ISTITUTO NAZIONALE DI FISICA NUCLEARE
Laboratori Nazionali di Frascati

# FRASCATI PHYSICS SERIES

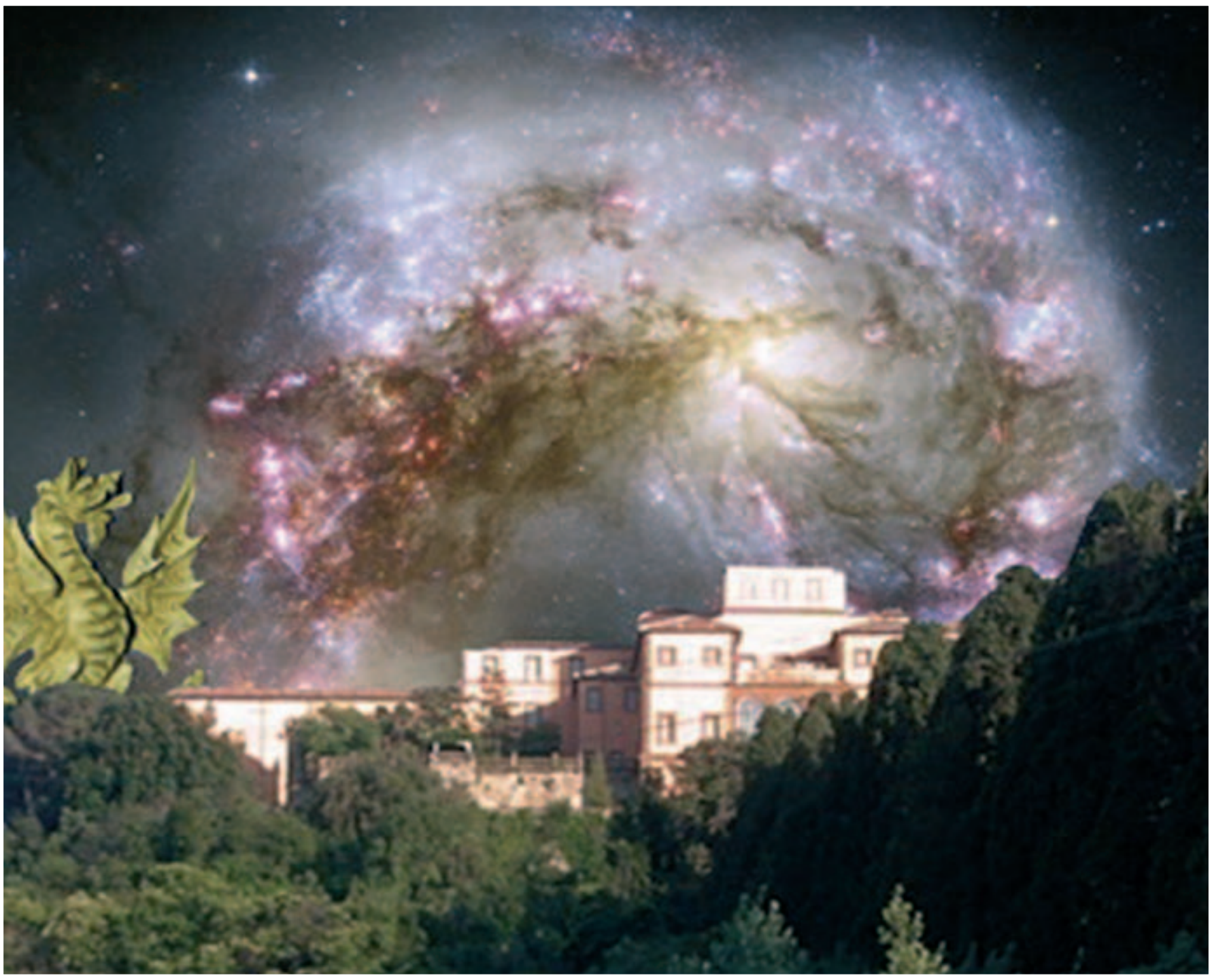

Proceedings of the Fifth International Workshop on

# SCIENCE WITH THE NEW GENERATION HIGH ENERGY GAMMA-RAY EXPERIMENTS

Editors

A. Lionetto, A. Morselli

45

2007

Proceedings of the fifth International Workshop on

# SCIENCE WITH THE NEW GENERATION HIGH ENERGY GAMMA- RAY EXPERIMENTS

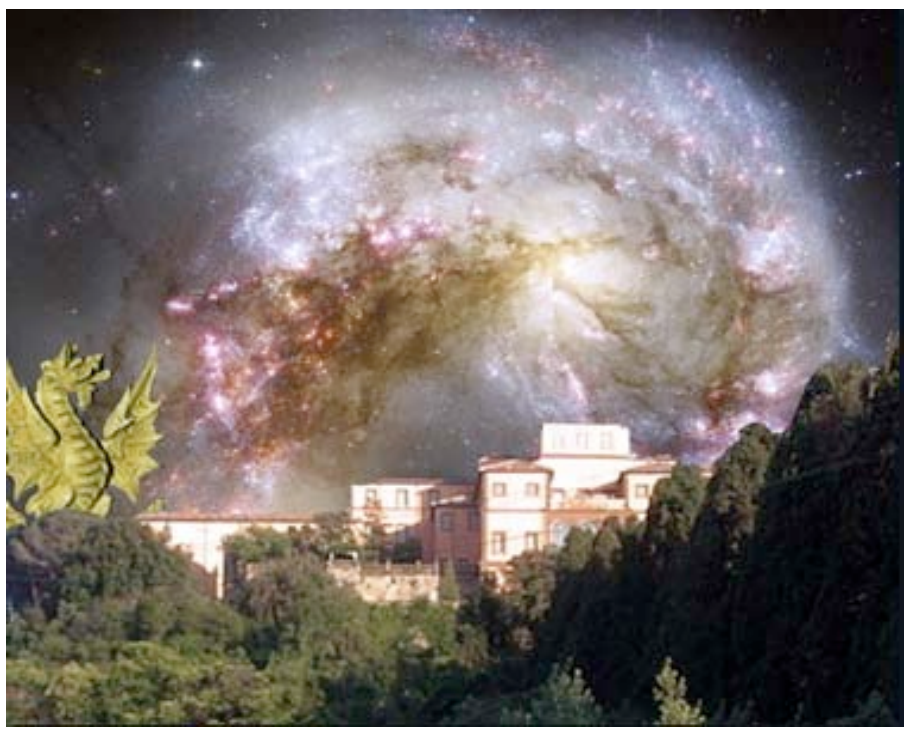

FRASCATI PHYSICS SERIES

FRASCATI PHYSICS SERIES

Proceedings of the Third International Workshop on

**FRONTIER SCIENCE**

FRASCATI PHYSICS SERIES

Volume XLV

Proceedings of the fifth International Workshop on

**SCIENCE WITH THE NEW GENERATION HIGH ENERGY GAMMA- RAY EXPERIMENTS**

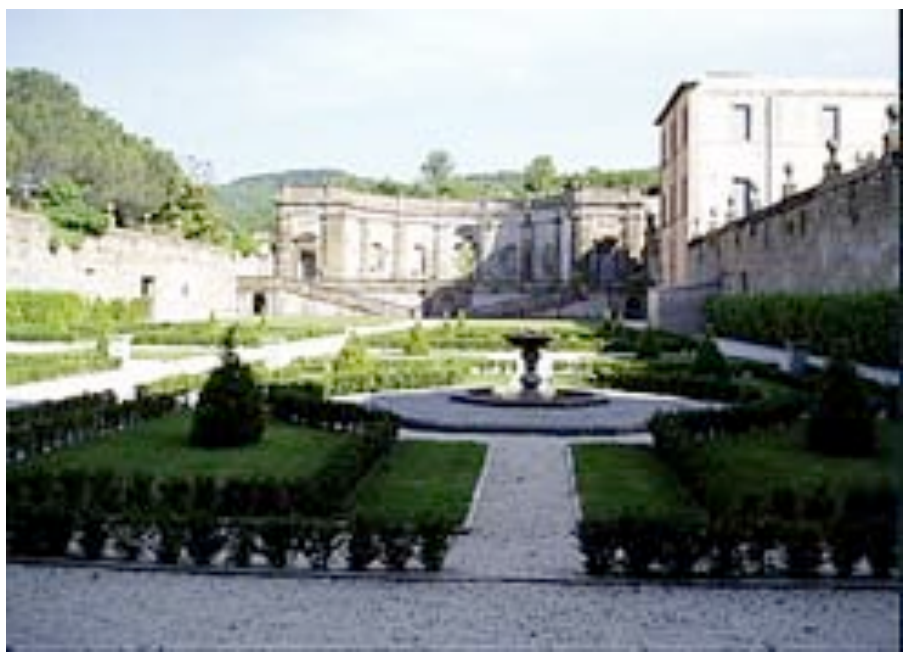

Editors

A. Lionetto, A. Morselli

Villa Mondragone, Monteporzio

June 18 –20, 2007

*Organising Committee*

| | |
|---|---|
| A. Morselli | *(INFN & University Roma Tor Vergata) – Chair* |
| D. Bastieri | *(INFN & Padova University)* |
| V. Buttaro | *(INFN & University Roma Tor Vergata)* |
| A. De Angelis | *(INFN & University Udine)* |
| A. Ferrari | *(CIFS & University Torino)* |
| N. Giglietto | *(INFN & Politecnico Bari)* |
| A. Lionetto | *(INFN & University Roma Tor Vergata)* |
| F. Longo | *(INFN & University Trieste)* |
| N. Omodei | *(INFN Pisa)* |
| E. Orazi | *(INFN & University Roma Tor Vergata)* |
| G. Tosti | *(INFN & University Perugia)* |

*International Scientific Advisory Commitee*

| | |
|---|---|
| F. Aharonian | *(DIAS (Dublin) & MPIK (Heidelberg))* |
| G. Barbiellini | *(Trieste University & INFN, Italy)* |
| G. Barreira | *(LIP Lisboa, Portugal)* |
| R. Bellazzini | *(INFN Pisa, Italy)* |
| E. Bloom | *(SLAC, USA)* |
| S. Digel | *(SLAC, USA)* |
| P. Giommi | *(INAF-ASI, Italy)* |
| G. Kanbach | *(Max Planck Institut fuer Physik, Germany)* |
| P. Lubrano | *(INFN Perugia, Italy)* |
| J. McEnery | *(NASA GSFC, Washington, USA)* |
| P. Michelson | *(Stanford University, USA)* |
| R. Paoletti | *(Siena University & INFN Pisa, Italy)* |
| L. Peruzzo | *(Padova University, Italy)* |
| P. Picozza | *(INFN & Roma Tor Vergata Unversity)* |
| S. Ritz | *(NASA GSFC, Washington, USA)* |
| F. Ryde | *(Stockholm University, Sweden)* |
| A. Saggion | *(Padova University & INFN, Italy)* |
| P. Spinelli | *(Bari University & INFN, Italy)* |
| M. Teshima | *(Max Planck Institut fuer Physik, Germany)* |
| M. Tavani | *(INAF Roma, Italy)* |
| D. Torres | *(ICREA/IEEC-CSIC, Barcelona Spain)* |

# PREFACE

This Conference is the fifth of a series of Workshops on High Energy Gamma- ray Experiments, following the Conferences held in Perugia 2003, Bari 2004, Cividale del Friuli 2005, Elba Island 2006.

This year the focus was on the use of gamma-ray to study the Dark Matter component of the Universe, the origin and propagation of Cosmic Rays, Extra Large Spatial Dimensions and Tests of Lorentz Invariance.

High energy gamma rays give a great chance to study physics beyond the standard model of the fundamental interactions. They are an important probe to better understanding dark matter. Weakly interacting massive particle (WIMP) are the most favorite candidates for dark matter and their nature can be explored studying gamma rays coming from WIMP pair annihilations.

This approach is complementary to the information that will come from the measurement of the antiproton and positron spectrum by the next generation cosmic-ray experiments. Mapping gamma rays coming from the interaction of primary p and He can also give a deep insight on cosmic-ray production and propagation mechanisms.

Finally many theories of physics beyond the standard model predict the existence of large extra space-time dimensions at an energy scale as low as 1 TeV and a possible high energy break-down of the Lorentz invariance.

The existence of extra dimensions can imply an enhancement of the expected gamma ray flux while a test of the Lorentz invariance can be done through correlated measurements of the difference in the arrival time of gamma-ray photons and neutrinos emitted from active galactic nuclei or gamma-ray bursts. In this Workshop all these topics had been covered both from the theoretical and experimental point of view.

An update on the current and planned research for space-borne and ground-based experiments dedicated to the observation of the gamma-ray sky was given with particular enphasis on the succesfull launch of AGILE.

We warmly thank the session chairpersons and all the speakers for their contribution to the scientific success of the Conference.

The Conference was sponsored and supported by the Department of Physics of the University of Roma “Tor Vergata, the Italian Istituto Nazionale di Fisica

Nucleare (INFN) Section of Roma Tor Vergata and the National Laboratory of Frascati.

We wish to thank the Organising Committee and International Scientific Advisory Commitee members for their valuable scientific advice and support all along the course of the conference organization.

Special thanks go to all the people involved in the Conference organization: We are particularly grateful to Liù Catena and all the Villa Mondragone Staff, Marta Solinas and Gabriella Ardizzoia for their valuable help in preparing and dealing with all the logistics and for the day-by-day assistance at Villa Mondragone and to Vincenzo Buttaro for the creation and updating of the web page.

Finally, our special thank goes to Luigina Invidia, of the Ufficio Pubblicazioni of the Frascati Laboratories, for the technical editing of these Proceedings.

Andrea Lionetto, Aldo Morselli, Piergiorgio Picozza

October 2007

The book is available in electronic format at
**http://www.roma2.infn.it/SciNeGHE07/**

# 5th Workshop on
# Science with the New Generation High Energy Gamma-ray Experiments
# SciNeGHE07

**The light of the dark: solving the mysteries of the Universe**

*Villa Mondragone, Frascati, Rome, (Italy)*
*June 18 - 20, 2007*

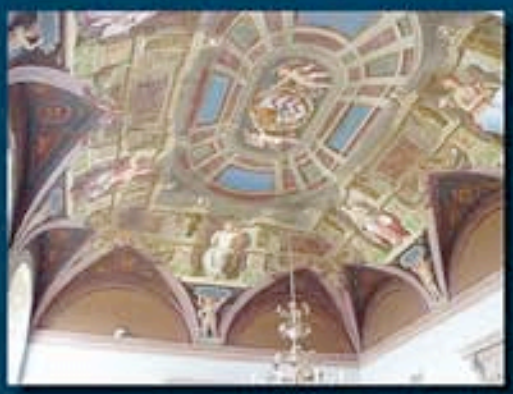

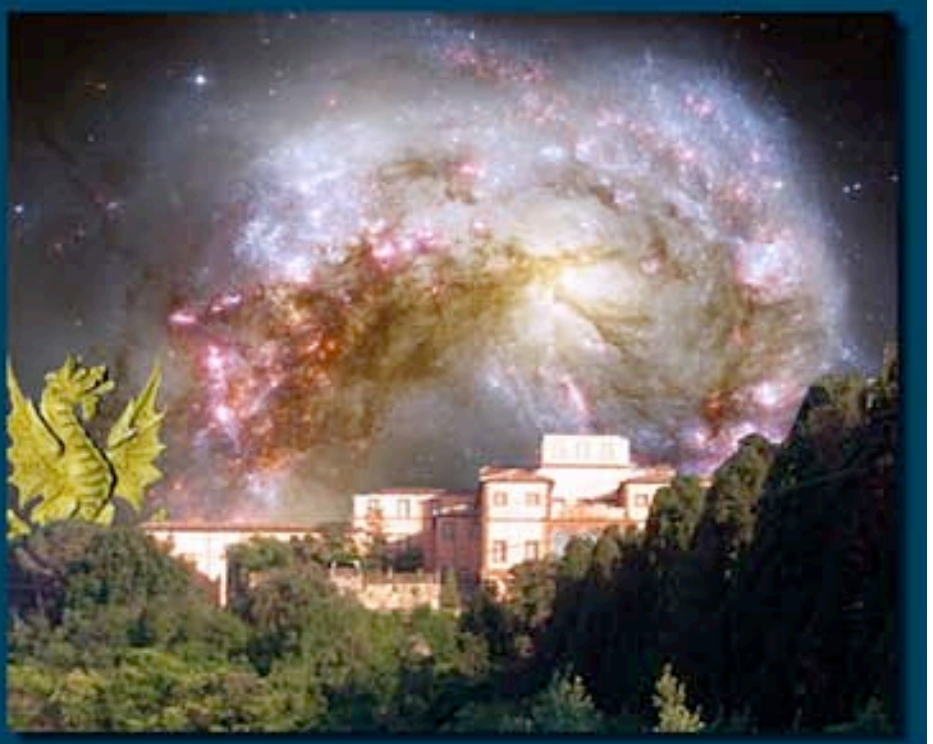

This will be the fifth of a series of Workshops on High Energy Gamma- ray Experiments, following the Conferences held in Perugia 2003, Bari 2004, Cividale del Friuli 2005, Elba Island 2006.
This year the focus will be on the use of gamma-ray to study the Dark Matter component of the Universe, the origin and propagation of Cosmic Rays, Extra Large Spatial Dimensions and Tests of Lorentz Invariance.
An update on the current and planned research for space-borne and ground-based experiments dedicated to the observation of the gamma-ray sky will be given.
Among the participants there are both hardened veterans of the first dedicated gamma-ray missions (like SAS-2 COS-B, CGRO) and young students entering the fascinating field of gamma-ray astrophysics participating in the new generation of high energy gamma-ray astrophysics experiments like GLAST, AGILE, MAGIC, VERITAS, HESS and ARGO.

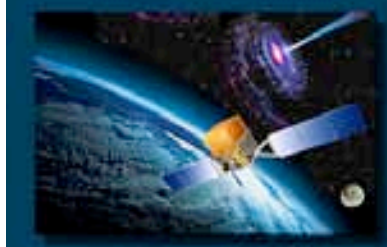

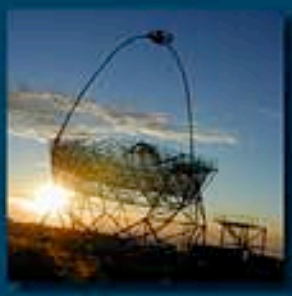

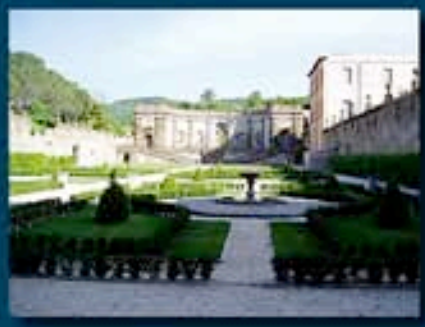

## ORGANISING COMMITTEE

A. Morselli (INFN & University Roma Tor Vergata) – Chair
D. Bastieri (INFN & Padova University)
V. Buttaro (INFN & University Roma Tor Vergata)
A. De Angelis (INFN & University Udine)
N. Giglietto (INFN & Politecnico Bari)
A. Lionetto (INFN & University Roma Tor Vergata)
F. Longo (INFN & University Trieste)
N. Omodei (INFN Pisa)
E. Orazi (INFN & University Roma Tor Vergata)
G. Tosti (INFN & University Perugia)

## INTERNATIONAL SCIENTIFIC ADVISORY COMMITEE

F. Aharonian (DIAS (Dublin) & MPIK (Heidelberg))
G. Barbiellini (Trieste University & INFN, Italy)
G. Barreira (LIP Lisboa, Portugal)
R. Bellazzini (INFN Pisa, Italy)
E. Bloom (SLAC, USA)
S. Digel (SLAC, USA)
P. Giommi (INAF-ASI, Italy)
G. Kanbach (Max Planck Institut fuer Physik, Germany)
P. Lubrano (INFN Perugia, Italy)
J. McEnery (NASA GSFC, Washington, USA)
P. Michelson (Stanford University, USA)
R. Paoletti (Siena University & INFN Pisa, Italy)
L. Peruzzo (Padova University, Iyaly)
P. Picozza (INFN & Roma Tor Vergata Unversity)
S. Ritz (NASA GSFC, Washington, USA)
F. Ryde (Stockholm University, Sweden)
A. Saggion (Padova University & INFN, Italy)
P. Spinelli (Bari University & INFN, Italy)
M. Teshima (Max Planck Institut fuer Physik, Germany)
M. Tavani (INAF Roma, Italy)
D. Torres (ICREA/IEEC-CSIC, Barcelona Spain)

## CONTACT

Scientific Secretariat: Marta Solinas

Phone: +39 6 0672594102
Fax: +39 6 72594647
E-mail: marta.solinas@roma2.infn.it

Dipartimento di Fisica
Università degli Studi di Roma
Tor Vergata
Via Ricerca Scientifica, 1
00133 Rome - Italy

web: http://www.roma2.infn.it/SciNeGHE07

INFN Istituto Nazionale di Fisica Nucleare

Università degli Studi di Roma Tor Vergata Dipartimento di Fisica

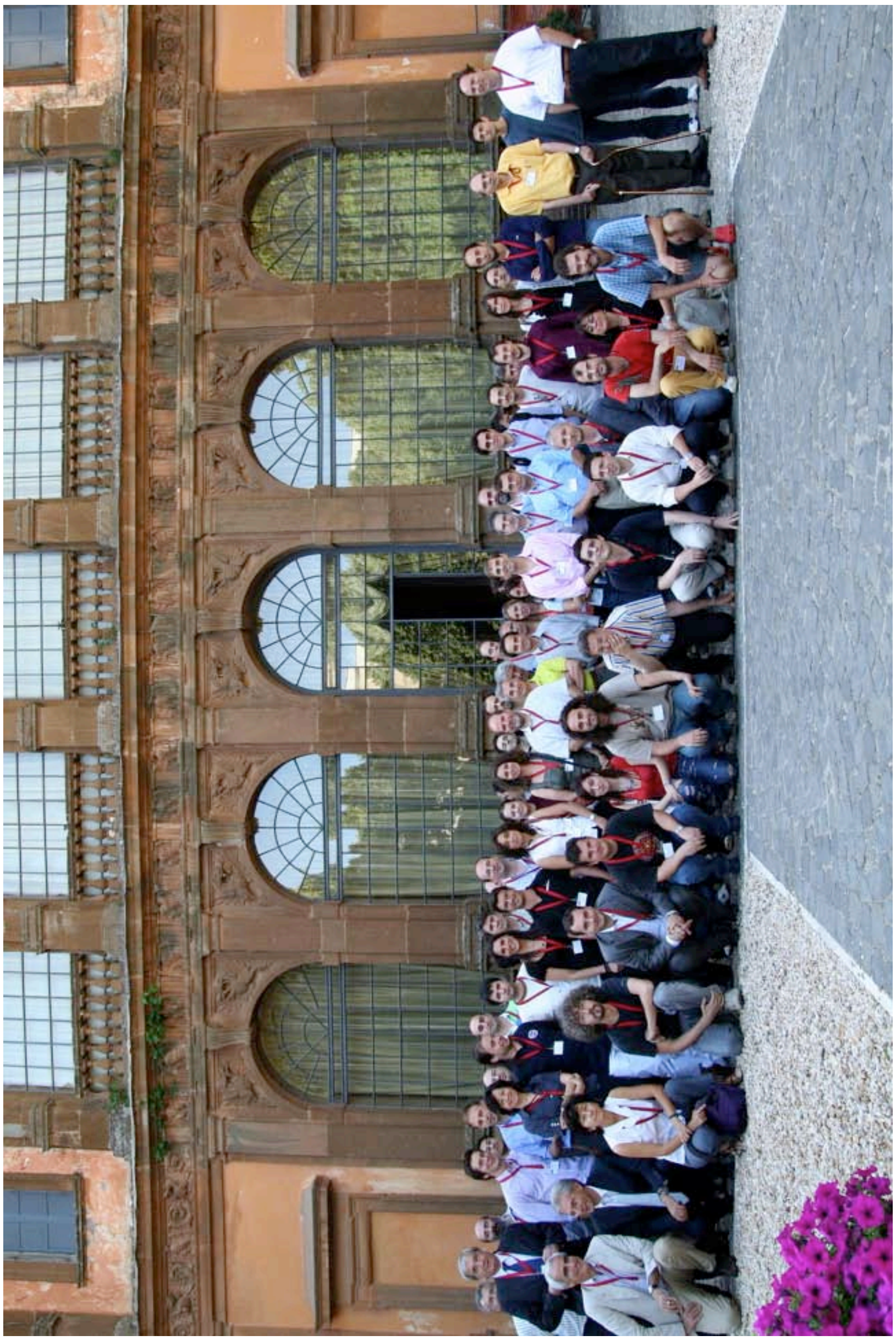

*Photo by: Pietro Oliva*

**TABLE OF CONTENTS**

# AGILE IN ORBIT

Tavani M.[a,b], Barbiellini G.[d,e], Argan A.[f], Bulgarelli A.[h], Caraveo P.[f], Chen A.[c,f], Cocco V.[b], Costa E.[a], De Paris G.[a], Del Monte E.[a], Di Cocco G.[h], Donnarumma I.[a], Feroci M.[a], Fiorini M.[f], Froysland T.[c,g], Frutti M.[a], Fuschino F.[h], Galli M.[i], Gianotti F.[h], Giuliani A.[f], Labanti C.[h], Lapshov I.[a], Lazzarotto F.[a], Lipari P.[j,k], Longo F.[d,e], Marisaldi M.[h], Mastropietro M.[a], Mattaini E.[f], Mauri F.[l], Mereghetti S.[f], Morelli E.[h], Morselli A.[b,g], Pacciani L.[a], Pellizzoni A.[f], Perotti F.[f], Picozza P.[b,g], Pontoni C.[d], Porrovecchio G.[a], Prest M.[d], Pucella G.[a], Rapisarda M.[m], Rossi E.[h], Rubini A.[a], Soffitta P.[a], Traci A.[h], Trifoglio M.[h], Trois A.[f], Vallazza E.[d], Vercellone S.[f], Zambra A.[c,f], Zanello D.[j,k]
and
Giommi P.[n], Antonelli A.[n], Pittori C.[n]

[a] *INAF-IASF Roma, via del Fosso del Cavaliere 100, I-00133 Roma, Italy*

[b] *Dip. Fisica, Universitá Tor Vergata, via della Ricerca Scientifica 1, I-00133 Roma, Italy*

[c] *CIFS, villa Gualino - v.le Settimio Severo 63, I-10133 Torino, Italy*

[d] *INFN Trieste, Padriciano 99, I-34012 Trieste, Italy*

[e] *Dip. Fisica, Universitá di Trieste, via A. Valerio 2, I-34127 Trieste, Italy*

[f] *INAF-IASF Milano, via E. Bassini 15, I-20133 Milano, Italy*

[g] *INFN Roma 2, via della Ricerca Scientifica 1, I-00133 Roma, Italy*

[h] *INAF-IASF Bologna, via Gobetti 101, I-40129 Bologna, Italy*

[i] *ENEA Bologna, via don Fiammelli 2, I-40128 Bologna, Italy*

[j] *INFN Roma 1, p.le Aldo Moro 2, I-00185 Roma, Italy*

[k] *Dip. Fisica, Universitá La Sapienza, p.le Aldo Moro 2, I-00185 Roma, Italy*

[l] *INFN Pavia, via Bassi 6, I-27100 Pavia, Italy*

[m] *ENEA Frascati, via Enrico Fermi 45, I-00044 Frascati (RM), Italy*

[n] *ASI Science Data Center, ESRIN, I-00044 Frascati (RM), Italy*

**Abstract**

AGILE is an Italian Space Agency mission that will explore the gamma-ray Universe with a very innovative instrument combining for the first time a gamma-ray imager (sensitive in the range 30 MeV - 50 GeV) and a hard X-ray imager (sensitive in the range 18-60 keV). An optimal angular resolution and very large fields of view are obtained by the use of state-of-the-art Silicon detectors integrated in a very compact instrument. AGILE was successfully launched on April 23, 2007 from the Indian base of Sriharikota and was inserted in a low-particle background equatorial orbit. AGILE will provide crucial data for the study of Active Galactic Nuclei, Gamma-Ray Bursts, unidentified gamma-ray sources, Galactic compact objects, supernova remnants, TeV sources, and fundamental physics by microsecond timing.

## 1 INTRODUCTION

The space program AGILE (*Astro-rivelatore Gamma a Immagini LEggero*) is a high-energy astrophysics Mission supported by the Italian Space Agency (ASI) with scientific and programmatic participation by INAF, INFN and several Italian universities [25]. The industrial team includes Carlo Gavazzi Space, Thales-Alenia-Space-Laben, Oerlikon-Contraves, and Telespazio.

The main scientific goal of the AGILE program is to provide a powerful and cost-effective mission with excellent imaging capability simultaneously in the 30 MeV-50 GeV and 18-60 keV energy ranges with a very large field of view [26, 7, 27].

AGILE was successfully launched by the Indian PSLV-C8 rocket from the Sriharikota base on April 23, 2007 (Fig. 1). The launch and orbital insertion were nominal, and a quasi-equatorial orbit was achieved with the smallest inclination (2.5 degrees) ever achieved by a high-energy space mission.

The AGILE instrument design is innovative and based on the state-of-the-art technology of solid state Silicon detectors and associated electronics developed in Italian laboratories [3, 4, 5, 6, 16]. The instrument is very compact (see Fig. 2) and light ($\sim$ 120 kg) and is aimed at detecting new transients and monitoring gamma-ray sources within a very large field of view (FOV $\sim$ 1/5 of the whole sky). The total satellite mass is equal to 350 kg (see Fig. 3).

AGILE is expected to substantially advance our knowledge in several research areas including the study of Active Galactic Nuclei and massive black holes, Gamma-Ray Bursts (GRBs), the unidentified gamma-ray sources, Galactic transient and steady compact objects, isolated and binary pulsars, pulsar wind nebulae (PWNae), supernova remnants, TeV sources, and the Galactic Center. Furthermore, the fast AGILE electronic readout and data processing

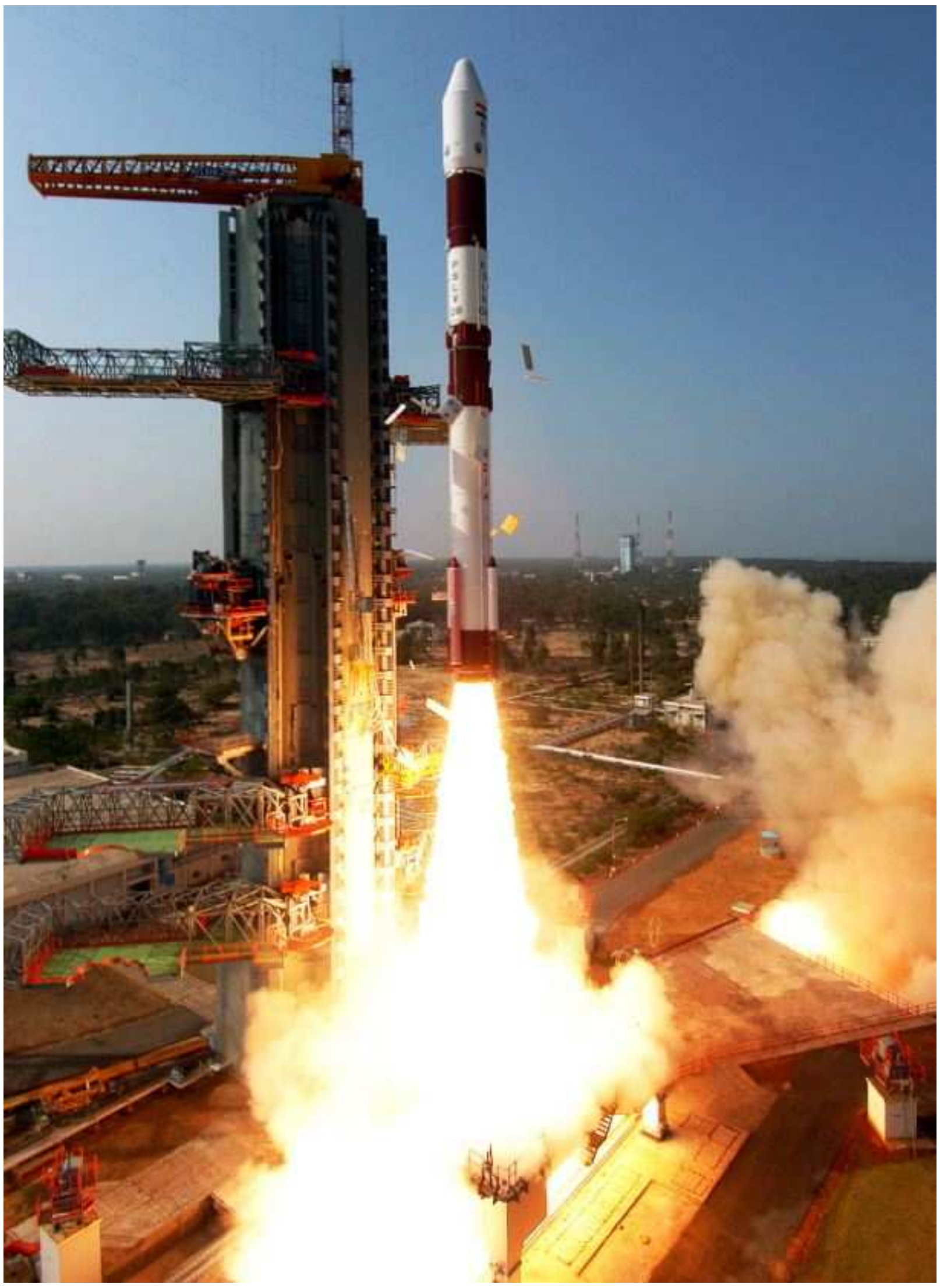

Figure 1: The launch of the AGILE satellite by the Indian PSLV-C8 rocket from the Sriharikota base on April 23, 2007.

(resulting in detectors' deadtimes smaller than $\sim 200\ \mu$sec) allow for the first time a systematic search for sub-millisecond gamma-ray/hard X-ray transients that are of interest for both Galactic objects (searching outburst durations comparable with the dynamical timescale of $\sim 1\,M_{\odot}$ compact objects) and quantum gravity studies [30] .

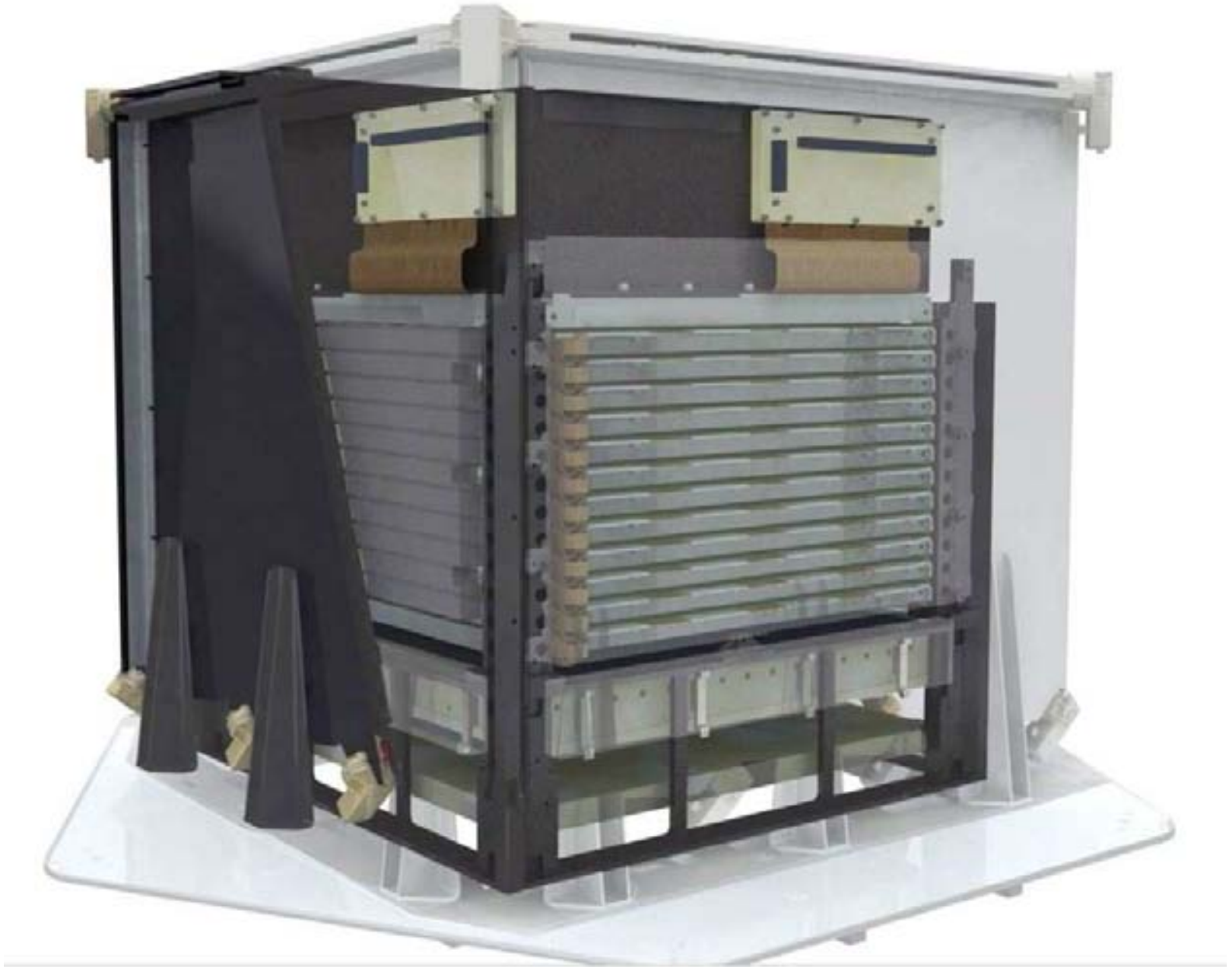

Figure 2: The AGILE scientific instrument showing the hard X-ray imager, the gamma-ray Tracker, and Calorimeter. The Anticoincidence system is partially displayed, and no lateral electronic boards and harness are shown for simplicity. The AGILE instrument "core" is approximately a cube of about 60 cm size and of weight equal to 100 kg.

The AGILE Science Program will be focused on a prompt response to gamma-ray transients and alert for follow-up multiwavelength observations. AGILE will provide crucial information complementary to several space missions (Chandra, INTEGRAL, XMM-Newton, SWIFT, Suzaku) and it will support ground-based investigations in the radio, optical, and TeV bands. Part of the AGILE Science Program will be open for Guest Investigations on a competitive basis. Quicklook data analysis and fast communication of new transients will be implemented as an essential part of the AGILE Science Program.

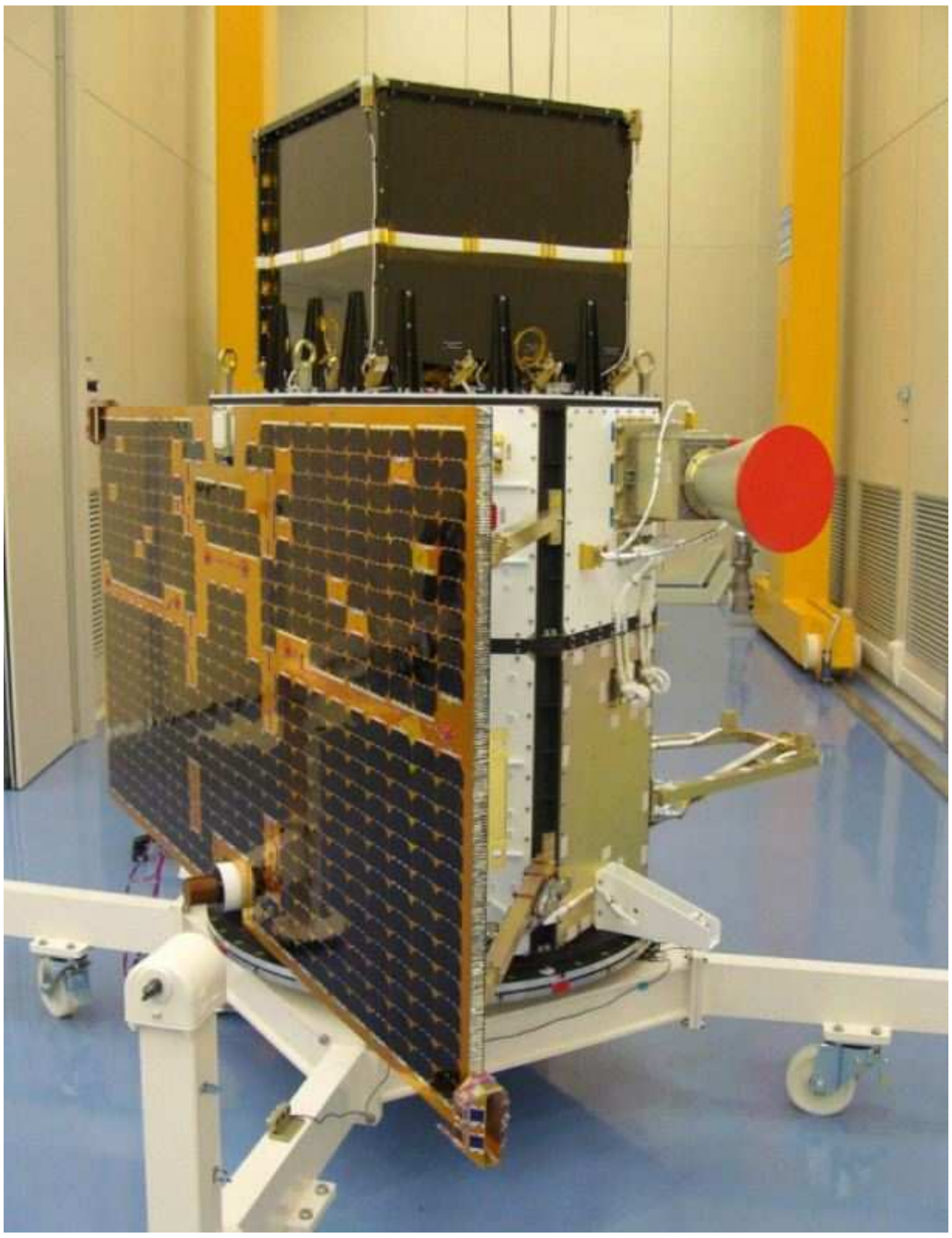

Figure 3: The integrated AGILE satellite in its final configuration before being covered by the thermal blanket. The total satellite mass is equal to 350 kg.

## 2 The Scientific Instrument

The AGILE scientific payload is made of three detectors combined into one integrated instrument with broad-band detection and imaging capabilities. The Anticoincidence and Data Handling systems complete the instrument. Table 1 summarizes the instrument scientific performance.

**The Gamma-Ray Imaging Detector (GRID)** is sensitive in the energy range $\sim$ 30 MeV–50 GeV, and consists of a Silicon-Tungsten Tracker, a Cesium Iodide Calorimeter, and the Anticoincidence system[1]. It is characterized by a very fine spatial resolution (obtained by a special arrangement of Silicon microstrip detectors and analog signal storage and processing) and by the smallest ever obtained deadtime for gamma-ray detection ($\lesssim$200 $\mu$s). The GRID is designed to achieve an optimal angular resolution (source location accuracy $\sim 15'$ for intense sources), an unprecedentedly large field-of-view ($\sim$ 2.5 sr), and a sensitivity comparable to that of EGRET for sources within 10-20 degree off-axis (and substantially better for larger off-axis angles).

**The hard X-ray imager (Super-AGILE)** is a unique feature of the AGILE instrument. This imager is placed on top of the gamma-ray detector and is sensitive in the 18-60 keV band. It has an optimal angular resolution (6 arcmin) and a good sensitivity over a $\sim$ 1 sr field of view ($\sim$10-15 mCrab on-axis for a 1-day integration). The main characteristic of AGILE will be then the possibility of simultaneous gamma-ray and hard X-ray source detection with arcminute positioning and on-board GRB/transient source alert capability.

**A Mini-Calorimeter operating in the "burst mode"** is the third AGILE detector. It is part of the GRID, but also is capable of independently detecting GRBs and other transients in the 350 keV - 100 MeV energy range with optimal timing capabilities.

Fig. 3 shows the integrated AGILE satellite and Fig. 2 a schematic representation of the instrument. We briefly describe here the main detecting units of the AGILE instrument; more detailed information will be presented elsewhere [28].

- **The Silicon-Tracker** (ST) providing the gamma-ray imager is based on photon conversion into electron-positron pairs. It consists of a total of 12 trays with a repetition pattern of 1.9 cm (Fig. 4). The first 10 trays are capable of converting gamma-rays by a Tungsten layer. Tracking of charged particles is ensured by high-resolution Silicon microstrip detectors that are configured to provide the two orthogonal coordinates for each element (point) along the track. The fundamental Silicon detector unit is a tile of area $9.5 \times 9.5$ cm$^2$, microstrip pitch equal to 121 $\mu$m, and thickness 410 $\mu$m. The AGILE ST readout system is capable of detecting and storing the energy deposited by the penetrating particles for half of the Silicon microstrips. This implies an alternating readout system characterized by "readout" and "floating" strips. The analog signal produced in the readout strips is read and stored for further processing. The fundamental element is a Silicon tile with a total of 384 readout channels (readout pitch equal to 242 $\mu$m) and 3 TAA1 chips required to process independently the analog signal from the readout strips. Each Si-Tracker layer is made of $4 \times 4$ Si-tiles, for a total geometric area of $38 \times 38$ cm$^2$. The first 10 trays are equipped with a Tungsten layer of 245 $\mu$m (0.07 $X_0$) positioned in the lower part of the tray. The Silicon detectors providing the two orthogonal coordinates are positioned at the very top and bottom of these trays. For each tray there are $2 \times 1,536$ readout microstrips. Since the ST trigger requires at least three planes to be activated, two more trays are inserted at the bottom of the Tracker without the Tungsten layers. The total readout channel number for the GRID Tracker is then $36,864$. Both digital and analog information (charge deposition in Si-microstrip) is read by TAA1 chips. The distance between mid-planes equals 1.9 cm (optimized by Montecarlo simulations). The ST has an *on-axis* total radiation length near 0.8 $X_0$. Special trigger logic algorithms implemented on-board (Level-1 and Level-2) lead to a substantial particle/albedo-photon background subtraction and a preliminary on-board reconstruction of the photon incidence angle. Both digital and analog information are crucial for this task. Fig. 5 shows a typical gamma-ray event detected by the GRID during the pre-launch satellite tests. Fig. 6 shows the first gamma-ray event detected by the GRID in orbit. The positional resolution obtained by the ST is excellent, being below 40 $\mu$m for a large range of particle incidence angles [8].

[1] In contrast with previous generation instruments (COS-B, EGRET), AGILE does not require gas operations and/or refilling, and does not require high-voltages.

- **Super-AGILE** (SA), the ultra-compact and light hard-X-ray imager of AGILE [14] is a coded-mask system made of a Silicon detector plane and a thin Tungsten mask positioned 14 cm above it (Fig. 7). The detector plane is organized in four independent square Silicon detectors ($19 \times 19\,\mathrm{cm}^2$ each) plus dedicated front-end electronics based on the XAA1.2 chips (suitable in the SA energy range). The total number of SA readout channels is 6,144. The detection cabability of SA includes: (1) photon-by-photon transmission and imaging of sources in the energy range 18-60 keV, with a large field-of-view (FOV $\sim$ 1 sr); (2) a good angular resolution (6 arcmin); (3) a good sensitivity ($\sim$ 10 mCrab between 18-60 keV for 50 ksec integration, and $\lesssim$ 1 Crab for a few seconds integration). The hard-X-ray imager is aimed at the simultaneous X-ray and gamma-ray detection of high-energy sources with excellent timing capabilities (a few microsecond deadtime for individual detectors). The AGILE satellite is equipped with an ORBCOMM transponder capable of trasmitting GRB coordinates to the ground within 1-2 min.

- **The Mini-Calorimeter** (MCAL) is made of 30 Thallium activated Ce-

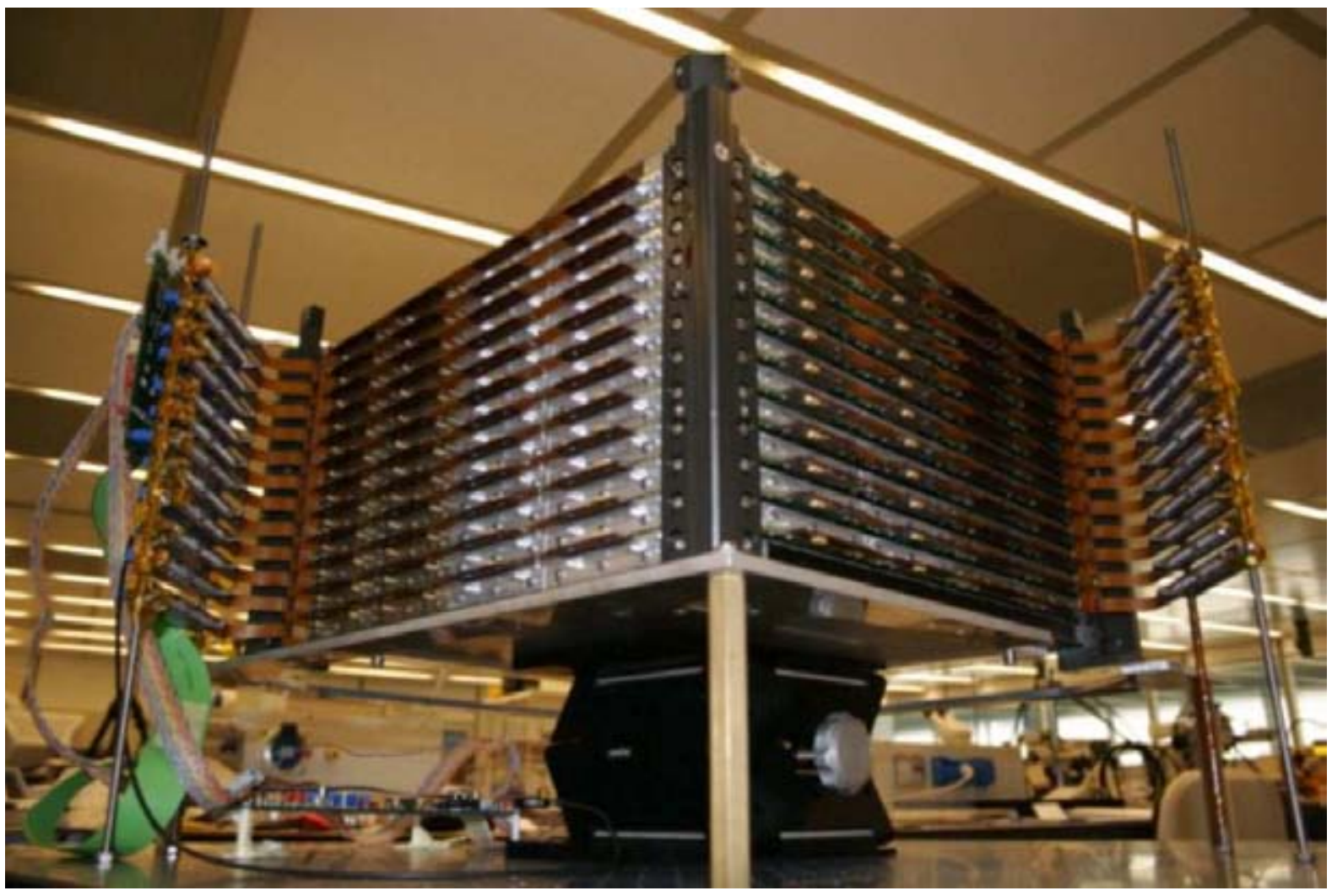

Figure 4: The assembled AGILE Silicon Tracker developed by the INFN laboratories of Trieste before being integrated with the rest of the instrument (June 2005).

sium Iodide (CsI(Tl)) bars arranged in two planes, for a total (on-axis) radiation length of 1.5 $X_0$ [20] . The signal from each CsI bar is collected by two photodiodes placed at both ends. The MCAL aims are: *(i)* obtaining additional information on the energy deposited in the CsI bars by particles produced in the Silicon Tracker (and therefore contributing to the determination of the total photon energy); *(ii)* detecting GRBs and other impulsive events with spectral and intensity information in the energy band $\sim 0.35 - 100$ MeV. An independent burst search algorithm is implemented on board with a wide dynamic range for the MCAL independent GRB detection.

- **The Anticoincidence (AC) System** is aimed at both charged particle background rejection and preliminary direction reconstruction for triggered photon events. The AC system completely surrounds all AGILE detectors (Super-AGILE, Si-Tracker and MCAL). Each lateral face is segmented in three plastic scintillator layers (0.6 cm thick) connected with photomultipliers placed at the bottom. A single plastic scintillator layer (0.5 cm thick) constitutes the top-AC whose signal is read by four light photomultipliers placed externally to the AC system and supported by the four corners of the structure frame. The segmentation of the AC

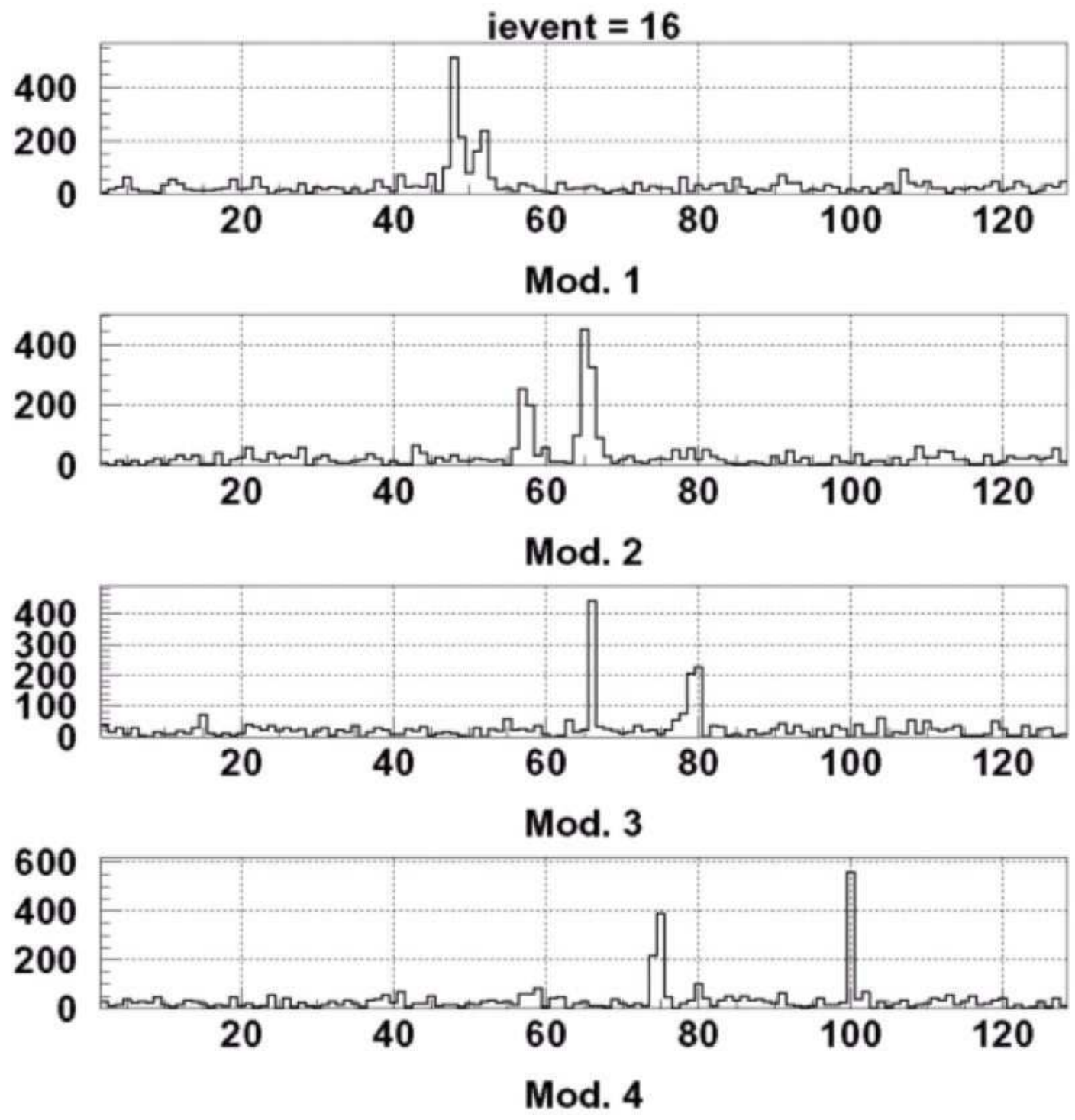

Figure 5: A typical gamma-ray event produced by cosmic-rays and detected by the AGILE Tracker with its characteristic pattern of released energy in clusters of hit readout Silicon microstrips.

System and the ST trigger logic contribute to produce the very large field of view of the AGILE-GRID.

The Data Handling (DH) and power supply systems complete the instrument. The DH is optimized for fast on-board processing of the GRID, Mini-Calorimeter and Super-AGILE data. Given the relatively large number of readable channels in the ST and Super-AGILE (∼40,000), the instrument requires a very efficient on-board data processing system. The GRID trigger logic for the acquisition of gamma-ray photon data and background rejection is structured in two main levels: Level-1 and Level-2 trigger stages. The Level-1 trigger is fast ($\lesssim 5\mu$s) and requires a signal in at least three out of four contiguous tracker planes, and a proper combination of fired TAA1 chip number signals and AC signals. An intermediate Level-1.5 stage is also envisioned (lasting $\sim 20\ \mu$s), with the acquisition of the event topology based on the identification of fired TAA1 chips. Both Level-1 and Level-1.5 have a hardware-oriented veto logic providing a first cut of background events. Level-2 data processing includes a GRID readout and pre-processing, "cluster data acquisition" (analog and digital information). The Level-2 processing is asynchronous (estimated duration $\sim$ a few ms) with the actual GRID event processing. The GRID deadtime turns out to be $\lesssim$ 200 $\mu$s and is dominated by the Tracker readout.

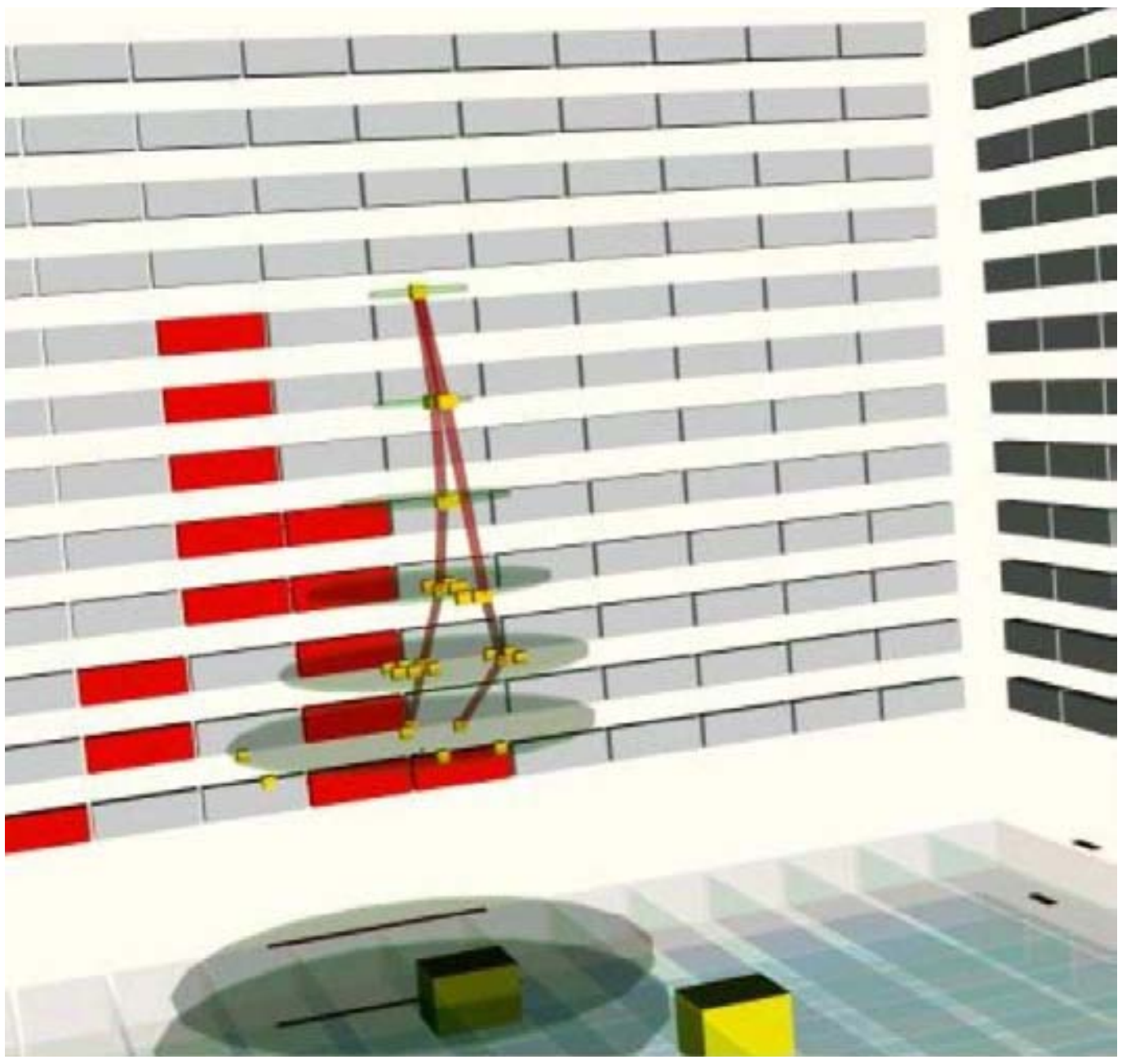

Figure 6: The first gamma-ray event detected by AGILE in space (May 10, 2007).

The charged particle and albedo-photon background passing the Level-1+1.5 trigger level of processing is simulated to be $\lesssim$ 100 events/sec for the nominal equatorial orbit of AGILE. The on-board Level-2 processing has the task of reducing this background by a factor between 3 and 5. Off-line processing of the GRID data with both digital and analog information is being developed with the goal to reduce the particle and albedo-photon background rate above 100 MeV to $\sim$0.01 events/sec.

In order to maximize the GRID FOV and detection efficiency for large-angle incident gamma-rays (and minimize the effects of particle backsplash from the MCAL and of "Earth albedo" background photons), the data acquisition logic uses proper combinations of top and lateral AC signals and a coarse on-line direction reconstruction in the ST. For events depositing more than 200 MeV in the MCAL, the AC veto may be disabled to allow the acquisition of gamma-ray photon events with energies larger than 1 GeV.

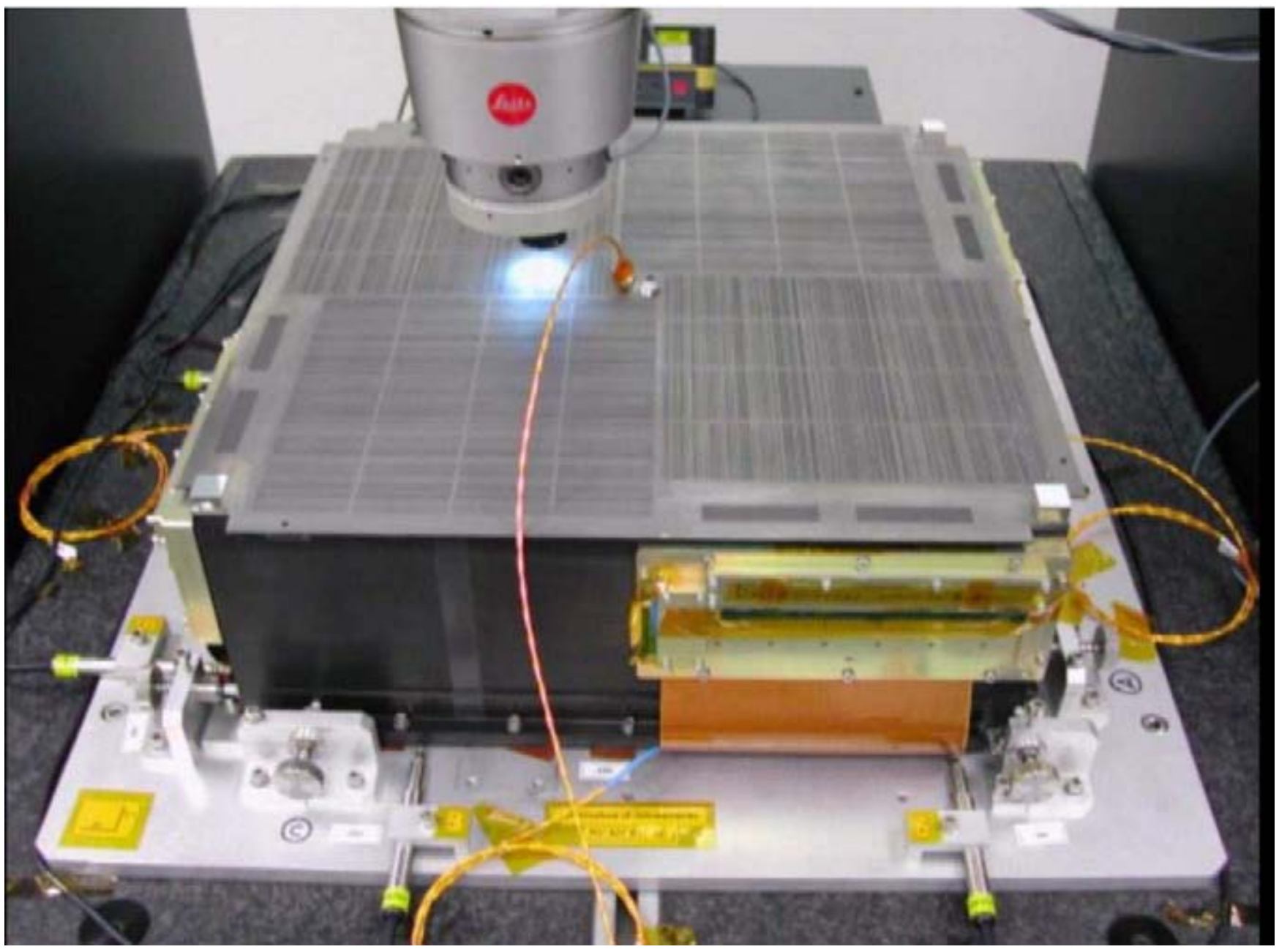

Figure 7: The Super-AGILE detector during metrology measurements (March 2005).

A special set of memory buffers and burst search algorithms are implemented to maximize data acquisition for transient gamma-ray events (e.g., GRBs) in the ST, Super-AGILE and Mini-Calorimeter, respectively. The Super-AGILE event acquisition envisions a first "filtering" based on AC-veto signals, and pulse-height discrimination in the dedicated front end electronics (based on XA1 chips). The events are then buffered and transmitted to the CPU for burst searching and final data formatting. The four Si-detectors of Super-AGILE are organized in sixteen independent readout units, resulting in a $\sim 5\ \mu$s global deadtime[23].

In order to maximize the detecting area and minimize the instrument weight,

the GRID and Super-AGILE front-end-electronics is partly accommodated in special boards placed externally on the Tracker lateral faces. Electronic boxes, P/L memory (and buffer) units are positioned at the bottom of the instrument within the spacecraft body.

Table 1: **AGILE Scientific Performance**

| | |
|---|---|
| **Gamma-ray Imaging Detector (GRID)** | |
| Energy range | 30 MeV – 50 GeV |
| Field of view | $\sim$ 2.5 sr |
| Flux sensitivity ($E > 100$ MeV, $5\sigma$ in $10^6$ s) | $3\times10^{-7}$ (ph cm$^{-2}$ s$^{-1}$) |
| Angular resolution at 400 MeV (68% cont. radius) | 1.2 degrees |
| Source location accuracy (high Gal. lat., 90% C.L.)) | $\sim$15 arcmin |
| Energy resolution (at 400 MeV) | $\Delta$E/E$\sim$1 |
| Absolute time resolution | $\sim 2\,\mu$s |
| Deadtime | $100 - 200\,\mu$s |
| **Hard X–ray Imaging Detector (Super-AGILE)** | |
| Energy range | 18 – 60 keV |
| Single (1-dim.) detector FOV (FW at zero sens.) | $107^\circ\times68^\circ$ |
| Combined (2-dim.) detector FOV (FW at zero sens.) | $68^\circ\times68^\circ$ |
| Sensitivity (at 18-60 keV, $5\sigma$ in 1 day) | $\sim$ 10 mCrab |
| Angular resolution (pixel size) | $\sim$ 6 arcmin |
| Source location accuracy (S/N$\sim$10) | $\sim$2-3 arcmin |
| Energy resolution (FWHM) | $\Delta$E< 8 keV |
| Absolute time accuracy | $\sim 4\,\mu$s |
| **Mini-Calorimeter** | |
| Energy range | 0.35 – 100 MeV |
| Energy resolution ( at 1.3 MeV ) | 13% FWHM |
| Absolute time resolution | $\sim 3\,\mu$s |
| Deadtime (for each of the 30 CsI bars) | $\sim 20\,\mu$s |

## 3 Science with AGILE

This section summarizes the main features of the AGILE scientific capability. The AGILE instrument has been designed and developed to obtain:

- **excellent imaging capability in the energy range 100 MeV-50 GeV**, improving the EGRET angular resolution by a factor of 2, see Fig. 9;
- **a very large field-of-view** for both the gamma-ray imager (2.5 sr) and the hard X-ray imager (1 sr), (FOV larger by a factor of $\sim$6 than that of EGRET);

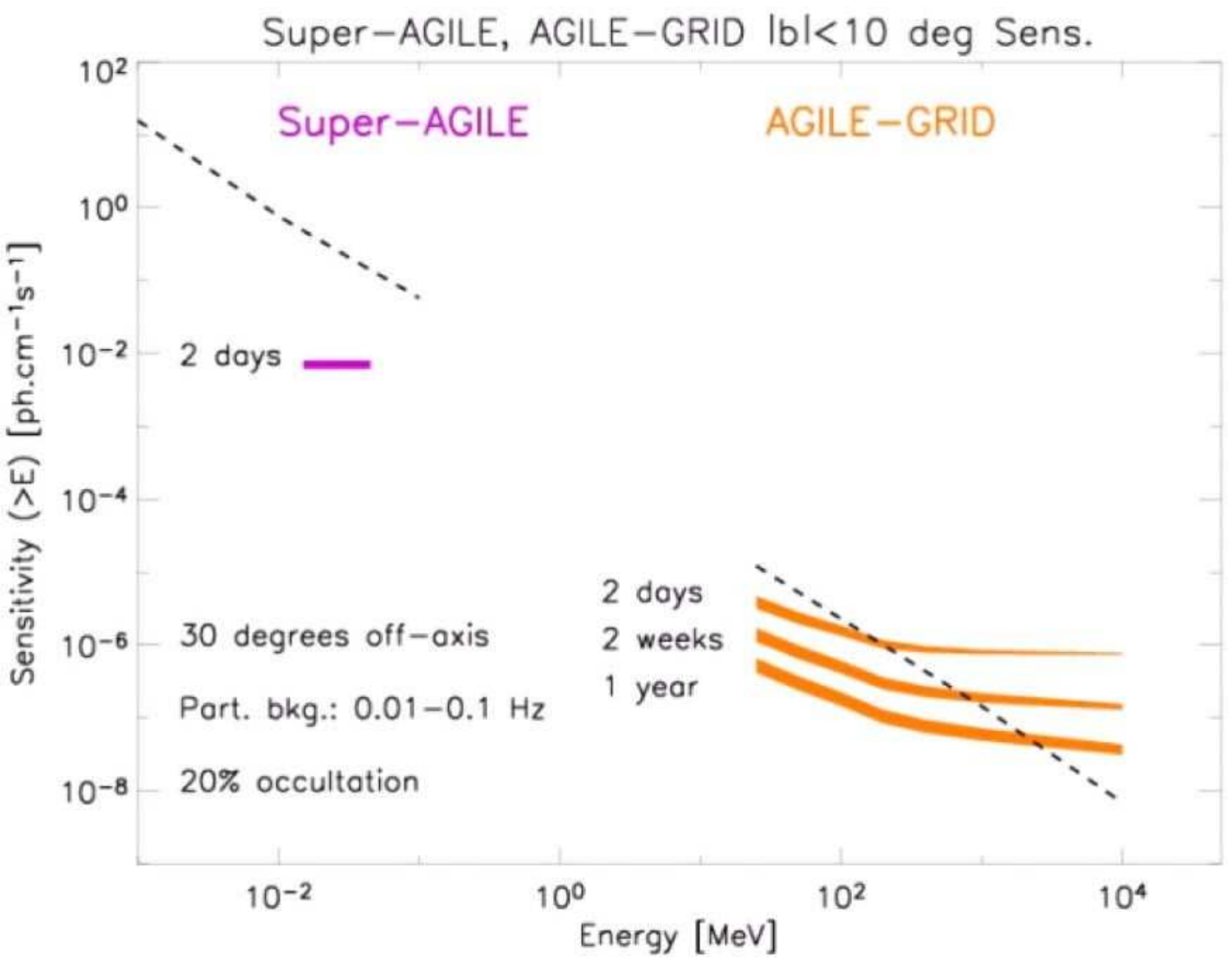

Figure 8: Simulated integrated flux sensitivity of AGILE (GRID and SA) as a function of energy for a 30-degree off-axis source in the Galactic plane. The Crab spectrum is shown by the dotted line.

- **excellent timing capability**, with overall photon absolute time tagging of uncertainty below $2\,\mu$s and very small deadtimes ($\lesssim 200\,\mu$s for the GRID, $\sim 5\,\mu$s for the sum of the SA readout units, and $\sim 20\,\mu$s for each of the individual CsI bars);

- **a good sensitivity for point sources**, comparable to that of EGRET in the gamma-ray range for sources within 20 degrees off-axis (except a central region of smaller effective area), and very flat up to 50-60 degrees off-axis (see Fig. 8). Depending on exposure and the diffuse background , the flux sensitivity threshold can reach values of $(10-20)\times 10^{-8}$ ph.cm$^{-2}$ s$^{-1}$ above 100 MeV. The hard X-ray imager sensitivity is between 10 and 20 mCrab at 20 keV for a 1-day integration over a field of view near 1 sr;

- **good sensitivity to photons in the energy range ∼30-100 MeV**, with an effective area above 200 $cm^2$ at 30 MeV;

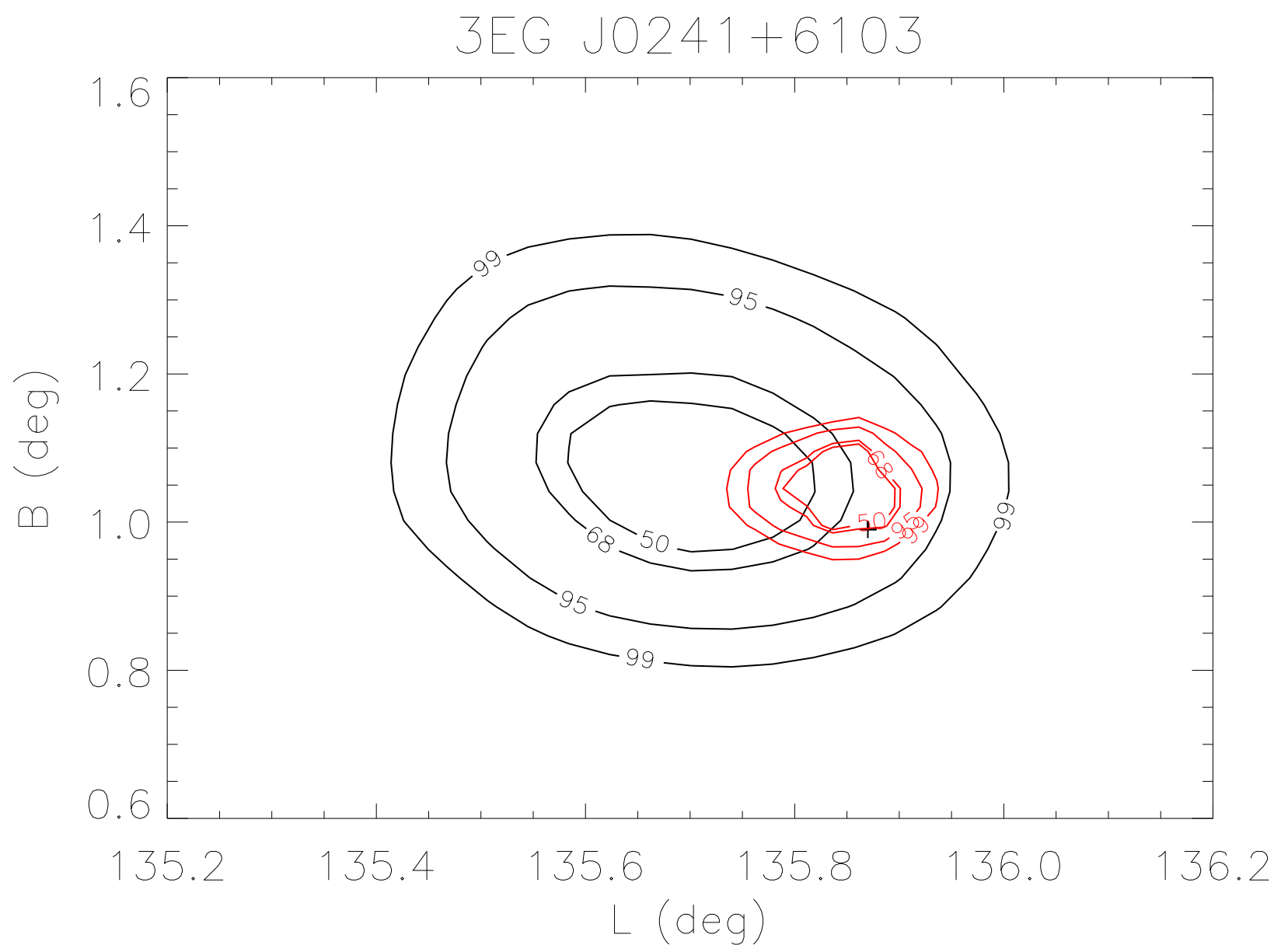

Figure 9: Simulated positioning of a gamma-ray source in the Galactic plane obtained by the AGILE Likelihood and diffuse gamma-ray background processing. The (more compact) red contours show the 50, 68, 95, 99 percent confidence levels obtained for the AGILE-GRID; the black contours show the confidence levels obtained with the EGRET likelihood analysis.

- **a very rapid response to gamma-ray transients and gamma-ray bursts**, obtained by a special quicklook analysis program and coordinated ground-based and space observations.
- **accurate localization (∼2-3 arcmins) of GRBs and other transient events** obtained by the GRID-SA combination (for typical hard X-ray transient fluxes above ∼1 Crab); the expected GRB detection rate is $\sim 1 - 2$ per month;
- **long-timescale continuous monitoring (∼2-3 weeks) of gamma-ray and hard X-ray sources**;
- **satellite repointing after special alerts** (∼1 day) in order to position the source within the Super-AGILE FOV to obtain hard X-ray data for gamma-ray transients detected by the GRID in the external part of its FOV or by other high-energy missions (e.g., GLAST).

The combination of simultaneous hard X-ray and gamma-ray data will provide a formidable combination for the study of high-energy sources. We briefly address here some of the relevant features of the expected scientific performance.

### 3.1 Angular resolution

Fig. 9 shows and example of the expected positioning of a gamma-ray source in the Galactic plane. The GRID configuration will achieve a PSF with 68% containment radius better within $1° - 2°$ at E > 300 MeV allowing a gamma-ray source positioning with error box radius near $10' - 20'$ depending on source spectrum, intensity, and sky position. Super-AGILE operating in the 15-45 keV band has a spatial resolution of 6 arcminute (pixel size). This translates into a positional accuracy of 1-3 arcmins for relatively strong transients at the Crab flux level.

### 3.2 Large FOV monitoring of gamma-ray sources

A crucial feature of AGILE is its large field of view for both the gamma-ray and hard X-ray detectors. Fig. 10 shows typical gamma-ray FOVs obtained for a sequence of AGILE pointings. Relatively bright AGNs and Galactic sources flaring in the gamma-ray energy range above a flux of $10^{-6}\,\mathrm{ph\,cm^{-2}\,s^{-1}}$ can be detected within a few days by the AGILE Quicklook Analysis. We conservatively estimate that for a 2-year mission AGILE is potentially able to detect hundreds of gamma-ray flaring AGNs and other transients. The very large FOV will also favor the detection of GRBs above 30 MeV. Taking into account the high-energy distribution of GRB emission above 30 MeV, we conservatively estimate that ∼1 GRB/month can be detected and imaged in the gamma-ray range by the GRID. Super-AGILE may be able to detect about 30% of the sources detected by INTEGRAL [10]; about 10 hard X-ray sources per day are expected to be detected by SA for typical pointings of the Galactic plane.

### 3.3 Fast reaction to strong high-energy transients

The existence of a large number of variable gamma-ray sources (extragalactic and near the Galactic plane, e.g., [24]) makes necessary a reliable program for quick response to transient gamma-ray emission. Quicklook Analysis of gamma-ray data is a crucial task to be carried out by the AGILE Team. Prompt communication of gamma-ray transients (requiring typically 1-3 days to be detected with high confidence for sources above $10^{-6}\,\mathrm{ph\,cm^{-2}\,s^{-1}}$) will be ensured. Detection of short timescale (seconds/minutes/hours) transients (GRBs, SGRs, and other bursting events) is possible in the gamma-ray range. A primary responsibility of the AGILE Team will be to provide positioning of short-timescale transient as accurate as possible, and to alert the community though dedicated channels.

### 3.4 Accumulating exposure on Galactic and extragalactic sky areas

The AGILE average exposure per source will be larger by a factor of $\sim 4$ for a 1-year sky-survey program compared to the typical exposure obtainable by EGRET for the same time period. Deep exposures for selected regions of the sky can be obtained by a proper program with repeated overlapping pointings. This can be particularly useful to study selected Galactic and extragalactic sources.

### 3.5 High-Precision Timing

AGILE detectors have optimal timing capabilities. The on-board GPS system allows to reach an absolute time tagging precision for individual photons better than $2\,\mu$s. Depending on the event characteristics, absolute time tagging can achieve values near $1-2\,\mu$s for the Silicon-Tracker, and $3-4\,\mu$s for the Mini-Calorimeter and Super-AGILE.

Instrumental deadtimes will be unprecedentedly small for gamma-ray detection. The GRID deadtime will be lower than $200\,\mu$s (improving by almost three orders of magnitude the performance of previous spark-chamber detectors such as EGRET). Taking into account the segmentation of the electronic readout of MCAL and Super-AGILE detectors (30 MCAL elements and 16 Super-AGILE elements) the MCAL and SA effective deadtimes will be less than those for individual units. We obtain $\sim 2\,\mu$s for MCAL, and $5\,\mu$s for SA. Furthermore, a special memory will ensure that MCAL events detected during the Si-Tracker readout deadtime will be automatically stored in the GRID event. For these events, precise timing and detection in the $\sim$ 1–200 MeV range can be achieved with temporal resolution well below $100\,\mu$s. This may be crucial for AGILE high-precision timing investigations.

### 3.6 AGILE and GLAST

AGILE and GLAST [15] are complementary missions in many respects: we briefly outline here some important points, postponing a more general discussion [29]. The GLAST gamma-ray instrument covers a broad spectrum and is especially optimized in the high energy range above 1 GeV . On the other contrary, AGILE is optimized in the range below 1 GeV with emphasis to the simultaneous hard X-ray/gamma-ray imaging with arcminute angular resolution. The GLAST large gamma-ray effective area allows deep pointings and good imaging within a few arcminutes for strong gamma-ray sources. AGILE has a gamma-ray effective area near 100 MeV smaller by a factor of $\sim 4$ compared to the upper portion of the LAT instrument on board of GLAST. However, it can reach arcminute positioning of sources because of Super-AGILE. Furthermore, the GLAST sky-scanning mode adopted during the first phase of the mission, and the AGILE fixed pointing strategy offer a way for joint investigations. In the overlapping pointed regions, strong time variability of gamma-ray sources (1-2 day timescale) can be very effectively studied simultaneously by the two missions. AGILE with its hard X-ray imager can also provide additional and very useful information.

The most relevant feature of AGILE is its unique combination of a large-FOV hard X-ray imager together with a gamma-ray imager. AGILE will be able to reach arcminute positioning of sources emitting in the hard X-ray range above 10 mCrab. Furthermore, AGILE can react to transients, and can point at gamma-ray sources detected by other missions to position the source within the FOV of both the GRID and SA. Interesting gamma-ray transients detected by GLAST might then be pointed by AGILE for a broad-band study of their temporal and spectral properties.

## 4 Scientific Objectives

Currently, nearly 300 gamma-ray sources above 30 MeV were detected (with only a small fraction, 30%, identified as AGNs or isolated pulsars) [17, 32, 33, 34]. AGILE fits into the discovery path followed by previous gamma-ray missions (SAS-2, COS-B, and EGRET) and be complementary to GLAST. We summarize here the main AGILE's scientific objectives.

• **Active Galactic Nuclei**. For the first time, simultaneous monitoring of a large number of AGNs per pointing will be possible. Several outstanding issues concerning the mechanism of AGN gamma-ray production and activity can be addressed by AGILE including: (1) the study of transient vs. low-level gamma-ray emission and duty-cycles [35]; (2) the relationship between the gamma-ray variability and the radio-optical-X-ray-TeV emission; (3) the correlation between relativistic radio plasmoid ejections and gamma-ray flares; (4) hard X-ray/gamma-ray correlations. A program for joint AGILE and ground-based monitoring observations is being planned. On the average, AGILE will achieve deep exposures of AGNs and substantially improve our knowledge on the low-level emission as well as detecting flares. We conservatively estimate that for a 3-year program AGILE will detect a number of AGNs 2–3 times larger than that of EGRET. Super-AGILE will monitor, for the first time, simultaneous AGN emission in the gamma-ray and hard X-ray ranges.

• **Gamma-ray bursts**. A few GRBs were detected by the EGRET spark chamber [21]. This number appears to be limited by the EGRET FOV and sensitivity and probably not by the intrinsic GRB emission mechanism. GRB detection rate by the AGILE-GRID is expected to be at least a factor of $\sim$ 5 larger than that of EGRET, i.e., $\geq$5–10 events/year). The small GRID deadtime ($\sim$ 500 times smaller than that of EGRET) allows a better study of the initial phase of GRB pulses (for which EGRET response was in many cases inadequate). The remarkable discovery of ‘delayed’ gamma-ray emission up to $\sim$ 20 GeV from GRB 940217 [18] is of great importance to model burst acceleration processes. AGILE is expected to be efficient in detecting photons above 10 GeV because of limited backsplashing. Super-AGILE will be able to locate GRBs within a few arcminutes, and will systematically study the interplay between hard X-ray and gamma-ray emissions. Special emphasis will be given to the search for sub-millisecond GRB pulses independently detectable by the Si-Tracker, MCAL and Super-AGILE.

• **Diffuse Galactic and extragalactic emission**. The AGILE good angular resolution and large average exposure will further improve our knowledge of cosmic ray origin, propagation, interaction and emission processes. We also note that a joint study of gamma-ray emission from MeV to TeV energies is possible by special programs involving AGILE and new-generation TeV observatories of improved angular resolution.

• **Gamma-ray pulsars**. AGILE will contribute to the study of gamma-ray pulsars (PSRs) in several ways: (1) improving timing and lightcurves of known gamma-ray PSRs; (2) improving photon statistics for gamma-ray period searches; (3) studying unpulsed gamma-ray emission from plerions in supernova remnants and studying pulsar wind/nebula interactions, e.g., as in the Galactic sources recently discovered in the TeV range[1]. Particularly interesting for AGILE are the $\sim$ 30 new young PSRs discovered [19] in the Galactic plane by the Parkes survey.

• **Search for non-blazar gamma-ray variable sources in the Galactic plane**, currently a new class of unidentified gamma-ray sources such as GRO J1838-04 [24].

• **Galactic sources, micro-quasars, new transients**. A large number of gamma-ray sources near the Galactic plane are unidentified, and sources such as 2CG 135+1/LS I 61 +61 303 can be monitored on timescales of months. Cyg X-1 will be also monitored and gamma-ray emission above 30 MeV will be intensively searched. Galactic X-ray jet sources (such as Cyg X-3, GRS 1915+10, GRO J1655-40 and others) can produce detectable gamma-ray emission for favorable jet geometries, and a TOO program is planned to follow-up new discoveries of *micro-quasars.*

• **Fundamental Physics: Quantum Gravity**. AGILE detectors are suited for Quantum Gravity studies [30]. The existence of sub-millisecond GRB pulses lasting hundreds of microseconds [9] opens the way to study QG delay propagation effects by AGILE detectors. Particularly important is the AGILE Mini-Calorimeter with independent readout for each of the 30 CsI bars of small deadtime ($\sim 20\,\mu$s) and absolute timing resolution ($\sim 3\,\mu$s). Energy dependent time delays near $\sim 100\,\mu$s for ultra-short GRB pulses in the energy range 0.3–3 MeV can be detected. If these GRB ultra-short pulses originate at cosmological distances, sensitivity to the Planck’s mass can be reached [30].

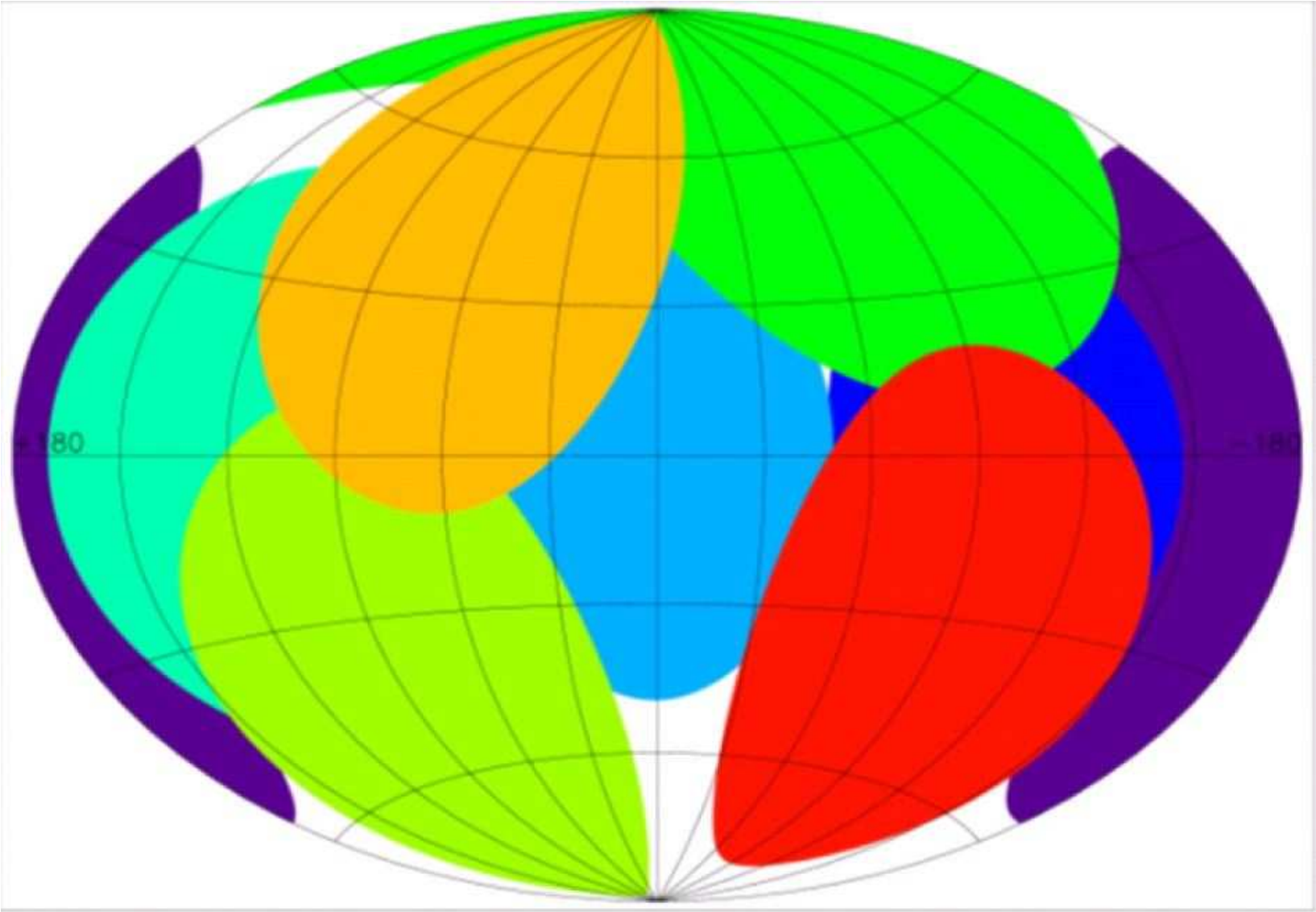

Figure 10: An example of a sequence of AGILE pointings showing the very large gamma-ray field of view (in Galactic coordinates).

## 5 The Mission

### 5.1 Satellite operations

The AGILE spacecraft is of the MITA class and is developed by Carlo Gavazzi Space (CGS) as prime contractor and by Oerlikon-Contraves. The spacecraft provides a 3-axis stabilization with an accuracy near 0.5–1 degree. The final satellite pointing reconstruction is required to reach an accuracy of $\sim$1 arcmin by a set of two Star Sensors. A GPS transceiver will also ensure an on-board timing accuracy within 2 microseconds. The AGILE scientific instrument generates under normal conditions a telemetry rate of $\sim$ 50 kbit/s. The satellite downlink telemetry rate is 512 kbit $s^{-1}$, that is adequate to transmit at every passage over the ground station all the satellite and scientific data.

The fixed solar panels configuration and the necessity to have them always exposed to the Sun imposes some constraints on the AGILE pointing strategy. However, in practice the AGILE large FOV does not sensibly limit the accessible sky: only the solar and anti-solar directions are excluded from direct pointings. The AGILE Pointing Plan is being finalized and will be ready in advance before the start of Cycle-1 (first year). AGILE might react to transient events of great importance occurring outside the accessible FOV. For transients detected by the AGILE-GRID and not by Super-AGILE, a minor re-pointing (20-30 degrees) is envisioned to allow the coverage of the gamma-ray transient also by the X-ray imager within 1 day. A drastic re-pointing strategy (Target-of-Opportunity, TOO) is foreseen for events of major scientific relevance detected by other observatories.

### 5.2 The Orbit

The AGILE orbit is quasi equatorial, with an inclination of 2.5 degrees from the Equator and average altitude of 540 km. A low earth orbit (LEO) of small inclination is a clear plus of the mission because of the reduced particle background as verified in orbit by all instrument detectors. Also a low-inclination orbit optmizes the use of the ASI communication ground base at Malindi (Kenya).

## 6 The AGILE Science Program

AGILE is a Small Scientific Mission with a science program open to the international scientific community. The AGILE Mission Board (AMB) oversees the scientific program, determines the pointing strategy, and authorizes Target of Opportunity (TOO) observations in case of exceptional transients. A substantial fraction of the gamma-ray data will be available for the AGILE Guest Observer Program (GOP) that will be open to the international community on a competitive basis. The AGILE Cycle-1 GOP is scheduled to start in December, 2007.

### 6.1 Data Analysis and Scientific Ground Segment

AGILE science data (about 300 Mbit/orbit) are telemetered from the satellite to the ASI ground station in Malindi (Kenya) at every satellite passage (about 90 minutes). A fast ASINET connection between Malindi and the Telespazio Satellite Control Center at Fucino and then between Fucino and the ASI Science Data Center (ASDC) ensures the data transmission every orbit. The AGILE Mission Operations Center is located at Fucino and will be operated by Telespazio with scientific and programmatic input by ASI and the AGILE Science Team through the ASDC.

Scientific data storage, quicklook analysis and the GOP will be carried out at ASDC. After pre-processing, scientific data (level-1) will be corrected for satellite attitude data and processed by dedicated software produced by the AGILE Team in collaboration with ASDC personnel. Background rejection and photon list determination are the main outputs of this first stage of processing. Level-2 data will be at this point available for a full scientific analysis.

Gamma-ray data generated by the GRID will be analyzed by special software producing: (1) sky-maps, (2) energy spectra, (3) exposure, (4) point-source analysis products, and (5) diffuse gamma-ray emission. This software is aimed to allow the user to perform a complete science analysis of specific pointlike gamma-ray sources or candidates. This software will be available for the GOP. Super-AGILE data will be deconvolved and processed to produce 2-D sky images through a correlation of current and archival data of hard X-ray sources. GRB data will activate dedicated software producing lightcurves, spectra and positioning both in the hard X-ray (18 – 60 keV) and gamma-ray energy (30 MeV – 30 GeV) ranges. The AGILE data processing goals can be summarized as follows:

- **Quicklook Analysis (QA)** of all gamma-ray and hard X-ray data, aimed at a fast scientific processing (within a few hours/1 day depending on source intensity) of all AGILE science data.
- **web-availability of QA results** to the international community for alerts and rapid follow-up observations;
- **GRB positioning and alerts through the AGILE Fast Link**, capable of producing alerts within 1-2 minutes since the event;
- **standard science analysis of specific gamma-ray sources open to a Guest Observer Program**;
- **web-availability of the standard analysis results of the hard X-ray monitoring program by Super-AGILE**.

### 6.2 Multiwavelength Observations Program

The scientific impact of a high-energy Mission such as AGILE (broad-band energy coverage, very large fields of view) is greatly increased if an efficient program for fast follow-up and/or monitoring observations by ground-based and space instruments is carried out. The AGILE Science Program overlaps and be complementary to those of many other high-energy space Missions (INTEGRAL, XMM-Newton, Chandra, SWIFT, Suzaku , GLAST) and ground-based instrumentation.

The AGILE Science Program will involve a large astronomy and astrophysics community and emphasizes a quick reaction to transients and a rapid communication of crucial data. Past experience shows that in many occasions there was no fast reaction to $\gamma$-ray transients (within a few hours/days for unidentified gamma-ray sources) that could not be identified. AGILE will take advantage, in a crucial way, of the combination of its gamma-ray and hard X-ray imagers.

An AGILE Science Group (ASG) aims at favoring the scientific collaboration between the AGILE Team and the community especially for coordinating multiwavelength observations based on AGILE detections and alerts. The ASG is open to the international astrophysics community and consists of the AGILE Team and qualified researchers contributing with their data and expertise in optimizing the scientific return of the Mission. Several working groups are operational on a variety of scientific topics including blazars, GRBs, pulsars, and Galactic compact objects. The AGILE Team is also open to collaborations with individual observing groups.

Updated documentation on the AGILE Mission can be found at the web sites http://asdc.asi.it, and http://agile.iasf-roma.inaf.it.

## ACKNOWLEDGMENTS

The AGILE program is developed under the auspices of the Italian Space Agency with co-participation by the Italian Institute of Astrophysics (INAF) and by the Italian Institute of Nuclear Physics (INFN). Research partially supported under the grant ASI-I/R/045/04. We acknowledge the crucial programmatic support of the ASI Directors of the *Observation of the Universe* Unit (S. Di Pippo) and of the *Small Mission* Unit (F. Viola, G. Guarrera), as well as of the AGILE Mission Director L. Salotti.

# THE AGILE SILICON TRACKER

GUIDO BARBIELLINI,[a,b], MARCO BASSET[b], LUCA FOGGETTA[c], FERNANDO LIELLO[a,b], FRANCESCO LONGO[a,b], ELENA MORETTI[a,b], CRISTIAN PONTONI[a], MICHELA PREST[c], ERIK VALLAZZA[b]

[a] *Dipartimento di Fisica, Università di Trieste, via Valerio 2, Trieste, Italy*

[b] *INFN, Sezione di Trieste, via Valerio 2, Trieste, Italy*

[c] *INFN, Sezione di Milano, via Celoria 16, Milano, Italy*

**Abstract**

The AGILE small scientific satellite for the detection of $\gamma$-ray cosmic sources in the energy range 50 MeV-30 GeV is made of three different detectors. The main detector is a silicon-tungsten tracker. Its structure and assembly are described in this paper. ).

## 1 The AGILE detector

The AGILE satellite (fig. 1) was launched by the PSLV rocket from the Sriharikota base in India on April 23, 2007[1]. The scientific instrument is made by four detectors: an **anticoincidence system**, made of 12 lateral panels and 1 top panel of plastic scintillators read out by photomultipliers, the X-ray detector **Superagile** with a silicon microstrip detection plane made of 16 silicon microstrip tiles (strip pitch is 121 $\mu$m) read by XAA1 chips[2], a little **calorimeter** made of 30 CsI scintillating bars in orthogonal directions, read out by photodiodes[3] and the silicon-tungsten **tracker**.

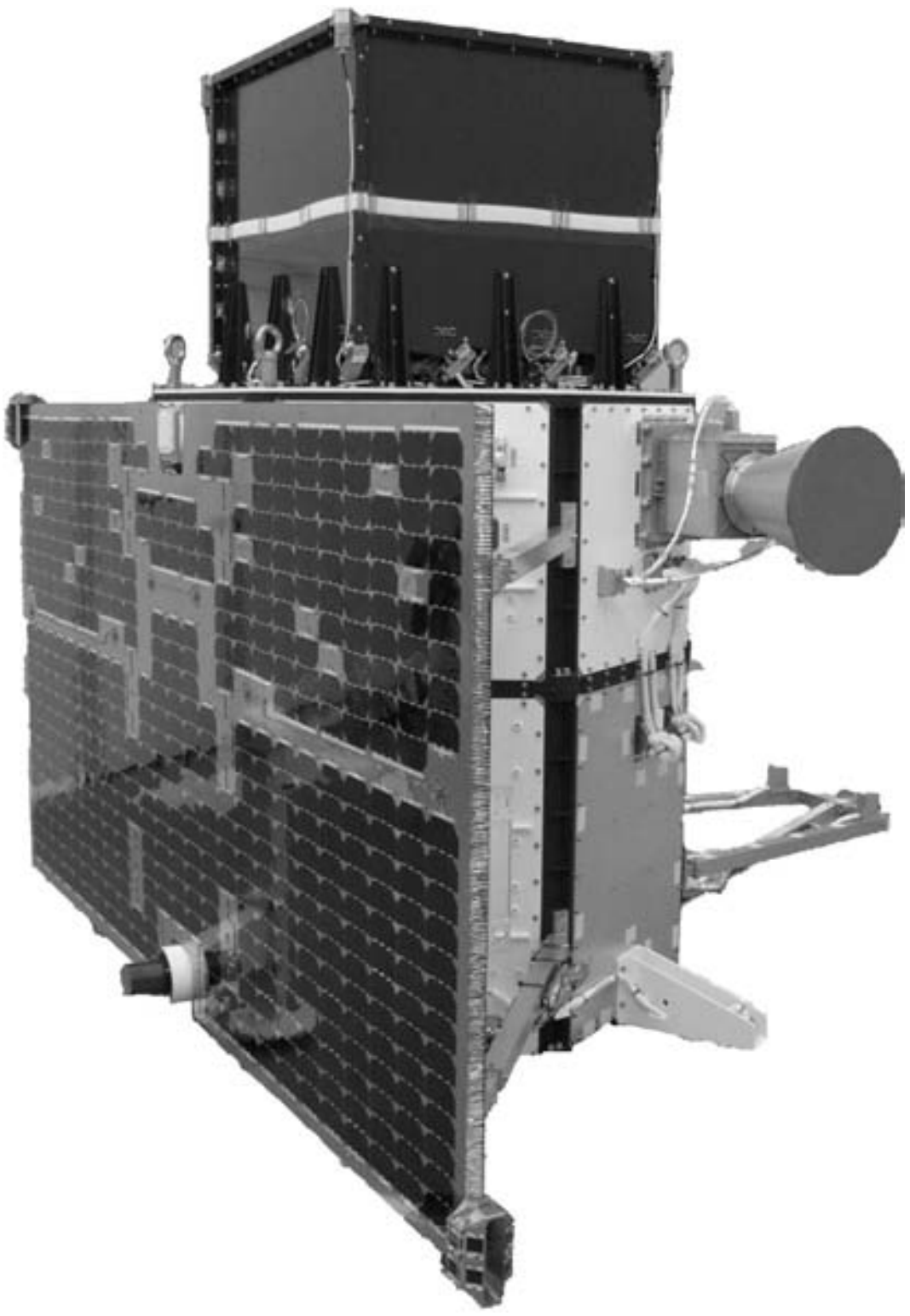

Figure 1: Picture of the integrated satellite.

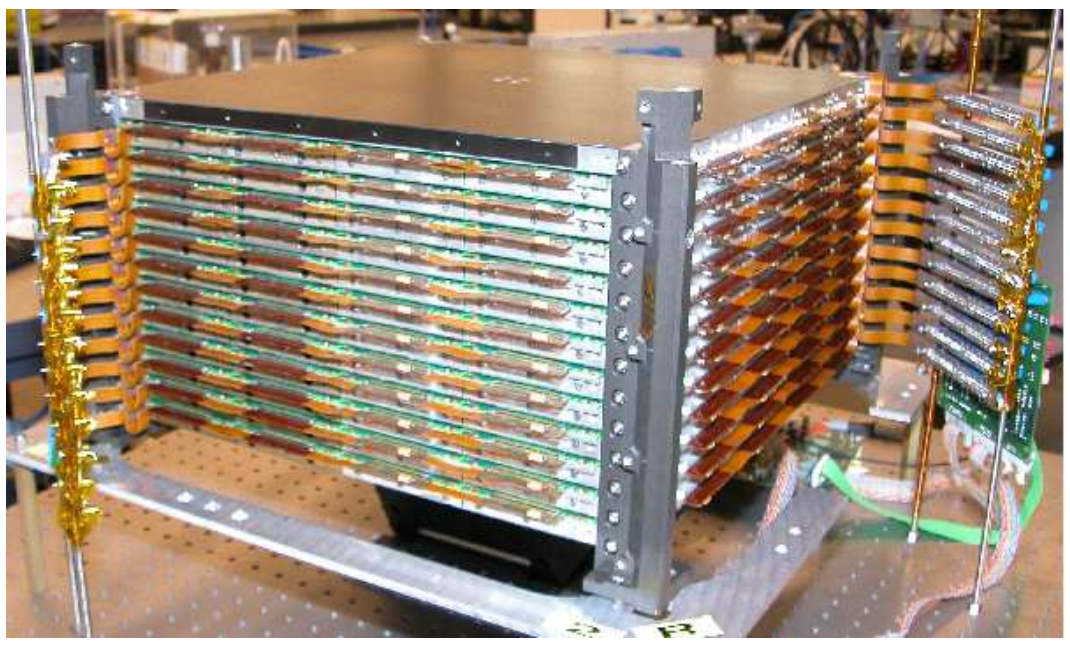

Figure 2: The silicon tracker before the integration with the payload and the flight DAQ (June 2005)

## 2 The AGILE Silicon Tracker

The silicon tracker (see fig. 2) consists of 12 planes with two views of 16 silicon microstrip tiles (Hamamatsu), organized in 4 modules ("ladders") of 4 tiles. The two views are positioned orthogonally one with respect to the other in order to obtain a $x$-$y$ imaging system and the planes are organized in 13 mechanical trays. The first 10 planes have a 245 $\mu$m (corresponding to about 0.07 $X_0$) thick tungsten layer for the photon conversion.

Each tile consists of a 9.5×9.5 cm$^2$ AC-coupled 410 $\mu$m thick 384 silicon strip detector. The 4 ladder tiles are connected in series strip by strip through 25 $\mu$m thick aluminium ultrasonic wire bondings. The strip physical pitch is 121 $\mu$m, while the readout one is 242 $\mu$m with one floating strip in order to reduce the number of electronics channels and thus the power consumption while maintaining a good spatial resolution.

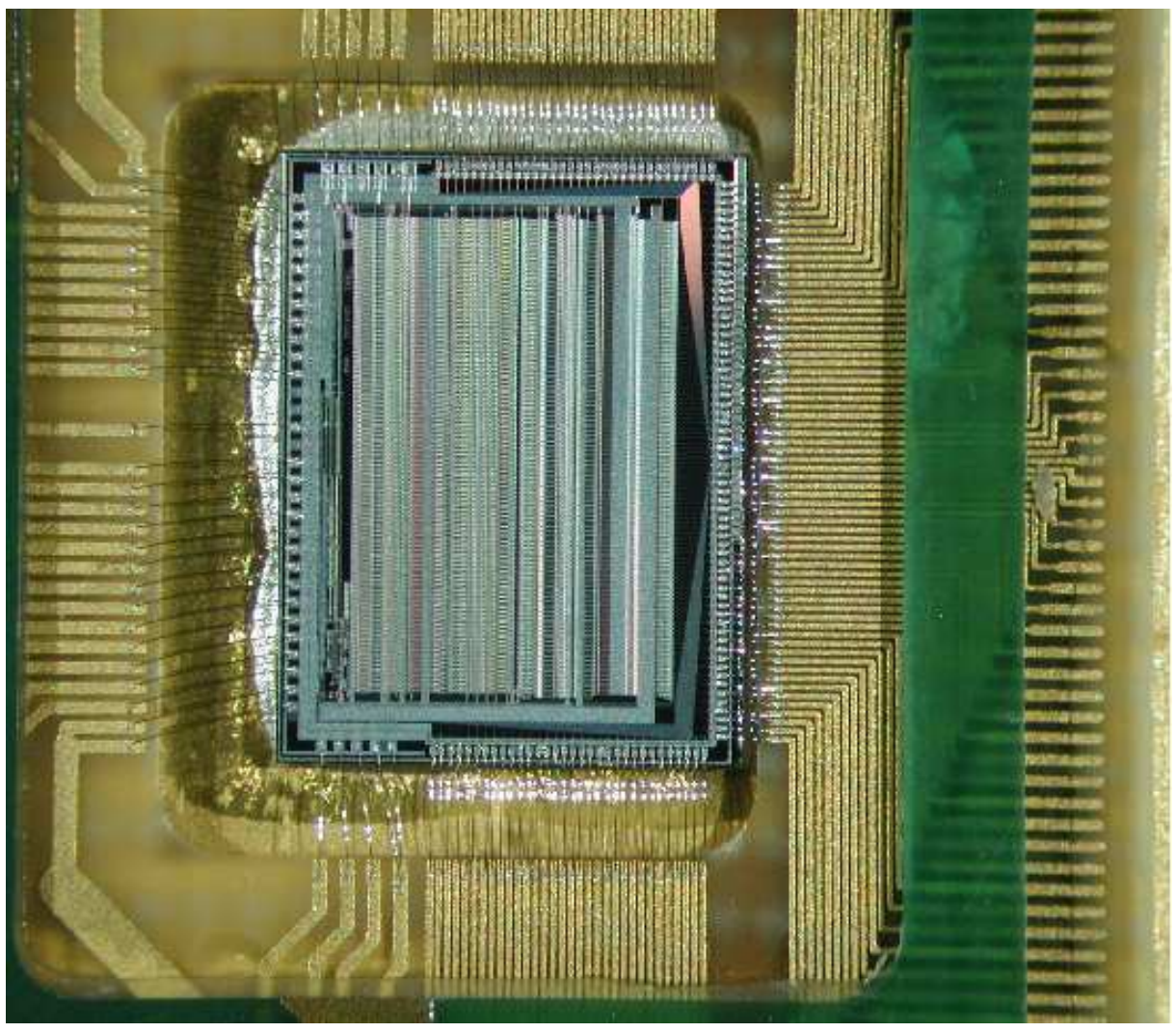

Figure 3: Picture of the TAA1 frontend chip

The readout ASIC is the TAA1 (fig. 3), a 128 channel, low noise, low power, self triggering ASIC designed by Ideas (Norway) and produced by AMS (Austria) with a 0.8 $\mu$m N-well BiCMOS, double poly, double metal on epitaxial layer technology. To limit the power consumption of the Tracker, the ASIC is operated in a very low power configuration ($< 400$ $\mu$W per channel). The die is $5.174 \times 6.919$ mm$^2$ and $\sim 600$ $\mu$m thick. The 128 input pads have a 100 $\mu$m pitch and have been bonded to the PCB (Printed Circuit Board) lines with a 17.5 $\mu$m thick aluminium wire, while a 25 $\mu$m one has been used to bond the 41 output, control and power 200 $\mu$m pitch pads.

The HDI (High Density Interconnection) is the circuit that interconnects the ladder ASICs and the detectors to the readout electronics. It consists of a ten layer FR4 PCB designed by INFN Trieste (Italy) and produced by ILFA GmbH (Hannover, Germany). One of the layers is a 38 cm long kapton foil extending outside the PCB (fig. 4) which is used both as a mechanical support for the silicon detectors and to provide the bias to the detectors themselves. The four silicon tiles are glued (see fig. 5) on the HDI and then bonded to form one silicon ladder (fig. 6). The ladders are then glued on the opposite sides of the support trays.

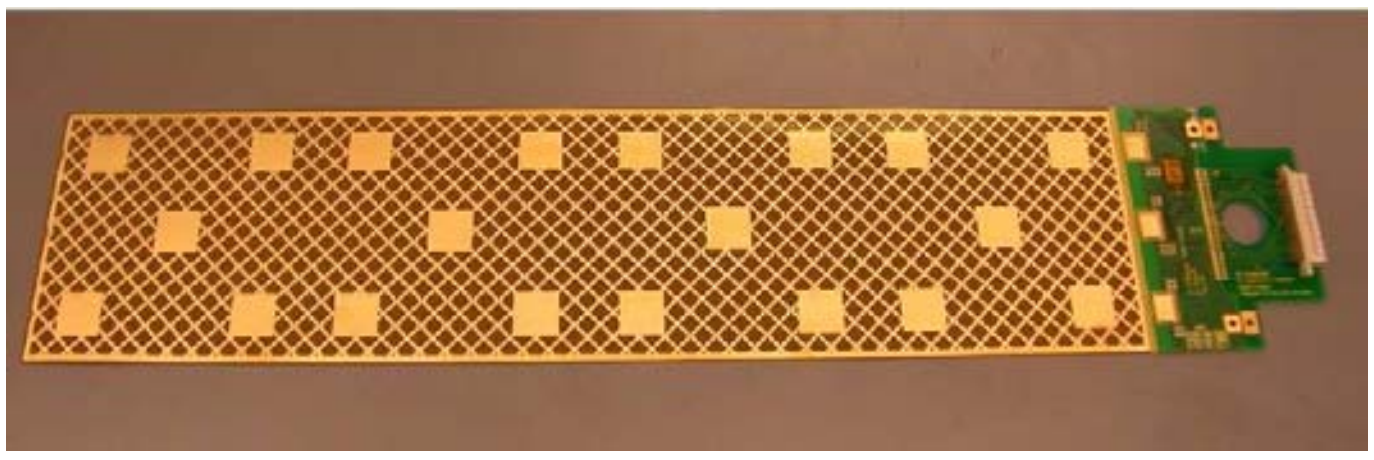

Figure 4: Picture of the HDI PCB. The 38 cm long kapton cable for the bias is clearly visible; the HDI output is connected to the temporary adapter for testing.

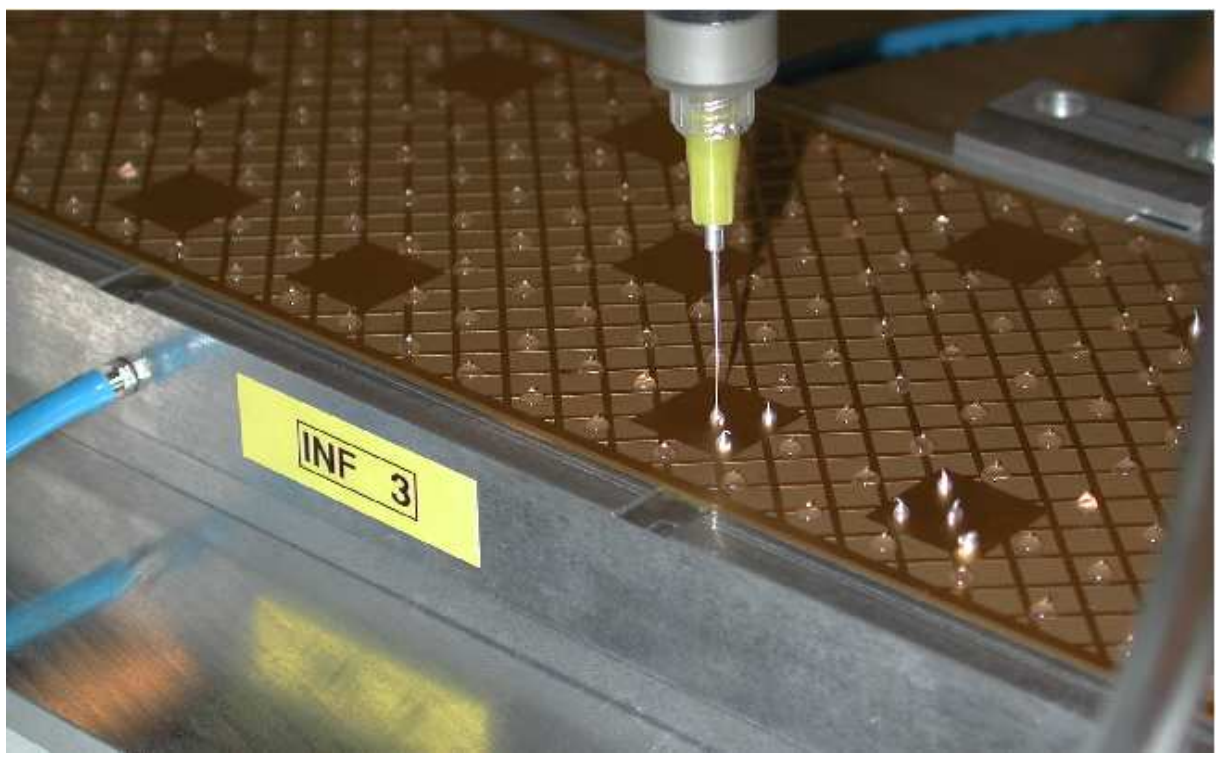

Figure 5: Glue dispenser during the preparation of the kapton layer for the silicon ladder.

Each tray (Oerlikon-Contraves, fig. 7 is made of a 15 mm thick aluminium honeycomb core with a 0.5 mm carbon fiber layer per side. The tungsten foil is glued on to the bottom part of the first 10 trays.

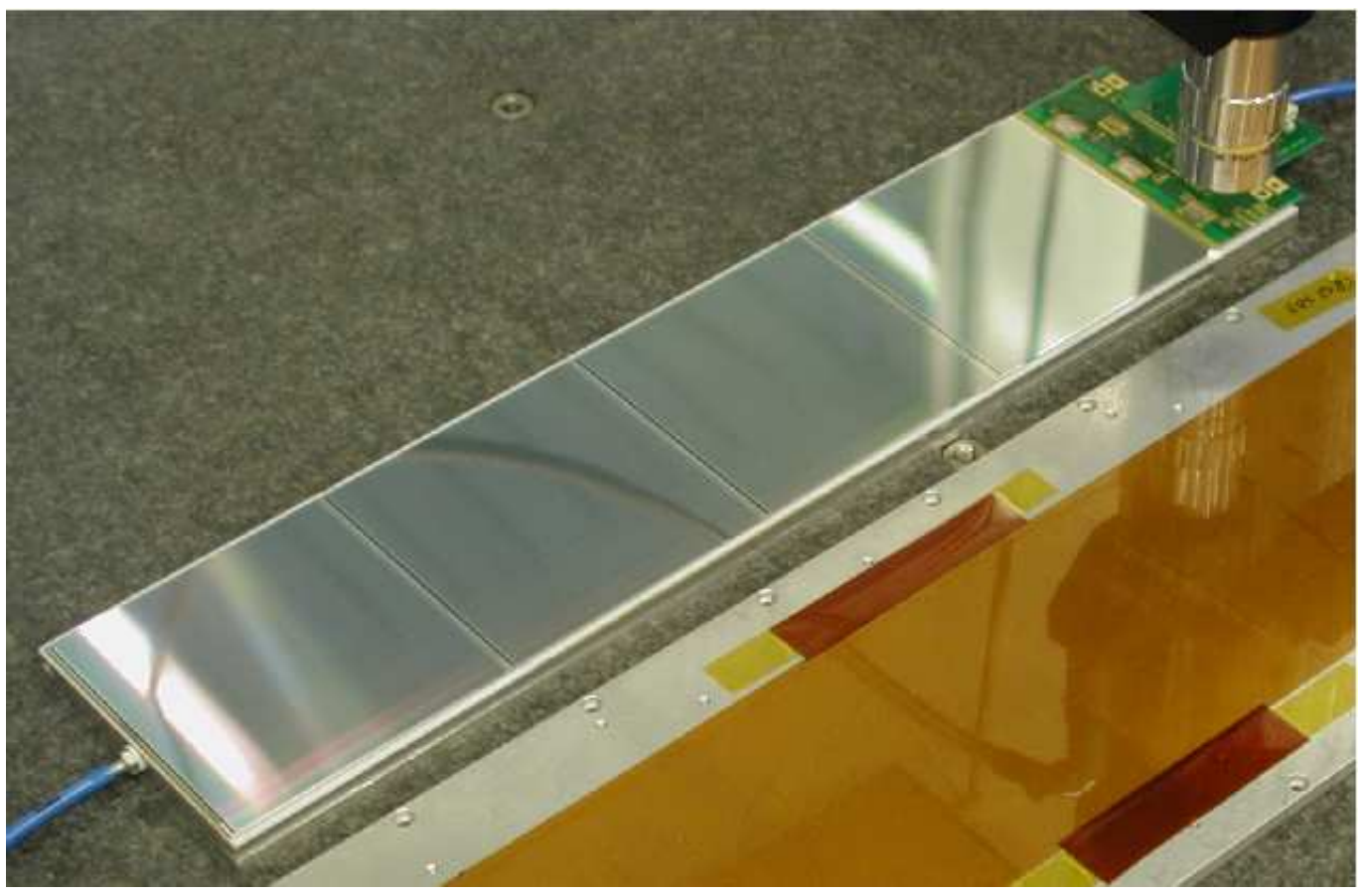

Figure 6: A ladder during the mechanical measurements

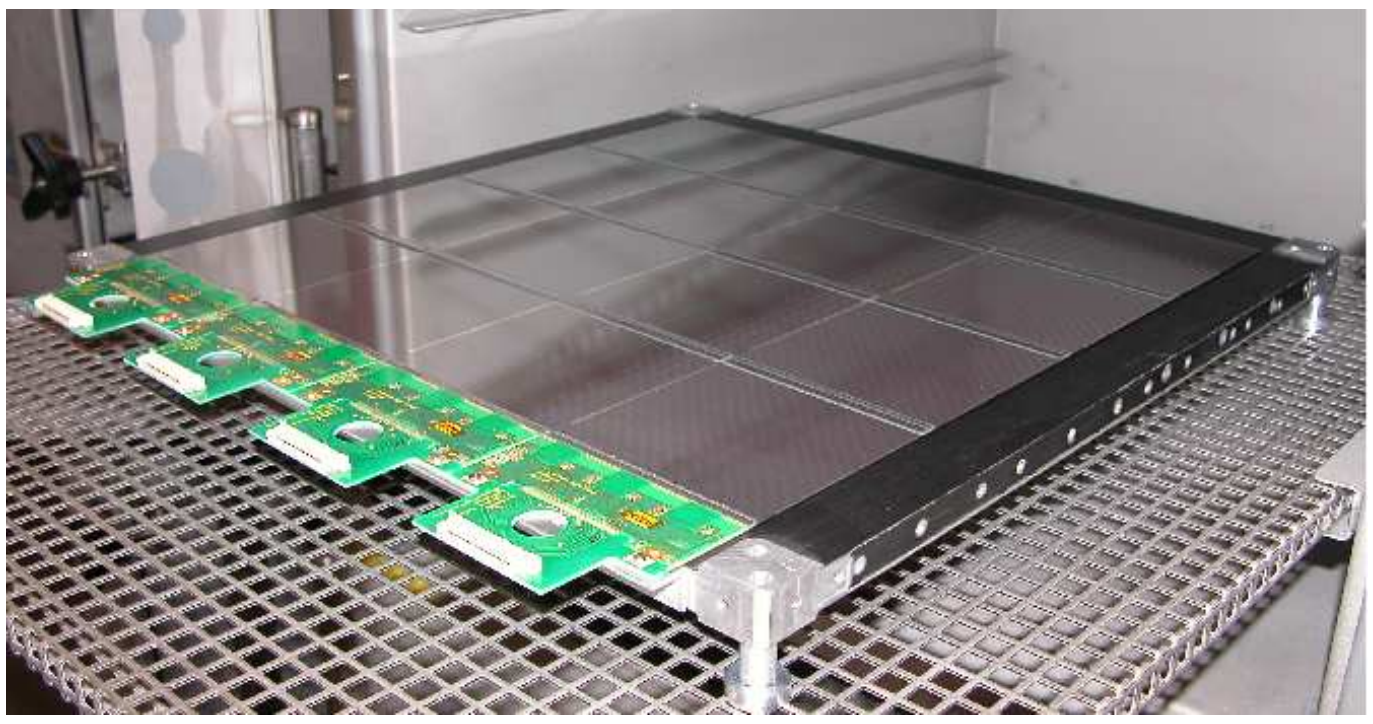

Figure 7: A tray with the four ladders glued on one of the views.

## 3 The assembly steps

The prototype and flight model assemblies have been performed by Mipot S.p.A. (Cormons, Italy).
The assembly steps can be divided into four groups:

- Assembly of the passive components and of the ASICs on the HDIs (fig. 4). The following operations are performed on the PCBs, once they have passed the incoming inspection: tin soldering of the passive components (MIL-883 resistors and capacitors); gluing and bonding of a temporary adapter for the connection to the test equipment; gluing (Epotek H20E), bonding and conformal coating (RTV615) of the ASICs;

gluing of the FR4 covers; final 72 hour test in a thermo-vacuum chamber with temperatures ranging from -30°C to +50°C; each cycle lasts 2 hours.

- Assembly of the silicon ladder. Four tiles are glued head-on (AV138) and the so obtained silicon ladder is glued on the HDI kapton (Epotek H20E on the gold dots for the elctrical connection to the bias voltage and DC3145 on the rest) (fig. 4c, fig. 6). The head-on gluing between the first silicon tile and the HDI is made with three segments of AV138 separated by RTV615, to avoid breaking caused by eventual thermal gradients. The strips are then bonded together between the four silicon tiles and between the HDI and the first silicon tile. The bondings are protected with an epoxy dam (2216) filled with RTV615 (fig. 4c).
- Tray assembly. Four ladders are glued first on the bottom side of a tray and then on the top view (fig. 7). The adapters are taken away and the cable to connect the frontend to the readout electronics (MLC, MultiLayer Connection) is fixed with screws and glued. The HDI and the MLC are bonded with 25 $\mu$m Al wire and the bondings are protected as in the silicon ladder case.
- The 13 trays are assembled together with the mechanical supports, obtaining the Silicon Tracker (fig. 2).

After the prototype performance studies, the final flight version of the Tracker has been assembled in one year. A large number of tests (e.g. mechanical measurements, thermo-vacuum cycling tests, electrical and functional tests) have been performed on every Tracker element at every assembly step.

## 4 Conclusions

The silicon tracker, core of the AGILE GRID detector, was assembled and tested in one year. The Hamamatsu silicon detectors and the IDEAS front-end electronics have proved to fullfill the design requirements. The capability of AGILE to detect of gamma-rays over a large field of view with optimal spatial resolution of electron-positron pairs is based on the excellent performance of this detector.

# SUPERAGILE: TWO MONTHS IN ORBIT

MARCO FEROCI [a], *on behalf of the AGILE Team*

[a] *Istituto di Astrofisica Spaziale e Fisica Cosmica, Istituto Nazionale di Astrofisica, via Fosso del Cavaliere 100, Roma, Italy*

**Abstract**

SuperAGILE is the hard X-ray imager onboard the AGILE mission, launched into an equatorial orbit on April 2007. Here I provide a short description of the experiment and report its status before lunch, as derived from the on-ground tests and calibrations, and after the first two months of operation in space.

## 1 Introduction

AGILE [1], a small mission of the Italian Space Agency (ASI), was injected by a PSLV rocket from India on $23^{rd}$ April 2007 in a $\sim 2°.5$ inclination orbit at $\sim$545 km altitude. The AGILE scientific payload is composed of a Gamma Ray Imaging Detector (GRID) - based on a silicon tracker, a CsI "mini" calorimeter and a plastic anticoincidence - and a hard X-ray ($\sim$18-60 keV) monitor, SuperAGILE. The simultaneous observation of the gamma-ray sources with an instrument operating in hard X-rays, an energy range where higher sensitivity and better angular resolution can be achieved, can improve the chance of detecting and identify the gamma ray sources, with the benefit of correlating the simultaneous emission in two distant energy bands. This was the main scientific driver for including SuperAGILE in the AGILE payload. However, with a good sensitivity and a very large field of view, SuperAGILE also operates as an independent wide field monitor of the hard X-ray sky, allowing the simultaneous monitoring of several known Galactic sources and possibly the discovery of new transient sources.

## 2 The SuperAGILE Experiment onboard AGILE

SuperAGILE [2] is an experiment based on the coded mask technique, imaging the X-ray sky in the energy range ∼18-60 keV. The field of view is larger than one steradian, and the on axis angular resolution is 6 arcminutes. Each of the four detection units performs one-dimensional imaging, in two orthogonal directions in the sky, with redundancy for both coordinates. The primary scientific goal of SuperAGILE is to provide a rapid and accurate identification of the gamma-ray sources observed with the GRID. This is a task that, based on the current knowledge, can be accomplished on a few classes of sources (gamma-ray pulsars, blazars, gamma-ray bursts). However, with an on-axis sensitivity of ∼15 mCrab (for a 50 ks net exposure) and field of view of $2\times(107^\circ\times68^\circ)$ (zero response), SuperAGILE will also operate as a (quasi) all sky monitor of the X-ray sky. In Table 1 the main instrument parameters are summarized. A more detailed description of the SuperAGILE experiment may be found in [2].

Table 1: Main instrumental parameters of the SuperAGILE experiment.

| Instrument Parameter | Value |
|---|---|
| Detector | 410 $\mu$m thick Si $\mu$-strip |
| Energy Range | 18-60 keV |
| Energy Resolution | ∼8 keV FWHM |
| Geometric Area | 1344 $cm^2$ |
| 1D Field of View | $2\times(107^\circ\times68^\circ)$ |
| 2x1D Field of View | $68^\circ\times68^\circ$ |
| Mask Transparency | 50% |
| Angular Resolution | 6 arcmin |
| Mask Element Size | 242 $\mu$m |
| Mask-Detector Distance | 142 mm |
| Timing Resolution | 2 $\mu$s |
| Timing Accuracy | 5 $\mu$s |
| Point Source Location Accuracy | 1-2 arcmin |
| Point Source Sensitivity | 15 mCrab (1D) |
| Data Transmission | Event-by-Event, 32-bit/event |

## 3 The Ground Calibrations

The SuperAGILE experiment underwent a series of ground calibration campaigns. The naked Detection Plane was calibrated on June 2005 at IASF

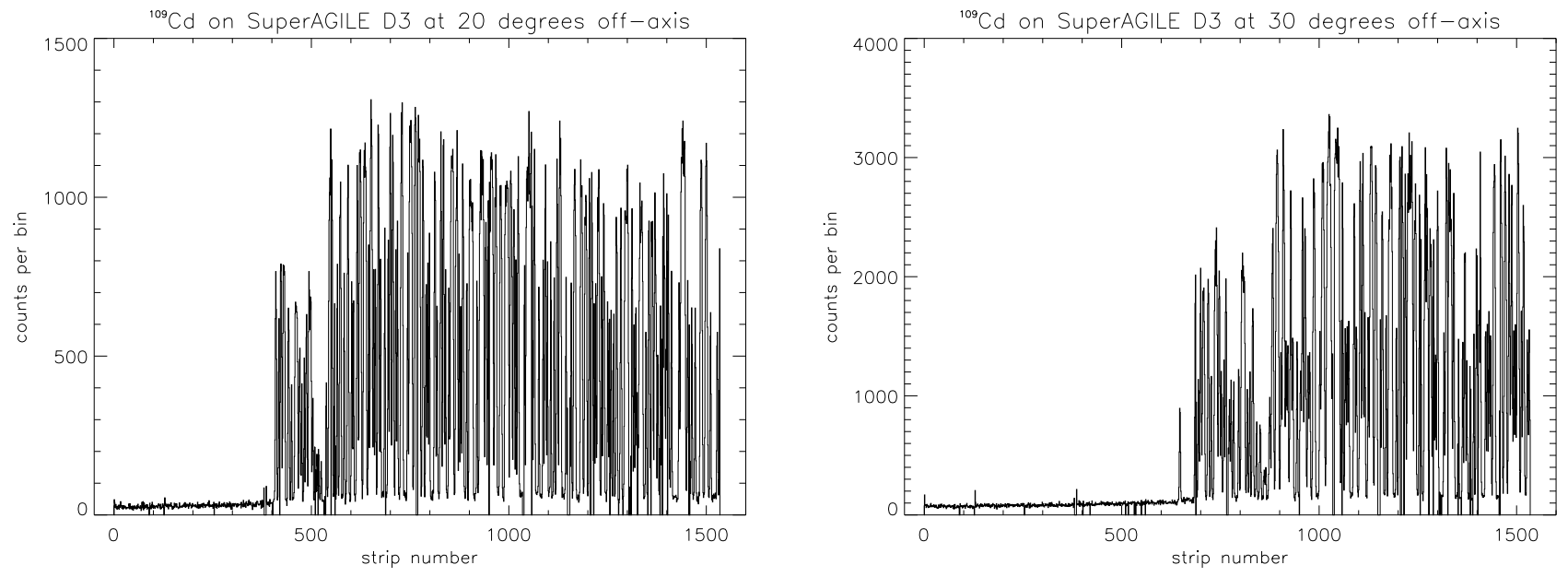

Figure 1: The detector images taken with one SuperAGILE detector during the ground calibration when illuminated with a 22 keV source at 20 and 30 degrees (left and right panels, respectively).

Rome, by scanning the detectors with a pencil beam obtained by collimating the beam from an X-ray tube, and illuminating it with omnidirectional radioactive sources. On August 2005 the complete experiment was again scanned with the pencil beam (simulating an infinite distance source) and illuminating with radioactive sources at ~2 m distance, correcting for the beam divergence in order to derive the response from an infinite distance source. Finally, the experiment was again calibrated after its integration in the satellite with radioactive sources at finite distance on January 2007 at the Carlo Gavazzi Space premises in Tortona (AL, Italy). The results of the 2005 calibrations may be found in [3] and [4].

The results of the 2007 calibration will be reported in a dedicated paper. Here we show a few snapshots, providing a feeling of the data quality. In Figure 1 the detector images obtained with a 22 keV source at 20 and 30 degrees off-axis are shown. The plots clearly show the obscuration effect of the collimator, as well as the open/closed pattern of the coded mask (shadowgram). The smaller structure appearing at the separation between the obscured section of the image and the mask shadowgram is due to support structures partially obstructing the SuperAGILE field of vew. When the de-convolution with the code of the mask and the finite-distance correction are applied to these data the source is imaged as shown in Figure 2, showing the 6 arcminutes point spread function of the experiment (here slightly increased by the finite extension of the radionuclide in the radioactive source, not subtracted here).

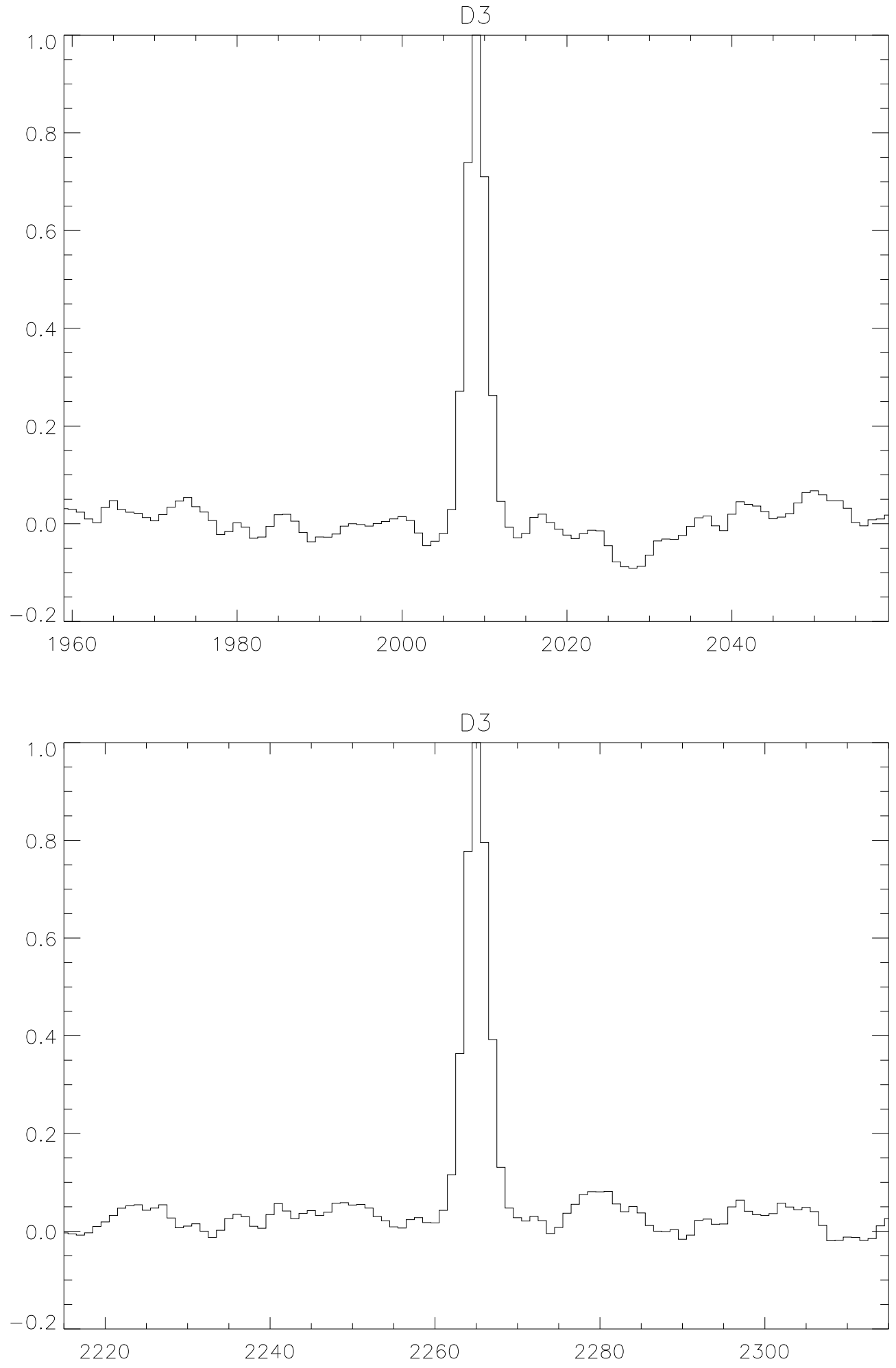

Figure 2: Source image obtained by the detector images shown in Figure 1, after the correction for the finite distance and the mask de-convolution process.

## 4 First News from Above

After the April 2007 launch, the in-flight commissioning phase of SuperAGILE started on May $3^{rd}$. Within the first two months of commissioning only 31 useful ground contacts (effectively, one week of satellite operation) were allocated for sending SuperAGILE commands. The commissioning was thus not yet completed at the time of this conference. The main activities carried out were: gradual switch-on of the experiment; health check-out (successful!); setting of ∼half of the digital parameters; configuration of the analog thresholds of most of the channels; configuration of the timing with the veto systems. All of this activities were carried out, upon our request, pointing the experiment boresight to a "blank" field (that is, a field where no X-ray sources are expected at a flux level above the SuperAGILE sensitivity), to be able to set the instrument parameters without contamination of the count rate by celestial sources (except for those contributing to the diffuse X-ray background). The first light of the experiment is then expected, after completion of the check-out and setting activities, on the field of Vela.

The type of activities carried out in this first phase of the SuperAGILE in flight operation does not allow to have results other than of engineering type. In Figure 3 we show the count rate recorded by one of the four SuperAGILE detectors during a typical orbit. The plots show the rate in 16s integration bins of the 4 "daisy chains", the basic electronic units composing each of the 4 SuperAGILE detectors. In practice, this is the count rate over 384 strips (1/4 of one detector), in 16 seconds. The plots show the main features registered along one orbit: the "de-occultation" of the experiment field of view by the Earth (the step-like feature at the beginning of the orbit), and the effect of the dead-time induced by the very high count rate of the AGILE anticoincidence during the passage through the South Atlantic Anomaly (the sudden decrease of counts towards the end of the orbit).

The Earth occultation of the field of view is best seen in the panel bottom left, due to the lower energy threshold of this daisy chain at the time of the measurement (at this stage, the energy threshold equalization process had not been completed yet). This effect allows to derive the net effect of the diffuse X-ray background, by comparing the on-Earth to the off-Earth energy spectra. Indeed, in this comparison also the Earth albedo plays a role when the instrument is looking at the Earth, so the difference is expected to be attenuated. This comparison is shown in Figure 4.

In Figure 5 we show an effect of the passage through the South Atlantic Anomaly. Here the "raw" SuperAGILE count rate is reported, as it is recorded before the energy and anticoincidence cuts are applied. The plot shows a sequence of more than 80 consecutive orbits. At each orbit the satellite crosses the South Atlantic Anomaly at a different depth, due to the precession of the satellite orbit over the geomagnetic Earth. At a different depth in the South

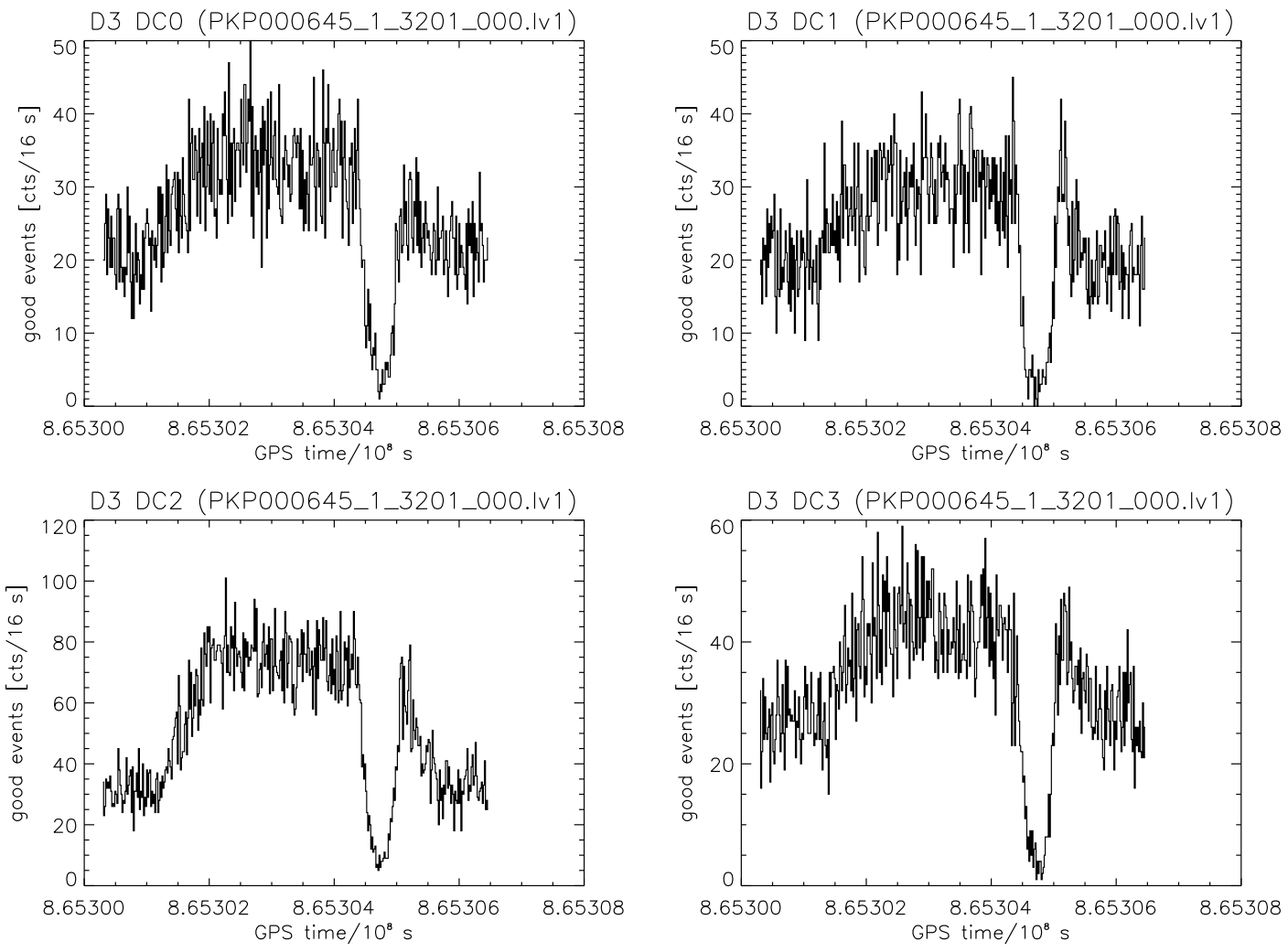

Figure 3: The count rate (cts/16s) of one SuperAGILE detection unit over a typical orbit. The detector count rate here is shown divided in its four sections.

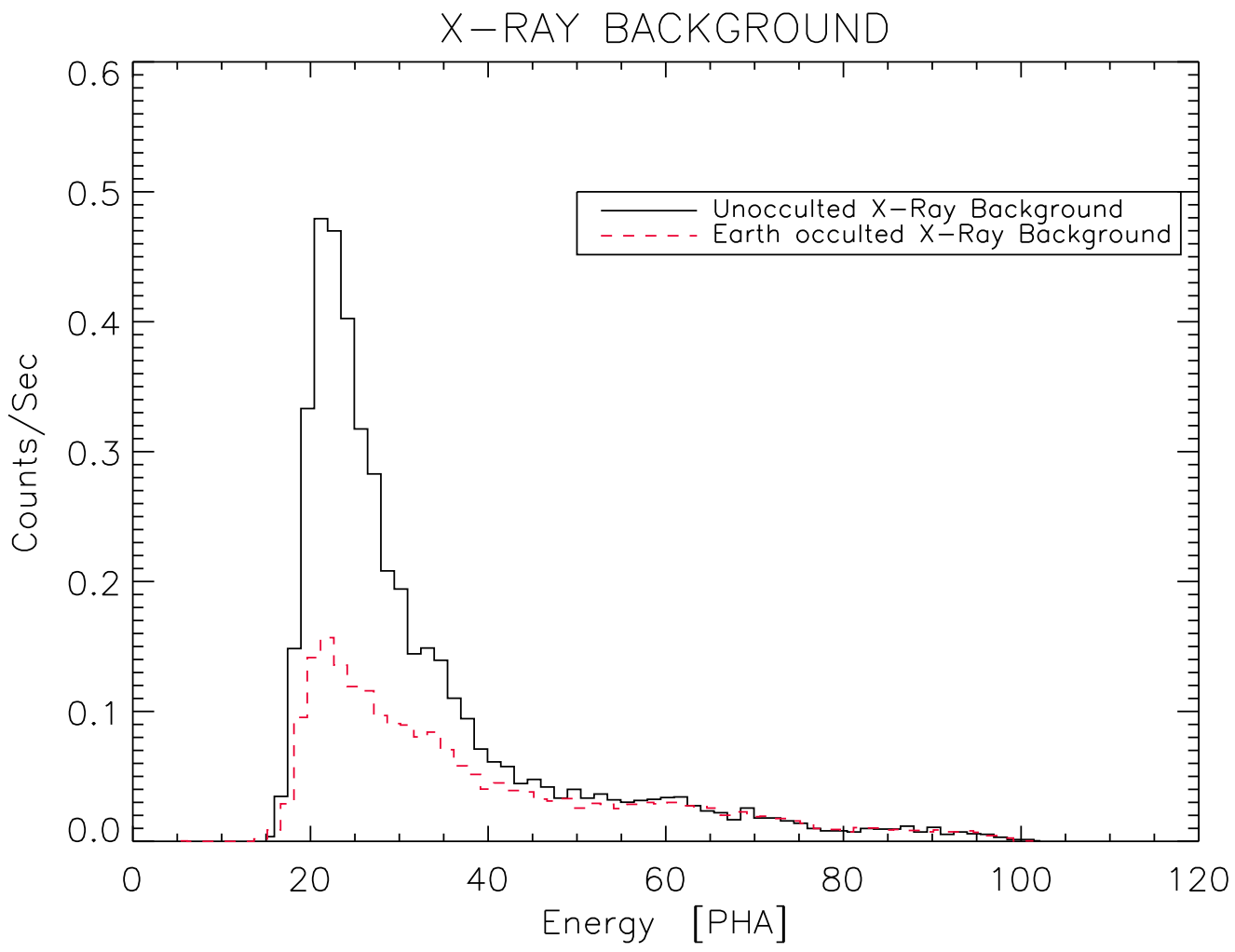

Figure 4: Energy spectrum of the background recorded by one SuperAGILE detector with and without the Earth occulting its field of view.

Atlantic Anomaly corresponds a different particle flux (here the dominant contribution to the event rate) and the net effect is a modulation of the count rate when seen over a multiple orbit scale.

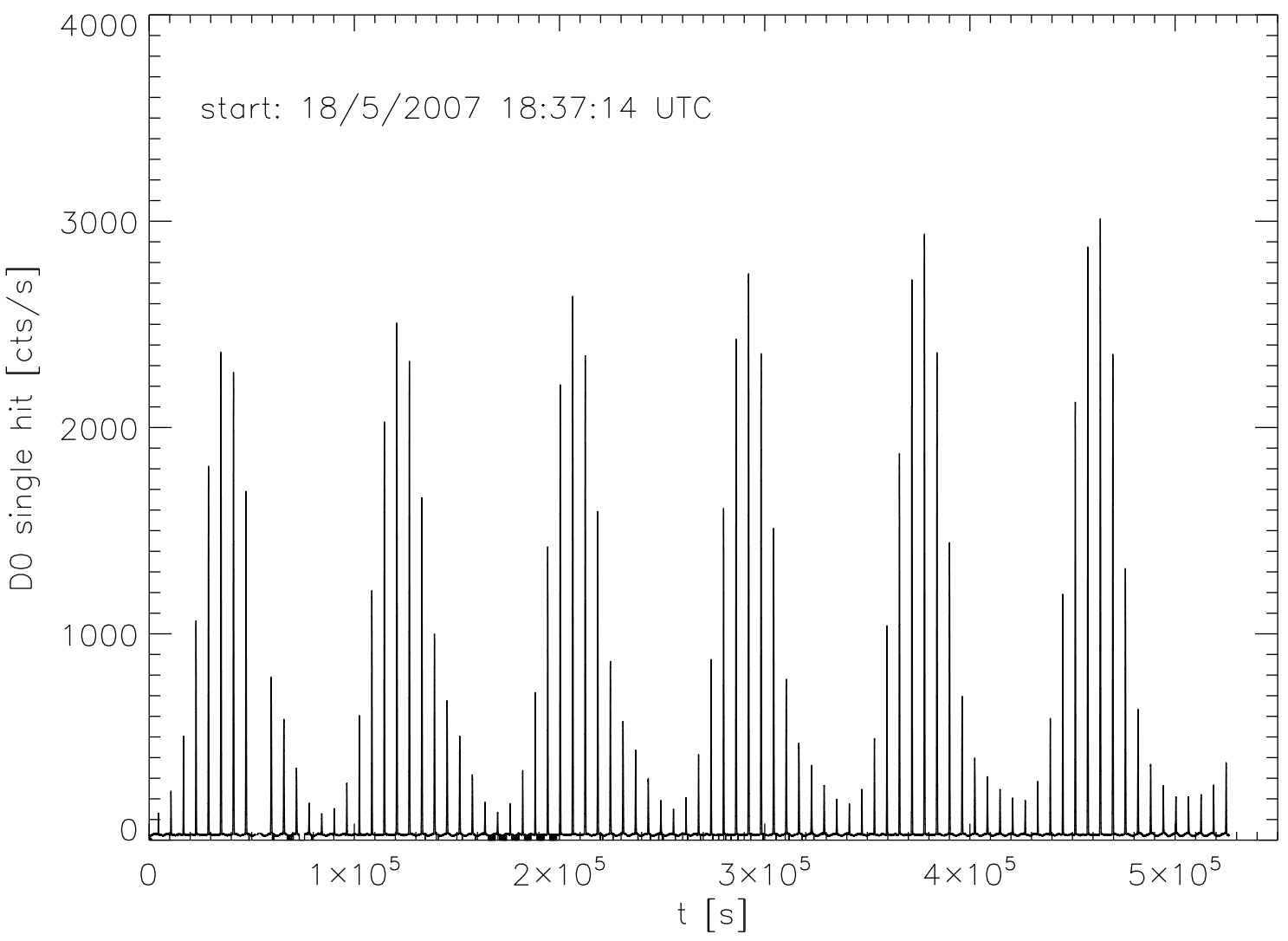

Figure 5: "Raw" count rate of one SuperAGILE detector recorded over 86 consecutive orbits. The rate here is different from that shown in Figure 3 in the fact that this is the rate before the energy upper threshold and anticoincidence cuts are applied. For this reason, the rate increase during the crossing of the South Atlantic Anomaly is clearly visible (each vertical line is an orbit, here).

## 5 Conclusions and Perspectives

As explained in the previous sections, despite the two months elapsed between this conference and the launch of the AGILE satellite, the SuperAGILE commissioning phase is not yet completed. Then, the near future will be spent in completing the experiment setting and then the Science Verification Phase will begin. The current scheduled period for this phase is from July to September 2007. During this phase SuperAGILE is expected to have its first light pointing to fields with usually bright X-ray sources, the Vela field and the Cignus region. Then, the in-flight calibration plan foresees a raster scan using the Crab nebula, that is observing this standard source at different off-axis angles, in order to map the experiment response at several locations over the field of view. Due to the tight Sun-angle constraints of AGILE (solar panels at $90\pm1$ degrees from the Sun), this scan will span a very long (not continuous) period, starting at the beginning of August and ending on November.

During these pointings the main goal will be to study and calibrating the imaging response, especially for what concerns the attitude correction, both onboard and in the on-ground data analysis, with the real aspect precession. In fact, the AGILE attitude varies by more than one SuperAGILE pixel every second. The data from the Star Trackers allow to know the real pointing at the arcminute level every tenth of a second. These data are then used to correct the arrival direction of each individual photon, and almost-fully exploit the SuperAGILE capabilities demonstrated by the on ground calibrations. This will be particularly important to fulfill the scientific objectives concerning the prompt response to cosmic gamma-ray bursts (see [5] and [2] for details).

# ORIGIN AND PROPAGATION OF COSMIC RAYS (SOME HIGHLIGHTS)

Igor V. Moskalenko

*Hansen Experimental Physics Laboratory and Kavli Institute for Particle Astrophysics and Cosmology, Stanford University, Stanford, CA 94305, U.S.A.*

**Abstract**

The detection of high-energy particles, cosmic rays (CRs), deep inside the heliosphere implies that there are, at least, three distinctly different stages in the lifetime of a CR particle: acceleration, propagation in the interstellar medium (ISM), and propagation in the heliosphere. Gamma rays produced by interactions of CRs with gas, radiation, and magnetic fields can be used to study their spectra in different locations. Still, accurate direct measurements of CR species inside the heliosphere (such as their spectra and abundances) are extremely important for the understanding of their origin and propagation. In this paper, an emphasis is made on very recent advances and especially on those where GLAST and PAMELA observations can lead to further progress in our understanding of CRs.

## 1 Introduction

Cosmic rays and $\gamma$-rays are intrinsically connected: $\gamma$-ray emission is a direct probe of proton and lepton spectra and intensities in distant locations of the Galaxy. Diffuse emission accounts for ~80% of the total $\gamma$-ray luminosity of the Milky Way and is a tracer of interactions of CR particles in the ISM [1]. On the other hand, direct measurements of CR species can be used to probe CR propagation, to derive propagation parameters, and to test various hypotheses [2]. Luckily, two missions of the present and near future are targeting these issues and present unique opportunities for breakthroughs. The Payload for Antimatter-Matter Exploration and Light-nuclei Astrophysics (PAMELA) [3] has been launched in June 2006 and is currently in orbit. During its projected 3 yr lifetime it will measure light CR nuclei, antiprotons, and positrons in the energy range 50 MeV/n – 300 GeV/n with high precision. The Gamma-ray Large Area Space Telescope (GLAST) [4] is scheduled for launch in early 2008. It has significantly improved sensitivity, angular resolution, and much larger field of view than its predecessor EGRET and will provide excellent quality data in the energy range 20 MeV – 300 GeV.

## 2 Cosmic-ray accelerators

A supernova (SN) – CR connection has been discussed since the mid-1930s when Baade and Zwicky proposed that SNe are responsible for the observed CR flux. The first direct evidence of particle acceleration up to very high energies (VHE) came from observations of synchrotron X-rays from the supernova remnant (SNR) SN1006 [5]. More recently, observations of TeV $\gamma$-rays [6] confirm the existence of VHE particles. Still, definitive proof that SNRs are accelerating protons is absent. Recent observations of the SNR RX J1713 by HESS suggest that its spectrum is consistent with the decay of pions produced in $pp$-interactions [6], while the spectrum from inverse Compton scattering (ICS) does not seem to fit the observations. However, a calculation of ICS and synchrotron emission using a one-zone model [7] and a new calculation of the interstellar radiation field (ISRF) shows that a leptonic origin is also consistent with the data [8]. Another interesting case study is the composite SNR G0.9+0.1 near the Galactic center [9] which can also be fitted using the leptonic model [8]. In this instance, the major contribution comes from ICS off optical photons while the $\gamma$-ray spectrum exhibits a "universal" cutoff in the VHE regime due to the Klein-Nishina effect. If this modelling is correct, GLAST observations can be used to probe the ISRF in the Galactic center. Observations of SNRs by GLAST will be vital in distinguishing between leptonic or hadronic scenarios as their predictions for the spectral shape in the GeV energy range are distinctly different.

A new calculation of the ISRF shows that it is more intense than previously thought, especially in the inner Galaxy where the optical and infrared photon density exceeds that of the cosmic microwave background (CMB) by a factor of 100 [8]. For a source in the inner Galaxy, properly accounting for inverse Compton energy losses flattens the electron spectrum in the source compared to the case of pure synchrotron energy losses [10]. This effect leads to a flatter intrinsic $\gamma$-ray spectrum at the source. On the other hand, the intense ISRF also leads to $\gamma\gamma$-attenuation which starts at much lower energies than for the CMB alone [11]. For VHE $\gamma$-ray sources located in the inner Galaxy, the attenuation effects should be seen for energies $\sim$30 TeV [12].

A new class of VHE $\gamma$-ray sources and thus CR accelerators, close binaries, has been recently found by HESS [13]. The observed orbital modulation due to the $\gamma\gamma$-attenuation on optical photons of a companion star testifies that the VHE $\gamma$-ray emission is produced near the compact object. Such an effect has been predicted long ago in a series of papers [14], where the light curves were calculated for binaries which were suspected to be VHE emitters at that time, like Cyg X-3. The phase of the maximum of the emission depends on the eccentricity of the orbit and its orientation; for orbital parameters of LS 5039 it is about 0.7 [14], in agreement with observations. For a recent discussion of the orbital modulation in LS 5039, see [15]. GLAST observations in the GeV–TeV range will be the key to understanding the emission mechanism(s).

## 3 Propagation of cosmic rays and diffuse gamma-ray emission

### 3.1 Particle propagation near the sources

Diffuse $\gamma$-ray emission in the TeV energy range has been recently observed by HESS [16] from the Galactic center. The emission clearly correlates with the gas column density as traced by CS. If this emission is associated with a relatively young SNR, say Sgr A East, observation of the individual clouds will tell us about CR propagation there. A simple back-of-the-envelope calculation shows that if the SNR age is <10 kyr and the shock speed is $<10^4$ km/s, the shell size should be <100 pc, while the emission is observed from distant clouds up to 200 pc from the Galactic center. The emission outside the shock, therefore, has to be produced by protons which were accelerated by the shock and left it some time ago. The spectrum of such particles can be approximated by a $\delta$-function in energy which depends on the SNR age; the resulting $\gamma$-ray spectrum is essentially flatter than expected from a power-law proton spectrum in the shell (Figure 1) [17, 18]. Observations of individual clouds in the GLAST energy range will be a direct probe of this model and thus of proton acceleration in SNRs.

Milagro has recently observed the diffuse emission at 12 TeV from the Cygnus region [19]. The observed emission, after subtraction of point sources correlates with gas column density. The VHE $\gamma$-ray flux is found to be larger than predicted by the conventional and even optimized model tuned to fit the GeV excess [25]. This may imply that freshly accelerated particles interact with the local gas, but other possibilities such as ICS or unresolved point sources can not be excluded on the base of Milagro observations alone.

These observations show that the diffuse emission exists even at VHE energies and variations in brightness are large. A contribution of Galactic CRs to the diffuse emission in this energy range is still significant. Observations of the diffuse emission may be used to study CR propagation and their penetration into molecular clouds. GLAST is ideally suited to address these issues.

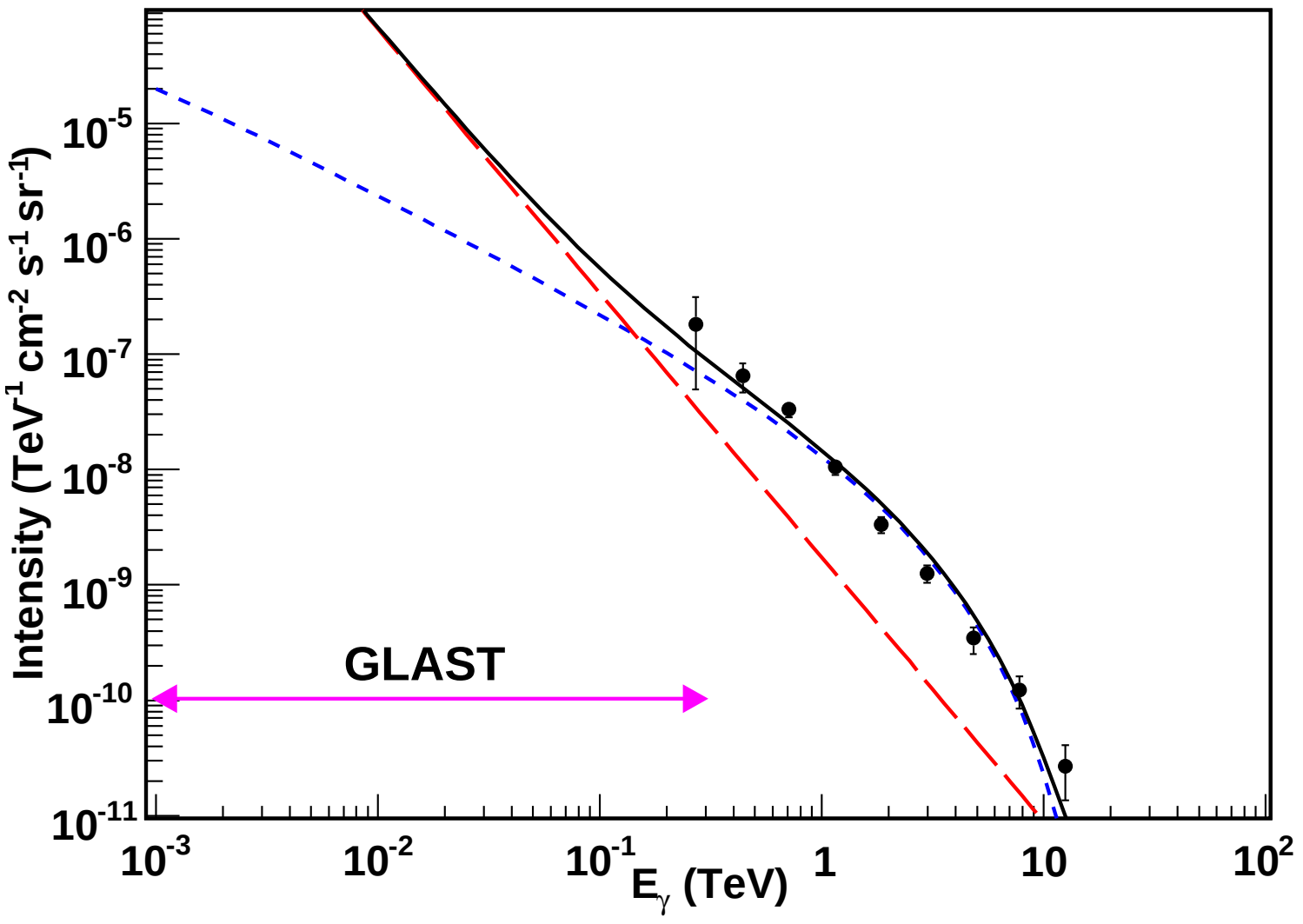

Figure 1: The spectrum of $\gamma$-rays [18] from the gas clouds outside of the SNR shell (monoenergetic protons of 25 TeV, dots) and from the shell (power-law with index $-2.29$, dashes); normalisations are arbitrary. The solid line is the total spectrum. Data: HESS observations of the Galactic center ridge [6].

### 3.2 Galactic cosmic rays

Propagation in the ISM changes the initial spectra and composition of CR species due to spallation, energy losses, energy gain (e.g., diffusive reacceleration), and other processes (e.g., diffusion, convection). The destruction of primary nuclei via spallation gives rise to secondary nuclei and isotopes which are rare in nature (i.e., Li, Be, B), antiprotons, pions and kaons that decay producing secondary leptons and $\gamma$-rays. Studies of stable secondary nuclei (Li, Be, B, Sc, Ti, V) allow the ratio (halo size)/(diffusion coefficient) to be determined and the incorporation of radioactive secondaries ($^{10}$Be, $^{26}$Al, $^{36}$Cl, $^{54}$Mn) is used to find the diffusion coefficient and the halo size separately. For a recent review on CR propagation see [2].

Measurement of the B/C ratio with a single instrument and in a wide energy range is long overdue. The best data >0.8 GeV/n to-date are those taken by the HEAO-3 experiment more than 25 years ago, while modern spacecraft, e.g., ACE, provide high quality data at low energies 150–450 MeV/n [20]. The sharp maximum in the B/C ratio observed at ∼1 GeV/n is difficult to explain in a physical model and has been long debated; it may well be an instrumental artefact. On the other hand, the high energy tail of the B/C ratio is sensitive to the rigidity dependence of the diffusion coefficient and thus its accurate measurement can be used to distinguish between models of CR propagation in the ISM. The PAMELA has the capability of measuring the B/C ratio in the energy range 100 MeV/n – 250 GeV/n and will address both issues.

The majority of CR antiprotons observed near the Earth are secondaries produced in collisions of CRs with interstellar gas. Because of the kinematics of this process, the spectrum of antiprotons has a unique shape distinguishing it from other CR species. It peaks at ∼2 GeV decreasing sharply toward lower energies. Because of their high production threshold and the unique spectral shape antiprotons can be used to probe CR propagation in the ISM and the heliosphere, and to test the local Galactic average proton spectrum (for a discussion and references see [21]). Because the $pp$ (and $\bar{p}p$) total inelastic cross section is ten times smaller than that of carbon, the ratio $\bar{p}/p$ can be used to derive the diffusion coefficient in a much larger Galactic volume than the B/C ratio. The CR $\bar{p}$ spectrum may also contain signatures of exotic processes, such as, e.g., WIMP annihilation. However, currently available data (mostly from BESS flights [22]) are not accurate enough, while published estimates of the expected flux differ significantly; in particular, the reacceleration model underproduces antiprotons by a factor of ∼2 at 2 GeV [21]. Secondary CR positrons are produced in the same interactions as antiprotons and are potentially able to contribute to the same topics. Accurate measurements of the CR $e^+$ spectrum may also reveal features associated with the sources of primary positrons, such as pulsars. During its lifetime, PAMELA will measure CR antiprotons and positrons in the energy range 50 MeV – 250 GeV with high precision [23]. Independent CR $\bar{p}$ measurements below ∼3 GeV will be provided by the new BESS-Polar instrument scheduled to fly in December of 2007 [21].

The diffuse emission is a tracer of interactions of CR particles in the ISM and is produced via ICS, bremsstrahlung, and $\pi^0$-decay. The puzzling excess in the EGRET data above 1 GeV [24] relative to that expected has shown up in all models that are tuned to be consistent with local nucleon and electron spectra [25]. The excess has shown up in all directions, not only in the Galactic plane. If this excess is not an instrumental artefact, it may be telling us that the CR intensity fluctuates in space which could be the result of the stochastic nature of supernova events. If this is true, the local CR spectra are not representative of the local Galactic average. Because of the secondary origin of CR antiprotons, their intensity fluctuates less than that of protons and $\bar{p}$ measurements can be used instead to derive the *average* local intensity of CR protons. Interestingly, a model based on a renormalized CR proton flux (to fit antiprotons) and a CR electron flux (using the diffuse emission itself), the so-called optimized model, fits the all-sky EGRET data well [25] providing a feasible explanation of the GeV excess. The GLAST observations of the diffuse emission will be able to resolve this puzzle. On the other hand, accurate measurements of CR antiprotons by PAMELA can be used to test the CR fluctuation hypothesis.

## 4 Cosmic rays in the heliosphere

Interestingly, GLAST will be able to trace the CRs in the heliosphere as well. The ICS of CR electrons off solar photons produces $\gamma$-rays with a broad distribution on the sky contributing to a foreground that would otherwise be ascribed to the Galactic and extragalactic diffuse emission [26]. Observations by GLAST can be used to monitor the heliosphere and determine the electron spectrum as a function of position from distances as large as Saturn's orbit to close proximity of the Sun, thus enabling unique studies of solar modulation. A related process is the production of pion-decay $\gamma$-rays in interactions of CR nuclei with gas in the solar atmosphere [27]. The albedo $\gamma$-rays will be observable by GLAST providing a possibility to study CR cascade development in the solar atmosphere, deep atmospheric layers, and magnetic field(s). The original analysis of the EGRET data assumed that the Sun is a point source and yielded only an upper limit [28]. However, a recent re-analysis of the EGRET data [29] has found evidence of the albedo (pion-decay) and the extended ICS emission. The maximum likelihood values appear to be consistent with the predictions.

GLAST will also be able to measure CR electrons directly. It is a very efficient electron detector able to operate in the range between $\sim$20 GeV and 2 TeV [30]. The total number of detected electrons will be $\sim 10^7$ per year. Accurate measurements of the CR electron spectrum are very important for studies of CR propagation and diffuse $\gamma$-ray emission. There is also the possibility to see the features associated with the local sources of CR electrons.

The Moon emits $\gamma$-rays [28, 31] due to CR interactions in its rocky surface. Monte Carlo simulations of the albedo spectrum using the GEANT4 framework show that it is very steep with an effective cutoff around 3 GeV and exhibits a narrow pion-decay line at 67.5 MeV [32]. The albedo flux below $\sim$1 GeV significantly depends on the incident CR proton and helium spectra which change over the solar cycle. Therefore, it is possible to monitor the CR spectrum at 1 AU using the albedo $\gamma$-ray flux. Simultaneous measurements of CR proton and helium spectra by PAMELA, and observations of the albedo $\gamma$-rays by GLAST, can be used to test the model predictions. Since the Moon albedo spectrum is well understood, it can be used as a standard candle for GLAST. Besides, the predicted pion-decay line at 67.5 MeV and the steep spectrum at higher energies present opportunities for in orbit energy calibration of GLAST.

I thank Troy Porter for useful suggestions. This work was supported in part by a NASA APRA grant.

# MEASUREMENTS OF COSMIC RAYS FLUXES WITH PAMELA

ROBERTA SPARVOLI [a,b] AND VALERIA MALVEZZI [a]
FOR THE PAMELA COLLABORATION

[a] *INFN, Sezione di Roma "Tor Vergata", via della Ricerca Scientifica 133, Roma, Italy*

[b] *Dipartimento di Fisica, Università di Roma "Tor Vergata" and INFN, via della Ricerca Scientifica 133, Roma, Italy*

**Abstract**

On the $15^{th}$ of June 2006 the PAMELA experiment, mounted on the Resurs-DK1 satellite, was launched from the Baikonur cosmodrome and it has been collecting data since July 2006. PAMELA is a satellite-borne apparatus designed to study charged particles in the cosmic radiation, to investigate the nature of dark matter, measuring the cosmic-ray antiproton and positron spectra over the largest energy range ever achieved, and to search for antinuclei with unprecedented sensitivity. In this paper we will present the status of the apparatus after one year in orbit; furthermore, we will discuss the PAMELA cosmic-ray measurements capabilities.

## 1 Introduction

PAMELA (*Payload for Antimatter/Matter Exploration and Light-nuclei Astrophysics*) is a satellite-borne experiment which has been designed to study charged cosmic rays, in particular having been optimized to reveal the rare antiparticle component of the cosmic radiation. Its principal aim is the measurement of the energy spectra of antiprotons and positrons with high precision and over a wide range, but also other more common components like protons, electrons and light nuclei will be thoroughly investigated. This will allow to look for evidences of the existence of dark matter, to check the correctness of cosmic-ray propagation models and also to test for the possible presence of antinuclei by direct detection.

The instrument was launched into space on June $15^{th}$ 2006 from the Baikonur cosmodrome in Kazakhstan. The apparatus is installed inside a pressurized container attached to the Russian Resurs-DK1 earth observation satellite. The satellite orbit is elliptical and semi-polar, with an altitude varying between 350 km and 600 km, at an inclination of 70°. The mission is foreseen to last for at least three years.

In this paper we will present a status report of the mission after one year of flight, and we will show preliminary results about particle identification and cosmic ray fluxes measurements.

## 2 The scientific case

The primary scientific goal of the PAMELA experiment is the study of the antimatter component of the cosmic radiation, in order: 1. to search for antinuclei, in particular antihelium; 2. to search for evidence of annihilations of dark matter particles by accurate measurements of the antiproton and positron energy spectra; 3. to test cosmic-ray propagation models through precise measurements of the antiproton and positron energy spectra and precision studies of light nuclei and their isotopes.

Concomitant goals include: 1. a study of solar physics and solar modulation during the $24^{\mathrm{th}}$ solar minimum; 2. a study of trapped particles in the radiation belts. The semipolar orbit allows PAMELA to investigate a wide range of energies for the different antiparticles, particles and nuclei. Three years of data taking will provide unprecedented statistics in these energy ranges with no atmospheric overburden, consenting to greatly reduce the systematic errors of the balloon measurements, and to explore for the first time the $\bar{p}$ and $e^+$ energy spectra well beyond the present limit of experimental data (~40 GeV).

## 3 The PAMELA apparatus

The core of the PAMELA instrument, as sketched in figure 1, is a permanent magnet spectrometer equipped with a silicon tracker. The tracking system consists of six 300 $\mu$m thick silicon sensors segmented into micro-strips on both sides. The mean magnetic field inside the magnet cavity is 0.43 T with a value of 0.48 T measured at the centre[1].

A sampling electromagnetic calorimeter, composed of W absorber plates and single-sided, macro strip Si detector planes is mounted below the spec-trometer. A scintillation shower tail catcher and a neutron detector made of $^{3}He$ counters enveloped in polyethylene moderator complete the bottom part of the apparatus[2].

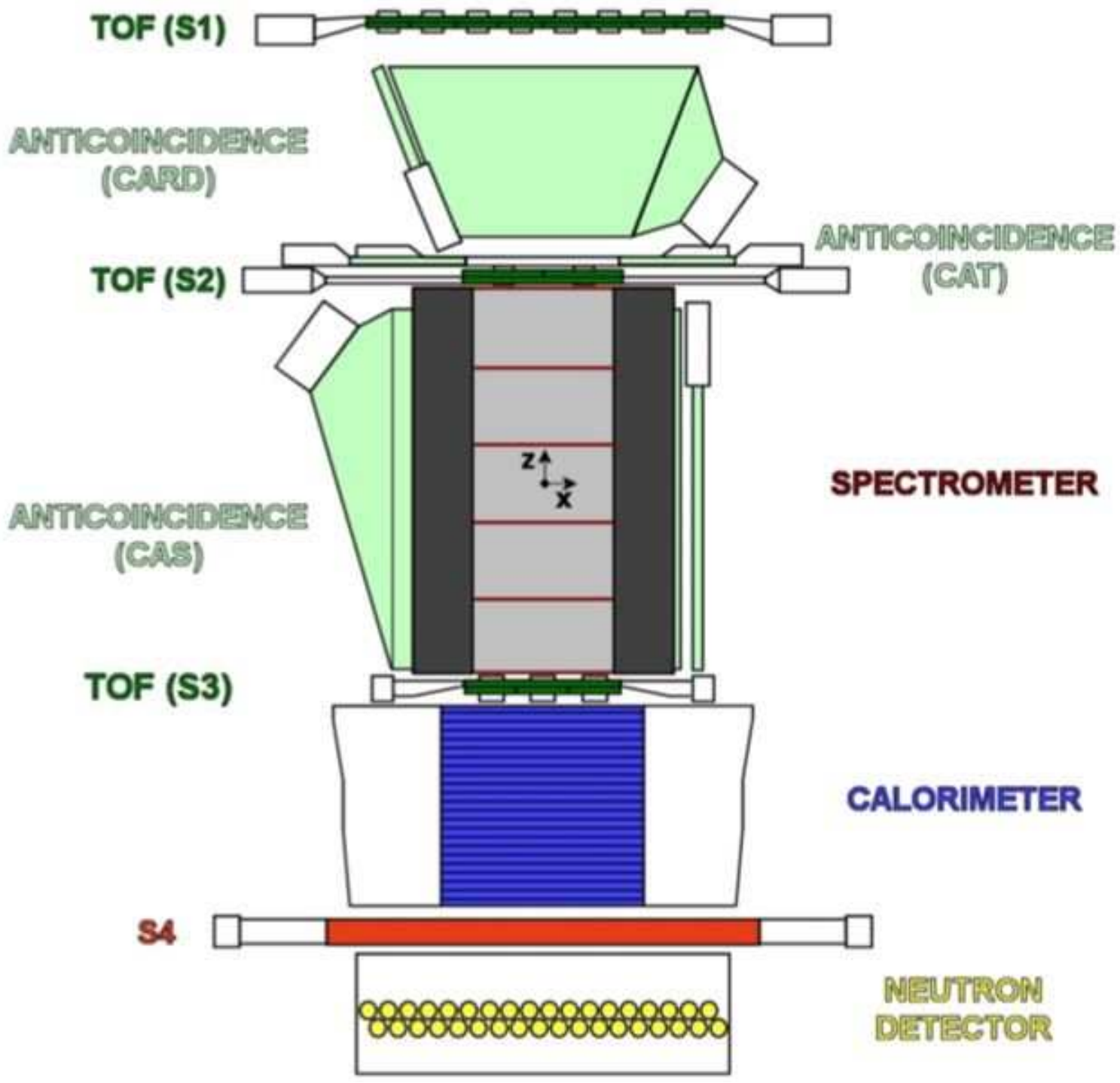

Figure 1: Schematic overview of the PAMELA apparatus. The detector is approximately 1.3 m high, has a mass of 470 kg and an average power consumption of 355W. The magnetic field lines inside the spectrometer cavity are oriented along the y direction. The average magnetic field is 0.43 T.

A Time-of-Flight (ToF) system, made of three double-layers of plastic scintillator strips, and an anticoincidence system complement the apparatus. Particles trigger the instrument when crossing the ToF scintillator paddles. The ToF system also measures the absolute value of the particle charge and flight time crossing its planes. In this way downgoing particles can be separated from up-going ones[3].

Particles not cleanly entering the PAMELA acceptance are identified by the anticounter system[4]. Then, the rigidities of the particles are determined by the magnetic spectrometer. Thus, positively and negatively charged particles can be identified. The final identification (i.e. positrons, electrons, antiprotons, etc.) is provided by the combination of the calorimeter and neutron detector[5] information plus the velocity measurements from the ToF system and ionization losses in the tracker system at low momenta.

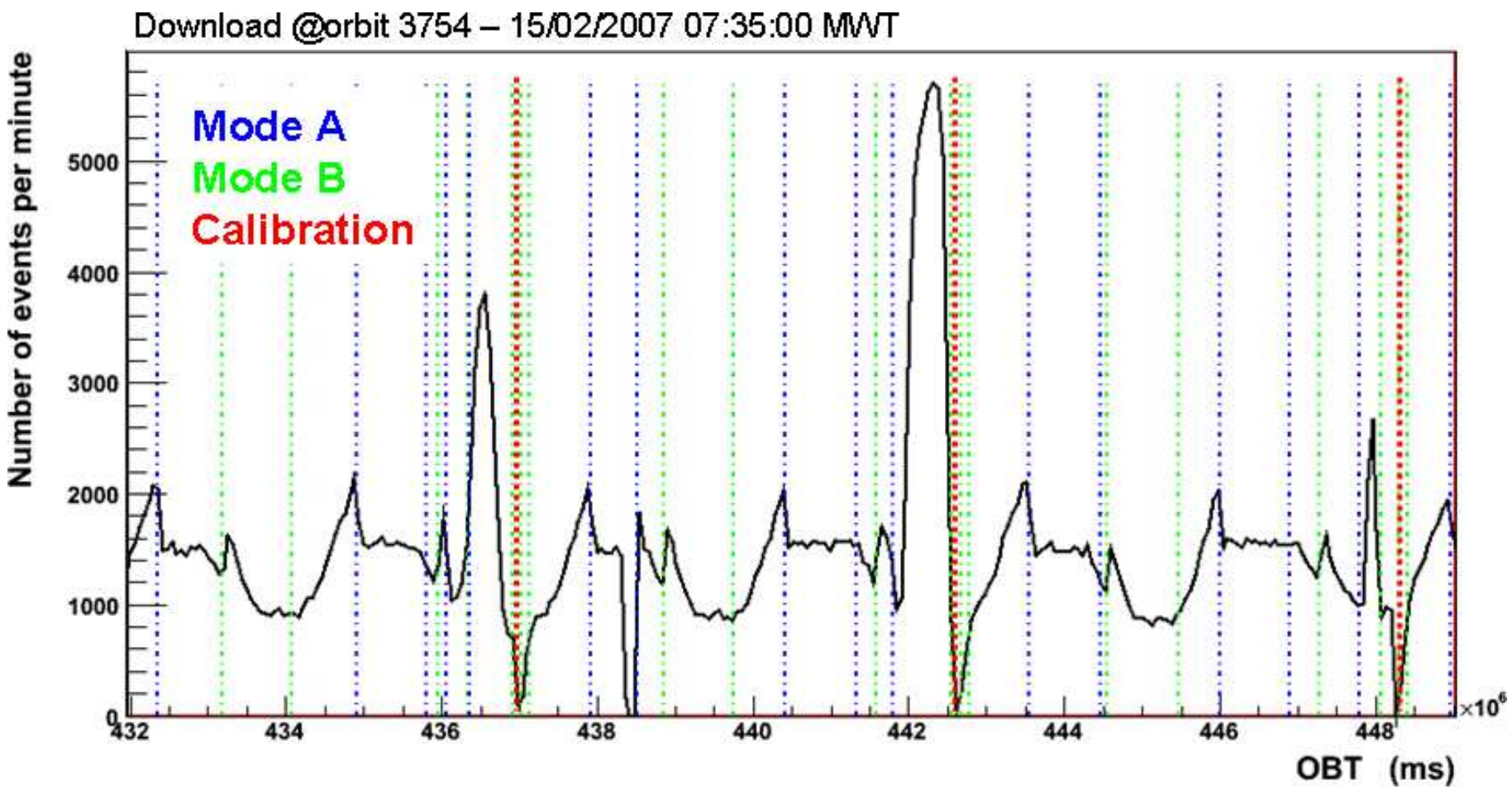

Figure 2: Sequence of acquisition modes during a typical orbit of PAMELA. Mode A refers to high-radiation regions and it is typical for polar regions and SAA, whilst mode B holds for low radiation regions like the equator. The calibration is done at every ascending node.

Electrons and protons are distinguished by comparing the particle patterns and energy losses inside the calorimeter. Additional hadron-rejection power is provided by the neutron detector and this increases as the energy increases.

The detector is approximately 120 cm high, has a mass of about 470 kg and a power consumption of 355 W. A very detailed description of the PAMELA detector along with an overview of the entire mission can be found in [6].

## 4 PAMELA in-flight operations

PAMELA is hosted by the Russian earth-observation satellite Resurs-DK1 that, onboard a Soyuz vehicle, was successfully launched in space on June $15^{th}$ 2006 from the Baikonur (Kazakhstan) cosmodrome.

The Resurs-DK1 satellite is manufactured by the Russian space company TsSKB Progress to perform multispectral remote sensing of the earth's surface and acquire high-quality images in near real-time. The satellite has a mass of 6.7 tonnes and a height of 7.4 m. During launch and orbital manoeuvres, the PC is secured against the body of the satellite. During data-taking it is swung up to give PAMELA a clear view into space.

On June $21^{st}$ 2006 PAMELA was switched on for the first time. After a few weeks of commissioning, during which several trigger and hardware configurations were tested, PAMELA has been in a nearly continuous data taking mode since July $11^{th}$. Until June 2007, the total acquisition time has been $\sim$ 300 days, for a total of $\sim$610 million collected events and 5.4 TByte of down-linked raw data.

All in-flight operations are handled by the PSCU (PAMELA Storage and Control Unit). The PSCU manages the data acquisition and other physics tasks and continuously checks for proper operation of the apparatus. The data acquisition is segmented in *runs*, defined as continuous period of data taking with constant detector and trigger configurations. The duration of a *run* is determined by the PSCU according to the orbital position. Two acquisition modes are implemented, for high- (radiation belts and polar regions - MODE A) and low- (equatorial region - MODE B) radiation environments. The run configuration, in both acquisition modes, and the criterion to switch between low- and high-radiation environments can be varied from ground. The main PAMELA trigger conditions are defined by coincident energy deposits in the scintillator ToF layers. The high-radiation trigger environment uses only information from the S2 and S3 scintillators while the low-radiation one also from the S1 scintillators. Figure 2 shows the PAMELA trigger rate as a function of orbital position for a typical orbit. The coloured lines define the regions where the different acquisition modes are alternated. From the figure it can been seen that the average trigger rate of the experiment is $\sim$ 25 Hz, varying from $\sim$ 20 Hz at the equatorial region to $\sim$ 30 Hz at the poles. Furthermore also the region where the satellite crosses the inner proton radiation belt, i.e. the South Atlantic Anomaly, can be clearly seen as a significant increase in the trigger rate. Figure 3 presents instead a map of PAMELA triggers as integrated over many orbits of PAMELA. The colours are proportional to the trigger rate, as indicated by the palette. Again the South Atlantic Anomaly is clearly seen.

The control of the experiment from ground is performed via two different type of commands: *macrocommands*, which talk directly to the PAMELA PSCU, and *telecommands*, which are commands sent instead to Resurs-DK1 handling main power lines. Hundreds of parameters (current and voltage values, thresholds, switching on and off of parts of the subdetectors, trigger configurations, ...) are modifiable through macrocommands therefore the system is extremely flexible so to be able to meet any unknown and unpredictable situation in flight.

The average fractional live time of the experiment exceeds 70%. During the entire PAMELA observational time to far some error conditions (approximately one per week) occurred, mainly attributable to anomalous electronics conditions in the detector electronics. Every time the PSCU was able to recover the system functionality and continue the acquisition. The thermal profile of the instrument has been very stable and no power-off due to over-temperature occurred. Furthermore, no radiation dose effects have been observed in the PAMELA sub-detectors. Indeed, all sub-detectors are behaving nominally.

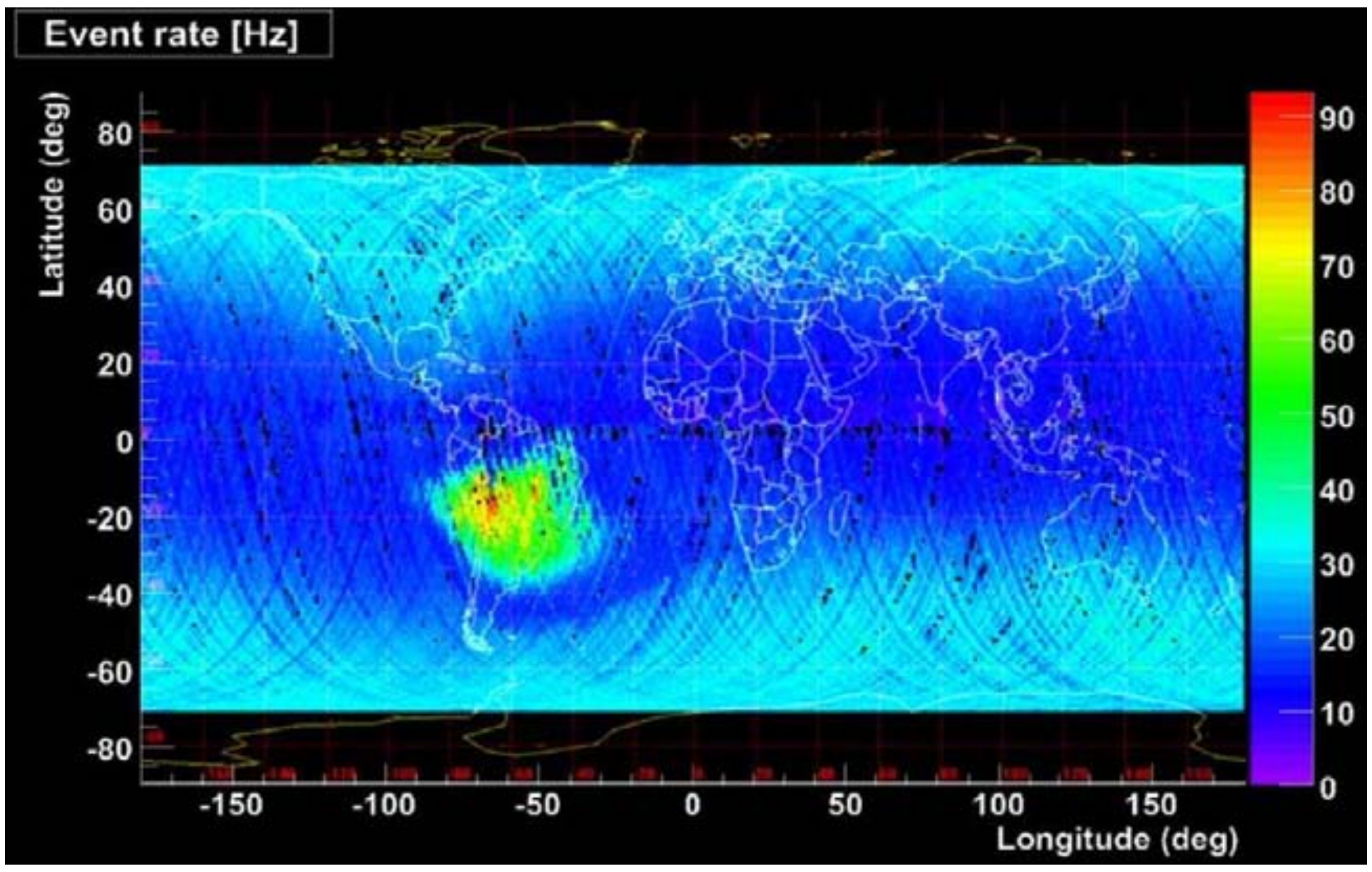

Figure 3: Map of the PAMELA trigger rate summed over hundreds of orbits. The colour along the orbit lines is proportional to the rate intensity, as indicated by the palette on the right.

### 4.1 Ground-data processing

The ground segment of the Resurs-DK1 system is located at the Research Center for Earth Operative Monitoring (NTs OMZ) in Moscow, Russia. The reception antenna at NTs OMZ is a parabolic reflector of 7 m diameter (see figure 4, left), equipped with an azimuth elevation rotation mechanism, and has two frequency multiplexed radio channels. The Resurs-DK1 radio link towards NTs OMZ is active between 4 and 6 times a day.

About 14 GB of PAMELA data are transferred to ground every day. The PAMELA data processing scheme is shown in figure 4 (right). Data received from PAMELA are collected by a dataset archive server which provides a secure connection from the Ground Station to the PAMELA Ground Segment. The downlinked data are then transmitted to a server dedicated to data processing for instrument monitoring and control, and are also written onto magnetic tapes for long-term storage. A pre-processing software calculates the downlink session quality (the error probability per bit), removes from the files all transport headers and footers and prepares the data for unpacking. At the same time information about received data files is stored into a MySQL data base; the access to the data base is made through a web interface, thus giving the possibility to the operators to easily check up data and faulty downlink sessions. In case of low quality transmission, a given downlink can be assigned for retransmission up to several days after the initial downlink.

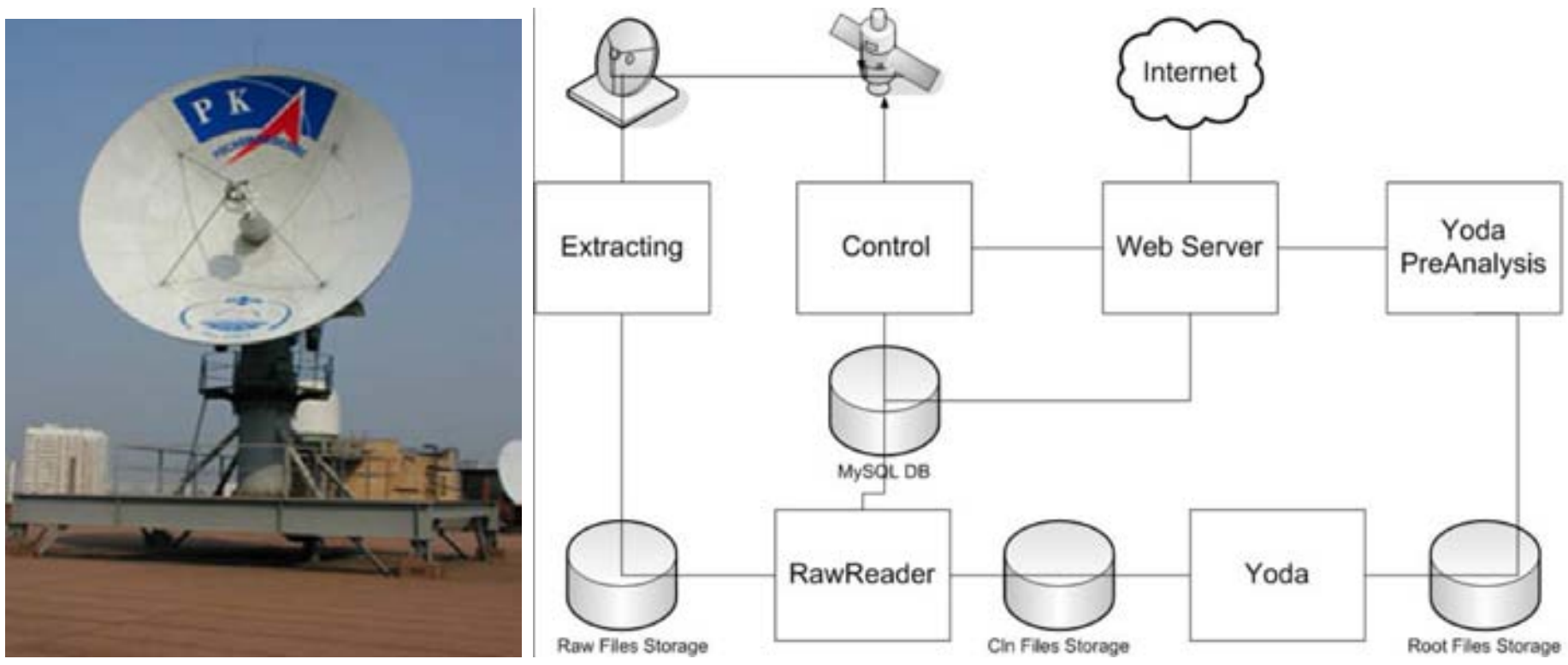

Figure 4: Left: main antenna in NTs OMZ. Right: Data processing scheme at PAMELA Groung Segment in NTs OMZ.

Data are then processed by a 1st Level software, a ROOT[1] based C++ program. This software unpacks all different structures creating the various trees (event, calibration, housekeeping, orbital information, etc). A quicklook applicative software monitors the status of housekeeping and physics data in order to allow local and remote (web based) assessment of the status of the mission. Short term programming and telecommand/macrocommand issuing is based on the result of the quicklook: for instance, in case of Solar Particle Event the number of allocated on-board memory may be increased.

After this level of data analysis, both raw and 1st Level processed data are moved through a normal internet line to the main storage centre in Eastern Europe, which is located at MEPHI (Moscow, Russia). From here, the GRID infrastructure is used to move raw data to the main storage and analysis centre of the PAMELA Collaboration, located at CNAF (Bologna, Italy), a specialized computing centre of INFN (Italian National Institute of Nuclear Physics). Here data are accessible to all various institutions within the PAMELA collaboration.

[1] http://root.cern.ch

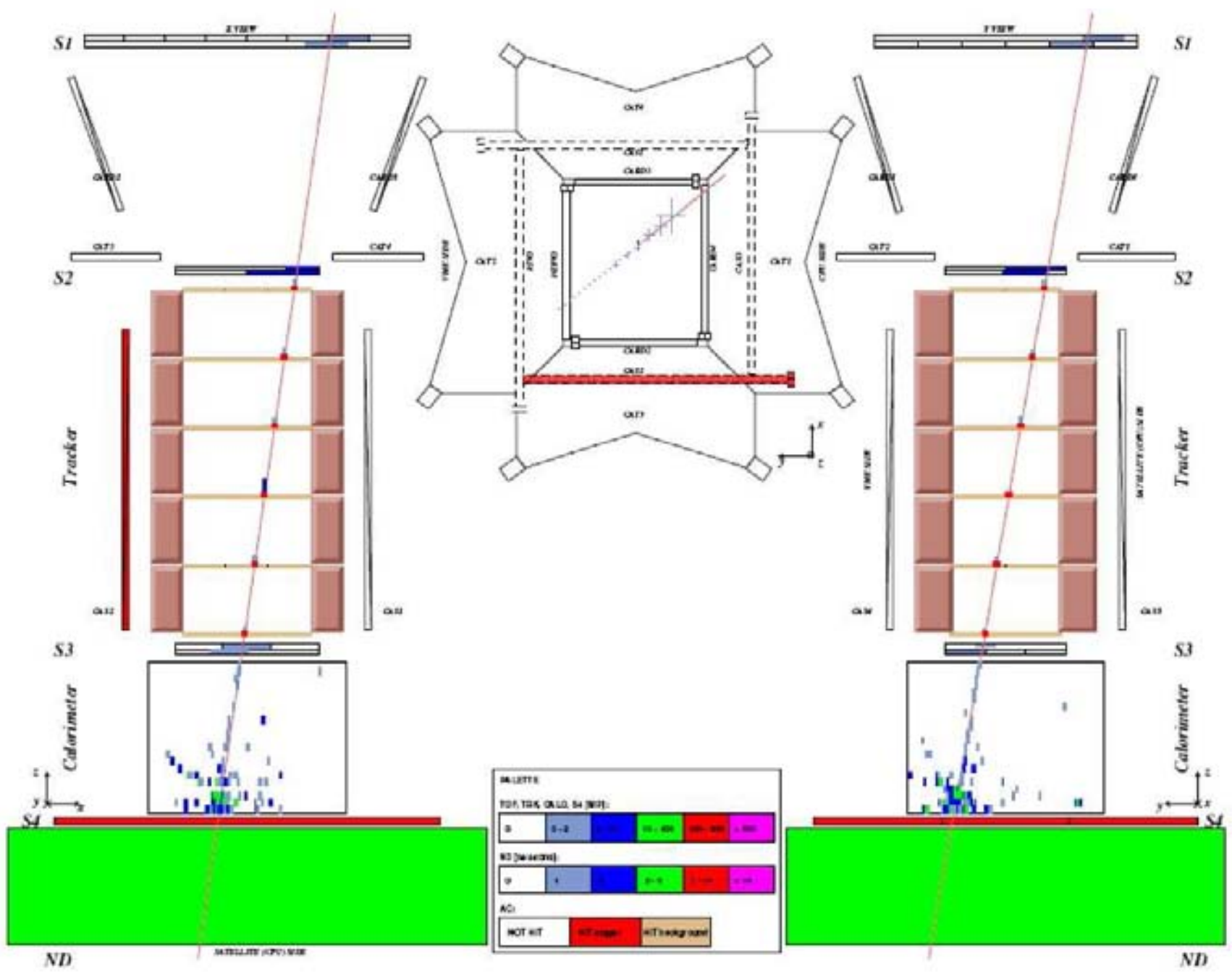

Figure 5: The event display a $\sim$ 41 GV interacting antiproton. The bending (x) and non-bending (y) views are shown on the left and on the right, respectively (plane 19 of the calorimeter x-view was malfunctioning.). A plan view of PAMELA is shown in the center. The signal as detected by PAMELA detectors are shown along with the particle trajectory (solid line) reconstructed by the fitting procedure of the tracking system.

## 5 Particle selection and preliminary cosmic ray measurements

The central components of PAMELA are a permanent magnet and a tracking system composed of six planes of silicon sensors, which form a magnetic spectrometer. This device is used to determine the rigidity $R = pc/Ze$ and the charge of particles crossing the magnetic cavity. The rigidity measurement is done through the reconstruction of the trajectory based on the impact points on the tracking planes and the resulting determination of the curvature due to the Lorentz force. The direction of bending of the particle (i.e. the discrimination of the charge sign) is the key method used to separate matter from antimatter.

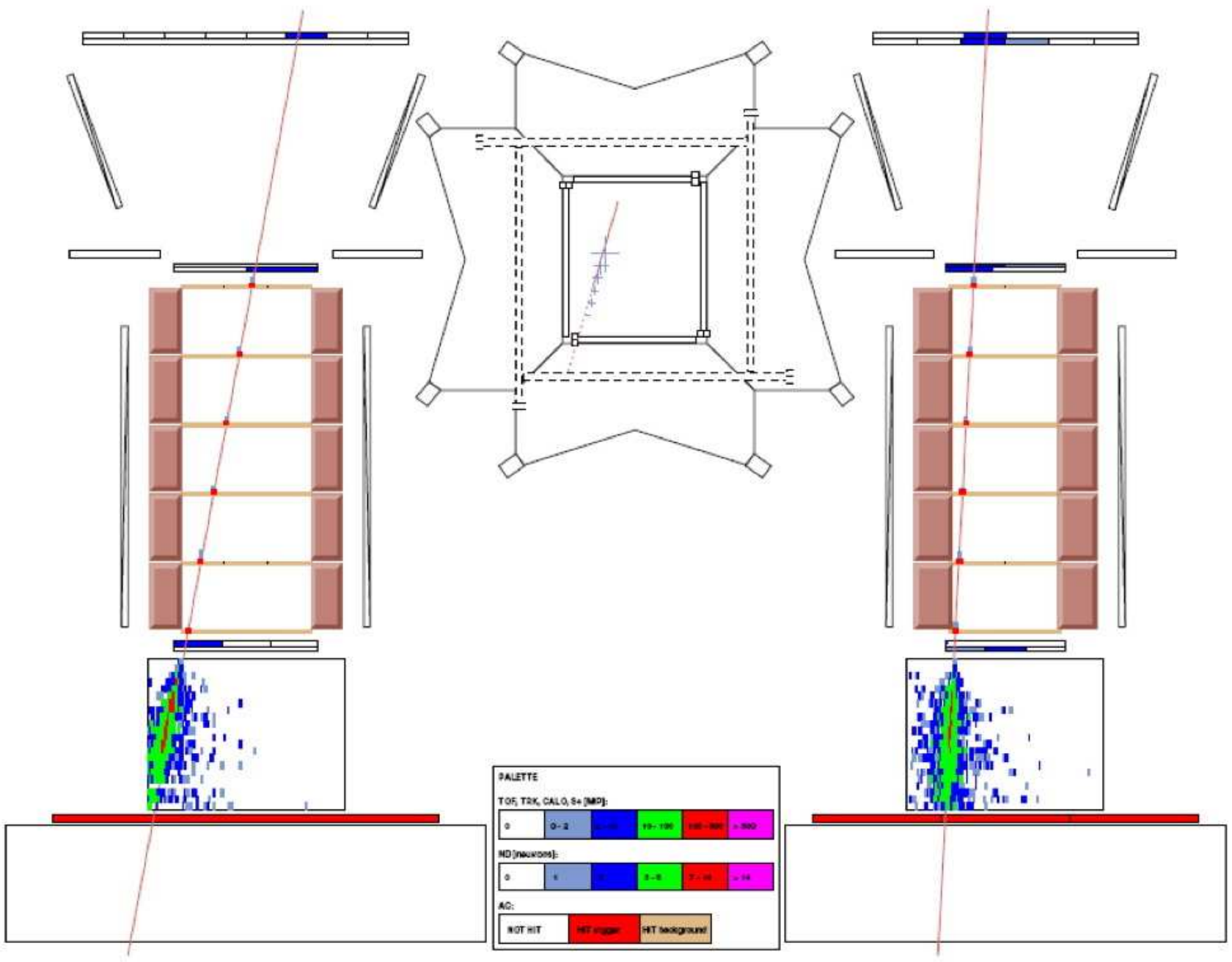

Figure 6: The event display a $\sim$ 70 GV positron. The bending (x) and non-bending (y) views are shown on the left and on the right, respectively (plane 19 of the calorimeter x-view was malfunctioning.). A plan view of PAMELA is shown in the center. The signal as detected by PAMELA detectors are shown along with the particle trajectory (solid line) reconstructed by the fitting procedure of the tracking system.

Figures 5 and 6 show respectively $\sim$ 41 GV negatively-charged interacting hadron identified as an antiproton and a a $\sim$ 70 GV positively-charged electromagnetic particle identified as a positron, crossing PAMELA and triggering the acquisition. The particle path, reconstructed by the tracking system, is extrapolated to the other detectors. It is clearly seen that the track is consistent with the energy deposits both in the scintillator ToF system and in the calorimeter. A different signature in the neutron detector - due to the different nature (hadronic or leptonic) of the shower produced in the calorimeter - can be clearly noticed. Indeed, additional hadron-rejection power is provided by the neutron detector and this increases as the energy increases.

Protons and electrons dominate the positively and negatively charged components of the cosmic radiation, respectively. Hence, positrons must be iden-tified from a background of protons and antiprotons from a background of electrons. To achieve its scientific goals, the PAMELA system must separate electrons from hadrons at a level of $10^5 - 10^6$. Much of this separation must be provided by the calorimeter, i.e.: electrons must be selected with an acceptable efficiency and with as small a hadron contamination as possible.

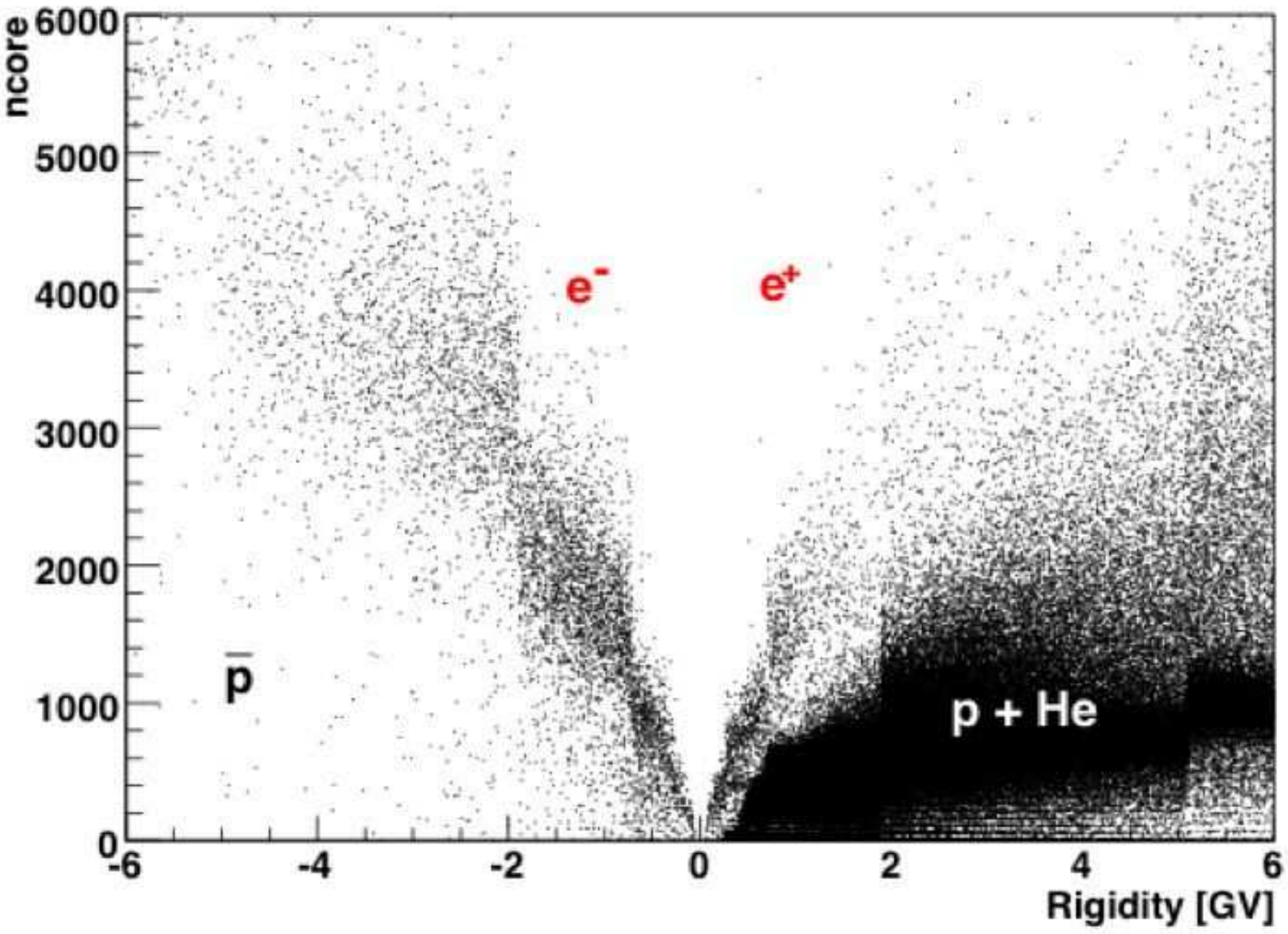

Figure 7: Distribution of the topological variable "ncore" as function of the rigidity for a sample of events. Electrons, positrons, anti–protons and protons/helium bands can be distinguished.

A complex set of variables is used to separate electromagnetic and hadronic showers in the calorimeter. These variables characterize the shower inside the calorimeter taking into account its starting point, its longitudinal and transverse profiles, its topological development and the energy release compared to the one given by the tracking system.

Figure 7 represents the particle discrimination capabilities using a topological variable as function of the rigidity as measured by the tracking system. This variable takes into account the number of hits along the track till the plane of maximum calculated using the momentum of the particle and assuming an electromagnetic shower. Hence this variable grows when the shower inside the calorimeter started in the first planes and is collimated along the track while it tends to be smaller for non interacting or hadronic particles. Electron and positrons bands can be clearly distinguished from anti–protons and protons plus helium bands.

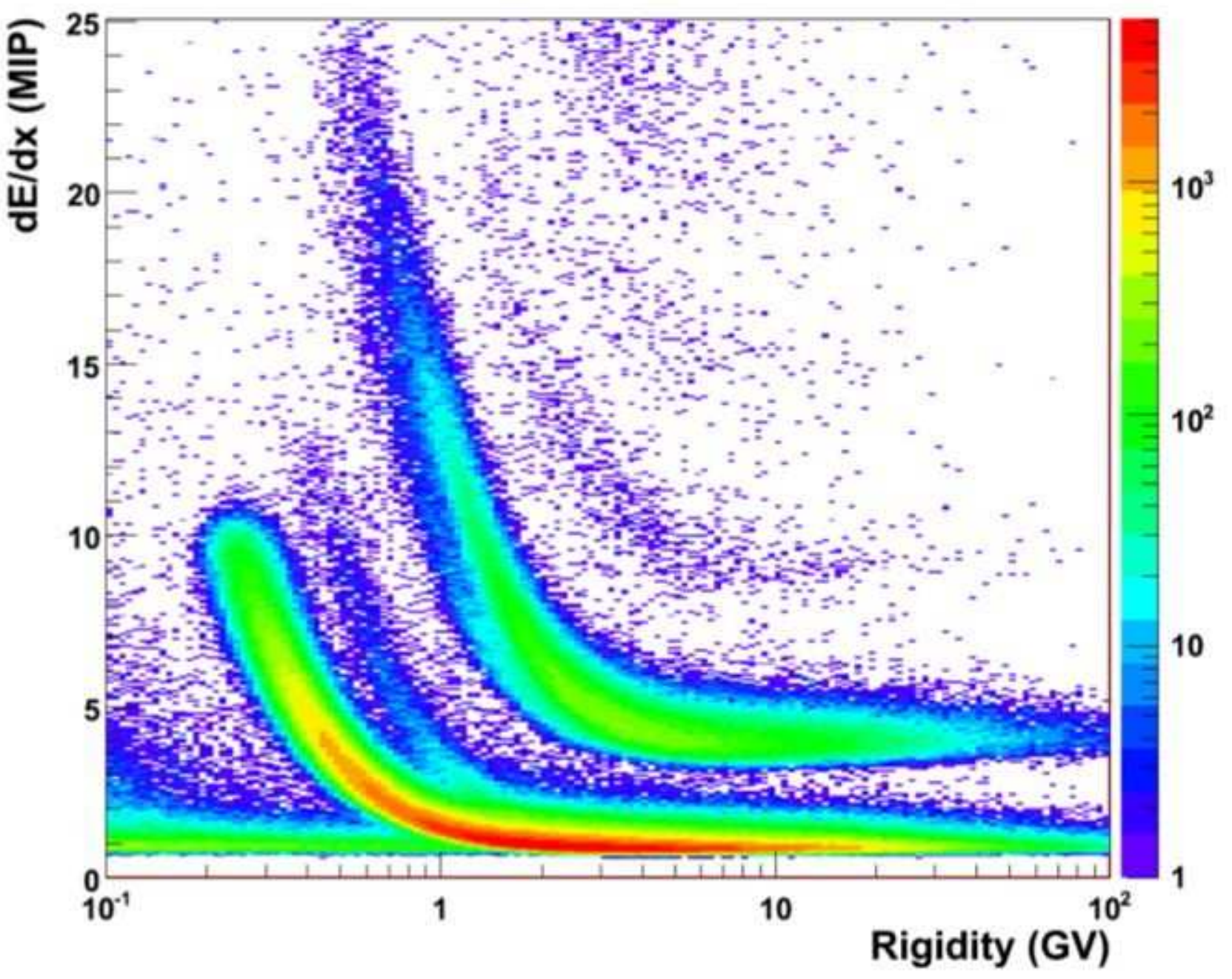

Figure 8: Distribution of the mean rate of energy loss in the silicon sensors of the tracking system, as a function of the particle rigidity, for a sample of positively charged cosmic rays. Moving from bottom-left to top-right, the following particle species can be recognized: $e^+$, p and d, $^3$He and $^4$He, Li, Be, B and C.

The PAMELA instruments is optimized for the detection of positrons and antiprotons, nevertheless three different sub-detectors (ToF, tracker and calorimeter) are able to identify light nuclei, with different efficiencies, resolutions and Z ranges. Such particle identification, together with the particle momentum measured by the spectrometer, will allow to reconstruct their energy spectra. Accurate simulations are in progress to evaluate systematic uncertainties resulting from the various correction factors needed to evaluate fluxes such as uncertainties in the determination of the geometry factor, spallation loss within the instrument, and the tracking efficiency as function of Z.

The tracking system can be used to determine the absolute value $Z$ of the charge, by multiple measurements of the mean rate of energy loss in the silicon sensors. Figure 8 shows the $Z$ discrimination capability of the tracking system. The spectrometer can contribute with a good charge resolution at least up to Be (when the single-channel saturation of the silicon sensors reduces the performances), and it is also able to perform isotopic discrimination for H and He at low rigidities.

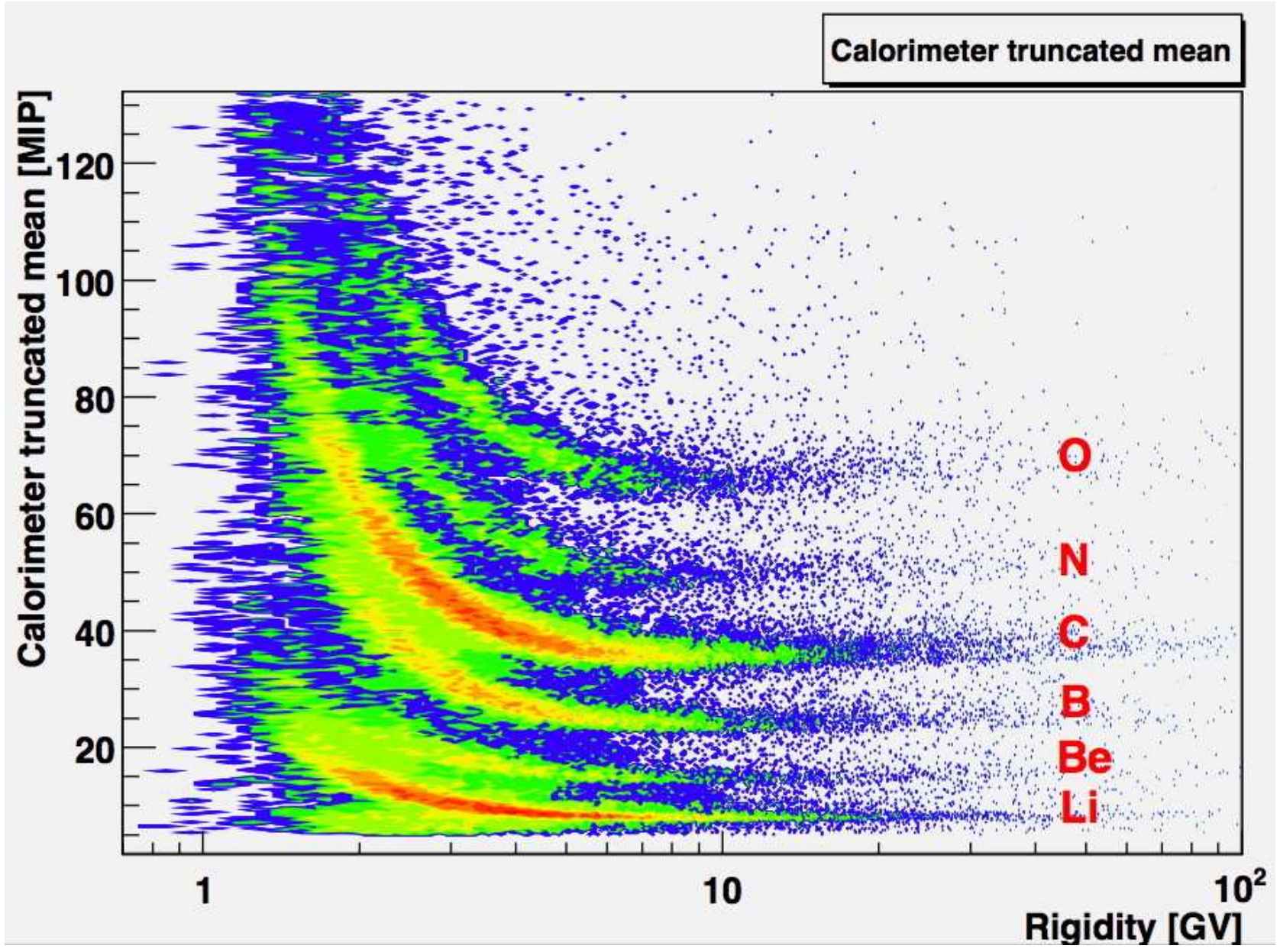

Figure 9: The ionization loss in the calorimeter with truncated mean using three points and at least four dE/dx measurements versus rigidity. A rough cut was used to throw away Helium nuclei which appear as a diffused region at about 2 GV and 5/6 MIP energy release.

The dE/dx measurement on the calorimeter silicon planes can be used to determine the charge of the incident nuclei too. The charge of the particles can be measured in the calorimeter by considering the energy released in the first plane of the detector which is not covered by tungsten plates. In presence of a reconstructed track, is possible to localize with precision the hit strips and to collect the charge.

More sophisticated methods can be used for nuclei not interacting in the first calorimeter layers. By determining the interaction plane, it is possible to use all the multiple energy losses in the planes preceding the interaction to derive the charge of the incident particle. An iterative algorithm taking into account the energy release along the track as given by the tracking system is able to recognize the interaction point inside the calorimeter. We used the three points with the smallest energy measurements to determine the average energy release, requiring the presence of at least four dE/dx measurements before the interaction. Figure 9 shows the charge bands for different nuclei from Lithium to Oxygen obtained with this method. Obviously the charge separation increases with the number of planes required, but the efficiency of the measurement decreases.

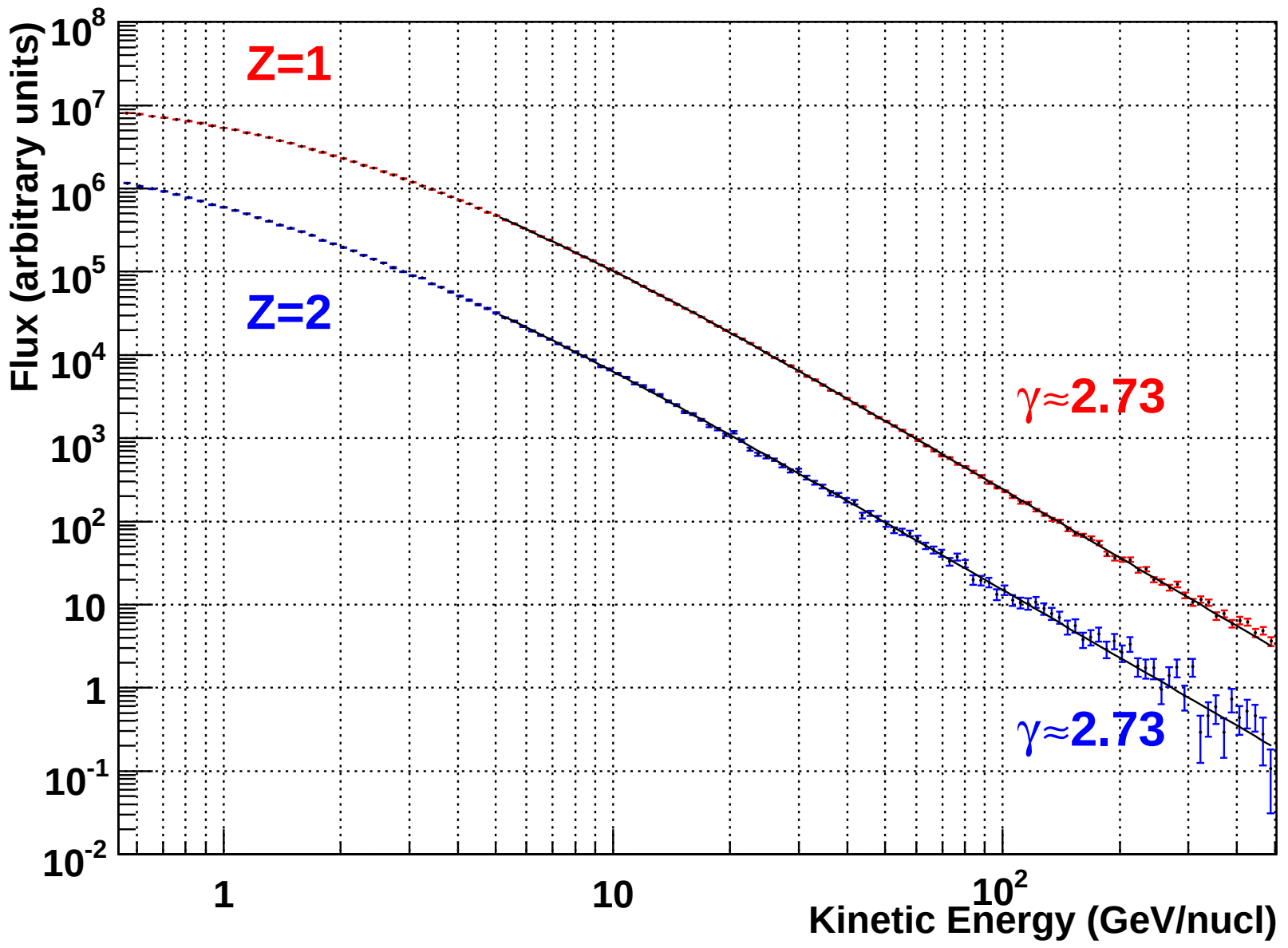

Figure 10: Preliminary reconstructed fluxes of H and He obtained by the combination of the particle identification given by all PAMELA detectors, and the momentum reconstruction performed by the magnetic spectrometer. Power-law fits are superimposed to the data: $\gamma$ indicates the spectral index of the power-law.

Finally figure 10 shows preliminary reconstructed fluxes of H and He - still in arbitrary units - obtained by the combination of the particle identification given by all PAMELA detectors, and the momentum reconstruction performed by the magnetic spectrometer. Fluxes are still in arbitrary units since we have not completed the estimation of the full detector efficiency and the possible systematic yet. We fitted the spectra with power-laws, obtaining spectral indexes consistent with expectations.

## 6 Conclusions

The PAMELA satellite experiment was successfully launched on the $15^{th}$ of June 2006. Detectors did not suffer any damage due to the launch and the experiment has been continuously taking data since then. Individual detectors are performing nominally; the instrument in-flight performance as well as its particle identification capabilities are consistent with design and ground tests, allowing for precise measurement of cosmic-ray spectra over a wide energy range.

Several thousand events have been identified as positrons and hundreds of events as antiprotons. Besides selection of charge one particles, PAMELA is able to identify light nuclei particles, up at least to Oxygen, using the ionization losses in the calorimeter, ToF and tracker systems. More than 80000 particles heavier than helium reached PAMELA instrument in its first year of life.

# LIGHT-NUCLEI IDENTIFICATION WITH PAMELA TIME-OF-FLIGHT: CALIBRATION AND PRELIMINARY IN-FLIGHT RESULTS

R. CARBONE [a,b] FOR THE PAMELA COLLABORATION

[a] *Dipartimento di Fisica, Università di Napoli "Federico II", Napoli, Italy*

[b] *INFN, Sezione di Napoli, Napoli, Italy*

**Abstract**

The determination of light nuclei fluxes in a wide energy range is one of the scientific objectives of the PAMELA experiment. The Time-of-Flight system is one of the key elements for the particle identification: its charge discrimination capabilities for light nuclei, determined during a test beam calibration, will be presented and compared with preliminary in-flight results.

## 1 The PAMELA experiment

PAMELA is a satellite-borne experiment [1] built to detect charged particles in cosmic rays with particular attention to antiparticles. On June 15th 2007 PAMELA has reached the first year in orbit. All the detectors are working nominally and analysis is in progress to achieve many scientific goals: principally the search for antimatter in primary radiation, the search for dark matter sources but also the measurement of fluxes and ratios of the different components of the cosmic radiation and the study of interactions of the radiation itself with Sun and Magnetosphere.

To reach all these scopes, PAMELA is composed by several instruments perfectly integrated; one of the most important elements is the *Time-of-Flight* system (ToF). The experimental setup allows us to operate precise momentum measurements, $dE/dx$ and charge spectrum reconstruction as well as investigation of many other topics. Details about the science of PAMELA, as well as the detector and its general status, will be described in another paper in this conference [2].

## 2 The *Time-of-Flight* system

The ToF system of PAMELA [3, 4] is composed by 24 paddles; each paddle is made by a plastic scintillator connected to a couple of photomultiplier tubes (PMT) by means of plastic light guides. The 24 paddles are arranged in 6 layers to build 3 planes (S1, S2 and S3) with $x$ and $y$ view.

The ToF electronic system [5] converts the 48 PMT pulses into time-based and charge-based measurements.

The instrument has to fulfill several scientific goals:

- to provide a fast trigger signal to the whole experiment
- to measure the time of flight of the passing particles, from which a measurement of $\beta$ can be derived
- to reject albedo particles events
- to support the magnetic spectrometer in the tracking phase
- to reconstruct $dE/dx$ spectrum for light nuclei up to Oxygen.

In this paper we will focus our attention on the last point, describing the method currently used in data analysis to identify light nuclei with the $ToF$ system.

## 3 Charge discrimination with $ToF$

Consider a charged particle coming from above which hits two paddles of different planes; in this case we get four independent time measurement related to the position of the hit point, the velocity of the passing particle, the flight path and some peculiar constants. Combining these relations, and joining together all the constant values (depending, for example, on geometry or on intrinsic characteristics of electronics), we have that the time of flight within the two hits is function of only $\beta$, according to:

$$T_{ToF} = K_1 + \frac{K_2}{\beta} \tag{1}$$

where $K_1$ and $K_2$ are calibration constants; for flight data they are continuously monitored and recalculated.

According to *Bethe-Block*, if we plot the energy loss inside the counter as a function of $\beta$, different families of nuclei separates into different bands which can be fitted. $Z$ values can be extrapolated from the curves.

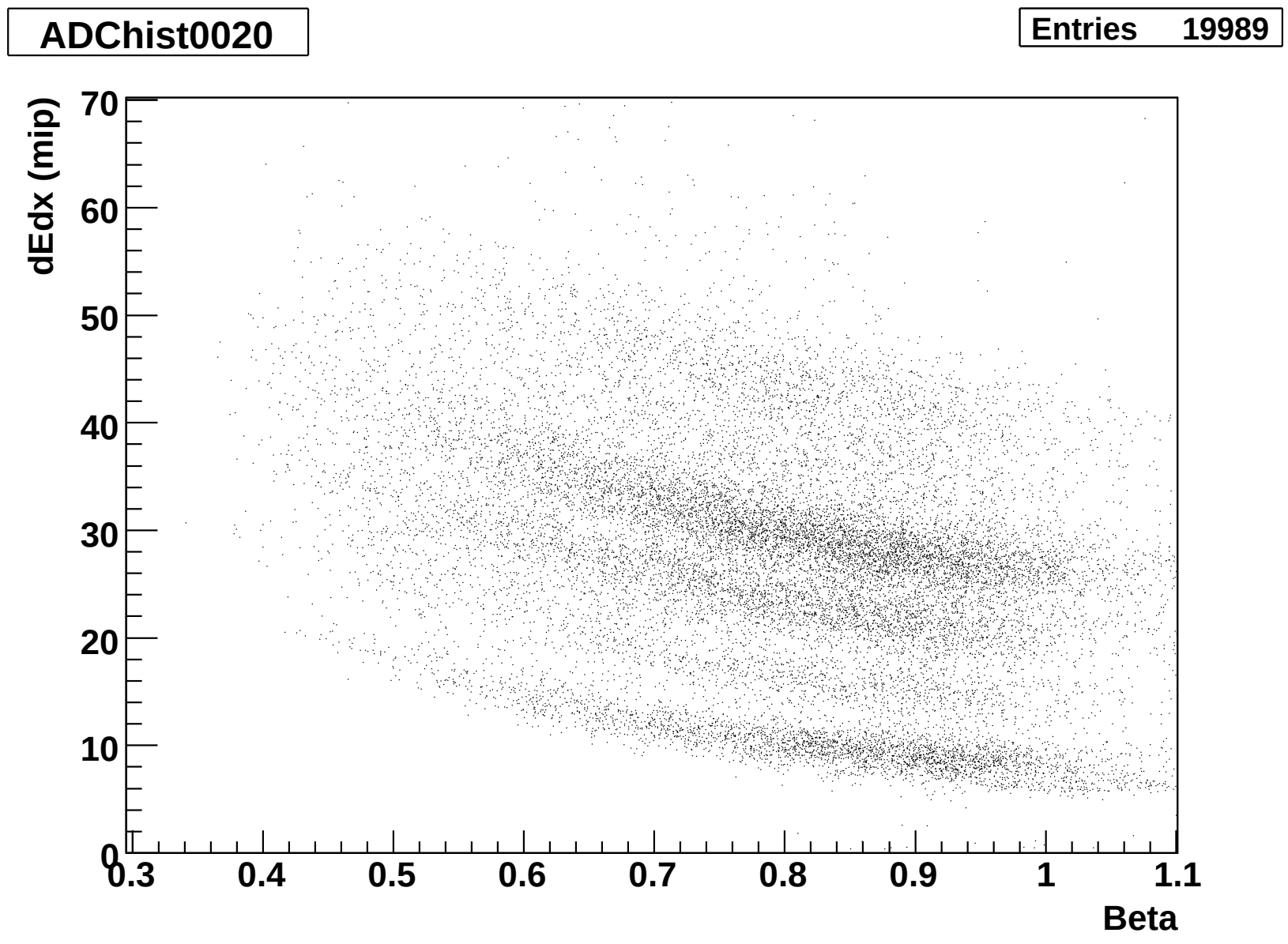

Figure 1: Plot of $dE/dx$ versus $\beta$ for a PMT of S2 for a sample of nuclei, with $Z \geq 3$, as reconstructed by ToF system alone. Band structures, corresponding to different families of ions, are visible up to Oxygen.

To evaluate energy loss we have, first of all, to convert digital data from ADC to informations on charge-integral of PMT pulses in units of $pC$ using the calibration relations of Front-End boards taking into account also some eventual not perfect linearity of the response of the electronics. The energy loss inside each counter has to be evaluated in units of *minimum ionizing particle* (*mip*); to obtain this estimation we have to use informations from the magnetic spectrometer to select particles of fixed $Z$, in particular p and He ions. Working on this selected sample we have to normalize for the angle of incidence (which is to correct for the different width of scintillator passing by the particle) and correct for the position of the hit point along the paddle (which is to normalize for attenuation of the signal using a double exponential fit).At this point we can plot *mip*s versus $\beta$ and select a sample of only nuclei above Helium with a graphical cut. In addiction, we have to eliminate those events for which the electronics saturates. Plotting the remaining sample, we finally reach the plot of energy loss as a function of $\beta$ (see also figure 1) in which we can fit different bands for different families (the procedure is represented in figure 2 for Lithium band) and normalize for velocity to obtain $Z$ distribution.

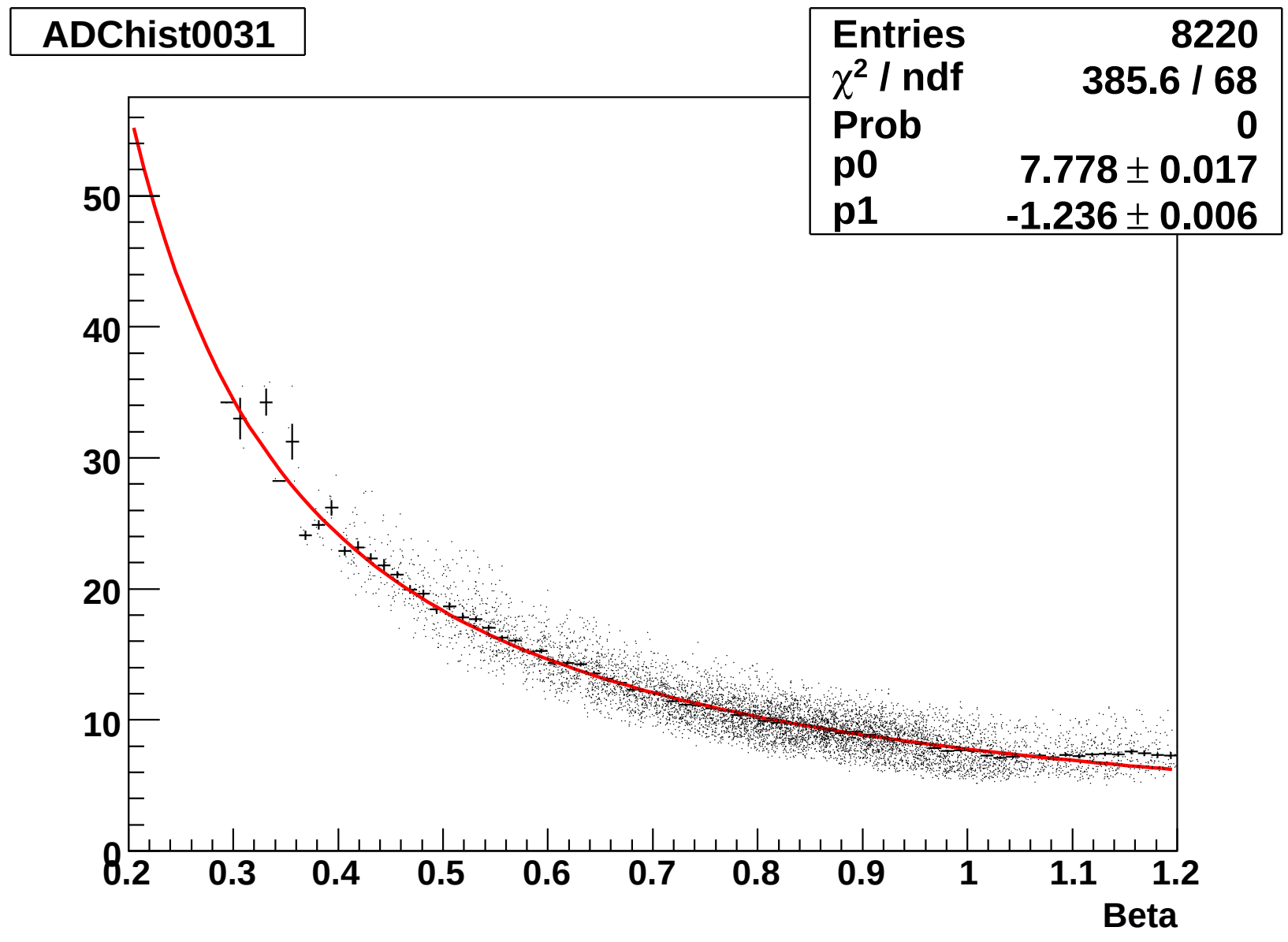

Figure 2: In this figure of $dE/dx$ versus $\beta$ for a single PMT, only events reconstructed as Lithium in 4 paddles or more are plotted. The band structure is fitted by a *Bethe-Block* curve.

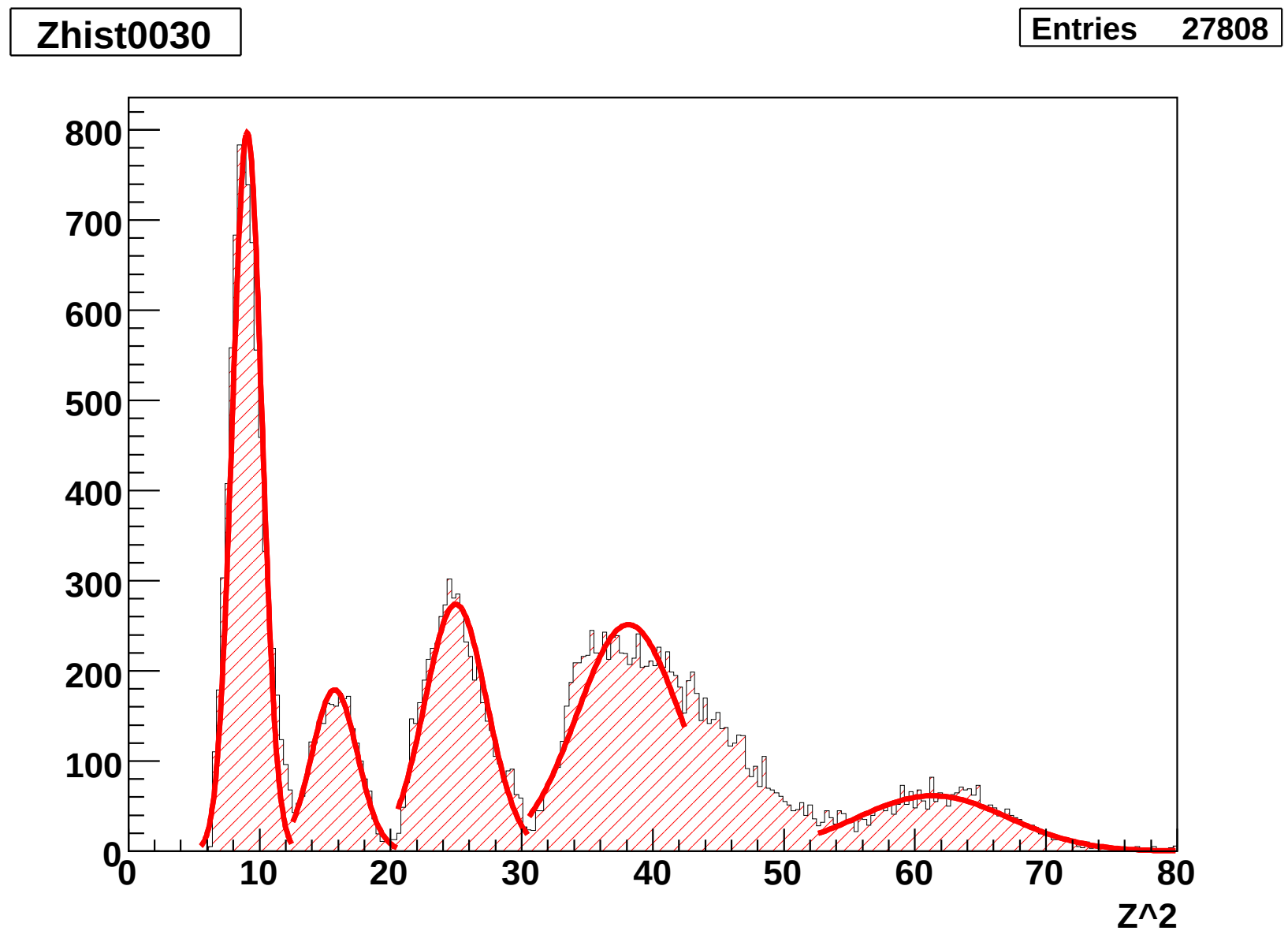

Figure 3: Preliminary distribution of $Z$ squared for light nuclei, with $3 \leq Z \leq 8$, as reconstructed by the ToF system alone, for a photomultiplier of the S2 plane. Where possible, gaussian fit is superimposed to the peaks.

In figure 3 there is an example of $Z^2$ distribution derived by means of the described method: this plot is not significant for the estimation of relative abundances of different elements and is very preliminary for the evaluation of resolution of charge measurements, but it is useful to verify the validity of the method. Moreover, it confirms the capability of the instrument to detect ions at least up to Oxygen, as we expected from beam test calibration [6].

# FUTURE DARK MATTER SEARCHES WITH THE GLAST-LAT

JAN CONRAD [a,b]
REPRESENTING THE GLAST-LAT COLLABORATION

[a] *Fysikum, Stockholms Universitet, AlbaNova, S-10691 Stockholm, Sweden*

[b] *Institutionen för fysik, Kungliga Tekniska Högskolan, AlbaNova, S-10691 Stockholm, Sweden*

**Abstract**

The Large Area Telescope, one of two instruments on the Gamma-ray Large Area Space Telescope mission, scheduled for launch by NASA in early 2008, is a high-energy gamma-ray telescope, sensitive to an energy range from approximately 20 MeV to more than 300 GeV and exceeding the sensitivity of its predecessor EGRET by nearly two orders of magnitude. Annihilation of Weakly Interacting Massive Particles (WIMP), predicted in many extensions of the Standard Model of Particle Physics, may give rise to a signal in gamma-ray spectra from many cosmic sources. In this contribution we give an overview of the searches for WIMP Dark Matter performed by the GLAST-LAT collaboration, in particular we discuss the sensitivity of GLAST to thermal WIMP dark matter in the galactic center and to high energy (50 -150 GeV) gamma-ray lines resulting from WIMP pair annihilation into two gamma-rays.

## 1 LAT searches for Dark Matter

The Gamma ray Large Area Space Telescope (GLAST) [1] [2] [3] is due for launch in early 2008. With its large effective area, excellent energy and angular resolution in an energy range of 20 MeV to 300 GeV, the Large Area

Telescope (LAT) on board of GLAST is an ideal detector to search for high energy gamma emission from the annihilation of pairs of Weakly Interacting Massive Particles (WIMP) in the universe.

The GLAST-LAT collaboration pursues complementary searches for Dark Matter each presenting its own challenges and advantages. In table 1 we summarize the most important ones.

The center of our own galaxy is a formidable astrophysical target to search for a Dark Matter signal, the reason being that simulations of Dark Matter halos predict high densities at the center of the galaxy and since the WIMP annihilation rate is proportional to the density squared, significant fluxes can be expected. On the other hand, establishing a signal requires identification of the high energy gamma-ray sources which are in (or near) the center [6] and also an adequate modeling of the galactic diffuse emission due to cosmic rays colliding with the interstellar medium. The latter is even more crucial for establishing a WIMP annihilation signal from the galactic halo.

Due to the $2\gamma$ production channel, a feature in the spectrum from the various astrophysical sources would be the gamma-ray line placed at the WIMP mass. This is a “golden” signal, in the sense that it would be difficult to explain by an astrophysical process different from WIMP annihilation. Also it would be free of astrophysical uncertainties, since the background can be determined from the data itself. However, since the $2\gamma$ channel is loop-suppressed, the number of photons will be very low.

In the following sub-sections, the above mentioned searches will be described in more detail. The sensitivity calculations presented here are work in progress, the reader is referred to a shortly forthcoming paper [4] for updated results.

## 2 Sensitivity to signal from the Galactic Center

The $\gamma$-ray flux for a generic WIMP, $\chi$, at a given photon energy $E$ is given by

$$\phi_\chi(E) = \frac{\sigma v}{4\pi} \sum_f b_f \frac{dN_{\gamma,f}}{dE} + \sum_g b_{\gamma g} n_\gamma \delta\left(E - m_\chi(1 - m_Z^2/4m_\chi^2)\right) \int_{l.o.s} dl \frac{1}{2} \frac{\rho(l)^2}{m_\chi^2} \tag{1}$$

This flux depends on the WIMP mass $m_\chi$, the total annihilation cross section times WIMP velocity $\sigma v$ and the sum of all the photon yields $dN_{\gamma,f}/dE$ per annihilation channel weighted by the corresponding branching ratio $b_f$. In addition, there is the “line” yield due to annihilation into 2 photons or a photon and a $Z$ boson. The fragmentation and/or the decay of the tree-level annihilation states gives rise to photons, predominantly via the generation of neutral pions and their decay into $2\gamma$. The simulation of the photon yield is performed with Pythia 6.202 [8] implemented in the DarkSUSY package [9]. Apart from the annihilation cross-section and photon yields, the WIMP annihilation flux also depends on the WIMP density in the galactic halo, $\rho(l)$,integrated along the line of sight. For the results presented here, we assume a Navarro-Franck-White (NFW) profile [10]. The GLAST detector response has been simulated using the fast simulation program *ObsSim*, developed by the GLAST LAT collaboration [11].

Table 1: *Summary of the different searches for Particle Dark Matter undertaken by the GLAST-LAT collaboration. For reference we include the contributions to the 1st International GLAST Symposium describing the respective analyses [5]*

| Search | advantages | challenges | GLAST Symp. |
|---|---|---|---|
| Galactic center | good statistics | Difficult source id, uncertainties in diffuse background | Morselli et al. |
| Satellites | Low background, good source identification | low statistics | Wang et al. |
| Galactic halo | Large statistics | Uncertainties in diffuse background | Sander et al. |
| Extra galactic | Large statistics | Uncertainties in diffuse background, astrophysical uncertainties | Bergstrom et al. see also [7] |
| Spectral lines | No astrophysical uncertainties "golden" signal | low statistics | Edmonds et al. |

Fixing the halo density profile, a dominant annihilation channel and the corresponding yield, we perform a scan in the $(m_{\rm wimp}, \sigma v)$ plane in order to determine the GLAST reach. The EGRET source given in [6] is also simulated at the Galactic Center and we perform a standard $\chi^2$ statistical analysis to check for each WIMP model compatibility with EGRET data at the $5\sigma$ level. For models consistent with EGRET, we calculate a second $\chi^2$ to check if GLAST is able to disentangle the WIMP contribution from the galactic diffuse gamma-ray emission, which is modeled using the GALPROP package [15]. In particular, we consider two different models: the conventional model [16] and the "optimized" model [17]. With respect to the former the EGRET measurement of gamma-ray diffuse emission [18] reveals an excess in the multi-GeV energy range. The latter has been adjusted to explain the excess in the framework of conventional physics, such as cosmic-ray intensity fluctuations. The "optimized" model is thus more conservative since it assumes that the GeV excess is not made by Dark Matter whereas the "conventional" model allows for a Dark Matter explanation. Figure 1 shows the results of the scans for annihilation into $b\bar{b}$ only.

## 3 GLAST senstivity to $\gamma$-ray lines

In some dark matter halo models, an annulus around the galactic center could give a signal to noise ratio as much as 12 times greater than at the galactic center [19]. To estimate the line sensitivity we therefore consider an galactic centered annulus with inner radius of 25 degrees and outer radius of 35 degrees, excluding the region within 10 degrees from the galactic plane to estimate the photon background for WIMP annihilation into narrow lines. LAT line energy sensitivities were calculated at $5\sigma$ significance for the case when the line energy is known (e.g. supplied by discovery at the Large Hadron Collider) and for unknown line energy including a trial factor. We use ObsSim to generate LAT resolved annihilation lines. ObsSim was run for 55 days (uniform all-sky coverage for the LAT in scanning mode) with a narrow ($\sigma(E)/E_0 = 10^{-3}$) Gaussian energy distribution centered on 25, 50, 75, 100, 125 and 150 GeV. A 5 year all-sky diffuse background was generated using the previously mentioned optimized model for the galactic diffuse background radiation. The region of interest, the galactic annulus excluding the area closest to the galactic plane, was fit with an exponential background in the energy range between 40 and 200 GeV. For each line the input background was bootstrapped with a Monte Carlo signal 1000 times and fit to an exponential plus a double Gaussian (giving a good fit to the line) as well as an exponential only to calculate a $\Delta\chi^2$ between the best fit background only and the best fit signal plus background hypotheses. This series of 1000 bootstraps was rerun varying only the average number of MC signal counts until average $\Delta\chi^2$ is larger than 25 (corresponding to a $5\sigma$ detection). The average number of signal counts needed at each energy was then converted to a sensitivity using average exposures integrated over the annulus. A bin width of FWHM/E = 8% was used based on the FWHM energy resolution of the GLAST-LAT. The flux needed for a $5\sigma$ detection is shown in figure 2. Assuming a dark matter density and a gamma-ray yield, sensitivity as function of velocity averaged cross-section can be obtained. The results presented here should be considered preliminary that current refinement of the used instrument response parameterizations is ongoing. This will, among other things allow us to study line sensitivity in the whole specified energy range (upto 300 GeV). Also, low number of counts will require more refined statistical methods, which are currently under study, see for example [20].

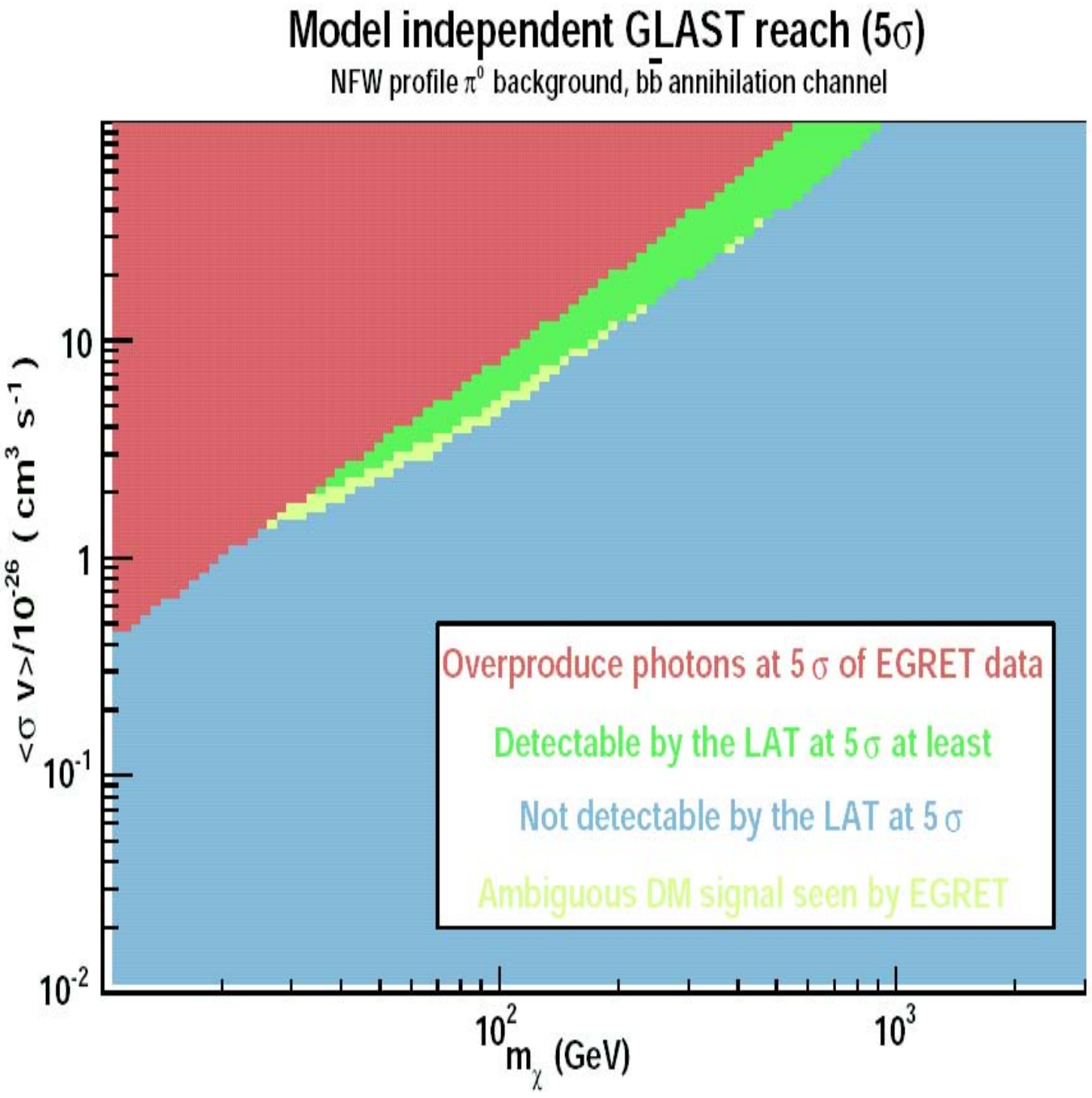

Figure 1: Cross Section times WIMP velocity versus the WIMP mass for the $b\bar{b}$ annihilation channel. The red region might be excluded by the EGRET data around the galactic center. It should however be noted that there is considerable controversy as to if the EGRET data above 1 GeV are reliable, see e.g. [12], [13][14]. Models in blue region are not detectable by GLAST and green region correspond to models detectable by GLAST for both conventional and optimized astrophysical background. The background uncertainties are reflected in the yellow regions as for these models, only one of the astrophysical background allows a detection by GLAST.

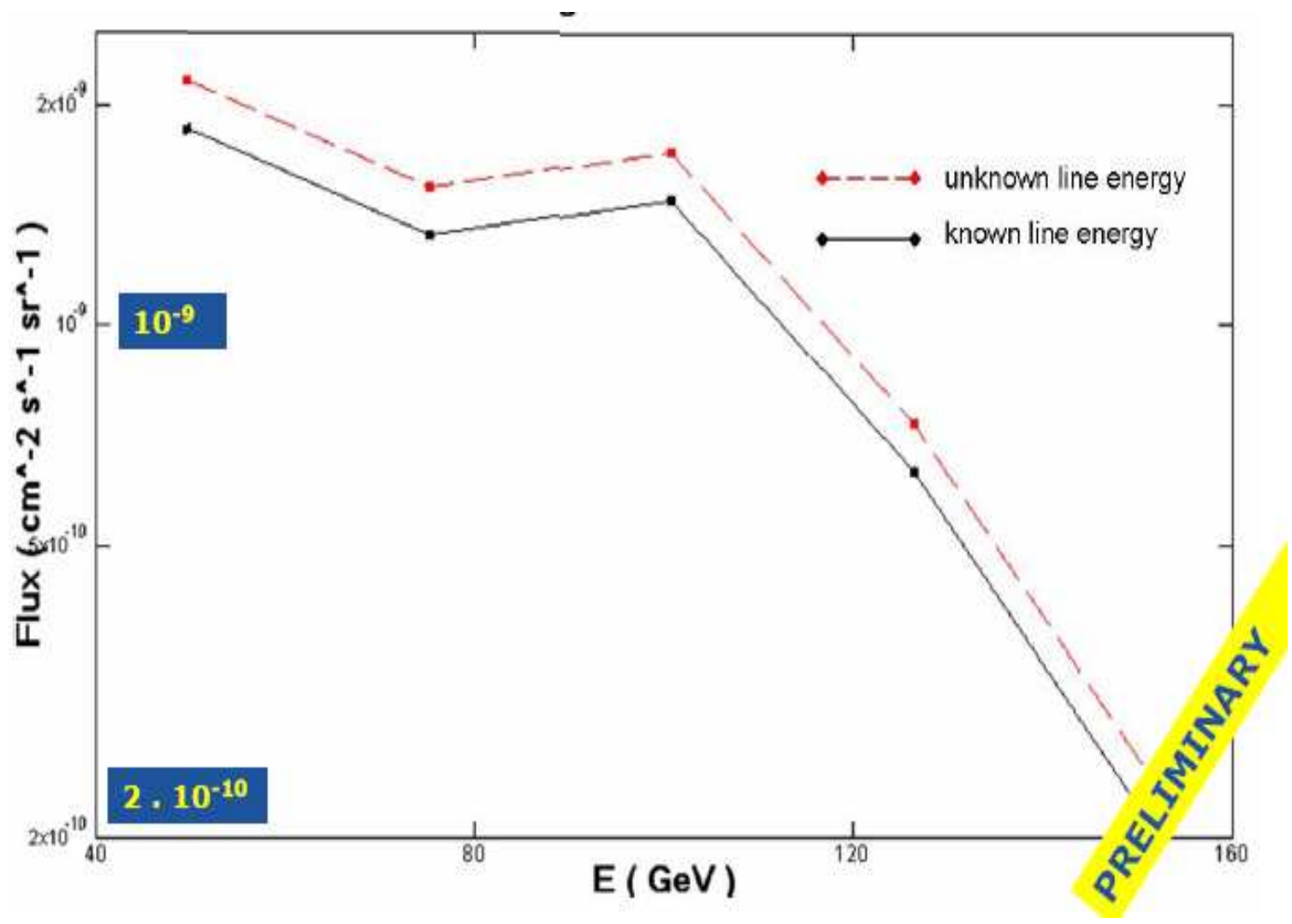

Figure 2: $5\sigma$ line flux sensitivity achievable within 5 years GLAST-LAT operation as function of energy (WIMP mass). Solid line: known WIMP mass, dashed: unknown WIMP mass

## 4 Conclusions

In this note we summarize the searches for particle Dark Matter to be performed with the GLAST-LAT instrument. Several complementary astrophysical sources will be examined, each presenting its own advantages and challenges. Those discussed in this note regard the WIMP annihilation signal from galactic center and the golden signal for presence of particle dark matter, namely gamma-ray lines from direct annihilation of WIMPS. For the galactic center, cosmologically interesting regions of the parameter space ($\sigma v \sim 10^{-26}$) are within the reach. Line flux sensitivity for an annulus around the galactic center are presented. These can be converted into a sensitivity for annihilation cross-section assuming a dark matter distribution. GLAST is now integrated on the space-craft and undergoing final testing. The launch is foreseen for early 2008.

## 5 Acknowledgements

The author would like to thank all members of the GLAST working group for Dark Matter and New Physics which contributed to this note. Support from Vetenskapsrådet is acknowledged.

# GLAST LAT COSMIC RAY DATA ANALYSIS AT GROUND

M. BRIGIDA [a,b], G.A. CALIANDRO [a,b], C. FAVUZZI [a,b], P. FUSCO [a,b], F. GARGANO [b] , N. GIGLIETTO [a,b], F. GIORDANO [a,b], F. LOPARCO [a,b], M.N. MAZZIOTTA [b], C. MONTE [a,b], S. RAINÒ [a,b], P. SPINELLI [a,b]

[a] *Dipartimento Interateneo di Fisica "M. Merlin", Università degli Studi di Bari e Politecnico di Bari, via G. Amendola 173, 70126 Bari, Italy*

[b] *INFN, Sezione di Bari, via E. Orabona 4, 70126 Bari, Italy*

**Abstract**

The Large Area Telescope (LAT) on board the Gamma Ray Large Area Space Telescope (GLAST) is able to perform gamma-ray astronomy in the energy range from 20 $MeV$ to 300 $GeV$. The LAT has a modular structure, consisting of 16 identical towers. Each tower is composed by a tracker, a calorimeter and a data acquisition module. A plastic scintillator anticoincidence system covers all the towers. The integrated LAT has been tested as a cosmic-ray observatory at "General Dynamics" in Gilbert, Arizona. The cosmic ray data samples collected at ground have been used to study the performance of the LAT tracker. Preliminary experimental data and a first comparison with Monte Carlo simulations for the tracker are shown.

## 1 The GLAST Large Area Telescope (LAT)

The Gamma-ray Large Area Space Telescope is a satellite-based observatory to study the high energy gamma-ray sky. The two instruments on board GLAST are the Large Area Telescope (LAT) which is an imaging, wide field-of-view, high-energy gamma-ray telescope, that covers the energy range from 20 $MeV$ to more than 300 $GeV$, and the GLAST Burst Monitor (GBM) which will provide spectra and timing in the energy range from 8 $keV$ to 30 $MeV$ for Gamma-Ray Bursts (GRB).

The LAT is a pair-conversion telescope equipped with a precision tracker (TKR) and a calorimeter (CAL), consisting of a $4 \times 4$ array of identical towers, supported by a low-mass grid and surrounded by a segmented anticoincidence (ACD) shield that provides most of the rejection against the charged cosmic-ray background. Each tower is composed by a *tracker module*, consisting of a vertical stack of 18 x-y single-sided silicon strip detectors tracking planes interleaved with tungsten foil converters, a *calorimeter module* composed by 96 CsI(Tl) crystals, with a total depth of 8.6 radiation lengths, and a *data acquisition module (DAQ)* [1].

## 2 Observatory data taking and muon data analysis

The verification strategy of the Science Requirements for the LAT consists of a combination of simulations, beam tests [2], and cosmic ray induced ground-level muon tests [3]. Ground-level muon data allow to test and validate the Monte Carlo (MC) simulation of the instrument response. Data taking with cosmic ray muons was performed at Stanford Linear Accelerator Center (SLAC) in 2005-2006 during the phase of Integration and Test (I&T) of the LAT [3]. Similar tests have been also performed starting from March 2007, with the two GLAST instruments completely integrated onto the spacecraft (Observatory) at "General Dynamics (GD) Advanced Information System" in Gilbert, Arizona.

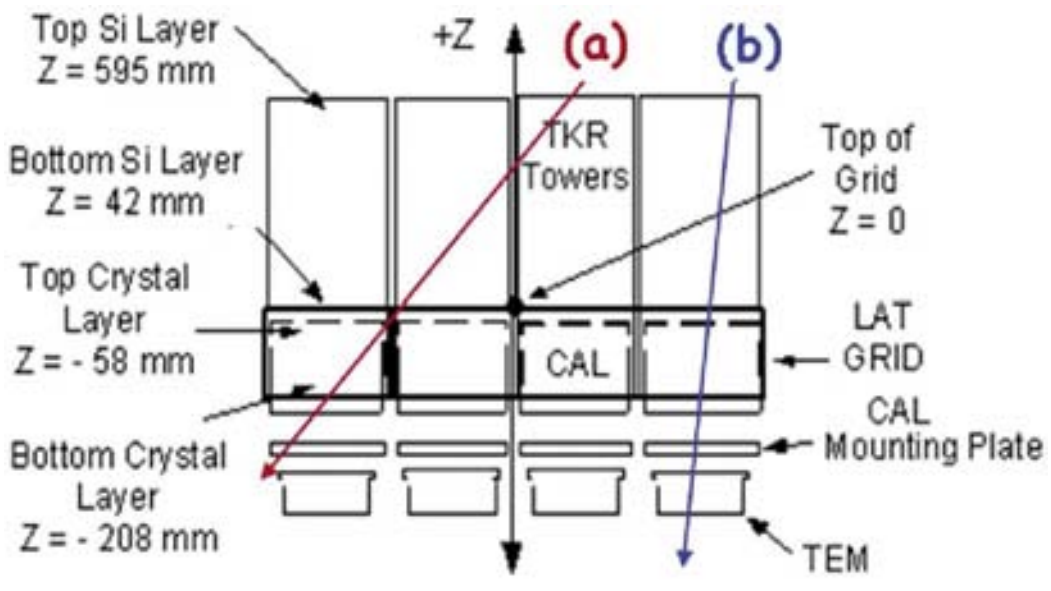

Figure 1: Geometrical selection criteria

### 2.1 Muon event selection

For our analysis we selected a data sample consisting of single muon tracks crossing the LAT from the top to the bottom. We selected single track events triggered by both TKR and ACD. We required that the particles crossed the whole LAT (track (a) in figure 1) or a single tower (track (b) figure 1) from the TKR top layer to the CAL bottom layer and that the energy deposition in the CAL layers was consistent with that of a minimum ionizing particle.

### 2.2 Monitoring of the TKR performance

We monitored the TKR performance analyzing both I&T and observatory data and comparing them with a Monte Carlo simulation developed for the detector in the I&T configuration. In figure 2 the distributions of the TKR hits (fired strips) and clusters (groups of adjacent fired strips) for single tower events are shown. The cluster distributions of cosmic ray data (both I&T and observatory) are in agreement with MC predictions, while the hit distributions are slightly different for the three samples. The discrepancies between I&T and observatory data are due to different settings of the DAQ: observatory data have been taken with the hardware in flight configuration, while I&T data were taken with the hardware configured for ground analysis [3]. For what concerns the differences between data (both observatory and I&T) and simulations, it is worth to point out that electronics simulation is still being developed and ground-level muon data can provide good inputs to improve the Monte Carlo codes.

In figure 3, the dependence of the total number of hit strips in the TKR on the zenith angle $\theta$ is shown for samples of muons crossing the whole LAT (like track (a) in figure 1). As expected, the total number of hits increases linearly

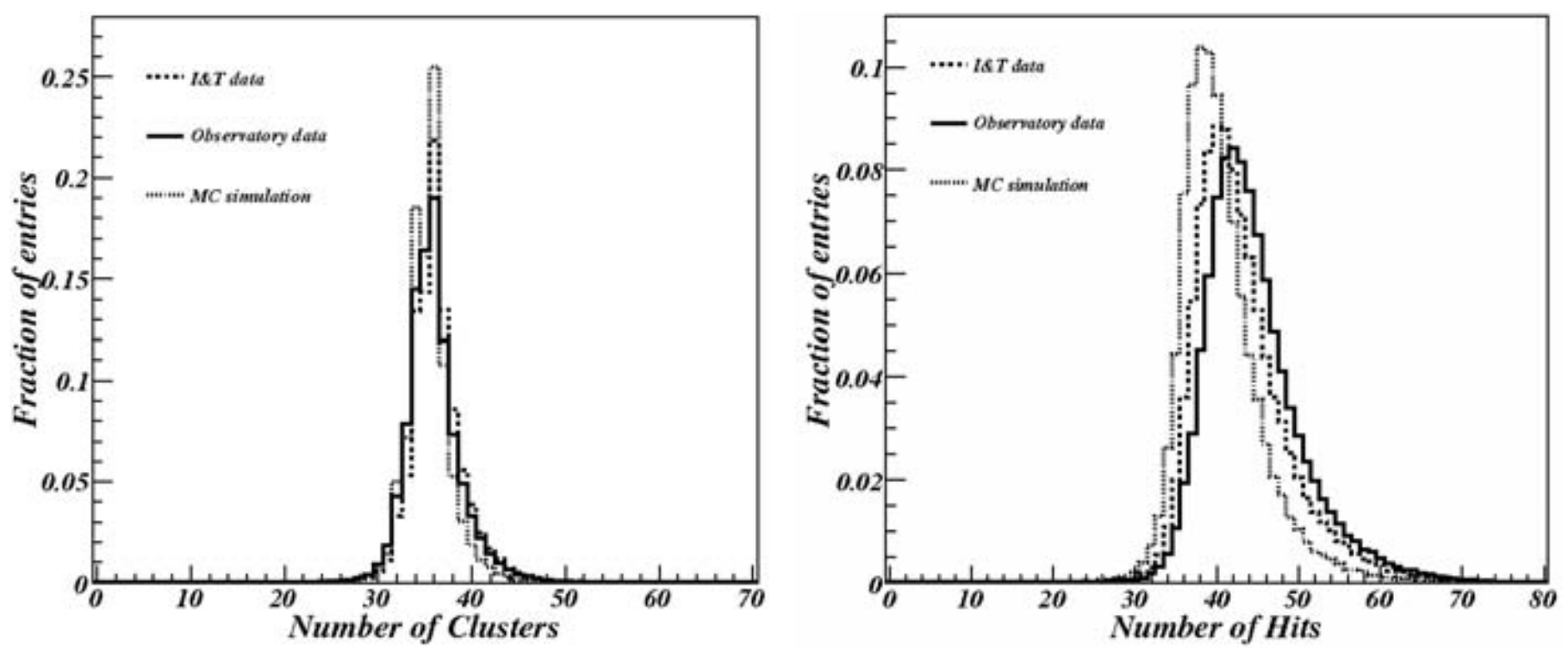

Figure 2: TKR cluster (left) and hit (right) distributions for single tower events.

with *sec* $\theta$, i.e. it is proportional to the track length.

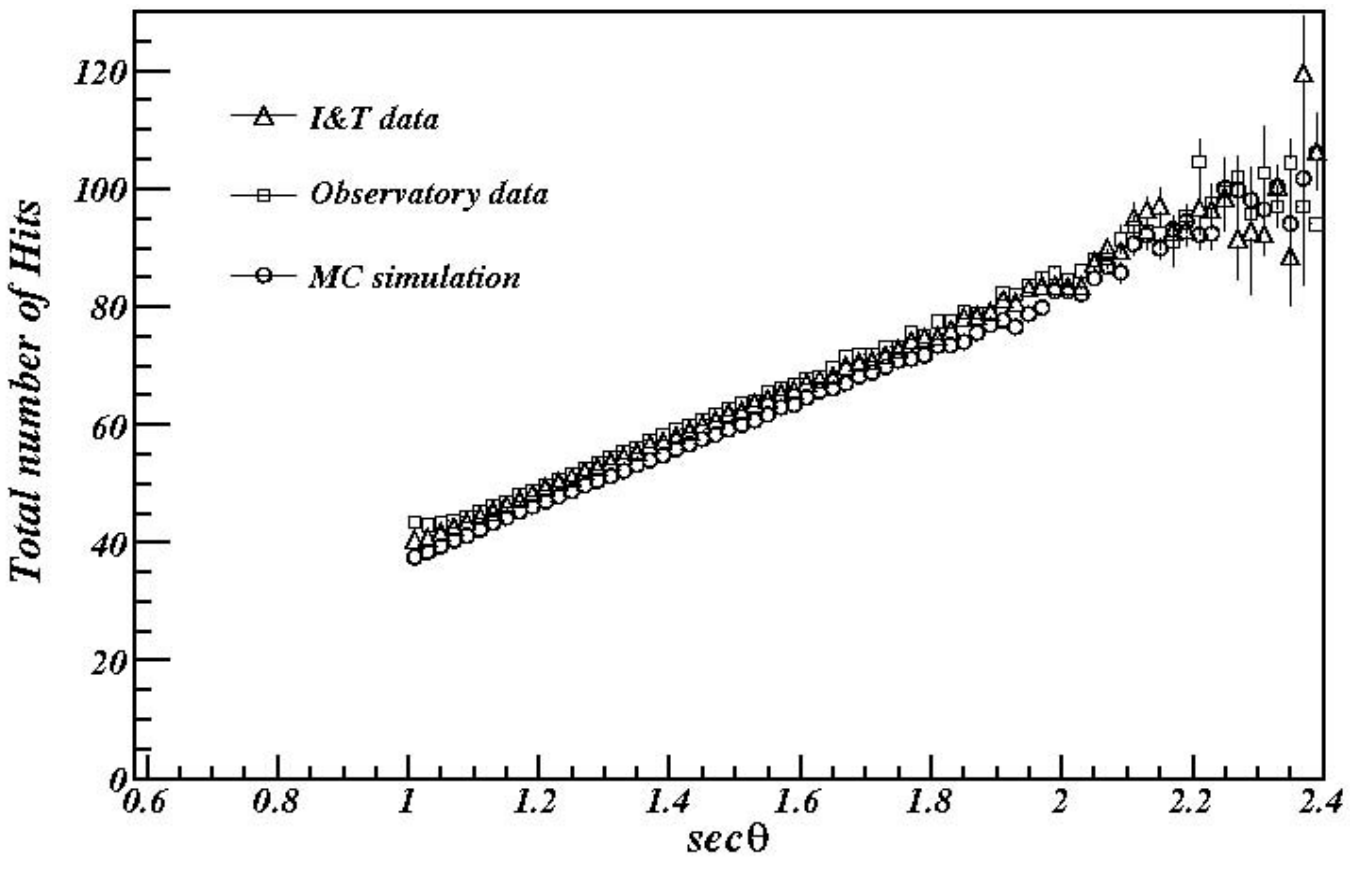

Figure 3: Total number of hit strips in the TKR vs *sec* $\theta$.

## 3 Conclusions

The LAT performance have continuosly been monitored during the phase of the instrument integration and the monitoring will continue during the pre-launch phase up to the launch date (actually scheduled for January 31, 2008). The MC simulation is still being tuned, taking also into account the beam test data collected at CERN and at GSI in 2006 [2].

# SEARCHING FOR POINT-LIKE DARK MATTER SOURCES

Torsten Bringmann [a]

[a] *SISSA/ISAS and INFN, via Beirut 2-4, I-34013 Trieste, Italy*

**Abstract**

The indirect search for dark matter is still hampered by relatively large uncertainties about the dark matter distribution and thus the absolute strength of the expected annihilation signal. In addition, at least for traditional sources like the galactic center, one often has to face a considerable astro-physical background. This points more and more towards an unavoidable need for clear spectral signatures that could unambiguously identify a potential dark matter annihilation signal. After stressing that gamma rays are particularly promising in this respect, I give an updated overview over possible point-sources of dark matter annihilations and the prospects for current and future experiments to detect them.

## 1 Introduction

While there has accumulated overwhelming evidence that non-baryonic, cold dark matter (DM) provides the building block of the observed structures in the universe, comparably little is known about its detailed properties. Shedding light on the nature of the DM thus remains a major challenge for cosmology and (astro-)particle physics. A theoretically particularly well-motivated class of DM candidates, however, are weakly interacting massive particles (WIMPs) [1]. They arise in almost all extensions to the standard model of particle physics (SM) as the lightest of a set of new, massive particles and are stable due to some internal symmetry; thermally produced in the early universe, they automatically acquire a relic density consistent with the needed amount of DM. Typically, the WIMP mass falls into the range $50\,\mathrm{GeV} \lesssim m_\chi \lesssim 10\,\mathrm{TeV}$.

A possible way for the *indirect detection* of DM makes use of the fact that WIMPs usually can pair-annihilate into SM particles and thereby potentially leave an imprint in cosmic rays of various kinds. The great advantage of gamma rays – as compared to other cosmic ray species – is that to them the galactic halo looks almost entirely transparent; gamma rays thus point directly to their sources and no further assumptions have to be made about their propagation like in the case of charged particles. The same is in principle true also for neutrinos; from an observational point of view, however, they are less interesting since the sensitivities for their detection are significantly below those for gamma-rays (this situation only changes for very nearby sources like the center of the earth or the sun). On the experimental side, this theoretical interest is met by a new generation of space- (AGILE and GLAST [2]) and ground-based (ARGO, CANGAROO, HESS, MAGIC, MILAGRO, VERITAS [3]) experiments with impressive performances that will allow to put non-trivial limits on the relevant parameter space of popular WIMP candidates. Upcoming and planned experiments like AMS-02, CTA, GAW [4] will show further significant improvements with respect to both the gamma-ray sensitivity and the range of observable energies.

The expected gamma-ray flux (in units of $\mathrm{GeV}^{-1}\mathrm{cm}^{-2}\mathrm{s}^{-1}\mathrm{sr}^{-1}$) from a source with DM density $\rho$ is given by

$$\frac{d\Phi_\gamma}{dE_\gamma}(E_\gamma, \Delta\psi) = \frac{\langle\sigma v\rangle_{\mathrm{ann}}}{4\pi m_\chi^2} \sum_f B_f \frac{dN_\gamma^f}{dE_\gamma} \times \frac{1}{2}\int_{\Delta\psi} \frac{d\Omega}{\Delta\psi} \int_{\mathrm{l.o.s}} d\ell(\psi)\rho^2(\mathbf{r})\,. \tag{1}$$

Here, $\langle\sigma v\rangle_{\mathrm{ann}}$ is the total annihilation cross section, $B_f$ the branching ratio into channel $f$ and $N_\gamma^f$ the number of resulting photons. The right part of the above expression essentially counts the number of DM particle *pairs* (times $m_\chi^2$) along the line of sight of a detector with an opening angle $\Delta\psi$. Since it depends strongly on the largely unknown distribution of DM, it results in a considerable uncertainty in the overall *normalization* of the expected annihilation flux. The first part of the above expression for the flux, on the other hand, depends only on the underlying particle physics model and can be determined to a much better accuracy. It is here that the energy dependence of the signal enters, illustrating the need for clear *spectral signatures* in order to eventually be able to discriminate DM from more conventional astrophysical sources.

This first part of Eq. (1) is discussed in more detail in Section 2. In Section 3, an overview over the most promising point-sources of DM annihilations is given, before turning in Section 4 to detection prospects for current and planned experiments. Section 5, finally, concludes.

## 2 Dark matter annihilation spectra

Even though DM does not directly couple to photons, annihilating DM particles can give rise to gamma-rays in various ways . Principally, one can distinguish between three different kinds of contributions to the total annihilation signal:

- secondary photons
- line signals
- final state radiation (FSR)

*Secondary photons* arise from the fragmentation and further decay of other annihilation products, mainly through the decay of neutral pions, i.e. $\pi^0 \rightarrow \gamma\gamma$. The resulting photon distribution can be obtained from Monte Carlo packages like PYTHIA [5]. Except for the case of $\tau$ lepton final states, which are usually sub-dominant and thus of less practical importance, these spectra are very similar and show an $E_\gamma^{-1.5}$ behaviour, up to an exponential cutoff at $m_\chi$ that appears due to kinematic reasons. In fact, taking into account that neither the absolute amplitude nor $m_\chi$ are *a priori* known, these spectra are almost indistinguishable (see, e.g., [6]). Secondary photons usually dominate the total annihilation spectrum at low and intermediate energies, i.e. $E_\gamma \lesssim 0.5\, m_\chi$. Unless the cutoff can be clearly observed with good statistics, it is quite difficult to distinguish these rather featureless spectra from typical astrophysical sources, which very often also show a power law-like behaviour.

Sharp *line signals* result from two-body final states where at least one of the particles is a photon; possible annihilation products are thus $\gamma\gamma$, $Z\gamma$ or $H\gamma$. While the observation of such a line feature at $E_\gamma = m_\chi$ (or $E_\gamma = m_\chi[1 - m_\chi^2/(4m_{Z,H}^2)]$, respectively), with a typical width $\mathcal{O}(10^{-3})$ due to the WIMP velocity in the galactic halo, would be a smoking-gun signature for DM annihilations, these processes are necessarily loop-suppressed and thus generically give rise to rather low fluxes. However, there exist WIMP examples where such line signals may be detectable even for current detector performances [7], especially when powerful enhancement mechanisms due to non-perturbative effects are at work [8]. In contrast to secondary photons, the corresponding annihilation rates are highly model-dependent and could thus in principle provide valuable information about the nature of the annihilating DM particles, especially when more than just one line is observed.

Whenever DM annihilates into pairs of charged SM particles (which is a generic situation for WIMPs), the same process with an additional photon in the final state unavoidably also takes place. If the produced charged SM particles are much lighter than the annihilating DM particles, this *final state radiation* (FSR) actually dominates the total gamma-ray spectrum at high energies, $E_\gamma \gtrsim 0.5\, m_\chi$ [9]. Prominent examples include the supersymmetric neutralino when it is a pure Higgsino or Wino [10], Kaluza-Klein DM in theories with universal extra dimensions [11], or scalar DM [12]; in fact, FSR turns out to be important even for more general neutralino compositions [13]. An important feature of FSR photons is furthermore that they provide much more pronounced spectral signatures than secondary photons. This could not only help to unambiguously identify the DM origin of an observed signal, but also to actually distinguish between different DM candidates and thus to provide important information about the underlying nature of the annihilating DM particles (see, e.g., [14] for an illustrative example).

## 3 Point source candidates

For sources of DM annihilation with an angular extent smaller than the resolution of the detector, the line-of-sight integral in expression (1) takes the following form:

$$\int_{\Delta\psi} \frac{d\Omega}{\Delta\psi} \int_{\mathrm{l.o.s}} d\ell(\psi)\rho^2(\mathbf{r}) \simeq \left(D^2\Delta\psi\right)^{-1} \int d^3\, r\rho^2(\mathbf{r})\,, \tag{2}$$

where $D$ is the distance to the point-source. Obviously, promising point-sources of DM annihilations should thus i) be as close as possible, ii) exhibit high DM concentrations and iii) lie in directions where a small contamination from astrophysical backgrounds is expected. Possible DM gamma-ray point-sources that have been discussed include:

- The galactic center
- Milky Way satellites (dwarf spheroidals) and external galaxies
- intermediate mass black holes
- DM substructures ("clumps") in the MW halo

The *galactic center* (GC) has traditionally been regarded as the most promising source of gamma rays. Unfortunately, the observed spectrum from that direction – well fitted by a $E_\gamma^{-2.2}$ power law and most likely of astro-physical rather than DM origin [15] – provides a major challenge to detect any DM signature on top of the background flux [16]. While still challenging, however, this situation improves when including the effect of FSR photons [17]. Note, further, that there is more room for a DM induced gamma-ray flux at energies between the EGRET ($\lesssim$ 30 GeV) and HESS ($\gtrsim$ 100 GeV) observations; this window will be closed by GLAST. In any case, it is intriguing to realize that DM annihilations can give rise to gamma-ray fluxes of the same strength as the actually observed one, for reasonably optimistic assumptions about the DM distribution near the GC (see, e.g., [11]).

*Dwarf spheroidals* (dSphs) that are satellites to the Milky Way (MW) are also very interesting potential targets for indirect DM searches. Since they are DM dominated, the expected background from more conventional astrophysical sources is expected to be rather low; at the same time, they are still in he galactic neighborhood. As a rule of thumb, the expected photon flux from these sources is at the per-cent level as compared to that from the GC (under similar astrophysical assumptions); for a recent analysis see, e.g., [18]. *Extra-galactic objects* that have been proposed as possible DM induced gamma-ray sources include Andromeda (M31), the Large Magellanic Cloud and the Coma Cluster. Taking into account the considerable astro-physical uncertainties about the DM distribution, detectional prospects are not not too different from the case of MW dSphs.

As realized in [19], a population of *intermediate mass black holes* (IMBHs) in the galactic halo could be bright sources of gamma rays from DM annihilations. In fact, for typical WIMP masses of $m_\chi = 150\,\mathrm{GeV}$, GLAST is expected to see $\sim 15$ of them - while for TeV-scale DM particles, at most a handful should be visible [6].

Numerical $N$-body simulations predict that a large fraction of DM is not distributed homogeneously but in the form of substructures, or *clumps*, down to the free-streaming scale for WIMPs [20] as determined by the scale of their kinetic decoupling in the early universe [21]. While this certainly has the effect of boosting any annihilation signal with respect to a smooth DM distribution, it is unlikely that the smallest structures can be resolved separately [22]. If, however, by chance, one of the larger clumps would be sufficiently nearby and thus detectable by GLAST, this could be a unique chance to study in detail the observed annihilation signal.

## 4 Experimental prospects

Gamma-ray observations can be performed both in space and on earth. The sensitivity of space-based telescopes is bounded by rather small effective areas and there is usually an upper bound on the photon energy that can be resolved ($\sim 300\,\mathrm{GeV}$ for GLAST, $\sim 1\,\mathrm{TeV}$ for AMS-02); a big advantage that allows the discovery of previously unknown sources, on the other hand, is typically a great field of view. Earth-based telescopes are complementary in that they have to face a lower bound on the observable energy since, below $\sim 30$ GeV, no electromagnetic showers are produced in the atmosphere; due to an effective area of several $\mathrm{km}^2$ they can, however, achieve much better sensitivities than space-based telescopes. Their rather small field of view makes them ideal for pointed observations.

For potential DM sources with known locations (GC, dSphs, external galaxies), one would thus use earth-based telescopes (note that their energy threshold roughly coincides with the lower range of masses for typical WIMP candidates), while for objects like IMBHs or DM clumps, which have no known counterpart in other wavelengths, one needs space-based experiments to locate them; once detected, they could be studied more thoroughly with ground-based telescopes, hoping, in particular, to see some of the pronounced spectral signatures discussed in Section 2.

A nice overview over the point-source sensitivities for various existing and planned gamma-ray experiments can be found in Ref. [23] (the planned CTA, still missing in that comparison, is expected to reach sensitivities about one order of magnitude better than the HESS telescope). For the upcoming GLAST satellite, whole-sky sensitivity maps are also available – both for the detection of point-sources and for the discrimination of a DM from an astrophysical source [6]. Translating these sensitivities into the needs for the potential sources discussed in the last section, and making reasonably optimistic assumptions about the local DM distributions, one finds that already operating or shortly upcoming experiments start to probe the parameter space of standard WIMP candidates for *all* of these cases (with the possible exception of micro-halos). For very pessimistic astrophysical assumptions (cores instead of steep central DM profiles, only small amplification effects from substructure, no baryonic compression in the GC, etc.), however, not even the projected sensitivity of large-scale experiments like the CTA is sufficient to see a signal from *any* of the sources discussed above.

## 5 Conclusions

A new generation of experiments is about to probe the high-energy gamma-ray sky with an accuracy that could make the indirect detection of DM feasible in the foreseeable or even near future. Despite these exciting prospects, however, one should keep in mind that they rely on optimistic (though not overly optimistic!) astrophysical assumptions; in the worst case scenarios, with the most pessimistic DM profiles consistent with simulations and actual observations, none of the standard DM candidates may give rise to visible gamma-ray fluxes.

Given the great astrophysical uncertainties, an obvious strategy consists of reducing them by both higher-resolution simulations and improved observations (taking, e.g., into account more stars in order to determine the gravitational potential to a better accuracy). On the other hand, one will most likely always be left with a considerable uncertainty in the absolute strength of DM annihilation signals; it is therefore of great importance to focus on distinctive spectral signatures that can clearly be attributed to a DM origin – such as the discussed line signals or FSR spectra.

Finally, let us stress that such a strategy will be even more successful if complemented with the search for indirect DM signals both at other frequencies and in other cosmic ray species. Such a complementary approach, including also bounds from direct DM searches as well as collider data from the LHC in the near future, seems to be the most promising way to finally reveal non-trivial information about the nature of the DM.

**Acknowledgments**

I would like to thank the organizers for inviting me to this very nice – yet maybe a bit too tightly packed – workshop with its pleasant atmosphere at Villa Mondragone.

# COSMOLOGICAL WIMPS, HIGGS DARK MATTER AND GLAST

A. Sellerholm [a], J. Conrad [a,b], L. Bergström [a] J. Edsjö [a]
Representing the GLAST-LAT collaboration

[a] *Department of Physics, Stockholm University, Stockholm, Sweden*

[b] *Department of Physics, Royal Institute of Technology (KTH), Stockholm, Sweden*

**Abstract**

Measurement of the extragalactic background (EGBR) of diffuse gamma-rays is perhaps one of the most challenging tasks for future gamma-ray observatories, such as GLAST. This is because any determination will depend on accurate subtraction of the galactic diffuse and celestial foregrounds, as well as point sources. However, the EGBR is likely to contain very rich information about the high energy-gamma ray sources of the Universe at cosmological distances.

We focus on the ability of GLAST to detect a signal from dark matter in the EGBR. We present sensitivities for generic thermal WIMPs and the Inert Higgs Doublet Model. Also we discuss the various aspects of astrophysics and particle physics that determines the shape and strength of the signal, such as dark matter halo properties and different dark matter candidates. Other possible sources to the EGBR are also discussed, such as unresolved AGNs, and viewed as backgrounds.

## 1 Introduction

The nature of dark matter (DM) is still one of the most challenging mysteries in present day cosmology and is so far completely unknown. The upcoming Gamma Ray Large Area Space Telescope (GLAST) [2] will survey a previous unexplored window to the high energy $\gamma$-ray universe, playing a crucial role in the indirect detection of weakly interactive massive particle (WIMP) DM through their self annihilation products, resulting in photons. GLAST will pursue different searches for DM, including point source surveys, such as the galactic center, and diffuse emission studies [3] .

In this paper we focus on the diffuse signal from cosmological, extragalactic WIMPs and prospects for GLAST to detect such a signal. We do this for a typical, thermal WIMP and a specific particle physics scenario with an extended Higgs sector.

Also we examine a recent claim that the EGBR is compatible with a 60 GeV cosmological WIMP [4] but where we suspect that the effect of cosmology has not been taken into account in the calculations.

Any sensitivity calculation is dependent on the background of the signal and we present an estimate of possible astrophysical contributions to the EGBR.

### 1.1 DM candidates

There exists many extensions of the standard model of particle physics that contain suitable WIMP DM candidates. Usually these are neutral, stable particles with masses and interaction strengths that give the observed, present day relic abundance. Probably the most studied of such particles is the neutralino, the lightest neutral particle that arises in supersymmetric extensions of the standard model (see, e.g., [5]) and is often used as the archetype for fermionic DM . The mass range of the neutralino is usually from around 50 GeV to a few TeV.

The lightest Kaluza Klein excitation (often the first excitation of the hyper charge gauge boson) gives an archetype for vector bosonic DM with mass in the range of about 0.5 TeV to a few TeV, see for example [16] and references therein. Below we shall also discuss an archetype for scalar DM.

**The Inert Doublet Model** (IDM) [12] is a minimal extension of the standard model – an added second Higgs doublet $H_2$, with an imposed unbroken discrete $Z_2$ symmetry that forbids its direct coupling to fermions (i.e. $H_2$ is *inert*). In the IDM the mass of the particle that plays the role of the standard model Higgs can be as high as about 500 GeV and still fulfill present experimental precision tests. Furthermore, conservation of the $Z_2$ parity implies that the lightest inert Higgs particle ($H^0$) is stable and hence a good DM candidate. One of the interesting features of the IDM is that it offers very high annihilation branching ratios into $\gamma\gamma$ and $Z\gamma$ final states, compared to the branching ratios into quarks, yielding the continuum spectra [13]. The range of WIMP masses is just in the range where GLAST will be sensitive. A spectrum from a cosmological IDM WIMP can be seen in figure 1.

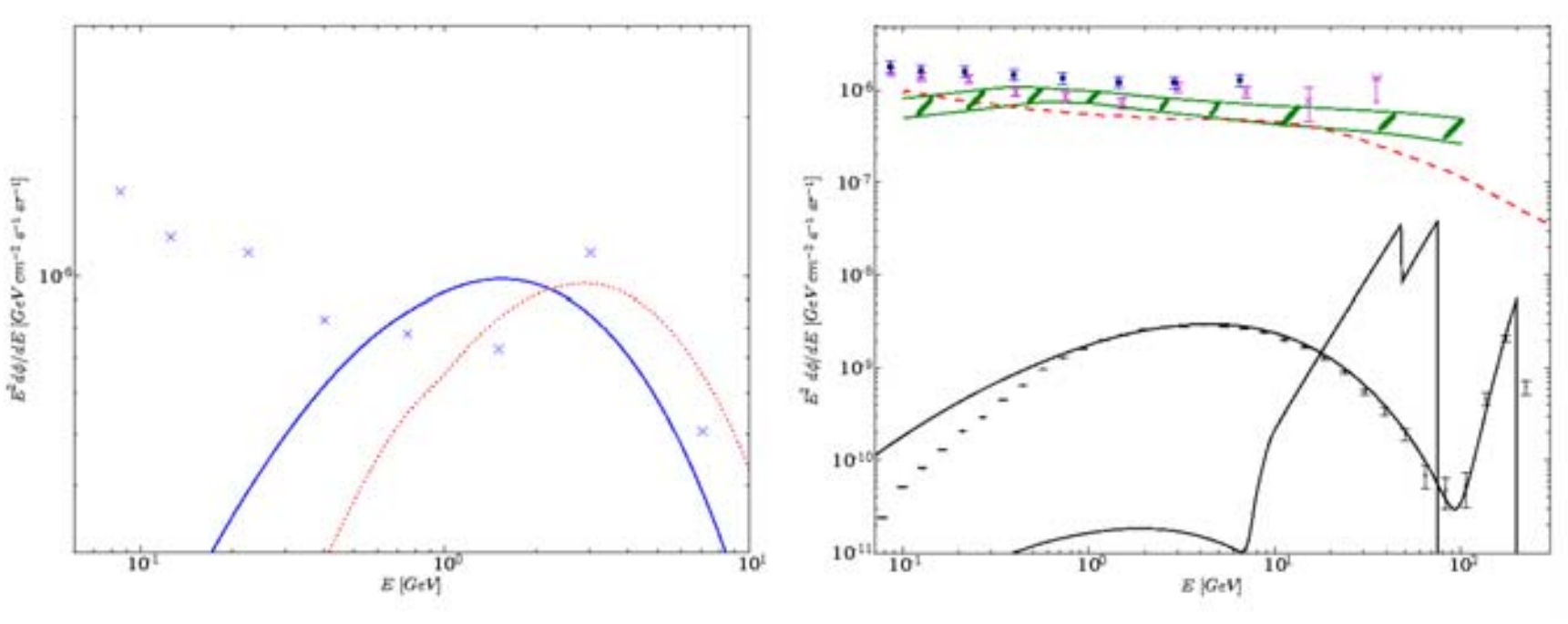

Figure 1: Left panel: Comparing the photon spectra at emission (red dotted) to the cosmological signal (blue solid), from a 60 GeV WIMP, see text. Right panel: EGRET data points (squares from [15], crosses from [14]). The green hatched area represents the upper and lower limits of astrophysical sources contributing to the EGBR, as taken from [17]. The red, dashed line is the unresolved blazar model used in our sensitivity calculation. Also we are showing two examples of cosmological WIMP spectra; a 75 GeV IDM WIMP and a 200 GeV WIMP of the kind used for the GLAST sensitivities, the latter also shows the response of GLAST.

## 2 The cosmological WIMP signal

The diffuse photon-signal originating from DM annihilating throughout the Universe can be calculated in several ways. Here we follow the procedure of [6], where the number of photons per unit effective area, time and solid angle in the redshifted energy range $E_0$ to $E_0 + dE_0$, is given by:

$$\frac{d\phi}{dE_0} = \frac{\sigma v}{8\pi} \frac{c}{H_0} \frac{\bar{\rho}_0^2}{M_\chi^2} \int dz\,(1+z)^3 \frac{\Delta^2(z)}{h(z)} \frac{dN_\gamma(E_0(1+z))}{dE} e^{-\tau(z,E_0)}. \quad (1)$$

In the following we will discuss the various quantities contributing to eq. (1).

### 2.1 High energy $\gamma$-ray environment

Any extragalactic $\gamma$-ray signal is strongly affected by absorption in the intergalactic medium, especially at high energies. The absorption is parameterized by the parameter $\tau$, the optical depth. The dominant contribution to the absorption in the GeV-TeV energy range is pair production on the extragalactic background light emitted in the optical and infrared range. For the optical depth, as function of both redshift and observed energy, we use the results of [7]. Newer calculations of optical depth are now available, see [8]. These results imply a slightly lower optical depth at low redshifts and slightly higher at high redshifts, which in turn slightly enhances or suppresses the WIMP signal, respectively.

### 2.2 Particle physics

The preferred particle physics model enters the differential gamma-ray flux via the cross section, $\sigma$, the WIMP mass $M_\chi$ and the differential gamma-ray yield per annihilation $dN/dE$, which is of the form:

$$\frac{dN_\gamma}{dE} = \frac{dN_{\rm cont}}{dE}(E) + 2b_{\gamma\gamma}\delta(E - M_\chi) + b_{Z\gamma}\delta(E - M_Z^2/4M_\chi)) \qquad (2)$$

The first term in eq. (2) is the contribution from WIMP annihilation into the full set of tree-level final states, containing fermions gauge or Higgs bosons, whose fragmentation/decay chain generates photons. These processes give rise to a continuous energy spectrum. The second and third terms correspond to direct annihilation into final states of two photons and of one photon and one Z boson, respectively. Although of second order (one loop processes), these terms can give rise to significant amounts of monochromatic photons.

Since the emission spectrum of the continuum and line signal are very different in shape, the result of the integration over redshift, in eq. (1), is quite different. The continuum spectra becomes slightly broadened and the peak is red shifted to lower energies. As a rule of thumb one can keep in mind that the total energy emitted ($E^2 d\phi/dE$), as a function of energy, peaks at about $E = M_\chi/20$ for the intrinsic emission continuum spectrum and about $E = M_\chi/40$ for the cosmological spectrum. In figure 1, a comparison between cosmological and emission spectra can be seen.

The line signal is different since all photons are emitted at the same energy, $E = M_\chi$ (in the case of a $2\gamma$-final state) and are observed at the energy $E_0 = E(1+z)^{-1}$, depending on at which redshift the WIMPs annihilated. At high redshifts the universe becomes opaque to high energy photons and the signal goes down dramatically at lower $E_0$. This results in the characteristic spectral feature of a sharp cut-off at the WIMP mass, with a tail to lower energies as seen in figure 1.

The left panel of figure 1 shows a comparison of the photon spectra at emission and the cosmological spectra from a 60 GeV WIMP only taken into account annihilation into $b\bar{b}$. The crosses are the reanalyzed EGBR of EGRET by [4], where a 60 GeV WIMP has been included in the galactic foreground emission model. Without doing any analysis we note that, in contrasts to claims made in [4], the cosmological spectra, from a 60 GeV WIMP, does not peak at the characteristic 3 GeV bump in the EGBR measurement, where the observed emission spectrum peaks.

### 2.3 Halos

The question of how dark matter is distributed on small, galactic and sub-galactic scale is still a matter of debate. However, N-body simulations show that large structures form by the successive merging of small substructures, with smaller objects usually being denser [9]. The density distribution in DM halos, from simulations, are well fitted by simple analytical forms, where the most common one is the NFW profile, [10].

Since the annihilation rate is proportional to the dark matter density squared, any structure in the DM distribution will significantly boost the annihilation signal from cosmological WIMPs. To take this effect into account we again follow the calculations in [6]. The quantity $\Delta^2(z)$ in eq. 1 describes the averaged squared over-density in halos, as a function of redshift.

Clumping the DM into halos typically yields a boost of $10^4 < \Delta^2(z=0) < 10^6$ depending on the choice of halo profile and the model of halo concentration parameter dependence of redshift and halo mass, where we use results from [11]. This freedom of choices introduces about a factor of ten each to the uncertainty in the normalization of the cosmological WIMP signal. However, this can be compared to the uncertainty in the signal from point sources, where only the choice of density profile can change the normalization by several orders of magnitude, which is the case with WIMP signals from the galactic center. Another difference, compared to point sources, is that the astrophysics of the halo concentration parameter can change the shape of the $\gamma$-ray spectrum, which is solely determined by particle physics in the case of point sources.

The largest contribution to $\Delta^2(z)$ comes from small halos formed in an earlier, denser universe. However, our understanding of halos at the low mass end is limited due to finite resolution of the N-body simulation. Therefore we have to use a cut-off mass, below which we do not trust our toymodels for the halo concentration parameters. We put this cut-off at $10^5 M_\odot$. Lowering the cut-off might boost the signal even further but will also introduce further uncertainties.

Also within larger halos, N-body simulations indicates that there should exist smaller, bound halos that have survived tidal stripping. These halos are indicated to have masses all the way down to $10^{-6}\, M_\odot$. Although not as massive as the primary halos the substructure halos arise in higher density environments which makes them denser than their parent halo. The phenomena of halos within halos seems to be a generic feature since detailed simulations reveals substructures even within sub halos [9].

## 3 Astrophysical contributions to the EGBR

The "standard" model for explaining the EGRB is that it consists of diffuse emission from unresolved, $\gamma$-ray point sources such as blazars, quasars, star-burst galaxies and starforming galaxies. Contributions from unresolved blazars, consistent with the EGRET blazar catalogue, could account for about 20% of the measured EGRB at 1 GeV. Taking into account predictions of starburst and starforming galaxies one gets about the measured values of the EGRB at 1 GeV [17]. However, these models under-predict the $\gamma$-ray flux at higher energies, arguing for new, hard $\gamma$-ray sources.

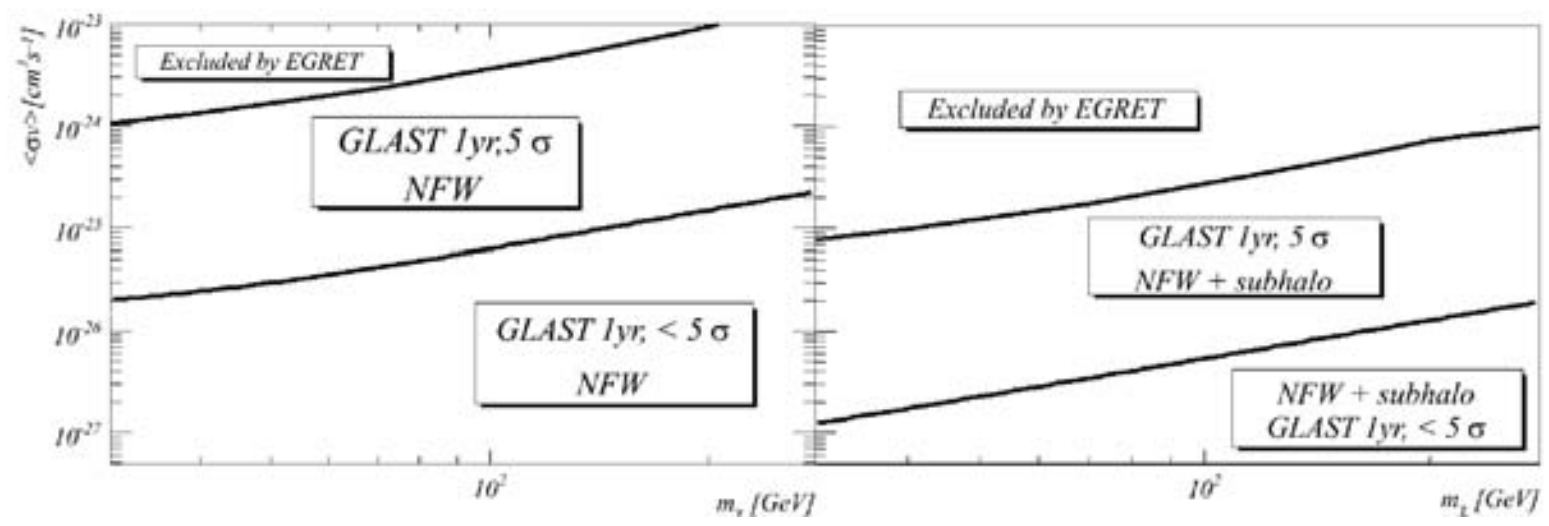

Figure 2: GLAST 1-year, $5\sigma$ sensitivity for generic, thermal WIMPs annihilation into $b\bar{b}$ and a branching of $10^{-3}$ into two-photon lines. See text for details.

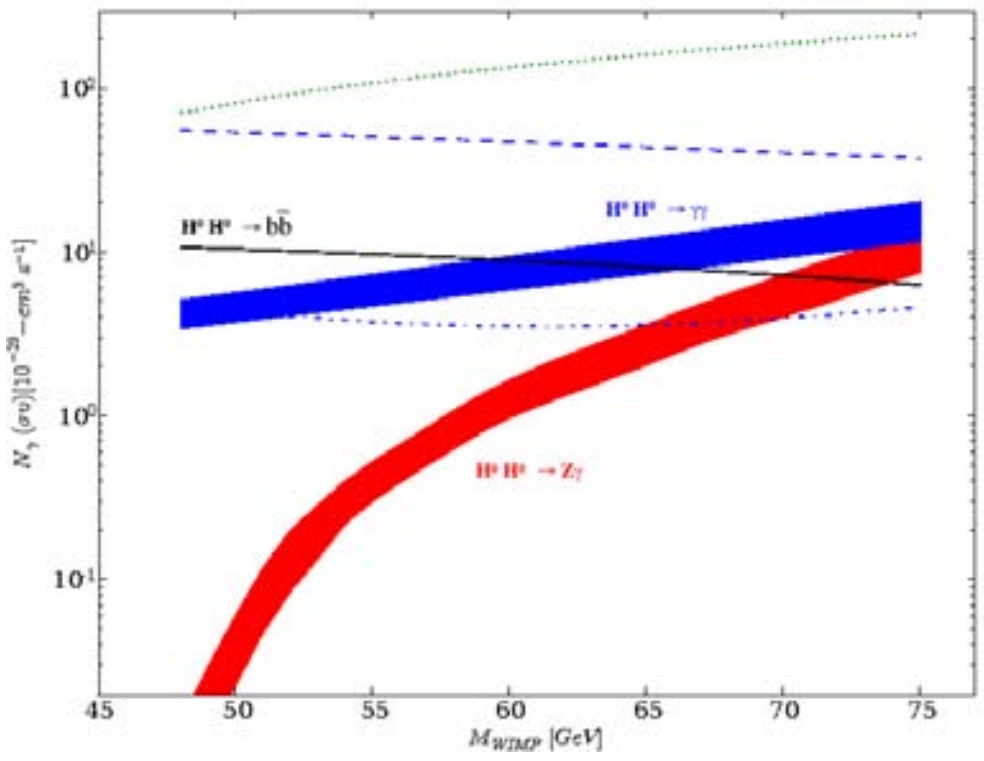

Figure 3: The parameter space of IDM shown as photon flux vs. WIMP mass for different annihilation channels [13]. The dashed line is the 1 year GLAST $5\sigma$ line sensitivity with a NFW profile and the dot dashed line the sensitivity when including substructures. Points in the parameter space that could be resolved by GLAST are the $\gamma\gamma$ fluxes above the sensitivity lines. The green, dotted line marks the region already excluded by EGRET, assuming an NFW profile with substructures.

The background used in our sensitivity calculations consists only of unresolved blazars [6] where GLAST's increased sensitivity to point sources have been taken into account. In figure 1, this background can be compared to other background models as well as EGRET measurements of the EGBR. Note however that the backgrounds from [17] are not treated together in a consistent way, for instance with respect to the optical depth. However, this has been done for the unresolved blazar model and the cosmological WIMPs.

## 4 GLAST sensitivity

Fast detector simulations [18] were done for a generic model of WIMPs annihilating into $2\gamma$ and into $b\bar{b}$ for WIMP masses between 30 and 280 GeV. A $\chi^2$ analysis was performed, assuming that the background consists of unresolved blazars, to obtain a sensitivity plot in $< \sigma v >$ vs $M_\chi$. Also, to the background we added an irreducible contribution from charged particles, at the level of 10 % of the blazar background. It should be noted that for the calculations presented here we somewhat optimistically assume that we have an ideal extraction of the EGBR as well as a perfect understanding of the conventional astrophysical backgrounds.
The WIMP signal was computed using a NFW profile and including the effect of substructures, assuming that they constitute 5% of the mass and have four times the concentration parameter of the parent halo. The result, viewed in figure 2, shows that GLAST is sensitive to total annihilation cross-sections of the order $10^{-24} - 10^{-27}$ cm$^3$ s$^{-1}$, depending on the exact halo model, for low masses and about an order of magnitude higher cross section for higher masses. In figure 3 the line sensitivity of GLAST to IDM cosmological WIMPs can be seen. Since the IDM offers so much higher ratio between the line branching and continuum branching the sensitivity was calculated only for the lines as $n_{\gamma,\chi}/\sqrt{n_{bkg}}$. The result is quite dependent on the choice of halo model. For the plain NFW profile GLAST cannot reach a $5\sigma$ level within one year but when adding substructures GLAST is sensitive to almost the entire parameter space of the IDM, at a $5\sigma$ level.

## 5 Conclusion

We have shown that studying the EGBR with GLAST could offer an interesting way of indirect detection of WIMP DM. Also since the cosmological signal differs in many ways from other point like sources of DM, it will be a useful compliment to such surveys. The level of sensitivity of GLAST still depends on many unknowns, many of them associated with the fact that we do not know enough about the nature of DM. But under our assumptions we find that GLAST will be sensitive to a wide range of interesting WIMP candidates. However, both in the case of the generic and the IDM WIMP the signal needs a light boost for GLAST to be able to cover the most interesting region which could be achieved by adding substructures in the halos.

# GAMMA-RAYS, ANTIMATTER AND THE COSMOLOGICAL EVOLUTION OF THE UNIVERSE

M.SCHELKE[a], R.CATENA[b] F.DONATO[a,c] N.FORNENGO[a,c] A.MASIERO[d,e]
M.PIETRONI[e]

[a] *INFN, Sezione di Torino, via P. Giuria, Torino, Italy*

[b] *Deutsches Elektronen–Syncrotron DESY, 22603 Hamburg, Germany*

[c] *Dipartimento di Fisica, Università di Torino, via P. Giuria, Torino, Italy*

[d] *Dipartimento di Fisica, Università di Padova, via Marzolo, Padova, Italy*

[e] *INFN, Sezione di Padova, via Marzolo, Padova, Italy*

**Abstract**

There exist a number of viable cosmological models which predict a pre-Big-Bang-Nucleosynthesis Hubble rate enhanced compared to the case of standard cosmology. The increased expansion rate has important consequences for the freeze out of the Dark Matter particles. Consequently, the enhancement of the pre-BBN Hubble rate can be constrained by the upper bounds on the Dark Matter annihilation cross section derived from the searches for indirect DM signals.

## 1 Introduction

In this paper we are going to report on and extend the analyses of our papers Ref. [1] and [2]. The basic idea, that we are going to explain further in Section 2, is that by combining the constraint on Dark Matter relic abundance and annihilation cross section we can derive constraints on the enhancement of the Hubble rate in the very early Universe. As a Dark Matter candidate we consider a generic weakly interacting massive particle (WIMP), $\chi$, which is produced in thermal equilibrium in the early Universe. We shall assume that the WIMP makes up the main part of the Cold Dark Matter (CDM) in the Universe. From cosmological observations we have that the CDM density falls in the following narrow interval (at $2\sigma$ C.L.) [3]

$$0.092 \leq \Omega_{\rm CDM} h^2 \leq 0.124 \tag{1}$$

where as usual $\Omega$ denotes the ratio between the mean density and the critical density and $h$ is the present Hubble constant in units of 100 $\rm km\,s^{-1}\,Mpc^{-1}$.

## 2 The relic density calculation

In this section we present the main idea on which this paper is based: If the expansion rate in the Universe differs from that of standard cosmology then this affects the freeze out of the CDM particles. This has been discussed in Ref. [4] for the expansion rate of a scalar–tensor gravity model, in Refs. [5] for the case of a Quintessence model with a kination phase and in [6] for braneworld models. See also Refs. [7] for the case of anisotropic expansion and other models of modified expansion. To understand the idea we must solve the Boltzmann equation, which controls the relic density of the WIMP DM candidate, assumed to be in thermal equilibrium in the early Universe.

$$\frac{dn}{dt} = -3Hn - \langle\sigma_{\rm ann} v\rangle (n^2 - n_{\rm eq}^2) \tag{2}$$

where $n$ is the WIMP number density, $n_{\rm eq}$ its equilibrium value and $t$ the time. Whereas $\langle\sigma_{\rm ann} v\rangle$ is the thermally averaged value of the WIMP annihilation cross section times the relative velocity and $H(t)$ is the Hubble expansion rate. This is the Boltzmann equation as it also looks in the standard case, but in our case we imagine that the Hubble rate of standard cosmology has been substituted by the expansion rate of a modified cosmology, *i.e.* $H(t) = A(t) H_{\rm std}(t)$. The analytic solution of the modified Boltzmann equation is

$$\Omega_\chi h^2 \sim {\rm const} \cdot \left( \int_{x_{\rm f}}^{\infty} \frac{h_{\rm eff}(x)\, \langle\sigma_{\rm ann} v\rangle}{A(x)\, \sqrt{g_{\rm eff}(x)}\, x^2}\, dx \right)^{-1} \tag{3}$$

where $h_{\rm eff}$ and $g_{\rm eff}$ are respectively the entropy–density and energy–density effective degrees of freedom. We have also introduced the usual definition $x = m_\chi/T$, where $m_\chi$ is the WIMP mass and $T$ the temperature. The integration starts at the WIMP freeze–out temperature and ends today. We note that the standard solution is recovered when $A(t) = 1$.

The Boltzmann equation and its solution describes how the relic WIMP density is determined by the competition between the expansion of the Universe and the particle processes that create and destroy the WIMP. A large WIMP annihilation cross section makes the particle processes able to compete with the dilution of the expansion for a longer time, and consequently reduces the relic WIMP density. On the contrary, an increase in the Hubble rate, *i.e.* $A > 1$, will make a given WIMP freeze out earlier than in standard cosmology, with the consequence of an increased relic WIMP density. Assuming that the WIMP makes up the mayor part of the CDM, we have very strong observational constraints on its relic density, Eq. (1). For the density constraint to be fulfilled, cosmologies with an enhancement of the Hubble rate in the early Universe select WIMP candidates with a larger annihilation cross section than does the standard cosmology scenario. The larger the Hubble rate, the larger the cross section. In the references mentioned in the beginning of this Section, it has been shown that the WIMP annihilation cross section favoured by models of modified cosmologies can be many orders of magnitude larger than the cross section favoured in the standard case. This without violating observables related to BBN, Large Scale structure, General Relativity tests etc. In our papers Ref. [1] and [2], we went one step further. We used the fact that the WIMP annihilation cross section (today) is constrained from above by the results of indirect DM searches. Combining this constraint with the constraint on the relic CDM density, we have derived constraints on the enhancement of the Hubble rate in the early Universe. In these proceedings we report on our previous results and we also show some new calculations for the prospects for the GLAST satellite.

## 3 Models with enhanced pre-BBN expansion

As we saw in the previous section, cosmological models with an enhanced Hubble expansion rate in the early Universe introduce interesting effects in the Dark Matter sector. In order not to be in conflict with the successful predictions of BBN and the formation of the Cosmic Microwave Background Radiation, we must require that the Hubble rate recover the standard evolution no later than at the time of the BBN.

$$H = A(T) H_{\rm std} \quad T >\sim 1 \text{ MeV} \quad ; \quad H = H_{\rm std} \quad T <\sim 1 \text{ MeV} \tag{4}$$

There exist a number of interesting cosmological models with $A(T) > 1$. In this paper we discuss quintessence models with a kination phase, the RSII braneworld model and scalar-tensor models. The enhancement function, $A(T)$, of these models can be described by the following parametrization:

$$A(T) = 1 + \eta \left(\frac{T}{T_{\rm f}}\right)^{\nu} \tanh\left(\frac{T - T_{\rm re}}{T_{\rm re}}\right) \tag{5}$$

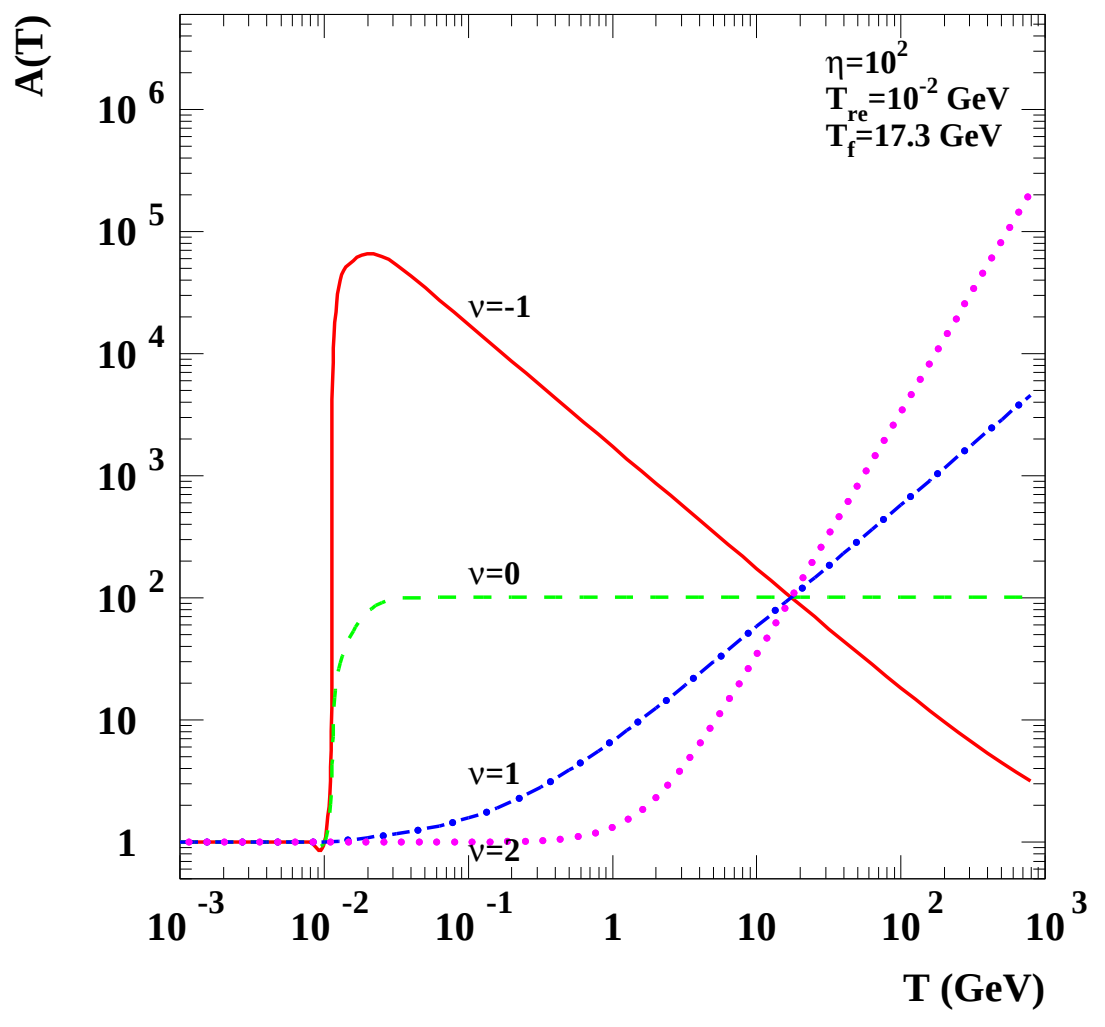

Figure 1: The enhancement function of Eq. 5 as a function of the temperature. Note that the time is running from right to left. The function is shown for 4 values of the slope parameter $\nu$ and for a specific choice of the other parameters. Figure from Ref. [1].

for temperatures $T > T_{\rm re}$ and $A(T) = 1$ for $T \leq T_{\rm re}$. By $T_{\rm re}$ we denote the temperature at which the Hubble rate "*re–enters*" the standard rate. The hyperbolic tangent has been chosen to assure that $A(T)$ goes continuously to "1", and $H \rightarrow H_{\rm std}$ before the "*re–entering*" temperature. As we showed in Ref. [1], the derived constraint is insensitive to the value of the reentering temperature, $T_{\rm re}$, as long as this is much lower than the other temperature parameter, $T_{\rm f}$. We choose $T_{\rm re} = 10^{-3}$ GeV (except for in Fig. 1) , which is always much lower than $T_{\rm f}$ for which we use the temperature where the DM particle freezes out in standard cosmology. $T_{\rm f}$ is not a free parameter, but is determined by the WIMP mass and annihilation cross section. The enhancement function $A(T)$ of Eq. (5) is shown in Fig. 1 for some specific choice of the parameters. The slope parameter $\nu$ is determined by the cosmological model: The RSII braneworld model [8] can be describe by $\nu = 2$, quintessence models with a kination phase by $\nu = 1$, while some specific scalar–tensor model can be approximated by $\nu = -1$. For further details about the parameters $\nu$ and $\eta$ in different cosmologies, we refer to our papers [1] and [2]. We note that $A(T_f) = \eta$ (for $\eta \gg 1$ and $T_f \gg T_{re}$). In Sec. 5 we derive the bound on $\eta$.

## 4 Upper bounds on the cross section

As we found in Section 2, we can derive constraints on the expansion rate in the early Universe if we can constrain the WIMP annihilation cross section. What matters is clearly the WIMP annihilation cross section in the early Universe. In this paper we are going to assume that the cross section is dominantly temperature–independent (or s–wave). This mean that we directly can use the constraints on the WIMP annihilation cross section today, which can be derived from the data of indirect searches for Dark Matter. Some examples of deviation from this case was discussed in our paper [1].

Let us first consider the upper bound on the WIMP annihilation cross section derived from the HESS data of the $\gamma$–ray signal from the Galactic Center (GC). The HESS Collaboration has reported in Refs. [9] on a power–law shaped spectrum of very high energy $\gamma$–rays from a point–like source in the GC. The GC hosts different astrophysical candidates for the observed spectrum. Also annihilating WIMP would give rise to a $\gamma$-ray signal. As it was discussed in *e.g.* [10] it is very difficult to explain all the HESS GC point-like signal by annihilating CDM. As also done in Ref. [10] we therefore try to explain the HESS signal by adding a WIMP signal to a power law $kE_\gamma^\Gamma$ assumed to come from some astrophysical source. The power–law parameters $k$ and $\Gamma$ and the WIMP annihilation cross section are taken as free parameters. For each DM halo model and each choice of WIMP mass, we make a $\chi^2$ analysis to find the maximum allowed value of the WIMP annihilation cross section, which is in agreement with the HESS data. For further details of this analysis see Ref. [2].

In Fig. 2 we show the HESS upper limit on the WIMP annihilation cross section as a function of the WIMP mass. In this paper we show the result only for the case of a NFW halo profile. As we have shown in Ref. [2] the bound can differ by many orders of magnitude from one halo profile to another. In Fig. 2 we also show the upper bound from searches for an indirect WIMP signal in the cosmic antiproton flux. The observed flux can be explained by the standard production from spallation of nuclei on the diffuse Milky Way gas, but the error–bars also leave a little room for a signal of exotic origin. In the Figure we show the bound found by using the best–fit values for the galactic propagation and diffusion parameters. For more details on the derivation of the antiproton bound we refer to our paper Ref. [1] and references therein. As a new feature we have included in Fig. 2 the prospects for the GLAST satellite–based $\gamma$–ray telescope. The GLAST sensitivity has been estimated in *e.g.* Ref. [11]. We use their result for a WIMP model independent framework. They have added the $\gamma$–ray signal from a dominant $b\bar{b}$ WIMP annihilation channel to a standard

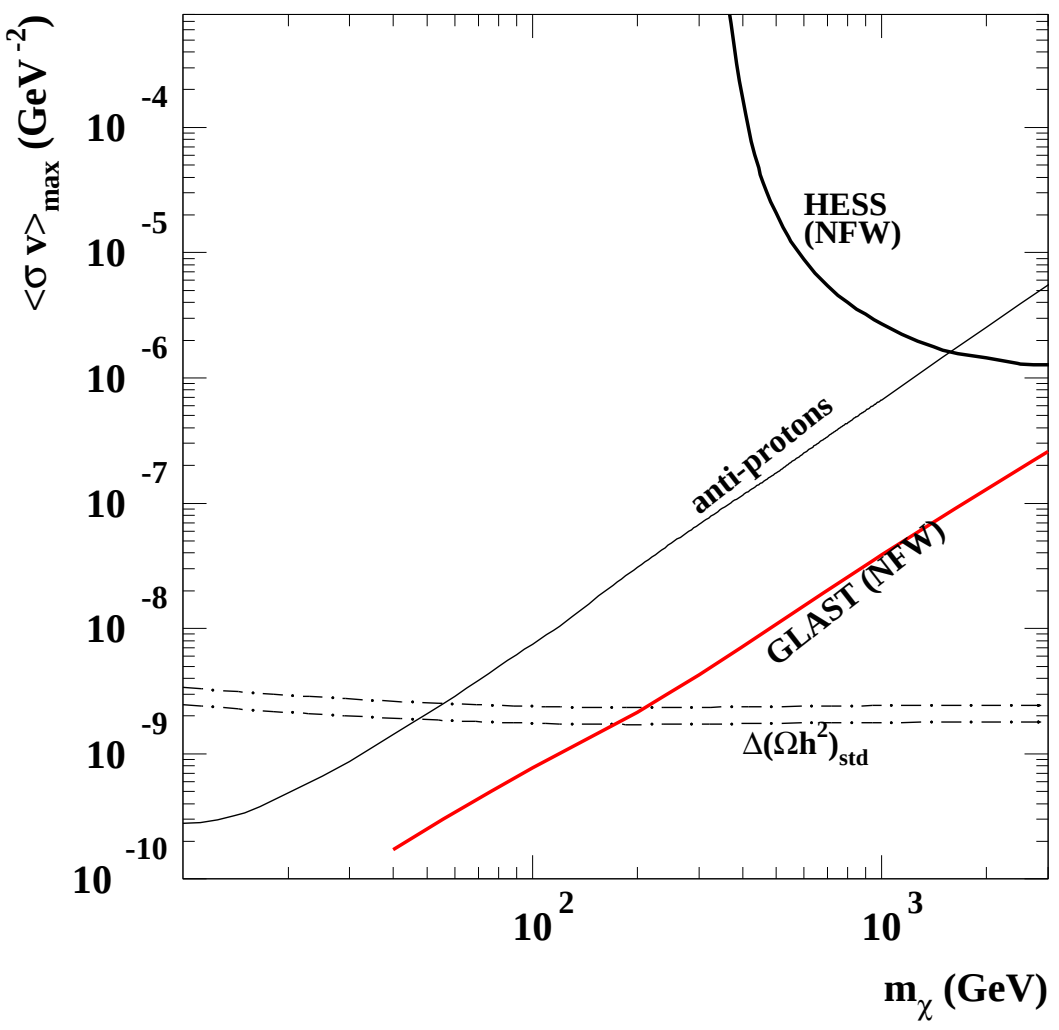

Figure 2: Upper bounds on the WIMP annihilation cross section as a function of the WIMP mass. The upper bound is shown as derived from different indirect searches for DM. See the text for further information.

astro–physical $\pi^0$ background. The resulting $3\sigma$ GLAST reach is shown in our Fig. 2, assuming a NFW halo profile and 4 years of data taking. For further details we refer to Ref. [11] and references therein. WIMP's with an annihilation cross section below the GLAST line in Fig. 2 cannot be detected by GLAST.

Finally, Fig. 2 also show the small band of the WIMP annihilation cross section derived applying the CDM relic density constraint, Eq. (1), and assuming the standard cosmological model and a temperature–independent WIMP annihilation cross section. Models, that stay above this "density–band" would be underabundant in standard cosmology. The idea, that we explained in Section 2, is now that in models with an enhanced pre-BBN Hubble rate we can fill all the space between the lower and the upper limit of the cross section and still fulfil the density constraint. We just have to adjust the enhancement of the expansion rate in order to obtain the correct amount of DM. The larger the cross section the larger the enhancement.

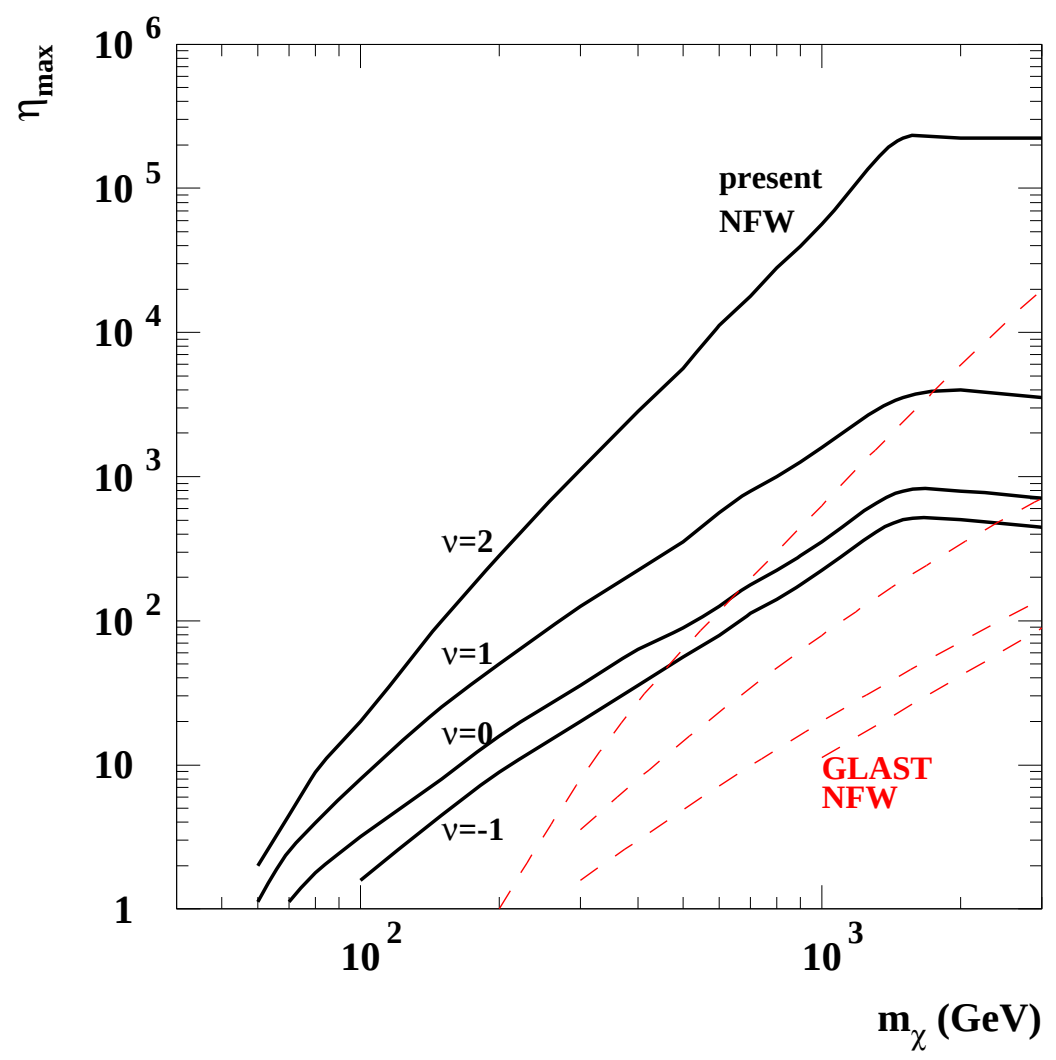

Figure 3: The upper bound of the enhancement parameter $\eta$ as derived by combining the upper limit on the WIMP cross section, Fig. 2, and relic density. The result is show for 4 values of the slope parameter $\nu$ for the present limit as well as for the GLAST prospects. A NFW halo profile has been assumed.

## 5 Present and future constraints on the expansion rate

In this Section we are going to find the upper limit of the enhancement parameter $\eta = A(T_{\rm f})$ for different choices of cosmology, $\nu$. It is interesting to note that this means that we constrain the expansion rate in the very early Universe, at $T_{\rm f}$ which is much earlier than BBN. Furthermore, once we know $\eta$, we know $A(T)$ at any temperature for a given cosmological model and $T_{\rm f}$.

For each WIMP mass we take the upper bound on the WIMP annihilation cross section, $\langle\sigma_{\rm ann}v\rangle_0$, as it can be found from Fig. 2. For the current bound we combine the antiproton and the HESS bound, taking always the most constraining of the two. For the future prospects we use the GLAST sensitivity limit. From the WIMP mass and cross section we can determine $T_{\rm f}$, the WIMP freeze–out temperature in standard cosmology. For each cosmological model, $\nu$, we then solve the Boltzmann Equation while making a scan in the enhancement parameter $\eta$. The upper bound on $\eta$ is the one where the solution of the Boltzmann Equation gives a WIMP density equal to the observational upper bound in Eq. (1). We note again that this method is based on the assumption that the WIMP annihilation cross section is temperature–independent. We show the result in Fig. 3 for four different choices of the slope–parameter $\nu$. The lower mass limit of the curves are not exact, but merely due to the step-size in the scan. We see that the derived constraints are very strong for low WIMP masses. We see that the exact limit depend strongly on the cosmological model, being strongest for a scalar–tensor model, $\nu = -1$. As we showed in Ref. [2], the derived limits depend also very strongly on the halo profile chosen. The strongest constraints were found for steep halo profiles like the Moore profile.

## 6 Conclusions

In this paper we have shown that indirect searches for Dark Matter can be used to derive strong constraints on the possible enhancement of the Hubble expansion rate in the very early Universe.

# DIFFUSE GAMMA-RAY AND NEUTRINO EMISSIONS OF THE GALAXY

C. EVOLI[a], D. GAGGERO [b,c], D. GRASSO[c], L. MACCIONE [a,d]

[a] *SISSA, via Beirut, 2-4, I-34014 Trieste, Italy*

[b] *Università di Pisa, Dipartimento di Fisica Largo B. Pontecorvo 3, I-56127 Pisa, Italy*

[c] *INFN, Sezione di Pisa, Largo B. Pontecorvo 3, I-56127 Pisa, Italy*

[d] *INFN, Sezione di Trieste, INFN, Sezione di Trieste, Via Valerio, 2, I-34127 Trieste, Italy*

## Abstract

We present recent results concerning the $\gamma$-ray and neutrino emissions from the Galactic Plane (GP) as should be originated from the scattering of cosmic ray (CR) nuclei with the interstellar medium (ISM). By assuming that CR sources are distributed like supernova remnants (SNRs) we estimated the spatial distribution of primary nuclei by solving numerically the diffusion equation. For the first time in this context we used diffusion coefficients as determined from Montecarlo simulations of particle propagation in turbulent magnetic fields. Concerning the ISM, we considered recent models for the $H_2$ and HI spatial distributions which encompass the Galactic Centre (GC) region. Above the GeV we found that the angular distribution of the simulated $\gamma$-ray emission along the GP matches well EGRET measurements. We compare our predictions with the experimental limits/observations by MILAGRO and TIBET AS$\gamma$ (for $\gamma$-rays) and by AMANDA-II (for neutrinos) and discuss the perspectives for a km3 neutrino telescope to be built in the North hemisphere.

## 1 Introduction

Several orbital observatories and especially EGRET [1, 2], found that, at least up to 10 GeV, the Galaxy is pervaded by a $\gamma$-ray diffuse radiation. While a minor component of that emission is likely to be originated by unresolved point-like sources, the dominant contribution is expected to come from the interaction of galactic cosmic rays (CR) with the interstellar medium (ISM). Since the spectrum of galactic CRs extends up to the EeV, the spectrum of the $\gamma$-ray diffuse galactic emission should also continue well above the energy range probed by EGRET. That will be soon probed by GLAST [3] up to 300 GeV and by air shower arrays (ASA) (e.g. MILAGRO [4, 5] and TIBET [6]) above the TeV.

Above the GeV, the main $\gamma$-ray emission processes are expected to be the decay of $\pi^0$ produced by the scattering of CR nuclei onto the diffuse gas (hadronic emission) and the Inverse Compton (IC) emission of relativistic electrons colliding onto the interstellar radiation field (leptonic emission). It is unknown, however, what are the relative contributions of those two processes and how they change with the energy and the position in the sky (this is so called *hadronic-leptonic degeneracy*). Several numerical simulations have been performed in order to interpret EGRET as well as forthcoming measurements at high energy (see e.g. [7, 8] ). Generally, those simulations predict the hadronic emission to be dominating between 0.1 GeV and few TeV, while between 1 and 100 TeV a comparable, or even larger IC contribution may be allowed.

The 1-100 TeV energy range, on which we focus here, is also interesting from the point of view of neutrino astrophysics. In that energy window neutrino telescopes (NTs) can look for up-going muon neutrinos and reconstruct their arrival direction with an angular resolution better than $1^\circ$. Since hadronic scattering give rise to $\gamma$-rays and neutrinos in a known ratio, the possible measurement of the neutrino emission from the GP may allow to get rid of the hadronic-leptonic degeneracy.

In this contribution we discuss the main results of a recent work where we modelled the $\gamma$-ray and neutrino diffuse emissions of the Galaxy due to CR hadronic scattering [9]. Our work improves previous analysis under several aspects which concern the distribution of CR sources; the way we treated CR diffusion which accounts for spatial variations of the diffusion coefficients; the distribution of the atomic and molecular hydrogen.

## 2 The spatial structure of the ISM

In order to assess the problem of the propagation of CRs and their interaction with the ISM we need the knowledge of three basic physical inputs, namely: the distribution of SuperNova Remnants (SNR) which we assume to trace that of CR sources; the properties of the Galactic Magnetic Field (GMF) in which the propagation occurs; the distribution of the diffuse gas providing the target for the production of $\gamma$-rays and neutrinos through hadronic interactions.

### 2.1 *The SNR distribution in Galaxy*

Several methods to determine the SNR distribution in the Galaxy are discussed in the literature (see e.g. that based on the surface brightness - distance ($\Sigma-D$) relation [10]). Here we adopt a SNR distribution which was inferred from observations of pulsars (for core-collapsed SNe) and old progenitor stars (for type-Ia SNe)[11]. Respect to surveys based on the $\Sigma-D$ relations, this approach is less plagued from sistematics and it agrees with the observed distribution of radioactive nuclides like of $^{26}$Al which are known to be correlated with SNRs. A similar approach was followed in [12] where, however, the contribution of type I-a SNRs (which is dominating in the inner 1 kpc) was disregarded.

### 2.2 *Regular and random magnetic fields*

The Milky Way, as well as other spiral galaxies, is known to be permeated by large-scale, so called *regular*, magnetic fields as well as by a random, or turbulent, component. The orientation and strength of the regular fields is measured mainly by means of Faraday Rotation Measurements (RMs) of polarised radio sources. From those observations it is known that the regular field in the disk of the Galaxy fills a *halo* with half-width $z_r \simeq 1.5$ kpc and that, out of galactic bulge, it is almost azimuthally oriented. Following [13, 14] we adopt the following analytical distribution in the halo:

$$B_{\rm reg}(r,z) = B_0 \exp\left\{-\frac{r-r_\odot}{r_B}\right\} \frac{1}{2\cosh(z/z_r)} \ , \qquad (1)$$

where $B_o \equiv B_{\rm reg}(r_\odot, 0) \simeq 2\ \mu$G is the strength at the Sun circle. The parameters $r_B$ is poorly known. Fortunately, we found that our final results are practically independent on its choice. In the following we adopt $r_B \simeq r_\odot \simeq 8.5$ kpc.

More uncertain are the properties of the turbulent component of the GMF. Here we assume that its strength follows the behaviour

$$B_{\rm ran}(r,z) = \sigma(r)\ B_{\rm reg}(r,0)\ \frac{1}{2\cosh(z/z_t)} \ . \qquad (2)$$

where $\sigma(r)$ provides a measure of the turbulence strength. Here we assume $z_t =$ 3 kpc. From polarimetric measurements and RMs is known GMF are chaotic on all scales below $L_{\rm max} \sim 100$ pc. The power spectrum of the those fluctuations is also poorly known. While in [9] we considered both a Kolmogorov ($B^2(k) \propto k^{-5/3}$) and a Kraichnan ($B^2(k) \propto k^{-3/2}$) power spectra in the following we consider only a Kolmogorov spectrum.

### 2.3 *The gas distribution*

The model which we consider here is based on a suitable combination of different analyses which have been separately performed for the disk and the galactic bulge. For the $H_2$ and HI distributions in the bulge we use a detailed 3D model recently developed by Ferriere et al. [15] on the basis of several observations. For the molecular hydrogen in the disk we use the well known Bronfman's et al. model [16]. Since in [16] $r_\odot = 10$ kpc was adopted, we correct the gas density and the scale height given in that paper to make them compatible with the value $r_\odot = 8.5$ kpc which we use in this work. Furthermore, we accounted for a radial dependence of the $H_2$-CO conversion factor $X$. Here we assume $X = 0.5 \times 10^{20}$ cm$^{-2}$ K$^{-1}$ km$^{-1}$s for $r < 2$ kpc and $X = 1.2 \times 10^{20}$ cm$^{-2}$ K$^{-1}$ km$^{-1}$s for $2 < r < 10$ kpc. For the HI distribution in the disk, we adopt Wolfire et al. [17] analytical model. We also assume that the ISM helium is distributed in the same way of hydrogen nuclei.

## 3 CR diffusion

The ISM is a turbulent magneto-hydro-dynamic (MHD) environment. Since the Larmor radius of high energy nuclei is smaller than $L_{\rm max}$, the propagation of those particles takes place in the spatial diffusion regime. We solved the diffusion equation following the approach developed in [18]. In the energy range considered in our work energy losses/gains can be safely neglected. Since we assume cylindrical symmetry the only physically relevant component of the diffusion tensor is the perpendicular diffusion coefficient $D_\perp$. We verified that Hall diffussion can be neglected at the energies considered in our work. It is crucial to know how $D_\perp$ changes as a function of the energy and of the turbulence strength (which, as we mentioned, may depend on the position). Here we adopt expression of $D_\perp(E, \sigma)$ which have been derived by means of Montecarlo simulation of charge particle propagation in turbulent magnetic fields [19]. Respect to other works, where a mean value of the diffusion coefficient has been estimated from the observed secondary/primary ratio of CR nuclear species (see e.g. [8]), our approach offers the advantage to provide the diffusion coefficients *point-by-point*. We solved the diffusion equation using the Crank-Nicholson method by imposing $N(E) = 0$ at the edge of the MF turbulent halo ($r = 30$ kpc, $z = 3$kpc) and by requiring that it matches the observed CR spectra at the Earth position for most abundant nuclear species.

## 4 Mapping the $\gamma$-ray and $\nu$ emission

Under the assumption that the primary CR spectrum is a power-law and that the differential cross-section follows a scaling behaviour (which is well justified at the energies considered here), the $\gamma$-ray (muon neutrino) emissivity due to hadronic scattering can be written as

$$\frac{dn_{\gamma\ (\nu)}(E;\ b, l)}{dE} \simeq f_N\ \sigma_{pp}\ Y_\gamma(\alpha) \int \mathrm{d}s\ I_p(E_p;\ r, z)\ n_H(r, z)$$

Here $I_p(E_p;\ r, z)$ is the CR proton differential flux at the position $r, z$ as determined by solving the diffusion equation; $\sigma_{pp}$ is the $pp$ cross-section; $Y_\gamma \simeq 0.04$ and $Y_{\nu_\mu + \bar{\nu}_\mu} \simeq 0.01$ are, respectively, the $\gamma$-ray and muon neutrino yields as obtained for a proton spectral slope $\alpha = 2.7$ [20]; the factor $f_N \simeq 1.4$ represents the contribution from the other main nuclear species both in the CR and the helium in the ISM; $s$ is the distance from the Earth; $b$ and $l$ are the galactic latitude and longitude.

In the following we assume that the turbulent component of the GMF has a Kolmogorov spectrum (in [9] we also considered a Kraichnan spectrum), which implies $D_\perp \propto E^{1/3}$, and that the turbulence strength trace the SNR radial distribution being normalised to the value $\sigma(r_\odot) = 1$.

In order to verify the consistency of our findings with EGRET observations, we extrapolated our results down to few GeV's. By doing that we assume that the energy dependence of $D_\perp$ does not change going from the TeV down to few GeVs. In fig. 4 we compare the longitudinal $\gamma$-ray flux profile as obtained with our model with EGRET measurements in the $4-10$ GeV energy range [21, 2]. Clearly, as we used an analytical expression for the gas distribution, not all fea-

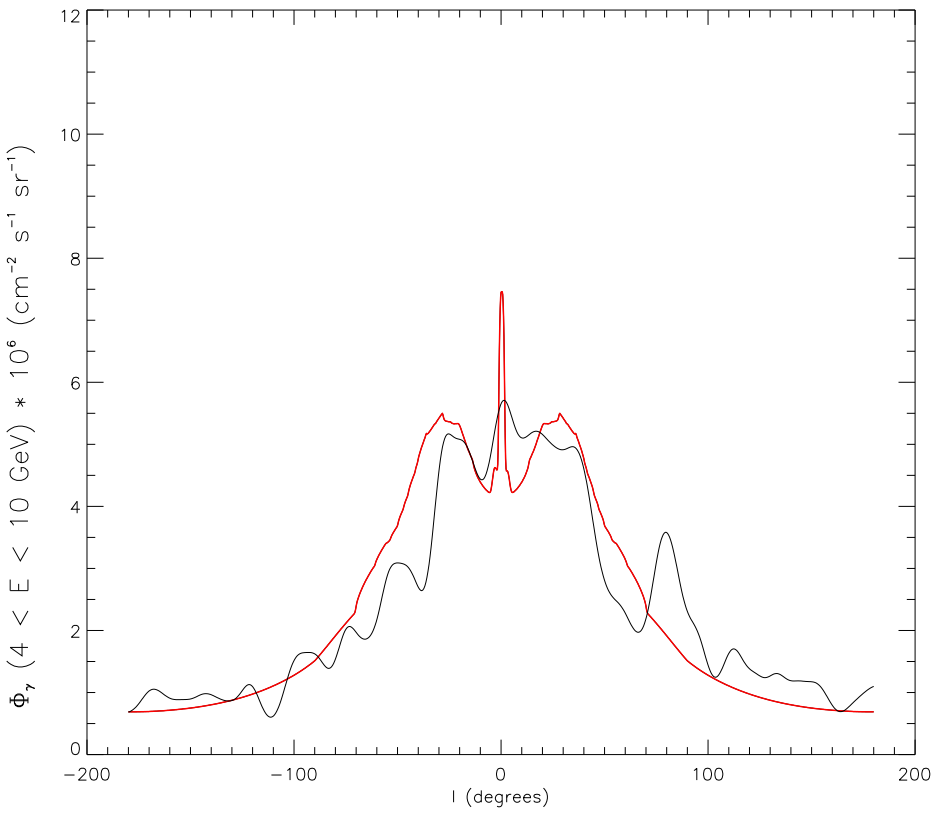

Figure 1: The simulated $\gamma$-ray flux profile along the GP integrated between 4 and 10 TeV (red, continuos line) is compared with that measured by EGRET (black, dashed line). Both simulated and are averaged over $|b| < 1°$.

tures in the EGRET diffuse sky map can be reproduced. It is evident, however,

Table 1: Our predictions for the mean $\gamma$-ray flux are compared with some available measurements. Since measurement's errors are much smaller than theoretical uncertainties they are not reported here.

| sky window | $E_\gamma$ | $\Phi_\gamma(>E_\gamma)$ $(\mathrm{cm}^2\ \mathrm{s\ sr})^{-1}$ | |
|---|---|---|---|
| | | our model | measurements |
| $\|l\| < 0.3°$, $\|b\| < 0.8°$ | 1 TeV | $\simeq 2 \times 10^{-9}$ TeV | $\simeq 1.6 \times 10^{-8}$ [22] |
| $20° < l < 55°$, $\|b\| < 2°$ | 3 TeV | $\simeq 5.7 \times 10^{-11}$ | $< 3 \times 10^{-10}$ [6] |
| $73.5° < l < 76.5°$, $\|b\| < 1.5°$ | 12 TeV | $\simeq 2.9 \times 10^{-12}$ | $\simeq 6.0 \times 10^{-11}$ [5] |
| $140° < l < 200°$, $\|b\| < 5°$ | 3.5 TeV | $\simeq 5.9 \times 10^{-12}$ | $< 4 \times 10^{-11}$ [4] |

that our simulations provide a good description of EGRET measurements on large scales. No tuning, neither of the SNR radial distribution nor of the $X_{CO}$ conversion factor, seems to be required to match the emission from the GP.

## 5 Discussion

Reassured by our good description of EGRET data, we modelled the hadronic emission above the TeV both for the $\gamma$-rays an the neutrinos. In Fig.2 we show two significant $\gamma$-ray flux profiles above the TeV. In Tab. 1 we compare our results with MILAGRO and TIBET AS$\gamma$ measurements. We found that, while over most of the sky the predicted diffuse flux is significantly below the experimental upper limits, in the Cygnus region and in the GC Ridge [22] the observed flux exceed the theoretical expectations. We interpret those excesses to be originated by local concentrations of CR which are likely to be correlated to molecular gas cloud complexes and that cannot be accounted by means of the analytical distributions used in this work.

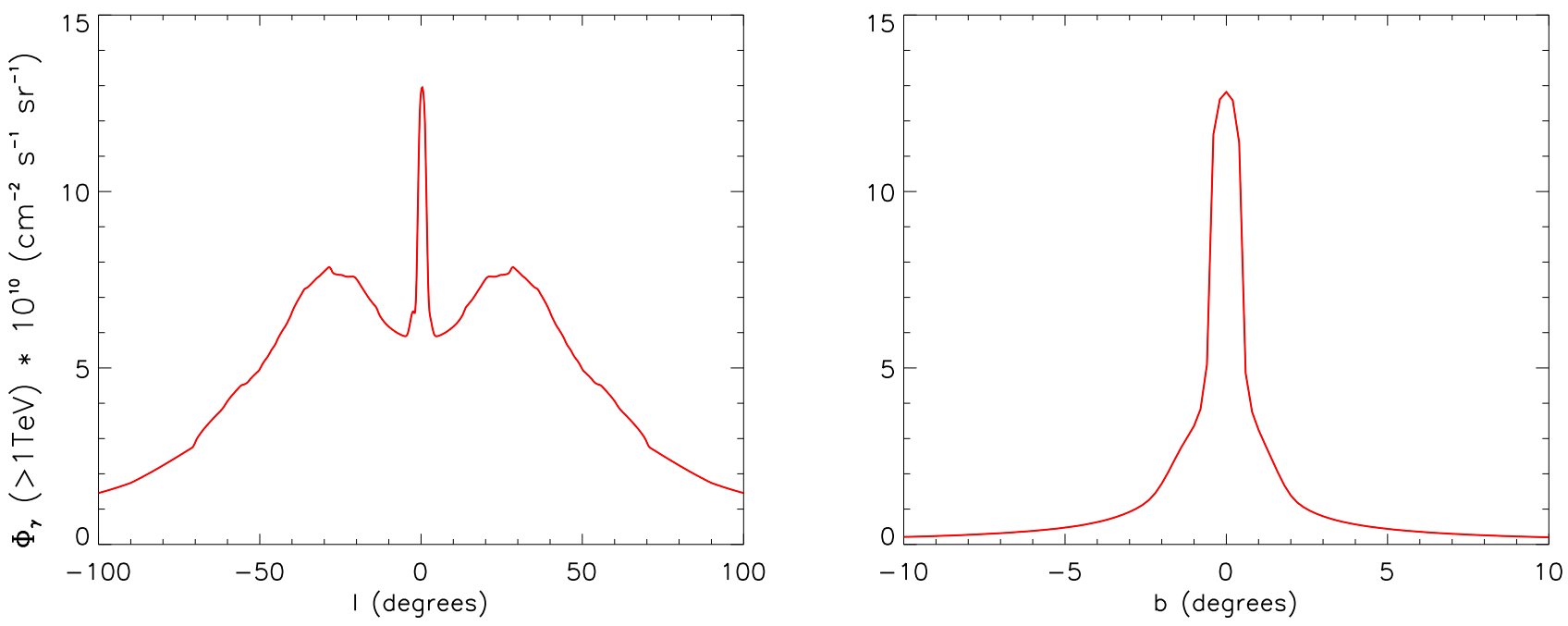

Figure 2: The $\gamma$-ray flux profiles along the GP (left panel) and along $l = 0$ (right panel) for $E > 1$ TeV, averaged over $1° \times 1°$ angular bins. The neutrino flux can be obtained by dividing this diagram by 3.1.

Concerning neutrinos, the only available upper limit on the neutrino flux from the Galaxy has been obtained by the AMANDA-II experiment [23]. Being located at the South Pole, AMANDA cannot probe the emission from the GC. In the region $33° < l < 213°$, $|b| < 2°$, and assuming a spectral index $\alpha = 2.7$, their present constraint is $\Phi_{\nu_\mu+\bar{\nu}_\mu}(> 1\ \mathrm{TeV}) < 3.1\times10^{-9}\ (\mathrm{cm}^2\ \mathrm{s\ sr})^{-1}$. According to our model the expected flux in the same region is $\Phi_{\nu_\mu+\bar{\nu}_\mu}(> 1\ \mathrm{TeV}) \simeq 4.2\times10^{-11}\ (\mathrm{cm}^2\ \mathrm{s\ sr})^{-1}$. That will be hardly detectable even by IceCube. Since a neutrino telescope placed in the North hemisphere may have better changes, we investigated this possibility in details. Assuming that such an instrument will be placed at the same position of ANTARES [24] and have a 40 times larger effective area we estimated the expected signal and the background along the GP. Unfortunately, we found that the detection of the smooth component of the diffuse emission may require more than 10 years of data taking. However, as we mentioned in the above, several observations suggest that the CR and gas distributions may be more clumpy than what considered in this work. This may lead to a significant enhancement of the neutrino flux from some regions as may be the case for the GC ridge [20] and the Cygnus region [25, 26].

# SEARCH FOR NEW PHYSICS AT THE LHC

ANNA DI CIACCIO [a,b]

*on behalf of the ATLAS Collaboration*

[a] *Dipartimento di Fisica, Università di Roma "Tor Vergata", via della Ricerca Scientifica, Roma, Italy*

[b] *INFN, Sezione di Roma "Tor Vergata", via della Ricerca Scientifica, Roma, Italy*

**Abstract**

The Large Hadron Collider, with its unique energy in the center of mass of $\sqrt{s} = 14$ TeV and an ultimate peak luminosity of $10^{34} cm^{-1} s^{-1}$ will start its operation soon, allowing to understand the physics at the TeV Scale. Several models have been proposed to describe the new physics, among them the most known are: Supersimmetry[1], Extra-dimensions[2], Technicolors[3] and Little Higgs[4]. The LHC experiments will be able to test these models and might find something totally unexpected. In any case the LHC outcome will certainly represents a major milestone in the history of particle physics. In this paper the search for new physics in the first phase of the LHC running will be presented. The strategy for the commissioning of the ATLAS detector will be reviewed and the expectations for the discovery of the Higgs boson, supersymmetry and more exotic particles up to 10 $fb^{-1}$ of integrated luminosities will be discussed.

## 1 Introduction

Presently, both the Large Hadron Collider and the experiments are in the commissioning phase and getting ready for the first collisions at low luminosity, foreseen in 2008. There are still a few uncertainties in the schedule depending on how the commissioning of the machine will actually evolve. Therefore we assume here that the integrated luminosity collected by the end of 2008 will range between 100 $pb^{-1}$ and 10 $fb^{-1}$ per experiment, and we discuss the LHC physics potential in this range.

Over the first year of operation, huge event samples should be available from known Standard Model (SM) processes, which will allow to complete the commissioning of the detectors, to fully debug the software and also to look for possible deviation from SM prediction, as possible indication of new physics. It will take certainlly some time before the accelerator ramps up in luminosity and the detectors are fully understood and optimally calibrated. Nevertheless we know that possible new physics phenomena could have large cross sections and striking topologies, that even a limited amount of collected data and a non-ultimate detector performance could lead to exciting results. In the following section, the strategy for commissioning the detector and the prospect for early discoveries will be presented.

## 2 ATLAS detector commissioning strategy

The ATLAS experiment, approved in January 1996, is a general purpose detector. The construction is basically finished, the installation in the cavern is well advanced and the commissioning phase using cosmic rays is in progress. A detailed description of the detector can be found in ref.[5]. It is the biggest among the LHC experiments, with its diameter of $\sim 22m$, a length of $\sim 44m$ and a total weight of $\sim 7000$ tons. The main features are summarized here :

- precision electromagnetic calorimetry for electrons and photons ;
- a large acceptance hadronic calorimetry for jets and missing transverse energy measurements ;
- a high-precision muon momentum measurement with the possibility at the highest luminosity of using the external muon spectrometer alone ;
- triggering and measurements of particles at low $p_T$ thresholds.

One important question to address concernes the detector performance at the beginning of the data taking. Based on quality checks during the construction, on the known precision of the hardware calibration and alignment systems, on test-beam measurements and on simulation studies we expect that the initial uniformity of the electromagnetic calorimeters (ECAL) should be at the level of 1% for the ECAL liquid-argon calorimeter. Prior to data taking, the jet energy scale may be established to about 10% from a combination of test-beam measurements and simulation studies. The tracker alignment in the transverse plane is expected to be known at the level of $20\mu m$ in the best case from surveys, from the hardware alignment systems and possibly from some studies with cosmic muons and beam halo events. This performance should be significantly improved as soon as the first data will be available and, thanks to the huge event rates expected at the LHC, the ultimate statistical precision should be achieved after a few days/weeks of data taking. Table 1 shows the data samples expected to be recorded by ATLAS for some example physics processes and for an integrated luminosity of $10 fb^{-1}$. The trigger selection efficiency has been included.

| Channel | Recorded events for $10fb^{-1}$ |
|---|---|
| $W \rightarrow \mu\nu$ | $7 \times 10^7$ |
| $Z \rightarrow \mu\mu$ | $1.1 \times 10^7$ |
| $t\bar{t} \rightarrow \mu + \mathrm{X}$ | $0.08 \times 10^7$ |
| QCD jets $p_T > 150 GeV$ | $\sim 10^7$ assuming 10% of trigger bandwidth |
| minimum bias | $\sim 10^7$ assuming 10% of trigger bandwidth |
| $\tilde{g}\tilde{g}, m(\tilde{g}) \sim 1\ TeV$ | $10^3 - 10^4$ |

Table 1: Number of expected events in ATLAS for some example physics processes for an integrated luminosity of $10 fb^{-1}$. The hypothetical production of gluinos with a mass of 1 $TeV$ is also considered.

During the first year(s) of data taking, the big event samples will allow to calibrate the detectors, tune the software and understand SM physics. We stress that this is possible even if the integrated luminosity collected during the first year is a factor of hundred smaller, i.e. $100pb^{-1}$. More in details, the following goals can be addressed with the first data:

- Commissioning and calibration of the detectors in situ. Understanding the trigger performance in an unbiased way, with a combination of minimum-bias events, QCD jets collected with various thresholds, single and dilepton samples. $ZZ \rightarrow ll$ is a gold-plated process for a large number of studies, e.g. to set the absolute electron and muon scales in the ECAL and tracking detectors respectively, whereas $t\bar{t}$ events can be used for instance to establish the absolute jet scale and to understand the b-tagging performance.

- Perform extensive measurements of the main SM physics processes, e.g. cross sections and event features for minimum-bias, QCD dijet, W, Z, $t\bar{t}$ production, etc. These measurements will be compared to the predictions of the MonteCarlo (MC) simulations, which will already be quite constrained from theory and from studies at the Tevatron and HERA energies. Typical initial precisions may be 10-20% for cross section measurements, and a few GeV on the top-quark mass, and will likely be limited by systematic uncertainties after just a few weeks of data taking.

- Prepare the road to discoveries by measuring the backgrounds of possible new physics channels. Processes like W/Z+jets, QCD multijet and $t\bar{t}$ production are the main backgrounds for a large number of new searches and need to be understood in all details.

As an example of initial measurement with limited detector performance, fig. 1 shows the reconstructed top-quark signal in the channell $t\bar{t} \to bjjbl\nu$, as obtained from a simulation of the ATLAS detector. The event sample corresponds to an integrated luminosity of $300pb^{-1}$, which can be collected in a week of data taking at L = $10^{33}cm^{-1}s^{-1}$. A very simple analysis was used to select these events, requiring an isolated electron or muon with $p_T \geq 20$ GeV, the event $E_T^{miss} > 20\ GeV$ and only four jets with $p_T \geq 40\ GeV$. The invariant mass of the three jets with the highest $p_T$ is plotted. No kinematic fit is made, and no b-tagging of some of the jets is required, assuming conservatively that the b-tagging performance would not have been well understood yet. Figure 1 shows that, even under these conditions, a clear top signal should be observed above the background after a few weeks of data taking. In turn, this signal can be used for an early validation of the detector performance. For instance, if the top mass is wrong by several $GeV$, this would indicate a problem with the jet energy scale. Furthermore, top events are an excellent sample to understand the b-tagging performance of ATLAS.

## 3 Search for discoveries

Only after a full understanding of the SM processes one can start a serious work to extract a convincing discovery signal from the data. Some examples of new physics in the first year(s) of operation are briefly discussed below: namely a possible $Z' \to e^+e^-$ signal, a SUSY signal, and the Standard Model Higgs boson.

### 3.1 $Z'$

A particle of mass 1-2 TeV decaying into $e^+e^-$ pairs, such as a possible new gauge boson $Z'$, is probably the easiest signal to be discovered at the LHC, for several reasons. First, if the branching ratio into leptons is at least at the percent level as for the $Z$ boson, the expected number of events after all experimental cuts is relatively large, e.g. about ten for an integrated luminosity as low as $300pb^{-1}$ and a particle mass of 1.5 $TeV$. Second, the dominant background, dilepton Drell-Yan production, is small in the $TeV$ region, and even if it is a factor of two or three larger than expected today (which is unlikely for such a theoretically well-known process), it would still be negligible compared to the signal. Finally, the signature will be very clear, since it will appear as a resonant peak on top of a negligeable background. These expectations are not based on ultimate detector performance, since they hold also if the calorimeter response is understood to a conservative level of a few percent.

### 3.2 Supersymmetry

Finding a convincing signal of SUSY in the early phases of the LHC operation is not straightforward, since good calibration of the detectors and detailed understanding of the numerous backgrounds are required. As soon as these two pre-requisites are satisfied, observation of a SUSY signal should be relatively easy and fast. This is because of the huge production cross sections, and hence event rates, even for squark and gluino masses as large as 1 $TeV$, as can be seen in Table 1, due to the clear signature of such events in most scenarios. Therefore, by looking for final states containing several high $p_T$ jets and large $E_T^{miss}$, which is the most powerful and model independent signature if R parity is conserved, the ATLAS experiment should be able to discover squarks and gluinos up to a masses of 1.5 $TeV$, after one month of data taking at L = $10^{33} cm^{-1} s^{-1}$, as shown in fig. 2.

Although detailed measurements of the SUSY particle masses will likely take several years, it should nevertheless be possible to obtain a first determination of the SUSY mass scale quickly after discovery. This is illustrated in fig. 3, which shows the striking SUSY signal on top of the SM background, expected at a point in the minimal SUGRA parameter space where squark and gluino masses are about 1 $TeV$. The plotted variable, called "effective mass" ($M_{eff}$), is defined as the scalar sum of the event $E_T^{miss}$ and of the transverse energies of the four highest jets. More precisely, the position of the peak of the $M_{eff}$ signal distribution (see fig.3) moves to larger and or smaller values with increasing/decresing squark and gluino masses. Therefore a measurement of the signal peak position should also provide a first fast determination of the mass scale of SUSY. The expected precision is about 20% for an integrated luminosity of $10 fb^{-1}$, at least in minimal models like mSUGRA.

A crucial detector performance issue for an early SUSY discovery is a reliable reconstruction of the event $E_T^{miss}$, which could be contaminated by several instrumental effects (calorimeter non-linearities, cracks in the detector, etc.). Final states with non-genuine $E_T^{miss}$ can be rejected by requiring the event primary vertex to be located close to the interaction centre (which also helps to suppress the background from cosmic and beam-halo muons), no jets pointing to detector cracks, and that the missing $p_T$ vector is not aligned with any jet. Concerning the physics backgrounds (e.g. $Z \to \nu\nu$ +jets, $t\bar{t}$ production, QCD multijet events), most of them can be measured by using control samples. For instance, $Z \to ll$ +jet production provides a normalization of the $Z \to \nu\nu$ +jets background. More difficult to handle is the residual background from QCD multijet events with fake $E_T^{miss}$ produced by the above-mentioned instrumental effects. An important element in the ability to calibrate these backgrounds using the theoretical MC predictions to extrapolate from the signal-free to the signal-rich regions, is the reliability of the MC themselves. Their level of accuracy and their capability to describe complex final states, such as the multijet topology typical of new phenomena like SUSY, have improved significantly over the past few years[6].

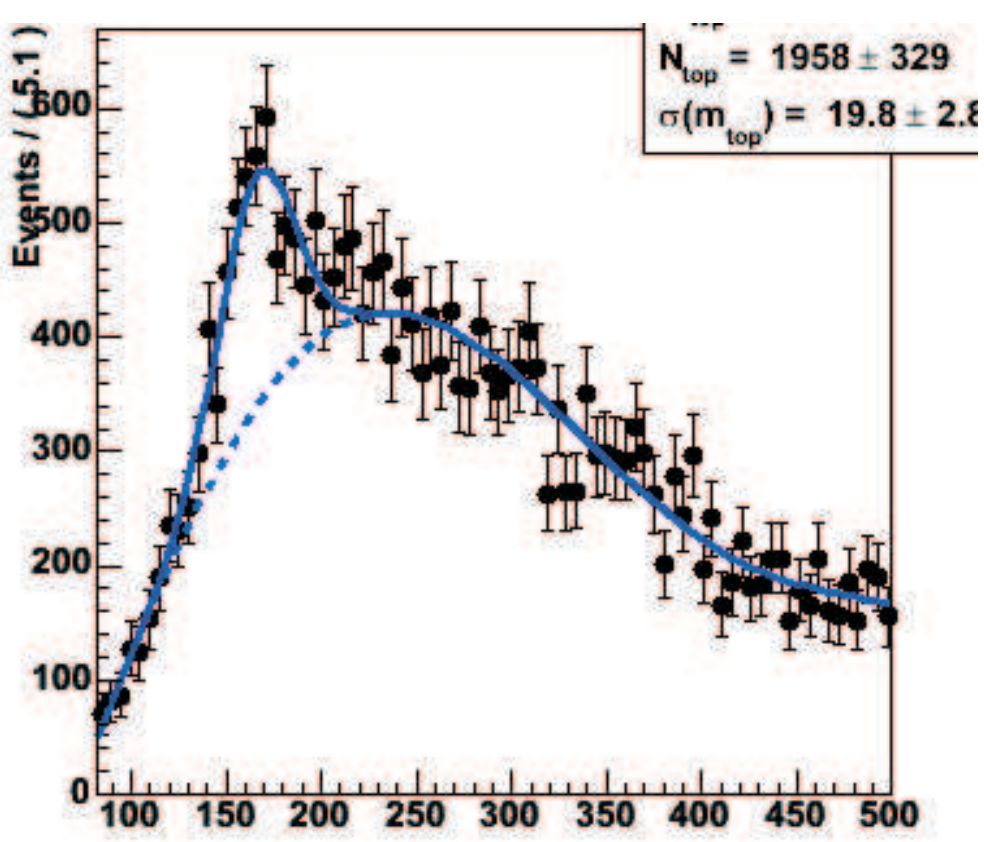

Figure 1: Tree jet invariant mass distribution for events $t\bar{t} \rightarrow bjjbl\nu$, simulated in ATLAS, as obtained with the selection explained in the text with an integrated luminosity of $300pb^{-1}$. The dots with error bars show a clear top signal plus the background. The dashed line shows the $W$ + 4jets background alone (ALPGEN MC).

### 3.3 Standard Model Higgs Boson

The possibility of discovering a SM Higgs boson at the LHC during the first year(s) of operation depends very much on the Higgs boson mass, as shown in fig. 4. If the Higgs boson mass is larger than 180 GeV, discovery may be relatively easy thanks to the gold-plated $H \rightarrow 4l$ channel, which is essentially background-free. The main requirement in this case is an integrated luminosity of at least $5 - 10pb^{-1}$, since the signal has a cross section of only a few $fb$. The low-mass region close to the LEP limit (114.4 $GeV$) is much more difficult. The expected sensitivity for a Higgs mass of 115 $GeV$ and for the first good (i.e. collected with well calibrated detectors) $10pb^{-1}$ is summarized in Table 2.

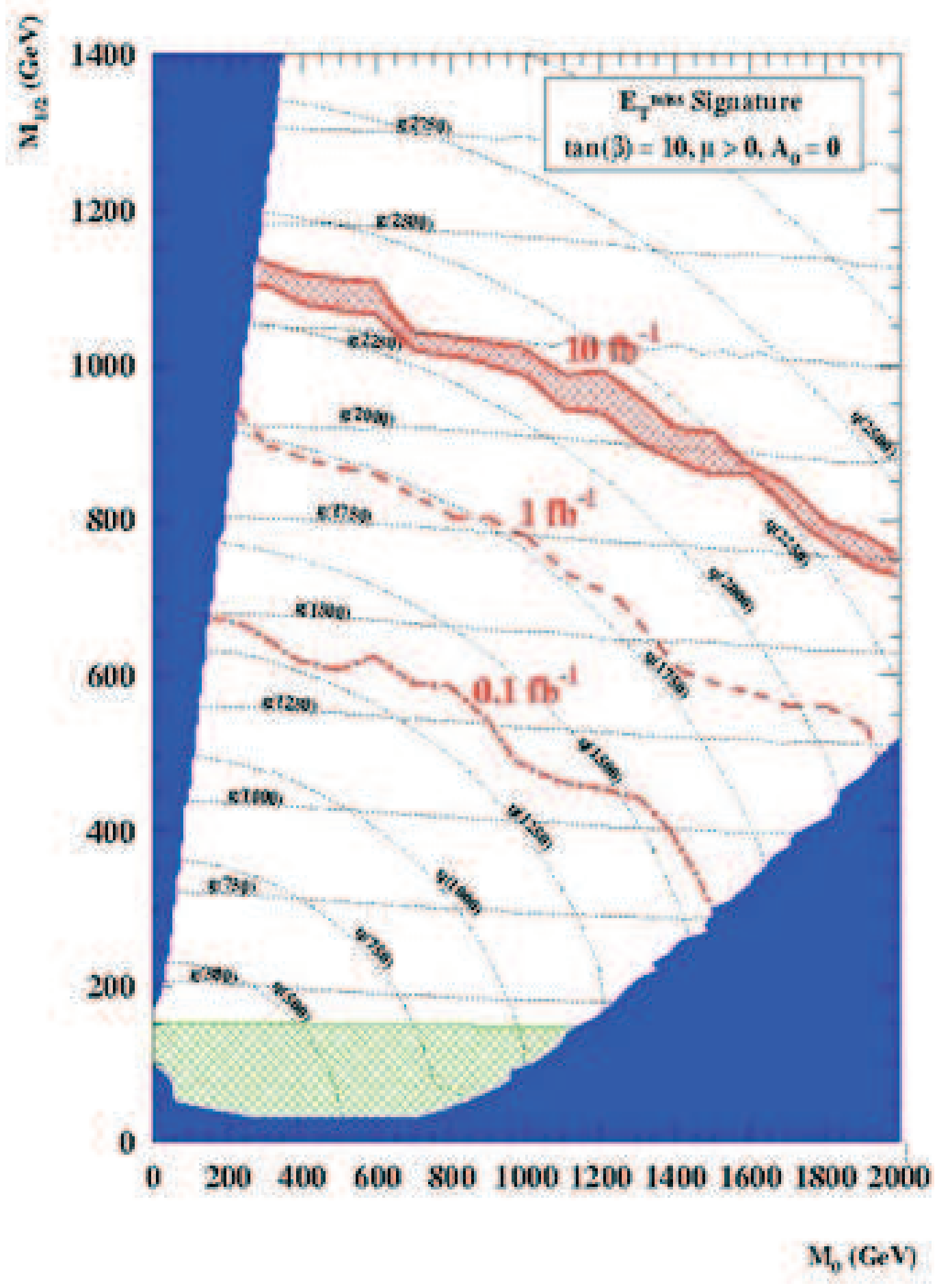

Figure 2: Discovery potential for squarks and gluinos in mSUGRA models, parametrized in terms of the universal scale $m_0$ and the universal gaugino mass $m_{1/2}$ as a function of the integrated luminosity.

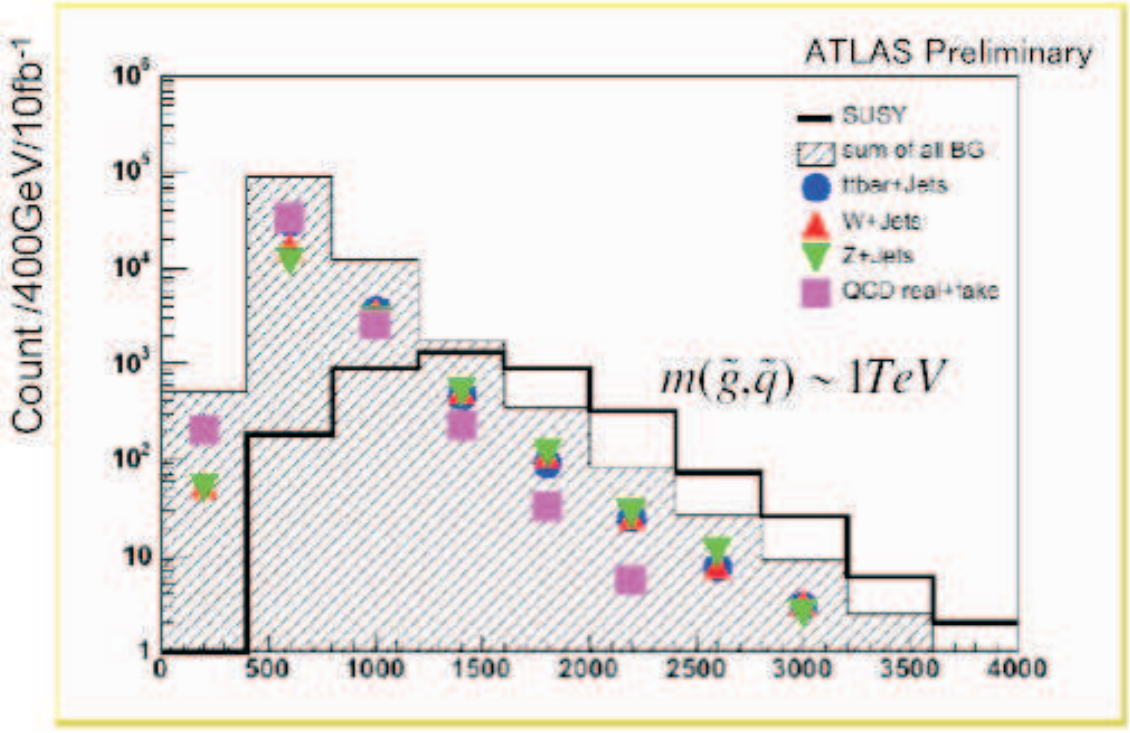

Figure 3: Expected distribution of the "effective mass" for the SUSY signal in mSUGRA, as obtained from a simulation of the ATLAS detector. The histogram shows the total SM background, which includes $t\bar{t}$, W+jets, Z+jets and QCD jets.

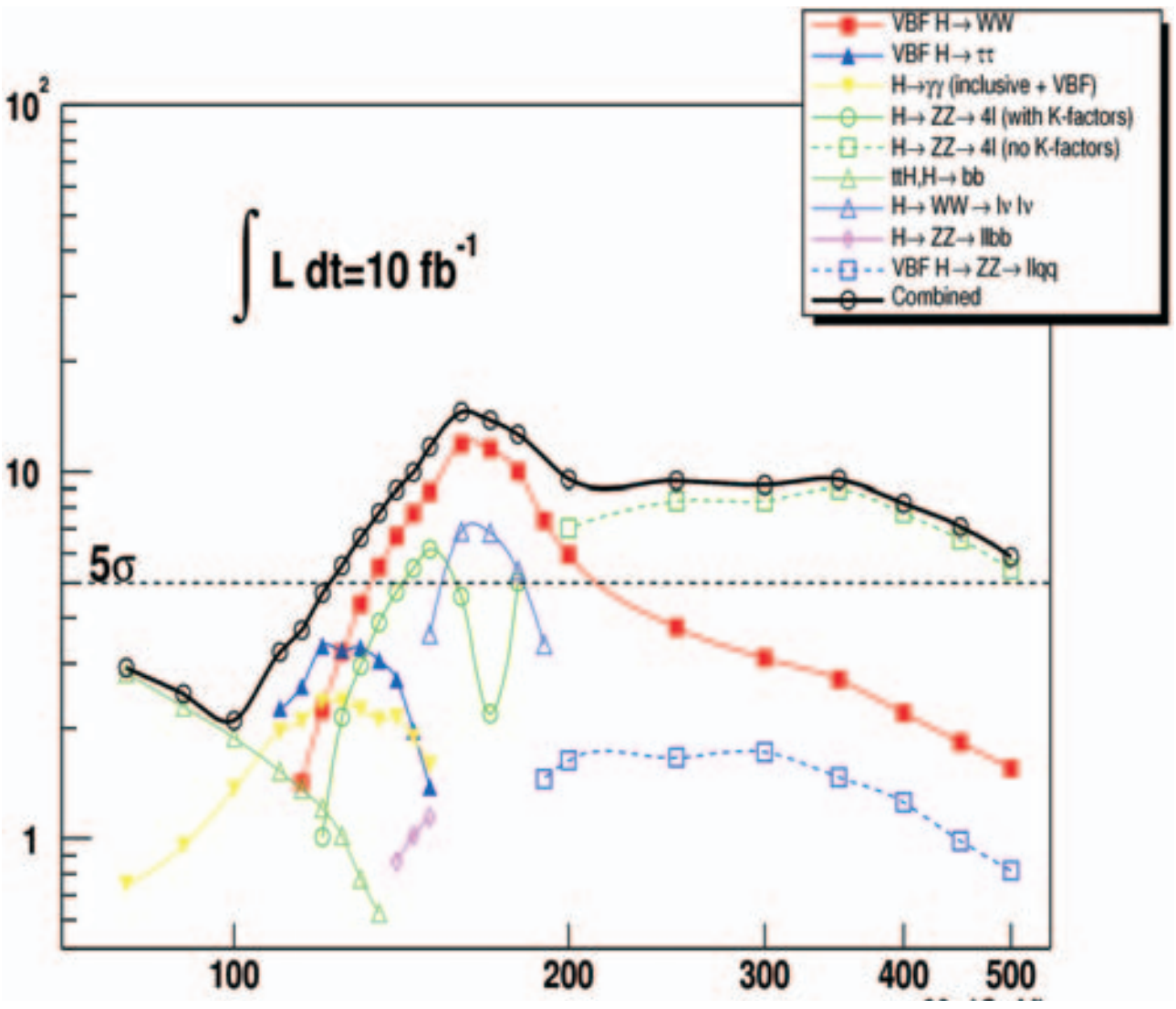

Figure 4: Expected signal significance for a SM Higgs boson in ATLAS as a function of the mass for an integrated luminosity of 10 $fb^{-1}$. The horizontal line indicates the minimum significance (5 $\sigma$) needed for discovery.

| | $H \to \gamma\gamma$ | $t\bar{t}H \to t\bar{t}bb$ | $qqH \to qq\tau\tau \to l$ +X |
|---|---|---|---|
| $S$ | 130 | 15 | $\sim 10$ |
| $B$ | 4300 | 45 | $\sim 10$ |
| $S/\sqrt{B}$ | 2.0 | 2.2 | $\sim 2.7$ |

Table 2: Expected number of signal(S) and background(B) events and signal significance for a Higgs boson mass of 115 $GeV$ for an integrated luminosity of $10pb^{-1}$.

The combined significance of about $4\sigma$ per experiment is more or less equally shared among three channels: $H \to \gamma\gamma$ , $t\bar{t}H$ production with $H \to b\bar{b}$ , and Higgs production in vector-boson fusion followed by $H \to \tau\tau$. It will not be easy to extract a convincing signal with only $10fb^{-1}$, because the significances of the individual channels are small, and because an excellent knowledge of the backgrounds and optimal detector performances are required. Finally, all three channels demand relatively low trigger thresholds (at the level of 20-30 $GeV$ on the lepton or photon $p_T$), and a control of the backgrounds to a few percent. These requirements are especially challenging during the first year(s) of operation. Therefore, the contribution of the both experiments, ATLAS and CMS, and the observation of all three channels, will be crucial for an early discovery.

## 4 Conclusions

The LHC offers the opportunity of major discoveries already at the very beginning. The conditions for discovery require a very good understanding of the detector performance and of the Standard Model and QCD processes. The performance of the LHC at the beginning is the next crucial issue.

With an integrated luminosity between 100 $pb^{-1}$ and 10 $fb^{-1}$, a $Z'$ boson with M > 1 $TeV$ can be discovered, as well a possible SUSY signal and the Higgs boson with M > 130 $GeV$. More difficult is instead the discovery of a light Higgs Boson.

# DIRECT DARK MATTER SEARCH

R. BERNABEI[a,b], P. BELLI[b], F. CAPPELLA[c], R. CERULLI[d], C. J. DAI[e], H. L. HE[e], A. INCICCHITTI[c], H. H. KUANG[e], J. M. MA[e],F. MONTECCHIA[a,b], F. NOZZOLI[a,b], D. PROSPERI[c,f], X. D. SHENG[e], Z. P. YE[e]

[a] *Dip. di Fisica, Università di Roma "Tor Vergata", via della Ricerca Scientifica, Rome, Italy*

[b] *INFN, Sezione di Roma Tor Vergata, via della Ricerca Scientifica, Rome, Italy*

[c] *INFN, Sezione di Roma, P.le A. Moro, Rome, Italy*

[d] *INFN - Laboratori Nazionali del Gran Sasso, I-67010 Assergi (Aq) - Italy*

[e] *IHEP, Chinese Academy, P.O. Box 918/3, Beijing 100039 - PR China*

[f] *Dip. di Fisica, Università di Roma "La Sapienza", P.le A. Moro, Rome, Italy*

## Abstract

The DAMA/NaI experiment at the Gran Sasso National Laboratory of the INFN. has pointed out - by a model independent approach - the presence of Dark Matter particles in the galactic halo at 6.3 $\sigma$ C.L. over seven annual cycles. Some of the many possible corollary model dependent quests for the candidate particle either have been carried out or are in progress; many of the related aspects are still under investigation. At present the second generation DAMA/LIBRA set-up is in data taking. Many searches for other rare processes have been and are under investigation as well.

## 1 Introduction

A large number of possibilities exists for candidates as Dark Matter (DM) particles in the Universe either produced at rest or non relativistic at decoupling time. DM particles at galactic scale can be directly investigated through their interaction on suitable deep underground target-detectors. The detector's signal can be induced either by recoiling nucleus or, in case of inelastic scattering, also by successive de-excitation gamma's; moreover, other possibilities, which directly involve practically only ionization/excitation phenomena in the detector, are open (see e.g. later) as well as the excitation of bound electrons in scatterings on nuclei giving rise contemporaneously to recoiling nuclei and electromagnetic radiation, etc.

The DAMA/NaI experiment was realized having the main aim to investigate in a model independent way the presence of DM particles in the galactic halo. For this purpose, we planned to exploit the effect of the Earth revolution around the Sun on the DM particles interactions in a suitable low background set-up placed deep underground. In fact, as a consequence of its annual revolution, the Earth should be crossed by a larger flux of DM particles around roughly June 2nd (when its rotational velocity is summed to the one of the solar system with the respect to the Galaxy) and by a smaller one around roughly December 2nd (when the two velocities are subtracted). This annual modulation signature – originally suggested in the middle of '80 in ref. [1] – is very distinctive since a seasonal effect induced by DM particles must simultaneously satisfy all the following requirements: (i) the rate must contain a component modulated according to a cosine function; (ii) with one year period; (iii) a phase roughly around $\sim$ June 2nd; (iv) this modulation must only be found in a well-defined low energy range, where DM particles can induce signals; (v) it must only apply to those events in which just one detector of many actually "fires", since the probability that DM particles would have multiple interactions is negligible; (vi) the modulation amplitude in the region of maximal sensitivity must be $\lesssim$ 7% for usually adopted halo distributions, but it can be significantly larger in case of other possible scenarios such as e.g. those of refs. [2, 3].

The DAMA/NaI experiment was located deep underground in the Gran Sasso National Laboratory of I.N.F.N.. It has been part of the DAMA project, which also includes several other low background set-ups, such as: i) DAMA/LXe [4]; ii) DAMA/R&D [5]; iii) the new second generation larger mass NaI(Tl) radiopure set-up DAMA/LIBRA; iv) DAMA/Ge detector for sample measurements which is installed in the LNGS Ge facility [6]. Detailed descriptions of the DAMA/NaI set-up, of its upgrading occurred in 2000 and of its performances have been given e.g. in [7, 8, 9, 10]. Thanks to its radiopurity and features, DAMA/NaI has also investigated other approaches for DM particles and several other rare processes [11].

## 2 DAMA/NaI model-independent result

The DAMA/NaI set-up has pointed out the presence of a modulation satisfying the many peculiarities of a DM particle induced effect, reaching an evidence at 6.3 $\sigma$ C.L. over seven annual cycles [7, 8]. In particular, the residual rate of the *single-hit* events in the cumulative (2-6) keV energy interval has a modulated cosine-like behaviour at 6.3 $\sigma$ C.L. and the fit for this cumulative energy interval offers modulation amplitude equal to $(0.0200 \pm 0.0032)$ cpd/kg/keV, a phase $t_0 = (140 \pm 22)$ days and a period $T = (1.00 \pm 0.01)$ year, all parameters kept free in the fit. The period and phase agree with those expected in the case of an effect induced by DM particles in the galactic halo ($T = 1$ year and $t_0$ roughly at $\simeq 152.5^{th}$ day of the year). The $\chi^2$ test on the (2–6) keV residual rate disfavours the hypothesis of unmodulated behaviour giving a probability of $7 \cdot 10^{-4}$ ($\chi^2/d.o.f. = 71/37$). The same data have also been investigated by other approaches as e.g. a Fourier analysis.

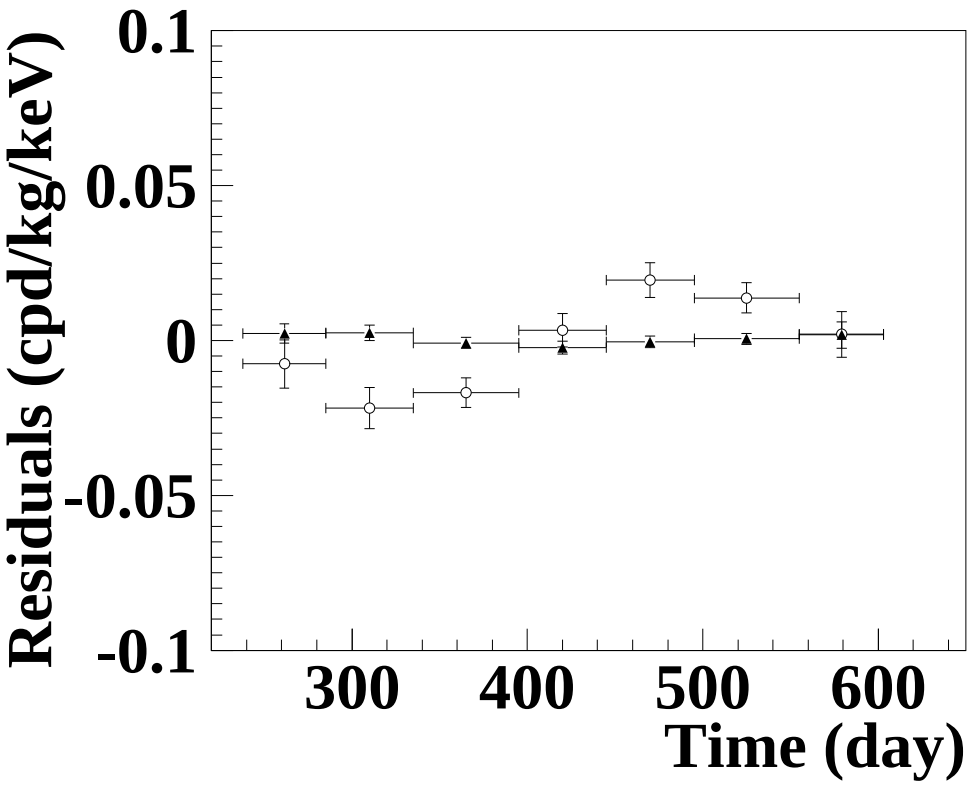

Figure 1: Experimental residual rates over seven annual cycles for *single-hit* events (open circles) – class of events to which DM particle events belong – and over the last two annual cycles for *multiple-hits* events (filled triangles) – class of events to which DM particle events do not belong – in the (2–6) keV cumulative energy interval. They have been obtained by considering for each class of events the data as collected in a single annual cycle and using in both cases the same identical hardware and the same identical software procedures. The initial time is taken on August $7^{th}$. See ref. [8].

A careful quantitative investigation of all the known possible sources of systematics and side reactions has been regularly carried out and published at time of each data release. No systematic effect or side reaction able to account for the observed modulation amplitude and to satisfy all the requirements of the signature has been found; for a detailed quantitative discussion see e.g. ref. [7, 8, 9].

As a further relevant investigation, the *multiple-hits* events collected during the DAMA/NaI-6 and 7 running periods (when each detector was equipped with its own Transient Digitizer with a dedicated renewed electronics) have been studied and analysed by using the same identical hardware and the same identical software procedures as for the case of the *single-hit* events (see Fig. 1). The fitted modulation amplitudes are: $A = (0.0195 \pm 0.0031)$ cpd/kg/keV and $A = -(3.9 \pm 7.9) \cdot 10^{-4}$ cpd/kg/keV for *single-hit* and *multiple-hits* residual rates, respectively. Thus, evidence of annual modulation with proper features [7, 8] is present in the *single-hit* residuals (events class to which the DM particle-induced signals belong), while it is absent in the *multiple-hits* residual rate (event class to which only background events belong), offering an additional strong support for the presence of DM particles in the galactic halo further excluding any side effect either from hardware or software procedures or from background.

## 3 Corollary model-dependent quests

Corollary investigations can also be pursued on the nature of the DM particle candidate [7, 8, 3, 12, 13]. This latter investigation is instead model-dependent and - considering the large uncertainties which exist on the astrophysical, nuclear and particle physics assumptions and on the parameters needed in the calculations - has no general meaning (as it is also the case of exclusion plots and of DM particle parameters evaluated in indirect detection experiments).

Low and high WIMP mass candidates interacting with ordinary matter via: i) mixed SI&SD coupling; ii) dominant SI coupling; iii) dominant SD coupling; iv) preferred SI inelastic scattering; have been considered for the WIMP class of DM particles in refs. [7, 8] for several (but still a limited number with respect to the possibilities) given scenarios. The general solution for these candidates is a 4-dimensional allowed volume in the space ($\xi\sigma_{SI}$, $\xi\sigma_{SD}$, $\theta$, $m_W$): examples of slices of such a volume in some of the many possible scenarios can be found in refs. [7, 8]. The DAMA/NaI allowed regions are well compatible with theoretical expectations for neutralino in MSSM (see e.g. [14]). Also the inelastic DM particle scenario (where heavy nuclei are favoured with the respect to lighter ones) of ref. [2] has been analyzed obtaining an allowed volume in the 3-dimensional space ($\xi\sigma_p$, $\delta$, $m_W$) for the considered scenarios [7, 8].

Galaxy hierarchical formation theories, numerical simulations, the discovery of the Sagittarius Dwarf Elliptical Galaxy (SagDEG) in 1994 and more recent investigations suggest that the dark halo of the Milky Way can contain non thermalized substructures. The effect of the inclusion in the considered halo models of the contribution from the SagDEG has been included in the analyses. A detailed discussion and examples of the modification of allowed volumes/regions following such an inclusion are given in ref. [12].

It has also been investigated the implication of the effect pointed out by A. B. Migdal in the 40's whose presence has so far been usually neglected in the direct searches for WIMP Dark Matter candidates. This effect consists in the ionization and the excitation of bound atomic electrons induced by the recoiling atomic nucleus. In ref. [13] the theoretical arguments have been developed and examples of the effect of the inclusion of this well known existing physical effect on the corollary quests of ref. [7] have been given. Additional allowed windows in the GeV region are open and several candidates and support from astrophysical consideration are available.

An additional corollary quest for the candidate particle considering a light ($\simeq$ keV mass) bosonic candidate, either with pseudoscalar or with scalar coupling, as DM component in the galactic halo has also been carried out [3]. For these candidates, the direct detection process is based on the total conversion in NaI(Tl) crystal of the mass of the absorbed bosonic particle into electromagnetic radiation. Thus, in these processes the target nuclei recoil is negligible and is not involved in the detection process. In ref. [3] the theoretical arguments have been developed and the obtained allowed regions for these very interesting candidates [15] have been given.

Other corollary quests are also available in literature, such as e.g. in refs. [16, 17, 14, 2, 18], and many other scenarios can be considered as well.

## 4 Some comparisons in the field

No experiment is available so far – with the exception of DAMA/LIBRA – whose results can be directly compared in a model independent way with that of DAMA/NaI. Thus, claims for contradictions have intrinsically no scientific meaning. Some discussions can be found e.g. in ref.[7, 8] and in proceedings.

In particular, as regards some claimed model-dependent comparisons presented so far we just mention – among the many existing arguments – that the other experiments available so far: i) are insensitive to the annual modulation signature; ii) use different exposed materials; iii) release just a marginal exposure (orders of magnitude lower than the one by DAMA/NaI) after several/many years underground; iv) exploit strong data selection and strong and often unsafe rejection techniques of their huge counting rate, becoming at the same time insensitive to several DM candidates; v) generally quote in an incorrect/partial/not updated way the DAMA/NaI result; vi) consider a single model fixing all the astrophysical, nuclear and particle Physics assumptions as well as all the theoretical and experimental parameters at a single questionable choice [1]. Thus, e.g. for the WIMP case they do not account for the existing

[1] We note that the naive and partial "prescription" of ref. [19] on some aspects for a single

uncertainties on the real coupling with ordinary matter, on the spin-dependent and spin-independent form factors and related parameters for each nucleus, on the spin factor used for each nucleus, on the real scaling laws for nuclear cross sections among different target materials; on the experimental and theoretical parameters, on the effect of different halo models and related parameters on the different target materials, etc. For example, large differences are expected in the counting rate among nuclei fully sensitive to the SD interaction (as $^{23}$Na and $^{127}$I) with the respect to nuclei largely insensitive to such a coupling (as e.g $^{nat}$Ge, $^{nat}$Si, $^{nat}$Ar, $^{nat}$Ca, $^{nat}$W, $^{nat}$O) and also when nuclei in principle all sensitive to this coupling but having different unpaired nucleon (e.g. neutron in case of the odd spin nuclei, such as $^{129}$Xe, $^{131}$Xe, $^{125}$Te, $^{73}$Ge, $^{29}$Si, $^{183}$W and proton in the $^{23}$Na and $^{127}$I). Moreover, in case the detection of the DM particles would involve electromagnetic signals (see, for example, the case of the light bosons discussed above, but also electromagnetic contribution in WIMP detection arising e.g. from known effect induced by recoiling nuclei), all the other experiments do lose the signal in their data selection and "rejection" procedures of the electromagnetic contribution to the counting rate.

In addition, the other experiments present many critical points e.g. regarding the energy threshold, the energy scale determination and the multiple selection procedures, on which their claimed "sensitivities" for a "single" set of assumptions and parameters' values are based. In particular, critical items in the used "rejection" procedures are the related stabilities and efficiencies, the systematics in the evaluations of the rejection factors (ranging from $10^{-4}$ to $10^{-8}$), stabilities and monitoring of the spill-out factors, ...

Discussions at some extent are also reported e.g. in ref. [7, 8, 3, 12]; all those general comments also hold in the substance for more recent claims, such as e.g. the one by XENON10. In particular, we remind that in this latter experiment the physical energy threshold is unproved by source calibrations; despite of the small light response (2.2 phe/keVee) an energy threshold of 2 keVee is claimed and no mention is done about the energy resolution at 2 keVee, notwithstanding whatever exclusion plot also critically depends on the used values for these quantities. Moreover, the claimed background rate in the "fiducial" volume is quite hard to be justified considering the design and all the involved materials of this detector. In addition, the used gas is natural xenon, particular WIMP case cannot be defined – on the contrary of what appears in some papers – as a "standard theoretical model". Such a paper summarized a single oversimplified approach adopted at that time. Its use as "unique" reference is obviously incorrect, since it did not account at all for the level of knowledge on all the involved astrophysics, nuclear and particle physics aspects and parameters, for the many possibilities open on the astrophysical, nuclear and particle physics aspects and for the different existing approaches as e.g. the annual modulation signature, which requires – among others – time/energy correlation analysis.

that is with an unavoidable content of Kripton. Many cuts are applied, each of them introduces systematics, which – in addition – can vary along the data taking and cannot be correctly evaluated with the needed precision just in a short period of calibrations; etc.

For completeness, it is also worth to note that no results obtained with different target material can intrinsically be directly compared even for the same kind of coupling, although apparently all the presentations generally refer to cross section on the nucleon. The situation is much worse than the one in the field of double beta decay experiments when different isotopes are used.

As regards the indirect searches, a comparison would always require the calculation and the consideration of all the possible DM particle configurations in the given particle model, since it does not exist a biunivocal correspondence between the observables in the two kinds of experiments. However, the present positive hints provided by indirect searches are not in conflict with the DAMA/NaI result.

Finally, it is worth to note that – among the many corollary aspects still open – there is f.i. the possibility that the particle dark halo can have more than one component; some example have already been considered in literature.

## 5 The new DAMA/LIBRA and beyond

In 1996 DAMA proposed to realize a ton set-up and a new R&D project for highly radiopure NaI(Tl) detectors was funded and carried out for several years in order to realize - as an intermediate step - the second generation highly radiopure DAMA/LIBRA experiment (successor of DAMA/NaI). Thus, as a consequence of the results of this second generation R&D, the new experimental set-up DAMA/LIBRA (Large sodium Iodide Bulk for RAre processes), $\simeq$ 250 kg highly radiopure NaI(Tl) scintillators, was funded and realized. DAMA/LIBRA started operations in March 2003 and the first release of results will, most probably, occur at end 2008. Further R&D's are in progress.

# PRIMORDIAL ANTIMATTER IN THE CONTEMPORARY UNIVERSE

Cosimo Bambi

Dipartimento di Fisica, Università degli Studi di Ferrara
via Saragat 1, 44100 Ferrara, Italy

**Abstract**

In some baryogenesis scenarios, the universe acquires a non-vanishing average baryonic charge, but the baryon to photon ratio is not spatially constant and can be even negative in some space regions. This allows for existence of lumps of antimatter in our neighborhood and the possibility that very compact antimatter objects make a part of cosmological dark matter. Here I discuss the peculiar signatures which may be observed in a near future.

One can conclude from simple considerations that there is much more matter than antimatter around us [1]. However, the origin of matter–antimatter asymmetry in the universe is unknown: the Standard Model of particle physics is certainly unable to explain it and new physics is necessary [2]. Assuming a homogeneous and isotropic universe, from the Big Bang Nucleosynthesis (BBN) [3] and the Cosmic Microwave Background Radiation (CMBR) [4] one can determine the baryon to photon ratio $\beta$

$$\beta = \frac{n_B - n_{\bar{B}}}{n_\gamma} \approx 6 \cdot 10^{-10} \tag{1}$$

where $n_B \gg n_{\bar{B}}$. On the other hand, the freeze-out abundances in a homogeneous baryo-symmetric universe would be $n_B/n_\gamma = n_{\bar{B}}/n_\gamma \sim 10^{-18}$ [5].

However, Eq. (1) may not be the end of the story. One can indeed distinguish three main types of cosmological matter–antimatter asymmetry:

1. **Homogeneous matter dominated universe**. Here $\beta$ is constant and the universe is 100% matter dominated. This is certainly the most studied case (see e.g. Refs. [6, 7]) but it is not very interesting for astrophysical observations, because there is only one observable quantity, $\beta$, which cannot contain much information on high energy physics.

2. **Globally B-symmetric universe**. Such a possibility appears quite reasonable and "democratic": the universe would consist of equal amount of similar domains of matter and antimatter. However, it seems observationally excluded or, to be more precise, the size of the domain where we live should be at least comparable to the present day cosmological horizon [8]. So, even in this case observations cannot determine nothing but $\beta$.

3. **Inhomogeneous matter dominated universe**. In this case the universe has a non-vanishing baryonic charge, but $\beta$ is not spatially constant and can even be negative in some space regions. Lumps of antimatter can be scattered throughout the universe.

Here I will discuss possible observational signatures of the third case: even if at first glance such a picture may appear strange, just because we are used to think about ordinary matter around us, there are no theoretical and experimental reasons to reject it. At present, the source of CP violation responsible for the observed B-asymmetry in the universe is unknown, so generation of lumps of antimatter is not so exotic as one may naively think. Moreover, compact antimatter objects can easily survive in a matter dominated universe up to the present days. The talk is based on a work made in collaboration with Alexander Dolgov [9]. The reference baryogenesis mechanism is the one in [10]. The phenomenology of other scenarios can be found in Refs. [11, 12].

## 1 Baryogenesis framework

Let us now briefly review the baryogenesis framework suggested in Ref. [10]. The basic ingredient is the Affleck-Dine mechanism [13], where a scalar field $\chi$ with non-zero baryonic charges have the potential with flat directions, that is directions along which the potential energy does not change. Due to the infrared instability of light fields in de Sitter spacetime [14], during inflation $\chi$ can condense along the flat directions of the potential, acquiring a large expectation value. In the course of the cosmological expansion, the Hubble parameter drops down and, when the mass of the field exceeds the universe expansion rate, $\chi$ evolves to the equilibrium point and the baryonic charge stored in the condensate is transformed into quarks by $B$-conserving processes. Since here CP is violated stochastically by a chaotic phase of the field $\chi$, then during the motion to the equilibrium state the matter and antimatter domains in a globally symmetric universe would be created. An interesting feature of the model is that regions with a very high $\beta$, even close to one, could be formed.

If the scalar field $\chi$ is coupled to the inflaton $\Phi$ with an interaction term of the kind $V(\chi, \Phi) = \lambda|\chi|^2(\Phi - \Phi_1)^2$, the "gates" to the flat directions might be open only for a short time when the inflaton field $\Phi$ was close to $\Phi_1$. In this case, the probability of the penetration to the flat directions is small and $\chi$ could acquire a large expectation value only in a tiny fraction of space. The universe would have a homogeneous background of baryon asymmetry $\beta \sim 6 \cdot 10^{-10}$ generated by the same field $\chi$, which did not penetrate to larger distance through the narrow gate, or by another mechanism of baryogenesis, while the high density matter, $\beta > 0$, and antimatter, $\beta < 0$, regions would be rare, although their contribution to the cosmological mass density might be significant or even dominant. In the simple model of Ref. [10], such high density bubbles could form clouds of matter or antimatter and more compact object like stars, anti-stars or primordial black holes. In the non-collapsed regions, primordial nucleosynthesis proceeded with large $|\beta|$, producing nuclei heavier than those formed in the standard BBN [15].

## 2 Phenomenology

In what follows I will not dwell on possible scenarios of antimatter creation, but simply consider phenomenological consequences of their existence in the present day universe, in particular in the Galaxy. Some considerations on the cosmological evolution of lumps of antimatter in a baryon dominated universe can be found in Refs. [9, 12].

### 2.1 Indirect detection

The presence of anti-objects in the Galaxy today should lead to the production of the gamma radiation from matter–antimatter annihilation. Hence we would expect $\sim 100$ MeV $\gamma$ from the decay of $\pi^0$ mesons produced in $p\bar{p}$ annihilation, with an average of 4 $\gamma$ per annihilation, and $2\,\gamma$ from $e^+e^-$ annihilation with $E = 0.511$ MeV, if $e^+e^-$ annihilate at rest. In addition to the slow background positrons, there should be also energetic secondary positrons produced by pion decays from $p\bar{p}$ annihilation. Astronomical observations are seemingly more sensitive to $p\bar{p}$ annihilation because the total energy release in $p\bar{p}$ annihilation is 3 orders of magnitude larger than that in $e^+e^-$ annihilation and the galactic gamma ray background at 100 MeV is several orders of magnitude lower than the one at 0.5 MeV. On the other hand, $e^+e^-$ annihilation gives the well defined line which is easy to identify.

For compact anti-objects like anti-stars, one find that the size of the anti-object, $R$, is much larger than the proton or electron mean free path inside the anti-object, $\lambda_{free} \sim 1/(\sigma_{ann}\, n_{\bar{p}})$, where $\sigma_{ann}$ is the annihilation cross section for $p\bar{p}$ or $e^+e^-$ (they have similar order of magnitude) and $n_{\bar{p}}$ is the antiproton number density in the anti-object. In this case, the annihilation takes place on the surface, all the protons and electrons that hit the surface of the anti-object annihilate and the annihilation cross section is given by the geometrical area of the anti-object, that is $\sigma = 4\pi R^2$. The gamma ray luminosity of such a compact anti-object is

$$L_\gamma \approx 10^{27} \left(\frac{R}{R_\odot}\right)^2 \left(\frac{n_p}{\text{cm}^{-3}}\right) \left(\frac{v}{10^{-3}}\right) \text{erg/s}, \tag{2}$$

where $R_\odot \sim 7 \cdot 10^{10}$ cm is the Solar radius and $n_p v$ is the proton flux. With this luminosity, a solar mass anti-star would have the life time of the order of $10^{27}$ s (considering only matter–antimatter annihilation), if all the factors in Eq. (2) are of order unity. For an anti-star in the galactic disc, the $\gamma$ flux observable on the Earth would be

$$\phi_{Earth} \sim 10^{-7} \left(\frac{R}{R_\odot}\right)^2 \left(\frac{1\,\text{pc}}{d}\right)^2 \text{cm}^{-2}\,\text{s}^{-1}\,. \tag{3}$$

where $d$ is the distance of the anti-star from the Earth. Such a flux should be compared with the point source sensitivity of EGRET [16], at the level of $10^{-7}$ photons cm$^{-2}$ s$^{-1}$ for $E_\gamma > 100$ MeV, and of the near-future GLAST [17], which should be about two order of magnitude better, i.e. $\sim 10^{-9}$ photons cm$^{-2}$ s$^{-1}$. So, anti-stars should be quite close to us in order to be detectable point-like sources and their observation would result difficult if they were very compact objects, as e.g. anti-neutron stars. On the other hand, if such an anti-star lived in the galactic center, where $n_p \gg 1/\text{cm}^3$, its luminosity would be larger. Anomalously bright lines of 0.5 MeV are observed recently in the galactic center [18], galactic bulge [19] and possibly even in the halo [20]. Though an excess of slow positrons is explained in a conventional way as a result of their creation by light dark matter particles, such a suggestion is rather unnatural, because it requires a fine-tuning of the mass of the dark matter particle and the electron mass. More natural explanation is the origin of these positrons from primordial antimatter objects.

The existence of primordial antimatter in the Galaxy would increase the galactic diffuse gamma ray background as well. Standard theoretical predictions and observational data agree on a galactic production rate of $\gamma$ in the energy range $E_\gamma > 100$ MeV [9]

$$\Gamma_\gamma^{tot} \sim 10^{43}\ \text{s}^{-1}\,. \tag{4}$$

Requiring that annihilation processes on anti-stars surface cannot produce more than 10% of the standard galactic production rate (4), we obtain the following bound on the present number of anti-stars

$$N_{\bar{S}} \lesssim 10^{12} \left(\frac{R_\odot}{R}\right)^2 , \tag{5}$$

where, for simplicity, we assumed that all the anti-stars have the same radius $R$. However the constraint is not very strong: for solar type anti-stars, their number cannot exceed the one of ordinary stars!

Let us now consider the annihilation of antimatter from the anti-stellar wind with protons in the interstellar medium. Since the number of antiprotons reached a stationary value, the production rate of 100 MeV $\gamma$ in the Galaxy has to be proportional to $N_{\bar{S}}$. The luminosity of the Galaxy in 100 MeV $\gamma$ rays from anti-stellar wind would be $L_{\bar{S}} \sim 10^{44} W\, N_{\bar{S}}/N_S$ erg/s, where $W$ is the anti-stellar wind to solar wind flux ratio. Since from Eq. (4) we find that the total Galaxy luminosity in 100 MeV $\gamma$ is $L_\gamma^{tot} \sim 10^{39}$ erg/s, the related bound on the anti-star to star number ratio is $N_{\bar{S}}/N_S \lesssim 10^{-6}\, W^{-1}$, always assuming that the contribution from new physics cannot exceed 10% of $L_\gamma^{tot}$. A similar restriction can also be obtained from the 0.511 MeV line created by $e^+e^-$ annihilation with positrons from the anti-stellar wind.

On the other hand, if anti-stars were formed in the very early universe in the regions with a high antimatter density [10], such primordial stars would most probably be compact ones, like white dwarfs or neutron stars. The stellar wind in this case would be much smaller that the solar one, $W \ll 1$. Their luminosity from the annihilation on the surface should be very low, because of their small radius $R$, and their number in the Galaxy may be even larger than the number of the usual stars. This possibility is not excluded by the previous bounds. Such compact dark stars could make a noticeable part of the cosmological dark matter.

### 2.2 Direct detection

It is common belief that the abundances of most elements in the cosmic rays reflect relative abundances in the Galaxy. Hence, as the simplest working hypothesis we can assume that the antimatter–matter ratio in cosmic rays is more or less equal to the anti-star–star ratio $N_{\bar{S}}/N_S$, at least if the anti-stars are of the same kind as the stars in the Galaxy.

As for antiprotons and positrons, they cannot be direct indicators for the existence of primordial antimatter, because they can be produced in many astrophysical processes. For example, the observed $\bar{p}/p$ ratio is at the level of $10^{-4}$ and is compatible with theoretical predictions for $\bar{p}$ production by the high energy cosmic ray collisions with the interstellar medium. A possible contribution of $\bar{p}$ from primordial lumps of antimatter is not more than about 10% of the total observed $\bar{p}$ flux, so $N_{\bar{S}}/N_S \lesssim 10^{-5}$ and the number of anti-stars $N_{\bar{S}}$ has to be no more than $10^6$, since the number of ordinary stars in the Galaxy is $N_S \sim 10^{11}$.

On the other hand, the possibility of producing heavier anti-nuclei (such as anti-helium) in cosmic ray collisions is completely negligible and a possible future detection of the latter would be a clear signature of antimatter objects.

At present there exists an upper limit on the anti-helium to helium ratio in cosmic rays, at the level of $10^{-6}$ [21], leading to the constraint $N_{\bar{S}} \lesssim 10^5$. Such an upper limit can probably be lowered by 2 or 3 orders of magnitude in a near future, thanks to AMS [22] and PAMELA [23] space missions. I would like to stress that here we are not assuming that these possible anti-helium nuclei were produced by nuclear fusion inside anti-stars, but that original anti-helium abundance inside anti-stars is roughly equal to the helium abundance inside ordinary stars. This is certainly a conservative picture, since anti-stars were formed in high density regions of the early universe, where the primordial nucleosynthesis produced much more anti-helium and heavier anti-nuclei [15]. On the other hand, if anti-stars were compact ones from the very beginning, the stellar wind from them and the shortage of anti-supernova events would spread much less anti-helium than the normal stars.

### 2.3 More exotic events

The presence of anti-stars in the Galaxy could lead to extraordinary events of star–anti-star annihilation. As a matter of fact, the radiation pressure produced in the collision prevents their total destruction. Still the released energy can be huge.

The most spectacular phenomenon is a collision between a star and an anti-star with similar masses $M$. A simple estimate of the amount of the annihilated matter in such a collision is $m_{ann} \sim Mv^2$ [9], where $v$ is the typical value of the relative velocity and is about $10^{-3}$. The total energy release would be $E \sim 10^{48}\,\text{erg}(M/M_\odot)(v/10^{-3})^2$. Most probably the radiation would be emitted in a narrow disk along the boundary of the colliding stars. The collision time is $t_{coll} \sim R$ and for the solar type star this time is about 3 s. The energy of the radiation should be noticeably smaller than 100 MeV, because the radiation should degrade in the process of forcing the star bounce. This makes this collision similar to gamma bursts, but unfortunately some other features do not fit so well: the released energy should be much larger, about $10^{53}\sqrt{v}$ erg and it is difficult to explain the features of the afterglow.

## 3 Conclusion

Unfortunately there are no true conclusions because we are unable to make clear predictions. However this is the problem of all the baryogenesis models: the physics responsible for the matter–antimatter asymmetry in the universe is unknown and common approaches are based on the construction and investigation of toy-models which contain free parameters that we can only partially constrain with the observed asymmetry (1). Moreover, most baryogenesis scenarios are based on physics at very high energy, which will be hardly tested in a near future by man-made colliders. On the other hand, if we are lucky and able to get evidences of the existence of primordial antimatter object, the latter will tell us much interesting information on high energy physics (CP violation, B violation, etc.) and, maybe, even on cosmological open questions such as the nature of dark matter.

Gamma rays from $p\bar{p}$ annihilation may be observable with future or even with existing $\gamma$-telescopes. Quite promising for discovery of cosmic antimatter are point-like sources of gamma radiation; the problem is to identify a source which is suspicious to consist of antimatter. The 100 MeV gamma ray background does not have pronounced features which would unambiguously tell that the photons came from the annihilation of antimatter. The photons produced as a result of $p\bar{p}$ annihilation would have a well known spectrum but it may be difficult to establish a small variation of the conventional spectrum due to such photons. In contrast, the 0.511 MeV line must originate from $e^+e^-$ annihilation and it is tempting to conclude that the observed excessive signal from the Galaxy and, especially, from the galactic bulge comes from astronomical antimatter objects. If an anti-star happens to be in the galactic center, its luminosity from the surface annihilation of the background matter should be strongly enhanced due to the much larger density of the interstellar matter there. So the search of the antimatter signatures in the direction of the center is quite promising. There is also a non-negligible chance to detect cosmic antinuclei and not only light anti-helium but also much heavier ones, especially if anti-stars became early supernovae.

# SEARCHING FOR KALUZA-KLEIN DARK MATTER SIGNATURES IN THE LAT ELECTRON FLUX

A.A. MOISEEV [a], EDWARD A. BALTZ [b], J.F. ORMES [c], AND L.G. TITARCHUK [d]

on behalf of the GLAST LAT Collaboration

[a] *CRESST and Astroparticle Physics Laboratory at NASA/GSFC, Greenbelt, USA*

[b] *KIPAC, Menlo Park, USA*

[c] *University of Denver, Denver, USA*

[d] *NRL, Washington, USA*

**Abstract**

We present here the prospects for the GLAST Large Area Telescope (LAT) detection of the signature of the lightest Kaluza-Klein particle (LKP). It decays by direct annihilation into electron-positron pairs that may be detectable in the high energy electron flux. We discuss the LAT capability for detecting the high energy (20 GeV - 1 TeV) cosmic ray electron flux and we analyze the LAT sensitivity to detect LKP-produced electrons for various particle masses. We include an analysis of the diffusive propagation of the electrons in the galaxy.

## 1 Introduction

The nature of dark matter and dark energy is one of the most exciting and critical problems in modern astrophysics. A number of theoretical models predict the existence of dark matter in different forms; so the experimental detection will be crucial. A large number of experiments are ongoing and planned to detect dark matter, directly and indirectly, both at accelerators and in space, where they search for dark matter signatures in cosmic radiation (see [1] and references therein). In this paper we explore the capability of GLAST Large Area Telescope (LAT), scheduled for launch in the beginning of 2008, to detect the signature of dark matter in the high energy cosmic ray electron flux. We have previously demonstrated that LAT will be a powerful detector of cosmic ray electrons, and will provide measurement of their flux with high statistical confidence [2]. We should mention that LAT is not designed to distinguish electrons from positrons, so we refer to their sum as electrons for simplicity. LAT will detect $\approx 10^7$ electrons per year above 20 GeV with the energy resolution 5-20%. Such good statistics permits detection of spectral features, among which could be ones caused by exotic sources such as Kaluza-Klein particles which manifest higher spatial dimensions. Baltz and Hooper [1] investigated the possibility of the annihilation of the Lightest Kaluza-Klein Particles (LKP), which can be a stable and viable dark matter candidate, directly into electron-positron pairs. They estimated that electron-positron pairs are produced in approximately 20% of the annihilation cases, which makes the observations viable within the model assumptions.

There are some indications of spectral features in the electron spectrum observed by ATIC [3] and PPB-BETS [4] around 300-500 GeV, as well as in the positron spectrum measured by HEAT [5], encouraging us for measurements with the LAT.

## 2 LAT Capability to detect cosmic ray electrons

It was demosnstrated earlier that the LAT can efficiently detect cosmic ray electrons [2]. Being a gamma-ray telescope, it intrinsically is an electron spectrometer. The main problem is to separate the electrons from all other species, mainly protons. In order to keep the hadron contamination in the detected electron flux under 10%, the hadron-electron separation power must be $10^3 - 10^4$. At very high energy (above a few TeV) the diffuse gamma-radiation could be a potential background, but it will be effectively eliminated by the LAT AntiCoincidence Detector. LAT's onboard trigger accepts all events with the detected energy above $\approx$ 20 GeV, which is very important in order to have unbiased data sample. We took advantage of this LAT feature and explored the instrument sensitivity in the energy range above 20 GeV. It is also good to mention that there should be no problem with albedo and geomagnetic variation in this energy range.

We have developed a set of analysis cuts that select electrons and applied them to simulated LAT data. The approach was based on using the difference in the shower development between hadron-initiated and electron-initiated events.

In order to obtain the instrument response function for electrons we simulated the electron spectrum incident on LAT and applied our selections. In the energy range from 20 GeV to 1 TeV the effective geometric factor (for electrons) after applying our cuts is $0.2 - 2$ m$^2$ sr and energy resolution is 5-20% depending on the energy. We also applied the selections to the simulated cosmic ray flux (LAT background flux is used, see [6]) and determined the residual hadron contamination to be $\approx 3\%$ of the remaining electron flux.

In order to test the approach, we run an independent simulation of the incident flux and used the response function obtained to reconstruct the spectrum. For the simulation of the electron flux we used the diffusion equation solution given in [7]. Fig.1 shows our spectrum reconstruction for the simulated electron flux collected during 1 year of LAT observations. The flux originated from an "hypothetical" single burst-like source, $2 \times 10^5$ years old, at a distance of 100 pc. The diffusion coefficient D was assumed to be energy dependent as $D = D_0\,(1 + E/E_0)^{0.6}$ with $D_0 = 10^{28}$ cm$^2$s$^{-1}$. The expected spectral cutoff for this model is $\approx 1.2$ TeV. With the demonstrated precision in the spectrum reconstruction we should be able to recognize the specific features which can be associated with dark matter.

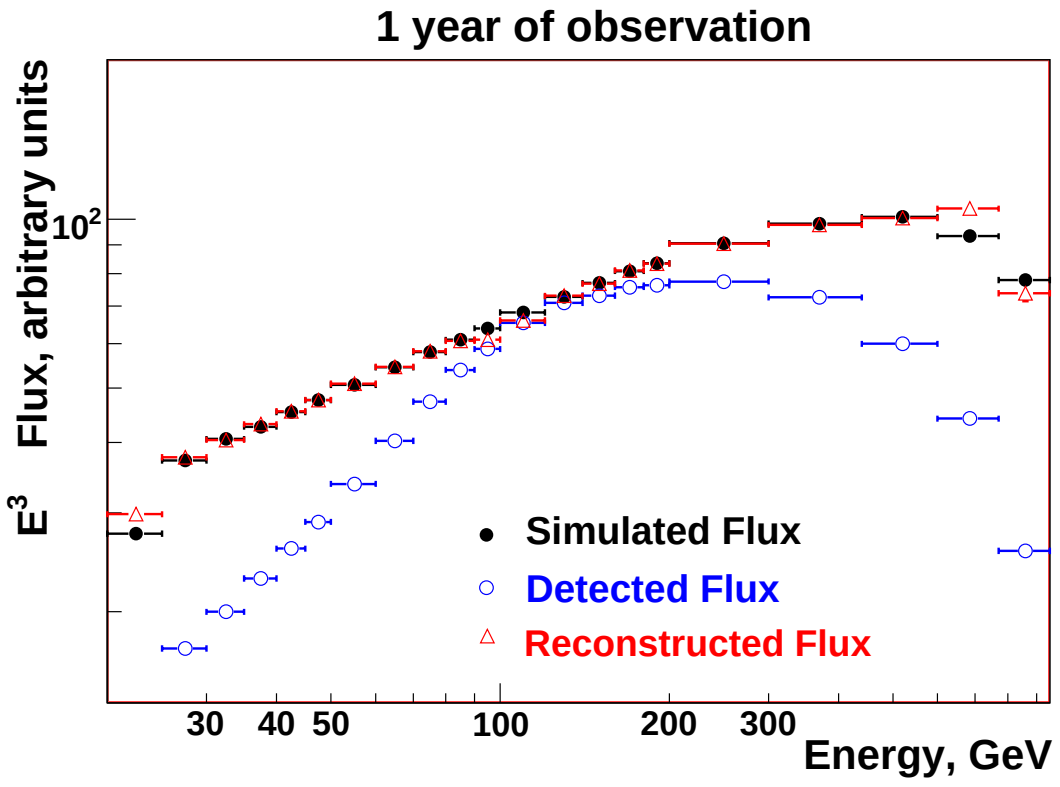

Figure 1: Simulated electron flux reconstruction for LAT

## 3 Diffusive propagation of the LKP signal

Now we want to include the contribution from the LKP annihilation into the electron flux, and see how it can be detected by the LAT. We consider this contribution as a continuous injection from a single point source of monoenergetic electrons, assuming dark matter clumpiness, and determine the effect of their propagation through space to the Earth. After that we will test LAT response to such a spectrum, varying the source parameters LKP mass and distance. We treat the propagation of electrons using the stationary diffusion equation which in the spherical symmetric case is presented as

$$\frac{D}{r^2}\frac{\partial}{\partial r}r^2\frac{\partial f}{\partial r} + \frac{\partial}{\partial \gamma}(Pf) = Q \tag{1}$$

where $f(r,\gamma)$ is a distribution of electron number over $\gamma = E/m_e c^2$ and radius $r$ at time $t$.

We assume that the continuous energy loss is dictated by synchrotron and inverse Compton losess

$$-d\gamma/dt = P(\gamma) = p_2\gamma^2 \tag{2}$$

where

$$p_2 = 5.2 \times 10^{-20}\frac{w_0}{1\mathrm{eV/cm^3}}\ \mathrm{s^{-1}} \tag{3}$$

and $\omega_0 \simeq 1$ eV/cm$^3$ (see [7] for the propagation details)

We choose the energy-dependent diffusion coefficient in the form

$$D(\gamma) = D_0(1+\gamma/\gamma_g)^\eta\ \mathrm{cm^2s^{-1}} \tag{4}$$

where $\gamma_g$ and $D_0$ are model parameters.

We derive the general solution of Eq. (1) for arbitrary source function in the factorized form

$$Q = \varphi(r)\psi(\gamma) \tag{5}$$

as well as a solution for a $\delta-$ function injection, i.e. for

$$Q(r,\gamma) = \delta(\gamma-\gamma_*)\delta(r-r_0)/4\pi r_0^2. \tag{6}$$

Using Eq. (1) it can be shown that the Green's function $G_{0,\gamma_*}(r,\gamma)$, as a solution for the delta-function source (see Eq. 6) is presented as

$$G_{0,\gamma_*}(r,\gamma) = \frac{1}{D_0}\frac{R_0[r,u(\gamma,\gamma_*)]}{\gamma^2} \tag{7}$$

where $R_0[r,u(\gamma,\gamma_*)]$ and $u(\gamma,\gamma_*)$ are determined by Eq. (8) and Eq. (9) respectively:

$$R_0(r,u) = \frac{1}{8u(\pi u)^{1/2}}\exp(-r^2/4u). \tag{8}$$

$$u(\gamma,\gamma_0) = \frac{D_0}{p_2}\int_\gamma^{\gamma_0}\frac{(1+\gamma/\gamma_g)^\eta d\gamma}{\gamma^2}. \tag{9}$$

Integral (9) can be presented in the analytical form in two cases: for $\gamma_g \to \infty$ or $\eta = 0$ it is (see [8])

$$u(\gamma, \gamma_0) = \frac{D_0}{p_2}\left(\frac{1}{\gamma} - \frac{1}{\gamma_0}\right). \tag{10}$$

and for $\eta = 0.5$ it is

$$u(\gamma, \gamma_0) = \frac{D_0}{p_2} \times$$

$$\left[\frac{1}{\gamma}\left(1 + \frac{\gamma}{\gamma_g}\right)^{1/2} - \frac{1}{\gamma_0}\left(1 + \frac{\gamma_0}{\gamma_g}\right)^{1/2} + \frac{1}{\gamma_g}\ln\frac{(\gamma_g/\gamma)^{1/2} + (1 + \gamma_g/\gamma)^{1/2}}{(\gamma_g/\gamma_0)^{1/2} + (1 + \gamma_g/\gamma_0)^{1/2}}\right]. \tag{11}$$

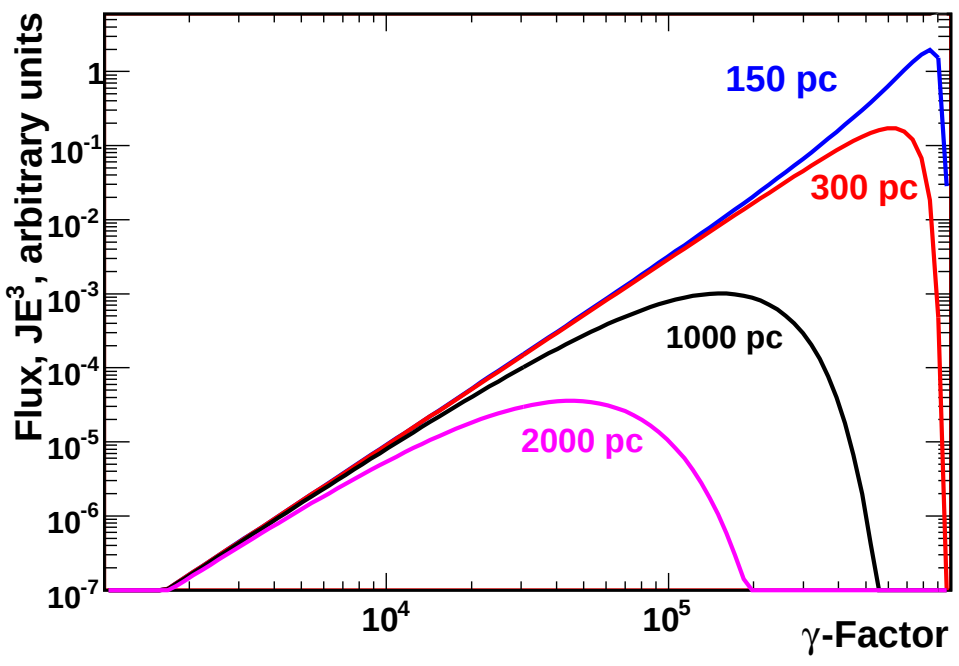

Figure 2: Propagation of the signal from LKP with the mass of 500 GeV, from different distances

Fig.2 illustrates the solution obtained and shows the propagation of the signal from annihilation of LKP, with mass of 500 GeV, for different distances. Due to losses during propagation, both the peak energy and the signal magnitude decrease with the increasing distance. There will be a superposition of contributions from dark matter clumps at different distances, which could reveal themselves as bumps in the spectrum, but the closest clump should determine the edge in the spectrum which should be clearly seen.

## 4 Prospects for LAT to observe LKP

Our analysis of the LAT capability for detection of electrons demonstrated the low residual hadron contamination in the resulting electron flux ($< 3\%$, see

Section 2). The contamination from gamma-radiation will also be negligible, so the dominant background in the search for LKP signature will consist of only "conventional" electron flux. Now we can apply the LAT capability for electron detection to one of the dark matter models, using the scenario given in [1] as an example. The LKP annihilation will be seen as a line (edge) in the electron spectrum of magnitude proportional to $m_{LKP}^{-6}$

$$\frac{dN_e}{dE_e} = \frac{Q_{line}(m_{LKP})}{b(E_e)}\theta(m_{LKP} - E_e)$$

$$\sim \langle\sigma v\rangle \left(\frac{\rho_0}{m_{LKP}}\right)^2 \left(\frac{1}{E_e^2}\right) \theta(m_{LKP} - E_e) \sim m_{LKP}^{-6} \tag{12}$$

where $\langle\sigma v\rangle$ is the total annihilation cross section of LKP, and $Q_{line}$ is the rate of electron and positron injection from direct LKP annihilation.

Using the numbers from [1]: boost factor $B = 5$ and $\rho_{local} = 0.4$ GeV cm$^3$, the magnitude of the signal after propagation is estimated as

$$\left(\frac{dN_e}{dE_e}\right) \approx \frac{9.5 \times 10^8}{m_{LKP}^6\,[\mathrm{GeV}]} m^{-2}\mathrm{s}^{-1}\mathrm{sr}^{-1}\mathrm{GeV}^{-1}. \tag{13}$$

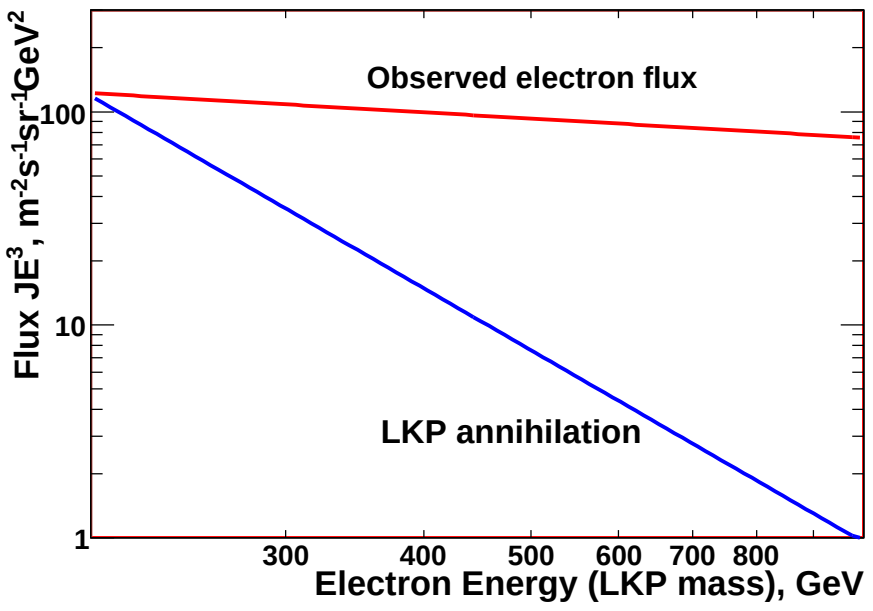

Figure 3: Expected electron flux from LKP annihilation, along with the observed electron flux

Fig.3 shows the expected signal from LKP vs. the $m_{LKP}$ plotted along with the "conventional" electron flux for the comparison. Of course, even in this optimistic model, the background dominates over the signal, but we now determine what will be the LAT sensitivity. Fig.4 shows the significance of LKP detection in the LAT-detected electron flux in one year of ovservation, and the observation time needed to detect LKP feature with $5\sigma$ confidence for a source at 100 pc. We can conclude that 600 GeV is probably the heaviest LKP

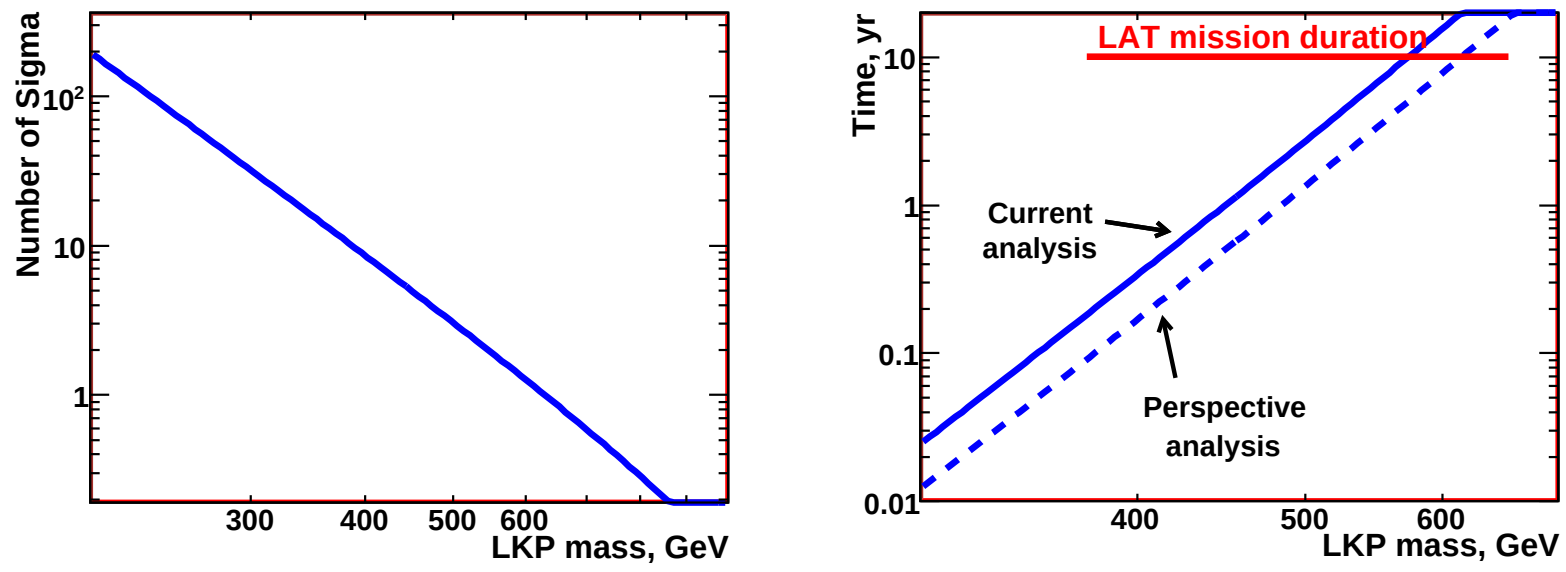

Figure 4: LKP detection in LAT electron spectrum. Left panel - detection significance for 1 year. Right panel - time needed to detect LKP feature with $5\sigma$ confidence

which could be observed within the constrains assumed. Taking into account that for thermal freeze-out, LKP mass in the range 600-700 GeV is preferred, the feasible window for LKP mass in the LAT search is rather narrow.

Now we illustrate our analysis by adding the signal from LKP (mass 300 GeV), from a single clump at a distantce of 100 pc, to the "conventional" electron flux shown in Fig.1, using the solution obtained in Section 3. The result is shown in Fig.6, with a clear signature of presence of a monoenergetic component. This is a very favorable situation, but to some extent consistent with references [3] and [4].

## 5 Conclusion

We analyzed the capability of the LAT to detect high energy cosmic ray electrons and applied it to the model of LKP direct annihilation into electron-positron pairs. Using the estimates for this model as given in [1] as an example to demonstrate the detection feasibility, we estimated that within this model, the LAT will be able to recognize the LKP-caused spectral edge in the electron spectrum up to a LKP mass of 600-700 GeV. The results obtained can be applied to any dark matter model where electrons are produced in order to estimate the LAT sensitivity. The important feature is that the dominating background in these measurements will be only the "conventional" electron flux.

We want to thank all LAT team members, and especially the members of LAT Dark Matter Science Working Group for their support and valuable suggestions. We are grateful to Robert Hartman and Jan Conrad for their comments on this paper.

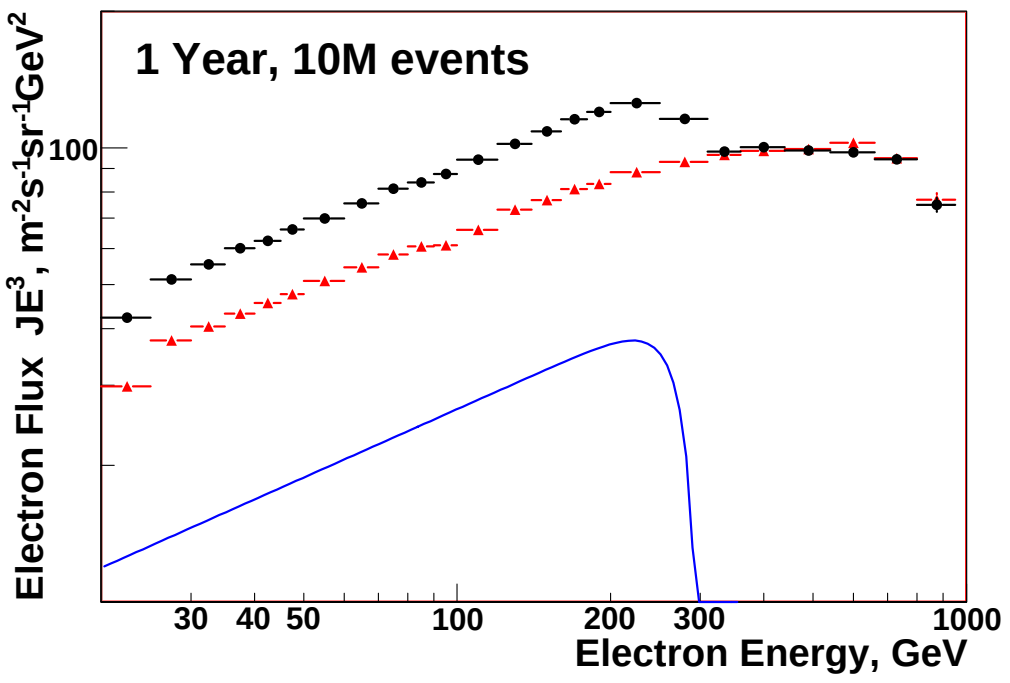

Figure 5: Illustration of the simulated reconstructed LAT electron spectrum with the presence of signal from LKP with the mass of 300 GeV, for one year of observations. Red dots - "conventional" electron flux, blue line - LKP signal, black dots - reconstructed spectrum with the presence of LKP signal

# RECENT RESULTS AND PERSPECTIVES FROM THE MAGIC EXPERIMENT

BARBARA DE LOTTO[a,b] FOR THE MAGIC COLLABORATION

[a] *Dipartimento di Fisica, Università di Udine, via delle Scienze 208, Udine, Italy*

[b] *INFN, Gruppo Collegato di Udine, Sezione di Trieste, Area di Ricerca, Padriciano 99, Trieste, Italy*

**Abstract**

MAGIC is currently the world's largest single dish imaging atmospheric Cherenkov telescope. During the first two Cycles of operation seventeen sources have been detected and four are in queue. In this talk I will review the most recent results, both on galactic and extragalactic observations. An overview on the ongoing upgrades and future perspectives will also be presented.

## 1 Introduction

The Major Atmospheric Gamma-ray Imaging Cherenkov (MAGIC) telescope was ready for commissioning in October 2003 and started regular data-taking in October 2004. Since then two cycles of observations have been completed, with about 2000 hours of dark time for physics and a mean efficiency of 85%. Also about 300 hours/year of moon time have been taken, and ∼200 hours for Target-of-Opportunity projects, which will likely increase with the increased number of collaborations (Suzaku, Swift, AGILE, GLAST, IceCube). With the current generation of ground based Cherenkov Telescopes (CT) and the upcoming launch of the new satellites, for the first time not only the search for sources, but also the detailed investigation of individual objects in the sky is possible.

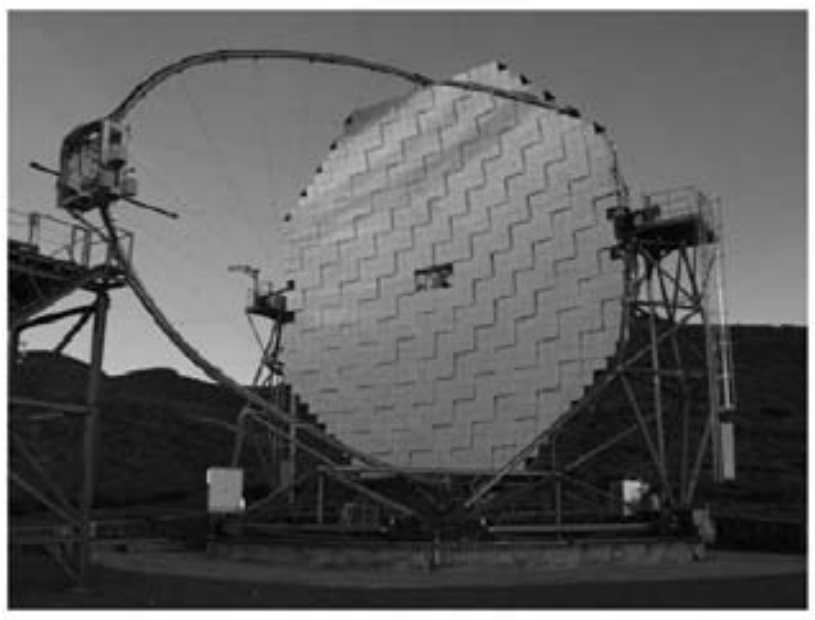

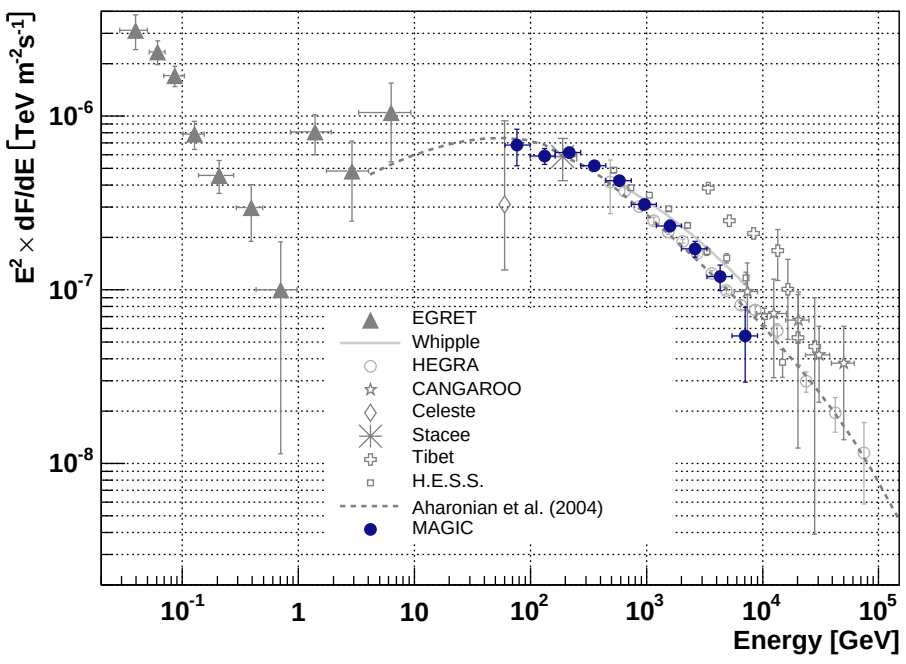

Figure 1: Left: the MAGIC telescope. Right: Spectral energy distribution (SED) of the $\gamma$-ray emission of Crab Nebula.

### 1.1 The MAGIC telescope

MAGIC [1] is currently the largest single-dish CT in operation, located at the Roque de los Muchachos in the Canary Island La Palma at $\sim$ 2200 m altitude. Its parabolic shaped reflector has a diameter of 17 m, resulting in a total mirror area of 236 m$^2$ formed by 974 Al-mirrors. The PMT camera is mounted at a distance of 17 m from the mirror and consists of 576 PMTs, resulting in a field of view of 3.5°.

The main goal of the MAGIC design was to reach low energy threshold while keeping the total weight as low as possible to allow fast rotation. To reach this, an Active Mirror Control system has been implemented to correct for deformations of the dish under varying gravitational load, camera sagging and other effects. The typical trigger rate during standard operation is approximately 250 Hz. The current performance reached by the MAGIC telescope is:

- angular resolution on a point-like source: $\sim$ 0.1°
- repositioning to any orientation: $<$ 40 s (with reduced motor power, a safety scheme implemented now) or $<$ 25 s (full motor power).

For small zenith angles, we reach:

- energy threshold (trigger): $\sim$ 50 GeV (trigger), $\sim$ 60 GeV (analysis)
- energy resolution: 22% above 150 GeV
- flux sensitivity (5$\sigma$ in 50 h): 2% of Crab flux

Additional information can be found at http://wwwmagic.mppmu.mpg.de/.

## 2 Recent Results: Galactic sources

Several results of observations of galactic sources have already been published [2], [3], [4], [5]; here we only mention the most recent ones: for more details see the dedicated talk by V. Vitale.

### 2.1 The Crab Nebula

Figure 1 (right) shows the spectral energy density distribution of the Crab Nebula [6] as measured by MAGIC, together with results from other experiments. Photons above 60 GeV are detected.

### 2.2 MAGIC J0616+225

The detection of a new source of very high energy $\gamma$-ray emission located close to the Galactic Plane, MAGIC J0616+225, which is spatially consistent with SNR IC 443, has been recently reported [7]. The statistical significance is 5.7 sigma in the 2006/07 data. The sky map of $\gamma$-ray candidate events (background subtracted) in the direction of MAGIC J0616+225 for an energy threshold of about 150 GeV in galactic coordinates is shown in Figure 2 (left). Overlayed are the multiwavelenght emission contours from other experiments.

### 2.3 Cignus X-1

Observations were performed in very high energy band (VHE, $E_\gamma \geq 100$ GeV) of the black hole X-ray binary Cygnus X-1, for a total of 40 hours between June and November 2006. Evidence for a signal at 4.0 standard deviations ($\sigma$) significance level was obtained for 154 minutes effective on-time on September 24, coinciding with an X-ray flare seen by satellites [8].

### 2.4 Cassiopeia A

The shell-type supernova remnant Cassiopeia A was observed between July 2006 and January 2007 for a total time of 47 hours. The source was detected above an energy of 250 GeV with a significance of 5.3 $\sigma$ and a photon flux above 1 TeV of $(7.3 \pm 0.7_{stat} \pm 2.2_{sys}) \times 10^{-13}$ cm$^{-2}$s$^{-1}$. The photon spectrum is compatible with a power law dN/dE $\propto$E$^{-\Gamma}$ with a photon index $\Gamma$=2.3$\pm$0.2$_{stat}\pm$0.2$_{sys}$. The source is point-like within the angular resolution of the telescope [9].

## 3 Recent Results: extragalactic sources

MAGIC detected up to now (July 2007) 9 extragalactic sources and all of them are well-established Active Galactic Nuclei (AGN). The list, with in-

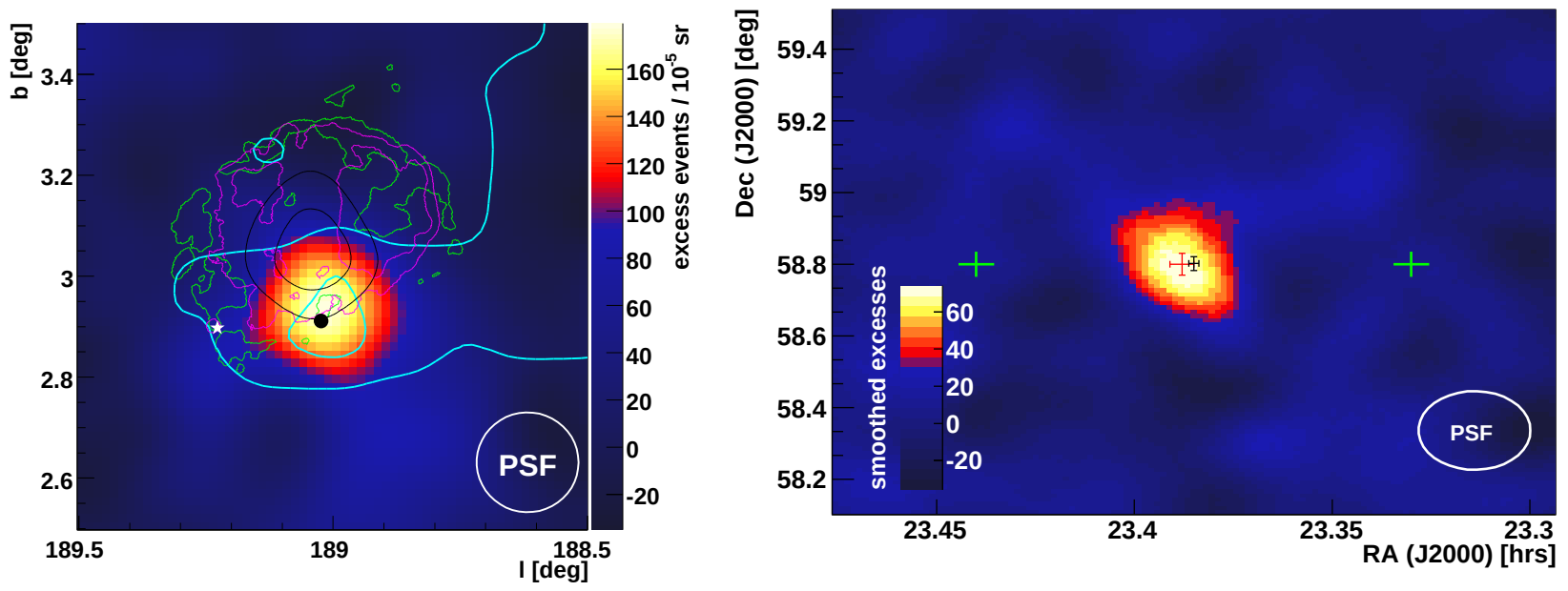

Figure 2: Skymap plot of the IC443 region (left) and Cas A (right).

creasing redshift value, is: Mkn 421 ($z = 0.030$) [10], Mkn 501 ($z = 0.034$) [11], 1ES 2344+514 ($z = 0.044$) [12], Mkn 180 ($z = 0.045$) [13], 1ES 1959+650 ($z = 0.047$) [14], BL Lacertae ($z = 0.069$) [15], 1ES 1218+304 ($z = 0.182$) [16], PG 1553+113 [17](unknown redshift, $0.42 > z > 0.09$ [18] [19]) and 1ES1011+496 ($z = 0.212$) [20]. Here we will highlight on the most recent results.

### 3.1 Mkn501 flare

Together with Mkn 421, Mkn 501 is one of the most studied VHE sources. It was observed with MAGIC between June and July 2005 [11].

We will focus here on the flare which was detected by MAGIC during the June/July '05 observation. In Figure 3 (left) the light curve for the night July 9 separated in different energy bands is shown. An unprecedented doubling time as short as less then 4 minutes was detected. The rapid increase in the flux level was accompanied by a hardening of the differential spectrum.

An energy dependent time delay of the flare peak emission can result from the dynamics of the source, such as gradual electron acceleration in the emitting plasma. A somewhat more speculative issue that blazar emission permits to explore is related to non-conventional physics, such as a violation induced in the Lorentz-Poincaré symmetry. The time analysis of the Mkn501 flare is ongoing.

### 3.2 BLLac

BL Lacertae is the historical prototype of a class of powerful $\gamma$-ray emitters: the "BL Lac objects"[21]. BL Lac objects are AGNs with their jet well aligned with the observer's line of sight. This class of object is further subdivided according to where the synchrotron emission peak lies: if it is in the sub-millimeter to optical band, the objects are classified as “Low-frequency peaked BL Lacs” (LBLs); if it is in the UV to X-ray band, they are referred at as “High-frequency peaked BL Lacs” (HBLs) [22, 23].

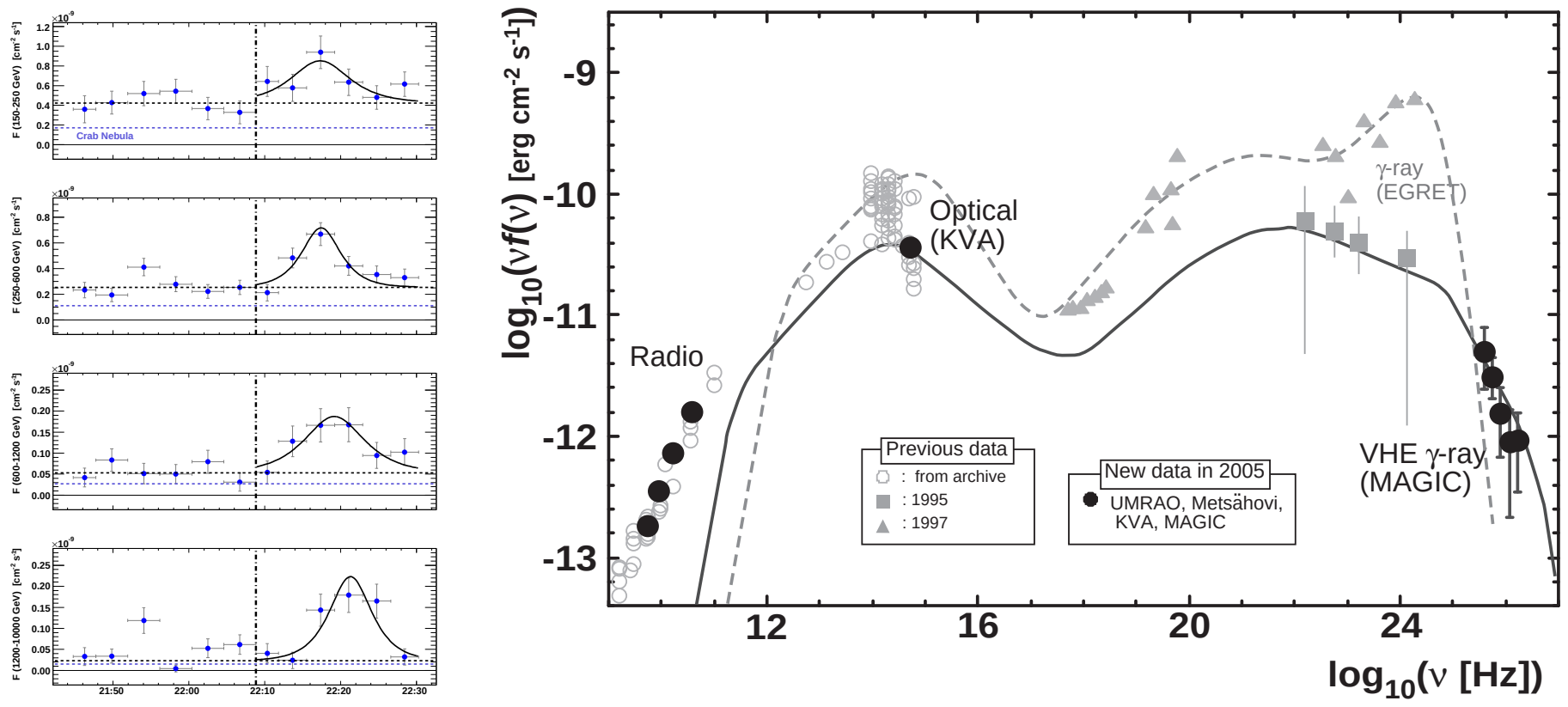

Figure 3: Left: Mkn501 light curve for the night July 9. Right: SED for BL Lacertae.

The Crimean Observatory claimed a detection at a $7.2\sigma$ level [24], but HEGRA could not detect any significant signal during the same period, obtaining only an upper limit [25], which contradicted the claim.

BL Lacertae was observed for 22 hours from August to December 2005 [15]. The observation showed at a level of $5.1\sigma$ a new VHE source of flux 3% CU above 200 GeV. The spectrum is compatible with a pure power law of index $3.6 \pm 0.5$. The SED is shown in Figure 3 (right) together with the results from other experiments. No significant variations of the VHE flux were detected. The source was also observed from July to September 2006 for 26 hours without any detection. There is a remarkable agreement of the observed trend with the optical activity of the source, that, together with the discovery of Mkn 180, reinforces the idea of a connection between optical activity and increased VHE emission.

BL Lac is the first member of LBL ever detected to emit in the VHE region. Given the very hard spectrum of the source, in agreement with LBL modelling, the low energy of MAGIC was necessary for its discovery.

### 3.3 1ES1011+496

Very recently we reported on the discovery of VHE $\gamma$-ray emission from the BL Lacertae object 1ES 1011+496. The observation was triggered by an optical outburst in March 2007 and the source was observed with the MAGIC telescope from March to May 2007. Observing for 18.7 hours we find an excess of $6.2\,\sigma$ with an integrated flux above 200 GeV of $(1.58\pm0.32)\cdot 10^{-11}$ photons cm$^{-2}$ s$^{-1}$. The energy spectrum of 1ES 1011+496 is shown in Fig. 4. It is well approximated by a power law with spectral index of $3.3 \pm 0.7$ after correction for EBL absorption. The redshift of 1ES 1011+496 has been detected, based on an optical spectrum that reveals the absorption lines of the host galaxy. The redshift of $z = 0.212$ makes 1ES 1011+496 the most distant source observed to emit VHE $\gamma$-rays up to date.

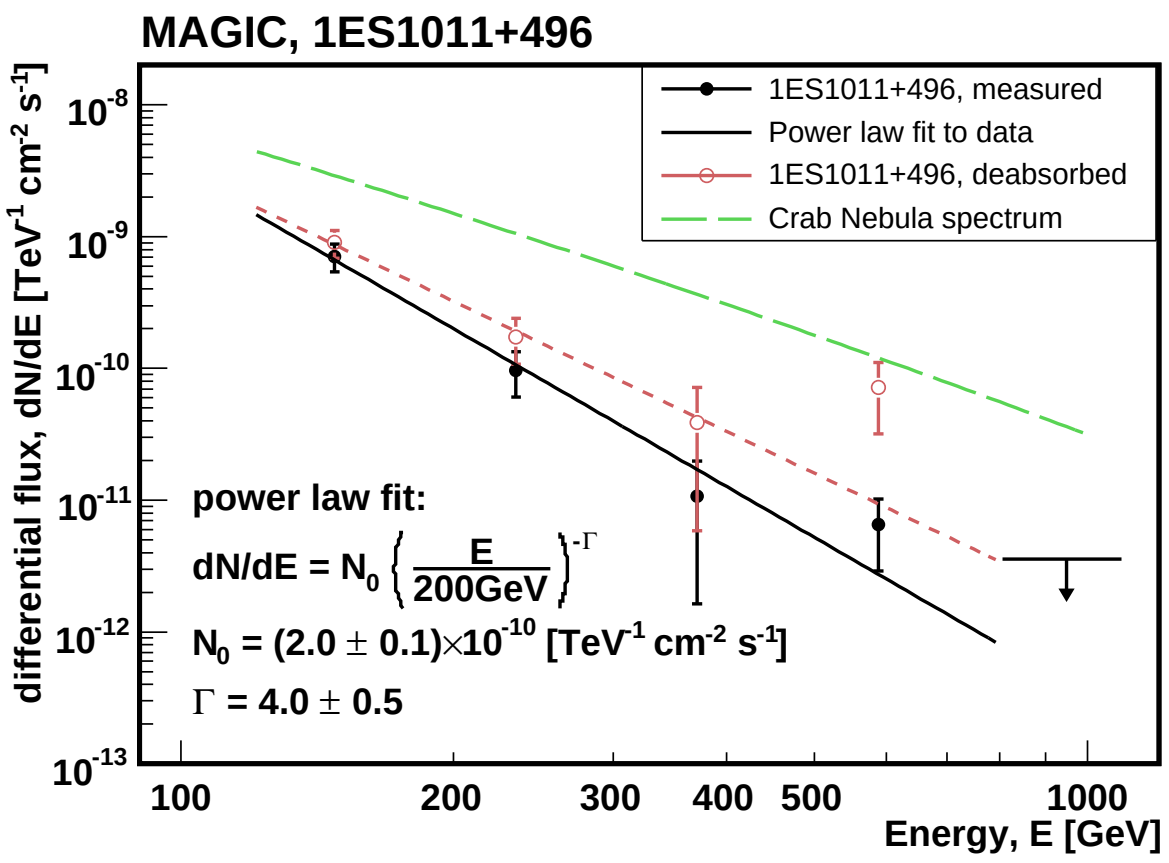

Figure 4: The measured spectrum for 1ES1011 (black filled circles), the power-law fit to the data (solid line), the deabsorbed spectrum (brown open circles), and the fit to the deabsorbed spectrum (dashed brown line).

## 4 Conclusion and Outlook

The performance of the MAGIC telescope complied with the design specifications. Since the end of commissioning, two full years of physics campaign followed and the data were almost completely analysed. MAGIC detected a total of eight galactic and nine extragalactic sources, discovering new populations and new features. Most of the new sources could be actually discovered because of the low energy threshold of MAGIC and its good sensitivity even below 100 GeV. Its good sensitivity, mainly due to the huge effective area common to all detectors exploiting the IACT, allowed also to resolve short term flux variability down to 2 minutes.

The construction of a second telescope, MAGIC II, close to the original one, has started. It incorporates some minor modifications suggested by the experience of running MAGIC for the last three years, as well as some basic changes.

Larger (1 $m^2$ surface) and lighter mirrors will be implemented, under the responsibility of the Italian INFN and INAF.

A huge improvement of the Data Acquisition System is foreseen: a new 2GHz sampler, based on a custom chip named Domino, is foreseen to reduce the Night Sky Background contribution on the signal. The heat dissipation required by standard FADC is causing troubles and the new design will reduce it by two orders of magnitude. The new readout system is also responsibility of the INFN.

A new camera design with a total number of channels of $\sim 1000$ is foreseen; a big improvement is expected by high quantum efficiency devices such as HPDs and and SiPMs.

MAGIC II is expected to be ready by September 2008, when stereo observations will be operational, allowing an increase in the sensitivity by at least a factor 2, and other improvements in the energy and direction reconstruction.

With the advent of MAGIC II, VHE $\gamma$-ray astrophysics will reach down a level of 1% CU. In the meanwhile, the AGILE results should come and GLAST should become fully operational, closing the current observational gap between $\sim 1$ and 60 GeV and extending observations of the electromagnetic radiation, without solution of continuity, up to almost 100 TeV.

# INDIRECT DARK MATTER SEARCHES WITH THE MAGIC TELESCOPE

M. DORO [a], H. BARTKO [b] G. BERTONE [c] A. BILAND [d] M. GAUG [e] M. MARIOTTI [a] F. PRADA [f] M. RISSI [d] M. SÁNCHEZ-CONDE [f] L.S. STARK [d] F. ZANDANEL [a] FOR THE MAGIC COLLABORATION [g]

[a] *University of Padova & INFN, Padova, Italy*

[b] *Max-Planck-Institut für Physik, München, Germany*

[c] *Institut d'Astrophysique de Paris, Paris, France*

[d] *Institute for Particle Physics ETH, Zurich, Switzerland*

[e] *Instituto de Astrofísica de Canarias (IAC), Tenerife, Spain*

[f] *IAA-CSIC, Granada, Spain*

[g] *http://wwwmagic.mppmu.mpg.de/collaboration/members/index.html*

### Abstract

The MAGIC telescope, currently the largest dish Imaging Atmospheric Cherenkov Telescope (IACT), is operating since 2004 in the Canary island of La Palma. The low energy threshold and the good sensitivity make MAGIC telescope well-suited for dark matter hunting in astronomical targets.
In this proceeding, after a brief introduction of the physics framework and a discussion on the possible interesting targets, a strategy for the observation will be given for the MAGIC telescope together with an outlook on the next future of the experiment.

## 1 Observation of dark matter with atmospheric Cherenkov telescopes

In the context of a ΛCDM scenario for the Universe, the WMAP experiment recently allowed to put strong constraints on a large number of parameters, including the total abundance of relic dark matter (DM) around $\Omega_{DM} = 0.24$ [1]. The ΛCDM scenario favours a WIMP-like DM particle, i.e. a weak and gravitational interacting particle but neutral for color and charge. A number of theoretical particles satisfies the requirements of the WMAP constraints together with other experimental limits (see [2] for a detailed discussion and references therein). Most of these candidates foresees a very outstanding new physics. We discuss here two very promising candidates: the *neutralino* particle ($\chi$) inside the mSUGRA extension of the Standard Model [2], and the *light Kaluza particle* (LKP) in multi-dimensional theories [3].
Both particles have decay channels that include gamma-rays in their final state of emission, both in continuous and line emissions, even if the estimation of the branching ratios depends on parameters which varies among different models. In the case of neutralino in the mSUGRA framework, the uncertainties relies on 4 continuous and 1 discrete parameters: The mixing angle $tan\beta$ which is the ratio of the vacuum expectation value for the two Higgs bosons constituting the neutralino, the universal trilinear mass term $A_0$, the universal scalar and gaugino masses $m_0$ and $m_{1/2}$ respectively and the sign of the Higgsino mass $\mu$. Different realizations for the neutralino are usually plotted in a $m_{1/2} - m_0$ bidimensional grid for different values of $\tan\beta$ (see Fig. 1). Currently, the $gaugino - like$ neutralino is best preferred, with mass of few hundreds GeV, where 45 GeV is the mass lower limit due to accelerator experiments and an upper limits can be placed as several TeV. Once the set of parameters is defined, the branching ratios are a consequence. Usually the annihilation of neutralinos, which is a Majorana particle, leads to the production of all standard models particles, and in particular quarks $b\,\bar{b}$, W and Z bosons, and $\tau$ leptons. Also a direct annihilation into two gammas is possible, even if loop-suppressed.

The LKP particle $B^{(1)}$ is the first Kaluza-Klein excitation of the photon. Also this particles decay into standard particles, with a preference for the leptonic channel and the quark channel. Also decay into neutrinos and Higgs bosons are foreseen. A tentative mass range can be fixed between 400 GeV and 1200 GeV.
Present dark matter searches are three-fold: direct detection through energy recoil in target materials where DM elastically scatters, direct production at accelerators, and indirect detection through products of the annihilation of DM which takes place at cosmic places. In this latter case one can use anti-particles, neutrino, protons and deuterons, and of course gamma-rays as tracers of the reaction.

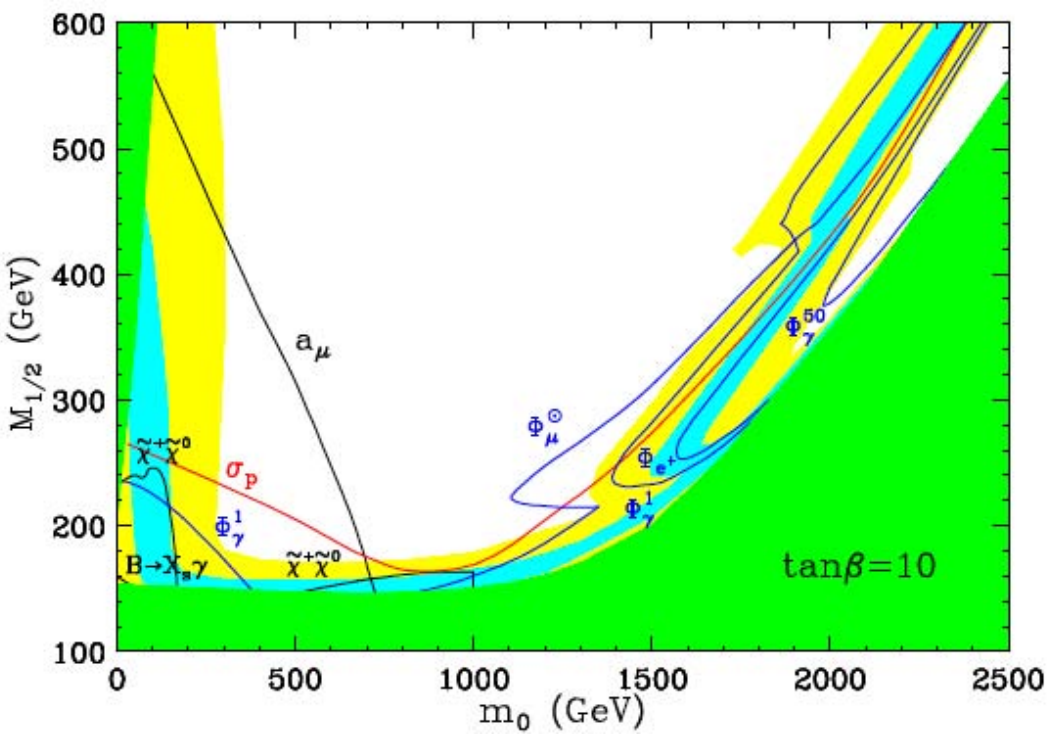

Figure 1: Possible neutralino realizations in a $m_{1/2} - m_0$ plane for a given value of $\tan\beta = 10$. The neutralino should be found in the shaded yellow and shaded blue region. Figure from [4].

### 1.1 Spectrum and fluxes from annihilations of DM

The spectrum of gamma-rays from annihilation of DM particles depends on the particle physics characteristics of the particle and its mass; nevertheless some basic features have generality: a) gammas from the fermionic channels have spectrum with slope close to $-1.5$, b) the cut-off in the LKP case is steeper than the exponential cut-off of the neutralino case, and if observed can constitute a discriminating factor between the two particles, c) a bump close to the cut-off can be found with final state radiation when W particles are created and d) all the spectrum look similar when the DM mass increases.
The gamma-rays flux for a given gamma spectrum $dN_\gamma/dE$ originating from annihilation regions is:

$$\frac{d\Phi(E)}{dE} = \frac{\sigma v}{2m_\chi^2}\frac{1}{d^2}\frac{dN_\gamma(E)}{dE}\int_{r_{cut}} \rho^2(r)\, r^2\, dr \tag{1}$$

where $\sigma v$ is the DM annihilation cross section times relative velocity, $d$ is the object distance, $dN/dE$ is the gamma spectrum and $\rho(r)$ the DM density profile.
The flux is factorized in a particle physics factor, and an astrophysical factor which described the DM profile around the target. Both factor play an important role and their interplay acts in a way that the uncertainties in the flux estimations can be of several orders of magnitude, which make strategies for observation very hard to define for most scenarios. We will show in Sec. 3.3 that in some cases the situation can be slightly improved.

## 2 Dark Matter searches with IACTs

The observation of the sky with IACTs happens through the detection of very huge showers of particles initiated by VHE cosmic particles impinging the Earth atmosphere. As a shower develops in the atmosphere, a larger and larger number of charged particles is produced, each having enough energy to boost the Cherenkov emission when the primary particle energy exceed the GeV. The shower creates then a very fast ($\sim$ 2 ns) flash of blue-UV light which is collected and focussed onto a pixellated camera. An off-line analysis allows to reconstruct the primary particle characteristics (energy, direction, type) on the basis of the image recorded in the camera, which is typically an ellipse pointing to the center.
The MAGIC telescope [5] is located at the Canary island of La Palma, at an altitude of 2,200 m.asl. Its world largest dish with a diameter of 17 m (236 $m^2$ area) allows a lower energy threshold of 60 GeV at zenith, and a sensitivity of less than 2.5% of Crab in 50 hours. Its energy range then overlaps with the energy range of the two DM candidates above described, making the telescope suited for their search. In case the signal from DM region in the sky is strong enough, IACTs will be able to detect DM, and through spectral analysis a clear sign of a particular DM candidate could be in turn obtained.

## 3 Annihilation sites of interest for the MAGIC telescopes

It is commonly believed that the entire Universe is pervaded by the presence of DM. Being gravitationally active, regions with over-densities are expected. Most of the dark matter is nevertheless concentrated in spherical halos around the galaxies. Nevertheless, as the annihilation rate depends on the square of the density $\Gamma \sim \rho_{DM}^2$ (see Eq. 1), the search can only be performed in "amplification sites" where DM concentrates. In the following sections, the interesting places for the MAGIC telescope are discussed.

### 3.1 The Milky Way centre

The centre of most galaxies is supposed to host a Super Massive Black Hole (SMBH) of the order of $10^9$ $M_\odot$. The SMBH is surrounded by a dense torus of accreting material, optically blind, and in some cases two jets of VHE particles and radiation come out from the SMBH in direction perpendicular to the thorus itself. The DM, driven by gravitational forces, accrete around the GCs and its effect can be clearly observed on the motion of stars around it, the gravitational lensing and other effects. The DM profile around the GC is almost well-known at distances beyond some kpc from the centre, where it is expected to be a power-law Navarro-Frenck-White (NFW) profile with very flat slope $-3$ [7]. In the inner region, namely the inner parsecs, experimental evidences and theorical models disagree in discriminating between *core* models, that means profile foreseeing a central plateau, and *cusp* models, which foresee the presence of a clump of slope between 0.5 and 1.5. The former case seems supported by experimental observations [8, 9], while the second one comes out from any N-body simulations.
The observation of the MW GC was performed by MAGIC in the period 2004-06 [6], together with other IACTs. A total of 24 hours of observation were carried on at large zenith angle, above 58 deg. The GC was observed with a significance of 7.3 sigma and 257 excess events.
The GC showed a steady emission within the observation time, with a pure power-law spectrum without cut-off ranging from 400 GeV to 20 TeV, with a slope of $-2.2 \pm 0.2$, see Fig. 2.

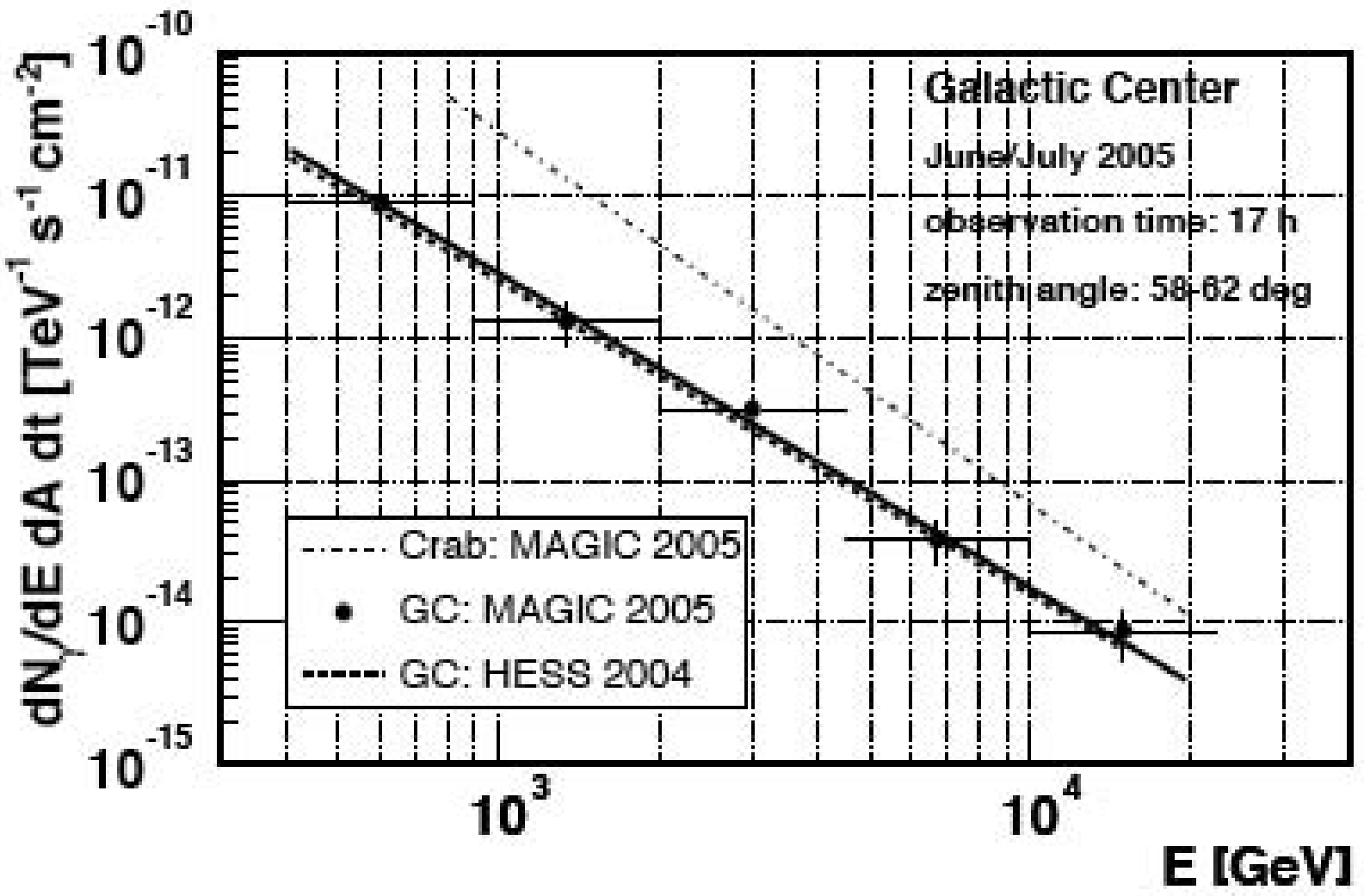

Figure 2: Gamma-ray spectrum of gamma–rays from the observation of GC with the MAGIC telescope during 2004-06.

The observed signal is a typical sign of an astrophysical emitter. In fact, located at the GC are two known emitters, namely the SMBH SgrA* and the SNR SgrA East, from where probably the observed signal comes from. This result anyhow does not indicate the absence of DM in the GC, but states that even if it exists, it is overcome by the stronger background astrophysical emission. Even if it exists, the DM signal cannot be then more than few percents of the total observed signal.

### 3.2 Dwarf Spheroidal Galaxies

A number of galaxies in the Universe is observed with a very low brightness compared to their mass content. The usual parameter for comparison is the ratio of there parameters in terms of the same units for the Sun $M_\odot/L_\odot$. Some of these object can have M/L larger than a hundred or even a thousand $M_\odot/L_\odot$ [10]. The common interpretation is that these objects are dominated by DM compared to barionic luminous matter or gas and dust.
Among these objects are the dwarf spheroidal galaxies (dSphs), which are small galaxies gravitationally bound to the Milky Way. Even if their identification is not straightforward as they are dark objects, already a number of them were clearly identified.
The DM distribution around these objects follows similar ideas with respect to the GC, with a flat distribution outside the central region, and a cusped or cored profile in the center. Studies on the motion of the stars were performed for some of these objects compatible with a core profile. A flux extimation is once more subject to large uncertainties. Studies on the observability of the Draco dSph shows furthermore that with the given pointing accurancy (PSF) of the MAGIC telescope, around 0.1 deg, it is not possible to discriminate between the two profiles because the inner region is seen as point-like and then the excess in the profile is smeared out, see Fig. 3 for details.
MAGIC telescope observed the Draco dSph during 2005-2007 for a total of about 15 hours without detecting any signal above the background from the region. With the obtained upper limits on the flux emission it is still not possible to put serious constraints on the SUSY parameters space and more observational time is required for this object. The results are currently under publication.

### 3.3 IMBHs

The underlying theory expressed in this section can be found in Ref. [11]. A different subject for DM observation raised the interest in the last few years. During the studies of the evolution of SMBHs which are believed to be present in the centre of most galaxies, one has to face the problem on how to create such huge objects within a very brief amount of time: in fact some SMBHs are observed in very young (1 Gyr) galaxy. A possible explanation resides in major event of merging of smaller seed black holes of intermediate mass (10-$10^5$ $M_\odot$) in the early Universe. A large number of these objects can in fact be produced either by collapse of first zero-metallicity Pop.3 stars (scenario I), or by collapse of giant molecular clouds (scenario II). Some of these objects could not have suffered major mergings and could have remained wandering objects in the galactic halo. Due to the low local barion content, adiabatic growth could be possible onto these objects so that the DM did have the possibility to form a peaked enhancement around the central black hole, constituting the so-called "mini–spike". The mini-spikes are twice interesting: from one side, due to the very large density, the signal in gammas annihilation could be very bright, brighter than the entire hosting galaxy itself. From the other side, the calculation of the flux is strongly dominated by the particle physics factor, while the astrophysical factor, namely the profile of DM, plays a less important role, as to say that the evolution of the object was largely affected only by the DM effective interaction. The dependence on the particle physics in the IMBH case is $\Phi \sim < \sigma v >^{2/7} \cdot m_\chi^{-9/7}$. According to the different evolution scenarios, a different number of these objects, ranging from 100-1000 for scenario I and II respectively, should be found in the galactic halo. In trying to figure out where IMBHs could be found, a search within the unidentified EGRET sources was performed. The EGRET experiment, onboard the CGRO satellite, discovered around 270 sources, out of which more than a hundred are still not recognized by other experiments. Some of them satisfies the IMBH model, being steady emitters, with spectral index as is expected by IMBH, with no counterpart at other wavelengths. MAGIC performed a selection between these sources to select the best candidates at large galactic latitudes ($>$ 20 deg) where no astrophysical emitters are expected so that the observation could be background–free. Observations are orgoing and foreseen for the next future.

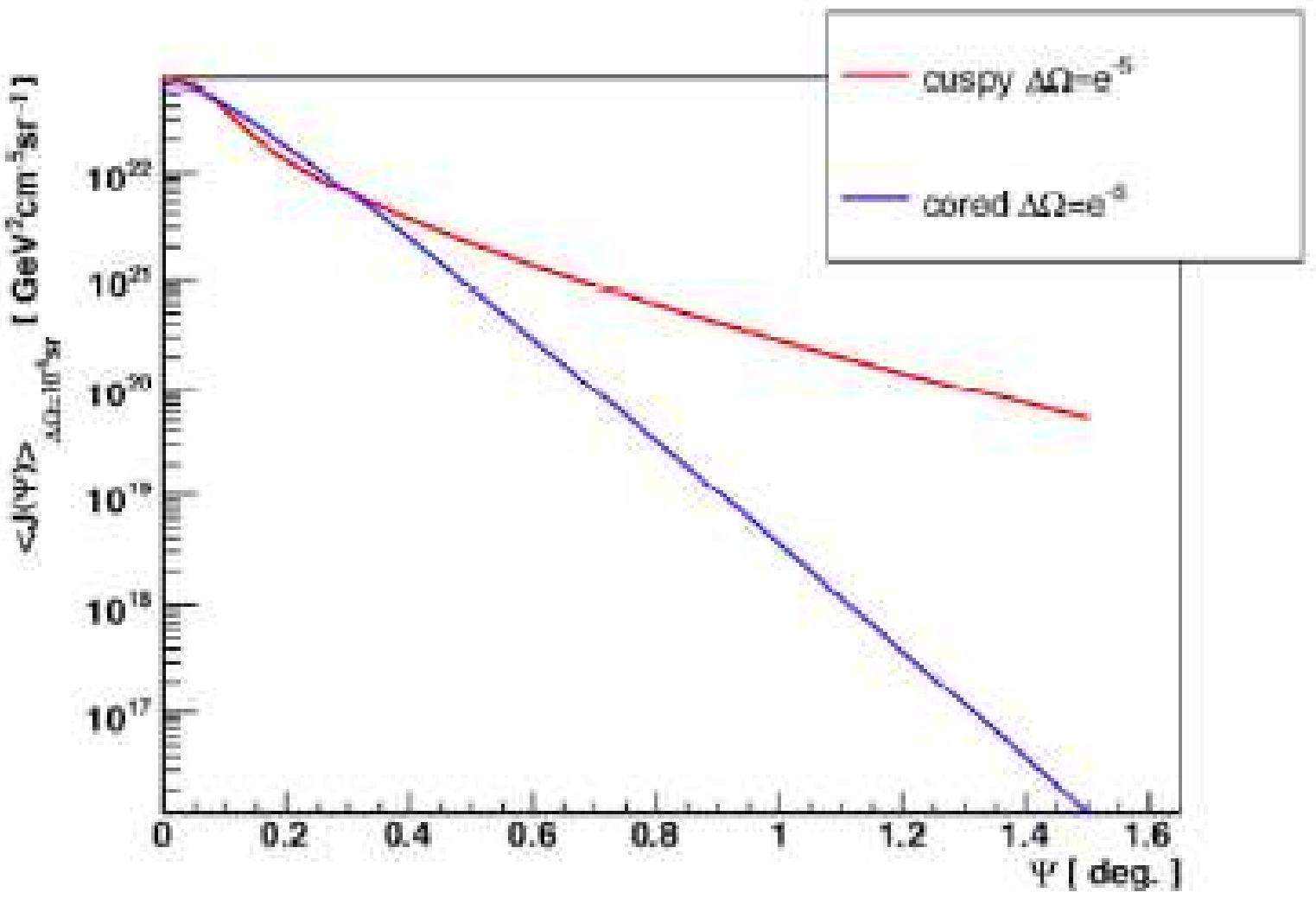

Figure 3: Convolution of cored and cuspy profiles with the MAGIC pointing accurancy for the Draco.

## 4 Outlook

The overlap between the DM candidates mass and the energy range to which IACT are currently sensitive make this technique suitable for DM observation. Nevertheless flux estimation are subject to large uncertainties of the orders of magnitude making observational plan hard to perform. The most interesting subjects of interest are the dwarf satellite galaxies of the Milky Way, where a large amount of dark matter is expected. A promising target are the mini-spikes of dark matter around intermediate mass black holes in the galactic halo. With the construction of a second clone telescope, the MAGIC overall performance will enhance in terms of pointing accurancy, background rejection and sensitivity due to the stereoscopic observation mode. The possibility to observe fainter sources with a better "eye" goes in the direction of an increased chance of observing a signal from annihilation of dark matter in the Universe.

# OBSERVATIONS OF GALACTIC SOURCES WITH MAGIC

VINCENZO VITALE [a], FOR THE MAGIC COLLABORATION

[a] *Dipartimento di Fisica, Università di Udine, Via delle Scienze, 208 - I-33100 UDINE, Italy*

**Abstract**

The MAGIC collaboration observed and studied galactic sources, both during the cycles I and II (respectively after spring 2005 and 2006). Many classes of sources were considered such as plerions, shell type supernova remnants, unidentified sources and also $\gamma$ ray binaries. Here are reported the investigations of Cassiopeia A, MAGIC J0616+225, the low energy study of Crab Nebula and the search for pulsar pulsed $\gamma$ emission.

## 1 Introduction

The MAGIC collaboration has built and operates the largest Imaging Atmospheric Cherenkov Telescope (IACT) currently in operation. The telescope and the scientific program for the study of extra-galactic sources, gamma-ray bursts and dark matter are described in details elsewere in these proceedings.

During the past years the VHE $\gamma$ ray astronomy made major advances and the detected sources went from 12 in 2003 to 71 in 2007, among which 51 are galactic or unidentified (21).

Here below some selected results of the MAGIC observation of galactic sources during cycle II are reported. Crab data are instead from cycle I. I focused on sources related to shell type or plerionic supernova remnant, candidate multi-TeV accelerators of charged particles. Results on $\gamma$ ray binaries, such as the evidence of $\gamma$ ray excess from Cygnus X-1 or the deep study of LS I +61 303, and other studies could not be reported for shortage of space.

## 2 Cassiopeia A

Cassiopeia A is a shell type supernova remnant, bright in the radio [9],[27] and X ray [7],[14] bands. The remnant consists of a patchy and irregular shell with diameter of 4', at an extimated distance of 3.4 kpc. A compact object was revealed with Chandra observations [21].The remnant results from the youngest known galactic supernova (A.D.1680). The progentor was probably a 15 to 25 $M_{sun}$ Wolf-Rayet star [15],[28], and the SN shock might expand into the progenitor wind bubble. Cass a was detected by the HEGRA Cherenkov Telescope System [1] with an exposure of 232 hours and a $\gamma$ ray flux of $(5.8 \pm 1.2_{stat} \pm 1.2_{sys})$ $10^{-13}$ph cm$^{-2}$ s$^{-1}$ above 1 TeV was reported (e.g. 3% of Crab flux). The spectral energy distribution was compatible with a power law with photon index of $(-2.5 \pm 0.4_{stat} \pm 0.1_{sys})$.

Cas A ($23.385^h$,58.800°) was observed with MAGIC and an exposure of 47 hours was available after quality data selection. The source is visible from the MAGIC site above 29° of zenith angle. Observations were performed also under moderate moonlight illumination, which did not substantially reduce the instrument performances. The data analysis was performed as reported in details in [3].

An excess of 157 $\gamma$ ray events, with a significance of $5.2\sigma$ was detected from the source. A sky map of the emission (Fig.1) was built and the c.o.g. of the emission was found at RA = $23.386 \pm 0.003_{stat} \pm 0.001_{sys}$h, and DEC = $58.81 \pm 0.03_{stat} \pm 0.02^{\circ}_{sys}$, consistent with the remnant position and the HEGRA source location. The X ray and radio diameter of Cas A (0.08°) are comparable with the MAGIC angular resolution and no source extension was found. The source differential energy spectrum (Fig.2) was obtained for $\gamma$ rays above 250 GeV up to 6 TeV, and is consistent with a power law with a photon index of $(-2.4 \pm 0.2_{stat} \pm 0.2_{sys})$. The differential flux is $(1.0 \pm 0.1_{stat} \pm 0.3_{sys})$ $10^{-12}$TeV$^{-1}$ cm$^{-2}$ s$^{-1}$ at 1 TeV. The $\chi^2$/d.o.f. of the fit was 2.83/3.

The MAGIC telescope detected the supernova remnant Cas A at level of 5.2 $\sigma$ and confirmed the previous discovery made by HEGRA. In the common energy range, above 1 TeV the results of MAGIC and HEGRA are in agreement, furthermore MAGIC measured the source energy spectrum down to 250 GeV. No spectral departures from a power law were found. The Cas A detection provides evidence of the acceleration of multi-TeV particle. Higher and lower energy measurements are needed in order to discriminate among leptonic or adronic $\gamma$ ray origin, as also a higher signal significance for such a weak source.

## 3 MAGIC J0616+225

The source MAGIC J0616+225 was discovered with the observations of the supernova remnant IC 443. This remnant is of shell type, asymmetric and has an angular diameter of 45' at an estimated distance of 1.5 kpc [16]. The source was studied in the radio band at various wave-lengths [24],[13]. Maser emission at 1720 MHz was reported from four regions of the shell [12], among which the most intense is in (l = -171.0, b = 2.9). The source was also well studied with X ray observatories [8],[23] [10], [26], [17] and a pulsar, CXOU J61705.3+222127, possibly associated to the remnant, was discovered. An EGRET source (3EG J0617+2238) was found to be coincident with the center of the shell [19]. Also for this reason IC 443 was searched for TeV $\gamma$ ray emission, but only upper limits [20], [22] were reported, before the MAGIC observation.

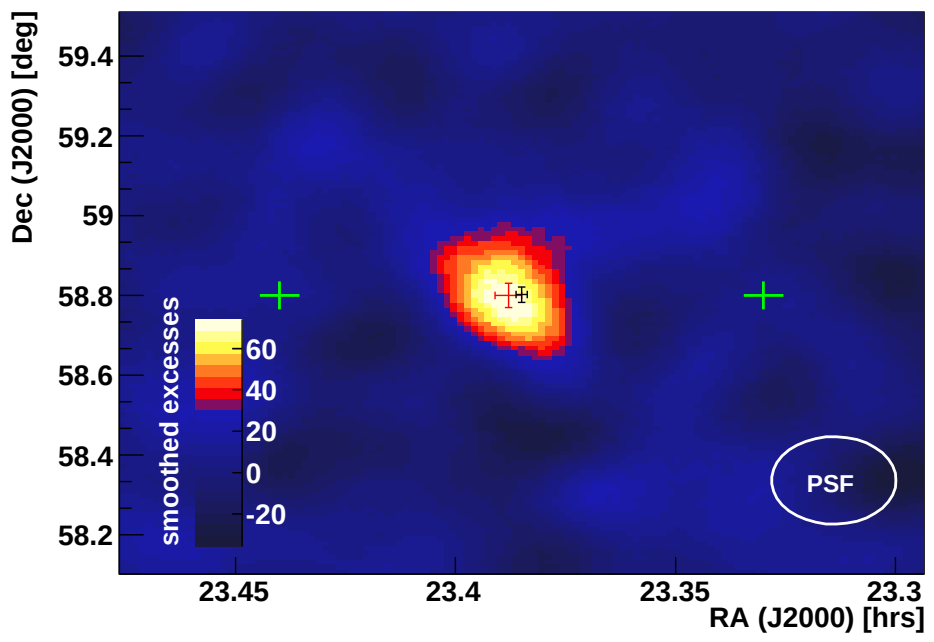

Figure 1: Sky Map of $\gamma$ ray emission from Cas A. The red cross is the MAGIC source location, the black cross the HEGRA source location, which is one $\sigma$ from the MAGIC one.

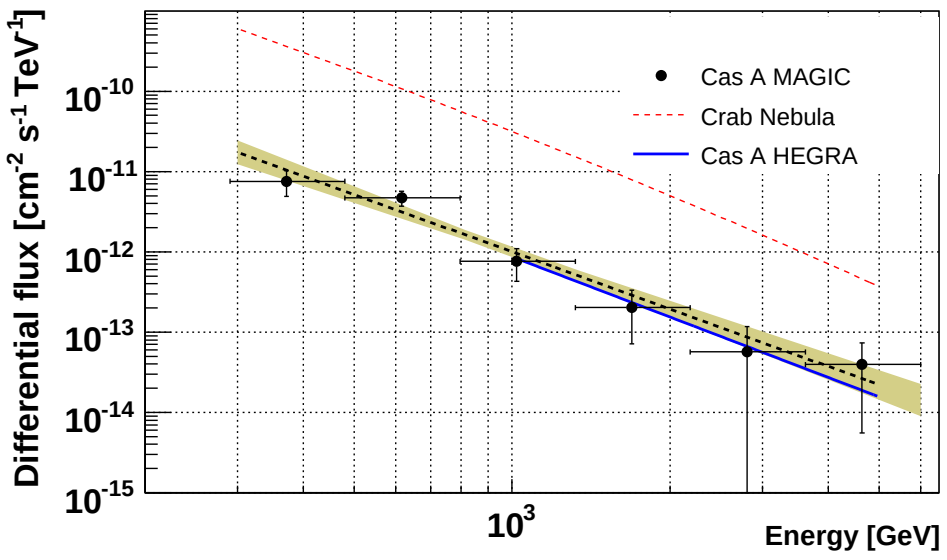

Figure 2: Differential energy spectrum of Cas A, above 250 GeV. The shaded area covers $1\sigma$ statistical error. The blue line is the HEGRA spectrum. The red line is the Crab Nebula spectrum

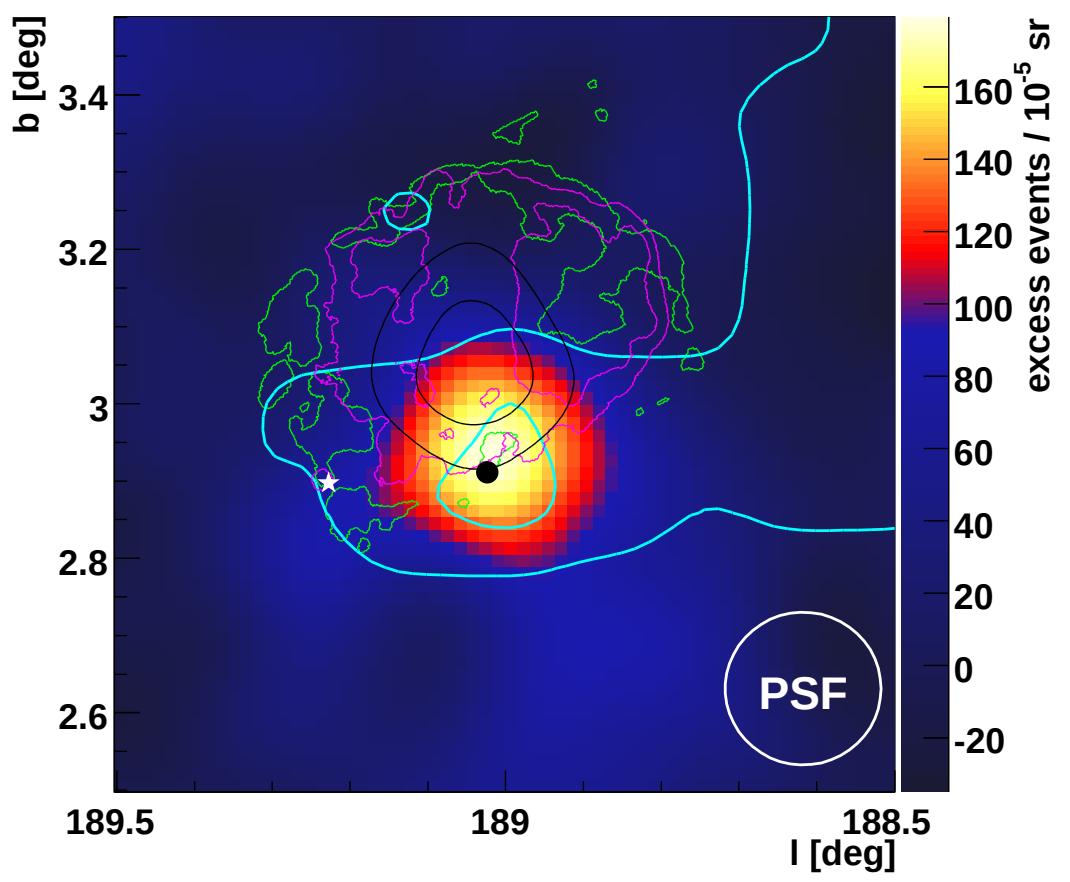

Figure 3: Sky map of $\gamma$ ray emission from MAGIC J0616+225, above 150 GeV. $^{12}$CO emission (cyan), 20cm VLA (green), Rosat X-ray (purple) and EGRET $\gamma$-ray contours are shown. The star indicates the pulsar CXOU J061705.3+222127 and the black dot the 1720 MHz maser emission position.

The data analysis was performed as described in details in [4]. A smoothed sky map of the $\gamma$ ray candidates was obtained (Fig.3). A $\gamma$ ray excess is visible in RA=$06^h 16^m 43^s$, DEC=+22° 31'48". This excess has a significance of $5.7\sigma$. The differential energy spectrum was calculated above 90GeV. It was fitted with a power law with photon index of (-3.1±0.3) and flux at 400 GeV of (1.0±0.2) $10^{-11}$ TeV$^{-1}$ cm$^{-2}$ s$^{-1}$ ($\chi^2$/d.o.f.= 1.1).

MAGIC J0616+225 is at the edge of the radio and X ray shell and of the EGRET source. A region of 1720 MHz maser emission and a large amount of molecular gas are instead co-located with the new discovered source. Therefore the $\gamma$ emission might originate from the interaction of particles accelerated in the shell with nearby dense molecular clouds. The MAGIC source is displaced from the pulsar CXOU J061705.3+222127 as is seen in Fig.3. Then PWN emission is disfavoured.

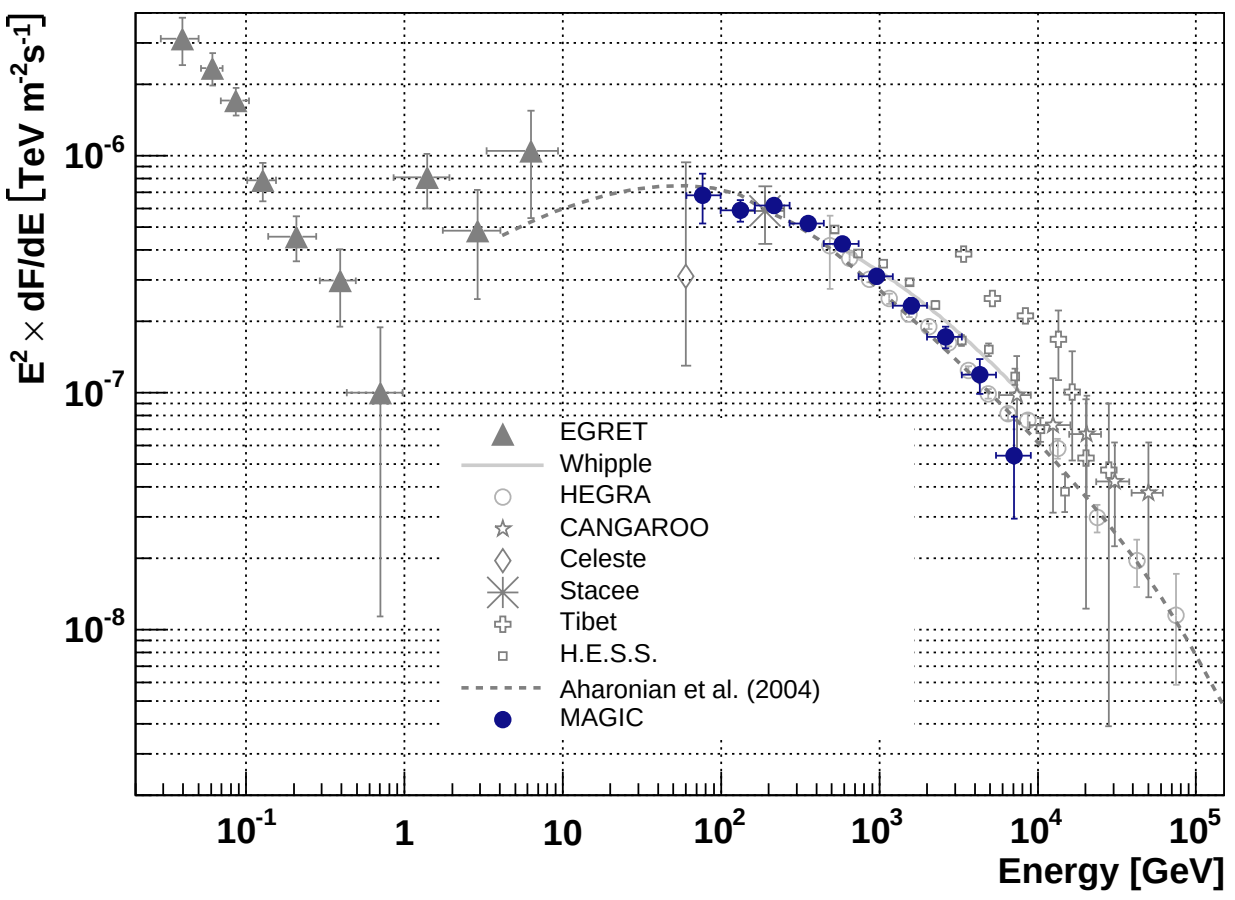

Figure 4: Spectral energy distribution of $\gamma$ ray emission of Crab Nebula. Dashed line is a theretical prediction [2]

## 4 Crab Nebula

Crab Nebula, detected in the TeV energy range by Wipple in 1989, is the first object firmly detected by IACTs. It is the standard candle for the VHE $\gamma$ ray astronomy also becouse of its large steady flux. The remanant originated by the galactic supernova of A.D. 1054. It is a plerion and is powered by the millisecond pulsar PSR B0531+21. Despite the deep study of this source, important questions on the emission mechanism are still open, in particular for the pulsar and the nebula at GeV energies and for the nebula at PeV energies. For a more detailed introduction see [5]. The search for pulsed emission above 10 GeV, is reported in the next section. These results are based on 19 hours of data taken between October and December 2005. The detailed discussion of the analysis is in [5]

The differential energy spectrum of the source was measured down to 60 GeV(Fig.4). This spectrum can be described with a curved power law $\frac{dF}{dE}$ = $A_0$ $(\frac{E}{300GeV})^{a+blog10(E/300GeV)}$, where $A_0$ = (6.0 $\pm$ $0.2_{stat}$) $10^{-10}$ $TeV^{-1}$ $cm^{-2}$ $s^{-1}$, a = -2.31$\pm$ $0.06_{stat}$ and b = -0.26$\pm$$0.07_{stat}$. The $\chi^2$/d.o.f. for the curved power law fit is 8/7, for comparison a simple power law fit has $\chi^2$/d.o.f. = 24/8 and yelds a photon index of (-2.48$\pm$$0.03_{stat}$$\pm$$0.2_{sys}$). The spectral energy distribution peak was found at 77$\pm$47 GeV. No source extension and no flux variation within the observation campain (14 nights between October and December 2005) were found. The excess location is consistent with the pulsar

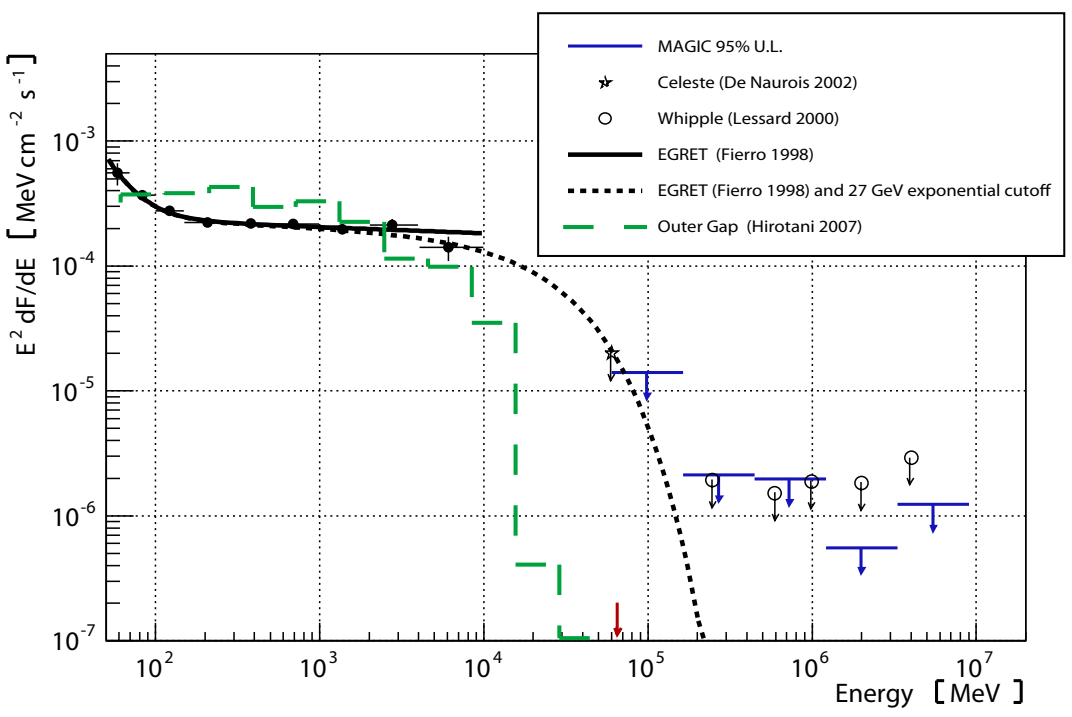

Figure 5: Upper limits on the pulsed $\gamma$ ray flux from Crab pulsar. Upper limits in differential energy bins are the blue points. The upper limit to the cutoff energy is indicated by the dashed line. The red arrow is the analysis threshold.

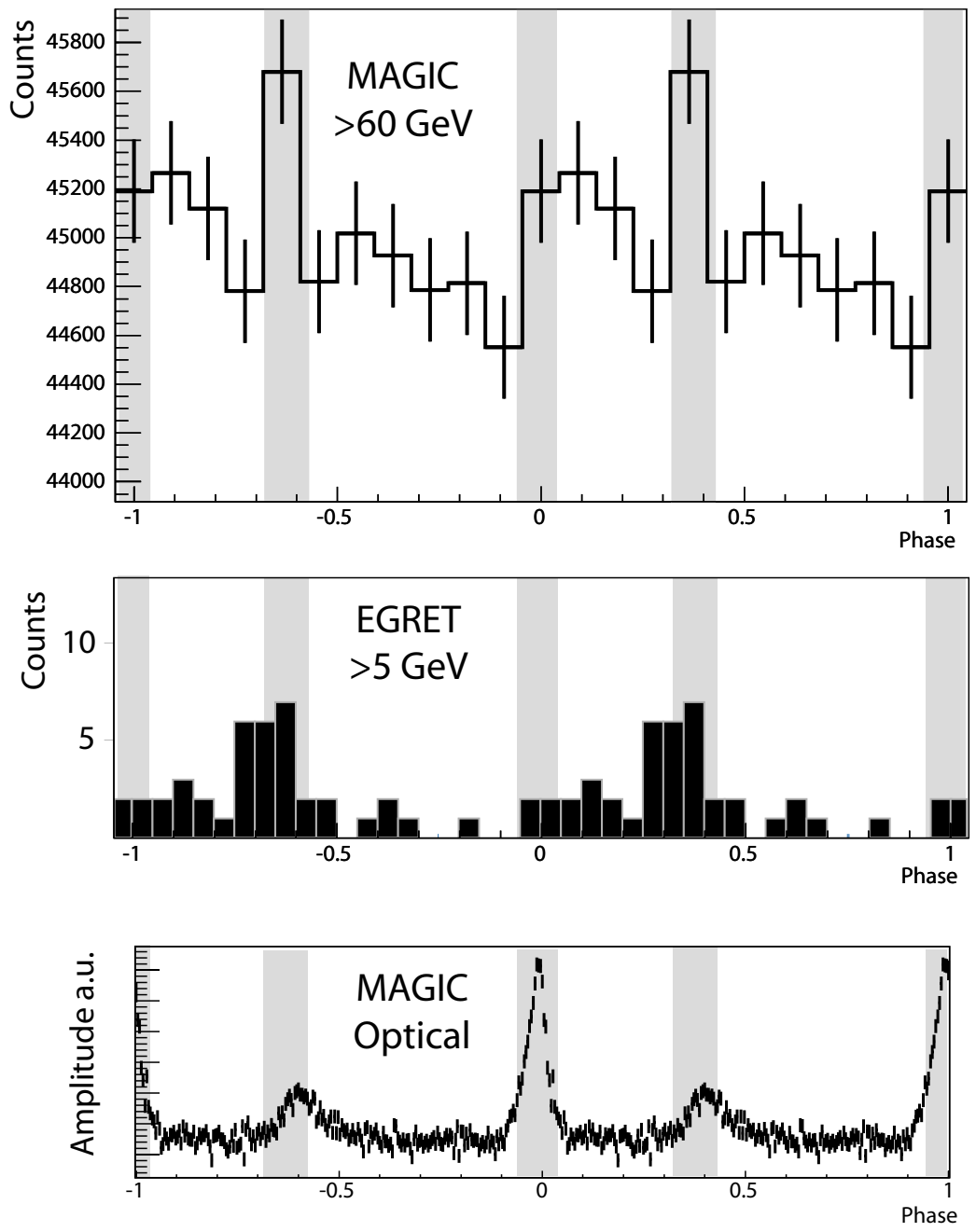

Figure 6: Pulse shape profiles of the Crab pulsar. The shaded regions indicate the EGRET measured positions of the peaks for $\gamma$ rays with energy above 100 MeV

position.

Before MAGIC the Crab observations between 10 and 200 GeV were sparse. The reported studies are the most detailed to date below 500 GeV. They support the generally accepted picture of emission which can be described in the SSC-model framework (for example [2])

## 5 Search for pulsed emission from PSR B1951+32 and PSR B0531+21(Crab)

The pulsar magnetosphere hosts particle acceleration, as also implied by the detected $\gamma$ ray emission. The main models of partcle acceleration in pulsars assume the acceleration to take place above the polar cap of the neutron star ([18]), or in the outer gap of the magnetosphere([11]). Also the region of interaction between the pulsar wind and the interstellar medium might host particle acceleration. PSR B1951+32 is one of the six pulsars which were detected with EGRET [25].

PSR B1951+32 was observed with MAGIC and was searched for steady VHE $\gamma$ emission. Only upper limits to the integral ($\phi$(E>140GeV) $< 1.5\ 10^{-11}$ ph cm$^{-2}$s$^{-1}$, with 95% confidence level) and to differential fluxes (see [6]) were obtained. Also the radio nebula CTB 80, thought to be associated to the pulsar, was searched for $\gamma$ emission and was not detected. The pulsed emission was searched as reported in details in [6]. Upper limits to the pulsed emission were obtained in various energy bands, ranging from $1.5\ 10^{-11}$ to $2.5\ 10^{-13}$ ph cm$^{-2}$s$^{-1}$. It was possible to set an upper limit also on the cutoff energy of the pulsed emission, which is $E_{cutoff} <$ 32GeV.

In the same way Crab pulsar (PSR B0531+21) was searched for pulsed emission, with MAGIC. Pulsed emission was searched in five energy bands (5) between 60 GeV and 9 TeV and no significant signal was found. The upper limit to the pulsed emission cutoff energy was calculated to be $E_{cutoff} <$ 27GeV. In Fig.6 is shown the pulse shape profile of the Crab pulsar. MAGIC data are for energies below 180 GeV. Also the optical pulse shape profile was measured with MAGIC. If the EGRET phase regions are defined as signal region the significance of the structures in the MAGIC phaseogram have significance of $2.9\sigma$. Then no pulsed emission was found in PSR B1951+32 or Crab pulsar, although a hint of signal is in the MAGIC phaseogram of Crab.

# OBSERVATIONS OF VERY HIGH ENERGY GAMMA-RAY GALACTIC SOURCES WITH H.E.S.S.

GIOVANNI LAMANNA[a]
FOR THE H.E.S.S. COLLABORATION

[a] *LAPP - Laboratoire d'Annecy-le-Vieux de Physique des Particules IN2P3/CNRS, 9 Chemin de Bellevue, 74941 Annecy-le-Vieux, France*

**Abstract**

The H.E.S.S array of imaging Cerenkov telescopes has discovered a number of previously unknown gamma-ray sources at very high energy (VHE) and has provided exciting results from the Galactic plane survey. In this communication a selected sample of highlights are presented.

## 1 The H.E.S.S. telescope system

The H.E.S.S. array, a system of four large (13 m diameter) imaging atmospheric Cerenkov telescopes, is operated since December 2003 by an international collaboration of about 100 physicists. Located in the Khomas highland of Namibia, the H.E.S.S. system covers a $5^o$ field of view, with a sensitivity which allows to detect sources with a flux of 1% of the Crab Nebula in 25 h of observation and an energy threshold between 100 and 700 GeV increasing with the observation zenith angle. The four telescopes provide multiple images of gamma-ray induced air showers in the Cerenkov light emitted by the shower particles, enabling the stereoscopic reconstruction of the shower geometry and the shower energy. The estimated energy resolution is 15% and $0.1^o$ is the angular resolution for individual gamma-ray corresponding to $1'$ location position of a VHE gamma-ray source. A personal selection of the highlights from H.E.S.S. is imposed by the lack of space: most recent published results from the galactic plane survey and dedicated source observations together with a summary on the studies of shell-type supernova remnants and pulsar wind nebulae will be the main topics of this letter.

## 2 The H.E.S.S. galactic plane survey

The Galactic plane survey was conducted in the summer of 2004 covering the region of -30$^o$ to 30$^o$ galactic longitude and -2.5$^o$ to 2.5$^o$ in galactic latitude, resulting in 15 new VHE gamma ray sources plus three previously known. Searching for counterparts in radio- and X-ray catalogs they resulted to be related to SNR, a significant fraction to PWNe and at least three "Dark accelerators" without counterpart known. A new observation campaign was conducted during the years 2005-2007 with the scanned region now reaching from -80$^o$ to 60$^o$ galactic longitude. The number of new sources is more than 15 with 6 new "Dark accelerators", others sources incrementing the known classes and some new results. Details on published results on the two campaigns can be found in [1] and [2] respectively.

### 2.1 HESS J1023-575

The discovery [3] of the source HESS J1023-575 is one of the most relevant highlight of the 2006 data taking: a clue to the investigation on the cosmic-rays origin. The detection of VHE gamma-ray emission associated with the young stellar cluster Westerlund 2 in the HII complex RCW 49 provides evidence that particle acceleration to extreme energies is associated with this region, a luminous massive star formation region already well studied at various wavelengths. The source (Fig. 1) has been observed for a total 14 h of data for a corresponding statistical significance of more than 9 $\sigma$ and clearly extended beyond the nominal PSF. The differential energy spectrum, extended about two order of magnitude in energy and with a minimum threshold of 380 GeV, can be described by a power law with index 2.53 $\pm$ 0.16 and an integral flux of 1.3 $\pm$ 0.3 $\times$ $10^{-11}$ cm$^{-2}$ s$^{-1}$. A variety of potential emission scenarios are suggested [4] for the interpretation of HESS J1023-575, a new type of astronomical object, profoundly distinguished from other source findings made during Galactic Plane Scan observations. Further investigation with H.E.S.S. will allow to discriminate among alternative interpretations.

## 3 Study of the shell-type supernova remnants (SNRs)

Two supernova shells already detected as gamma sources by CANGAROO, RX J1713.7-3946 [5] and RX J0852.0-4622 ("Vela Junior") [6](Fig. 2), are now firmly established VHE gamma-ray emitters and morphologically resolved by H.E.S.S.. The energy spectra follow a power law with index of about 2.3, constant across the entire remnants. For both sources, the gamma-ray shell intensity observed with H.E.S.S. is highly correlated with the X-rays one. This correlation would be natural if a common population of primary electrons were responsible for both emission regimes. Assuming that X-rays represents synchrotron radiation and that the gamma rays are generated in Inverse Compton scattering, as it is shown in Fig. 3, simple electronic models assuming an electron injection index of 2.5 and with a local magnetic field of B $\sim$ 10 $\mu$G, which accommodates both levels of spectra, fail to consistently fit the multi-wavelength data (e.g.: over shooting the radio flux [8]). In contrast models assuming higher magnetic field and adding gamma-rays from proton-interactions, achieve a good description of wide-band spectra [7]. This interpretation would support the hadronic origin of gamma rays even if a conclusive evidence is still lacking. The currently modest number of shell SNR resolved in VHE has been increased by the recent observation of RCW 86, a supernova remnant with a barrel-shaped shell, visible in X-rays, radio and optical waves. Hints for gamma-ray emission were seen with CANGAROO-II instrument, but no firm detection was claimed. A clear gamma-ray signal with more than 9 $\sigma$ has been detected by H.E.S.S.. A detailed analysis is in progress and preliminary results [9] have shown: a flux 5-10% of the Crab nebula, a 2.3-2.5 spectral index and a shell type morphology.

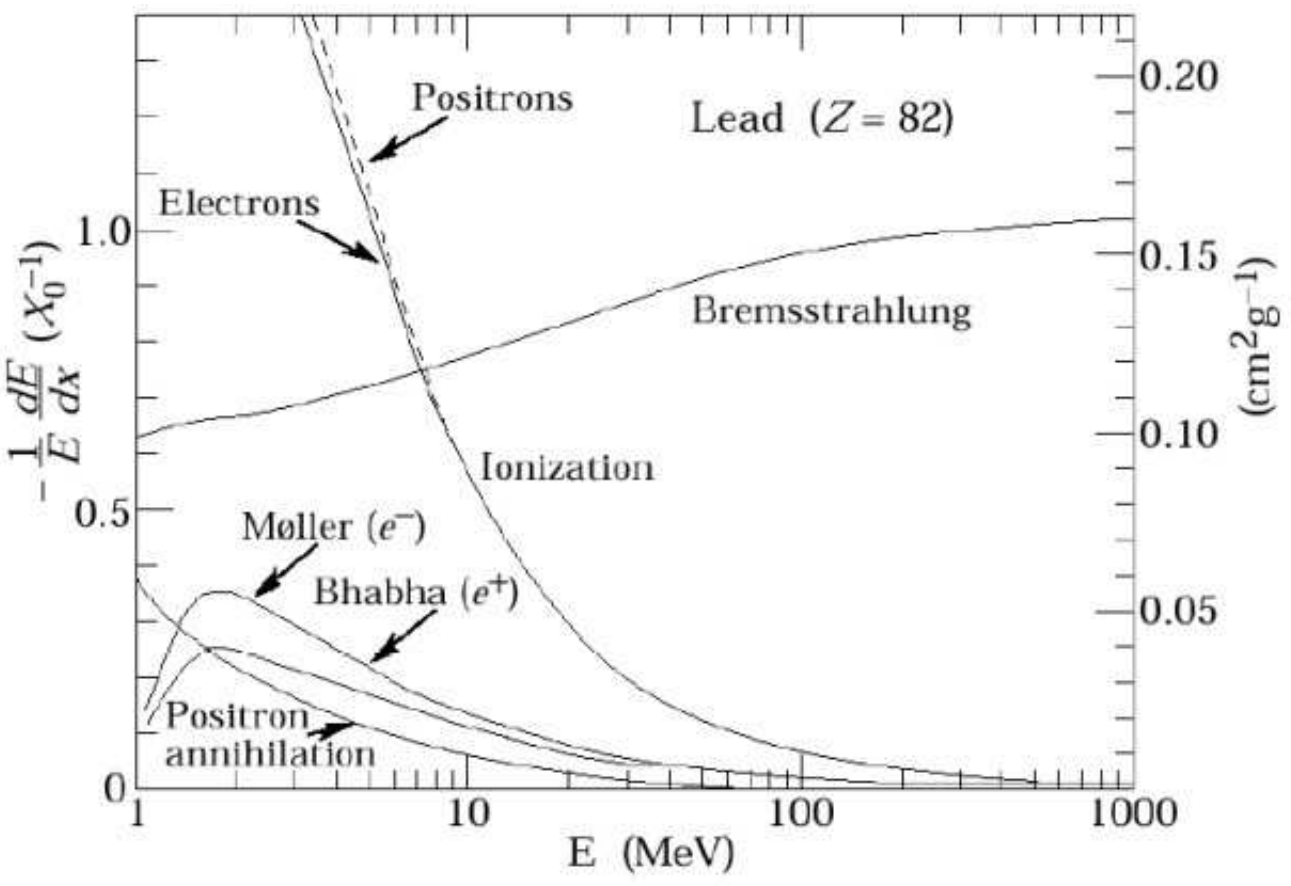

Figure 1: Left: H.E.S.S. gamma-ray sky map of the Westerlund 2 region. The WR stars WR20a and WR20b are marked as filled triangles, while the dashed circle is the extension of the luminous stellar cluster Westerlund2. Right: HESS J1023 5, 7, and 9 $\sigma$ significance contours overlaid on a radio image.

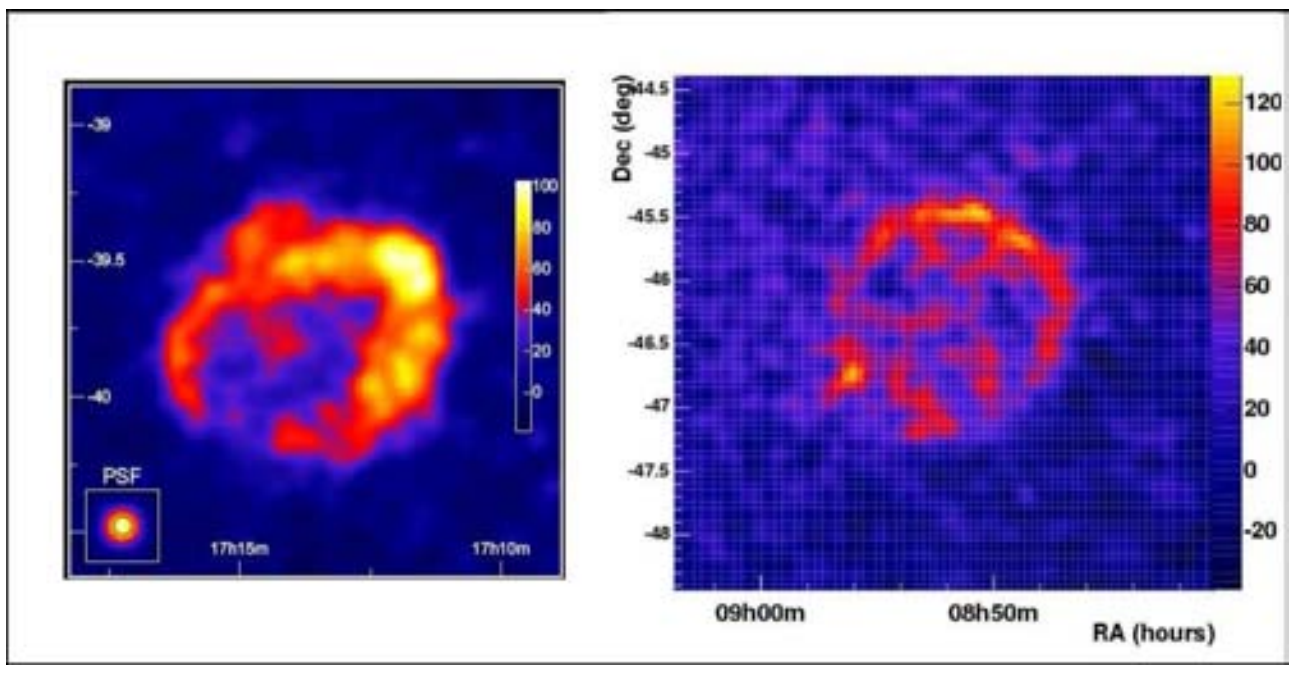

Figure 2: Shell supernova remnants seen in gamma rays by H.E.S.S.: RX J1713.7-3946 (*left*) and RX J0852.0-4622 (*right*).

### 3.1 HESS J1800-240 and HESS J1801-233

An other composite or mixed-morphology SNR, which is an ideal target for VHE observations, is W 28 (G6.4-0.1). The old-age W 28 SNR is thought to have entered its radiative phase of evolution. The shell-like radio emission peaks at the northern and northeastern boundaries where interaction with molecular cloud is established. The X-ray emission, which overall is well-explained by a thermal model, peaks in the SNR center but has local enhancements in the northeastern SNR/molecular cloud interaction region. On the south boundary several HII regions, including ultra-compact HII region W 28A2 are found. H.E.S.S. observations of W 28 have revealed VHE gamma-ray emission situated at its northeastern (HESS J1801-233) and southern boundaries (HESS J1801-240 with components A, B and C) (Fig. 4) [10]. A multi-wavelength analysis of W 28 has revealed a dense molecular cloud enveloping the southern region, and EGRET MeV/GeV emission centered on HESS J1801-233 and the northeastern interaction region. Overall, these results suggest that old-age SNRs are capable of multi-TeV particle accelerators and candidate hadrons diffusive shock accelerators.

## 4 Study of the Pulsar Wind Nebulae

Pulsar Wind Nebulae (PWN) are responsible for a significant fraction of the new VHE Galactic sources observed by H.E.S.S.. The purpose of the PWN study is a diagnostic of the spatial and spectral distribution of the high energy electrons responsible of the TeV gamma-ray production dominated by the Inverse Compton scattering off the well-known cosmic microwave background. HESS J1825-137 is a particularly interesting PWN candidate: it is a strong source extended over a fraction of degree [11]. It was detected during the first Galactic plane survey and then further observed. It is located south of the pulsar PSR B1823-13 which exhibits an X-ray nebula trailing extended over ∼5' in the direction of the VHE source but then much smaller in size. A natural explanation is that the X-ray generating electrons (via interaction with the nebula magnetic field ∼10 $\mu$G) have higher energies than those responsible via Inverse Compton scattering for the VHE gamma rays. The higher energy X-ray electrons cool faster and have a shorted range. More importantly, for the first time observations have revealed the energy dependent morphology of the source. This manifests itself as a steepening of the power-law spectral index with increasing distance from the pulsar, as would be expected from the radiative losses of high-energy electrons injected by the pulsar (see Fig. 5).

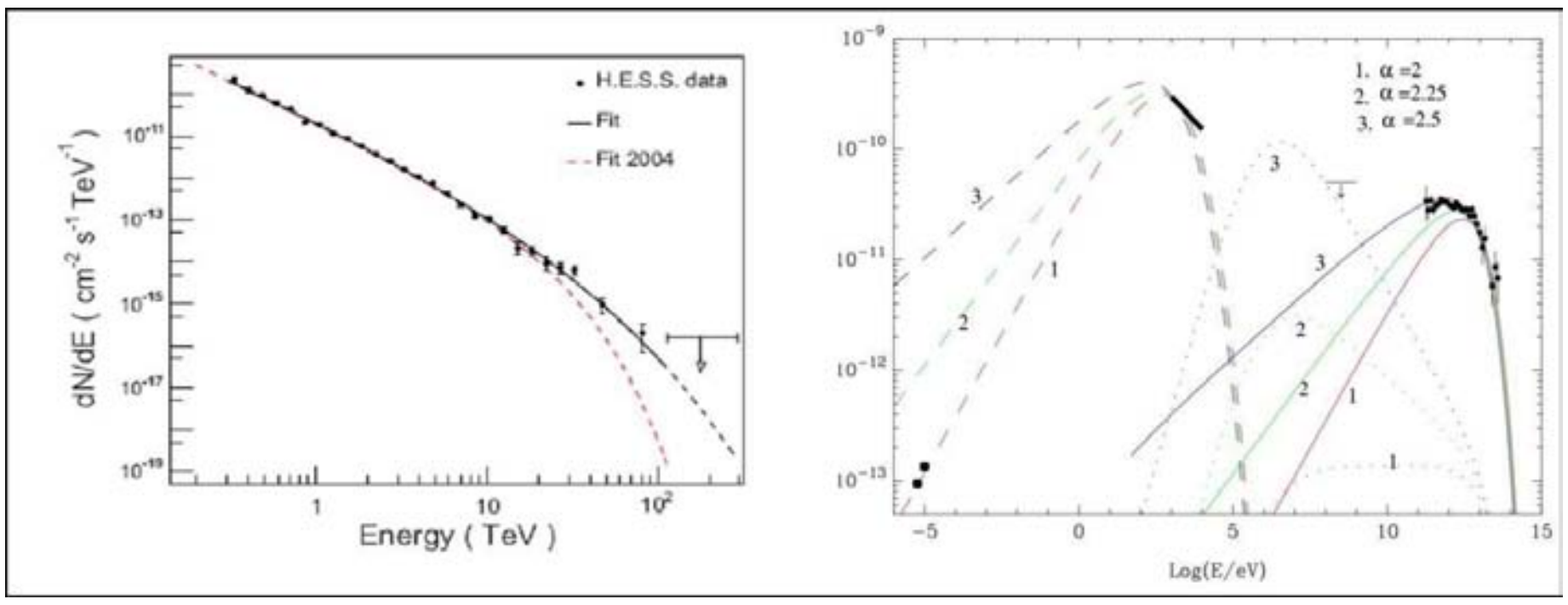

Figure 3: RX J1713.7-3946: energy spectrum (*left*); wide band spectra (together with X-rays, radio data and H.E.S.S. gamma rays measurements) for a magnetic field $\sim 10\ \mu$G and an electron injection index of 2.0, 2.25 and 2.5 (*right*).

Among the number of PWNe detected by H.E.S.S. and apart from the almost point-like Crab nebula, extensively studied through different consecutive observation campaigns [12], Vela X associated with the Vela pulsar (Fig. 6) is likely the most extended one (about a degree south of the pulsar) and significantly old (age ∼11 kyr). The energy spectrum is very hard reaching 50 TeV. The radio, X-rays and VHE gamma-rays emission regions of Vela X are markedly offset from the pulsar position. This may be due to the supernova explosion occurring in an inhomogeneous medium, and the resulting asymmetric reverse shock displacing the PWN in the direction away from the higher density medium. The displacement of the nebulae from the pulsar positions is a surprising constant of almost of extended PWN candidate sources. It is also the case of MSH 15-52, associated with the pulsar PSR B1509-58 inside the G 320.4-1.0 / RCW 89 shell. This elongated and single-sided nebula was revealed by H.E.S.S. as aligned in the same direction of the jet-like high-resolution X-ray Chandra image. More recently, studies of the “Kookaburra” region revealed two new TeV gamma-ray sources, one most likely associated with the pulsar PSR J1420-6048, the other one with the “Rabbit” feature presumably resulting from another pulsar. Both sources have relatively hard spectra index around 2.2, both are extended on the scale of about 10 pc and both are displaced by a similar amount from their pulsars.
Re-observations of the mentioned extended PWNe have been recently accomplished and analysis are in progress to further infer about the spectral and spatial distribution of the energetic electrons within the leptonic interpretation of the VHE emission.

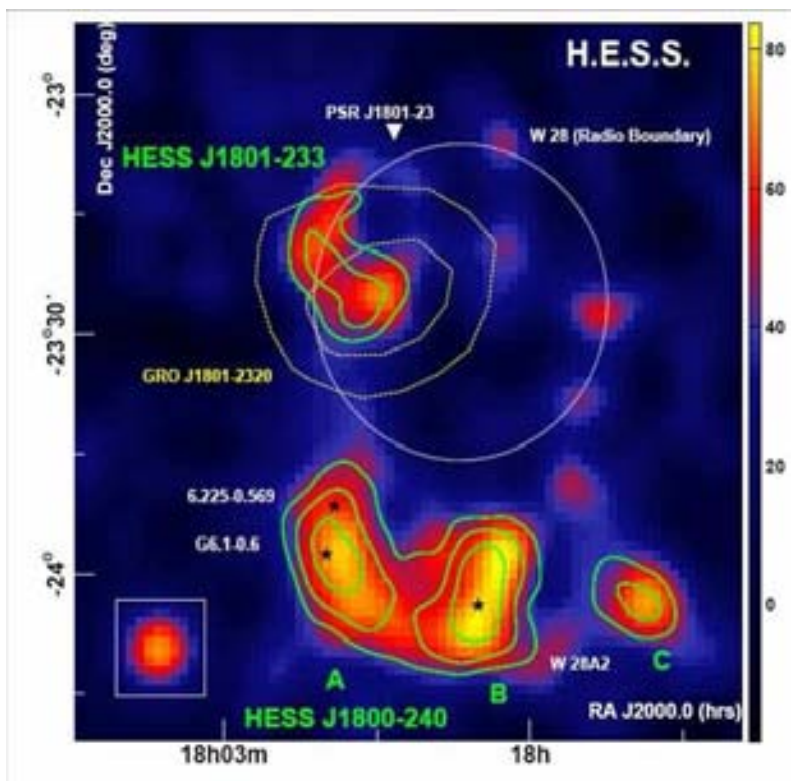

Figure 4: W 28 VHE gamma-rays excess map with 4, 5 and 6$\sigma$ contour levels for HESS J1801-233 and HESS J1801-240 A, B and C sources.

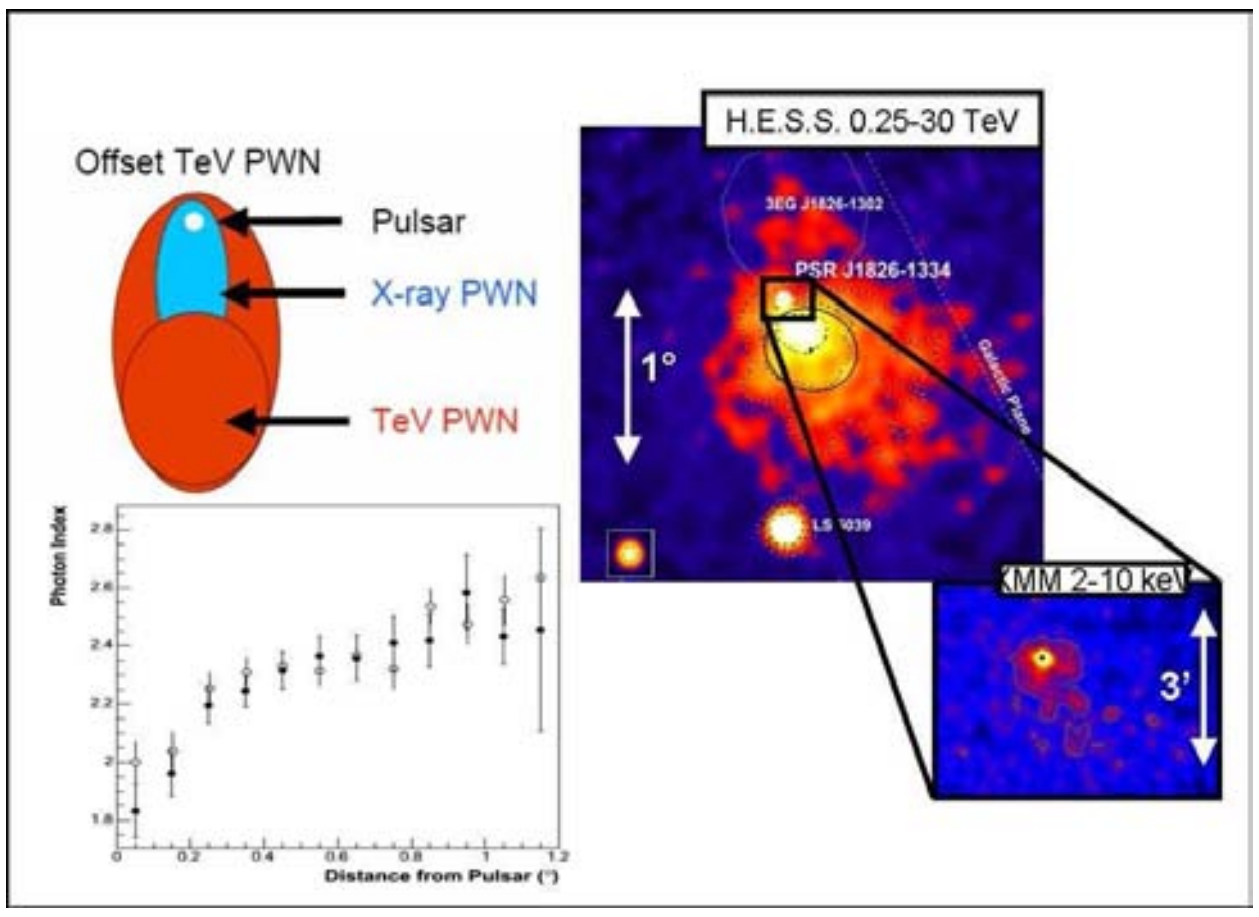

Figure 5: HESS J1825-137, a PWN candidate manifesting an offset from the associated pulsar position, a steepening of the power-law spectral index with increasing distance from the pulsar and a larger extension than in X-ray.

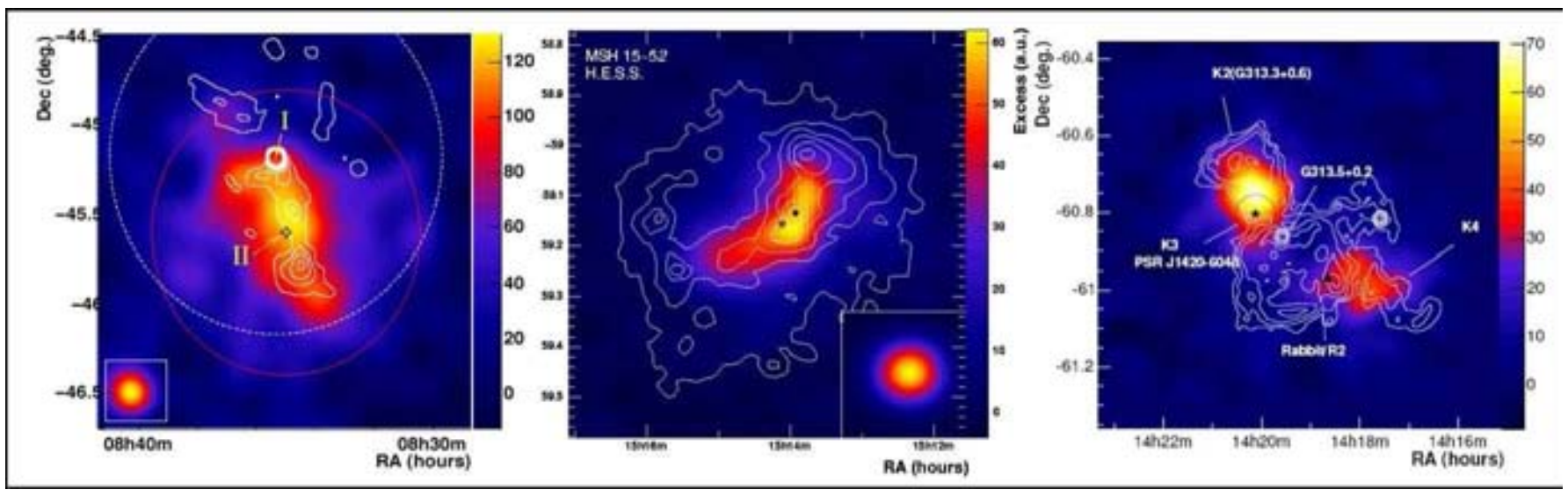

Figure 6: Smoothed gamma-ray excess map (from left to right) from Vela X, MSH 15-52 and the two sources in the Kookaburra region. White contours are X-ray corresponding to count rates contour lines.

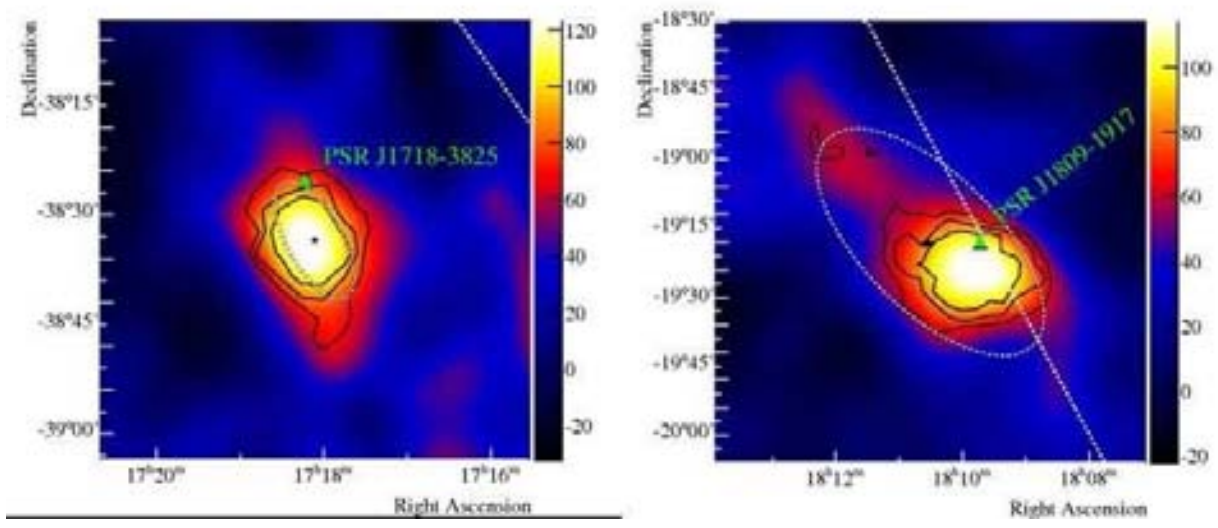

Figure 7: Two new PWNe candidates, HESS J1718-385 and HESS J1809-193, observed during the Galactic sky survey.

A systematic search for gamma-ray counterparts of known pulsars is addressed by the possibility that all pulsars have associated VHE gamma rays nebulae. Such a research has produced a bunch of new PWN candidates: e.g.: HESS J1718-385 and HESS J1809-193 (Fig. 7) [13]. They show that among pulsars with a spin-down energy flux above $10^{35}$ergs/s/kpc$^2$, a large fraction is visible as gamma-ray emitters, converting about 1% of their spin-down energy into 1-10 TeV gamma rays. This implies that about 10% of pulsar spin-down energy is fed into high-energy electrons. More observations of these sources and corresponding multi-wavelength investigations would provide important progress in the physics of PWNe.

# H.E.S.S. EXTRAGALACTIC $\gamma$-RAY SOURCES

BERRIE GIEBELS [a], FOR THE H.E.S.S. COLLABORATION[1]

[a] *Laboratoire Leprince-Ringuet, Ecole polytechnique, IN2P3/CNRS F-91140 Palaiseau, France*

**Abstract**

The observational field of extragalactic $\gamma$-ray astronomy has dramatically evolved in the past years, with the new generation of Atmospheric Čerenkov Telescopes (ACTs) such as H.E.S.S. and MAGIC coming online, and probing the radiative properties of Active Galactic Nuclei (AGN) with improved levels of sensitivity and spectral resolution. Light curves now show evidence for minute time-scale variability in the very high energy (VHE, $E > 100\,\mathrm{GeV}$) $\gamma$-ray regime, and quality spectra of objects up to $z \simeq 0.2$ are measured, allowing unprecedented constraints to the intrinsic behaviour of these objects, or to the Extragalactic Background Light (EBL) they propagate through.

## 1 The extragalactic VHE skyscape

The contents of Table 1, where all the currently known VHE-emitting blazars are listed, has to be compared with its counterpart written in April 2007 by [1], which had 6 entries. The new ones are all discoveries from H.E.S.S. and MAGIC, and the experiment to be credited can be identified by the first author's names of the associated journal paper(Aharonian and Albert, respectively). The threefold increase in extragalactic $\gamma$-ray emitters, all but one belonging to the BL Lac class, shows how ten-fold improvement in sensitivity translates into increased detections[2].

[1] http://www.mpi-hd.mpg.de/hfm/HESS/HESS.html

| Source | $z$ | Discovery & Confirmation |
|---|---|---|
| Mrk 421 | 0.031 | Punch et al. 1992, Petry et al. 1996 |
| Mrk 501 | 0.034 | Quinn et al.1996, Bradbury et al. 1997 |
| 1ES 2344+514 | 0.044 | Catanese et al. 1998, Tluczykont et al.2003 |
| Mrk 180 | 0.046 | Albert et al. 2006 |
| 1ES 1959+650 | 0.047 | Nishiyama et al. 1999, Holder et al. 2003, Aharonian et al. 2003 |
| BL Lac | 0.069 | Albert et al. 2007 arXiv:astro-ph/0703084 |
| PKS 0548-322 | 0.07 | Aharonian et al. 2007, In prep. |
| PKS 2005-489 | 0.07 | Aharonian et al. 2005 |
| PKS 2155-304 | 0.117 | Chadwick et al. 1999, Hinton et al. 2003 |
| H 1426+428 | 0.129 | Horan et al. 2002, Aharonian et al. 2002, Djannati et al. 2002 |
| ES 0229+200 | 0.140 | Aharonian et al. 2007, In prep |
| H 2356 | 0.167 | Aharonian et al. 2005 |
| 1ES 1218+30.4 | 0.182 | Albert et al. 2006 |
| 1ES 1101-232 | 0.186 | Aharonian et al. 2006 |
| ES 0347-121 | 0.188 | Aharonian et al. 2007, In prep. |
| 1ES 1101+496 | 0.212 | Albert et al. 2007 arXiv:0706.4435 |
| PG 1553+113 | 0.36 (?) | Aharonian et al. 2006, Albert et al. 2006 |

Table 1: The 17 blazars detected by ACTs as of the date this proceeding was written, ranked by increasing redshift. The references in the table are not given in this paper's reference list for convenience. Note that M87, a detected non-blazar $\gamma$-ray source, is not listed here.

These objects appearing mostly as point-like objects, the spectacular improvement in spatial $\gamma$-ray imaging that H.E.S.S. has achieved is not directly visible, otherwise than through the improved background rejection and hence sensitivity. Note also that the 4 closest objects in Table 1 are all northern hemisphere objects, making them easier targets not only because of their smaller luminosity distance but also because of the smaller attenuation due to their propagation through the EBL.

The propagation effects and the uncertainties thereupon complicate the estimation of what was intrinsically emitted. This is an annoyance for the understanding of the sources. The uncertainty in our understanding of these sources, notably the fact that they can probably not be standard candle-ized, is on the other hand an annoyance for the estimations of the EBL imprint on their spectra.

[2]The Galactic sources, far more numerous, are described in this context by G. Lamanna and V. Vitale, see these proceedings

## 2 Low state emissions

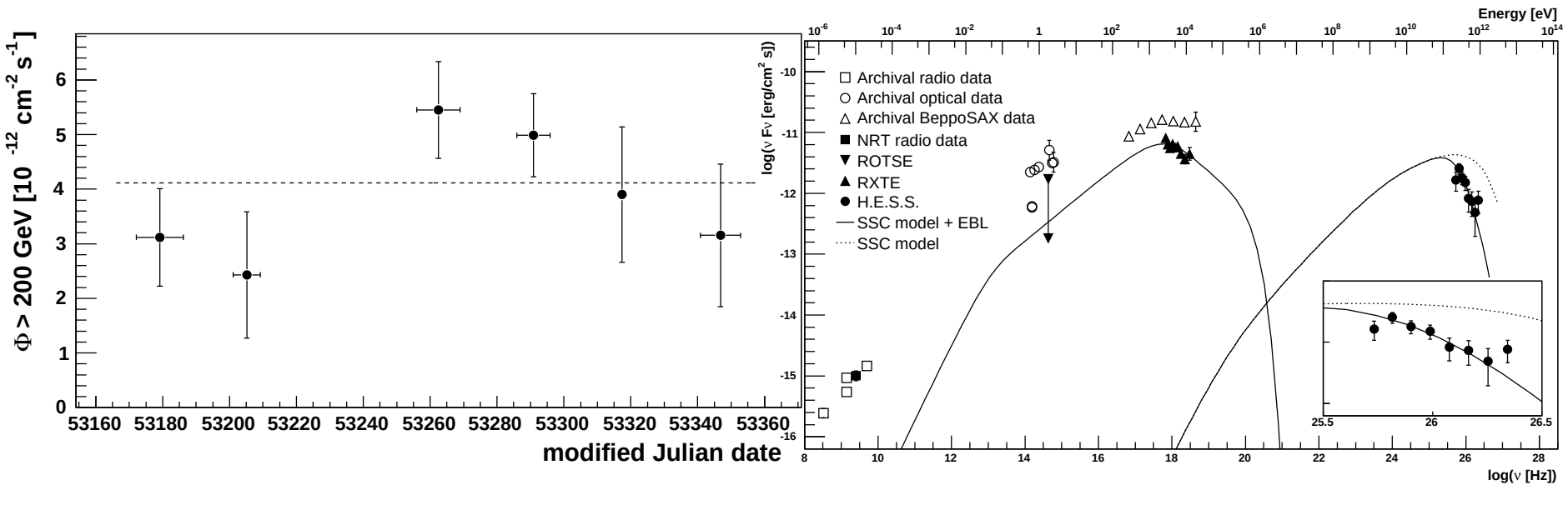

Figure 1: Left: light curve of H 2356-309 ($z = 0.165$) above 200 GeV in 6 different time windows from June to December 2004. No indication of variability was observed on nightly timescales either. Right: SED of the same object, with contemporaneous data shown as filled symbols. The line is a fit with a single-zone homogenous SSC model. The lack of variability occurs at a time when the source appears to be less intense than archival observations.

The mere detection of a blazar above a given significance level, usually 5 standard deviations (or $\sigma$), used to indicate an eruptive episode, because most of the time the source went undetected given the achieved sensitivity. It is now rather striking that detections of blazars often occur when simultaneous observations of their synchrotron component, where for obvious reasons archival data exist to compare with, indicate that the flux levels are close to the lower archival levels. Also for most of the recently discovered blazars with ACTs show relatively flat lightcurves.

The multi-wavelength campaign organized to observe PKS 2155-304 in 2003, thought to be in a high emission state after consecutive nightly detections by H.E.S.S. at the 20% Crab level, showed that the simultaneously measured optical and X-ray levels below those observed in its high state [2] by an order of magnitude. Interestingly, the small level of variability in the different wavelenghts ($\Delta F/F \approx 3$ in both X-rays and $\gamma$-rays) was not correlated during these observations, while they were actually seen to be correlated at higher fluxes in the 2004 campaign [3]. Similar $\gamma$-ray detections in low-state synchrotron states were found in the BL Lac object H 2356-309 (Fig.1) with H.E.S.S. [4], in 1ES 1959+650 [5] as well as 1ES 1218+30.4 .

## 3 Variability

The lightcurve of highly eruptive events are often searched for a variability timescale. The high $\gamma$-ray statistics allow shorter bins, and Figure 2 shows such an event that happened in July 2006 when the blazar PKS 2155-304 reached unprecedented luminosity levels [6], with variability that is among the fastest ever seen for such objects in all wavelenghts.

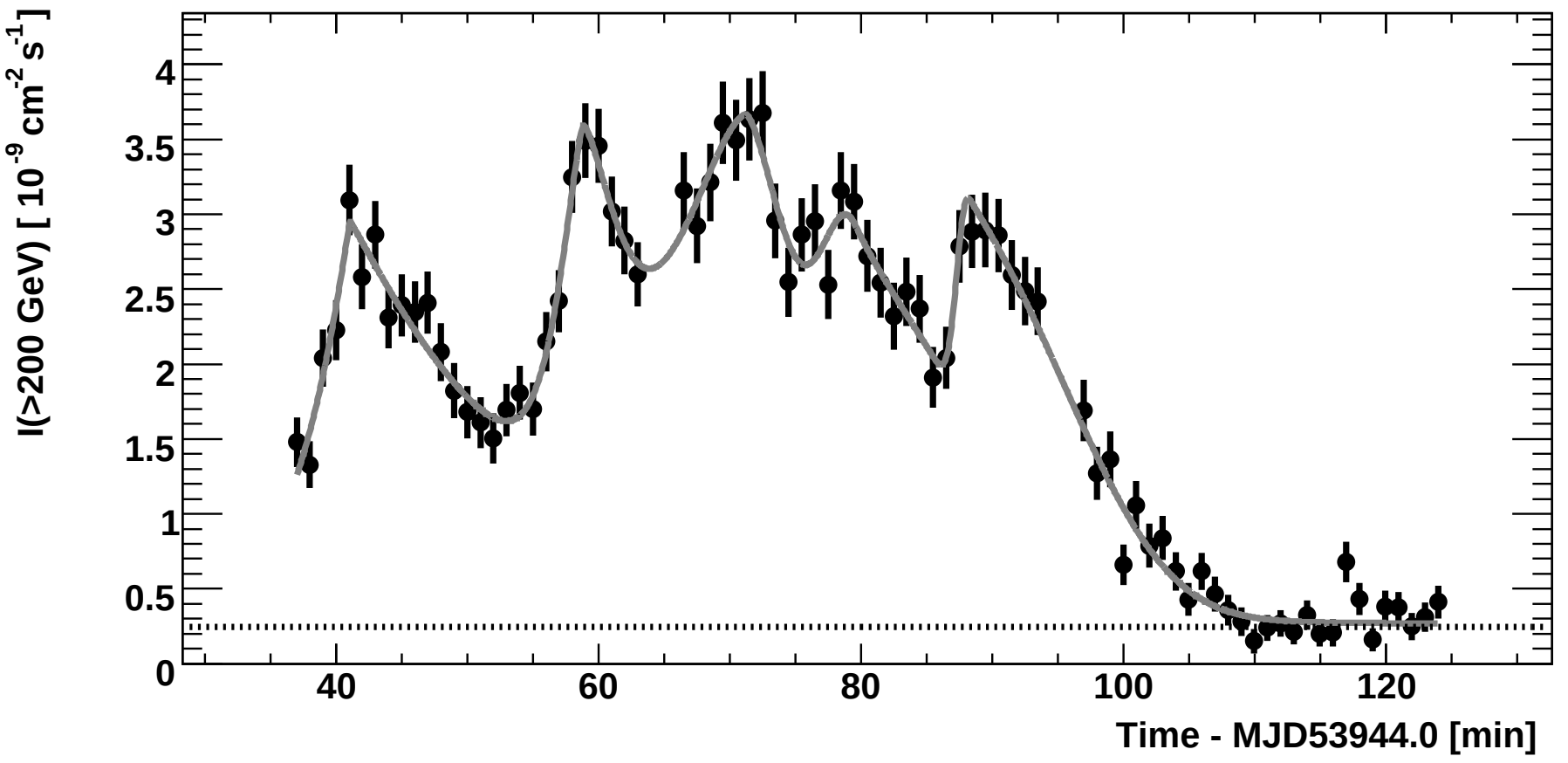

Figure 2: Light curve of PKS 2155-304 in a high state, on 24 July 2006. The sampling is 1', and the dotted line shows the 1 Crab level.

The fastest resolved transient in the light curve of Fig.2 has a rising timescale of $t_{\text{var}} = 173 \pm 28\,\text{s}$, which is a factor of 60 to 120 times smaller than the light crossing time of the Schwarzschild radius $R_{\text{S}}$ of the $1{-}2{\times}10^9 M_{\odot}$ black hole in the nucleus [7]. Doppler boosting of exactly this amount would then allow an emission region radius $R$ of the order of $R_{\text{S}}$, the smallest characteristic size in the system, but at the price of a rather large Lorentz bulk factor this presumes, greater than those typically derived for VHE $\gamma$-ray blazars ($\delta{\sim}10$) and come close to those used for GRBs, which would be a challenge to understand. Lower values of $\delta$ were derived with similar timescales for Mkn 421 [8], but the remarkable difference for that observation was the rather large spectral variability seen during an outburst,contrary to the outburst of PKS 2155-304.

The power density spectrum (PDS) of this light curve is similar to those derived in X-rays for this object, a featureless power-law Fourier spectrum of index -2 reaching the white noise level at $\sim 1.6 \times 10^{-3}$ Hz (600 s) [7], but with an order of magnitude more power than the archival X-ray PDS at similar frequencies.

## 4 Propagation effects

During their propagation from the source to the observer, it is well known that $\gamma$-rays can interact with ambient intergalactic photons of energy $E_p$ if $E_\gamma E_p \geq 2(m_e c^2)^2$ and create a pair of electrons. An observed spectrum $F(E)$ is then different from the intrinsic $F_i$ spectrum with $F_i(E) = F(E)/\exp -\tau(E)$, $\tau(E)$ being the optical depth. For 100 GeV − 10 TeV photons, the photon field in the 0.1 − 10 $\mu$m is probed and its intensity sets the level of attenuation that affects $F_i$.

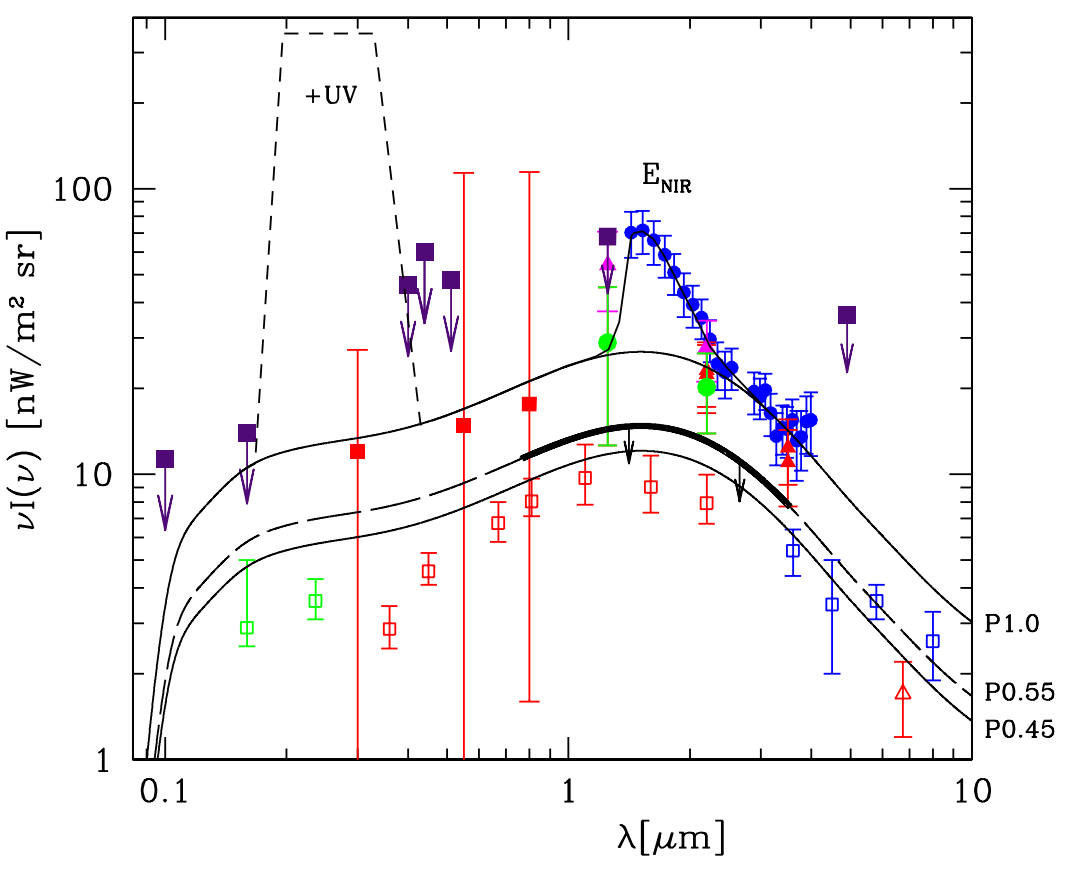

Figure 3: Spectra energy distribution of the Extragalactic Background Light, in the 0.1−10$\mu$m band where the H.E.S.S. data are most affected. The curves show the shapes used to constrain the EBL density, with the thick line indicating the range where the data constrain directly the shape. Details are in [9].

However the shape of $F_i$ is never known, so assumptions have to be largely model-dependent to derive constraints on the EBL. For instance, assuming that the intrinsic photon index $\Gamma_i$ of $F_i(E)$ cannot be larger than 1.5 in the VHE range, one can readily test some EBL models by constructing the intrinsic spectrum they would predict: the models violating the initial assumption on $\Gamma_i$ are then seriously in doubt.

Another possibility to derive the spectral energy distribution of the EBL consists in assuming a specific shape for it, and to leave the overall normalization as a free parameter only constrained by the fact that its imprint on the VHE spectrum should not yield a $\Gamma_i$ smaller than 1.5. Applying this to the object 1ES 1101-232, one of the most distant blazars listed in Table 1, [9] derived limits on the EBL by scaling a model designed to be in general agreement with the EBL spectrum expected from galaxy emission [10].

The rather low EBL level obtained this way is in good agreement with the expectations of standard galaxy evolution models, and is an indication that the Universe is more transparent to $\gamma$-rays than initially thought. A similar study on the object 1ES 1218+30.4 (which has also a very flat VHE light curve!), detected by the MAGIC experiment, yielded similar results, showing that the EBL level is remarkably close to the 'incompressible' leve derived from resolved source counts.

## 5 Non-blazars

A result awaiting a decisive confirmation was the detection in VHE $\gamma$-rays of the nearby ($z = 0.0008$) radio galaxy M87 by the HEGRA experiment. The large angle of the jet with the line of sight, estimated to be $\theta = 35°$, makes it an unlikely VHE source since Doppler boosting would be too low to make it bright enough. However the marginal detection of Cen A, an object similar to M87 but with $\theta = 60°$, by the EGRET experiment [11], shows the potential of this nearby unbeamed population of AGN.

Besides confirming M87 as a VHE emitter, variability on daily timescales was detected by H.E.S.S. (Fig.4), which is an order of magnitude faster than the monthly variability set by Chandra measurements in the X-rays. Using the causality argument with a 2-day timescale variability, the size of the $\gamma$-ray emitting region is smaller than $5 \times \delta R_{\rm S}$, excluding the elliptical galaxy itself, the entire extended kiloparsec jet, and dark matter annihilation.

## 6 Conclusions

So what have we learned in discovering all these new extragalactic sources? A clear global trend as of now is that more and more sources are discovered with little variability in their light curves, which could be interpreted as seeing them in their quiescent level[3]. This is corroborated with the fact that (quasi) contemporaneous observations of the synchrotron component show fluxes that are on the lower bound of archival observations when those are available. Observations of rather low X-ray fluxes are the most convincing fact, since $\gamma$-ray variability correlates best with X-rays (even though some ponctual exceptions exist). Having access to quiescent states will give insights on the duty cycle of blazars, as well as rise the question of the relationship with the more variable high flux emission region when these appear.

[3]Higher variability, or variance, has been claimed to be correlated with the flux in X-ray binaries and AGN [13], but this has yet to be established for blazars.

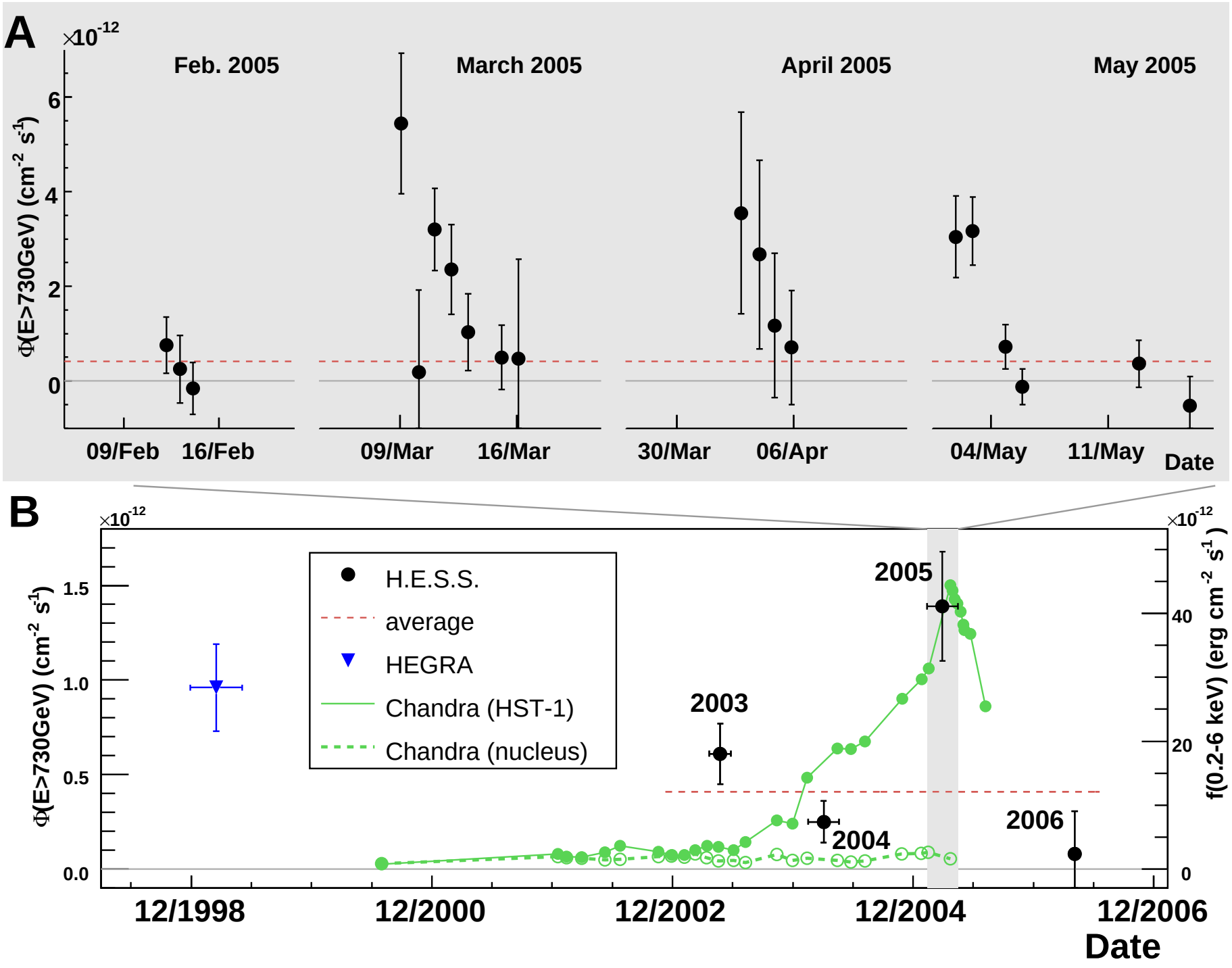

Figure 4: Light curve above 730 GeV of M87 as measured by H.E.S.S. [12]. The top panel are the nightly fluxes for each observed month of 2005. The yearly fluxes, along with the Chandra measurements of the knot HST-1, are in the bottom plot.

Individually, spectral measurements of the most distant blazars in VHE $\gamma$-rays have considerably constrained the intensity of the EBL in the $0.1 - 10\,\mu$m band, although in a model-dependent way. With little room left for additional components in the EBL, the the steep VHE spectra from relatively nearby objects can be deemed to be due to the intrinsic physical mechanisms at play in the accelerator, rather than EBL attenuation. The search for more distant blazars, with spectra able to constrain event further the EBL SED, doesn't seem over yet.

Very fast variability on minute timescales now leads to an increase in bulk Lorentz factors in the context of homogeneous 1-zone emitting models that are closer to the GRB context. Note that the outburst of PKS 2155-304 shown here is only a fraction of the multi-wavelength campaign that was triggered subsequently, with simultaneous observations in optical and X-ray on the following observations where the flux was also very high. These will be shown and published elsewhere. The monitoring of blazars with the current generation of ACTs could yield other surprises, the radically different spectral behaviour of PKS 2155-304 and Mkn 501 having yet to be understood.

# INDIRECT SEARCHES FOR DARK MATTER WITH HESS

J-F. GLICENSTEIN [a], FOR THE H.E.S.S.COLLABORATION [1]

[a] *DAPNIA, CEA-Saclay, F-91191 Gif-sur-Yvette, France*

**Abstract**

The High Energy Stereoscopic System (H.E.S.S.), an array of atmospheric Cherenkov telescopes, is used to perform searches for VHE (Very High Energy > 100 GeV) gamma ray emission from astrophysical objects. Recent searches for dark matter annihilation sources are reported in this paper. A strong signal from the high energy source HESS J1745-290 has been found in the vicinity of the Galactic Center. Limits on the contribution of WIMPs annihilation to this signal are given. Dark matter annihilations were also searched towards the Sagittarius (Sgr) dwarf elliptical galaxy. Constraints on high energy particle models from the negative results of this search are reported.

## 1 Introduction

The H.E.S.S. (High Energy Stereoscopic System) instrument is an array of four 107 $m^2$ imaging atmospheric Cherenkov telescopes installed in Namibia. It is operated by a collaboration of $\sim$ 100 astrophysicists mostly from Germany and France. Details on the H.E.S.S. collaboration and the operation of the array of telescopes are given on the H.E.S.S. collaboration homepage.

[1] address: http://www.mpi-hd.mpg.de/hfm/HESS/

Popular particle physics models such as the Minimal Supersymmetric Standard Model (MSSM) or Universal Extra Dimensions ("Kaluza-Klein" [15]) predict WIMP (Weakly Interacting Massive Particles) dark matter annihilations in galactic halos. These annihilations could give observable signals in Cherenkov telescopes (for a review see [6]). The flux $\mathrm{d}\Phi/\mathrm{d}E_\gamma$ of gamma rays is

$$\frac{\mathrm{d}\Phi}{\mathrm{d}E_\gamma} = \frac{1}{4\pi}\frac{\mathrm{d}N_\gamma}{\mathrm{d}E_\gamma}\frac{<\sigma v>}{M_\chi^2}\bar{J}\Delta\Omega. \tag{1}$$

It is the product of an astrophysical term $\bar{J}$ and a particle physics term. The former depends on the mass density profile $\rho$ of the dark halo

$$\bar{J} =< \int_{l.o.s} \rho^2 \mathrm{d}s > . \tag{2}$$

In equation 2, the average is taken over the solid angle $\Delta\Omega$ spanned by the Point Spread Function (PSF). The spatial resolution of H.E.S.S. is of the order of 5 arc minutes per event, giving $\Delta\Omega = 2\ 10^{-5}$. The particle physics term depends on the velocity averaged annihilation cross section $< \sigma v >$ and the WIMP mass $M_\chi$. A typical analytical $\mathrm{d}N_\gamma/\mathrm{d}E_\gamma$ photon spectrum [10] was used for the MSSM studies. The photon spectrum from the annihilation of the $B^1$ boson of the Kaluza-Klein model was obtained directly with PYTHIA [2].

The possible targets for WIMP annihilation searches can be ranked according to their values of $\bar{J}$. If the annihilation signal from a halo located at distance $D$ is "point-like", then $\bar{J} \propto M^2/D^5$ where $M$ is the (often measured) dark mass inside the PSF. The best astrophysical targets are thus the Galactic Center and nearby dwarf galaxies. The expected flux from galaxy clusters such as Virgo or Coma is smaller by at least 3 orders of magnitude. It is also possible to look for dark matter clumps. This paper reports limits on a dark matter contribution to the signal of the Galactic Center source HESS J1745-290 [1] and the observation of a satellite galaxy of the Milky Way, the Sgr dwarf elliptical galaxy [2]. H.E.S.S. has also observed several galaxy clusters. The observations of the center of the Virgo cluster (the M 87 galaxy) are reported in [5].

## 2 The Galactic Center source HESS J1745-290

H.E.S.S. has detected a bright, point-like source, HESS J1745-290, near the Galactic Center [4]. The size of the emission region is smaller than 15 pc. Diffuse emission has been detected along the galactic plane [3], also in the vicinity of the Galactic Center. Diffuse emission correlates well with the mass density of the molecular clouds of the Central Molecular Zone as measured by the CS line [16]. The position of the HESS J1745-290 source is compatible with that of the supermassive black hole Sgr A*. In the 2004 [4] and 2006 [1] H.E.S.S. papers, the position of the source was quoted with systematic errors of roughly 30 arc seconds. Recently, a careful study [17] allowed to lower the pointing errors down to the level of 8 arc seconds. The preliminary position of the source is located at an angular distance of 7.3 $''\pm 8.7$ $''$(stat) $\pm\, 8.5''$(syst) from the central galactic black hole Sgr A*. The new accurate position of HESS J1745-290 is incompatible with the centroid of the radio emission of the supernova remnant Sgr A East, but still compatible with sources such as the pulsar wind nebula G359.95-0.04.

[2]URL http://projects.hepforge.org/pythia6/

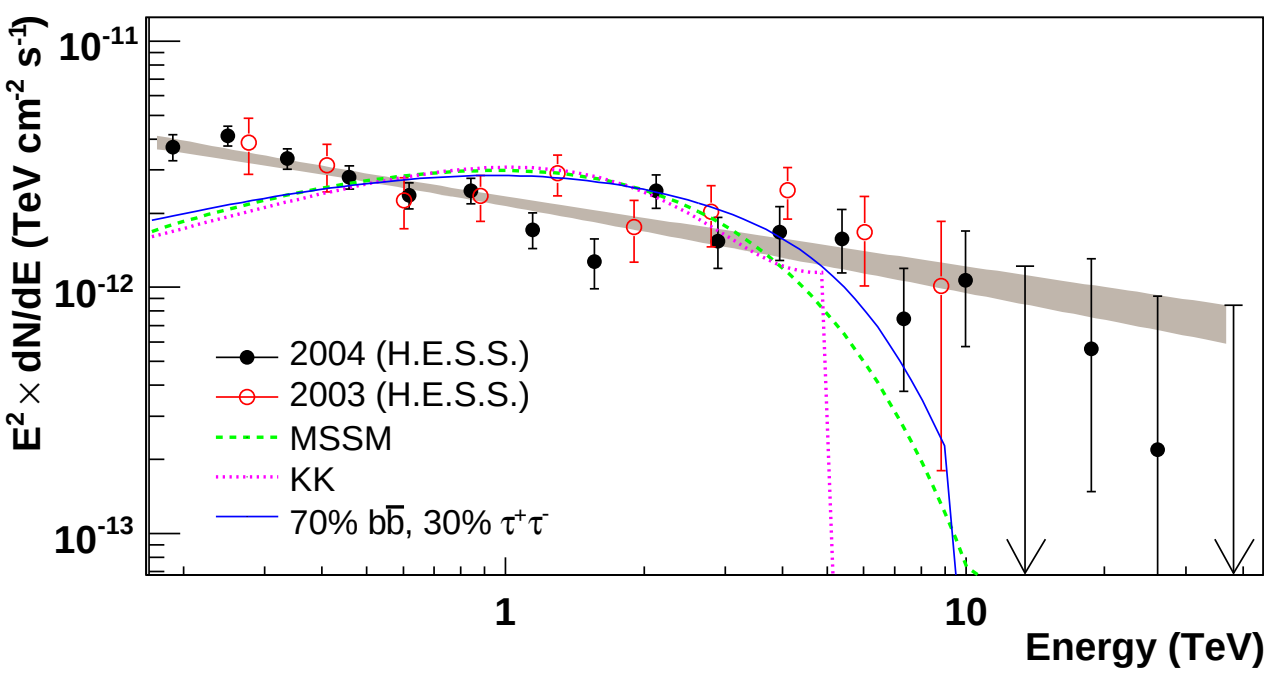

Figure 1: Spectrum of the HESS J1745-290 source. Both 2003 and 2004 spectra are shown. The shaded area shows the power-law fit to the data. The dashed and dotted lines show the spectra expected for the annihilation of respectively a typical MSSM-like 14 TeV neutralino, and for that of a 5 TeV Kaluza-Klein $B^1$ boson (see e.g [6]). The solid line shows the photon spectrum expected from 10 TeV DM particles annihilating into 30% $\tau^+\tau^-$ and 70% $b\bar{b}$, as proposed in [14].

The spectrum of HESS J1745-290 (figure 1) is well fitted by a power-law spectrum in the energy range (160 GeV-30 TeV) with a spectral index $\Gamma = 2.25 \pm 0.04\,(\text{stat}) \pm 0.1\,(\text{syst})$. No deviation from a power-law spectrum is seen, which implies a cut-off energy of more than 9 TeV (95% CL). The emission of HESS J1745-290 does not show any significant variability or periodicity at time-scales ranging from 10 minutes to 1 year [18].

The signal of HESS J1745-290 can been interpreted by a large variety of astrophysical models (see e.g. [9] and references inside). The possible non-standard interpretation of the HESS J1745-290 signal as annihilation of dark matter particles has been investigated by the H.E.S.S. collaboration [1]. The observed spectrum (figure 1) is not well fitted by expected annihilation spectra. Limits on the annihilation cross-section can be given by assuming that the observed signal is a blend of an astrophysical source and dark matter annihi-lations. The spectral shape is taken as the sum of a powerlaw and a known dark matter annihilation spectrum. The $< \sigma v >$ term in the annihilation flux formula 1 is allowed to vary. The dark matter annihilation flux involves also the astrophysical factor $\bar{J}$. For the Galactic Center, the values of $\bar{J}$ range from $J_0 = 3\ 10^{22}$ GeV$^2$ cm$^{-5}$ (core model) to $\sim 1.5\ 10^5\ J_0$ (Moore model). To derive the Galactic Center limits, the mass density of the dark halo was assumed to be described by a Navarro-Frenk-White [13] profile, for which $\bar{J} \sim 3000\ J_0$. The 95% CL upper limits on $< \sigma v >$ are in the range $10^{-24} - 10^{-23}$ cm$^3$ s$^{-1}$, for MSSM or Kaluza-Klein models with WIMP masses between 200 GeV and 20 TeV. This is one to two order of magnitude higher than the predictions of the MSSM or Kaluza-Klein models.

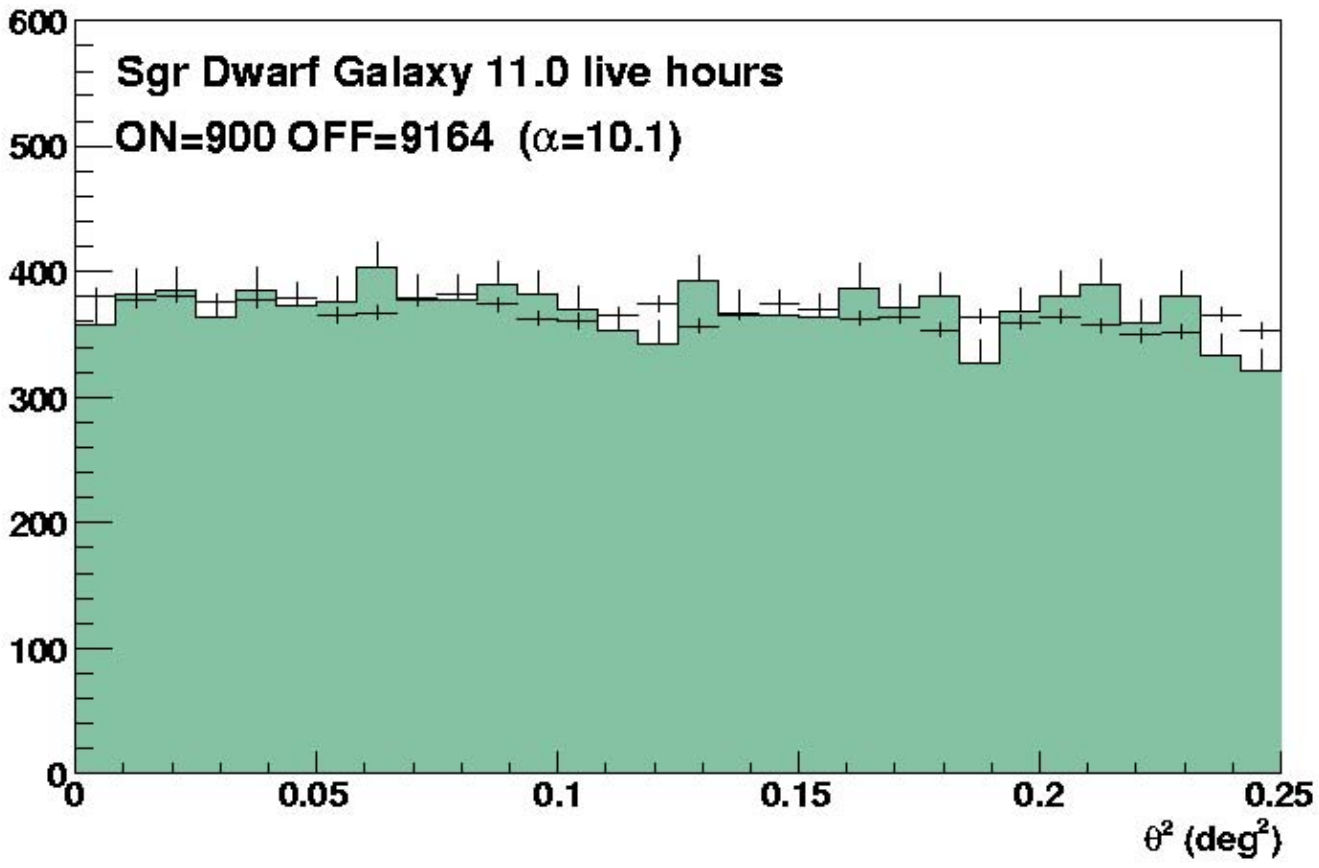

Figure 2: Distribution of $\theta^2$ for Sgr dwarf galaxy. $\theta$ is the angular distance between the direction of the gamma event and the target position. The target position is the M 54 globular cluster, which is also the center of the Sgr dwarf galaxy. The points with error bars are the data points. The filled histogram shows the expected background. The ON and OFF counts are obtained with the $\theta^2 < 0.02$ cut. No significant excess of photons is seen at the center of Sgr dwarf galaxy.

Searches for dark matter annihilations at the Galactic Center are limited at the present time by source confusion. Nearby dwarf galaxies are known for their large dark matter content. They are expected to be much less crowded with astrophysical sources than the Galactic Center.

## 3 Searches towards the Sagittarius dwarf galaxy

The Sgr dwarf galaxy is a satellite of the Galaxy. It is located in the galactic plane in the direction of the Galactic Center, at a distance of 24 kpc. This galaxy was discovered only recently [11]. It is being torn apart by the tidal force of the Galaxy. The visible mass profile of the Sgr dwarf galaxy is difficult to obtain because of the contamination of galactic foreground stars. The center of the Sgr dwarf galaxy is coincident with the globular cluster M 54 [12]. The interpretation of velocity dispersion measurements is difficult because of the tidal interaction with the Milky Way. The central velocity dispersion has been measured by several groups (see e.g. [19]).

The Sgr dwarf galaxy has been observed by H.E.S.S. in June 2006. After quality cuts, a total exposure of 11 hours was obtained. As shown on figure 2, no significant excess of signal was found at the location of M 54. Less than 56 photons (95% CL) with an energy of more than 250 GeV (corresponding roughly to the analysis threshold) are detected at a position of less than 8.4 arc-minutes from the center of M 54.

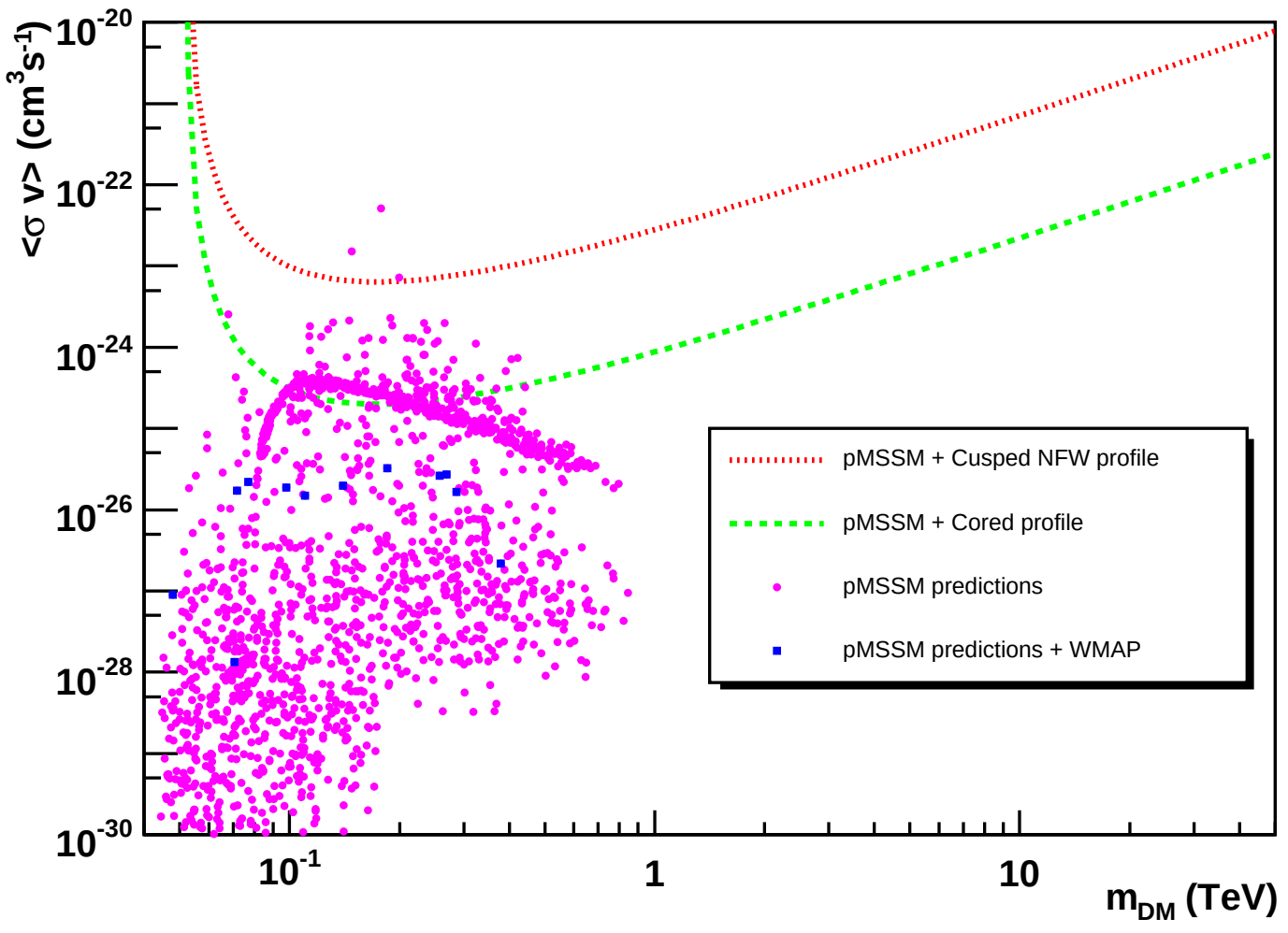

Figure 3: 95% CL upper limits on $<\sigma v>$ versus the neutralino mass obtained from the HESS J1745-290 signal. The purple points show the predictions of the (p)MSSM. The blue points show the effect of the relic density constraint from the measurements of the WMAP satellite. The values of $<\sigma v>$ excluded at the 95% CL lie above the green dashed line (cored model) or the red dotted line (NFW model).

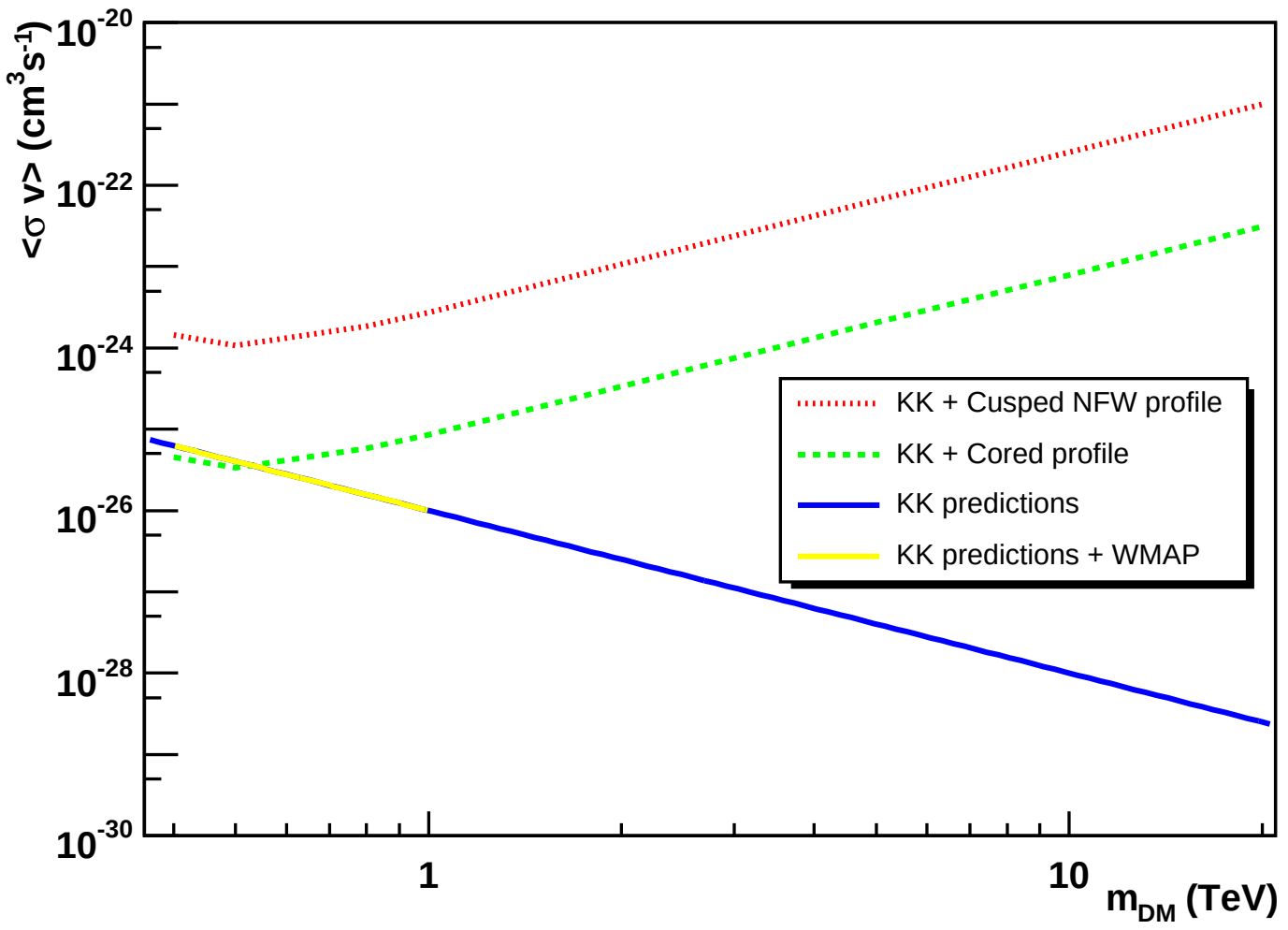

Figure 4: 95% CL upper limits on $< \sigma v >$ versus the WIMP mass obtained from the HESS J1745-290 signal. The predictions of the Kaluza-Klein model are shown by the blue line. The yellow segment corresponds to the WMAP constraint. The values of $< \sigma v >$ excluded at the 95% CL lie above the green dashed line (cored model) or the red dotted line (NFW model).

A dynamic model of the Sgr dwarf galaxy is needed to convert the upper limit on the number of photons into constraints on the WIMPs interactions. Dynamic models of the Sgr dwarf galaxy have been studied by Evans, Ferrer and Sarkar [8]. As in the Galactic Center study, our reference model has a NFW mass profile. The parameters of this model were taken from [8]. Another model (a cored isothermal model) also inspired from reference [8] was studied. Differences with the cored model of [8] include

1. A different velocity dispersion profile
   The velocity dispersion of the Sgr dwarf stars is supposed to be independent of position and equal to the measured value of the central velocity dispersion ([19]). In reference [8], the velocity dispersion data were the measured values of the Draco dwarf galaxy. As in [8], the velocity dispersion tensor was assumed to be isotropic.

2. A different mass density profile.
   The density profile of stars is taken from reference [12]. The cored profile model has a very small core due to a "cusp" in the mass density profile.

The value of $\bar{J}$ for the cored profile model turns out to be larger than that of the NFW model. Figure 3 shows the constraint obtained on the $< \sigma v >$ values of the MSSM models. The predictions of the MSSM (see [6]) were obtained with DarkSusy 4.1 [7] and are plotted as purple points. The blue squares are MSSM models which give a relic mass density of WIMPs compatible with the measurements of the WMAP satellite. Values of $< \sigma v > \geq 10^{-24}$ (cored mass profile) or $< \sigma v > \geq 10^{-22}$ (NFW mass profile) are excluded at the 95% CL. Figure 4 shows the constraints on $< \sigma v >$ for the Kaluza-Klein model. The prediction of the Kaluza-Klein model is shown as a blue line. The yellow segment gives a relic density compatible with the measurements of the WMAP satellite. Values of $< \sigma v > \geq 10^{-25}$ (cored mass profile) or $< \sigma v > \geq 10^{-23}$ (NFW mass profile) are excluded at the 95% CL for the Kaluza-Kein model.

## 4 Conclusion

The H.E.S.S. collaboration has studied two potential sources of dark matter annihilations: the Galactic Center and the Sgr dwarf galaxy. Constraints on the annihilation cross section have been given. These constraints are still one to two order of magnitude larger than the prediction (assuming an NFW profile). The energy threshold is around 250 GeV. In the phase 2 of H.E.S.S., starting in 2009, the analysis threshold will be lowered down to $\sim 80$ GeV or less. This will allow to explore more supersymmetric models.

# DETECTION AND MEASUREMENT OF GAMMA RAYS WITH THE AMS-02 DETECTOR

S. Di Falco [a]
on behalf of the AMS Collaboration

[a] *INFN & Università di Pisa, Largo B. Pontecorvo 3, Pisa, Italy*

**Abstract**

The AMS-02 detector will collect data on the International Space Station for at least three years. The gamma rays can be measured through e+e- pair conversion in the Silicon Tracker, as well as single photons directly detected in the Electromagnetic Calorimeter. AMS-02 will provide precise gamma measurements in the 1 GeV up to the few hundreds GeV range, which are particularly relevant for Dark Matter searches. In addition, the good angular resolution and identification capabilities of the detector will allow clean studies of the main galactic and extra-galactic sources and the observation of the high energy tail of some GRB.

## 1 The AMS experiment

The Alpha Magnetic Spectrometer (AMS) is a cosmic ray experiment which will be attached to the International Space Station (ISS) for three years, being launch ready in 2008. It is a major collaboration of European, Asian and American institutions, together with NASA. The main purpose of AMS measurements is to determine the characteristics of the incident particle such as its momentum, charge, velocity and mass, to the highest possible precision, to ensure its identification.

A precursor flight, AMS-01 [1], with a reduced acceptance and magnetic field and without electromagnetic calorimeter, has already succesfully flown on board of the Space Shuttle Discovery for 10 days in 1998: it has performed important measurement of the spectrum of primary and secondary protons, Helium, electrons and positrons and has improved the limit on the flux of cosmic antihelium.

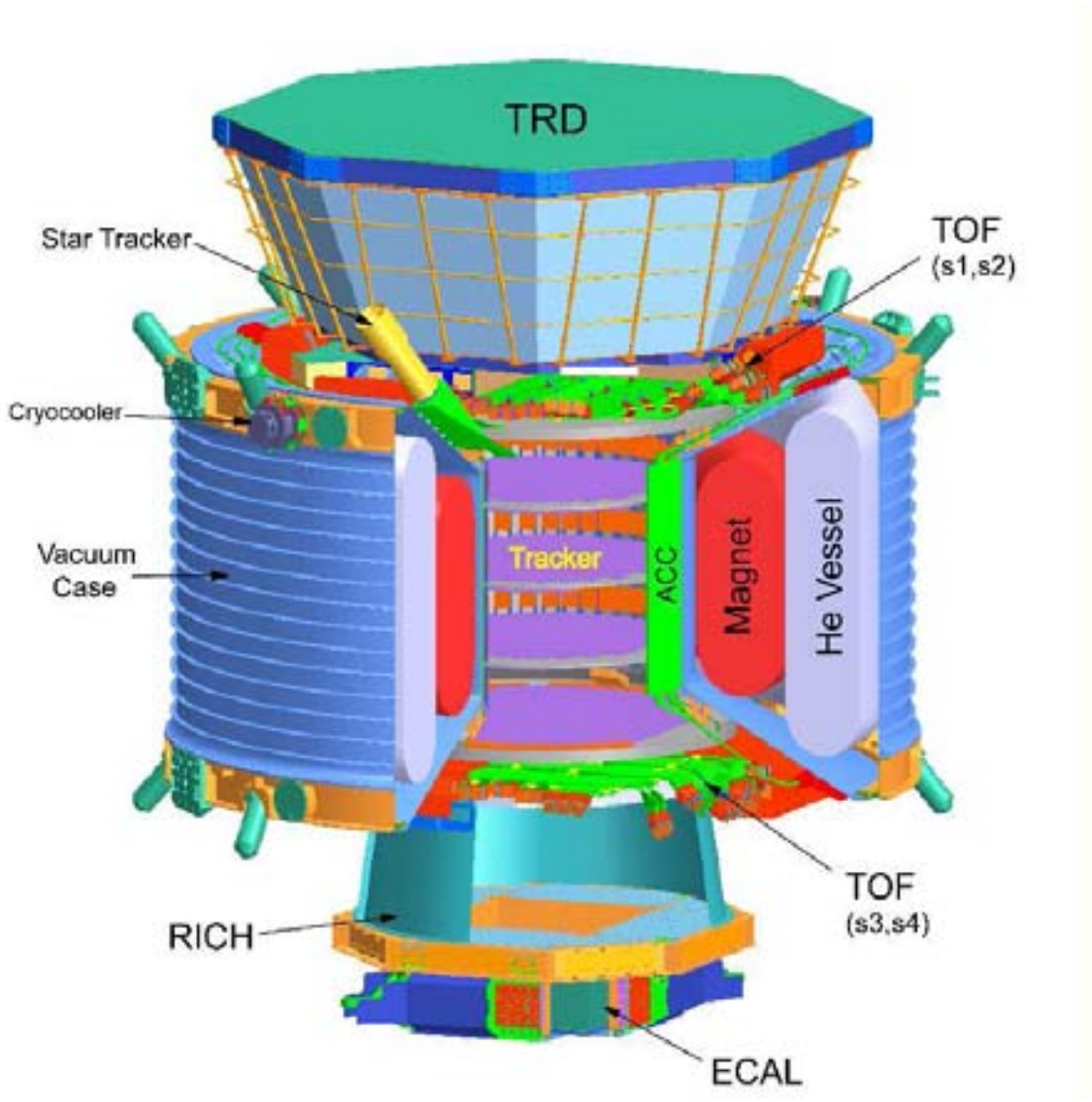

Figure 1: The AMS-02 detector.

The AMS-02 detector (fig. 1) has a cylindrical symmetry with a radius ∼2 m and an height ∼3 m for a total weight of about 7 tons. Its main components are: the *Superconducting Magnet*, producing a dipolar field with a bending power $BL^2 = 0.85\ Tm^2$; the *Transition Radiation Detector (TRD)* [2]; the *Silicon Tracker Detector (STD)* [3], capable to measure the particle rigidity up to ∼1 TV; the *Time Of Flight (TOF)* detector [4]; the *AntiCoincidence Counters (ACC)* used to inihibit the charged particle trigger provided by the TOF when the particles come from the side of the detector; the *Ring Imaging CHerenkov (RICH)* [5] and the *Electromagnetic CALorimeter (ECAL)* [6], a sampling calorimeter composed of scintillating fibers and lead with 9 super-layers alternated with the fibers along X or Y directions perpendicular to the detector axis, for a total of 17 radiation lengths.

The major scientific goals of AMS-02 include dark matter, and anti-matter searches, cosmic ray related astrophysics and, last but not least, gamma ray astronomy. Photons can be detected in AMS-02 following two complemen-tary methods: the *photon conversion* in an $e^+e^-$ pair and the *direct photon measurement* in the calorimeter.

## 2 Detection of photon conversions

The material in front of the first silicon tracker layer corresponds to $\sim 0.3\ X_0$, so there's a significant probability (∼25%) for a photon to convert in the upper part of the detector into an $e^+e^-$ pair that can be detected by the STD.

A photon conversion in AMS will have the following characteristics: no hits in the upper TRD layers; a couple of tracks well reconstructed in the STD; a vertex located in the TRD or in the upper TOF; the invariant mass of the 2 tracks close to 0; the extrapolated photon direction not crossing the mechanical structure of the detector; the total amount of hits in all the subdetector compatible with just 2 relativistic particles; the energy deposits in the calorimeter, if any, compatible with electromagnetic showers.

The last 3 requests have been inserted to suppress the background, mainly due to $\delta$ rays produced by protons or electrons before the STD.

The photon energy resolution is[7]:

$$\frac{\sigma_E}{E} = (3.2 \oplus 0.05 \cdot E(GeV))\% \qquad (1)$$

that is ∼3% below 30 GeV increasing to 10% at 200 GeV.

The angular resolution defined as the 68% containment angle is[8]:

$$\sigma_{68} = \left(-0.57 + 0.58 \cdot e^{1/E(GeV)}\right)^o \qquad (2)$$

i.e. $1^o$ at 1 GeV improving to $0.015^o$ at 200 GeV.

Considering as signal the photons coming from the galactic center the background to signal ratio is of the order of $10^4$ for protons and $10^2$ for electrons. The rejection power of the analysis cuts is better for $10^4$ for both protons and electrons. A further improvement can come, in case of source studies, from the excellent pointing capability obtained by the angular resolution.

## 3 Direct Photon Detection

Given the amount of material present in front of the calorimeter, a photon has a probability of ∼70% to reach the ECAL without converting.

**The AMS photon trigger.** Photons not converting before the calorimeter do not relaese enough energy in TOF scintillators to satisfy the TOF trigger conditions, so that a dedicated photon trigger is required. The *photon trigger*[10] has two levels: the *fast trigger*, produced before 200 ns, compares the dynode signal of the PMTs of the 6 central ECAL superlayers with a programmable threshold and requires at least 1 PMT dynode above threshold for at least 2 of the 3 superlayers of each view; the *level 1 trigger*, produced within 1 $\mu s$, checks that the particle inclination, obtained from the average position of the PMts above threshold in the different superlayers, is below $\sim 20^o$ [1]. The trigger efficiency on photons not converting before the calorimeter is 20% at 1 GeV, 90% at 2 GeV and better than 99% above 5 GeV. The total trigger rate is negligible with respect to the TOF trigger rate.

A photon directly interacting in the calorimeter can be recognized as an *electromagnetic shower* [2] in the calorimeter without any hits in the other detectors.

The main background is due to secondary particles, namely $\pi^0$, produced by proton interactions in the mechanical structure of the magnet and can be suppressed by asking that the reconstructed photon direction crosses all the others subdetectors.

According to the results of the ECAL test beam in 2002[11], confirmed by the ECAL flight model test beam of July 2007, the photon energy resolution is:

$$\frac{\sigma_E}{E} = \left(2.1 \oplus \frac{10.4}{\sqrt{E(GeV)}}\right)\% \tag{3}$$

that is ∼6% at 2 GeV improving to ∼2% above 100 GeV.

The angular resolution at $0^o$ defined as the 68% containment angle is:

$$\sigma_{68} = \left(0.2 + \frac{7.5}{\sqrt{E(GeV)}}\right)^o \tag{4}$$

i.e. $5^o$ at 1 GeV improving to $1^o$ at 100 GeV. Photon inclination also improves the angular resolution, thanks to the larger energy deposit: for example at $20^o$ it's $0.5^o$ at 100 GeV.

The selection efficiency for photons with an angle lower than $20^o$ is $\sim 10$ % at 1 GeV, $\sim 50$% at 2 GeV and $\sim 80$% above 10 GeV.

The low efficiency at low energy is partly due to the tight cuts used for the background rejection: the rejection factor on protons is about 10 at 1 GeV while is about $10^4$ above 10 GeV.

Also in this case further background suppression can come from the pointing to the source.

[1] This corresponds to the request that the photon crosses all the other AMS subdetectors.

[2] An energy deposit that satisfies the lateral and longitudinal profile, the collimation and the compactness expected for the interaction of an electromagnetic particle in the calorimeter[9].

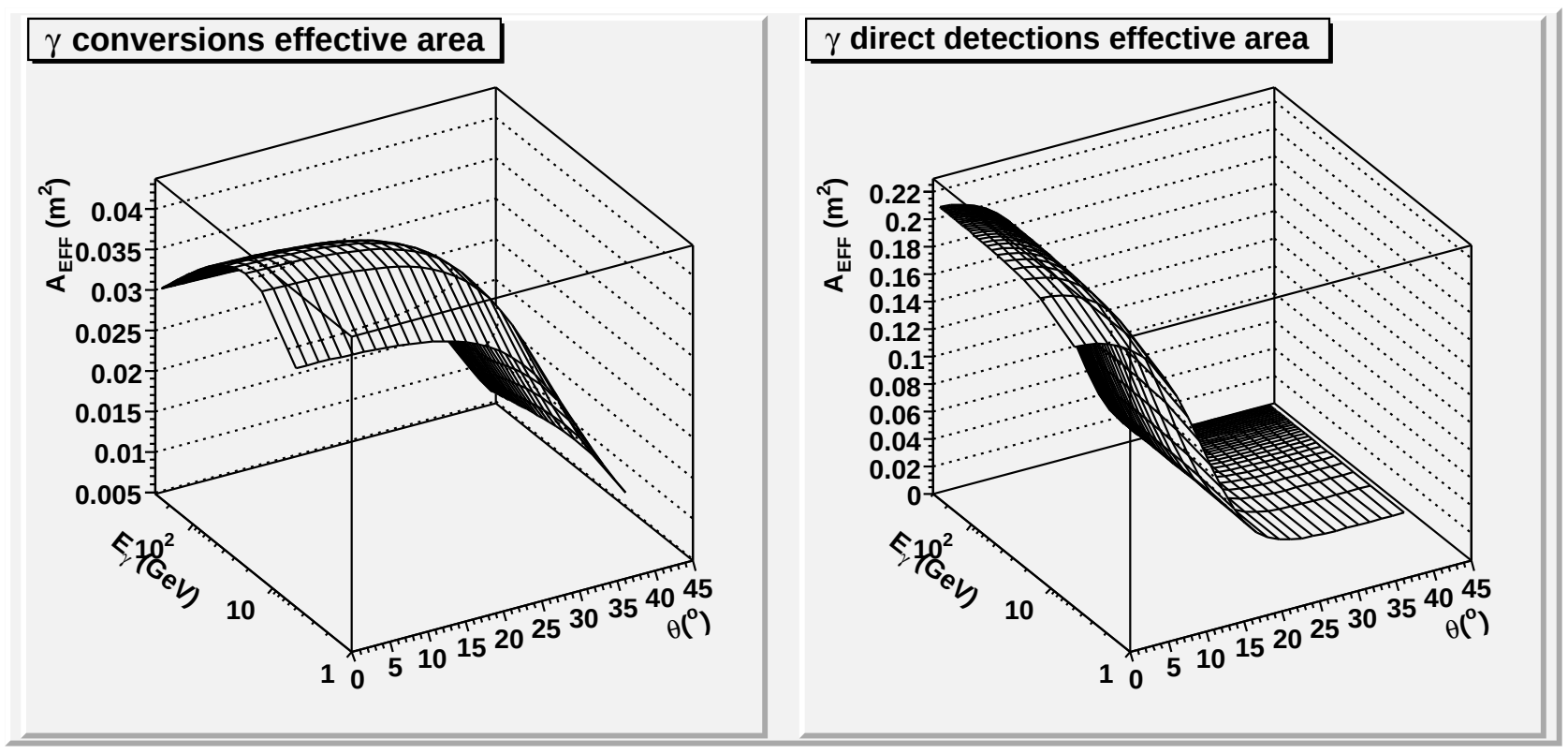

Figure 2: Effective area for $\gamma$ conversions (left) and direct detections (right).

## 4 Estimate of the number of photons detected

The number of photons of energy E observed from a source S seen under an angle $\theta$ for a time interval t can be expressed as:

$$N_S(E) = F_S(E) \cdot A_{EFF}(E, \theta) \cdot t \tag{5}$$

where $F_S$ is the photon flux from the source and $A_{EFF}$ is the detector effective area as function of the particle energy and inclination.

Integrating over the angle $\theta$ and approximating the integral with the sum on finite angle intervals I, one gets:

$$N_S(E) = F_S(E) \sum_I < A_{EFF}(E, \theta) >_I \cdot t_I \tag{6}$$

where the effective area is averaged upon the angle interval I and $t_I$ is the time during which the source is seen under an angle $\theta$ belonging to the interval I.

The effective areas for the two photon detection methods discussed in the previous paragraphs can be parametrized as:

$$A_{EFF}(E, \theta) = \frac{A_1(E)}{A_1(E_0)} A_2(\theta) \tag{7}$$

where $E_0$ is a fixed energy value taken as reference.

For the conversion method (fig. 2.a) $A_1(32\ GeV) = 0.604$ and:

$$
\begin{aligned}
A_1(E) &= e^{-0.5\cdot\left(\frac{log_{10}E(GeV)-1.39}{0.82}+e^{-\frac{log_{10}E(GeV)-1.39}{0.82}}\right)};\\
A_2(\theta) &= \frac{0.041}{1+e^{\frac{cos(\pi-\theta)+0.79}{0.057}}}
\end{aligned}
$$

For the direct detection method (fig. 2.b) $A_1(20\ GeV) = 0.065$ and :

$$
\begin{aligned}
A_1(E) &= 0.067 - 0.057\cdot e^{-2.6\cdot log_{10}E(GeV)};\\
A_2(\theta) &= \frac{0.23}{1+e^{\frac{cos(\pi-\theta)+0.96}{0.02}}}
\end{aligned}
$$

The maximum opening angle is $\sim 25^o$ for the direct detection and $\sim 45^o$ for the conversions.

For vertical particles the effective area for conversions has a maximum around 20 GeV at 0.04 $m^2$ while for direct detection it always increases with energy approaching 0.22 $m^2$ above 100 GeV.

The exposition time of a source under a given angle is obtained by simulating the Internation Space Station orbit: this is an ellipse with average radius of 386 Km, with an inclination of $51.57^o$ with respect to the ecliptic and precessing with a period of 71 days. It will be covered by the ISS in ~92 minutes.

Considering as an example the whole opening angle of the two detection methods, integrating over 15 precession periods ($\sim$ 3 years) one finds a total exposition time to a $1^o \times 1^o$ pixel around the galactic center of 120 days for the conversion method and 60 days for the direct detection [3].

Due to the limited opening angle and to the latitude limits of the ISS orbit, there are 2 regions of the galactic sky in which the direct photon measurement cannot be applied: for longitude $-80^o < l < -60^o$ and latitude $-40^o < b < -10^o$ and $110^o < l < 140^o, 10^o < b < 40^o$. On the contrary, the conversion method can cover the whole sky.

## 5 Expected number of detected photons

The *AMS Fast Simulator*[12] is a program appositely designed to calculate the number of photons from a source detected by AMS in one precession period using the effective areas of the 2 methods. Examples of its application are:

- the expected number of photons from the Crab nebula in 3 years is ~300 for the conversion method and $\sim$ 80 for the direct detection;

[3] The detector is assumed to be blind during the South Atlantic Anomaly crossing.

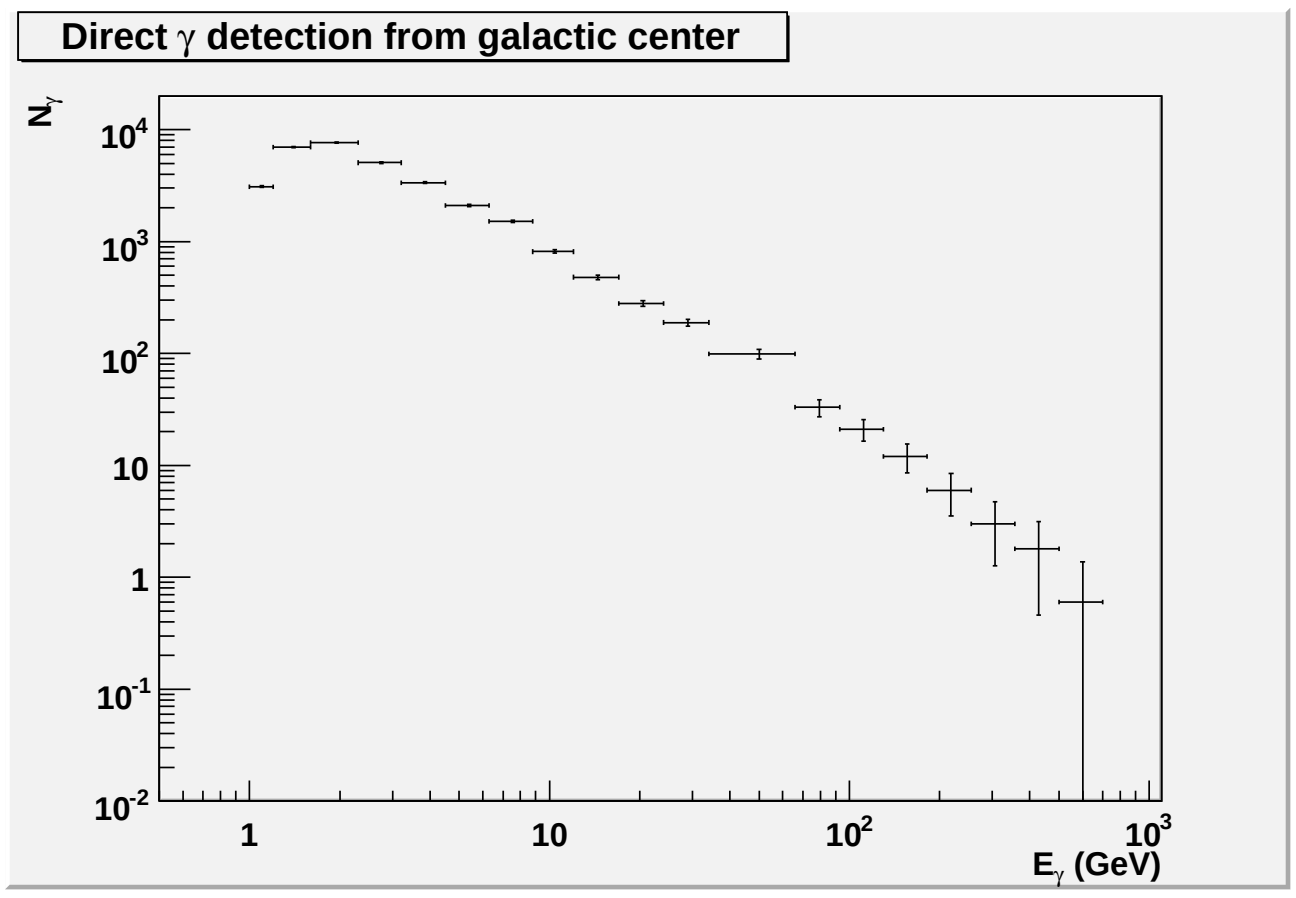

Figure 3: Exepected number of direct detections of photons coming from the galactic center region ($-30^o < l < -30^o, -5^o < b < 5^o$) in 3 years.

- $\sim$ 1500 photons will be detected from the Vela pulsar, allowing to discriminate between the polar cap and the outer gap emission models;
- the expected number of photons directly detected in 3 years from the galactic center region delimited by $-30^o < l < 30^o$ and $-5^o < b < 5^o$ is shown in fig. 3 [9] [4]: only 10 photons are expected above 200 GeV;
- the precise determination of the photon spectrum will allow to constrain the region of parameters for many models of Lightest Supersimmetric Particle and Lightest Kaluza-Klein Particle dark matter [14];
- due to the relatively high energy threshold only few tens of GRBs could be observed in 3 years. The most energetic GRB observed by EGRET, which lasted 80 s with an energy spectrum:

$$\frac{dN}{dE} = 9.6 \cdot (E(GeV))^{-2.2}\ m^{-2}s^{-1}GeV^{-1}, \tag{8}$$

  would produce 13 conversions and 45 direct detections if it whould happen at $0^o$ with respect to the AMS axis, but only 13 conversions if it would happen at $30^o$. Any spectral studies would be in any case quite difficult.

[4]This was calculated dividing the region in $1^o \times 1^o$ pixels whose diffuse emission has been obtained using the GALPROP program[13].

## 6 Summary

During its 3 years mission on ISS, the AMS experiment will be able to measure simultaneously and in most cases with unprecedented precision the spectra of the different components of cosmic radiation in the GeV-TeV range.

Thanks to a dedicated photon trigger and to the good number of radiation lengths of its calorimeter, a direct photon detection can be successfully used together with the usual photon conversions detection. Significative effective areas between 1 and 300 GeV will be available for both methods.

# THE LOCALIZATION OF GAMMA RAY BURSTS BY SUPERAGILE ON-BOARD OF AGILE

E. Del Monte[a], E. Costa[a], G. Di Persio[a], I. Donnarumma[a], Y. Evangelista[a,b], M. Feroci[a], M. Frutti[a], I. Lapshov[a], F. Lazzarotto[a], M. Mastropietro[c], E. Morelli[d], L. Pacciani[a], G. Porrovecchio[a], M. Rapisarda[e], A. Rubini[a], P. Soffitta[a], M. Tavani[a], A. Argan[a], A. Trois[a], C. Labanti[d], M. Marisaldi[d], F. Fuschino[d]

[a] *INAF IASF Roma Via Fosso del Cavaliere 100, I-00133 Roma, Italy*

[b] *Dip. di Fisica, Università di Roma "La Sapienza", Piazzale Aldo Moro 2, I-00185 Roma, Italy*

[c] *IMIP CNR, Via Salaria km 29.300, I-00016 Monterotondo Scalo (RM), Italy*

[d] *INAF IASF Bologna, Via Gobetti 101, I-40129 Bologna, Italy*

[e] *ENEA UTS Fusione Tecnologie, Via Enrico Fermi 45, I-00044 Frascati (RM), Italy*

## Abstract

SuperAGILE is the wide field of view, compact, light and low power consuming hard X-ray monitor of the AGILE space mission. AGILE is the first small scientific mission of ASI, is devoted to the study of the High Energy Astrophysics in the hard X-ray and gamma ray energy bands and was launched on 23rd April 2007. SuperAGILE is expected to image Gamma Ray Bursts and bright transients directly on-board, basing on the instrument ratemeters, integrated on various timescales from 64 ms up to 8 s. The accumulated image is deconvolved on-board by an algorithm introduced in the Payload Data Handling Unit and the event coordinates are rapidly transmitted to Earth by using the ORBCOMM satellite constellation. Shorter transients (down to the sub-ms level) can be detected but are not located. The on-board Gamma Ray Burst detection system has been extensively tested in laboratory by using radioactive sources. In this contribution we describe the SuperAGILE Gamma Ray Burst detection system and report about the laboratory testing.

## 1 The SuperAGILE scientific performances

SuperAGILE is the hard X-ray monitor of the AGILE satellite-borne mission of the Italian Space Agency (ASI), launched from the Satish Dawan Space Centre (India) on 23rd April 2007. AGILE is devoted to the High Energy Astrophysics and its payload is composed of two instruments: the Gamma Ray Imaging Detector (GRID), with about 2.5 sr Field of View (FOV) and sensitive in the 30 MeV – 50 GeV energy band, and SuperAGILE, with about 1 sr FOV and 15 – 45 keV energy band. In turn the GRID is composed of a Silicon Tracker, with tungsten converters and silicon microstrip detectors, and a Mini-calorimeter with CsI(Tl) scintillator bars. AGILE is flying on an equatorial orbit with about 550 km altitude, 2.5° inclination and 100 minutes period. A description of the AGILE mission may be found in [1].

SuperAGILE is a coded aperture instrument with four 1-D silicon microstrip detectors of $19 \times 19$ cm$^2$ surface each, 121 $\mu$m strip pitch and 410 $\mu$m thickness. The instrument FOV is composed of two orthogonal 1-D areas of $107° \times 68°$, overlapping in the central area of $68° \times 68°$ where both 1-D coordinates are encoded. The instrument pixel size is 6 arcmin with a source location accuracy of 1 – 2 arcmin for intense sources. The nominal energy band is 15 to 45 keV and the resulting sensitivity is 1 Crab (corresponding to $4 \cdot 10^{-1}$ photons/cm$^2$/s at 15 – 45 keV) at $5\sigma$ significance level in 10 s increasing up to 15 mCrab ($6 \cdot 10^{-3}$ photons/cm$^2$/s) at $5\sigma$ in 50 ks integration time. SuperAGILE is a small ($40 \times 40 \times 14$ cm$^3$), light (10 kg) and low power (12 W) instrument. Further information about SuperAGILE may be found in [2].

The Mini-calorimeter (a description can be found in [3]) is composed of 30 scintillator bars arranged in two planes, spatially aligned in orthogonal directions. This instrument cooperates with the Silicon Tracker in composing the GRID but it is designed also to detect photon transients in its 300 keV – 200 MeV energy band although it is not an imaging instrument.

## 2 The SuperAGILE Gamma Ray Burst detection and localization

Gamma Ray Bursts (GRB) are among the most important scientific objectives of AGILE. A detailed description of the GRB observational properties and theoretical models is far beyond the scope of this paper and the interested reader is addressed to dedicated reviews (for example [4] and [5]).

In brief, GRB are transient events at cosmological distance detected in X and gamma rays. One to three such events are detected every day and they do not show repetitions. The distribution of the GRB incoming direction in the Sky is isotropic (see for example [6]). If $T_{90}$ is defined as the interval of time in which 90 % of the photons are detected, the distribution of $T_{90}$ is bimodal, with the first peak at 0.6 s, the second peak at 35.5 and a third intermediate class required, as shown in [7]. The GRB spectrum can be described by the Band function (see [8] for details), composed of two power laws smoothly connected. GRB are often followed by X-ray (see [9] for details), optical and radio afterglows, starting just after the prompt event. For this reason a rapid communication of the GRB coordinates is of primary importance for the study of this class of sources.

The on-board detection of GRB is introduced in the AGILE Payload Data Handling Unit (PDHU) and is based on the SuperAGILE and Mini-calorimeter ratemeters. The signal and background are estimated by integrating the counting rate in two different timescales: the Signal Integration Time (SIT) and the Background Estimation Time (BET). Allowed values of the SIT are sub-ms, 1 ms, 16 ms, 64 ms, 256 ms, 1024 ms and 8192 ms. The BET can be 8, 16, 33, 66, 131, 262 and 524 s.

The trigger logic compares the $n_{sig}$ counts integrated in the SIT with the $n_{bkg}$ ones in the BET using two types of conditions: static logic (with fixed threshold counts $n_{thr}$),

$$n_{sig} \geq \alpha n_{bkg} + n_{thr} \tag{1}$$

where $\alpha$ is can be either 0 or 1, or adaptive logic (with threshold evaluated from the background standard deviation $\sigma$),

$$n_{sig} \geq n_{bkg} + n\sigma. \tag{2}$$

To increase the time resolution, using the SIT between 16 ms and 1024 ms the signal is estimated by a moving average updated every SIT/4 and the trigger condition is checked every SIT/4. Similarly, the signal of the 8192 s SIT is estimated with a moving average based on 512 ms intervals. For SIT of 16 ms and below the trigger condition is checked every 16 ms.

The trigger is applied independently to the four SuperAGILE detectors, each one divided in two energy bands for a total number of eight channels, and to the two Mini-calorimeter planes. A dedicated lookup-table is used to accept or reject the trigger depending on the channels coincidence. Another parameter of the burst search procedure is the delay between SIT and BET intervals, that can be selected between 0 and 128 s.

If a trigger is accepted with SIT ranging from 64 ms to 8192 ms, the accumulation of the four SuperAGILE detector images starts. The accumulation stops when the counts in the image are enough to produce a statistically significant localization or when the maximum integration time is reached. Both localization significance and maximum integration time can be selected. The background images, continuously accumulated during the burst search procedure, are then subtracted from the signal ones and the resulting images are decoded in order to extract the 1-D position of the transient and the peak counting rate. The background and signal 1-D images and the peak position and counting rate are formatted in a dedicated telemetry packet and transmitted to the AGILE ground station at Malindi. To avoid the delay due to the 100 minutes telemetry downlink period, a subset of the information from the imaging process is transmitted using the ORBCOMM satellite constellation and delivered to the Ground with a typical minutes delay.

The SuperAGILE localization procedure cannot be started using SIT values lower than 64 ms (sub-ms, 1 ms and 16 ms). The burst search trigger logic is the same for SuperAGILE and the Mini-calorimeter, that is not an imaging instrument. The burst search procedure can be configured in order to download the Mini-calorimeter data in photon-by-photon mode in case a SuperAGILE trigger is detected. Further inrmation about the Mini-calorimeter burst search procedure is in [10].

## 3 Laboratory testing

The SuperAGILE on-board GRB localization system was experimentally tested in order to verify the trigger and the imaging procedures. The test were performed on the integrated satellite using a $^{109}$Cd radioactive source (the X-ray line with the highest branching ratio is emitted at 22 keV), positioned inside a dedicated brass plate with place for two sources and two remote controlled shutters. The plate is suspended to an aluminum structure at about 2 m height above SuperAGILE. The shutter uncovers the source in less than 1 s and the source activity is high enough to produce statistically significant images at the source-SuperAGILE distance. A picture of the brass plate used in the test is shown in fig 1.

During the test, the source is unocculted using the remote-controlled shutter in order to produce a transient. This measurement allows to test the trigger capability, the start time (reconstructed from the trigger time) and the stop time (estimated depending on the statistics in the accumulated image). An example of a time series with the background counting rate (before 3450 s), the source counting rate (between 3450 s and 3600 s), the start time (vertical dashed line) and the stop time (vertical dotted line) is shown in fig. 2.

The detector images accumulated on-board are compared with the corresponding detector images accumulated off-line from the SuperAGILE photon-by-photon data simultaneously recorded. All the images are then decoded and the source position reconstructed on-board is verified against the position extracted from the photon-by-photon data analysed off-line. An example of the two 1-D reconstructed images of a transient is shown in fig. 3 and 4. By this measurement the on-board imaging reconstruction and source localization procedures are tested.

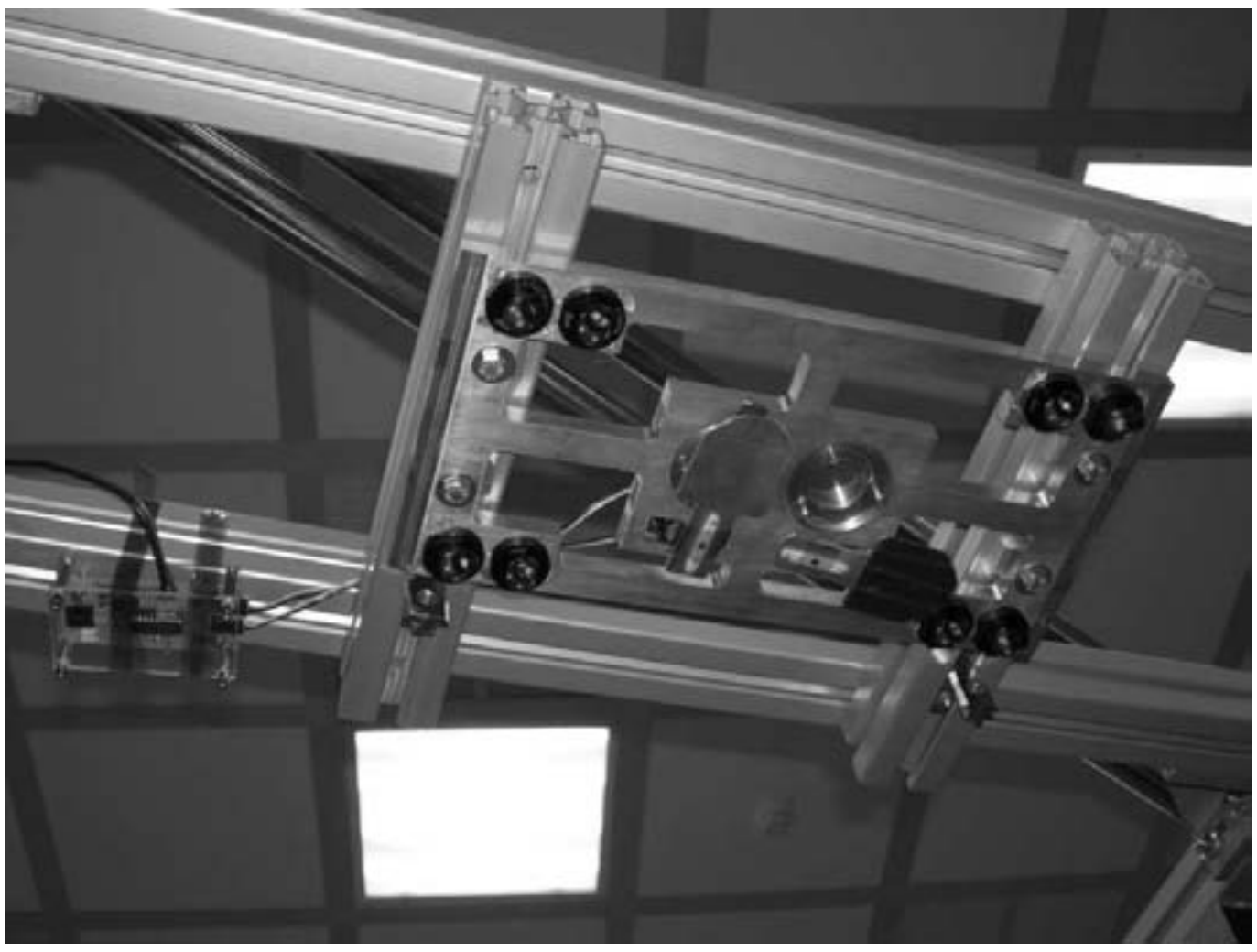

Figure 1: Picture of the brass plate equipped with the $^{109}$Cd radioactive source and the shutters, the left one closed and the right one open.

## 4 Scientific perspectives

About 15–20 GRB per year are expected in the SuperAGILE FOV and from 5 to 10 events in the GRID FOV, of which 1–3 per year are simultaneous. The Mini-calorimeter can independently detect 40 – 50 GRB per year. Five GRB have been detected by EGRET in the same energy band as the GRID (see [11] for details) but the lack of fine localization and redshift measurement did not allow to know their position and fluence. Similar events can be simultaneously detected by the GRID and localized by SuperAGILE, thus providing information about their position, possibly leading to the identification of the counterpart and distance thereof.

A detailed analysis of the AGILE sensitivity to GRB may be found in [12]. Due to the different energy band of SuperAGILE and Mini-calorimeter, the first instrument is more sensitive to soft GRB and X-Ray Flashes while the latter to the hard events.

## 5 Conclusions

The SuperAGILE wide field monitor on-board the AGILE satellite-borne mission include a GRB detection and localization system based on the instrument counting rate. The localization is performed within the SuperAGILE $68^\circ \times 68^\circ$ two 1-D FOV on timescales from 64 ms to 8192 ms and the reconstructed coordinates are provided to the Malindi ground station in some minutes by using the ORBCOMM satellite constellation. GRB can also be detected on timescales of sub-ms to 16 ms but without imaging capabilities. In case a trigger is detected, the Mini-calorimeter photon-by-photon data can be transmitted in telemetry to add the 300 keV – 200 MeV energy band.

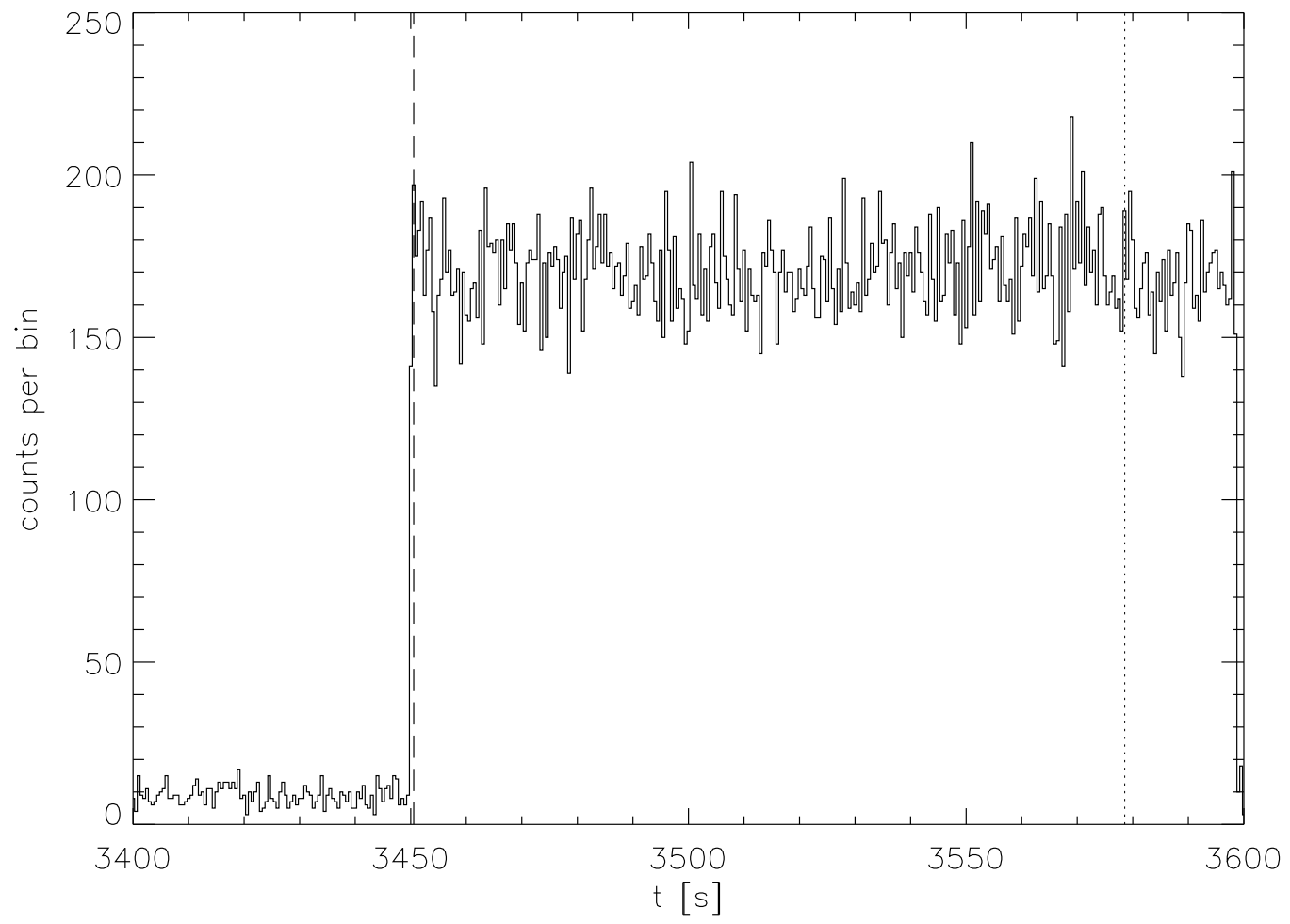

Figure 2: Example of a time series during the experimental tests: at the beginning the source is occulted, at 3450 s it is unocculted, the vertical dashed line is the start time and the vertical dotted line is the stop time, both from the on-board GRB trigger.

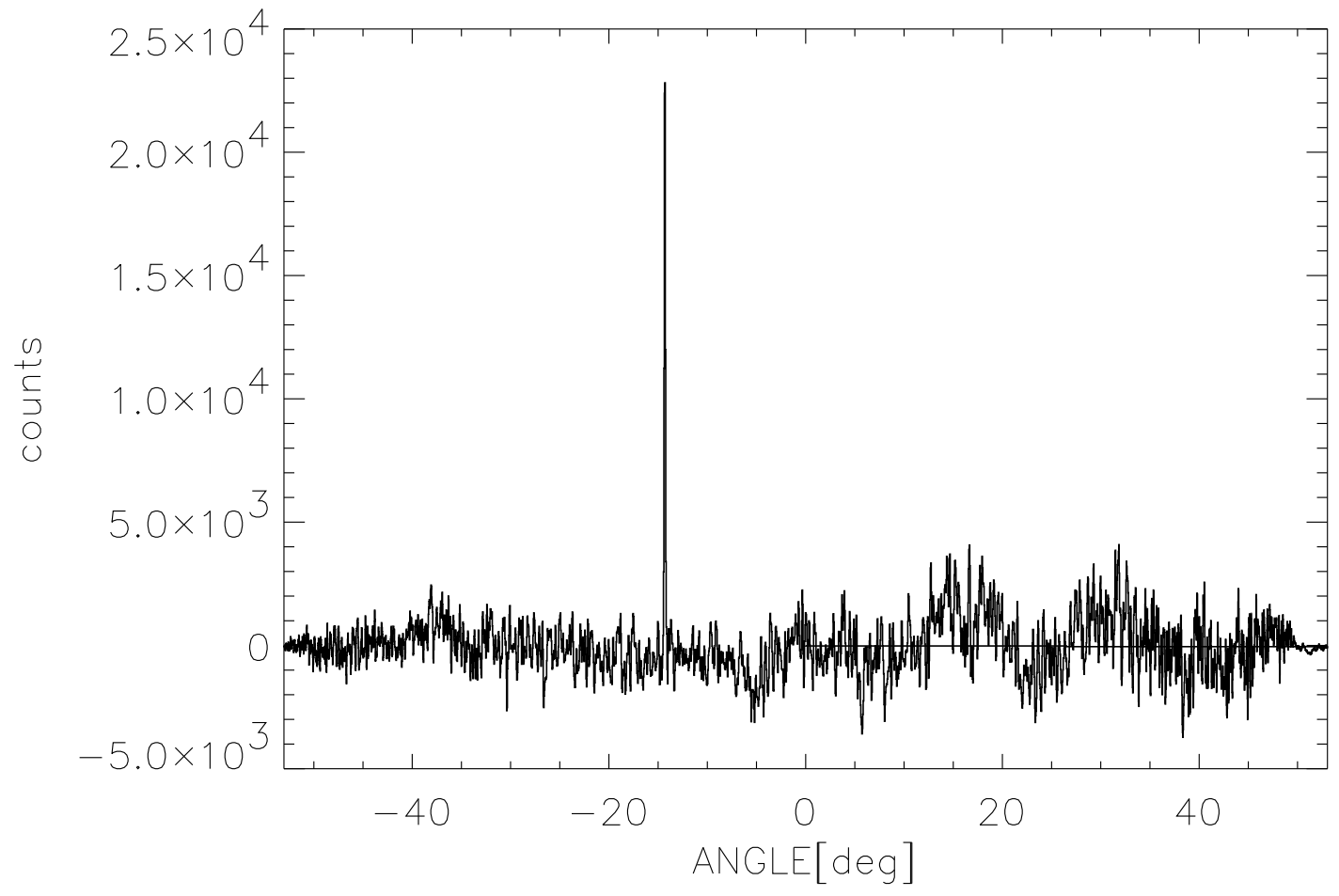

Figure 3: Example of the 1-D decoded image (x axis) from the same dataset as in fig. 2

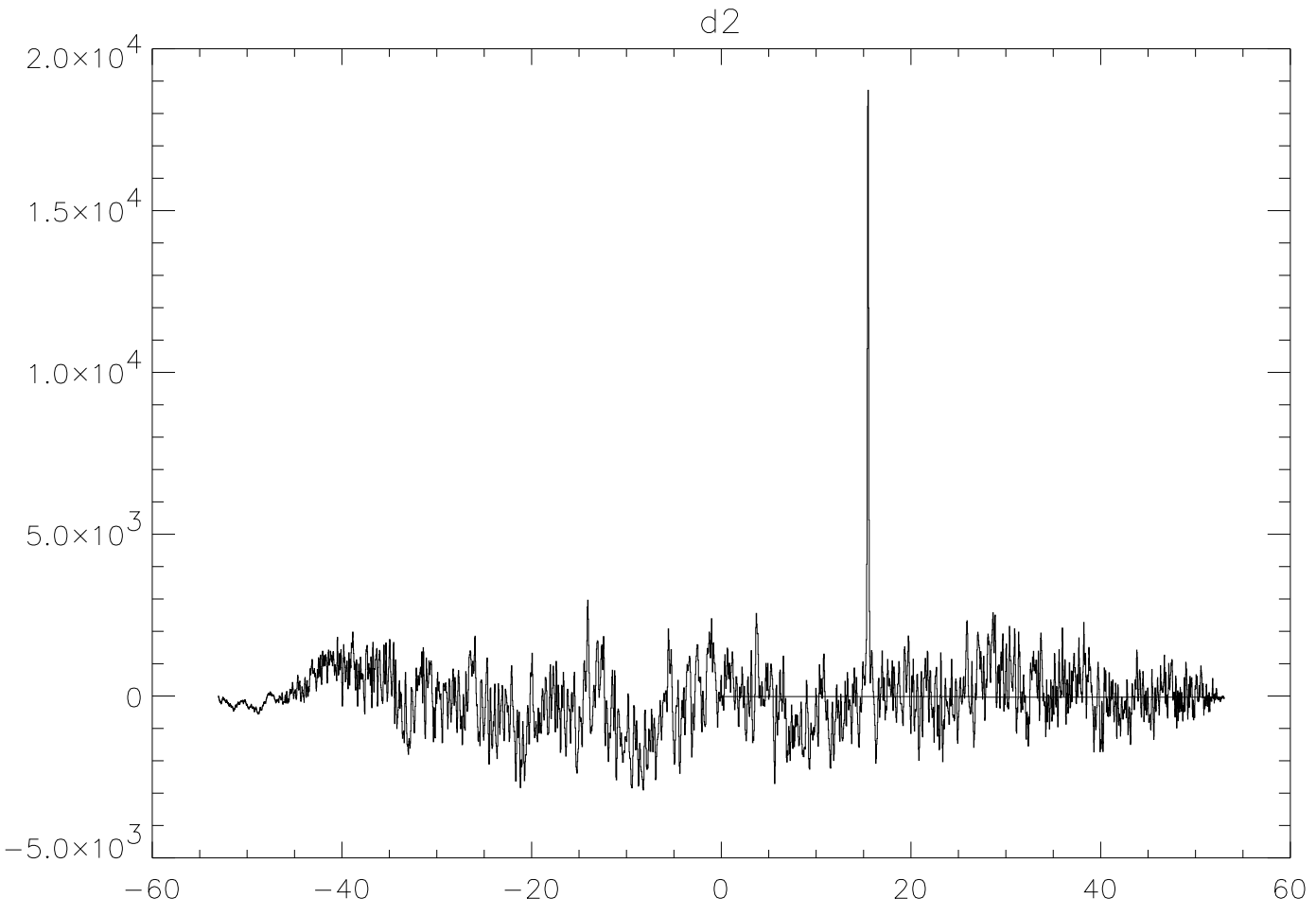

Figure 4: Example of the 1-D decoded image (z axis) from the same dataset as in fig. 2

# THE MINI-CALORIMETER OF THE AGILE SATELLITE

Martino Marisaldi [a], Claudio Labanti [a], Fabio Fuschino [a], Marcello Galli [b], Andrea Argan [c], Andrea Bulgarelli [a], Guido Di Cocco [a], Fulvio Gianotti [a], Marco Tavani [c], Massimo Trifoglio [a]

[a] *INAF, Istituto di Astrofisica Spaziale e Fisica cosmica, Via Gobetti 101, 40129 Bologna, Italy*

[b] *ENEA, Via Don Fiammelli 2, 40129 Bologna, Italy*

[c] *INAF, Istituto di Astrofisica Spaziale e Fisica cosmica, Via Fosso del Cavaliere 100, 00133 Roma, Italy*

## Abstract

AGILE is a small scientific mission of the Italian Space Agency devoted to X and gamma-ray astrophysics, succesfully launched on $23^{rd}$ April 2007. Among the AGILE scientific instruments, the Minicalorimeter is a gamma-ray detector based on scintillator detectors with solid state readout sensitive in the range 0.3-200 MeV. The Minicalorimeter can work both as a slave of the Silicon Tracker, to form the Gamma-Ray Imaging Detector, and as an independent all-sky monitor for gamma-ray bursts and transients detection. With just 5 W power consumption available, an energy threshold as low as 300 keV, an energy resolution of about 14% FWHM at 1 MeV, a position resolution of 6 mm for MIPs and a timing accuracy better than 1 $\mu$s has been achieved. In this paper the design and construction of the Minicalorimeter will be described, together with results of the on-ground calibration campaigns.

## 1 Introduction

AGILE is a small scientific mission of the Italian Space Agency devoted to X and gamma-ray astrophysics, succesfully launched on $23^{rd}$ April 2007 from Satish Dawan Space Centre (India). AGILE orbits in an equatorial Low Earth Orbit at 550 km altitude and 2.5° inclination. A review of the AGILE scientific instruments can be found in [1] and [2]. The AGILE payload is composed of a tungsten-silicon tracker (ST) sensitive in the gamma-ray energy range 30 MeV-50 GeV [3], SuperAGILE, an X-ray imager sensitive in the energy range 15-45 keV [4], the Minicalorimeter (MCAL) sensitive in the range 0.3 - 200 MeV, and an Anticoincidence shield. The Silicon Tracker, together with MCAL and the Anticoincidence shield make up the so called Gamma-Ray Imaging Detector (GRID). A Payload Data Handling Unit (PDHU) takes care of all subsystems data acquisition and performs dedicated tasks for transients detection. MCAL is made of 30 CsI(Tl) bar-shaped scintillator detectors with photodiode readout at both ends, arranged in two orthogonal layers. Each bar acts as an independent hodoscopic detector, so that energy and position of interaction can be derived from a proper composition of the signals readout at the bar's ends. A technical description of MCAL can be found in [5].

## 2 MCAL design and construction

MCAL is composed of a detection plane made of 30 CsI(Tl) scintillator detectors with the shape of a bar each one 15x23x375 mm in size, arranged in two orthogonal layers, for a total thickness of 1.5 radiation lengths. In a bar the readout of the scintillation light is accomplished by two custom PIN Photodiodes (PD) coupled one at each small side of the bar. For each bar the PDs signals are collected by means of low noise charge preamplifiers, and then conditioned in the Front End Electronics, (FEE). The circuits have been optimized for best noise performance, fast response, combined with low power consumption and a wide dynamic range. For each bar, the energy and the position of an interacting gamma-ray or ionizing particle can be evaluated combining the signals of the two PDs.

The detection plane is hosted in the upper part of MCAL main frame; the preamplifiers are arranged in four boxes on each side of the detection plane and at its same level. Below the detection plane is placed the FEE electronics board that has the same area of the whole detection plane. The overall mechanical envelope of MCAL constitutes the lower part of the whole AGILE payload.

### 2.1 The bar detectors

The active core of AGILE MCAL is made of the CsI(Tl) scintillating bars with two PIN PD readout. Each bar is wrapped with a reflective coating and ar-ranged inside a carbon fiber structure that provides rigidity and modularity to the detector. A detector bar is characterized by the charge per unit energy produced at PD level by a photon interacting in the crystal at a defined distance from the PD (signal output, expressed in $e^-$/keV), and by the relation governing the signal output as a function of the distance of interaction from the PD (light attenuation law). Each bar has been characterized independently with a collimated $^{22}$Na source at different positions before integration into the MCAL flight model. It has been found that an exponential model for the light attenuation law is quite a good representation along all the bar extension but the first few cm near the PDs, where border effects are responsible for a deviation from the above law up to 5-7%. The PDs signal $U(x)$, expressed in $e^-$, as a function of the distance of interaction $x$ from the PD will be described by a relation like $U(x) = EU_0e^{-\alpha x}$, where $E$ is the energy released in the bar, $U_0$ is the extrapolated signal output for interactions at the PD edge and $\alpha$ is defined as the light attenuation coefficient. On the whole set of 32 bars (30 flight detectors plus 2 spares), the average value for $U_0$ was found to be 21 $e^-$/keV with a standard deviation of 1 $e^-$/keV; while the average value for $\alpha$ was found to be 0.028 cm$^{-1}$ with a standard deviation of 0.002 cm$^{-1}$.

### 2.2 Operative modes

MCAL works in two possible operative modes:

- GRID mode: when a trigger is issued by the ST all the signals from the MCAL detectors are collected. The scientific objective of this operative mode is to contribute to the GRID event energy reconstruction and provide information for background events rejection.

- BURST mode: each bar behaves as an independent self triggering detector and generates a continuous stream of information of gamma events in the energy range 300 keV-100 MeV. In the data handling system these data are used to detect impulsive variations in count rates. The scientific objective of this operative mode is to provide high energy spectral coverage of Gamma-Ray Bursts (GRBs) and other intense transients.

Both operative modes can be active at the same time. Due to telemetry limitations BURST mode data are not sent on ground on a photon-by-photon basis unless a trigger for a transient is issued by a dedicated Burst Search logic. However BURST data are used to build a broad band energy spectrum (Scientific RateMeters, SRM) recorded and stored in telemetry every second. Scientific ratemeters are organized in 11 bands for each of the two MCAL detection planes and are expected to provide information on the high energy gamma-ray background in space and its modulation through orbital phases.

### 2.3 Front-end electronics

MCAL Front End Electronics (FEE) is physically divided in two different parts: the PD charge pre-amplifiers are mounted in four boxes very close to the PDs, while all the rest of the electronics is placed on a single board, 415x415 $mm^2$ in size which contains about 5000 components and that is placed below the detectors. The analogue chain of each PD is common for GRID and BURST operations up to the charge pre-amplifier and the shaping amplifier stage with shaping time of about 3 $\mu$s to cope with CsI(Tl) scintillation light decay time. The noise figure of the whole branch is contained in about 1000 $e^-$ rms. The amplified signals are then sent to two independent parallel processing branches, for GRID and BURST operative modes respectively.

## 3 On-ground calibration and scientific performance

MCAL has been tested both as a stand-alone system and as a subsystem of the fully integrated AGILE payload.

Since the bars light attenuation law can be considered exponential to a good extent, any event's position $x$ and energy $E$ can be calculated according to equations 1 and 2

$$x = A + B \ln \frac{U_B}{U_A} \tag{1}$$

$$E = C\sqrt{U_A U_B} \tag{2}$$

where $A$, $B$ and $C$ are constants based on the bar's parameters, $U_A$ and $U_B$ are the signals readout at the bar's PDs, expressed in ADC channels, corrected for the electronic chains offset. The parameters required for energy and position reconstruction have been obtained from stand-alone calibration tests performed on MCAL before integration with the other payload subsystems. Moreover these tests allowed to evaluate the instrument overall functionalities and performance.

The BURST mode has been tested using a collimated $^{22}$Na radioactive source. The source and the collimator were placed on a programmable positioning system able to automatically move the source within a defined set of coordinates. With this setup each MCAL bar was tested in a number of points ranging between 10 and 13, thus allowing reconstruction of the signal output along the bar. For each bar the signal output curve at an energy of 1.275 MeV as a function of the position of interaction was built. This curve was fitted with an exponential model to extract the current bar parameters (signal output and attenuation coefficient). This set of values was used as preliminar calibration parameters for position and energy reconstruction in data analysis software. Errors in these calibration parameters give rise to systematic errors, while the main contribution to energy and position resolution is given by the electronic noise.

After stand-alone calibrations, MCAL was integrated with all the other payload instruments and with the PDHU. Since this integration phase, data have been acquired directly from the PDHU in the form of telemetry packets. With the parameters obtained from stand-alone calibration tests the performance described in the following subsections have been obtained.

### 3.1 BURST mode performance

Source position in calibration datasets is properly reconstructed with a 1.8 cm standard deviation, at 1.275 MeV. When a bar is exposed to an uncollimated $^{22}$Na source the 1.275 MeV peak is properly reconstructed with a 0.076 MeV sigma, thus allowing a 14% FWHM energy resolution at 1.275 MeV. Figure 1 shows a background count spectrum obtained on ground during satellite integration activity. This spectrum was obtained considering MCAL as a single detector, i.e. summing together the contributions from different bars triggering at the same time. The low energy part of the spectrum is dominated by the $^{40}$K 1.460 MeV peak and by other features due to natural radioactivity, up to a few MeV. At higher energies the spectrum is dominated by two broad peaks at about 10 and 20 MeV. These peaks are due to cosmic muons crossing one or two MCAL planes, respectively. 10 MeV is just the expected energy loss by a minimum ionizing particle (MIP) in 1.5 cm of CsI.

### 3.2 GRID mode performance

Despite the bars parameters are the same for GRID and BURST modes, to evaluate the detectors gain in GRID mode an external trigger is required. Thus GRID mode energy calibration was performed after payload integration using cosmic ray muons as probes. Figure 2 shows the energy spectra due to muons for the two MCAL detection planes in GRID mode. Fitting the Landau peaks for each bar and comparison with the expected values for MIPs provide a good gain calibration for MCAL detectors in GRID mode.

Muon tracks can be easily selected in a GRID dataset requiring straight trajectories with low scattering angles and a track extended over most of the planes of the ST. Muons trajectories can then be extrapolated to MCAL to obtain the expected interaction position. After gain calibration, from the MCAL data the reconstructed position of interaction was calculated and compared to the extrapolated value. A good agreement was obtained, the distribution of the deviation with respect to the expected value having a 0.6 cm standard deviation, as shown in figure 3. After fine tuning of calibration parameters an even better position resolution is expected. This procedure will be used also during flight operations for MCAL gain calibration in GRID mode.

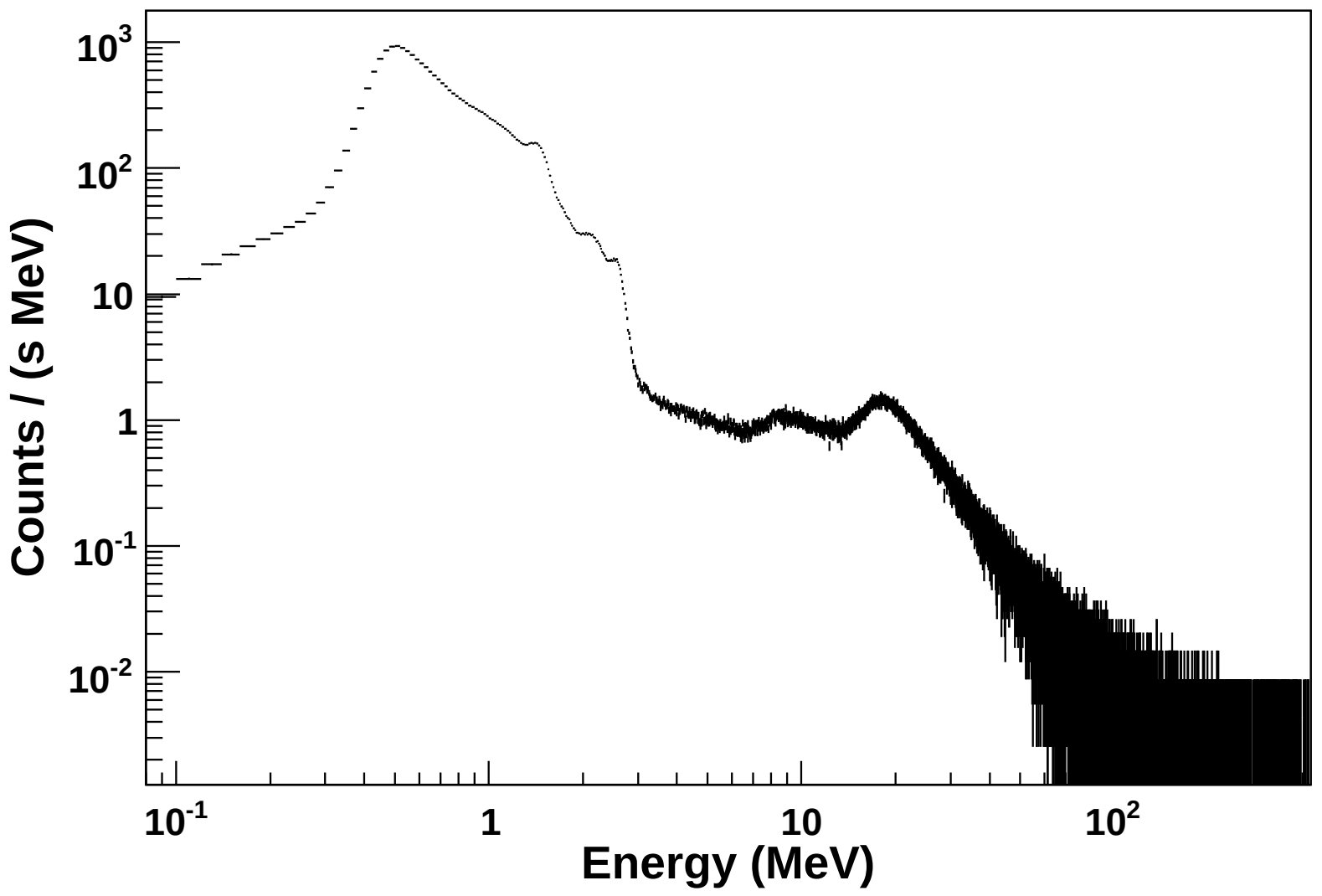

Figure 1: MCAL background count spectrum obtained in BURST mode during satellite integration. Spectral features due to radioactive isotopes in the environment and to cosmic muons can be observed.

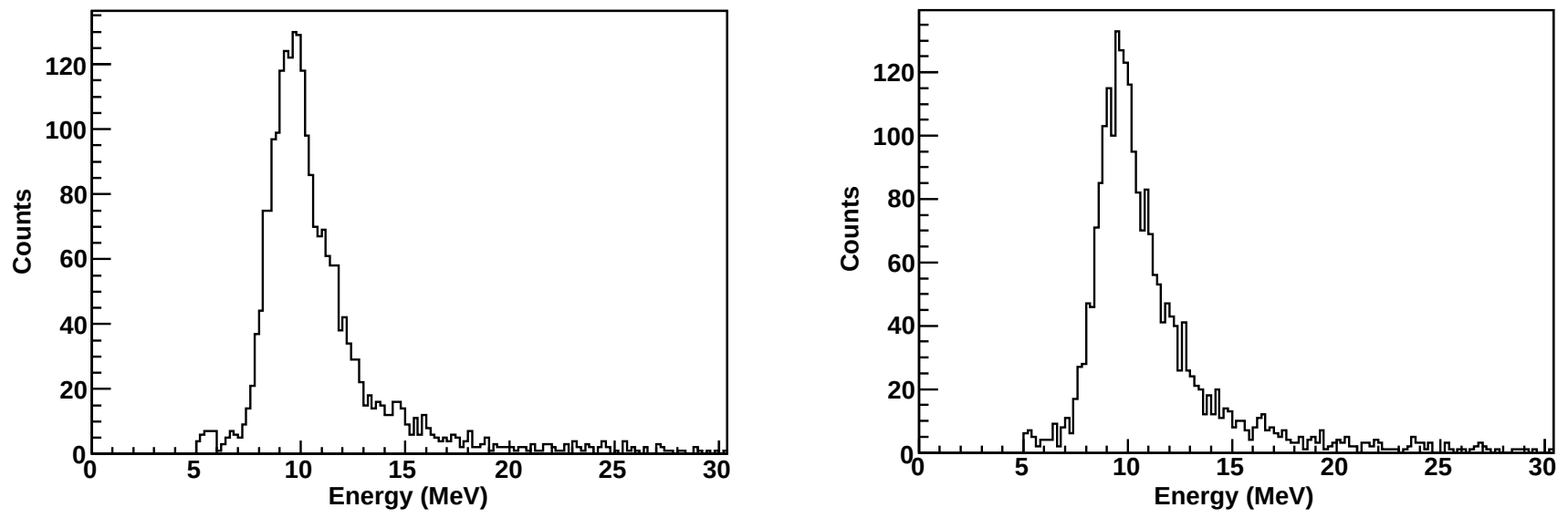

Figure 2: Reconstructed energy spectra for MCAL in GRID mode. Only muon events with incident angle less than 10 ° have been selected. Left panel: upper detection plane, close to the ST. Right panel: bottom detection plane, close to the electronics box. Measurements taken on ground during satellite integration.

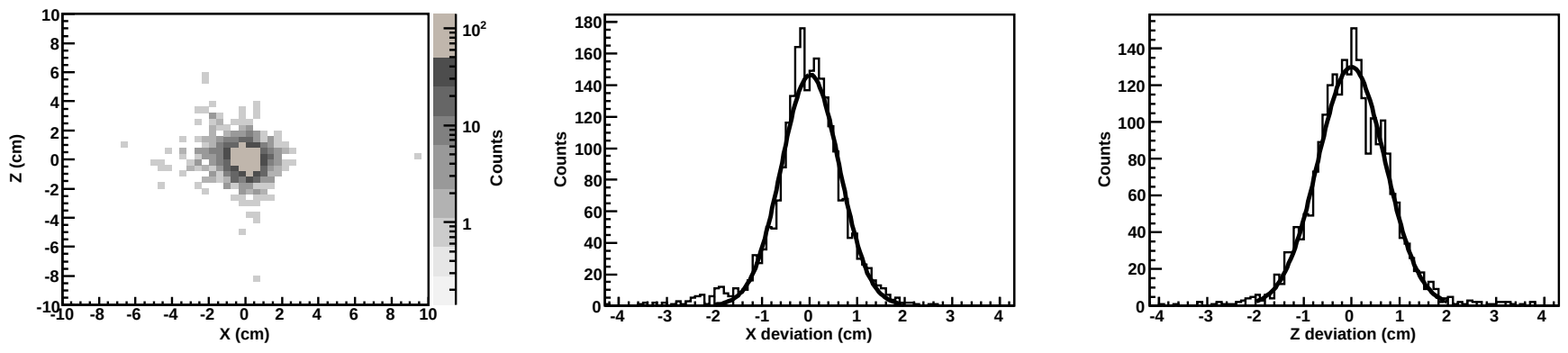

Figure 3: Deviation from extrapolated and reconstructed positions in GRID mode. Left panel: 2D deviation distribution. Central panel: deviation distribution along X direction. Right panel: deviation distribution along Z direction. Gaussian fits exhibit a 0.6 cm standard deviation in both directions. Measurements taken on ground during satellite integration.

## 4 Burst Search logic tests

Data produced in the BURST branch of the FEE are stored in a derandomizing FIFO and then sent to the PDHU where they are continuously processed by a Burst Search (BS) software algorithm for the detection of fast transients. The first task of the BS software is the formation of background ratemeters (RMs) evaluated integrating events on different time windows called Search Integration Times (SITs). To account for the different time profiles of a burst, the SITs range from sub-millisecond to several seconds. Furthermore, for each SIT several RMs are generated depending on the detector, the events energy and position of interaction. For MCAL 9 RMs are evaluated covering 3 ranges of energy respectively, from 0.35 to about 0.7 MeV, from 0.7 MeV to about 1.4 MeV, and above 1.4 MeV; the limits of the ranges can be varied via Telecommands. Events in the first two energy ranges contribute to generate different RMs depending on the place of interaction on MCAL; in this case MCAL is devided in four zones. The high energy events contribute to an unique RM. Even if the BS logic for the MCAL detector has been thoroughly investigated at software level, it was also tested experimentally with a dedicated hardware setup. A 10 cm thick lead collimator has been placed in front of a MCAL side and a radioactive source is moved across the collimator opening by means of a stepper motor. The motor speed can be adjusted to produce "bursts" of different duration between 32 ms and 1.9 s. With this setup it was possible to verify all the steps of the BS algorithm, from trigger generation to data transmission. A detailed description of the MCAL BS logic as well as of the validation tests carried out on ground can be found in [6]. Also SuperAGILE contributes to the BS in the energy range 15-45 keV, providing additional imaging and localization capabilities if the burst is detected inside the instrument's field of view [7].

## 5 Conclusions

The Mini-Calorimeter of the AGILE satellite, despite its low weight and power budget, is a powerful and versatile gamma-ray detector in the energy range 300 keV-200 MeV. The instrument can work both as a slave to the Silicon Tracker and as an independent detector for gamma-ray transient search. AGILE was launched on $23^{rd}$ April 2007 and at the time of writing (July 2007) is in its commissioning phase. MCAL was switched on in orbit on $2^{nd}$ May 2007, and since then it exhibits nominal behavior.

# GAMMA RAY ASTRONOMY WITH ARGO-YBJ: FIRST OBSERVATIONS

SILVIA VERNETTO [a,b]
FOR THE ARGO-YBJ COLLABORATION

[a] *IFSI Torino, INAF, Italy*

[b] *INFN, Sezione di Torino, Italy*

**Abstract**

The ARGO-YBJ experiment is the first EAS detector combining a very high mountain altitude (4300 m a.s.l.) to a "full coverage" detection surface. These features allow ARGO-YBJ to work in the typical energy range of Cherenkov telescopes, with an energy threshold of a few hundreds GeV. The low threshold and the large field of view (∼2 sr) make ARGO-YBJ suitable to monitor the gamma ray sky, searching for unknown sources and unexpected events, like Active Galactic Nuclei flaring episodes or high energy Gamma Ray Bursts.

In this paper we present the preliminary results on Gamma Ray Astronomy obtained with the events collected in the first months of data taking, in particular the detection of gamma rays from the Crab Nebula, the observation of a Markarian 421 outburst in July-August 2006, and finally a search for Gamma Ray Bursts emission in the GeV energy range using the scaler mode technique.

## 1 The detector

ARGO-YBJ is an air shower detector optimized to work with an energy threshold of a few hundreds GeV. It is located at the Yangbajing Cosmic Rays Laboratory (Tibet, China) at an altitude of 4300 m above the sea level.

It consists of a $74 \times 78$ m$^2$ "carpet" realized with a single layer of Resistive Plate Counters (92% of coverage), operated in streamer mode, surrounded by a partially instrumented " sampling ring", for a total active area of 6700 m$^2$ (see Fig. 1). The detector is logically divided into 154 clusters, 130 of them forming the central carpet and 24 the sampling ring. The cluster, consisting of a set of 12 RPCs, is the basic DAQ unit of the detector. Signals from each RPCs are picked-up by 10 electrodes of dimension $56 \times 62$ cm$^2$ (the "pads") which provide the space-time pattern of the shower front with a time resolution of $\sim$ 1 ns. Each pad is segmented into 8 strips ($62 \times 7$ cm$^2$) which count the number of particles hitting the pad itself.

In order to extend the measurable primary energy range, the RPCs are equipped with 2 large size pads ($140 \times 125$ cm$^2$), providing a signal of amplitude proportional to the number of particles.

The detector will be covered by a 0.5 cm thick layer of lead, in order to convert a fraction of the secondary gamma rays and to reduce the time spread of the shower front, increasing the angular resolution.

ARGO-YBJ operates in two independent acquisition modes: the "shower mode" and the "scaler mode".

In the "shower mode" all showers giving a number of fired pads $N_{pad} \geq N_{trig}$ in the central carpet during a time window of 420 ns are recorded. The spatial coordinates and the time of any fired pad are then used to reconstruct the position of the shower core and the arrival direction of the primary. To perform the time calibration of the 18480 pads, a software method had been developed[1].

Fig. 1 reports an example of a shower detected by the central carpet, showing the capability of ARGO-YBJ in providing a detailed view of the shower front.

The current trigger threshold $N_{trig} = 20$ corresponds to a primary gamma energy threshold of a few hundreds GeV, the exact value depending on the source spectrum and on the zenith angle of observation. With this trigger condition, the event rate is $\sim$4 kHz.

In the "scaler mode" the counting rates of each cluster are recorded every 0.5 s for 4 different levels of coincidence inside the cluster: $\geq 1$, $\geq 2$, $\geq 3$, $\geq 4$ pads, with a coincidence time window of 150 ns. The counting rates are respectively $\sim$40 kHz, $\sim$2 kHz, $\sim$300 Hz and $\sim$120 Hz per cluster, for the 4 coincidence levels.

This measurement allows the detection of secondary particles from very low energy showers (E>1 GeV) reaching the ground in a number not sufficient to trigger the detector operating in shower mode. In scaler mode the primary arrival directions are not reconstructed and the data are used to search for transient phenomena as Gamma Ray Bursts or Solar Ground Level Enhancements, and to study cosmic ray modulations due to the solar activity.

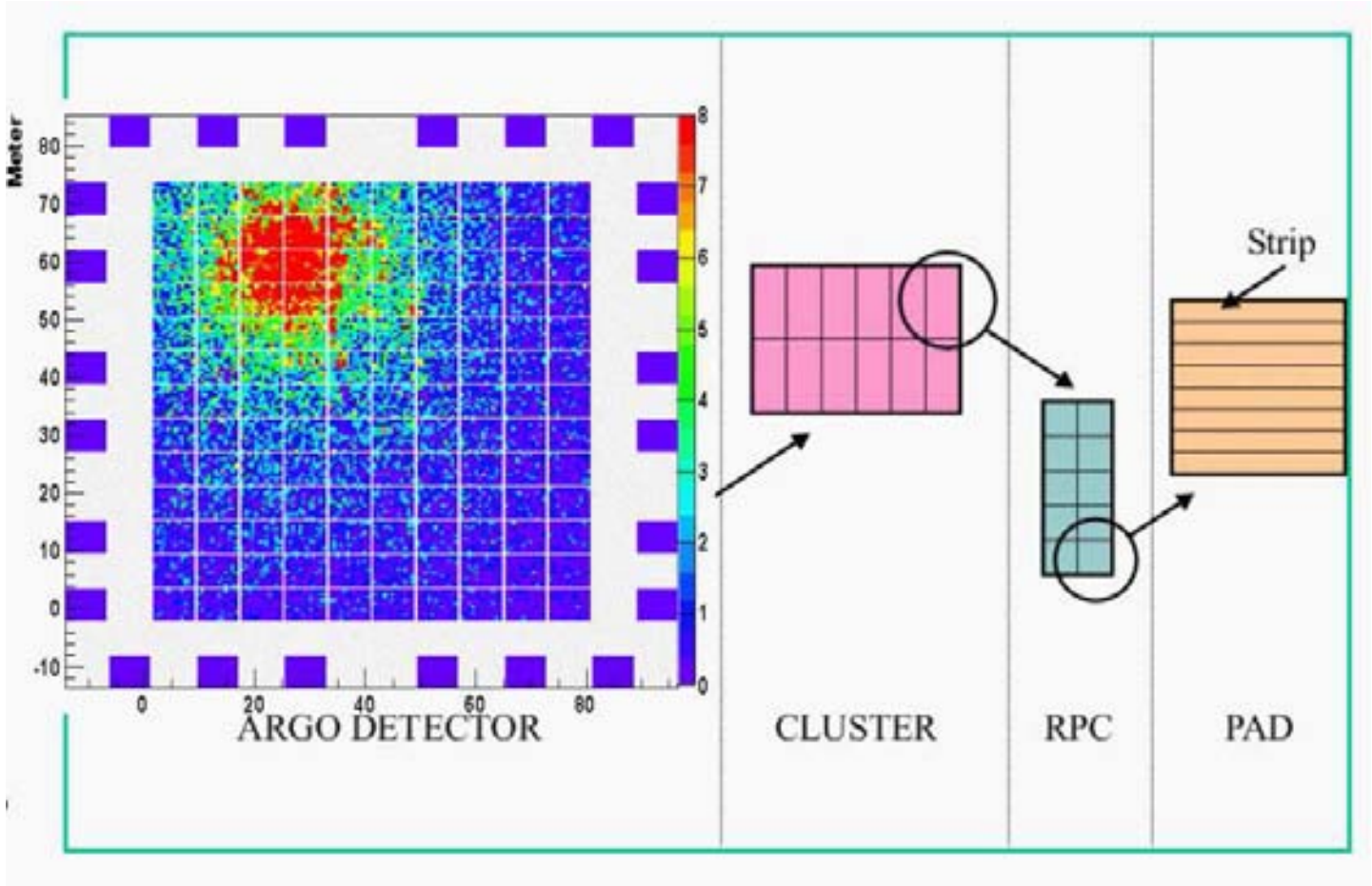

Figure 1: Layout of the ARGO-YBJ detector, with an example of a real shower superimposed to the central carpet. Each point represents a fired pad. Different colours indicate the number of fired strips per pad.

Since July 2006 the whole central carpet is in data taking. Preliminary results obtained with a subset of the data taken from July 2006 to March 2007 in shower mode and from December 2004 to March 2007 in scaler mode (with the detector surface increasing from $\sim$ 700 to $\sim$5600 m$^2$) are presented.

## 2 Detector angular resolution

The angular resolution has a statistical component, due to the fluctuations of the shower development and of the detector response, and a systematic one, arising from possible misalignment of the detector or systematic errors in the shower reconstruction.

Fig. 2 shows the angular resolution of ARGO-YBJ obtained by a Montecarlo simulation. The figure reports the value of $\Psi_{72}$, i.e. the angular difference between the true direction and the reconstructed one, as a function of the number of fired pads $N_{pad}$. $\Psi_{72}$ is the radius of the circular window around the true direction containing 71.5% of the reconstructed events, which in the case of a gaussian point spread function is the radius of the observation window around a point source that maximizes the signal to noise ratio, and it is equal to 1.58 $\sigma$, where $\sigma$ is the angular resolution. $\Psi_{72}$ has been obtained by selecting the events with the reconstructed shower core inside the central carpet, for both primary gamma rays and protons [2]. The accuracy in the direction determination increases with the number of pads. For $N_{pad} \geq 500$, $\Psi_{72}$ is less than 0.5°.

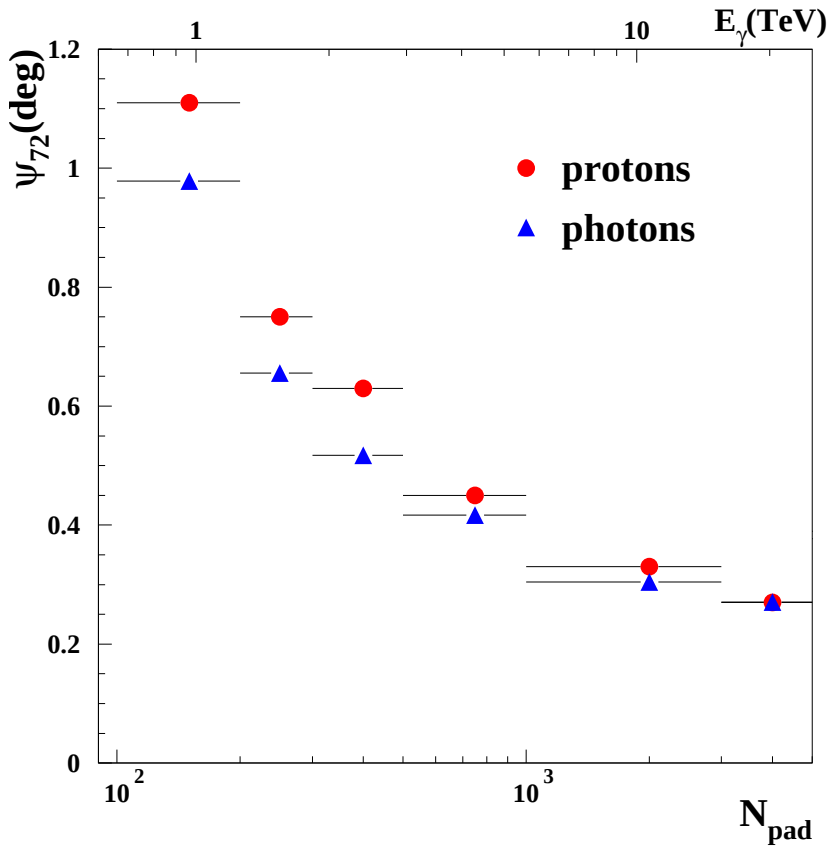

Figure 2: Angular resolution obtained by a Montecarlo simulation, comparing the true primary direction with the reconstructed one, for gamma rays and protons, as a function of $N_{pad}$. The upper scale gives the mean energy of gamma rays correspondent to a given $N_{pad}$, assuming a Crab-like spectrum.

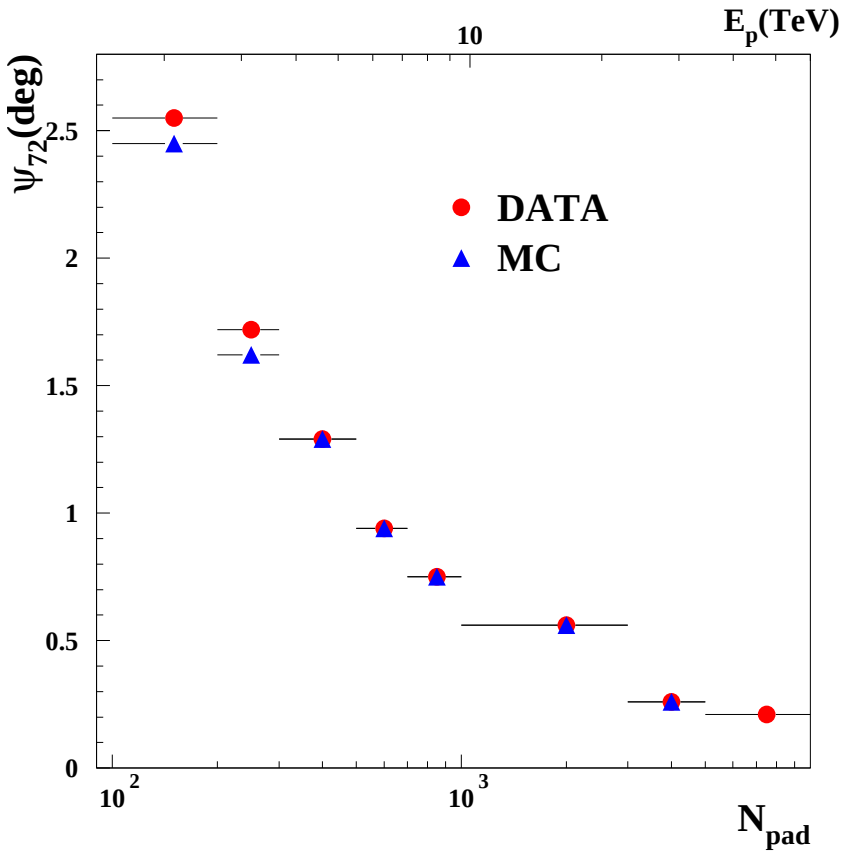

Figure 3: Measured angular difference in the shower direction reconstruction obtained with the "chessboard method", as a function of $N_{pad}$, compared with a Montecarlo simulation. The upper scale gives the mean energy of protons correspondent to a given $N_{pad}$, assuming a proton spectrum $\propto E^{-2.7}$.

A standard method to compare the simulated resolution with the real one, is the so called "even-odd" or "chessboard" method, consisting in splitting the detector in two parts (as the white and black squares of a chessboard) and comparing the arrival directions of showers independently obtained by the two detector subsets. Fig. 3 shows the opening angle $\Psi_{72}$ between the two different reconstructions as a function of the number of fired pads, compared with Montecarlo expectation. The agreement is excellent (note that the "even-odd" angular difference given in Fig. 3 is expected to be a factor ~2 larger than the one obtained with the full detector, given in Fig. 2).

The "even-odd" method allows to check only the statistical component of the angular resolution. To find out possible systematic effects it is common to study the profile and the position of the shadow that the Moon casts on the cosmic ray flux, the strongest "anti-source" of the sky.

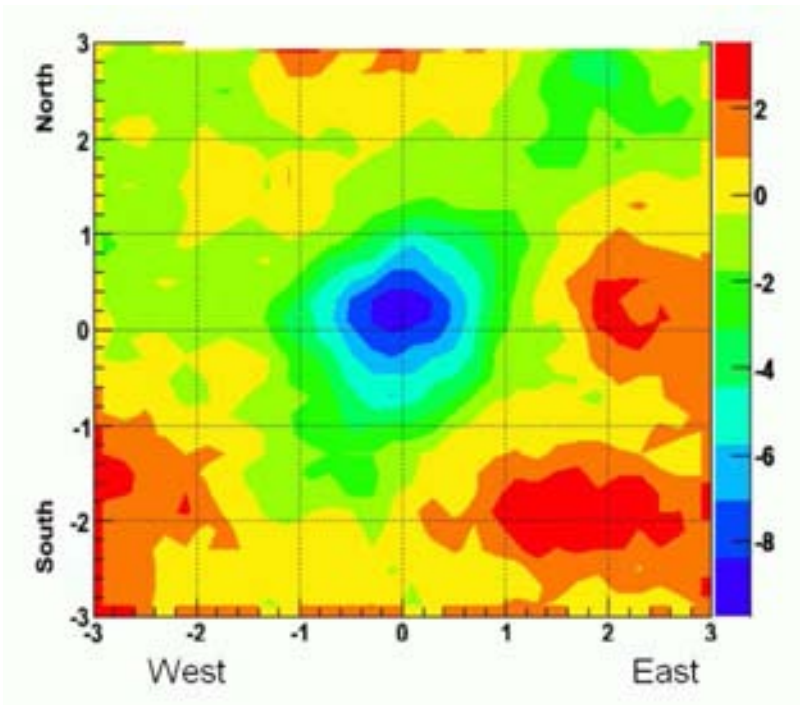

Figure 4: Map of the Moon region. The colour scale shows the significance of the shadow in standard deviations. The axes report the distance in degrees from the Moon position.

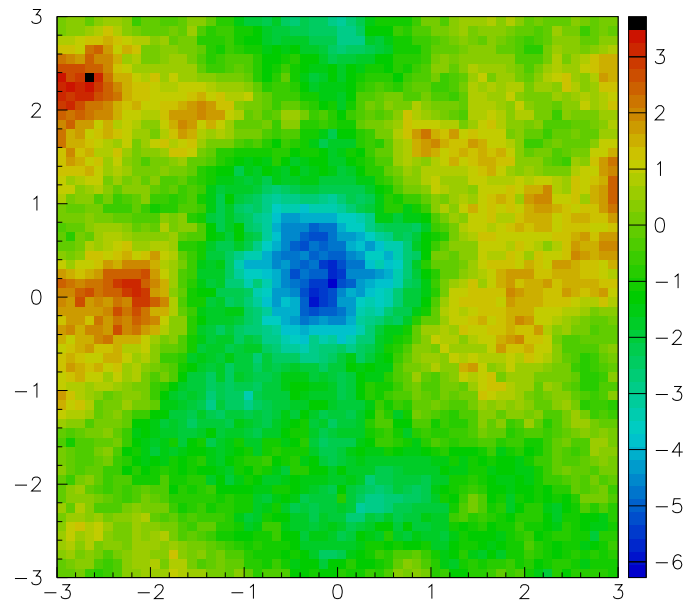

Figure 5: Map of the Sun region. The colour scale shows the significance of the shadow in standard deviations. The axes report the distance in degrees from the Sun position.

## 3 The Moon shadow

The Moon shadow is an important tool for ground-based detectors. The spread and the shape of the shadow at energies where the geomagnetic effect is small, provide a measurement of the angular resolution of the detector, and the position of the shadow allows to find out possible pointing biases.

To minimize the effects of the bending of cosmic rays in the geomagnetic field ($\sim 1.6°$ /E(TeV) westward) we consider only the events with $N_{pad} \geq 500$, corresponding to a median energy of ∼5 TeV. Fig. 4 shows the significance map of the Moon shadow, obtained during 558 hours of observation, selecting the events with zenith angle $< 50°$. A deficit of ∼10 standard deviations is visible, shifted by 0.04° toward the West and 0.14° toward the North with respect to the nominal Moon position [3].

The projection of the deficit along the North-South axis, where the magnetic deflection is expected to be negligible, can be fitted by a Gauss distribution with a width in good agreement with the expected angular resolution.

Using a Montecarlo that simulates the path of the cosmic rays in the geomagnetic field, the position of the Moon shadow is expected to be shifted toward the West by $\sim 0.3°$, due to the magnetic deflection. The observed position of the shadow shows a diplacement of ∼0.25° with respect to the expected position. The systematics causing this shift are currently under investigation.

Also the Sun casts a shadow on cosmic rays, but sometimes the effects of its highly variable magnetic field are so strong to hamper the observation of the shadow. However in 2006 the solar activity was at its minimum and the

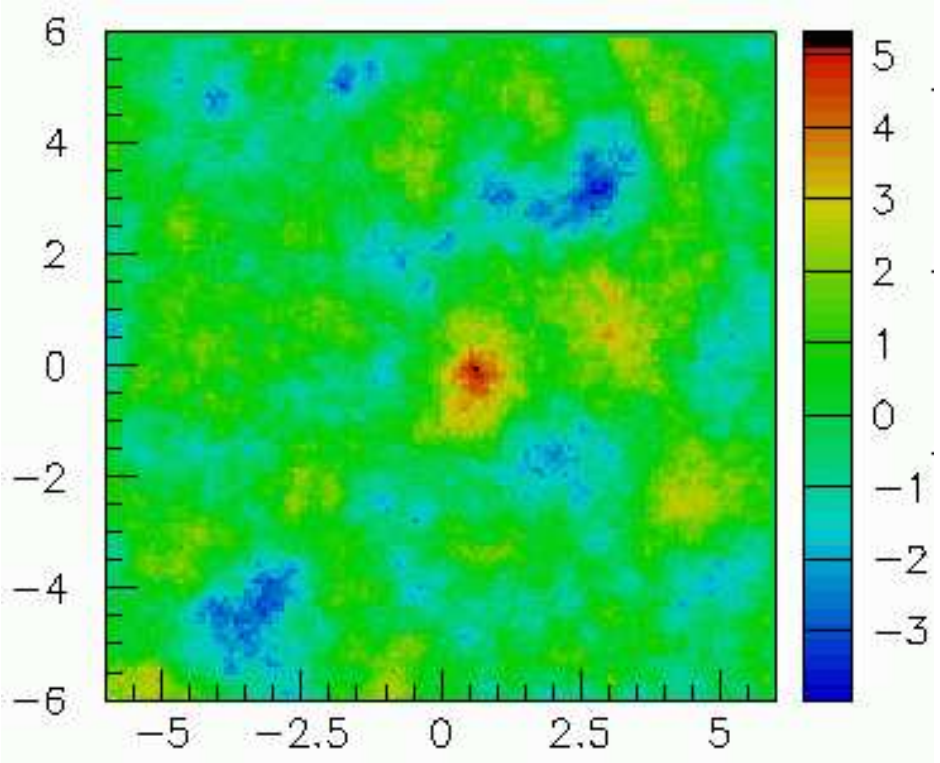

Figure 6: Map of the Crab Nebula region. The colour scale shows the significance of the signal in standard deviations. The axes report the distance in degrees from the source position.

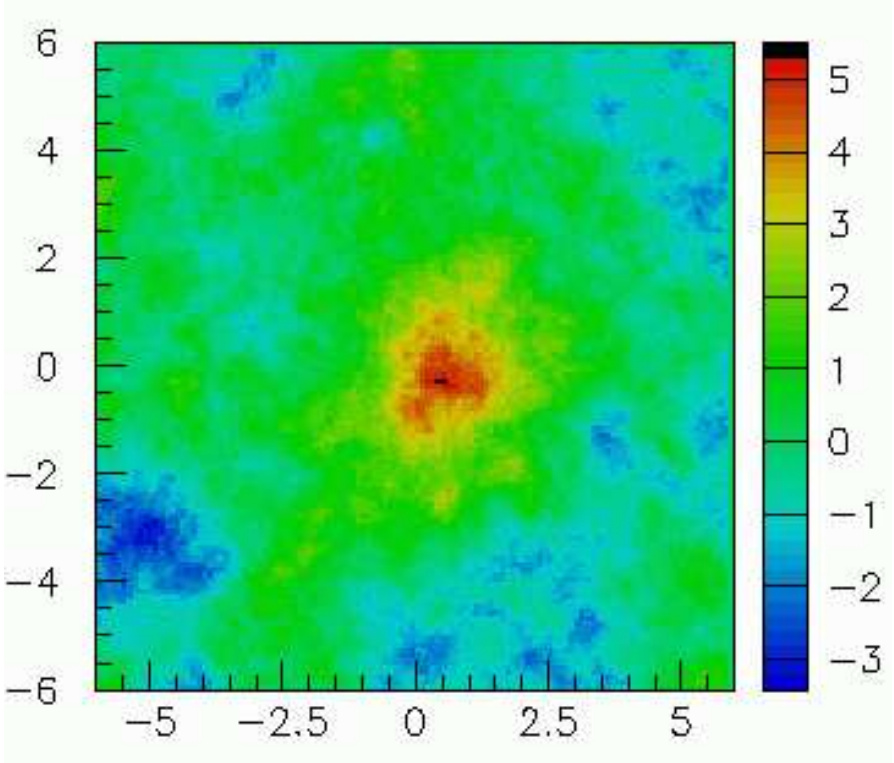

Figure 7: Map of the Mrk421 region during the outburst of July-August 2006. The colour scale shows the significance of the signal in standard deviations. The axes report the distance in degrees from the source position.

Sun shadow appeared with a statistical significance of more than 6 standard deviations, as shown in the map of Fig. 5 obtained in 208 hours of observation.

## 4 Gamma Ray sources

Among the steady TeV gamma ray sources, the Crab Nebula is the most luminous and it is used as a standard candle to check the detectors sensitivity.

Fig. 6 shows the map of the Crab Nebula region obtained by ARGO-YBJ using the events with $N_{pad} \geq 200$ and zenith angle $< 40°$ recorded in 290 hours of observation, equivalent to $\sim$50 transits of the source (one transit lasts 5.8 hours). The Crab is visible with a significance of more than 5 standard deviations.

In July-August 2006 the AGN Markarian 421 underwent an active period, with a rather strong increase of the X-ray flux[4]. As observed in many occasions during the past years, the X-ray flux increases are generally associated to increases in the TeV band, that can reach a flux several times larger than the Crab Nebula one. The 2006 summer outburst was not observable by Cherenkov telescopes, being the source high in the sky during the daytime.

ARGO-YBJ observed Mrk421 for 80 hours in July and August, during a debugging phase of the detector, just at the end of the installation of the central carpet.

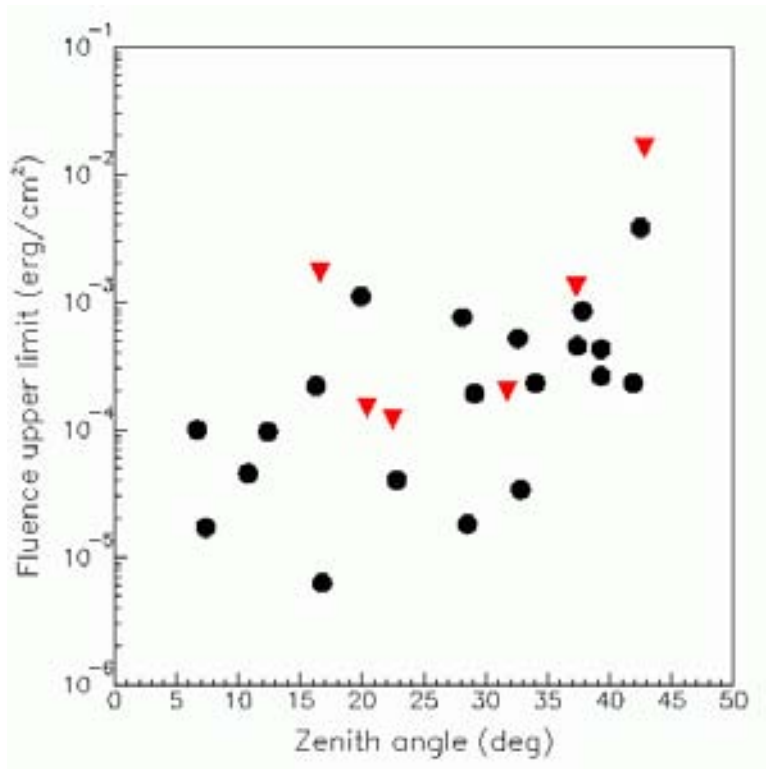

Figure 8: Fluence upper limits for 26 GRBs in the energy range 1-100 GeV, as a function of the zenith angle. The values shown by red triangles have been corrected for the absorption of gamma rays in the extragalactic space.

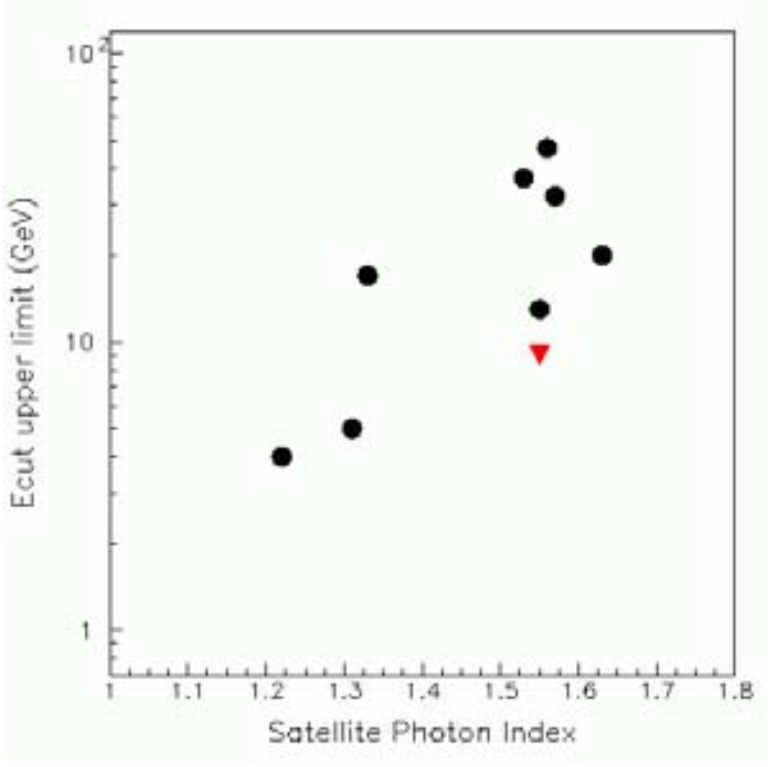

Figure 9: Upper limits to the maximum energy of GRBs spectra. The values shown by red triangles have been corrected for the absorption of gamma rays in the extragalactic space.

Fig. 7 shows the map of the region around the source, obtained with the events with $N_{pad} \geq 60$ and zenith angle $< 40°$. The significance of the signal is more than 5 standard deviations. Considering the low threshold used in this analysis, the detected emission is mostly at energies lower than 1 TeV. A more accurate evaluation of the energies is in progress.

The position of both the Crab Nebula and Mrk421 signals appears slightly shifted eastward with respect to the sources nominal position. The causes of this shift are under study.

## 5 Gamma Ray Bursts

The scaler mode technique offers a unique tool to study GRBs in the GeV energy range, where gamma rays are less affected by the absorption due to pair production in the extragalactic space[5].

The search has been done in coincidence with 26 GRBs detected by satellites (mainly by Swift). No excess has been found neither in coincidence with the low energy detection, nor in an interval of 2 hours around it [6].

Fig. 8 shows the fluence upper limits for the 26 GRBs in the energy range 1-100 GeV during the satellite time detection, as a function of the zenith angle, obtained assuming a power law spectrum with an index $\alpha$=-2.5. When the GRB distance is measured, the spectra are corrected with a factor that takes into account the extragalactic absorption[7].

Fig. 9 shows the upper limits to the maximum energy of the GRB spectra, obtained extrapolating with the same slope the power law spectrum measured by satellites. In some cases the limits are less than 10 GeV.

## 6 Conclusions

The central carpet of ARGO-YBJ has been installed and is taking data since July 2006. The data recorded in the first months of measurement have been analyzed in order to test the performance of the detector in the gamma ray astronomy field. The detection of the Moon and Sun shadows, together with a preliminary observation of a gamma ray flux from the Crab Nebula and Mrk421, show that the detector is properly working, with excellent angular resolution and sensitivity.

Working in "scaler mode" ARGO-YBJ has also performed a search for emission from GRBs in coincidence with 26 events observed by satellites, setting upper limits on the fluence between $6 \times 10^{-6}$ and $2 \times 10^{-2}$ erg cm$^{-2}$ in the 1-100 GeV energy range.

A further increase of the sensitivity to gamma rays is expected with the installation of a converter layer of lead above the detector, and after the implementation of an offline procedure to reject a fraction of the cosmic ray background, based on the different topological pattern of hadronic and electromagnetic showers.

# ARGO-YBJ: PRESENT STATUS AND FIRST INVESTIGATIONS IN COSMIC-RAY ASTROPHYSICS

P. CAMARRI [a,b]
ON BEHALF OF THE ARGO-YBJ COLLABORATION

[a] *Dipartimento di Fisica, Università di Roma "Tor Vergata", via della Ricerca Scientifica, Roma, Italy*

[b] *INFN, Sezione di Roma Tor Vergata, via della Ricerca Scientifica, Roma, Italy*

**Abstract**

The ARGO-YBJ experiment has been taking data for one year with a 6000 $m^2$ full-coverage carpet of resistive plate chambers. The status of the detector and its present performance are discussed, in connection with the first results of the experiment in cosmic-ray astrophysics.

## 1 Introduction

The ARGO-YBJ experiment was designed to study some major topics in astroparticle physics, and in particular:

- search for point-like $\gamma$-ray sources at a lower energy threshold of few hundreds of GeV;
- detection of $\gamma$-ray bursts;
- measurement of the $\bar{p}/p$ ratio at $\sim 1$ TeV energy;
- measurement of the cosmic-ray spectrum and composition close to the "knee".

Ground-based experiments aiming at performing precise measurements on the above-mentioned items must determine the arrival direction of primary cosmic rays at Earth. This is mainly dealt with by two different techniques: planar arrays for reconstructing the fronts of the extensive air showers produced by the incoming primaries, and air-Cherenkov telescopes detecting the Cherenkov light emitted by air showers. The ARGO-YBJ experiment [1] is based on a full-coverage planar array at high altitude, in order to lower the primary energy threshold with respect to standard grid arrays down to a few hundreds of GeV for gamma-initiated showers, and cover a dynamical range from few hundreds of GeV up to more than 1 PeV. The full-coverage technique is presently also being exploited by the Milagro experiment [2], which reconstructs the air-shower fronts using the Cherenkov-light emission in water of the incoming secondaries. The full-coverage technique allows continuous monitoring of a $\sim$ 2-sr angular sector of the sky.

Here the ARGO-YBJ detector layout and performance are described. In addition, the preliminary results obtained by the experiment on a few important issues in cosmic-ray astrophysics are reported.

## 2 The ARGO-YBJ detector

The ARGO-YBJ experiment is located at Yang Ba Jing (China), at 4300 m a.s.l. on the Tibet plateau ($90°31'50''$ E; $30°06'38''$ N). The experiment was designed so that the following requests were fulfilled:

- high altitude, closer to the shower maximum where the shower front is denser, allowing more precise reconstruction of the primary arrival direction;
- full coverage, so that smaller showers can be detected and a lower energy threshold can be reached.

The ARGO-YBJ detector is a full-coverage array of resistive plate chambers (RPCs) operated in streamer mode [3]. The sensitive part of the ARGO-YBJ RPCs is a 2-mm wide gas volume made of two 2-mm thick plastic-laminate resistive plates ($\rho \sim 10^{12}$ $\Omega$·cm). The RPCs operate in streamer mode with a 3-component gas mixture ($C_2H_2F_4$/Ar/i$C_4H_{10}$=75/15/10). The outer faces of the plates are painted with a thin graphite layer, so that the electric field across the gap is obtained by applying a 7.2 kV voltage and capacitive read-out metal strips ($62 \times 7$ cm$^2$ each) pick up the detector signal across the electrodes. In addition, two metal big pads pick up the analog signal on the opposite side of the gas volume. These additional pads are needed to extend the dynamical range of the detector up to few PeV, since the digital information coming from the read-out strips is saturated at energies $\gtrsim$ 100 TeV. A sketch of the transverse section of an ARGO-YBJ RPC is shown in Fig. 1. Twenty-four additional outer clusters form a "guard ring" allowing better reconstruction of the shower core for events not entirely contained in the central carpet; these clusters will be activated before the end of 2007.

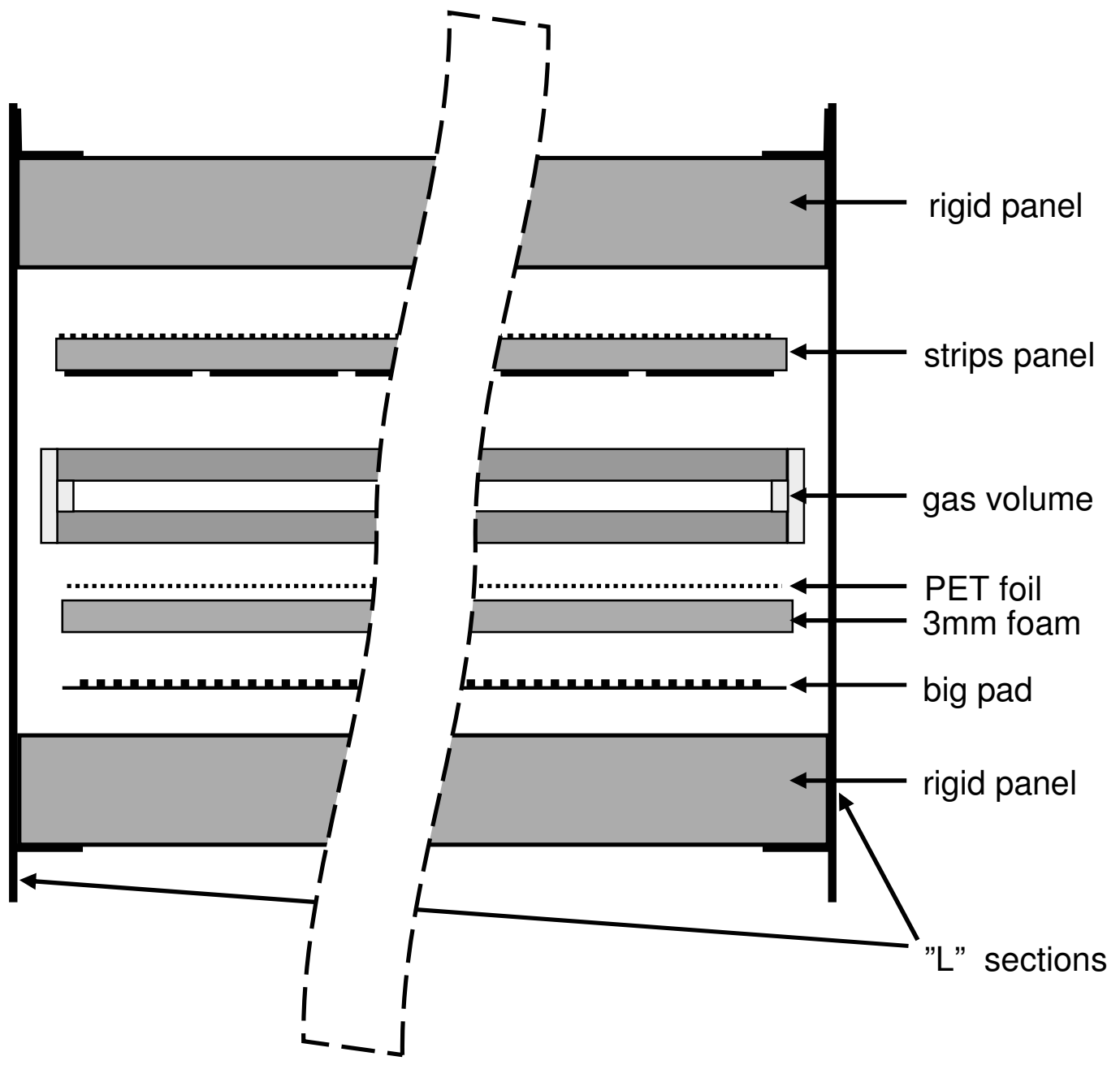

Figure 1: Scheme of the transverse section of an ARGO-YBJ resistive plate chamber.

The detector layout is shown in Fig. 2, where the details of the detector structure are evidenced. Each RPC has a surface of $1.26 \times 2.85$ m$^2$. The RPCs in the central carpet ($78 \times 74$ m$^2$) are arranged in a full-coverage array. A group of 12 close chambers is called a *cluster* ($7.64 \times 5.72$ m$^2$), and the logical acquisition unit is a group of 8 adjacent read-out strips in a chamber, called a *pad* ($0.62 \times 0.56$ m$^2$). The 130 clusters of the central full-coverage array have been taking data since July 2006.

The front-end electronics of the ARGO-YBJ RPCs [4] is based on a full-custom GaAs circuit. In one single die it includes 8 channels, each one com-posed of a 3-stage voltage amplifier, a variable-threshold comparator and a digital AC-coupled ECL driver. All the front-end boards were carefully tested before being inserted into the chambers. Fig. 3 shows: (a) a scheme of the front-end circuit for one channel; (b) the rise-time distribution for all the tested channels (the mean value is about 1 ns); (c) the power-absorption distribution per channel (the mean value is about 32 mW). So, the ARGO-YBJ RPCs provide single-pad time information with a 1-ns time resolution and a total dissipated power less than 5 kW on a $\sim 10^4$ m$^2$ surface.

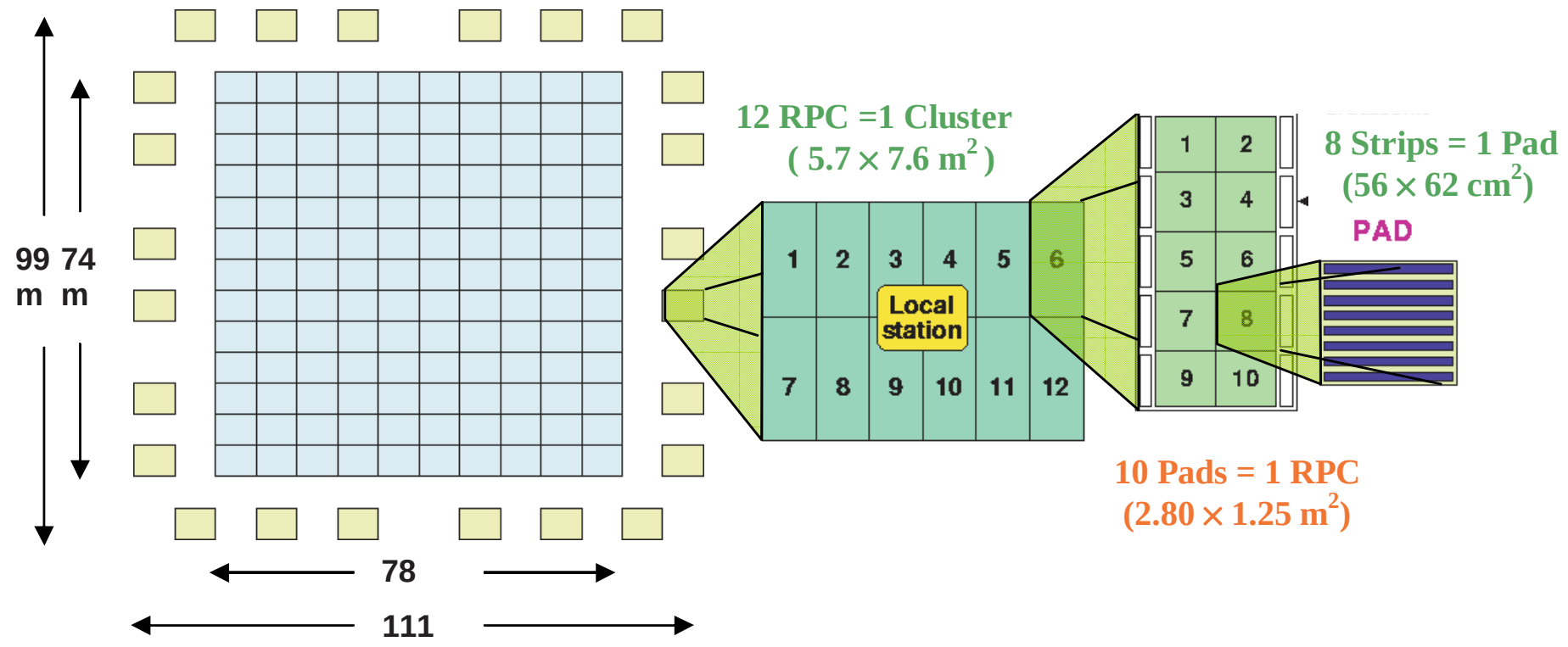

Figure 2: Scheme of the ARGO-YBJ detector layout. The details of the detector geometry and structure are also shown.

The ARGO-YBJ acquisition hardware and the trigger algorithm have already been described in other papers [5].

An example of a partially-contained shower detected in the ARGO-YBJ central carpet is shown in Fig. 4.

The monitoring data [6] show a good distribution for the current absorbed by the chambers, as shown in Fig. 5, with a mean value of about 3 $\mu$A. A more detailed study shows a good linear correlation between the RPC absorbed current and the local temperature if a suitable time delay is accounted for, as shown in Fig. 6. This kind of studies provides better understanding of the detector operation in the particular environmental conditions of the experimental site, and the increasing expertise coming from this will be used to keep the ARGO-YBJ detector operating in a stable and reliable way in the forthcoming years.

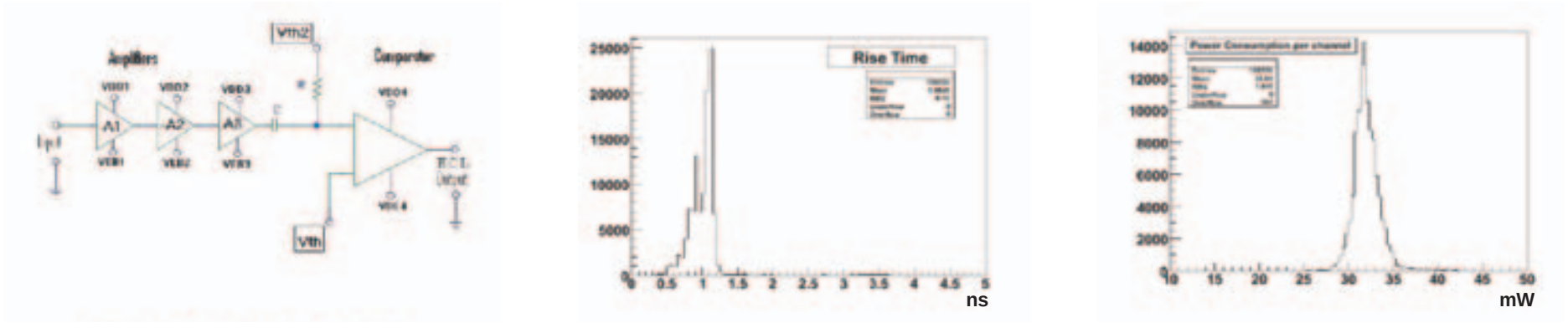

a. b. c.

Figure 3: (a) Scheme of the ARGO-YBJ RPC front-end circuit. (b) Rise-time distribution for all the tested front-end channels. (c) Power-absorption distribution per channel.

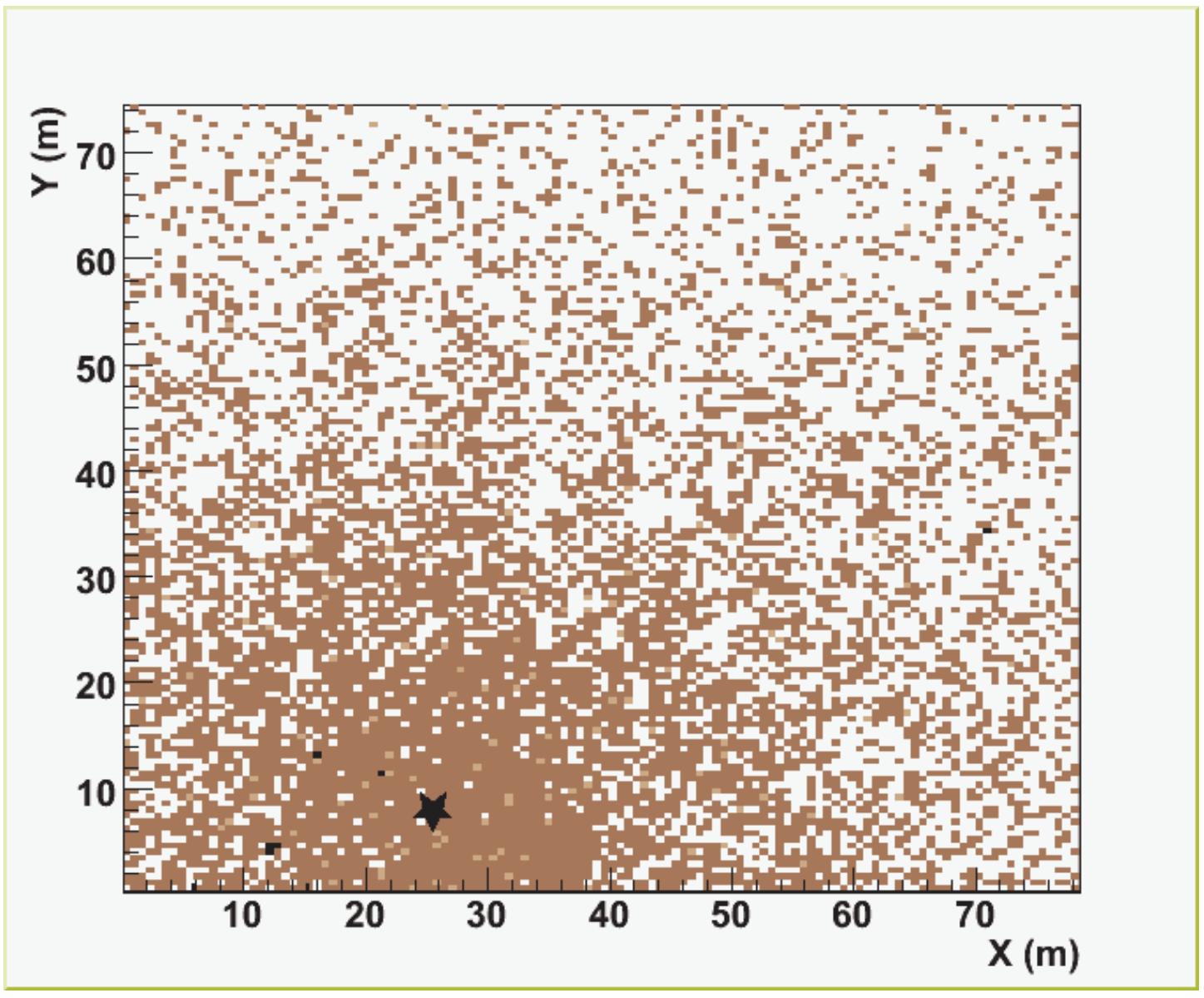

Figure 4: Hit map of an air shower on the ARGO-YBJ central carpet. The horizontal and vertical coordinates are measured with respect to the lower left corner of the carpet. The position of the shower core is shown.

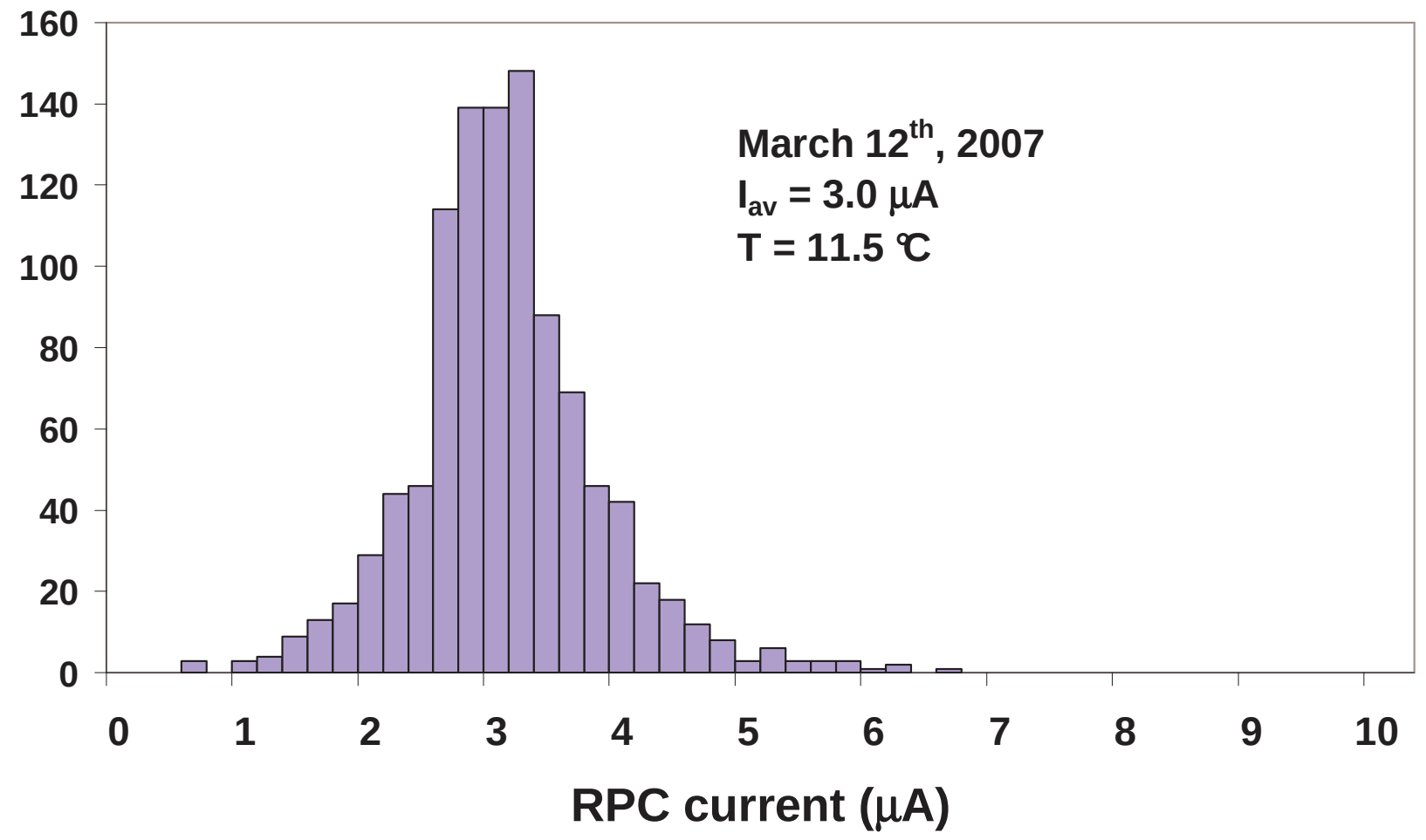

Figure 5: Distribution of the current absorbed by the ARGO-YBJ RPCs (March 2007 data).

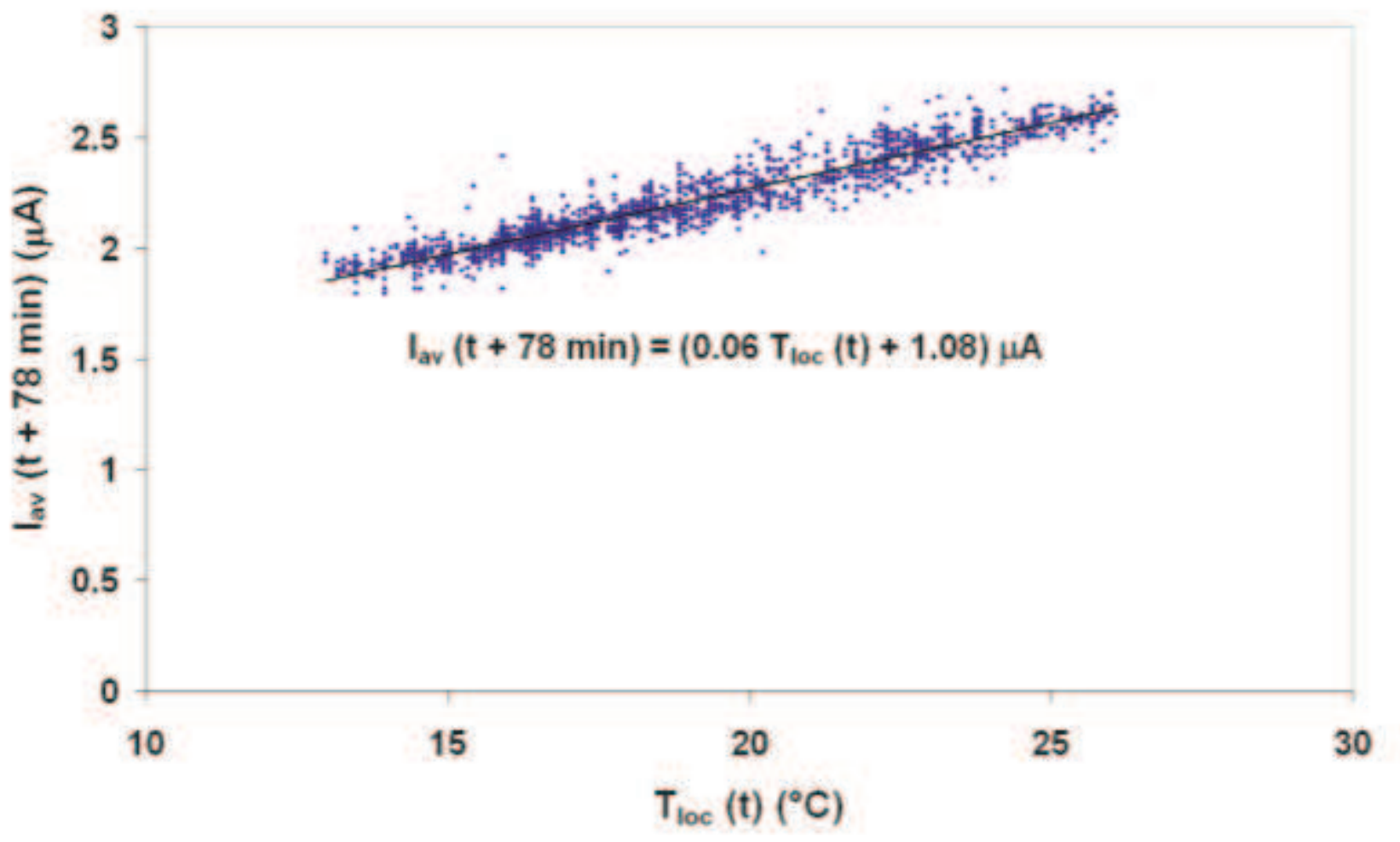

Figure 6: Correlation plot of the 78-minute delayed average RPC current and the local temperature for one ARGO-YBJ cluster near the carpet center (Cluster 91). The result of the linear fit is also shown.

## 3 First results in cosmic-ray astrophysics

The preliminary results from ARGO-YBJ concerning the detection of $\gamma$ rays from the Crab Nebula and from MRK 421 during its July 2006 flare were reported separately in this Workshop [7]. Here we will summarize the results of other important studies performed by the collaboration in cosmic-ray astrophysics.

A crucial investigation concerns the features of the shower fronts [8]. Two parameters can be measured: the *curvature*, which is the mean value of the time residuals with respect to a planar fit of the shower front; the *thickness*, which is the RMS of the residuals with respect to a conical fit (see Fig. 7). The experimental results for these two parameters as a function of the distance from the shower core for three different pad-multiplicity ranges (corresponding to three different mean primary energies) are shown in Fig. 8. The events were selected with a zenith angle less than 15°. A detailed Monte Carlo simulation shows that these parameters may be crucial to perform $\gamma$-hadron separation on a statistical basis.

Another study in progress is the measurement of the proton-air inelastic cross section at energy greater than 1 TeV [9]. For a given pad-multiplicity interval (corresponding to a given mean primary energy) the frequency of showers as a function of the zenith angle $\theta$ (for a fixed distance $X_{DM}$ between the detector and the shower maximum) is related to the probability distribution of the depth of the shower maximum $P(X_{max})$, with $X_{max} = h_0 \sec\theta - X_{DM}$, where $h_0$ is the observational vertical depth. For large $X_{max}$ values, $P(X_{max})$ has a decreasing exponential behaviour with attenuation length $\Lambda = \kappa\lambda_p$, where $\kappa$ (to be evaluated with a Monte Carlo simulation) is an adimensional parameter depending on the shower development and the detector response, and $\lambda_p(g/cm^2) \simeq 2.41 \cdot 10^4/\sigma_{p-Air}(mb)$. In order to select showers from primaries interacting deeper in the atmosphere, it was required that 70% of the fired strips were contained within 25 m from the reconstructed shower core. This event selection is independent of shower fluctuations for $\theta < 40°$, which is also the applied cut for the selected events. The experimental $\sec\theta - 1$ distributions for pad-multiplicity ranges $300 \leqslant N_{pad} \leqslant 1000$ (corresponding to a mean primary energy of (3.9±0.1) TeV) and $N_{pad} > 1000$ (corresponding to a mean primary energy of (12.7±0.4) TeV) are shown in Fig. 9 (upper and lower plot respectively). From these plots the values of $\kappa$ can be obtained by comparison with the Monte Carlo simulation. The result for $\sigma_{p-Air}$, after correcting for the contribution of primaries heavier than protons, is $(275 \pm 51)$ mb for the lower multiplicity range and $(282 \pm 31)$ mb for the upper one. These values are in good agreement with the results of other experiments. The possibility of performing a measurement of $\sigma_{p-Air}$ at energy ~1 PeV with a suitable selection of the shower development stage, also using the RPC analog read-out system, is being considered.

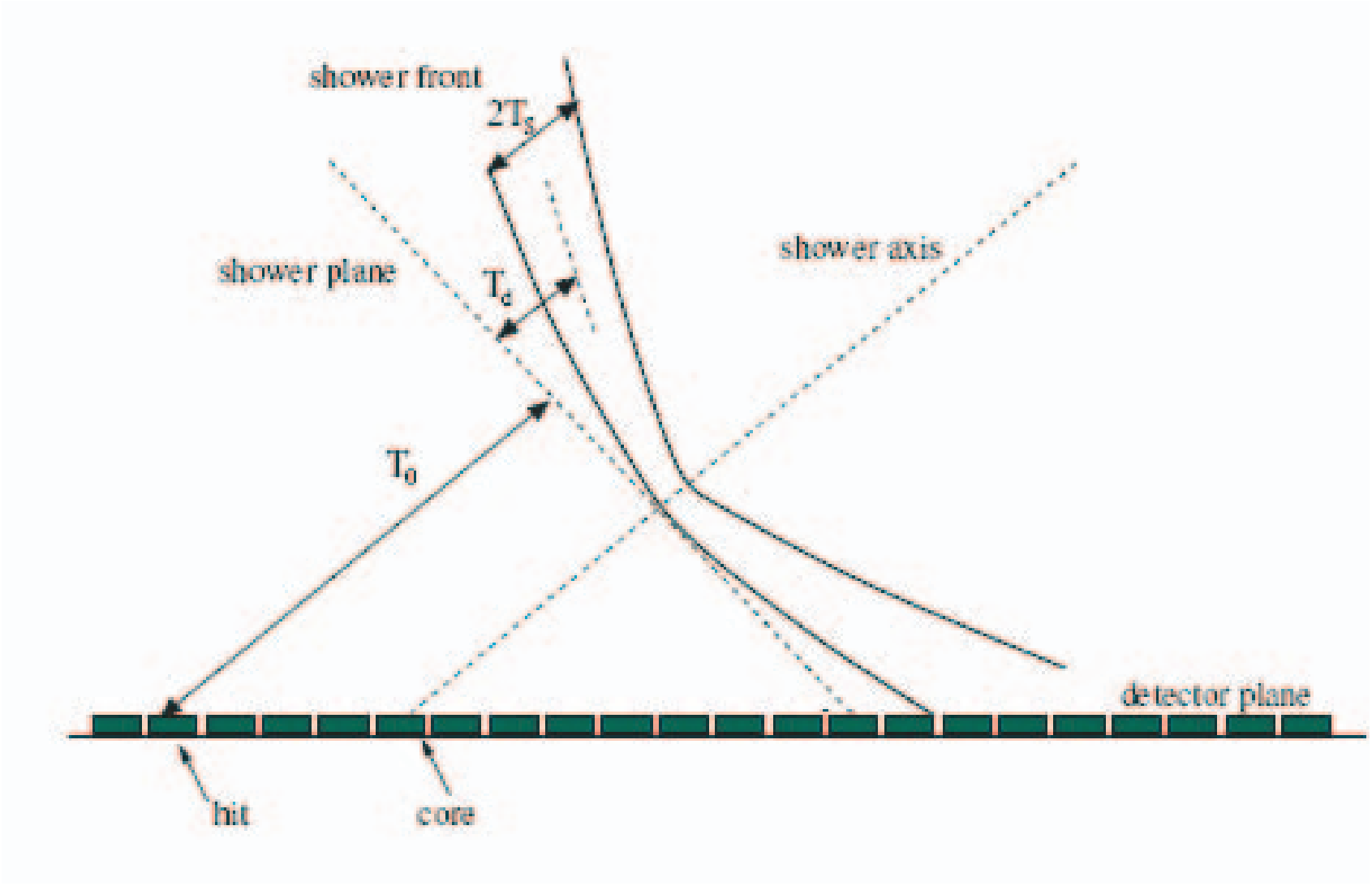

Figure 7: Scheme of the transverse view of an air-shower front hitting the ARGO-YBJ detector. The shower curvature $T_d$ and thickness $T_S$ at a given distance from the core are shown.

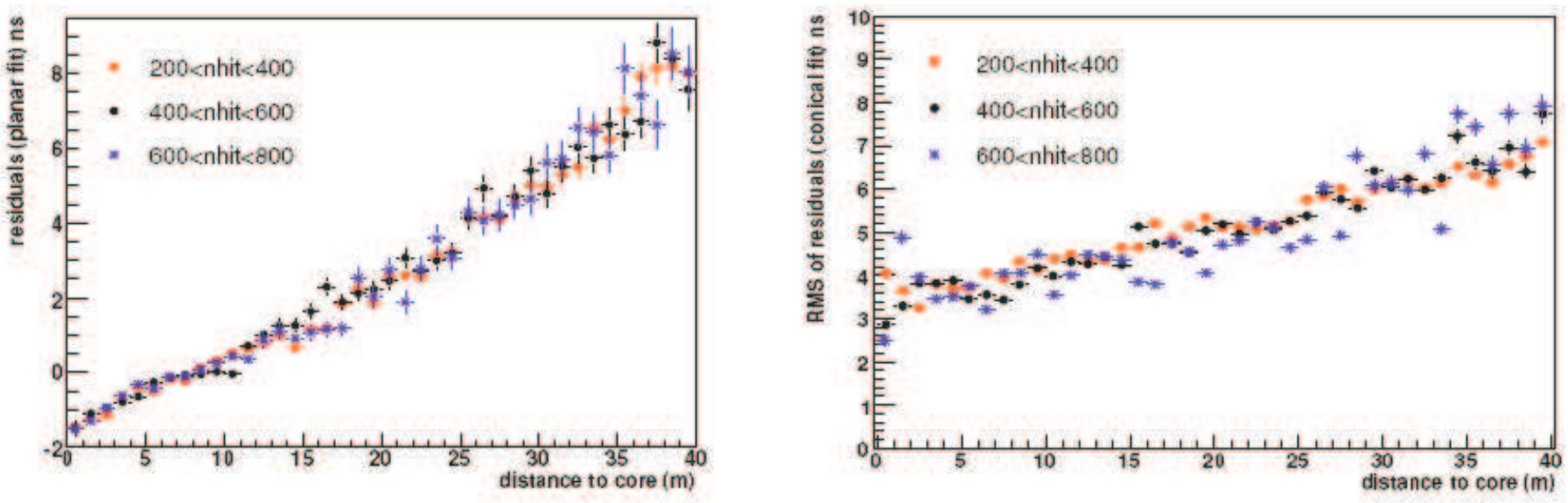

Figure 8: Left: shower curvature vs. the distance from the core, for three different pad-multiplicity ranges. Right: shower thickness vs. the distance from the shower core, for three different pad-multiplicity ranges.

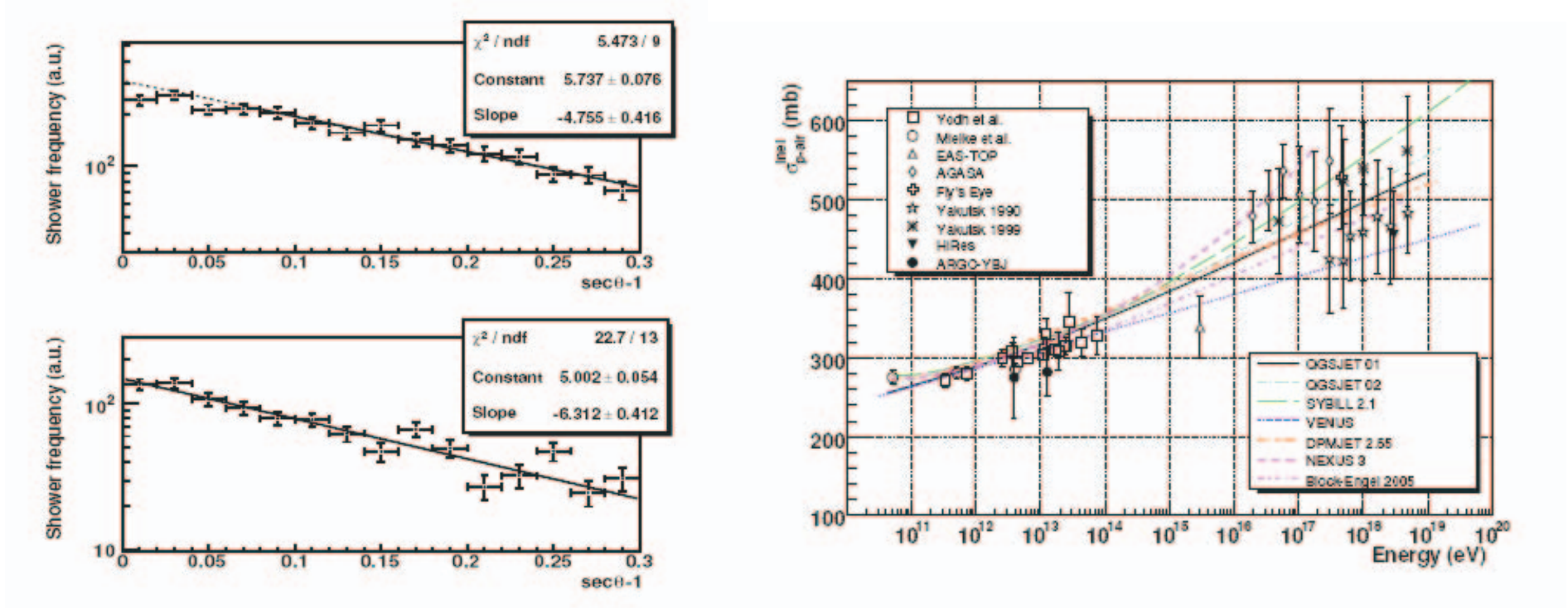

**a.** **b.**

Figure 9: **a.** Upper: experimental $\sec\theta$ distribution ($\theta < 40°$) for pad multiplicity $300 \leqslant N_{pad} \leqslant 1000$. Lower: experimental $\sec\theta$ distribution ($\theta < 40°$) for pad multiplicity $N_{pad} > 1000$. **b.** Inelastic p-air cross section measured by several experiments including ARGO-YBJ, with the prediction of different hadronic interaction models.

## 4 Conclusions

The performance of the ARGO-YBJ detector after the completion of the full-coverage central carpet (July 2006) is good. The first studies in $\gamma$-ray and cosmic-ray astronomy are giving excellent results and providing a good amount of information. The constant monitoring of the environmental and operational parameters is giving a better understanding of the detector and of the conditions which are needed in order to keep the operation stable for a long time. Before the end of 2007 the complete detector will be working, including the guard-ring clusters. The installation of the analog read-out system on all the clusters will be crucial to investigate the energy range from few hundreds of GeV up to few PeV.

# THE EFFECT OF SUBSTRUCTURES ON DARK MATTER INDIRECT DETECTION WITH $\gamma$-RAYS

Lidia Pieri [a,b], Gianfranco Bertone [c], Enzo Branchini [d]

[a] *INAF - Osservatorio Astronomico di Padova, Vicolo dell'Osservatorio 5, 35122 Padova, Italy*

[b] *INFN, Sezione di Padova, Via Marzolo 8, 35131 Padova, Italy*

[c] *Institut d'Astrophysique de Paris, UMR 7095-CNRS, Université Pierre et Marie Curie, 98bis boulevard Arago, 75014 Paris, France*

[d] *Department of Physics, Università di Roma Tre, Via della Vasca Navale 84, 00146 Rome, Italy*

## Abstract

Under the cosmological assumption of the Cold Dark Matter scenario of structure formation, and the particle physics scenario according to which the Dark Matter is composed by common candidates such as supersymmetric particles, the smallest bound structures have masses as low as $10^{-6}M_{\odot}$. N-body simulations show that these clumps may survive till present days. In this case, they are expected to boost up significantly the expected annihilation signal and might also be detected individually as bright spots in the $\gamma$-ray sky. In this work we perform an analysis of the prospects for indirect detection of these objects with GLAST-like experiments, exploring different prescriptions for the subhalos shape parameters currently allowed by numerical simulations. Our results confirm that while subhalos may contribute significantly to the diffuse Galactic annihilation signal, the possibility of detecting a single halo is very small, and it is restricted to high mass halos.

## 1 Introduction

The present-day description of the universe includes 26 % of cold dark matter (CDM), whose nature and distribution is unknown [1]. No dark matter (DM) particle has been observed so far, although hypotheses have been done on weakly interacting massive particles (WIMPs) coming from Supersymmetry or Universal Extra Dimension theories. The distribution of DM inside the halos is uncertain too. The smooth radial DM distribution is poorly constrained in the innermost regions around the halo center, where neither experiments nor numerical simulations have enough resolution to allow any conclusive modeling. The NFW density profile [2] is usually found to be consistent with numerical simulations. Extrapolated at small distances from the halo center, it predicts a $\rho(r) \propto r^{-1}$ behaviour.

In the hierarchical formation scheme of the CDM scenario, large systems as our Milky Way are the result of the merging and accretion of highly concentrated subhalos, the smallest of which ever accreted onto the present day halo have a mass of $10^{-6}M_{\odot}$, if the DM particle is a WIMP. In such dense areas, the DM annihilation into standard model particles is expected to give the bigger contibution.

References and further details can be found in [3]. In this paper we derive a prediction for the expected $\gamma$-ray flux deriving from the population of subhalos inside the MW, and we study its probability to be detected with a GLAST-like experiment. We assume different models for the subhalo concentration parameter, which result in more or less concentrated NFW subhalos.

## 2 $\gamma$-ray flux from subhalos

The $\gamma$-ray annihilation flux can be generally written as $\Phi_\gamma = \Phi^{\rm PP} \times \Phi^{\rm cosmo}$, which factorizes the particle physics and cosmological contributions. We define

$$\Phi^{\rm PP}(E_\gamma) = \frac{1}{4\pi}\frac{\sigma_{\rm ann}v}{2m_\chi^2} \times \sum_f B_f \int_E \frac{dN_\gamma^f}{dE_\gamma}dE. \tag{1}$$

and we adopt $m_\chi = 40$ GeV, $\sigma_{\rm ann}v = 3\times10^{-26}{\rm cm}^3{\rm s}^{-1}$, a 100% branching ratio in $b\bar{b}$, and integrate above 3 GeV. We refer to [4] for further details. On the other hand, $\Phi^{\rm cosmo}$ includes cosmological factors as well as geometrical details such as the angular resolution $\Delta\Omega$ of the instrument and the pointing angle $\psi$:

$$\Phi^{\rm cosmo}(\psi,\Delta\Omega) = \int_M dM \int_c dc \int\int_{\Delta\Omega} d\theta d\phi \int_{\rm l.o.s} d\lambda$$

$$[\rho_{sh}(M, R(R_\odot\lambda,\psi,\Delta\theta,\phi)) \times P(c)\times$$

$$\times \Phi^{\rm cosmo}_{halo}(M, c, r(\lambda, \lambda', \psi, \theta', \phi')) \times J(x, y, z|\lambda, \theta, \phi)] \tag{2}$$

where

$$\Phi^{\rm cosmo}_{halo}(M, c, r) = \int \int_{\Delta\Omega} d\phi' d\theta' \int_{\rm l.o.s} d\lambda'$$
$$\left[ \frac{\rho^2_\chi(M, c, r(\lambda, \lambda', \psi, \theta'\phi'))}{\lambda^2} J(x, y, z|\lambda', \theta'\phi') \right] ; \tag{3}$$

$\rho_\chi$ is the NFW density profile inside the halo, whose scale parameters are a function of the concentration parameter $c(M, z)$. $P(c(M, z))$ is the lognormal probability for a given value $c$ with width=0.24. J is the Jacobian determinant, $R$ is the distance from the MW center and $r$ is the distance from each halo center. $\rho_{sh}$ is the subhalo distribution inside the MW taken from [5]:

$$\rho_{sh}(M, R) = AM^{-2} \frac{\theta(R - r_{min}(M))}{(R/r_s^{MW})(1 + R/r_s^{MW})^2}, \tag{4}$$

where $r_s^{MW}$ is the scale radius of our Galaxy and $\theta(R - r_{min}(M))$ accounts for the effect of tidal disruption, according to the Roche criterion. We refer to [6] for the complete definition of symbols.

We use Eq.3 when deriving the contribution from each subhalo as well as the one coming from the smooth MW halo.

Eq.2 is used to derive the expected diffuse $\gamma$-ray foreground coming from unresolved halos. The resulting flux is then normalized not to exceed the EGRET extragalactic background far from the Galactic plane.

Ten Monte Carlo realizations of the closest and brightest subhalos have been realized in order to study the possibility of detecting an annihilation flux from a resolved subhalo.

As already pointed out in the Introduction, the $c(M, z)$ relation is not well established. We have used the following 6 different models when deriving the flux predictions for both the diffuse and the resolved halo flux:

- $B_{z_0}$ uses $c(M, z = 0)$ as computed in [7], extrapolated to $c(M = 10^{-6} M_\odot, z = 0) = 80$ [8]
- $B_{z_0,5\sigma}$ uses $c(M, z = 0)$ as computed in [7], extrapolated to $c(M = 10^{-6} M_\odot, z = 0) = 400$, corresponding to a 5 $\sigma$ density peak fluctuation.
- $B_{z_f}$ as $B_{z_0}$ but computed at the collapse redshift as derived from [7], extrapolated to $z_f(M = 10^{-6} M_\odot) = 70$ [8] through the relation $c(M, z = 0) = (1 + z_f) \times c(M, z_f)$
- $B_{z_f,5\sigma}$ as $B_{z_0,5\sigma}$ but computed at the collapse redshift

$\mathrm{ENS}_{z_0}$ as $\mathrm{B}_{z_0}$ but with $c(M, z=0)$ computed according to [9].

$\mathrm{ENS}_{z_f}$ as $\mathrm{B}_{z_f}$ but with $c(M, z=0)$ computed according to [9].

The $c(M, z=0)$ curves are shown in the upper panel of Fig.1. Further details and a more complete set of models can be found in [6].

## 3 Experimental sensitivity

We define the experimental sensitivity $\sigma$ as the ratio

$$\sigma \equiv \frac{n_\gamma}{\sqrt{n_{\mathrm{bkg}}}}$$

$$= \frac{\sqrt{T_{obs}} \int A_\gamma^{\mathrm{eff}}(E,\theta_i)[d\phi_\gamma^{\mathrm{signal}}/dEd\Omega]dEd\Omega}{\sqrt{\int \sum_{\mathrm{bkg}} A_{\mathrm{bkg}}^{\mathrm{eff}}(E,\theta_i)[d\phi_{bkg}/dEd\Omega]dEd\Omega}} \tag{5}$$

Here, $T_{obs}$ =1 yr, $A^{\mathrm{eff}} = 10^4 cm^2$ independent from both energy and incident angle, and the background is taken from [10] and [11]. The smooth MW and subhalo annihilation foregrounds have been considered as background for the detection of single halos. The lower panel of Fig.1 shows the sensitivity curves for the diffuse subhalo + MW foreground for the models described in Sec.2. Such a signal would be detectable only toward the Galactic Centre for the $z = 0$ models which are not affected from the normalization imposed by the EGRET data. Unfortunately, the astrophysical uncertainties in modeling the expected background in that region are very high.

We have then computed the 3-$\sigma$ detection probability for each one of the simulated halos, as the probability $P(c_{3\sigma})$ of having a concentration parameter $c_{3\sigma}$ whose corresponding flux would result in a 3-$\sigma$ level detection. The sum of the detection probabilities of all the simulated halos gives us the number of subhalos detectable at 3-$\sigma$ in 1 yr for the given model. The result is shown in Fig.2, where the number of detectable halos as a function of the subhalo mass is plotted for all the concentration parameter models. The sum of detectable halos integrated over the mass is greater than 1 in all cases but the $\mathrm{B}_{z_f,5\sigma}$ and $\mathrm{ENS}_{z_0}$. Yet, the only detectable subhalos would have a mass greater than $10^7 M_\odot$.

## 4 Conclusion

We have derived the expected $\gamma$-ray flux from the annihilation of DM in galactic subhalos. We have computed the smooth MW halo and unresolved subhalos components as well as the contribution from resolved halos, assuming a number of different models for the inner shape parameters of each subhalo. We have shown that detection of an annihilation signal with a GLAST-like satellite as a diffuse emission would be possible only toward the Galactic Centre, where astrophysical uncertainties would make it difficult to disentangle from the poorly known astrophysical background. On the other hand, even in the most optimistic models presented here, only a handful of subhalos with masses in the range $[10^7, 10^9]M_\odot$ could be detected individually. As pointed out in [6], other toy-models can be approached, though they are not supported by numerical simulations. Even in the most optimistic of those models, only very few high mass subhalos are visible.

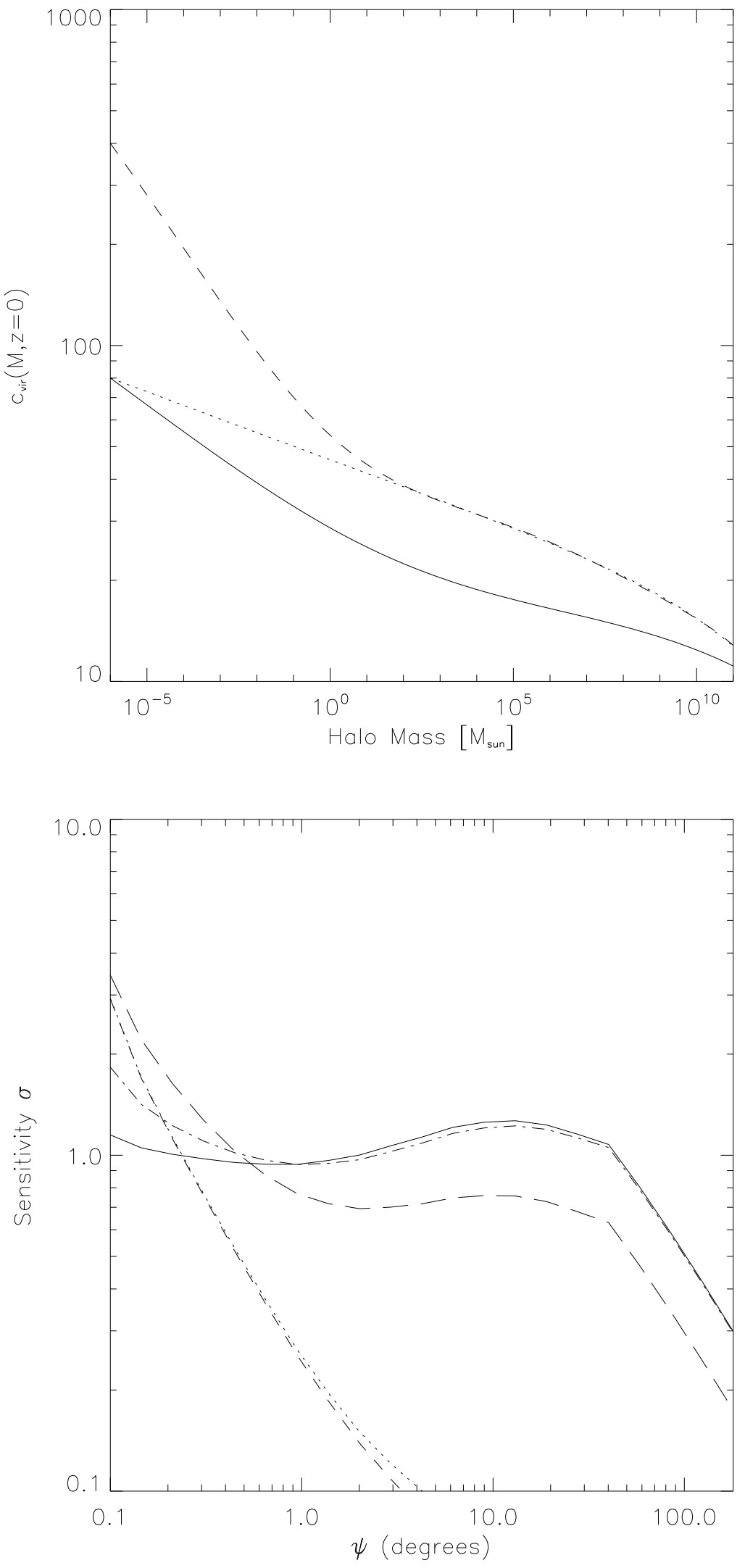

Figure 1: Upper panel: Concentration parameters as a function of halo mass at $z = 0$ computed for the $\mathrm{ENS}_{z_0}$(solid), $\mathrm{B}_{z_0}$(dashed) and the $\mathrm{B}_{z_0,5\sigma}$(dotted) model described in the text. Lower panel: Sensitivity curves for the smooth subhalo contribution obtained along l=0. Results for the $\mathrm{B}_{z_0}$(dotted), $\mathrm{ENS}_{z_0}$(short dashed), $\mathrm{B}_{z_0,5\sigma}$(long dashed), $\mathrm{B}_{z_f}$and $\mathrm{ENS}_{z_f}$(solid), $\mathrm{B}_{z_f,5\sigma}$(dot-dashed) models are shown.

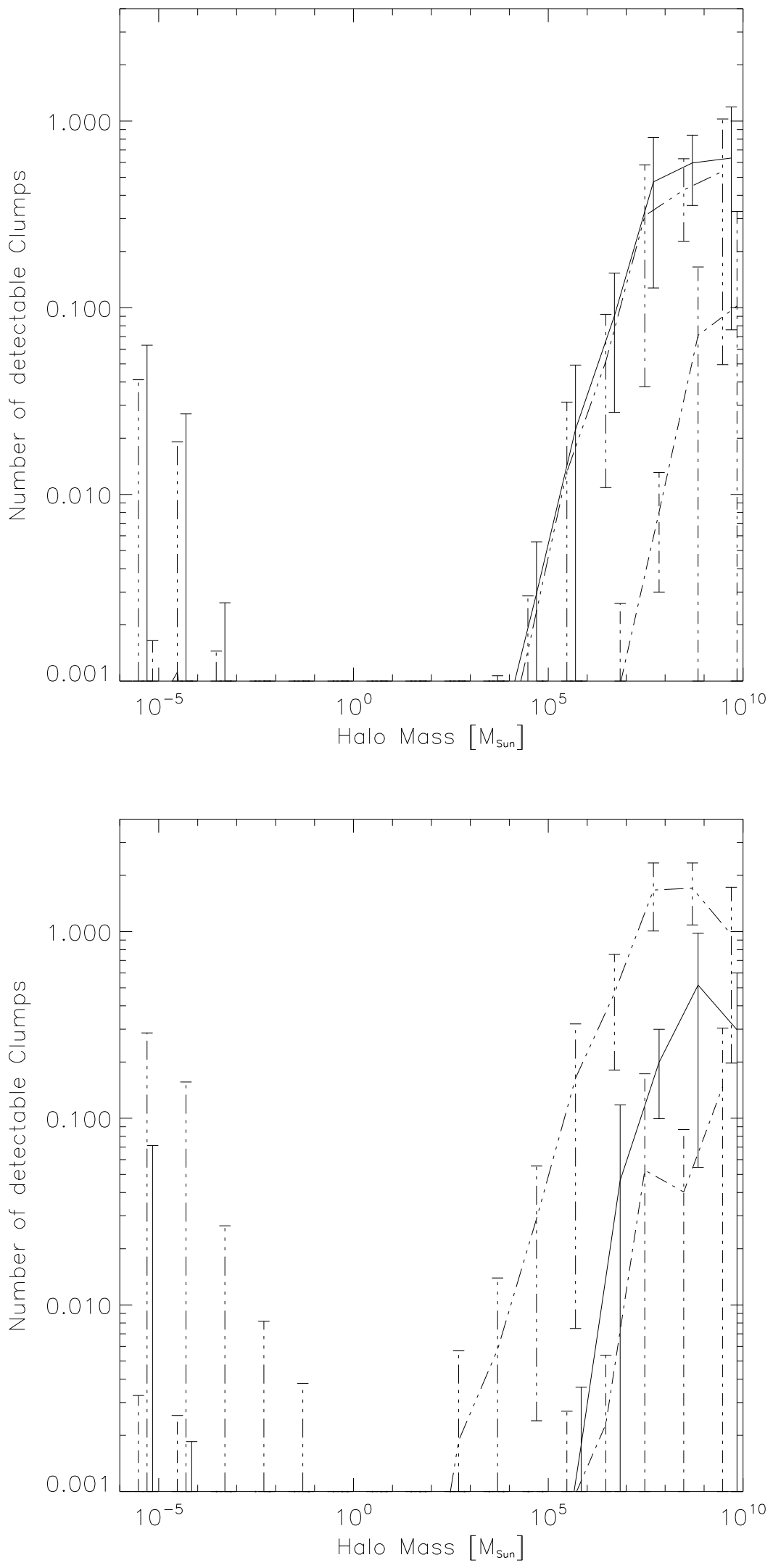

Figure 2: Upper panel: Number of detectable subhalos at 3 $\sigma$ in a 50 degrees cone of view towards the GC for the models $B_{z_0}$(solid), $B_{z_0,5\sigma}$(long dot-dashed) and $ENS_{z_0}$(short dot-dashed), given the lognormal distribution $P(c)$ for the concentration parameter. Lower panel: the same as the upper panel for the models $ENS_{z_f}$(solid), $B_{z_f}$(long dot-dashed) and $B_{z_f,5\sigma}$(short dot-dashed).

# NEW CRAB NEBULA LIMITS ON PLANCK-SCALE SUPPRESSED LV IN QED

LUCA MACCIONE [a,b], STEFANO LIBERATI [a,b],
ANNALISA CELOTTI [a], JOHN KIRK [c]

[a] *SISSA, via Beirut, 2-4, 34014 Trieste, Italy*

[b] *INFN, Sezione di Trieste, via A. Valerio, 2, 34127 Trieste, Italy*

[c] *Max-Planck-Institut fuer Kernphysik, Saupfercheckweg, 1, D-69117, Heidelberg, Germany*

## Abstract

We set constraints on $O(E/M)$ Lorentz Violation in QED in an effective field theory framework. One major consequence of such assumptions is the modification of the dispersion relations for electrons/positrons and photons, which in turn can affect the electromagnetic output of astrophysical objects. We consider the information provided by multiwavelength observations versus a full and self-consistent computation of the broad band spectrum of the Crab Nebula. We cast constraints of order $10^{-5}$ at 95% confidence level on the lepton Lorentz Violation parameters.

## 1 Introduction

Local Lorentz invariance (LI) is fundamental to both of the two pillars of our present physical knowledge: the standard model of particle physics and general relativity. Nonetheless the most recent progress in theoretical physics, in particular toward the construction of a theory of Quantum Gravity (QG), has led to a new perspective in which both the above mentioned theories are seen as effective ones to be replaced by a theory of some more fundamental objects at high energies. In this perspective it is easy to understand that even space-time fundamental symmetries (as local LI) could cease to be valid nearby the Planckian regime.

This intuition and several other issues of principle (see e.g. [1, 2]) for questioning LI at high energies, have been somewhat strengthened by specific hints of Lorentz violation (LV) that have come from tentative calculations in various approaches to QG (see [3] for more details).

In recent years most efforts for placing constraints on high energy deviations from LI have focused on modified dispersion relations for elementary particles. In fact, in most QG models LV is expressed through dispersion relations which can be cast in the general form (assuming rotation invariance to be preserved)

$$E^2 = p^2 + m^2 + f(E, p; \mu; M) \ , \tag{1}$$

where $c = 1$, $E, p$ are the energy and momentum of the particle, $\mu$ is some particle physics mass scale (possibly associated with some symmetry breaking/emergence scale) and $M$ denotes the relevant QG scale, which is usually assumed to be of order the Planck mass: $M \sim M_{\rm Pl} \approx 1.22 \times 10^{19}$ GeV. The function $f(E, p; \mu; M)$ can be generally expanded in powers of the momentum.

We investigate here the model of QED modified by the addition of non-renormalizable, dimension five, LV operators proposed in [4]. The effect of these extra terms is to modify the dispersion relations for the particles as follows. For the photon we have (the + and - signs denote right and left circular polarisation) $\omega_\pm^2 = k^2 \pm \xi k^3/M$, while for the fermion (with the + and - signs now denoting positive and negative helicity states) $E_\pm^2 = p^2 + m^2 + \eta_\pm p^3/M$, with $\xi, \eta_\pm$ constant parameters to be constrained. Hence standard processes (e.g. threshold reactions) are modified and new processes are open (e.g. Čerenkov emission in vacuum). The energy scale at which those effects become visible can be estimated as $k_{\rm th} \sim (m_e^2 M_{\rm Pl}/\eta)^{1/3} \approx 10\ \ {\rm TeV}\eta^{-1/3}$ [2].

Observations involving energies of $\sim 10$ TeV can potentially cast an $O(1)$ constraint on the above defined coefficients. This is indeed a high energy scale but it is well within the range of the observed phenomena in high energy astrophysics, which provides then the observations so far more effective in casting constraints on our test theory. But what is the theoretically expected value for the LV coefficients? In particular renormalisation group effects could in principle strongly suppress the low energy values of $(\xi, \eta_\pm)$ even if they are $O(1)$ at high energies. However, according to [3, 5], the running of the LV parameters is only logarithmic and, therefore, low energy constraints are robust.

In Section 2 we shall review the currents status of the constraints on this test theory, focusing on the role played so far by the Crab Nebula (CN). In Section 3 the current observations of the CN and their theoretical interpretation will be reviewed, and the role of the departures from LI in our theory will be discussed. Finally in Section 4 we shall present and discuss the constraints that the current observations allow to cast on the theory and the possible improvements expected from future experiments like GLAST.

These proceedings refer to the more detailed work [3], to which we point the interested reader.

## 2 Previous astrophysical constraints on QED with O(E/M) LV

The present status of the astrophysical constraints on our test theory is summarised in figure 1. Since we just sketch here the present status of the field, we point the reader to [2] and [3] for a more detailed discussion.

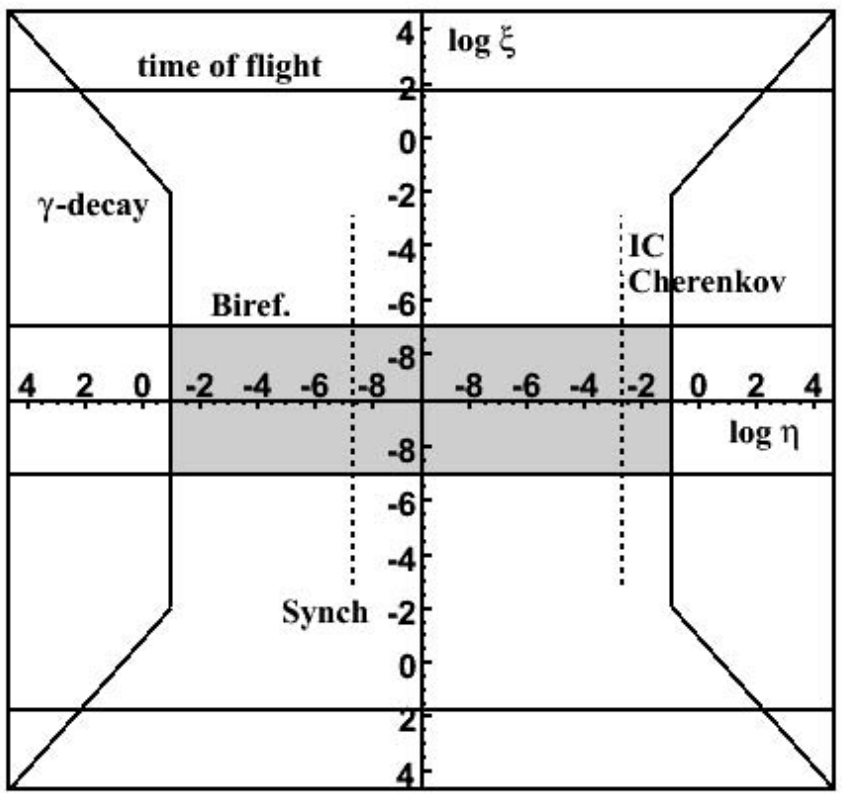

Figure 1: Present constraints on the LV coefficients for QED with dimension 5 Lorentz violation. The grey area is the allowed one and within it the region bounded by the two dashed vertical lines identifies the allowed range for at least one of the four lepton LV coefficients (assuming that a single population has to be simultaneously responsible for the synchrotron and Inverse Compton emission of the CN).

While the natural magnitude of the photon and electron coefficients $\xi, \eta_\pm$ would be $O(1)$ if there were one power of suppression by the inverse Planck mass, the coefficients are currently restricted to the region $|\xi| \lesssim 10^{-7}$ by birefringence [6] and $|\eta_\pm| \lesssim 10^{-1}$ by photon decay [2]. Birefringence is due to the fact that the LV coefficients for right and left circularly polarised photons are opposite, hence the linear polarisation of a photon is spoiled after long distance propagation, while the photon decay $\gamma \to e^+e^-$ is allowed in our test theory because the photon energy-momentum 4-vector is not null [2]. In the same way, in this theory also the emission of Čerenkov radiation in vacuum (VC) $e \to e\gamma$ by a superluminal charged particle is possible.

However, whereas the constraint on the photon coefficient is remarkably strong, the same cannot be said about the LV coefficients of leptons. A constraint on the lepton coefficients of comparable strength is given by the synchrotron limit found in [2], but this is not double sided and implies only that the LV coefficient of the population responsible for the CN synchrotron emission cannot be smaller than $-8 \times 10^{-7}$. Similarly the VC-IC bound $\eta < +3 \times 10^{-3}$ [2] constrains only one lepton population. These statements cannot be considered constraints on $\eta_{\pm}$, since for each of them one of the two parameters $\pm\eta_{+}$ (and $\pm\eta_{-}$) will always satisfy the bound.

It is clear however that these simple arguments do not fully exploit the available astrophysical information. A detailed comparison of the observations with the reconstructed spectrum in the LV case, where all reactions and modifications of classical processes are considered, can provide us with constraints on both positive and negative $\eta$, at levels comparable to those already obtained for the photon LV coefficient. We then move to reconsider such information concerning the astrophysical object that so far has proven most effective in casting constraints on the electron/positrons LV coefficients: the Crab Nebula.

## 3 The Crab Nebula

The CN is a source of diffuse radio, optical and X-ray radiation associated with a Supernova explosion observed in 1054 A.D. Its distance from Earth is about 1.9 kpc. A pulsar, presumably a remnant of the explosion, is located at the centre of the Nebula. It is thought to provide the Nebula with both the radiating particles and a magnetic field, and powers it with part of its spin-down luminosity of about $5 \times 10^{38}$ erg/s (for a recent review see [9]).

The Nebula emits an extremely broad-band spectrum (21 decades in frequency, see [3] for a comprehensive list of relevant observations), produced by two major radiation mechanisms that are related to the interactions of relativistic electrons with the ambient magnetic and radiation fields. The emission from radio to low energy $\gamma$-rays ($E < 1$ GeV) is thought to be synchrotron radiation from relativistic electrons, whereas IC scattering by these electrons is the favoured explanation for the higher energy $\gamma$-rays. The clear synchrotron nature of the non-thermal radiation from radio to low energy $\gamma$-rays, combined with a magnetic field strength of the order of $B \approx 100\ \mu$G implies, when exact Lorentz invariance is assumed, the presence of relativistic electrons with energies up to $10^{16}$ eV. The gyro period of these electrons is roughly equal to the timescale on which they lose energy by synchrotron, implying an acceleration rate close to the maximum estimated for shock-based mechanisms (e.g. [10]). However, though the maximum energy of the electrons is model dependent, the fact that photons with $E \gtrsim 10$ TeV have been detected from the CN is an unambiguous evidence of effective acceleration of particles beyond 100 TeV. Let us stress that this statement has to be considered robust also in our test theory, given that in this case energy-momentum conservation still holds.

From the theoretical point of view, the CN is one of the most studied objects. The current understanding of the whole environment is based on a spherically symmetric magneto-hydro-dynamics model presented in two seminal papers by Kennel & Coroniti [11], that accounts for the general features seen in the CN spectrum. We point the reader to [3] for a wider discussion of this model.

Because we consider a LV version of electrodynamics, it is interesting to study whether this introduces modifications into the model of the CN and, if so, what effects it produces. We now show how the processes at work in the CN would appear in a "LV world".

**Fermi mechanism** Several mechanisms have been suggested for the formation of the spectrum of energetic electrons in the CN. As discussed in [3], the power-law spectrum of high energy (> 1 TeV) particles is usually interpreted as due to first order Fermi mechanism operating at the ultra-relativistic termination shock front of the pulsar wind, since, in the simplest kinematic picture, this mechanism predicts a power law index of just the right value [9]. In [3] the possible modifications occurring to the Fermi mechanism due to LV have been discussed, including the interpretation of the high energy cut-off. In summary, if we phenomenologically model the cut-off at $E_c$ as an exponential, from the Fermi mechanism we expect a particle spectrum in the high energy region $E > 1$ TeV, of the form $n(E) \propto \gamma(E)^{-p} e^{-E/E_c}$ with $p \approx 2.4$ and $E_c \approx 2.5 \times 10^{15}$ eV. Then, we can safely deal with the electron/positron distributions inferred by [12], paying attention to replace the energy with the Lorentz boost factor in the expressions given by [12].

**Role of VC emission** In presence of LV the process of VC radiation can occur. Taking $\xi \simeq 0$ (see [6]), the threshold energy is given by $p_{\rm VC} = (m_e^2 M/2\eta)^{1/3} \simeq 11$ TeV $\eta^{-1/3}$. Just above threshold, this process has a time scale of order $10^{-9}$ s, so it is extremely efficient. The VC emission, due to its extreme rapidity above threshold, can produce a sharp cut-off in the acceleration spectrum. It can be verified that the modifications in the optical/UV spectrum produced by the VC radiation emitted by particles above threshold are negligible with respect to the synchrotron emission.

**Role of Helicity Decay (HD)** If $\eta_+ \neq \eta_-$, leptons can flip their helicities by emitting a suitably polarised photon. In order to understand whether the HD is effective, its typical time scale $\tau_{\rm HD} \sim 10^{-9}$ s $\times \Delta\eta^{-3}(p/10 \text{ TeV})^{-8}$ has to be compared with that of the spin precession of a particle moving in a magnetic field. The spin rotation will effectively prevent the helicity decay if the precession rate is faster than the time needed for HD. Therefore, we can estimate that the HD will become effective when the particle energy is above $p_{\rm HD}^{(\rm eff)} \gtrsim 930\ \text{GeV}\,(B/0.3\ \text{mG})^{1/8}\,|\Delta\eta|^{-3/8}$. Electrons and positrons with $E > p_{\rm HD}^{(\rm eff)}$ can be found only in the helicity state corresponding to the lowest value of $\eta_\pm$. Hence, the population of greater $\eta$ will be sharply cut off above threshold while the other will be increased.

**Synchrotron radiation** Synchrotron radiation by leptons cycling in a magnetic field is strongly affected by LV. In the LI case, as well as in the LV one [2], most of the radiation from an electron of energy $E$ is emitted around a critical frequency $\omega_c = 1.5\, eB\gamma^3(E)/E$. In the LV case, the electron group velocity is given by $v(E) \simeq 1 - m_e^2/2E + \eta E/M$. This introduces a fundamental difference between particles with positive or negative LV coefficient $\eta$. If $\eta$ is negative the group velocity of the electrons is strictly less than the (low energy) speed of light. This implies that, at sufficiently high energy, $\gamma(E)_- < E/m_e$, for all $E$. As a consequence, the critical frequency $\omega_c^-(\gamma, E)$ is always less than a maximal frequency $\omega_c^{\rm max}$ [2]. Then, if synchrotron emission up to some frequency $\omega_{\rm obs}$ is observed, one can deduce that the LV coefficient for the corresponding leptons cannot be more negative than the value for which $\omega_c^{\rm max} = \omega_{\rm obs}$. On the other hand, particles with positive LV coefficient can be superluminal and therefore, at energies $E_c \gtrsim 8\ \text{TeV}/\eta^{1/3}$, $\gamma(E)$ starts to increase more than $E/m_e$ and reaches infinity at a finite energy which corresponds to the threshold for soft VC emission. Therefore, also the critical frequency will be larger than the LI one and the spectrum will show a characteristic bump due to the enhanced $\omega_c$. Finally, it has been checked [3] that the modified energy loss rate does not affect the acceleration spectrum.

## 4 Constraints

Using the numerical tools developed in [3], we are able to study the effect of LV on the CN spectrum. The procedure requires first to fix most of the model parameters using radio to soft X-rays observations, which are not affected by LV [3]. The high energy cut-off of the wind lepton spectrum $E_c \simeq 2.5$ PeV and a spectral index of the freshly accelerated electrons $p = 2.4$ give the best fit to the data in the LI case [12].

It is clear that only two configurations in the LV parameter space are really different: $\eta_+ \cdot \eta_- > 0$ and $\eta_+ \cdot \eta_- < 0$, with $\eta_+$ assumed to be positive for definiteness. In fact, the one with both $\eta_\pm$ negative is the same as the $(\eta_+ \cdot \eta_- > 0,\ \eta_+ > 0)$ case, while that with the signs scrambled is equivalent to the case ($\eta_+ \cdot \eta_- < 0$, $\eta_+ > 0$). This is due to the fact that positron coefficients are related to electron coefficients through $\eta_\pm^{af} = -\eta_\mp^{f}$ [2].

Even though in many cases it is evident that too large LV parameters (at a level of e.g. $10^{-4}$) produce spectra incompatible with data, a $\chi^2$ analysis has been performed to quantify the agreement between models and data. The results are shown in figures 2, where the level curves of the reduced $\chi^2$ are drawn for the cases $\eta_+ \cdot \eta_- > 0$ and $\eta_+ \cdot \eta_- < 0$. Constraints at 90%, 95% and 99% Confidence Level (CL) correspond to respectively $\chi^2 > 8$, $\chi^2 > 10$, $\chi^2 > 13.5$, the minimum value of $\chi^2$ being $\sim 3.6$. According to [3] it is possible to conclude that the LV parameters for the leptons are both constrained, at 95% CL, to be $|\eta_\pm| < 10^{-5}$.

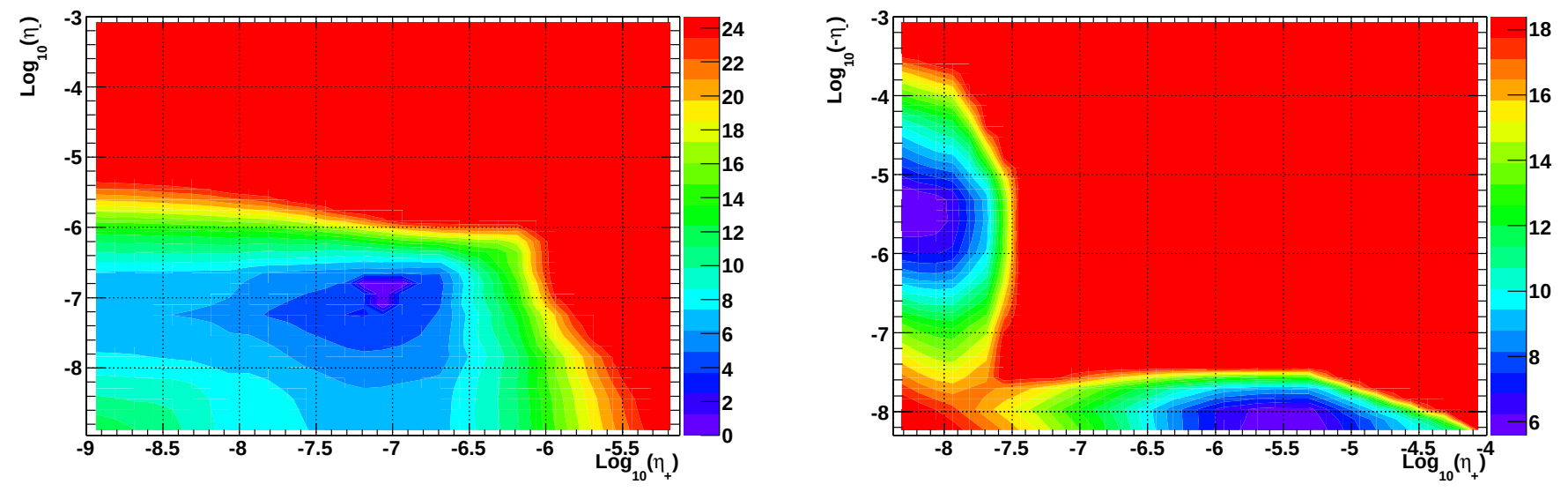

Figure 2: Contour plots of the reduced $\chi^2$ versus $\eta_+$ and $\eta_-$, in the case $\eta_+ \cdot \eta_- > 0$ (left) and $\eta_+ \cdot \eta_- < 0$ (right).

Although the best fit model is not the LI one, a careful statistical analysis shows that the best fit is statistically indistinguishable (with present day data) from the LI one at 95% CL [3].

## 5 Conclusions

We have shown how relaxing the LI assumption (within the framework set up in [4]) reflects into the electromagnetic output of astrophysical sources. We reproduced the observed synchrotron spectrum starting from the most accurate theoretical model of the CN [12], but taking into account the LV contributions of *all* the electron/positron populations and reconsidering all our LI "biases". In this way *both* $\eta_\pm$ can be constrained to limits at $|\eta_\pm| < 10^{-6}$ and $|\eta_\pm| < 10^{-5}$ at 90% and 95% CL respectively, by comparing the simulated spectra to the observational results. A significant step forward could be taken when the GLAST [13] observatory will be flying. An order of magnitude estimate of the improvement can be obtained considering its exposure, which is 30 times larger that EGRET's one. Therefore, assuming that GLAST will observe the CN at least as long as EGRET did, we can estimate that the errors associated with measurements in the 10 MeV-500 MeV band will be lowered by roughly a factor of 5 (statistical). Then, a constraint of order $10^{-6}$ at 99% CL ($10^{-7}$ at 95% CL) is realistic. Moreover, GLAST should permit to disentangle the LI model from the best fit LV one found in [3].

# CONSTRAINTS ON THE TEV SOURCE POPULATION AND ITS CONTRIBUTION TO THE GALACTIC DIFFUSE TEV EMISSION

SABRINA CASANOVA, [a,b], BRENDA L. DINGUS [c]

[a] *ICRANET, Piazzale Aldo Moro, I-00185 Rome, Italy*

[b] *MPI für Kernphysik, Saupfercheckweg 1, D-69117 Heidelberg*

[c] *LANL, 87545, Los Alamos, NM, USA*

**Abstract**

The detection by the HESS atmospheric Cerenkov telescope of fourteen new sources from the Galactic plane makes it possible to estimate the contribution of unresolved sources like those detected by HESS to the diffuse Galactic emission measured by the Milagro Collaboration. The number-intensity relation and the luminosity function for the HESS source population are investigated. By evaluating the contribution of such a source population to the diffuse emission we conclude that a significant fraction of the TeV energy emission measured by the Milagro experiment could be due to unresolved sources like HESS sources. Predictions concerning the number of sources which Veritas, Milagro, and HAWC should detect are also given. The new Milagro results from the Galactic scan confirm our predictions.

## 1 Introduction

Milagro,a water Cerenkov telescope surveying the northern sky at TeV energies has measured the diffuse Galactic $\gamma$-ray emission up to TeV energies. The emission is $(7.3 \pm 1.5 \pm 2.3) \times 10^{-11}\,\mathrm{photons\,s^{-1}\,cm^{-2}\,sr^{-1}}$ for $E > 3.5$ TeV in the region $40^o < l < 100^o$ [1]. Recently [2] have argued that no strong signal of pion decay is seen in the $\gamma$-ray spectrum, and therefore the pion decay mechanism would not be able to explain the excess above 1 GeV measured by the EGRET experiment [3]. Therefore, [2] claimed that the Milagro measurements of the diffuse flux at higher energies revealed a new excess at TeV energies. [2] investigated several possibilities for what might cause the TeV excess, among them possible dark matter decay, contribution from unresolved EGRET sources or sources that are only bright in the TeV range. An extrapolation of the EGRET sources within the range in Galactic longitude of the Milagro observation overestimates the diffuse TeV flux measured. However, there could be a population of sources undetectable by EGRET, but contributing to both the GeV and TeV excess diffuse emission. Recent observations by the TeV observatory, HESS, in fact point to a new class of hard spectrum TeV sources, whose average slope is about $E^{-2.3}$. The HESS detection of high energy $\gamma$ rays from fourteen new sources has improved significantly the knowledge of both the spatial distribution and the spectra and fluxes of VHE $\gamma$-ray galactic sources [4, 5]. These HESS results make it possible to estimate with unprecedented precision the contribution of unresolved sources to the Galactic diffuse emission recently extended to TeV energies by Milagro. Here, based on HESS results, knowing the sensitivity and the field of view of an experiment, we estimate the number of expected sources and their expected VHE $\gamma$-ray flux. We then evaluate the contribution of unresolved sources to the diffuse $\gamma$-ray emission for Milagro [6]. The new results obtained by Milagro during the scan of the Galactic Plane confirm our predictions [7, 9].

## 2 Number-intensity relation for HESS sources

In order to perform a study of the collective properties of HESS source population the number-intensity relation, $logN(> S) - logS$, is here used. The number-intensity relation has the advantage of using the flux data without any assumption on the distance and luminosity. In fact for many of HESS sources the location and luminosity are unknown. The major difficulties in the study of the collective properties of the HESS source population consist of the limited number of sources detected and the relatively small range of flux covered by the survey. Also, the HESS survey of the Galaxy was not performed with uniform sensitivity. In fact, whereas the sensitivity of the survey of the Galactic Plane in Galactic latitude is rather flat in the region between -1.5 and 1.5 degrees, its effective exposure and therefore its sensitivity is not uniform in longitude. Longer observation times were dedicated by HESS to locations in the Galactic plane close to where three sources, HESS J1747-218 and HESS J1745-290 (in the Galactic Center), and HESS J1713-397, were already known. The average sensitivity of the survey as a function of the longitude and the latitude are shown in Fig. 2 and Fig. 3 of [5], respectively. In some locations of the Galaxy the survey was done at peak sensitivity of 2 per cent of the Crab flux. From Fig. 3 of [5] one can deduce that in order for our sample to be complete, only sources detected with more than 6 per cent the Crab flux within $-2^o < l < 2^o$ can be included.

The number-intensity relation for the HESS sample of sources with integral fluxes bigger than $S_1 = 12 \times 10^{-12}\, cm^2\, s^{-1}$, which corresponds to 6 per cent of the Crab flux, is

$$N(> S) = (152 \pm 41)\, (\frac{S}{S_0})^{(-1.0 \pm 0.1)} \quad (1)$$

where $S_0 = 10^{-12}\, cm^2\, s^{-1}$ with reduced $\chi = 0.5$. The HESS-like sources are distributed in a thin disk, in a volume that is larger than the visibility limit. The $\gamma$-ray flux due to the HESS source population above 6 per cent of the Crab flux in the region $-30^o < l < 30^o$ and $-2^o < l < 2^o$ is

$$F(E > 200GeV) = \int_{S_1}^{S_2} N(> S)\, dS = 2.5 \times 10^{-10}\, photons\, \mathrm{s}^{-1}\, \mathrm{cm}^{-2}\,, \quad (2)$$

where $S_2 = 57.8 \times 10^{-12}\, cm^2\, s^{-1}$ is the maximum flux detected by HESS from a source.

For only a few of HESS sources a firm identification with counterparts at other wavelengths exists. Though there are some suggestions that many of the HESS sources might coincide with supernova remnants (SNRs) or pulsar wind nebulae (PWNe). In fact two of the HESS sources have SNRs as counterparts, and five of these most recently discovered HESS sources are associated with pulsar wind nebulae [10]. SNRs are an established source class in VHE $\gamma$ ray astronomy [11, 12, 13, 14]. PWNe formed from young pulsars with age less than a million years are considered as potential gamma-ray emitters [15]. Though a young age is not a sufficient condition for a pulsar to generate a PWN. The spin-down energy loss is the key parameter to determine whether a young energetic pulsar forms a PWN [16]. The ratio between $\gamma$-ray loud versus $\gamma$-ray quiet pulsars is uncertain. [16] suggests that all pulsars with $dE/dt > dE/dt_c = 3.4 \times 10^{36} erg/s$ are X-ray bright, manifest a distinct pulsar wind nebula (PWN), and are associated with a supernova event. By studying the Chandra data on the 28 most energetic pulsars of the Parkes Multibeam Pulsar Survey [15] [17] found that 15 pulsars with $\dot{E} > 3.4 \times 10^{36} ergs/s$ are X-ray bright, show a resolved PWN, and are associated with evidence of a supernova event. This means that about 2.5 per cent of the radio loud pulsar have a PWN and might emit $\gamma$-rays.

From the distributions of Galactic PSRs and SNRs there are 88 SNRs and 5324 PRSs in the region $-30^o < l < 30^o$ and $-2^o < b < 2^o$ , whereas in the Milagro region ($40^o < l < 100^o$ and $-5^o < b < 5^o$) there are 41 SNRs and 1358 PSRs. Assuming that supernova remnants and pulsars are the radio counterparts of high energy gamma ray sources, SNRs and PWNe, the Milagro region has 26 percent of the sources which are in the HESS region. Also, the Milagro Galactic diffuse emission is measured for a threshold energy of about 3.5 TeV, whereas the flux from unresolved sources calculated in Eq.(2) refers to the HESS threshold energy of 200 GeV. In order to estimate the contribution of unresolved sources to Milagro diffuse emission the flux in Eq.(2) has to be corrected for the different threshold energy. All HESS sources are fitted by a power law spectrum $\Phi(E,\Gamma) = \Phi_0\,(\frac{E}{1TeV})^{-\Gamma}$ and the average spectral index is $\Gamma = 2.32$. By correcting the flux in Eq.(2) for the Milagro threshold energy the flux which HESS-like sources contribute is

$$F(E > 3.5\,TeV,\, 40 < l < 100,\, -5 < b < 5) = 8.3\times10^{-12}\,photons\,\mathrm{s}^{-1}\,\mathrm{cm}^{-2}\mathrm{sr}^{-1}\,. \tag{3}$$

The contribution of HESS source population amounts to 10 per cent of the diffuse flux which Milagro measures above 3.5 TeV. This is a lower limit for the contribution of unresolved sources to Milagro diffuse emission, as only sources above 6 percent of the Crab flux were taken into account to estimate it.

The Galactic diffuse emission measured by Milagro can be used to constrain the minimum flux $S_{min}$ below which the logN-logS plot becomes flat in order not to overproduce the Milagro flux

$$\int_{S_{min}}^{S_2} dS\,\frac{dN}{dS} < 7.3\times10^{-11}\,\mathrm{photons\,s^{-1}\,sr^{-1}cm^{-2}}\,. \tag{4}$$

$S_{min}$ cannot be less than $1\times10^{-16}\,photons\,\mathrm{s}^{-1}\,\mathrm{sr}^{-1}\mathrm{cm}^{-2}$ in order not to violate the constraint in Eq.(4).

From the number-intensity relation for HESS source population it is possible to deduce the number of sources which HESS, VERITAS, Milagro and HAWC will detect. The number of SNRs and PWNs expected for HESS if its entire field of view is scanned with a uniform sensitivity of 2 per cent of the Crab flux above 200 GeV is about $43 \pm 10$. If VERITAS will survey the Northern sky reaching the level of 1 per cent of the Crab flux above 100 GeV, it should detect approximately $18\pm4$. For small fields of view experiments such as HESS and VERITAS the complete survey of the sky at the quoted sensitivity requires years of operation.

Milagro has been observing the Northern sky with a sensitivity equal to about 65 per cent the Crab flux at 3.5 TeV. At this level of sensitivity one should expect no detection of sources for Milagro and indeed no detection of sources was claimed in [1] with the data accumulated after the first three years of operation. Now after six years of data have been accumulated Milagro has reached a sensitivity equal to about 10 per cent of the Crab flux at 12 TeV median energy and should be able to detect $5 \pm 1$ sources. The proposed experiment HAWC after only one year of operation will have surveyed the Northern sky at 50mCrab sensitivity above 1 TeV and should have detected $5 \pm 1$ HESS-like SNRs and PWNe. Compared to the Milagro experiment which had no detection for the first three years of data the expectations of detection of SNRs and PWNe at VHE energy for HAWC after two/three years of operation are extremely promising.

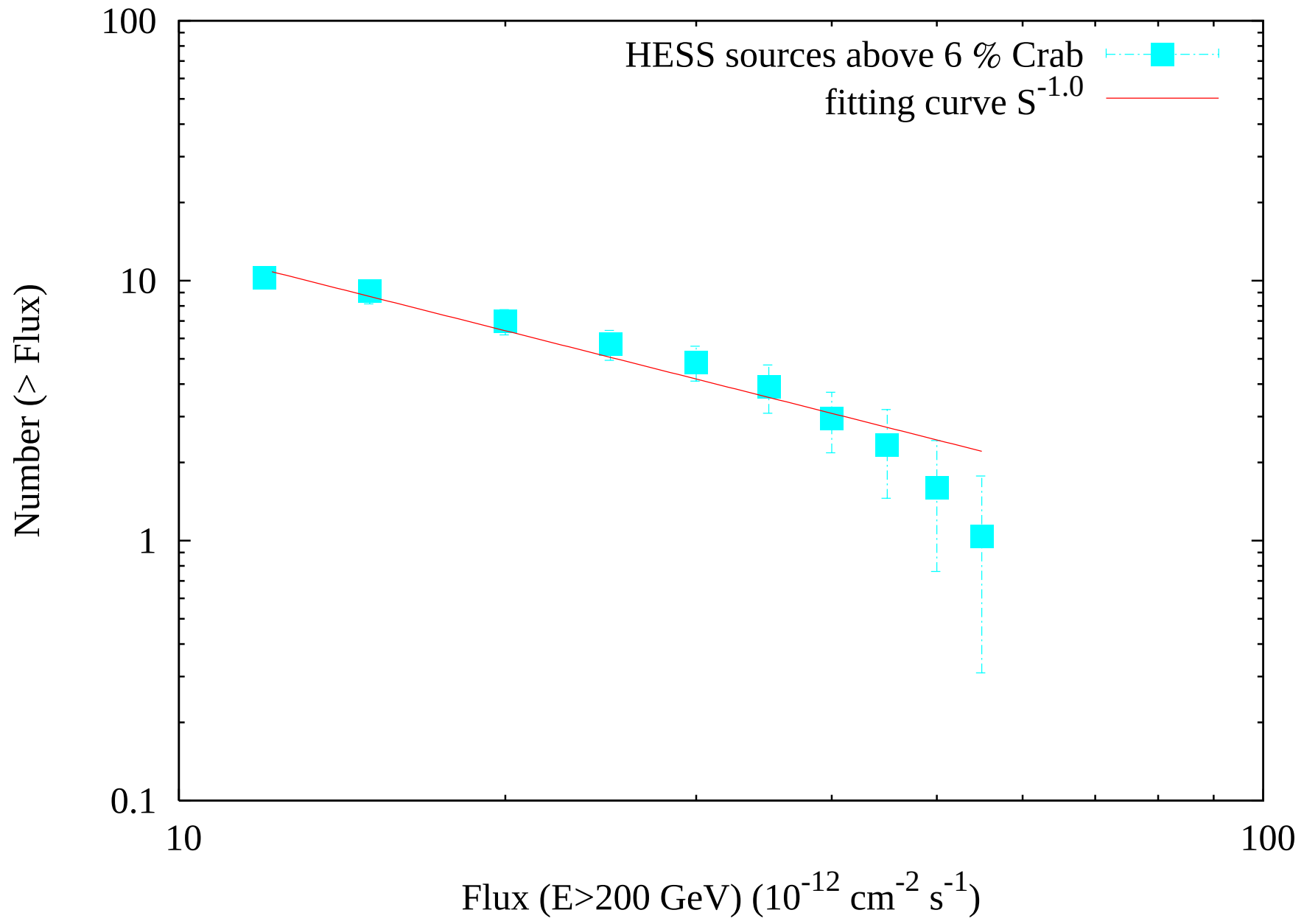

Figure 1: The number-intensity relation, $logN(> S) - logS$, for HESS source population. In order to have a complete sample, only sources detected above 6 per cent of the Crab flux are included. The population of sources is distributed in a thin disk.

## 3 Alternative method to estimate the contribution of unresolved sources to the diffuse emission

Thanks to the logN-logS relation we obtained a lower limit for the contribution of unresolved HESS-like sources to the diffuse emission measured by Milagro. Here we will assume the density of VHE $\gamma$-ray source candidates, SNRs and PWNe, follows the volume density of SNRs or of pulsars in the Galactic Plane as observed at radio wavelengths and we will then estimate their contribution to the Milagro diffuse emission.

Since the data on HESS sources are too sparse to constrain their luminosity function, we will leave it as a parametrised input. The luminosity function $\Phi(L)$ will be a power law with different indices $\alpha$ varying between -1 and -2

$$\Phi(L) = \frac{dN}{dL_\gamma} = c\,(\frac{L_\gamma}{L_{\gamma\,0}})^{\alpha}\,. \tag{5}$$

The assumed luminosity function will then be compared with the HESS source counts to fix the normalisation $c$. In Eq.(5) $L_{\gamma_0} = 1 \times 10^{34} erg/s$.

The range in luminosities for the HESS sources, for which the distance and thus the luminosity is known, varies between $10^{31} erg/s$ and $10^{36} erg/s$. In fact, most sources of $\gamma$-ray in the Galaxy are located close to the plane of the Galaxy, within a region which extends from $D_{min} = 0.3$ kpc up to $D_{max} = 30$ kpc [18]. The range in luminosity for the HESS sample can then be found from the HESS sensitivity (we assume 6 percent of the Crab flux) and the maximum flux detected by HESS, which are respectively $L_{\gamma_{min}} = \frac{\Gamma-1}{\Gamma-2}\, E_{th}\, 4\,\pi\, {D_{min}}^2\, f_{min} = 3 \times 10^{31} erg/s$ and $L_{\gamma_{max}} = \frac{\Gamma-1}{\Gamma-2}\, E_{th}\, 4\,\pi\, {D_{max}}^2\, f_{max} = 1 \times 10^{36} erg/s$, where $E_{th}$ is the detector threshold energy and $\Gamma$ is the spectral index if the $\gamma$-ray emission is a power law.

That HESS detects above 6 percent of the Crab flux constrains the normalisation factor $c$. If the slope of the luminosity function is $\alpha = -1.5$, the integral flux above 3.5 TeV due to HESS-like SNRs and PWNe is $7 \times 10^{-11}\, photons\, \mathrm{s}^{-1}\, \mathrm{cm}^{-2} \mathrm{sr}^{-1}$, which is comparable to the diffuse emission itself assuming a slope $\alpha = -1$. If the slope of the luminosity function $\alpha = -1.5$ the contribution of SNRs and PWNe to the diffuse emission measured by Milagro is about 5 percent and for $\alpha = -2$ this contribution becomes negligible, which is in disagreement with the lower limit of 10 percent for the contribution of HESS-like sources to the VHE diffuse emission previously found. Thus the slope of the luminosity function for HESS-like sources is constrained to be $-1 > \alpha > -1.5$.

## 4 Conclusions

The number-intensity relation and the luminosity function for the HESS source population were investigated using the assumption that HESS sources are distributed as PSRs and SNRs detected at radio wavelengths. In order for the chosen sample of sources to be complete only the HESS sources with fluxes above 6 percent of the Crab flux were taken into account to derive the number-intensity relation. The contribution of unresolved HESS-like sources to the diffuse emission measured by Milagro was also estimated. Using the logN-logS relation for the HESS sample of Galactic $\gamma$-ray emitters at least 10 per cent of the diffuse emission at TeV energies is estimated to be due to the contribution of unresolved HESS-like sources. This result is a lower limit for such a contribution because we have taken into account only sources detected above 6 per cent of the Crab flux and because HESS sensitivity gets worse for extended sources, meaning that some extended sources might have been missed by HESS. Using the logN-logS relation we have also predicted the number of HESS-like sources which VERITAS, HESS and HAWC should detect during their survey of the sky.

An alternative procedure to evaluate the contribution of unresolved HESS-like sources to Milagro diffuse emission gives the diffuse flux due to unresolved sources comparable to the diffuse emission itself. We also constrained the slope of the luminosity function. New observational results support the hypothesis that a population of unresolved sources contribute significantly to the emission at very high energy. Milagro has recently reported the discovery of TeV gamma ray emission from the Cygnus Region of the Galaxy, which exceeds the predictions of conventional models of gamma -ray production [19] from the same region in the Galaxy where the Tibet Array has detected an excess of cosmic rays [20]. Milagro's recently improved sensitivity has also better imaged the whole Northern sky and discovered seven new hot spots above 4.5 sigma, which contribute 25 per cent of the Milagro emission [7, 8, 9]. HESS has seen very high energy emission spatially correlated with giant molecular clouds located in the Galactic Center [21]. The energy spectrum measured by HESS close to the Galactic Center is $E^{-2.3}$, significantly harder than the $E^{-2.7}$ spectrum of the diffuse emission and equal to the average spectrum of the HESS source population. The emission from the Galactic Center might possibly unveil a cosmic ray accelerator.

The main uncertainty of our calculation consists in assuming that the distribution of $\gamma$-ray sources follows the distribution of either pulsars or SNRs observed in the radio. In particular, in order to predict how many PSRs observed in the radio have a PWN and are possible gamma ray emitters we used the result that the spin-down energy loss $dE/dt > dE/dt_c = 4 \times 10^{36} erg/s$ for a young energetic pulsar to form a PWN. In this respect we have ignored the existence of pulsars, such as Geminga, which are $\gamma$-ray loud, yet not observed in the radio. To draw more definitive conclusions about the very high energy $\gamma$-ray sky, new observations are of fundamental importance. New hints will be provide by both MAGIC and VERITAS, which already survey the Cygnus Region. Finally GLAST will investigate the window of energy between 10 MeV to 300 GeV, covering the energy gap left between EGRET and the ground-based low threshold gamma-ray observatories.

# UNIDENTIFIED GAMMA-RAY SOURCES AND THE MAXIMUM DURATION OF ASTRONOMICAL INCOMPREHENSION

PATRIZIA A. CARAVEO [a,b]

[a] *Istituto di Astrofisica Spaziale e Fisica Cosmica, INAF, Via Bassini, 15, 20133 Milano, Italy*

[b] *INFN, Sezione di Pavia, via Ugo Bassi, 6, Pavia, Italy*

**Abstract**

Unveiling the nature of a vast number of unidentified sources is the most compelling problem facing today's high-energy (MeV-to-GeV) gamma-ray astronomy. However, unidentified sources are not peculiar to high-energy gamma-ray astronomy, they have been an ever-present phenomenon in astronomy.

Owing to the intrinsic limitations of gamma-ray detection technique, however, the instruments' angular resolution has not yet reached the minimum level required to permit the transition from the unidentified limbo to the identified status, thus creating a continuing unidentified high-energy gamma-ray source problem.

## 1 Introduction

While trying to understand celestial objects, astronomers are frequently facing challenging sources (or classes of sources) which defy easy classification. The discovery of a new class of celestial objects can result in a rather quick major discovery, as was the case for pulsars and quasars, or in a long struggle. Trimble (2003) [17] has quantified the time needed to solve a number of astronomical puzzles finding values of "the maximum duration of astronomical incomprehension" ranging from the 25 years needed to unravel the mystery of GRBs to more than a century needed to understand coronal lines or Mira variables.

Unidentified gamma ray sources( UGOs for short), still happily with us since the publication of the first COS-B catalogue in 1977 [11], have long passed the quarter century mark, resisting all efforts to nail down their nature.

Indeed, as time (and gamma-ray missions) went by, the number of sources grew dramatically, making the source identification problem even more compelling.

As of today, while waiting for the new look of the gamma ray sky which will be provided by Agile and Glast, the current gamma ray source catalogue [10] lists 271 sources, 172 of which have no identification (neither certain nor probable). To make matters worse, the 99 identified sources belong to just two classes of celestial objects: pulsars (6 objects) and active galactic nuclei (all the remaining IDs), both classes having been recognized by virtue of their variability. A more diverse source portfolio would be most welcome, but no compelling evidence of different source classes has been collected so far.

## 2 Struggling towards identifications

Indeed, something unusual has already been found: the first and, so far, the only source successfully identified turned out to be the first example of a radio quiet pulsar, a variety of neutron star long posited to exist but so far very elusive. The phenomenology of Geminga seems custom-made for gamma ray astronomy. The source is indeed brightest at gamma-ray wavelength, detectable, but unremarkable, in X-rays and darn faint at optical wavelengths [3]. While X-ray photons provided the first evidence of the source periodicity, the optical counterpart yielded first the source proper motion and later its parallactic distance. Now Geminga is known to behave exactly like a pulsar, were it not for the lack of normal radio emission (sporadic emission has been reported, so far unconvincingly). Geminga-like objects are expected to make a significant contribution to the galactic gamma-ray source population, although their intrinsic low luminosity can account only for relatively nearby sources. Indeed, the success story of the identification of Geminga rests on the fortuitous combination of distance and flux values that happened to render the source just within reach of the best instruments available at the time of the "Chase".

The "Next Geminga" source provides an example of the difficulties one can encounter with a slightly more distant source. 3EG J1835+5918 is the brightest among the 172 unidentified gamma-ray sources and, luckily enough, is ideally positioned, well above the galactic plane in the Cygnus region. X-ray coverage of the gamma-ray error box unveiled a source for which no optical counterpart has been found, making it a convincing case for a radio quiet isolated neutron star, hence the name of "Next Geminga" given by the discoverers [14], [7], . Presumably about 4 times more distant than Geminga (still less than 1 kpc from us), 3EG J1835+5918 offers a clear example of the difficulties one faces when dealing with radio quiet neutron stars. Geminga-like luminosities yield flux values too faint to allow detection of the source pulsation in X-rays, while the optical emission is probably beyond reach of all current telescopes, thus hampering the possibility of measuring a proper motion which would clinch the identification. The only development foreseen at this point [8] is the detection of a periodicity in the gamma-ray data, i.e. the smoking gun for any INS. While this is certainly possible with the new generation of gamma-ray telescopes, unveiling pulsation directly in gamma-ray data has never been done before and would be a remarkable first.

Of course, compact objects in binary systems could also be of interest to gamma-ray astronomy. One such system is GT 0236 [15], also known as LSI 61°303, which has been considered a potential counterpart of the COS-B gamma ray source CG 135+01 since the discovery of its X-ray emission [2]. It is a peculiar binary system with a remarkable 26 day periodicity detected at radio wavelengths and also present at optical [12] and X-ray wavelengths [9]. Such periodicity offers a vary useful handle to a potential gamma-ray identification but, most unfortunately, no convincing evidence of orbital variability has yet been found in high energy gamma-ray data [16], [13]. Perhaps the orbital periodicity is superimposed to an erratic behavior, as happens to be the case in X-rays [5]. However, the detection of very high energy gamma rays varying at the correct orbital period [1] has recently revived the interest in the source, exactly 25 years after the original suggestion by [2].

## 3 Looking for diversity in the gamma-ray sky

The situation of today's gamma-ray astronomy is similar to that of X-ray astronomy at the end of the Uhuru mission, when the catalogue of Uhuru sources [6] listed 339 source, 206 of which without identification. At variance with gamma-ray astronomy, the sources identified belonged to different classes: pulsars, binary systems, stars, AGNs. The development of X-ray telescopes, based on grazing incidence techniques, allowed for a fantastic improvement in angular resolution, easing significantly the identification procedure.

Gamma-ray astronomy cannot count on such a dramatic improvement. Even if the performances of Agile and GLAST promise to be much better that that of the previous generation of gamma-ray telescopes, their source positioning will be, at best, in the several arcmin region, far too much to allow for an unambiguous identification, especially in crowded galactic regions .

In view of the limited angular resolution, gamma-ray source identification must rely, yet again, on additional pieces of information. While the photon arrival times will continue to be exploited searching for some kind of correlated variability, the number of sources expected for a mission such as GLAST, however, renders a Geminga-like multiwavelength approach impossible to pursue.

### 3.1 Towards a new identification strategy

A different strategy is being worked out by the GLAST Working Groups, by proposing the idea of going "from detection to association to identification", mostly (but not only) through a statistical approach. Since multiple positional coincidences with field objects will be the rule rather than the exception, a "figure of merit" approach has to be developed for ranking the proposed candidates, see e.g. [4]. First, for each source class a chance occurrence probability parameter should be computed in order to weight the relative abundance of a given source class. In general, counterparts belonging to a class of "certified" gamma-ray emitters will be ranked higher than a counterparts belonging to a new, yet unseen, class. Whereas population studies may offer insights on the existence of yet unknown gamma-ray sources classes, we ultimately need to single out prominent individuals of such classes and unequivocal identify them to establish confidence in the existence of new classes of high-energy gamma-ray emitters. Of course, the "figure of merit" should also takes into account the displacement of the proposed candidate from the best source position. Next, the energetic plausibility of the association should be evaluated. A proposed counterpart unable to meet the energetic requirement set by its gamma-ray flux and its supposed distance will be discarded. Finally, the general source phenomenology will be scrutinized to see if its parameters, such as gamma-ray emission efficiency, spectral shape, variability etc, are consistent with those of a given source class.

The use of time variabilities as an identification tool will be exploited also through a comprehensive program of coordinated observations devoted to different classes of objects. Multiwavelength campaigns will be the last resort used, when everything else has failed or there is no other way to confirm an identification. In order to avoid an a priori limitation on the GLAST potential for discovery, one should keep in mind that any figure of merit approach as well as any kind of monitoring campaign will be biased toward known source classes.

Being ready for the unknown is a challenging task but it is the only way to limit the duration of astronomical incomprehension in high-energy gamma-ray astronomy.

# FOSSIL AGN AS COSMIC PARTICLE ACCELERATORS

R.J. Protheroe [a], Gregory Benford [b]

[a] *Department of Physics, School of Chemistry & Physics, University of Adelaide, Adelaide, SA 5005, Australia*

[b] *Department of Physics and Astronomy, University of California, Irvine, CA 92697-4575, USA*

**Abstract**

Remnants of active galactic nucleus (AGN) jets and their surrounding cocoons leave colossal magnetohydrodynamic (MHD) fossil structures storing total energies $\sim 10^{60}$ erg. The original active galacic nucleus (AGN) may be dead but the fossil will retain its stable magnetic configuration resembling the reversed-field pinch (RFP) encountered in laboratory MHD experiments. Slow decay of the large-scale RFP field induces electric fields which can accelerate cosmic rays with an $E^{-2}$ power-law up to ultra-high energies. A similar mechanism, operting for fossil microquasars could contribute to Galactic cosmic rays and be responsible for some unidentified GeV and TeV gamma-ray sources.

## 1 Introduction

The energy spectrum of cosmic rays (CR) extends from below 1 GeV up to at least $10^{20}$eV, and at the highest energies is almost certainly extragalactic. Possible acceleration sites of these ultra-high energy (UHE) CR include hotspots of giant radio galaxies, the intergalactic medium, gamma ray bursts and blazar jets. UHE CR are subject to interaction with the cosmic microwave background radiation (CMBR) by pion photoproduction as was first noted by Greisen [1] and Zatsepin & Kuzmin [2], and the cut-off they predicted is referred to as the "GZK cut-off". See ref. [3] for a recent review of UHE CR.

Remnants of AGN jets and their surrounding cocoons may persist long after their parent AGN fade from view. These colossal MHD structures decay slowly and yet may retain their relatively stable self-organized configurations. Decay depends on the structure circuit resistance, and lifetimes could be quite long, given the large inductance of the circuit, an initial outward current along the jet and a return current back along an outer sheath or cocoon around the jet. On the immense scale of these fossil jets, the decay time from instability can be billions of years. However, decay of these colossal MHD structures on such time-scales result in electric fields capable of accelerating existing populations of lower energy cosmic rays up to ultra high energies with a flat spectrum extending from some minimum rigidity (momentum/charge) determined by fossil dimensions, magnetic field and decay time. A more extensive discussion and full details of our present work are given in ref. [4].

## 2 Evolution of AGN Fossil Magnetic Structures

Helical structures are common in AGN jets, arising during jet formation from rotation of magnetized plasma accreting toward the central black hole – this could be enhanced in the case of binary black hole systems. Azimuthal electric currents are therefore likely, yielding a magnetic field component along the jet direction. Laboratory MHD experiments show that reversed-field pinches are fairly stable, and are therefore likely configurations of "fossil jets", and Benford [7] has proved that for jet-built RFP structures, the same simple MHD stability conditions for a jet guarantee stability, even after the jet turns off. We discuss elsewhere [4] further details the stability of MHD structures, but here we concentrate on radio and X-ray observations of fossil jets. Recent detections of several ghost cavities in galaxy clusters [5] – often, but not always, radio-emitting – suggest that the cluster hot plasma stays well separated from the bulk of the relativistic plasma on a timescale of $\sim$100 Myr. This means that magnetic structures made stable while a jet is on can evolve into fossils that persist long after the building jet current has died away. These may be the relic radio "fossils", "ghost bubbles" or "magnetic balloons" found in clusters and made visible by contrast against the X-ray emission as seen in Hydra A [8]. Giant radio galaxies such as Cyg A are building such structures now. Such fossils have a massive inventory of magnetic energy that can be $\sim 10^{60}$ erg [6]. Such enormous amounts of energy can only come from the gravitational infall energy of a supermassive black hole, when jets convey a few percent of the energy outward.

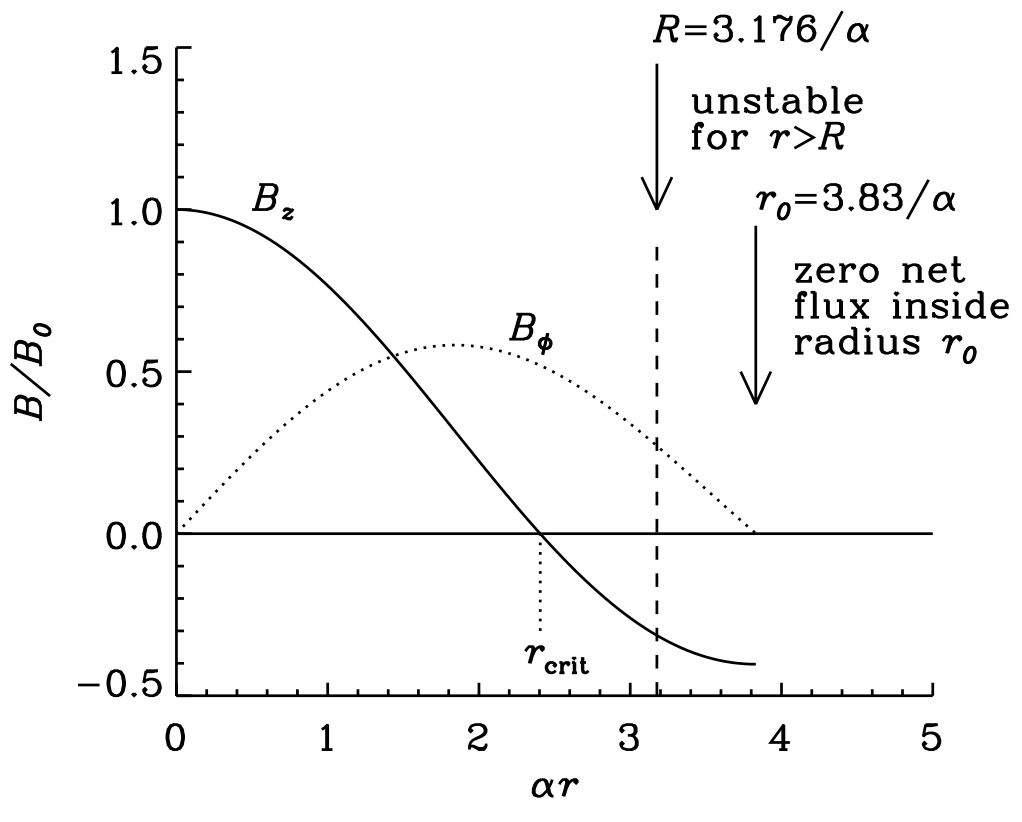

Figure 1: Magnetic field components of an RFP.

## 3 Particle acceleration in a reversed field pinch

The simplest idealization would be for an infinite cylindrical jet where the magnetic field in cylindrical coordinates $(r, \phi, z)$ is [9]

$$B_r = 0, \;\; B_\phi(r) = B_0 J_1(\alpha r), \;\; B_z(r) = B_0 J_0(\alpha r), \tag{1}$$

and this has been shown [10] to be stable for $\alpha r < 3.176$. This radius for stability which we take to be $R \equiv 3.176/\alpha$ is where a conducting wall with a large inertial mass would be present in an experimental situation, and provide part of the circuit along which a return current could flow, which we assume here to be the cocoon. The magnetic field is shown in Fig. 1. Notice the longitudinal field changes sign at $r_{\rm crit} = 2.405/\alpha$, the first zero of $J_0(\alpha r)$.

The current density and vector potential are everywhere proportional to the magnetic field,

$$\vec{j}(r) \;\; = \;\; \frac{\vec{B}(r)\alpha}{\mu_0} \;\; ({\rm A\ m^{-2}}), \quad \vec{A}(r,\phi,z) = \frac{1}{\alpha}\vec{B}(r,\phi,z). \tag{2}$$

Electric fields from reconnection are *emfs* induced according to Faraday's law, and so the electric field will be more extensive. The cold plasma (pressure is low) responsible for currents which maintain the magnetic structure cannot short out these electric fields, since they are inductively driven everywhere in the structure, allowing acceleration outside the reconnection zone. Assuming a flow of flux lines toward the reconnection region, the field will be changing everywhere and will induce an electric field. The simplest way of estimating this global electric field is by assuming an exponential decay of the magnetic field,

$$\vec{B}(t) = B_0 e^{-t/t_{\rm dec}}[J_1(\alpha r)\hat{\phi} + J_0(\alpha r)\hat{z}]. \tag{3}$$

Then,

$$\vec{\mathcal{E}} = -\frac{\partial \vec{A}}{\partial t} = 3.18 \times 10^{-5} \left(\frac{B_0}{10\ \mu\mathrm{G}}\right)\left(\frac{R}{100\ \mathrm{kpc}}\right)\left(\frac{t_{\rm dec}}{\mathrm{Gyr}}\right)^{-1} \times [J_1(\alpha r)\hat{\phi} + J_0(\alpha r)\hat{z}]e^{-t/t_{\rm dec}} \quad (\mathrm{V\ m^{-1}}) \tag{4}$$

We have simulated charged particle trajectories in the RFP magnetic field including the effect of energy change in the induced electric field. We inject particles uniformly and isotropically over the surfaces of disks of radius $R$ at both ends of the fossil jet of length $L$, and follow their motion until they escape. A typical example is shown in Fig. 2(a).

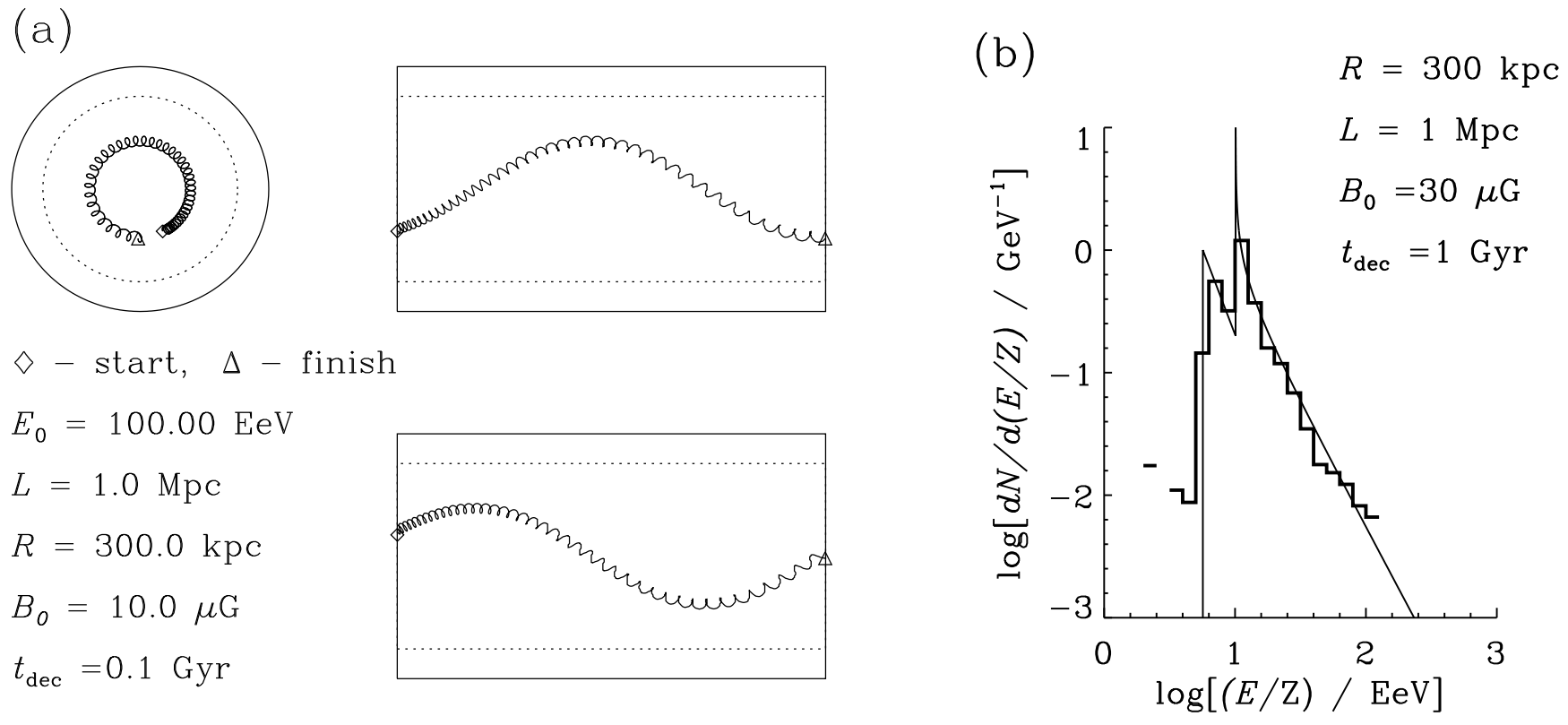

Figure 2: (a) Three orthogonal views showing a typical trajectory in the RFP fields, and critical radius (dashed). (b) Histogram: output spectrum for monoenergetic injection at $E_0/Z = 10^{18}$ eV, and fossil jet parameters as specified, and following particle trajectories as they undergo helical motion along field lines. Solid curve: shows analytic result from Fig. 3(b).

Ultra-relativistic particles of charge $Ze$ are injected with energy $E_0$ and their final energies are binned as shown in Fig. 2(b). Since the induced electric field is in the same direction as the magnetic field, energy is gained as particles move *along* field lines. Since the induced electric field is proportional to the magnetic field according to Eq. 4, positively charged particles will gain energy for pitch angles less than 90° and lose energy if their pitch angles are greater than 90°.

The energy gain on traversing the fossil jet length $L$ will actually depend on the pitch angle $\psi$ *of the helical magnetic field line* acting as the guiding centre. So, positive particles injected into the RFP with $r < r_{\rm crit}$ will gain energy while moving in the positive $z$ direction, and those injected with $r_{\rm crit} < r < R$ will gain energy while moving in the negative $z$ direction. The increase in energy of ultra-relativistic particles of charge $Ze$ is

$$E_{\rm gain} = E^0_{\rm gain} \frac{J_0(\alpha r)^2 + J_1(\alpha r)^2}{J_0(\alpha r)} \tag{5}$$

where

$$E^0_{\rm gain} \approx (10^{18} Z \text{ eV}) \left(\frac{B_0}{10\ \mu\text{G}}\right) \left(\frac{L}{\text{Mpc}}\right) \left(\frac{R}{100\ \text{kpc}}\right) \left(\frac{t_{\rm dec}}{\text{Gyr}}\right)^{-1} \tag{6}$$

and this is plotted in Fig. 3(a). Note that as $r \to r_{\rm crit}$, $p_{\rm gain} \to \infty$.

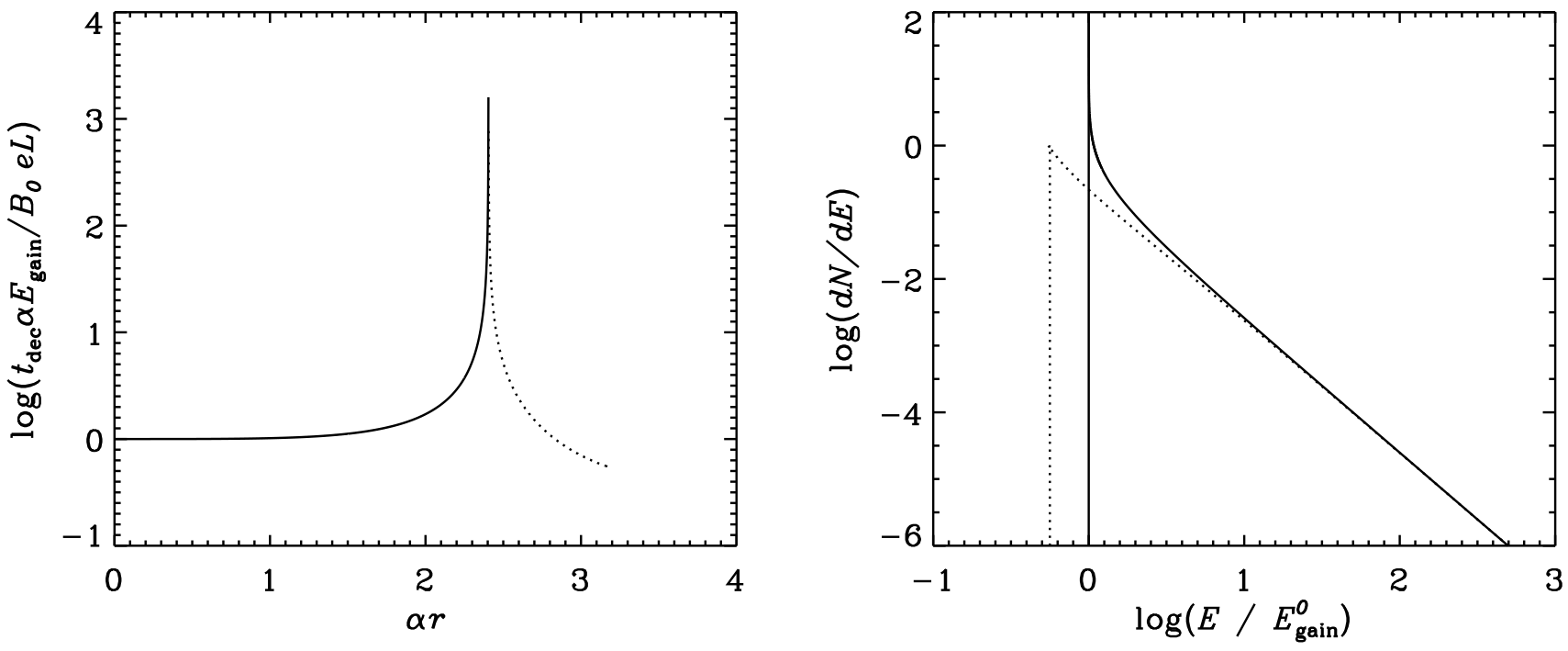

Figure 3: (a) Energy gain of particles injected at one end of the RFP of length $L$ and exiting at the other – solid curve for positive particles traveling in positive $z$ direction, dotted for positive particles traveling in negative $z$ direction. (b) Spectrum of accelerated particles – curves have same meaning as in part (a)

We can work out the energy spectrum as follows,

$$\frac{dN}{dE_{\rm gain}} = \frac{dN}{dr}\left[\frac{dE_{\rm gain}}{dr}\right]^{-1} \tag{7}$$

where $dN/dr$ is the distribution in radius of the injection points. For injection at one end of the RFP we would have uniform injection over the disk of radius $R$, giving

$$\frac{dN}{dr} = \frac{2r}{R^2} \qquad \text{for } 0 < r < R, \tag{8}$$

Differentiating Eq. 5 gives $dE_{\rm gain}c/dr$.

Thus, from Eqn. 7 we have $dN/dE_{\rm gain}$ as a function of the parameter $r$, and from Eqn. 5 we have $E_{\rm gain}$ as a function of the parameter $r$, and so we can plot $dN/dE_{\rm gain}$ vs. $E_{\rm gain}$, and this is shown in in Fig. 3(b). In Fig. 2(b) we have added the analytical spectrum and compared it with that obtained by following particle trajectories. The analytic slope, which asymptotically is $E^{-2}$, is consistent with the histogram. Note that the "double peaked" structure is due to separate contributions from injection at $r < r_{\rm crit}$ and $r > r_{\rm crit}$. The shape of the spectrum reflects a geometric property of the acceleration mechanism, as particles near $r_{\rm crit}$ being preferentially accelerated to become UHE CRs.

## 4 Discussion

For an extragalactic source distribution producing an $E^{-2}$ spectrum of protons, Lipari [11] estimates the local power requirement to be $\sim 10^{50}$ erg Mpc$^{-3}$ y$^{-1}$. Decaying magnetic fields with local filling factor $\eta_B$ lose energy at a rate

$$\dot{u}_B \quad \sim \quad 10^{53}\eta_B \left(\frac{B_0}{10\ \mu\text{G}}\right)^2 \left(\frac{t_{\rm dec}}{\text{Gyr}}\right)^{-1} \quad \text{erg Mpc}^{-3}\ \text{y}^{-1}$$

Magnetic fields from quasars can fill up to 5–20% of the intergalactic medium [12] – probably higher locally since our Galaxy is in a "Wall". Indeed, Gopal-Krishna & Wiita [13] estimate the fractional relevant volume that radio lobes born during the quasar era cumulatively cover is $\sim$0.5. Hence, our crude energetics arguments show fossil AGN structure decay could well be responsible for the observed UHE CR.

The spectrum of accelerated particles will cut off at some maximum momentum determined by either the finite thickness of the reconnection zone (recall that in the analytic approximation as $r \to r_{\rm crit}$, $E_{\rm gain} \to \infty$), or by the gyroradius increasing so that it is no longer much less than the radius of the fossil. From Fig. 1, we see that for $r < R$ the magnetic field is in the range $0.4B_0 < B < B_0$. Hence, the condition $r_L \ll R$ implies

$$E^{\rm max}_{\rm gain} \quad \ll \quad (10^{21} Z\ \text{eV}\ ) \left(\frac{B_0}{10\ \mu\text{G}}\right) \left(\frac{R}{100\ \text{kpc}}\right). \tag{9}$$

The spectrum of UHE CR observed at Earth would have contributions from nearby fossil jets at different distances, with different powers and each having different dimensions and magnetic fields, and hence a range of $E^0_{\rm gain}$ and $E^{\rm max}_{\rm gain}$. Given that several percent of the universe's volume may house such slowly decaying structures, these fossils may even re-energize ultra-high energy cosmic rays from distant/old sources, offsetting the GZK-losses due to interactions with photons of the cosmic microwave background radiation and giving evidence of otherwise undetectable fossils.

For an individual fossil, the cut-off is expected to be rigidity dependent, implying the observed composition would change from light to heavy close to the cut-off if one or two nearby AGN fossils dominate. However, if distant sources dominate nuclei will be photo-disintegrated by interactions with CMBR photons, and in this case the composition would remain light to the highest energies if distant sources or fossils dominated. Otherwise the composition could be mixed near the observed cut off.

We expect most of the fossil jets to be below the sensitivity of current radio telescopes, based on the work of Blundell & Rawlings [14], and it is impossible at the present time to make firm predictions for the expected UHE CR intensity at Earth. However, this may well change when the SKA (www.skatelescope.org/) is commissioned. Nevertheless, we have demonstrated that it is possible for this process to accelerate protons to UHE, and nuclei to a $Z$ times higher energy, and shown that the power requirements may reasonably be achieved given plausible volume filling factors.

In conclusion, remnants of jets and their surrounding cocoons may still be present around or close to galaxies which contain AGN which are now no longer active. These fossil jets are colossal MHD structures and may have total energies $\sim 10^{60}$ erg. We have shown that decay of such structures over timescales of ~Gyr induces large-scale electric fields which accelerate cosmic rays an $E^{-2}$ power-law up to ultra-high energies. Energetics arguments show that this provides a plausible mechanism for the origin of the UHE CR.

Finally, Heinz & Sunyaev [15] have shown that particles should be accelerated at the reverse shock of a micro-quasar jet colliding with the interstellar medium, and that this may give a contribution to the galactic cosmic rays up to about ~10 GeV. We mention here the possibility of particle acceleration by induced electric fields in a micro-quasar's decaying remnant magnetic bubbles after the micro-quasar's jets have switched off, as in the case of fossil AGN, if they form self organized magnetic structures such as the RFP. The minimum energy of accelerated particles would the be

$$E^0_{\rm gain} = (10^{12} Z \text{ eV}) \left(\frac{B_0}{0.1 \text{ mG}}\right) \left(\frac{L}{1 \text{ pc}}\right) \left(\frac{R}{1 \text{ pc}}\right) \left(\frac{t_{\rm dec}}{\text{Myr}}\right)^{-1} . \quad (10)$$

Such a mechanism might also apply to decaying pulsar wind nebulae, as well as fossil micro-quasars, and could be responsible for emission in unidentified EGRET and TeV gamma-ray sources, as well as contributing to galactic CR up to $\sim 10^{12} Z$ eV.

**Acknowledgements**

We thank Garang Yodh and Roger Blandford for stimulating discussions, and KIPAC and SLAC for hospitality in May 2007. RJP thanks the Australian Research Council for support through a Discovery Project grant.

# COSMOLOGICAL LARGE SCALE ANISOTROPIES IN THE HIGH-ENERGY GAMMA-RAY SKY

ALESSANDRO CUOCO [a]

[a] *Institut for Fysik og Astronomi, Aarhus Universitet Ny Munkegade, Bygn. 1520 8000 Aarhus Denmark*

**Abstract**

The interactions that characterize the propagation of $\gamma$ photons in the TeV energy range introduce a cosmological horizon at a distance of few hundreds Mpc, implying a correlation of the high-energy gamma-ray sky with the local Large Scale Structures. We provide detailed predictions of the expected anisotropies based on the map of the local universe from the PSCz astronomical catalogue. We then discuss the chances to detect the predicted signal with the forthcoming satellite observatory GLAST and the extensive air showers detectors Milagro, and HAWC.

## 1 Gamma Astronomy

The 0.1–10 TeV range represents one of the "last" photonic windows yet to be explored at large distances. Besides single sources, wide field of view instruments like the extensive air showers detectors (EAS) Milagro, Argo and the planned HAWC and satellite-based observatories like GLAST are sensitive to diffuse $\gamma$-ray emissions.

A particularly interesting emission is the extragalactic diffuse $\gamma$-ray background (in the following, cosmic gamma background, or CGB). The CGB is a superposition of all unresolved sources emitting $\gamma$-rays in the Universe and provides an interesting signature of energetic phenomena over cosmological time-scales. While a clear detection of this background has been reported by the EGRET mission [1], its origin is still uncertain, despite the fact that many models have been proposed. The most likely contribution is the one from unresolved blazars, i.e. beamed population of active galactic nuclei, with (probably sub-leading) components from ordinary galaxies, clusters of galaxies, and gamma ray bursts. However, exotic possibilities like dark matter annihilation have been proposed, that are compatible with existing data and constraints. It is extremely difficult to test such models as long as the only observable is the energy spectrum. Recently, it was proposed to use the peculiar small-scale anisotropy encoded in the MeV-GeV gamma sky to probe dark matter [2, 3] or astrophysical [4, 5] contributions to the CGB. We further study this topic, with particular emphasis on the large scale anisotropy in the energy range 0.1-10 TeV. The lower part of this range will be probed by the GLAST telescope [6], while the energy window above the TeV is in principle accessible to EAS detectors like Milagro [7] and Argo [8]. Different candidates to explain the CGB predict distinctive large scale features, even when similar energy spectra are expected. This is a consequence of the combined effect of a cutoff distance after which $\gamma$ of energy starting from about 100 GeV (the very-high energy regime, VHE) can travel undamped to us, and of the anisotropic distribution of matter in the local universe (i.e., within a few hundred Mpc from us), the local Large Scale Structures (LSS). We shall then use the redshift Point Sources Catalogue (PSCz) [9] as tracer of the real structures in the nearby universe, thus producing maps of the VHE gamma sky. For a more complete and detailed discussion of the present issues we refer the reader to the paper [10].

It is interesting to note that a similar horizon (and a similar correlation with LSS) is expected for cosmic rays particles of energy $\gtrsim 10^{19}$ eV (the so called ultra-high energy (UHE) regime) [11] (see also [12]). Indeed, possibly a fraction of the CGB could be associated to the $\gamma$ cascades produced by the energy losses from the propagation of UHE hadrons [13].

The observation of the extra-galactic TeV sky is limited by the presence of the much more intense galactic foreground. Indeed, the detection of a diffuse emission along the galactic plane has been recently reported by the MILAGRO collaboration with a median energy of $\sim$20 TeV [14]. Interestingly, the emission detected by MILAGRO significantly exceeds the flux expected from the models of diffuse emission in the galaxy, challenging our present understanding of the galactic radiation production mechanism [15]. This result clearly shows the importance and complementarity of large field of view observations for a full understanding of the astrophysical and/or exotic phenomena taking place in the gamma sky. The detection of the anisotropies of the extra-galactic gamma-sky, analogously, would complement the point sources study in the characterization of the extra-galactic high-energy gamma sky.

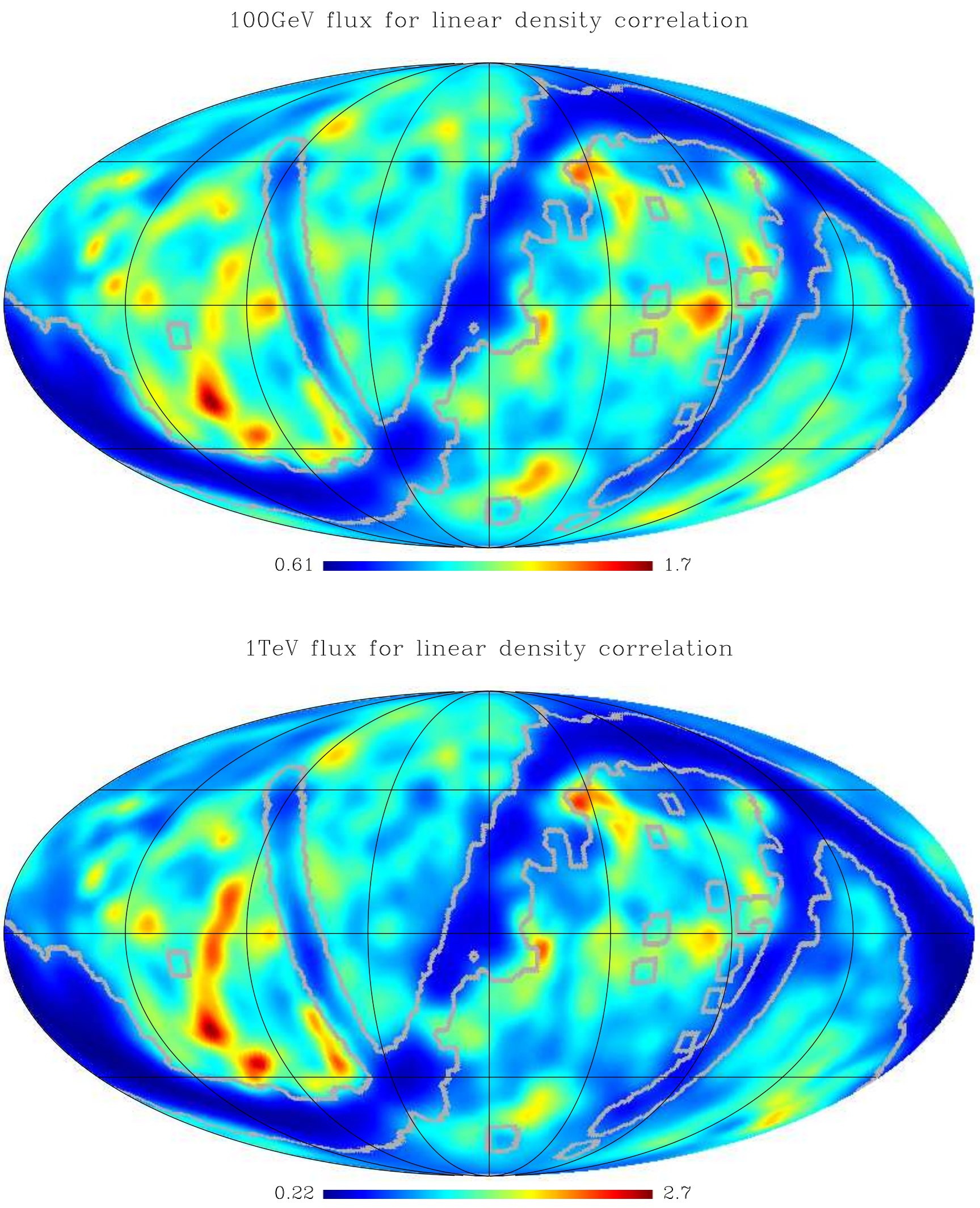

Figure 1: Equatorial density $\gamma$ sky maps from the PSCz catalogue for $E_{\rm cut} =$ 100 GeV and 1 TeV. The color scale is linear and the average flux outside the mask of the PSCz is normalized to 1 so that to represent adimensional maps. The mask of the PSCz survey is indicated by the thick grey contour.

## 2 Sky maps and forecast

In Fig. 1 we plot the resulting $\gamma$ maps from the PSCz catalogue in equatorial coordinates for $E_{\rm cut} = 100$ GeV and 1 TeV. For the case of the map with $E_{\rm cut} = 100$ GeV, modulo the "hole" due to the mask the pattern is quite isotropic, with some hot spots like e.g. from the Virgo and Perseus Clusters. Other structures which appear are the Shapley concentration and the Columba cluster (for a key of the local cosmological structures see [11]). Given the limited statistics of GLAST at high energies, the TeV map is of interest especially for the EAS gamma detectors like MILAGRO. We see in this case that the nearest structures, forming the Super-Galactic Plane, dominate. Of course, from the Northern emisphere (where all the present or planned EAS instruments are located) only the upper part of the map is visible. Here, the Virgo Cluster and the Perseus cluster offer the strongest anisotropy.

To obtain the maps the effects of the propagation of the particles through the relevant Infrared/Optical and Microwave backgrounds have been properly taken into account. Further details can be found in the paper [10].

What are the real chances to detect these anisotropies with the data from the forthcoming experiments? To answer the question we have performed an harmonic decomposition of the maps $f(\hat{\Omega}) = \sum_{lm} a_{lm} Y_{lm}(\hat{\Omega})$ and then assessed the predicted shot noise errors on the $a_{lm}$ due to the finite statistics collected by a given experiment. In particular, the errors read as $\sigma^2_{a_{lm}} = 4\pi f_{\rm sky}/N_\gamma \,(1 + N_{\rm CR}/N_\gamma)$ where $N_\gamma$ and $N_{\rm CR}$ are respectively the numbers of photons and background events collected and $f_{\rm sky}$ is the fraction of the sky accessible to the experiment (assumed with uniform acceptance over this region).

In Fig. 2 we report the coefficients $a_{lm}$'s up to $l_{\rm max} = 10$ calculated from the PSCz gamma maps of Fig.1, with the relative errors for a 4 year exposure of the GLAST mission and 10 years for the EAS Milagro and HAWC. Performing the analysis in terms of the harmonics coefficients $a_{lm}$ instead of angular power spectrum $C_l$'s has the advantage of exploiting the full information present in the map (for an angular scale of order $\theta = \pi/l$) without the limit imposed by cosmic variance.

GLAST should be able to detect some structures above 100 GeV at the $2\sigma$ level, while, on the contrary, instruments like MILAGRO may hardly find hints of structures at 1 TeV (gray band in the bottom panel of Fig. 2). We note that the intensity of the anisotropies increases sensibly from the 100 GeV energy band to 1 TeV, but, despite the increased signal and statistics collected, ground arrays have an hard task in detecting the CGB fluctuations. The signal detected by EAS arrays is infact buried under an heavy background of Cosmic Rays events that overwhelms the gamma signal typically by a factor of order $10^5$! Rejection capability helps in removing part of the background. Note that GLAST is expected to have an excellent background identification, so that only cosmic rays in the amount of $\sim 6\%$ of the gamma flux pass the cuts. On the other hand, EAS experiments have a poor rejection capability, which increases typically the gamma content of the diffuse flux by no more than one order of magnitude. Therefore even after gamma/hadron separation, the anisotropies of the gamma sky have to be identified against a *quasi*-isotropic background which is $\sim 10^4$ larger than the gamma flux.

The intensity of the anisotropies also potentially increases in some scenarios of DM gamma production. In this case, the gamma emission is expected to follow the square of the DM density distribution due to the peculiar production mechanism proceeding through DM-DM annihilation [2]. A detailed study of this issue is in progress [16].

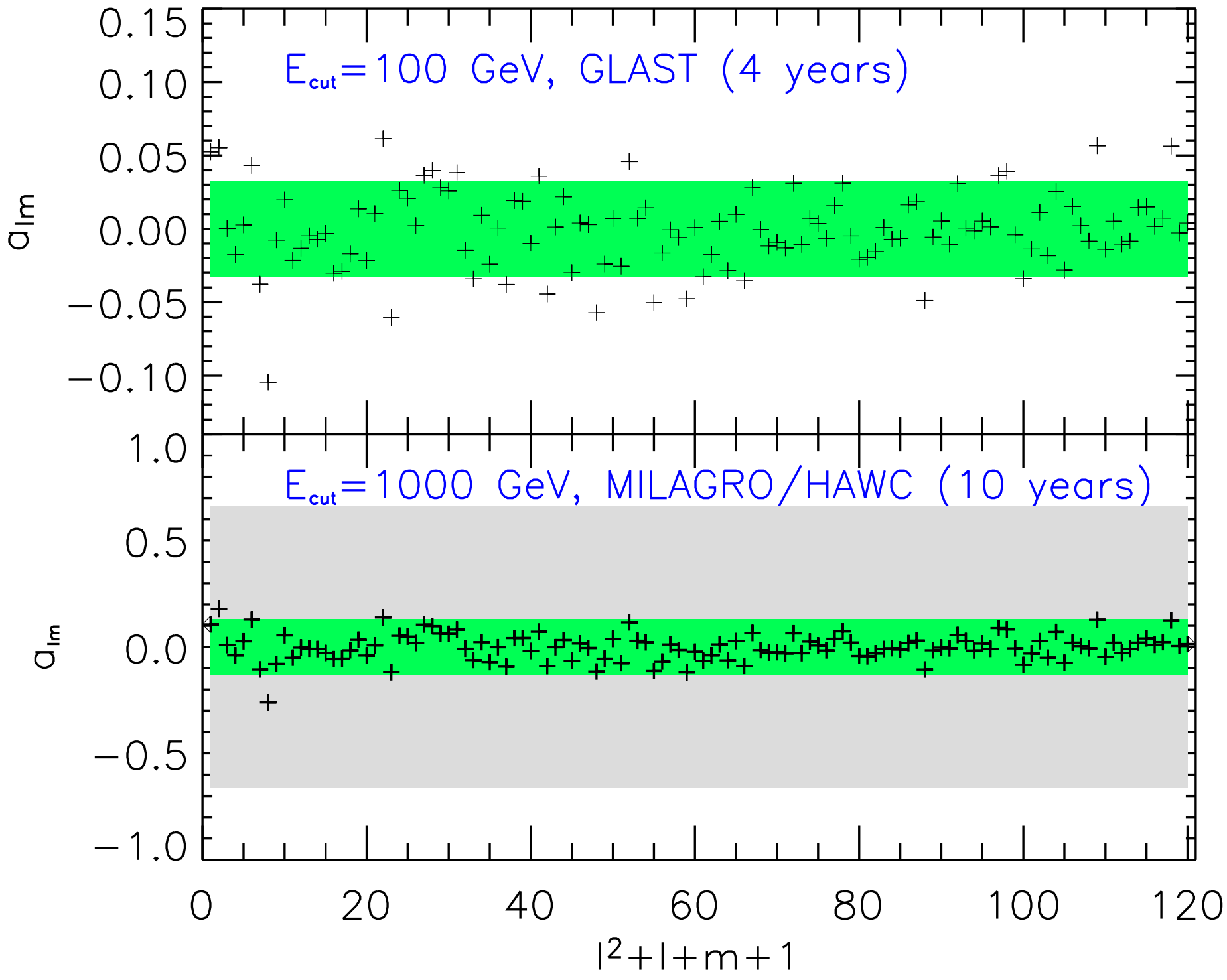

Figure 2: The coefficients $a_{lm}$ up to $l_{\max} = 10$ calculated from the PSCz gamma maps of Fig. 1. The shaded band shows the 1-$\sigma$ shot noise error; in the bottom panel the inner shaded region refers to HAWC, the outer one to MILAGRO.

It is further worth to notice that for an EAS detector the error on the $a_{lm}$'s scales as $\sqrt{N_{\rm CR}}/N_\gamma$. Therefore the reduction of the shot-noise error goes like $(t \cdot A_{\rm eff})^{-1/2}$ (both $N_{\rm CR}$ and $N_\gamma$ grow linearly with $t \cdot A_{\rm eff}$, the collecting time times the effective area of the experiment), or equivalently as $\sqrt{h_{\rm cut}}/g_{\rm cut}$ (where $g_{\rm cut}$ and $h_{\rm cut}$ are the fraction of $\gamma$'s and hadrons that survive after the trigger cuts): improving the exposure is equally important as improving the gamma/hadron separation capability. A simple inspection of Fig. 2 reveals that for a realistic detection of the features in the VHE sky one would need the improvement in effective area planned to be reached by instruments like HAWC [17] (see inner green band in the bottom panel of Fig. 2). An instrument like ARGO is expected to have performances in between MILAGRO and HAWC, and may have some chance especially if a significant improvement in hadron rejection can be made. Also, note that due to their altitude HAWC (planned at an altitude of ~4000 m [18]) and ARGO (at an altitude of 4300 m) have a significant acceptance of sub-TeV events. While the gamma/hadron separation is less efficient at lower energies, the higher statistics may help in revealing these structures. Indeed, the ARGO collaboration has recently presented the first preliminary results [19], showing a significant sensitivity increase and lower energy threshold with respect to MILAGRO (located at an altitude of 2630 m).

We also note that these estimates are somewhat conservative: summing the power at different $l$'s may favor the detection (see e.g. [5]), and cross-correlating directly with the maps we have produced would eventually rely on the whole information.

Finally, the ultimate limitation in detecting anisotropies in the gamma sky with EAS observatories is expected to come from the understanding of the intrinsic anisotropy in the CR background that are generally measured at the level of few$\times 10^{-4}$ and are then comparable to the expected intrinsic $\gamma$ anisotropy. One possible strategy to tackle this problem, could consist in reversing the gamma cut and thus enriching the sample in hadronic showers thus helping in identifying and removing non-gamma anisotropies.

## 3 Acknowledgements

It is a pleasure to thank S. Hannestad, T. Haugbølle, G. Miele, P. D. Serpico and H. Tu for the collaborations and the stimulating discussions on the topics here described.

# OBSERVATIONS OF GAMMA RAY BURSTS WITH THE MAGIC TELESCOPE

M. Gaug[a], D. Bastieri[b], N. Galante[c], M. Garczarczyk[c], F. Longo[d], S. Mizobuchi[b], V. Scapin[e] for the MAGIC Collaboration

[a] *Instituto de Astrofísica de Canarias, Via Láctea s/n, La Laguna, Spain*

[b] *Università degli Studi di Padova, Via Marzolo 8, Padova, Italy*

[c] *Max-Planck-Institut für Physik, Föhringer Ring 6, München, Germany*

[d] *INFN sez. Trieste, Via Valerio 2, Trieste, Italy*

[b] *Università degli Studi di Udine, Via delle Science 208, Udine, Italy*

## Abstract

The MAGIC telescope, taking regular data since more than two years, has been designed especially to observe Gamma Ray Bursts (GRB). Its low energy threshold at well below 100 GeV as well its fast reaction and repositionning capabilities make it an ideal instrument to follow up GRBs rapidly in the very high energy (VHE) energy range. In the past two years, the MAGIC collaboration has demonstrated in two cases that the telescope is able to observe the promt emission of GRBs, if rapid alerts from satellite experiments are provided. Due to the absorption of gamma-rays by the Meta-galactic Radiation Field (MRF), it is necessary that the GRB lies at a redshift below $z \sim 1$ in order to expect a signal around 100 GeV. Upper limits have been published for a total of 12 GRBs.

## 1 Introduction

Both the physical origin of GRBs as possible emission of gamma-rays in the VHE regime are still under debate, even 40 years after their discovery. Especially the detection of high-energy radiation will lead to a deeper understanding of the involved emission processes. Many attempts were therefore made in the past to observe GRBs in the GeV and TeV energy range, however without stringent evidence for the existance of a very high-energy emission component, neither during the prompt emission nor during the afterglow (see e.g. [1, 2, 3, 4, 5]). The only significant detections were made by the EGRET detector which could observe seven GRBs emitting gamma-rays with energies between 100 MeV and 18 GeV [6] without showing any apparent cut-off in the energy spectrum. Results from the TASC shower counter, part of the EGRET detector, jointly fit with BATSE data, indicate that the spectrum of at least one burst contained a very hard, luminous, long-duration component presumably due to ultra-relativistic hadrons with a differential photon flux spectral index of $\alpha = -1$ with no cut-off up to the TASC detector energy limit at 200 MeV [7].

Since the start of data taking of the second generation of Imaging Atmospheric Cherenkov Telescopes (IACT), a sufficiently good flux sensitivity and low energy threshold have been achieved to follow the GRB spectra up to energies of $\gtrsim$100 GeV. Nevertheless, as their small field of view does allow prompt observations only by - unprobable - serendipitous detection, they have to rely on external triggering, as the one provided by the automated satellite link to the Gamma-Ray Burst Coordinates Network (GCN), which broadcasts the coordinates of events triggered and selected by dedicated satellite detectors.

The detection of VHE emission from the GRB is important for comparing different theoretical models. In the framework of the popular fireball model, gamma-ray emission in the GeV-TeV range in the prompt and delayed phase is predicted by several authors. Possible emission mechanisms appear in leptonic and hadronic models. In the first case, inverse-Compton (IC) scattering by electrons in internal or external shocks (e.g. [9, 10]), IC in the afterglow shocks (e.g. [11, 12]) and IC by electrons responsible of optical flashes [13, 14] have been proposed. Even considering pure electron synchrotron radiation, measurable VHE gamma-ray emission for a significant fraction of GRBs is predicted [12]. Suggested hadronic models comprise proton-synchrotron emission, photon-pion production and neutron cascades (e.g. [15, 16, 17, 18])

The recent observations of X-ray flares in the early afterglow phase by the SWIFT satellite [19] suggested the possibility of correlated gamma-ray emission extending to the GeV and TeV energy range lasting for $10^3$ s to $10^4$ s where strong GeV-TeV flares from IC scattered photons in the forward shock were predicted [20].

However, due to the strong absorption of gamma-rays by the MRF [21, 22], it is necessary that the GRB lies at a redshift below $z \sim 1$ in order to expect a signal around 100 GeV.

## 2 The MAGIC Telescope

The Major Atmospheric Gamma Imaging Cherenkov (MAGIC) telescope [23], located on the Canary Island of La Palma (2200 m a.s.l, 28.45°N, 17.54°W), is currently the largest single dish (17 m diameter tessellated reflector dish) IACT with the lowest energy threshold. As recently reported [24], MAGIC is able to reconstruct energy spectra down to about 60 GeV if the observation is carried out at a low zenith angle. The faint Cherenkov light flashes are recorded by a pixellized camera, comprised of 577 photo-multiplier tubes. In its current configuration, MAGIC has an accuracy in reconstructing the arrival directions of gamma-rays (hereafter called point-spread-function PSF) of about 0.1°, slightly depending on the analysis. The telescope is focused at 10 km, i.e. the typical position of the shower maximum for energies at the threshold. The trigger collection area for gamma-rays is of the order of $10^5$ $m^2$, increasing further with the zenith angle of observation. Figure 1 shows the effective area for gamma-rays, after typical analysis cuts for three representative zenith angles under which the telescope observes the GRB. One can see the threshold energy depends sensibly on the zenith angle.

The incident light pulses are converted into optical signals and transmitted to a control house over 162 m of optical fiber. There, the signals are converted back and digitized by fast Flash ADCs (FADCs). At the beginning of this year, the FADC readout was upgraded from 300 MSamples/s to 2GSamples/s increasing further the sensitivity of the telescope. All data presented here were still taken with the old system, except for the very last burst shown. Due to its light carbon fibre structure, MAGIC is the only IACT with fast repositionning capabilities, designed especially for the observation of GRBs [25]. In case of a Target of Opportunity alert by GCN, an automated procedure takes only few seconds to terminate any pending observation, validate the incoming signal and start slewing toward the GRB position. In its current configuration, the repositionning speed of MAGIC is on average 42 s for GRB alerts. This allowed to set upper limits on the GRB prompt emission in some cases already [26].

Upon receipt of the alert, the telescope operators have to validate the alert whereupon the telescope enters in automatic fast-movement mode, moves directly to the transmitted GRB coordinates and starts observation immediately. Recently, additional security hardware has been installed which allows to make the procedure fully automatic and recover thus the couple seconds lost by the reaction time of the operators. As long as the GRB coordinates are visible below a zenith angle of 60°, the source is tracked for typically an hour until the observation of the previous source is recovered.

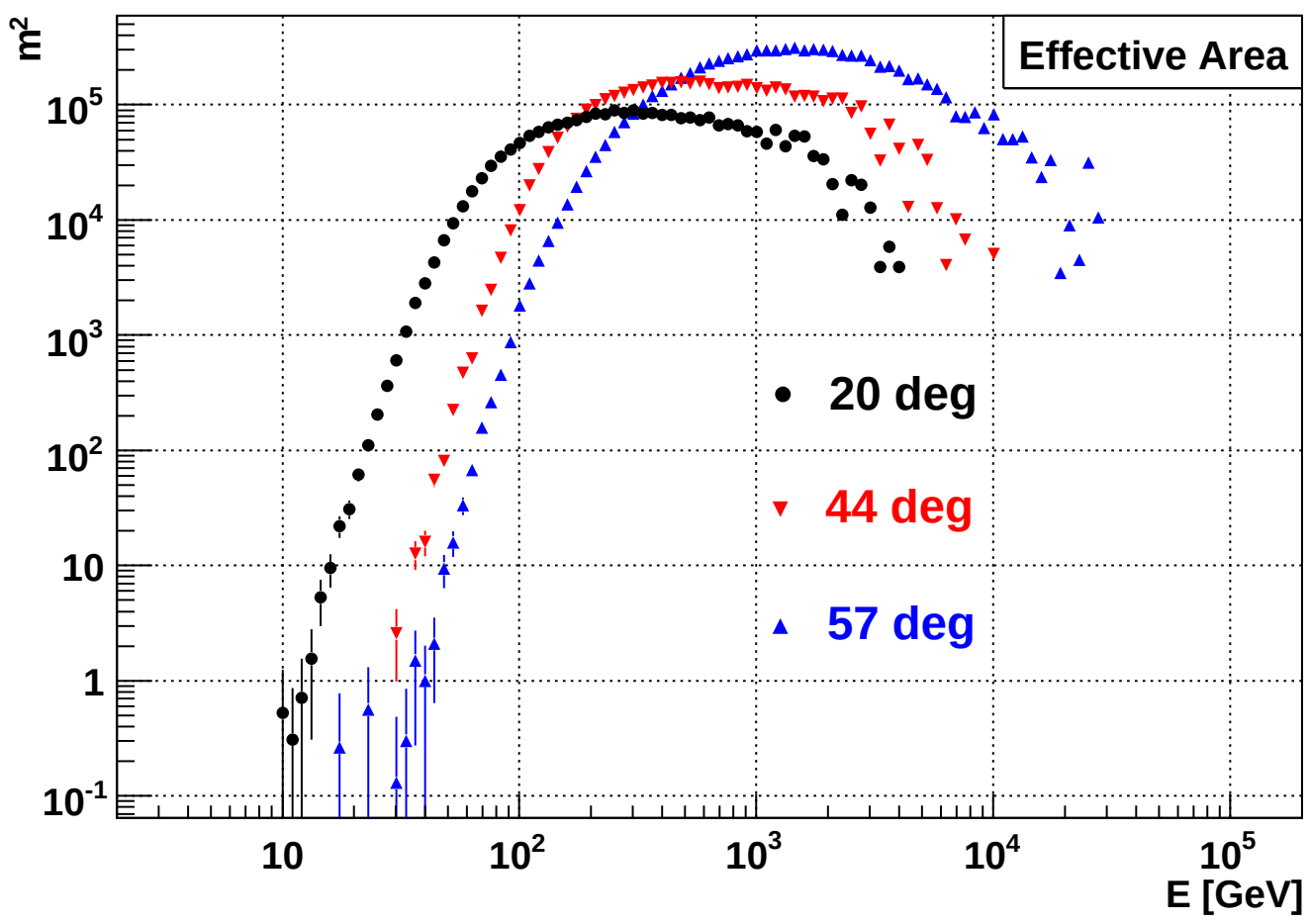

Figure 1: Effective area of MAGIC, after typical analysis cuts, for three different zenith angles

## 3 Blind test with the Crab Nebula

On 11$^{th}$ October 2005 at 02$^{h}$17$^{m}$37$^{s}$ UTC, the INTEGRAL satellite announced GRB051011 [28] at the position R.A. $= 5^{h}\ 34^{m}\ 47^{s}$, Dec. $= +21^{d}\ 54'\ 39''$ which later turned out to be triggered by gamma-ray emission from the Crab Nebula. As in the case of a real GRB, the MAGIC telescope immediately slewed to the source and started data taking at zenith angles between 33° and 42°. A total of 2814 seconds of data was gathered from the standard calibration source for IACTs, in a real blind test.

The analysis yielded a signal of about $14\,\sigma$ significance above 350 GeV corresponding to the sensitivity of the telescope. The analysis results shows that MAGIC can observe (arbitrarily chosen) benchmark fluxes of 5 Crab Units (C.U.) [1] with $5\,\sigma$ significance in 38 seconds above 300 GeV and in 90 seconds below 300 GeV.

[1] 1 Crab Unit (C.U.) is the measured Crab Nebula flux which can be fitted to $1.5 \cdot 10^{-6} \cdot E\,(\mathrm{GeV})^{-2.58}\mathrm{ph} \cdot \mathrm{cm}^{-2}\mathrm{s}^{-1}\mathrm{GeV}^{-1}$ above 300 GeV. At lower energies, the Crab Nebula flux is lower than the power-law fit.

## 4 Observations

In its first regular observation cycle, from April 2005 to May 2006, MAGIC collected nine GRBs, suitable for analysis. Since then, other GRBs were observed, out which four have been analyzed in a preliminary way. Table 1 summarizes the main properties of these thirteen bursts.

| GRB | Satellite | $T_{90}$ (s) | Fluence X-ray (erg cm$^{-2}$) | z | $E_{\rm thr}$ MAGIC (GeV) | U.L. Fluence (erg cm$^{-2}$) |
|---|---|---|---|---|---|---|
| GRB050421 | SWIFT | 10 | $1.8 \cdot 10^{-7}$ | - | 290 | $3.8 \cdot 10^{-8}$ |
| GRB050502a | INTEGRAL | 20 | $1.4 \cdot 10^{-6}$ | 3.8 | 190 | $6.2 \cdot 10^{-8}$ |
| GRB050505 | SWIFT | 60 | $4.1 \cdot 10^{-6}$ | 4.3 | 400 | $1.5 \cdot 10^{-7}$ |
| GRB050509a | SWIFT | 13 | $4.6 \cdot 10^{-7}$ | - | 290 | $7.7 \cdot 10^{-8}$ |
| GRB050713a | SWIFT | 70 | $9.1 \cdot 10^{-6}$ | - | 270 | $1.7 \cdot 10^{-7}$ |
| GRB050904 | SWIFT | 225 | $5.4 \cdot 10^{-6}$ | 6.3 | 95 | $1.1 \cdot 10^{-8}$ |
| GRB060121 | HETE-II | 2 | $4.7 \cdot 10^{-6}$ | - | 190 | $9.7 \cdot 10^{-8}$ |
| GRB060203 | SWIFT | 60 | $8.5 \cdot 10^{-7}$ | - | 210 | $3.9 \cdot 10^{-8}$ |
| GRB060206 | SWIFT | 11 | $8.4 \cdot 10^{-7}$ | 4.1 | 85 | $1.4 \cdot 10^{-7}$ |
| GRB060825 | SWIFT | 15 | $9.8 \cdot 10^{-7}$ | - | 115 | $3.6 \cdot 10^{-7}$ |
| GRB061028 | SWIFT | 106 | $9.7 \cdot 10^{-7}$ | - | 120 | $2.9 \cdot 10^{-7}$ |
| GRB061217 | SWIFT | 0.4 | $4.6 \cdot 10^{-8}$ | 0.8 | 400 | $1.0 \cdot 10^{-7}$ |
| GRB070412 | SWIFT | 40 | $4.8 \cdot 10^{-7}$ | - | 95 | $5.4 \cdot 10^{-7}$ |

Table 1: Summary of GRBs observed by MAGIC from April 2005 to March 2006. The last two columns denote the analysis energy threshold which depends strongly on the zenith angle of observation and the upper limit set on the first half an hour of MAGIC observation. The results shown for the last four bursts are still preliminary.

In two cases (GRB050713a and GRB060904), the observation took place while the prompt emission was still ongoing [26, 27]. Figure 2 shows the case of GRB050713a, where part of the prompt emission could be observed and upper limits in the VHE energy region be set. In six cases, the MAGIC observation overlapped with X-ray observations from space, namely in the case of GRB050421, GRB050713a, GRB050904, GRB060206, GRB060825 and GRB061028. During cycle-I, the X-ray telescope (XRT) onboard the Swift satellite detected two flares during the MAGIC observation window in case of GRB050421 and one flare in the case of GRB050904. During cycle-II, flaring activity was detected by XRT while observing GRB060825.

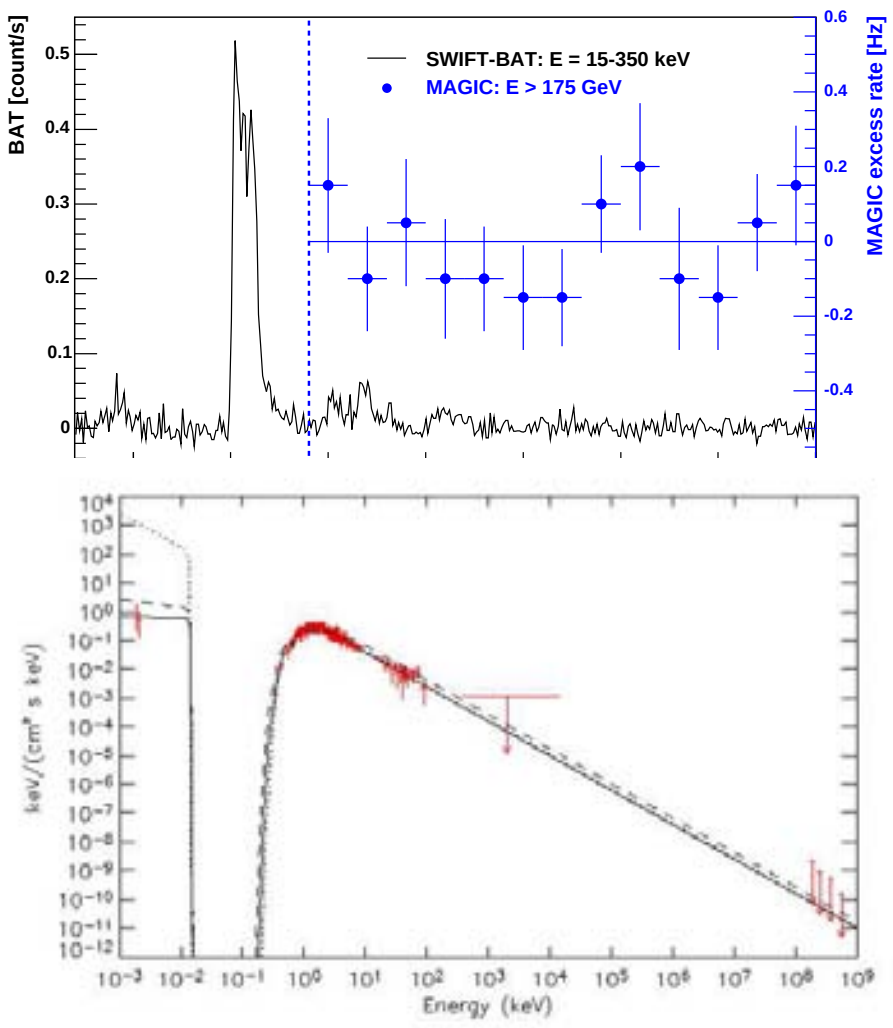

Figure 2: MAGIC observations of GRB050713a. Left: excess event rate, compared with Swift (BAT) observations of GRB050713a. The MAGIC observation started at 40 s from the BAT trigger, covering the second and third emission peak. Unfortunately, BAT triggered only on the main emission peak and not on the precursor at -60 s. In that case, MAGIC would have also fully observed the main peak. Right: Combined multi-wavelength SED of the early afterglow of GRB050713a. The three lines represent three emission models. Starting from 175 GeV, the upper limits set by MAGIC are in accordance with an extrapolated power law spectrum. In this particular case, the MAGIC sensitivity lay slightly above the expected synchrotron emission.

## 5 Analysis and Results

The recorded events are reconstructed and calibrated using the standard MAGIC analysis software [29]. The obtained signal charges are cleaned from spurious background from the light of night sky by requiring signals which exceed fixed reference levels and coincide in time with neighbouring channels. Standard Hillas parameters are calculated [30] and several pre-cuts applied which remove non-physical background images. The gamma-hadron separation is performed optimizing combinations of cuts by means of the Random Forest method [31]. The energy of the gamma-rays is estimated using the same approach, yielding an energy resolution of $\sim 30\%$ at 200 GeV.

Dedicated OFF data samples were selected for each GRB observation separately which match the ON data observation conditions with respect to zenith angle, background light intensity and atmospheric conditions. These OFF data samples were passed through the same analysis chain and used to determine the residual background due to hadronic showers.

Subsequently, signals are searched in the recorded prompt emission phase (if any), during X-ray flares, in the integrated first 1000 s of data, the entire data set and in time bins of 100 s. Up to now, no significant excess over background has been detected in any of these searches in any of the observed GRBs. Upper limits were derived using the method of [32] in different energy bins. Table 1 shows the obtained upper limits for the first half an hour of observation.

## 6 Conclusions

MAGIC is currently the fastest and most sensitive operative IACT at energies around 100 GeV and observes about 1 GRB per month. It has already shown its capability to observe part of the prompt emission phase of two GRBs. Due to the existing gamma-ray horizon, which limits the visibility of gamma-rays to distances less than $z \sim 1$ at energies $E > 100$ GeV, many of the obtained upper limits have only a small impact on the interpretation of GRB data. With the recent 2 GSamples/s FADC upgrade and other upgrades to further improve the repositionning speed, it is only a question of time until MAGIC will observe a sufficiently close-by burst to make a strong statement on the emission mechanisms of GRBs in the VHE regime.

## Acknoledgements

We would like to thank the IAC for excellent working conditions at the ORM at La Palma. The support of the German BMBF and MPG, the Italian INFN, the Spanish CICYT, ETH research grant TH 34/04 3, and the Polish MNiI grant 1P03D01028 are gratefully acknowledged.

# WHY TAU FIRST?

DANIELE FARGION [a,b], DANIELE D'ARMIENTO [a] PIER GIORGIO LUCENTINI DE SANCTIS [a].

[a] *Dipartimento di Fisica, Università di Roma "Sapienza", Pzle.A.Moro 2, Roma, Italy*

[b] *INFN, Sezione di Roma I, Pzle.A.Moro 2, Roma, Italy*

**Abstract**

Electron neutrino $\nu_e$ has been the first neutral lepton to be foreseen and discovered last century. The un-ordered muon $\mu$ and its neutrino $\nu_\mu$ arose later by cosmic rays. The tau $\tau$ discover, the heaviest, the most unstable charged lepton, was found surprisingly on (1975). Its $\nu_\tau$ neutrino was hardly revealed just on (2000). So why High Energy Neutrino Astronomy should rise first via $\nu_\tau$, the last, the most rare one? The reasons are based on a chain of three favorable coincidences found last decade: the neutrino masses and their flavor mixing, the UHECR opacity on Cosmic Black Body (GZK cut off on BBR), the amplified $\tau$ air-shower decaying in flight. Indeed guaranteed UHE GZK $\nu_\tau$, $\overline{\nu_\tau}$ neutrinos, feed by muon mixing, while skimming the Earth might lead to boosted UHE $\tau$,$\overline{\tau}$, mostly horizontal ones. These UHE $\tau$ decay in flight are spread, amplified, noise free Air-Shower: a huge event for an unique particle. To be observed soon: within Auger sky, in present decade. Its discover may sign of the first tau appearance.

## 1 The Cosmic multi-frequency spectra up to GZK edges

High energy neutrino astronomy at GZK [13],[16] limit is ready to be discovered. Its role may shine light in Universe understanding. Our present view of this Universe is summarized in radiation flux number spectra updated in Fig.1; its consequent energy fluency spectra in considered in Fig.2. The flux number spectra ranges from radio frequency to cosmic Black Body Radiation, BBR, toward Cosmic Rays up to Ultra High Energy ones, UHECR. For calibration also a BBR Fermi-Dirac for any eventual massless cosmic neutrino (at $1.9K$ temperature). Light neutrino with mass are probably non-relativistic and gravitationally clustered, not displayed here. The role in cosmology of relic neutrinos has been widely reviewed recently [8]. Infrared and the Optical photons fluxes are followed by UV and X cosmic ones. Last $\gamma$ Astronomy is at MeV-GeV-TeV band edge. In MeV region, Supernova Cosmic relic Neutrinos background may soon arise. Unfortunately the CR secondaries, $\pi^{\pm}$,$\mu^{\pm}$ are blurred as well as their $\nu_{\mu}$ and $\nu_e$ final atmospheric neutrinos: this cause to $\nu_{\mu}$ and $\nu_e$ astronomy to be also smeared and polluted. The inclined line on edge Fig. 1 tag the so-called Waxmann- Bachall limit,$\simeq E^{-2}$ as a minimal limit for GRB and GZK $\nu$ flux expectation. It is easy to note as this fluency is correlated to average cosmic radio background, as well as it is comparable to UHECR at ten EeV band and average GRB fluency. There are good reasons to foresee a WB GZK neutrino background too. The coexistence on many cosmic radiations makes the known windows on the Universe an exciting growing puzzle. Neutrino astronomy at different band may offer the key answers. Indeed while photons are neutral, un-deflected, offering Astronomy pictures, most of the Cosmic Ray are charged and smeared by galactic and cosmic magnetic fields. Therefore Cosmic Rays (CR) offer only an integrated, short-sighted Astrophysics. The presence of galactic magnetic fields are reminding us of the puzzling absence of magnetic monopoles in our Universe. The low energy multi-frequency spectra on left side (below TeV energy) is dominated by photons; at higher energies (TeV-PeVs) the photons are rarer and opaque to relic extra-galactic Infrared photons. Tens TeV photons arrive only from nearby Universe (hundred Mpc radius); at PeV energy, the cosmic Black Body Radiation (BBR) makes UHE photons bounded in our Local Group, (Mpc) size volume. Therefore only PeVs neutrinos may reach us from Universe edges. The right side Fig.1, the high energy one, is dominated by Cosmic Rays and its secondaries. The ruling dominance of solar photons and of its neutrinos is obviously hidden here to avoid confusions.Indeed beyond the MeV energies, where solar neutrino flux dominates, one expects a peculiar niche for the relic Supernova background, still on the edge of detection. At tens MeV atmospheric neutrino noise will pollute this SN signal. Hopefully upgraded underground SK (Super Kamiokande) detector might soon reveal the SN trace. At the same $10^7$ eV energy band, very rare and bright galactic supernova neutrino, as the famous SN1987A, might rarely blaze our Milky Way almost once a century. Future Megaton Neutrino detectors could observe even nearby Andromeda Supernova, making three times larger the previous rate. In the same energy range, much less power-full but more frequent neutrino burst, in all flavor, may rise from largest solar flares, once in a decade. They are better detectable by noise free anti-neutrino electron $\overline{\nu}_e$ component in Mega-ton detectors. Among the cosmic rays, the secondary Atmospheric Neutrinos arise, blurred and noisy as their parental CR. Therefore Neutrino muon and electron signals are largely polluted by abundant atmospheric secondaries. The charmed pions (rare parents of tau) are hardly produced by CR respect to pion and Kaon ($< 10^{-5}$) [1]. Energetic atmospheric $\nu_\mu$ cannot feed $\nu_\tau$ via neutrino mixing, because too short distance for known mass splitting. Then atmospheric HE $\nu_\tau$ are suppressed and its astronomy is noise free. Unfortunately TeVs Tau neutrino are still difficult to be disentangled from other neutral current neutrino events. But higher energy ones, in PeV-EeV band, may reveal themselves loudly [9]. As discussed below.

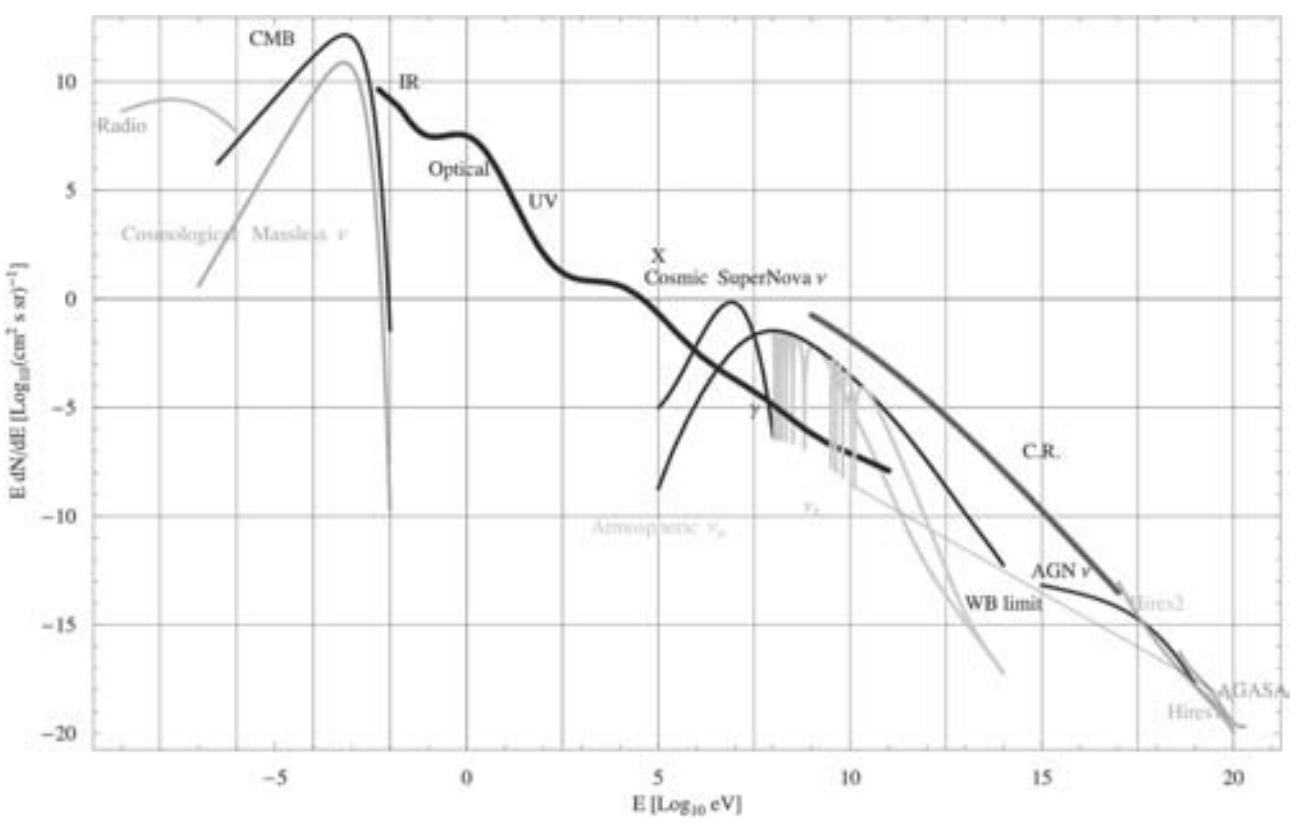

Figure 1: The flux number multi-frequency panorama of cosmic radiations. Solar and local galactic components $\gamma, \nu$ have been omitted. Both atmospheric $\nu_\mu$ and its parasite oscillated $\nu_\tau$ component are shown; the twin $\nu_\tau$ curves are showing both the vertical, crossing the Earth, (the one with a deep at 10 GeV) and the horizontal components in all energy band.

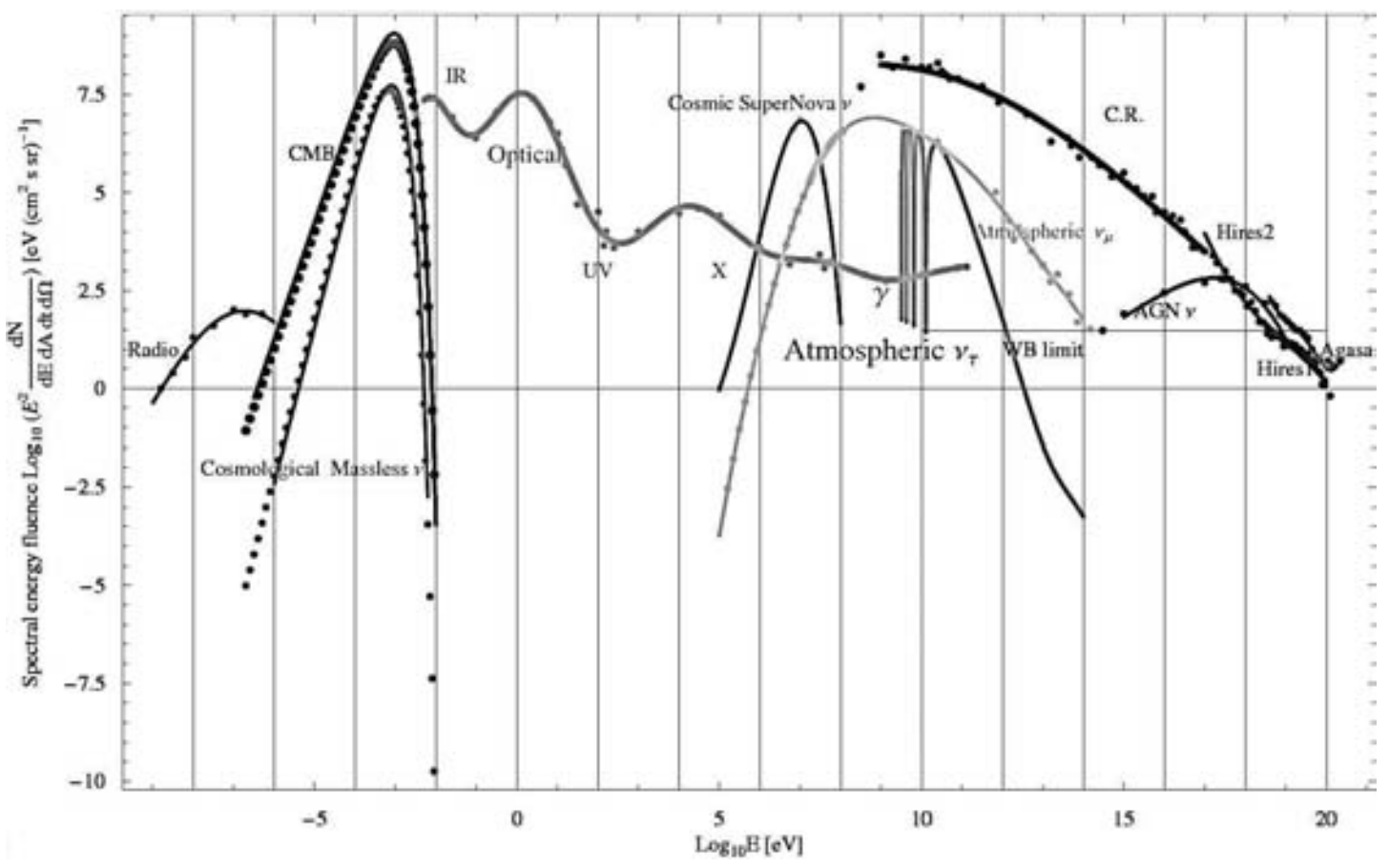

Figure 2: The energy spectra of the cosmic radiation consequent of the number flux in previous figure; theoretical and few observed data points are shown. Only the vertical $\nu_\tau$ are shown. Above TeV they are free of atmospheric noise. The GZK cut off at the extreme is the source of the GZK neutrino, at WB range, discussed in the article.

## 2 The Auger-Hires spectra: GZK cut-off, expected UHE $\nu$ Fluxes

Up to day the puzzles on CR and on UHECR remain unsolved: what are the sources, how they are accelerated, is there any GZK cut-off, why local UHECR sources are not yet observed? Agasa-Hires and Auger moved the problem answer randomly from one edges to another. BL-Lac connection with UHECR, found first by Agasa [12] and confirmed somehow by Hires in last few years [14], apparently fade away by Auger null results. Clustering events too . The early Agasa Galactic Anisotropy at EeV, hint for a timid, but relevant, new Galactic Neutron Astronomy, disappeared under Auger scrutiny. Moreover a surprising composition record in Auger UHECR data is unexpected: a turn toward heavy (Fe) nuclei at highest energy events. They may produce less neutrinos, if they are very local (but than, what are their arrival directions?). Otherwise being isotropic, they are call for a cosmic nature, possibly born at ZeV energy. In this view their photo-nuclear fragility (diffusion distance of few Mpc) imply once again a much abundant UHE GZK neutrino fluxes to be found. The puzzle grows. The presence of a drastic or at least a mild decrease in UHECR spectra edges arose from Hires and AUGER data. This in contrast with AGASA hint for the absence of a GZK cut off. The absence of source identifications within a GZK volume pose additional puzzles: are UHECR isotropic and homogeneous (as GRBs), spread along the whole Universe? How can they overcome the cosmic photon opacity (GZK puzzle)? To face this possibility we [6] did offer a decade ago the Z-Shower or Z-Burst model [6]. This model is based on UHE ZeV neutrinos primary, ejected from the cosmic sources as the courier, transparent to BBR photons, interacting at the end of the flight, with their relic non relativistic cosmic partners clustered in wide cloud as hot dark matter. They are the favorite target of the interaction via Z-boson resonance. The UHE Z produced and its decay in flight would lead to UHE nucleons traces of observed UHECR. This model got alternate attention and fortune, but its motivation (the need to overcome isotropic and homogeneity UHECR spectra) survived the last Auger test [18]. More models able to fit the spectra require a diffuse UHECR protons source around ZeV energy [3]; such an energy for the primary in complete agreement with the Z-shower model versions, with a tuned neutrino mass at $m_\nu = 0.08$ eV, well compatible with cosmological limits and atmospheric mass splitting [6],[7]. However, as noted above, last surprising Auger claim of a heavy UHECR composition is making all these conclusions questionable. To estimate a minimal GZK neutrino flux we note that the Auger UHECR at GZK knee $E = 3.98 \cdot 10^{19} eV$ is corresponding to a small fluency ($\Phi_{GZK} \simeq 6.6 eV \cdot cm^{-2} s^{-1} sr^{-1}$ ): at its average maxima $E \simeq 1.1 \cdot 10^{20} eV$, the flux is very suppressed ($\Phi_{GZK} \simeq 0.5 eV \cdot cm^{-2} s^{-1} sr^{-1}$ ); this flux must suffered severe losses along the whole cosmic volumes, into GZK secondaries, mostly few EeV GZK neutrinos. A simple estimate may be done based on this flux amplified by the Universe/GZK size ratio, a value of nearly two order of magnitude. The final total UHECR GZK fluence estimated in this and other ways is ($\Phi_{GZK} \simeq 50 eV \cdot cm^{-2} s^{-1} sr^{-1}$ ), whose main traces are electron pairs, $\nu$ pairs of all three flavors. This offer, following most authors, a neutrino (pair) GZK minimal energy spectra at EeV. $\Phi_{\nu_\tau + \overline{\nu_\tau}} \simeq 20 eV \cdot cm^{-2} s^{-1} sr^{-1}$ to assume for up-going taus. This value may be at worst a half of it, but not too far way. A different, convergent hint of a minimal fluency comes from the UHECR ankle threshold at $E = 3.98 \cdot 10^{18} eV$ : it may be mark the crossing from galactic to extragalactic components; it may also mark the electron pair losses; it may also be source of photo-pion production of UHECR escaping from their bright source. The consequent fluency may exceed ($\Phi_{EeVs} \simeq 25 eV \cdot cm^{-2} s^{-1} sr^{-1}$),compatible with previous fluency value. Therefore for sake of simplicity we assume around EeV energy a minimal flat ($\propto E^{-2}$) neutrino $\tau$ spectra (the sum of both two species), comparable with the WB one, at a nominal fluency $\Phi_{\nu_\tau + \overline{\nu_\tau}} \simeq 20 eV \cdot cm^{-2} s^{-1} sr^{-1}$. For a fluency 50% larger we derived [10], earlier estimate mainly for EUSO; we considered in detail the Earth opacity to UHE neutrinos for an exact terrestrial density profile: its column depth defined the survival for UHE $\nu_\tau$ at each zenith angle, the consequent $\tau$ probability to escape and to decay in flight considering the terrestrial finite size atmosphere. Our result for Auger now, for $\Phi_{\nu_\tau + \overline{\nu_\tau}} \simeq 20 eV \cdot cm^{-2} s^{-1} sr^{-1}$, are summarized in Fig. 3. It is evident that at EeV in rock matter (as the one in Auger territory), the expected rate exceed one event in three years. An enhancement, made by peculiar Ande screen, may amplify the rate from the West side (doubling the expected rate). Inclined hadronic showers the more their zenith angle is large, the higher their altitudes take place. At highest quota (twenty-forty km), the air density is low, the pair threshold increases, the Cherenkov and Fluorescent luminosity decrees drastically. Moreover the distance from the high altitude till the Auger telescope increase and the hadronic high altitude Showers (Hias)[9] [11] are not longer observable by present array telescopes: Auger is not able to reveal EeV hadronic air-shower above 75 degrees. Therefore, even within the poor Auger angular telescope resolution any the inclined event within $80^o - 100^o$, being long lived (because of the smaller air density ) air-showers must be indebt only to incoming UHE Neutrino. If upward, just tau ones.

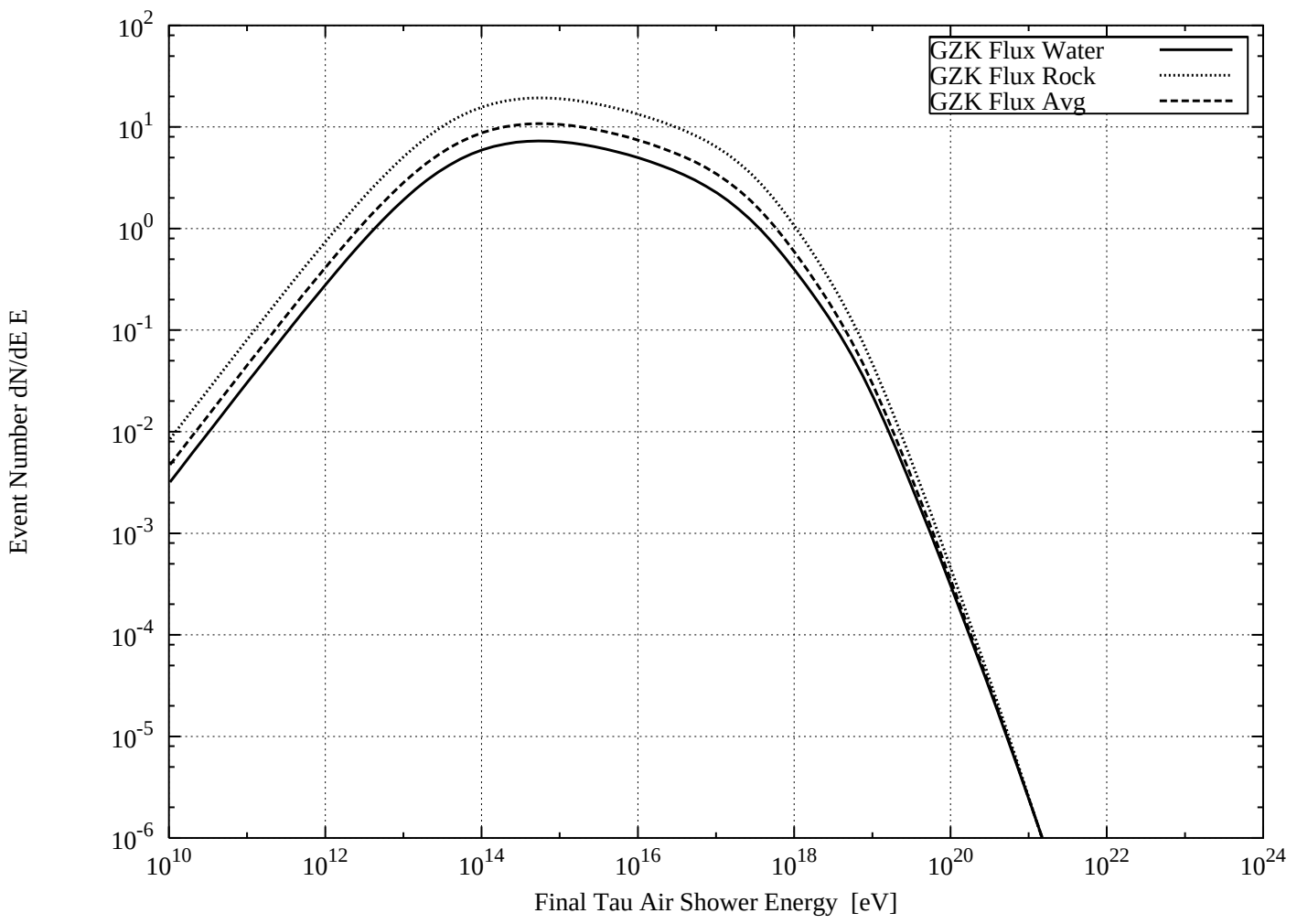

Figure 3: Our expected event rate spectra on Auger sky by Fluoresce Detector in three years of records assuming an arrival WB neutrino flux:$\Phi_{\nu_\tau+\overline{\nu_\tau}} = 20eV \cdot cm^{-2}s^{-1}sr^{-1}$. At EeV, where Auger FD the rate is $N_{eV} = 1.07$ in three years; at $3 \cdot 10^{17}$ eV it reaches $N_{eV} = 3.3$ ; at this energy the Auger acceptance is nearly a third of the area, reaching once again the unity. It means that within present three years, i.e. this decade, a Tau EeV event may rise in Auger sky within $2 - 0.3$ EeV . Additional event may occur as inclined showers on surface detectors mostly arriving within the Ande shadow, a tau amplifier (double-triple rate from West than East side) observable by FD and SD. Finally the extended horizontal and long tau air shower at high altitude (and low density) may be partially contained in Auger,increasing the area and the estimate above. Air shower Cherenkov reflection on clouds may also be observable.

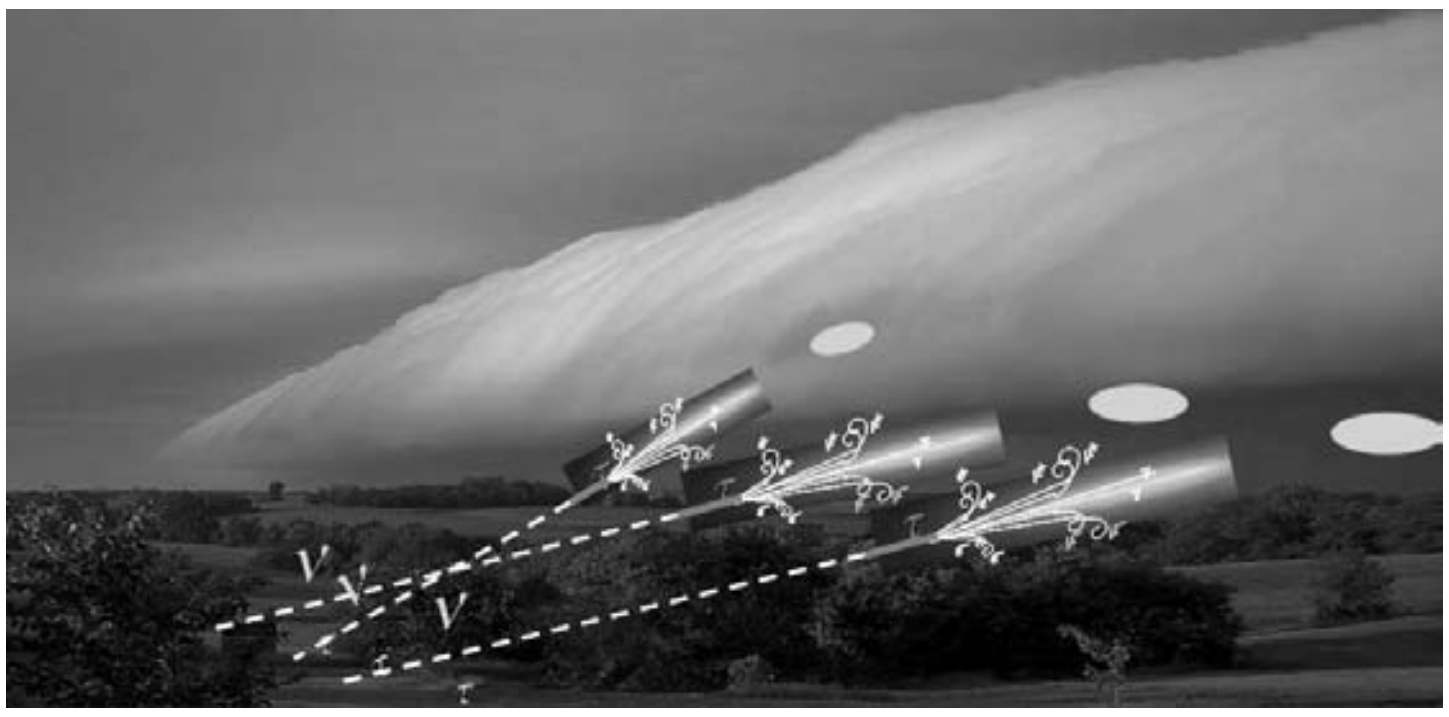

Figure 4: Upward Tau Air showering on the Auger clouds

## 3 Tau Air-Showers Rate: Young, in Ande Shadows, on Auger Sky

One of the most famous signature of young Neutrino air-shower is their curvature and its time structure: it may indicate the Tau neutrino origin.[4]. However it is not the unique and most powerful imprint. The Auger angular resolution and its limited statistics will not allow to reveal any Moon or Solar Shadows. At least in a decade. The Ande shadow [9],[15] however is at least a thousand times larger than moon; however on the horizons the UHECR rate decreases drastically, nearly three order of magnitude; nevertheless the West-East asymmetry would rise around $88^o$ horizons as a few hundred missing or asymmetric events, making meaningful its detection in one year. It *must* be observed soon by tuned trigger and angular resolution attention. Its discover is an important crosscheck of the Auger experiment. Within this Ande shadow horizons taus might be better born, nearly one-two any three year, mostly in FD (Fluorescence Detector), but also in SD (Surface Detector); without any care on thresholds it will rise more rarely. In this decade Auger may find up-going Tau in its whole area at the rate (see Fig. 3) of $N_{10^{18}eV} = 1.07$ event each three years ; at lower energy, $N_{3\cdot 10^{17}eV} = 3$ the Auger area detection is reduced ($\simeq 0.33$), leading to an important event rate $N_{E_\tau = 3\cdot 10^{17}eV} \simeq 1.1$. Because additional events are un-confined (Horizontal) air-shower, this increases the detection mass and its discover rate, almost doubling the expectation rate. Moreover the presence of an enhanced rate from Ande size on FD and SD may increase the West side rate. Finally a possible discover of FD could be amplified by final flash via Cherenkov reflection on clouds (see Fig. 4). Being cloudy nights a third or a fourth of the whole time, this time may be an occasion to exploit even if Moon arises. In conclusion, in partial disagreement to some earliest[2] and most recent Auger prospects [5] requiring one or two *decades* for a WB flux, we foresee,(in see also[17]), a sooner discover of GZK $\tau$ neutrino astronomy, possibly within two-three years from now. Auger may be even the first experiment in the world to detect a tau natural flavor regeneration processes. To reach and speed this goal we suggest: 1) to enlarge the telescope array facing towards the Ande. 2) To increase the array telescope angle of view, reducing the air-shower energy threshold, covering larger areas. 3) To tune the electronic trigger of FD to horizontal air-showers. 4) To map the UHECR Ande shadows at great angular resolution.

# GAW - AN IMAGING ATMOSPHERIC ČERENKOV TELESCOPE WITH LARGE FIELD OF VIEW

G. CUSUMANO[a], G. AGNETTA[a], A. ALBERDI[b], M. ÁLVAREZ[c], P. ASSIS[d], B. BIONDO[a], F. BOCCHINO[e], P. BROGUEIRA[d], J.A. CABALLERO[c], M. CARVAJAL[f], A.J. CASTRO-TIRADO[b], O. CATALANO[a], F. CELI[a], C. DELGADO[g], G. DI COCCO[h], A. DOMÍNGUEZ[c], J.M. ESPINO NAVAS[c], M.C. ESPIRITO SANTO[d], M.I. GALLARDO[c], J.E. GARCIA[f], S. GIARRUSSO[a], M. GÓMEZ[f], J.L. GÓMEZ[b], P. GONÇALVES[d], M. GUERRIERO[b], A. LA BARBERA[a], G. LA ROSA[a], M. LOZANO[c], M.C. MACCARONE[a], A. MANGANO[a], I. MARTEL[f], E. MASSARO[i], T. MINEO[a], M. MOLES[b], F. PEREZ-BERNAL[f], M.A. PÉRES-TORRES[b], M. PIMENTA[d], A. PINA[d], F. PRADA[b], J.M. QUESADA[c], J.M. QUINTANA[b], J.R. QUINTERO[f], J. RODRIGUEZ[f], F. RUSSO[a], B. SACCO[a], M.A. SÁNCHEZ-CONDE[b], A. SEGRETO[a], B. TOMÉ[d], A. DE UGARTE POSTIGO[b], P. VALLANIA [l]

[a] *IASF-Pa/INAF, Istituto di Astrofisica Spaziale e Fisica Cosmica di Palermo, Palermo, Italy*

[b] *Instituto de Astrofísica de Andalucía (CSIC), Granada, Spain*

[c] *Universidad de Sevilla, Dept. de Física Atómica, Molecular y Nuclear, DFAMN-US, Sevilla, Spain*

[d] *LIP, Laboratório de Instrumentação e Física Experimental de Partículas, Lisbon, Portugal*

[e] *OAPA/INAF, Osservatorio Astronomico di Palermo, Palermo, Italy*

[f] *Universidad de Huelva, Huelva, Spain*

[g] *IAC, Instituto de Astrofisica de Canarias, Tenerife, Spain*

[h] *IASF-Bo/INAF, Istituto di Astrofisica Spaziale e Fisica Cosmica di Bologna Bologna, Italy*

[i] *Universitá La Sapienza, Roma, Italy*

[l] *IFSI-To/INAF, Istituto di Fisica dello Spazio Interplanetario di Torino, Torino, Italy*

**Abstract**

GAW, acronym for Gamma Air Watch, is a Research and Development experiment in the TeV range, whose main goal is to explore the feasibility of large field of view Imaging Atmospheric Čerenkov Telescopes. GAW is an array of three relatively small telescopes (2.13 m diameter) which differs from the existing and presently planned projects in two main features: the adoption of a refractive optics system as light collector and the use of single photoelectron counting as detector working mode. The optics system allows to achieve a large field of view ($24° \times 24°$), suitable for surveys of large sky regions. The single photoelectron counting mode in comparison with the charge integration mode improves the sensitivity by permitting also the reconstruction of events with a small number of collected Čerenkov photons. GAW, which is a collaboration effort of Research Institutes in Italy, Portugal and Spain, will be erected in the Calar Alto Observatory (Sierra de Los Filabres - Andalucía, Spain), at 2150 m a.s.l.. The first telescope will be settled within Autumn 2007. This paper shows the main characteristics of the experiment and its expected performance.

## 1 Introduction

In the last fifteen years a new electromagnetic window (50 GeV - 20 TeV) has been opened thanks to the observation of the sky by ground-based Atmospheric Čerenkov Telescopes (ACT). The detection of the Čerenkov light produced by the electrophotonic showers, originated by the interaction of $\gamma$-ray photons in the atmosphere, has offered spectacular breakthroughs in this extreme observational energy domain. A remarkable number of sources has been firmly detected since the pioneer detection of Very High Energy (VHE) emission from the Crab Nebula [7]. Among the most exciting recent results we cite the discovery of many new sources in the Galactic plane, only some of them identified with known astronomical sources, and the detection of some Active Galactic Nuclei (AGN), three of them with redshifts in the range from 0.15 to 0.2. These are the three most distant extragalactic sources observed in this energy band and their detection is relevant for the evaluation of the Extragalactic Background Light and shows that the intergalactic space is more transparent to $\gamma$-rays than previously thought.

An increasing number of VHE experiments are now exploring the sky to capture emissions in this extreme energy band. They are producing observational results useful for the understanding of the physical processes responsible for the emission in AGNs and supernova remnants and helped us to study the cosmic-ray acceleration processes. The majority of telescopes have imaging capability (IACT), with the Čerenkov radiation focused onto a pixellated camera and the dominant cosmic-ray background is rejected by exploiting differences in the images from primaries $\gamma$ and proton air showers [6]. The main advantages of the atmospheric Čerenkov technique are: high sensitivity, good angular resolution, moderate energy resolution and a low energy threshold that is getting closer and closer to the observational energy window of the $\gamma$-ray space experiments (EGRET, AGILE, GLAST-LAT). IACT, however, have a reduced duty cycle of about 10% and a small Field of View (FoV) of a few degree (3-5 degrees). The duty cycle of ACTs depends upon the constrain of moonless and clean nights, while their limited FoV is tied to the optical telescope design adopted to collect and image the light. Present IACTs , using large mirror reflectors, cannot reach larger FoV because of the mirror optical aberrations, rapidly increasing with off-axis angles. Moreover, the increasing of the detector area necessary to cover large FoV would inevitably produce a strong reduction of the light collecting area of the primary mirror because of the shadow of the focal plane instrumentation onto the reflecting surface.

A large FoV, however, is an important instrumental requirement for several scientific goals. VHE astronomical events, in fact, can occur at unknown locations and/or random in time and a large FoV is then mandatory to increase their detection probability. Moreover, it is also very useful to perform sensitive surveys in and out of the Galactic Plane and to measure the celestial $\gamma$-ray diffuse emission as well. A further advantage offered by a large FoV is the increase of the effective area at the highest energies: the more distant is the core of the Čerenkov light pool from the telescope, the farther the image falls from the center of the focal plane detector. Large FoV will keep these events inside the detector area, increasing the sensitivity of the telescope for the most energetic electromagnetic radiation.

In this paper we describe the main characteristics of GAW, acronym for Gamma Air Watch, a R&D experiment that will test the feasibility of a new generation of IACT that incorporates high flux sensitivity with large FoV capability, stereoscopic observational approach and single photoelectron counting mode. A technical description of the GAW experiment can be found in other two papers presented at this conference [4, 1].

## 2 The experiment

GAW is conceived as an array composed by three identical IACTs located at the vertexes of a triangle, $\sim$ 80 m side. A detailed description of GAW is given in [5].

GAW is different from the present IACT experiments. A refractive optical system characterizes its optics system: the light collector is a non commercial single side flat Fresnel lens (ø 2.13 m) with focal length of 2.55 m and thickness of 3.2 mm. The Fresnel lens, an approximation of the refractive aspherical lens, provides a large FoV with imaging quality suitable to the coarse structure of the Čerenkov image and with the advantage of no central obstruction of the focal plane detector. The lens is made of UltraViolet (UV) transmitting acrylic with a nominal transmittance of $\sim$ 95% from 330 $nm$ to the near InfraRed. The material has a small refraction index derivative at low wavelength, thus reducing chromatic aberration effect. This will be further minimized by implementing diffractive optics design onto the side of the Fresnel lens containing the grooves. The lens design is optimized at $\sim$ 360 $nm$, and it is characterized by a quite uniform spatial resolution suitable to the requirements of the Čerenkov imaging up to 12° off-axis. The lens is made of 33 petals maintained in a rigid configuration by a spider structure. The optical system is designed and manufactured by the Fresnel Technologies, Fort Worth, Texas.

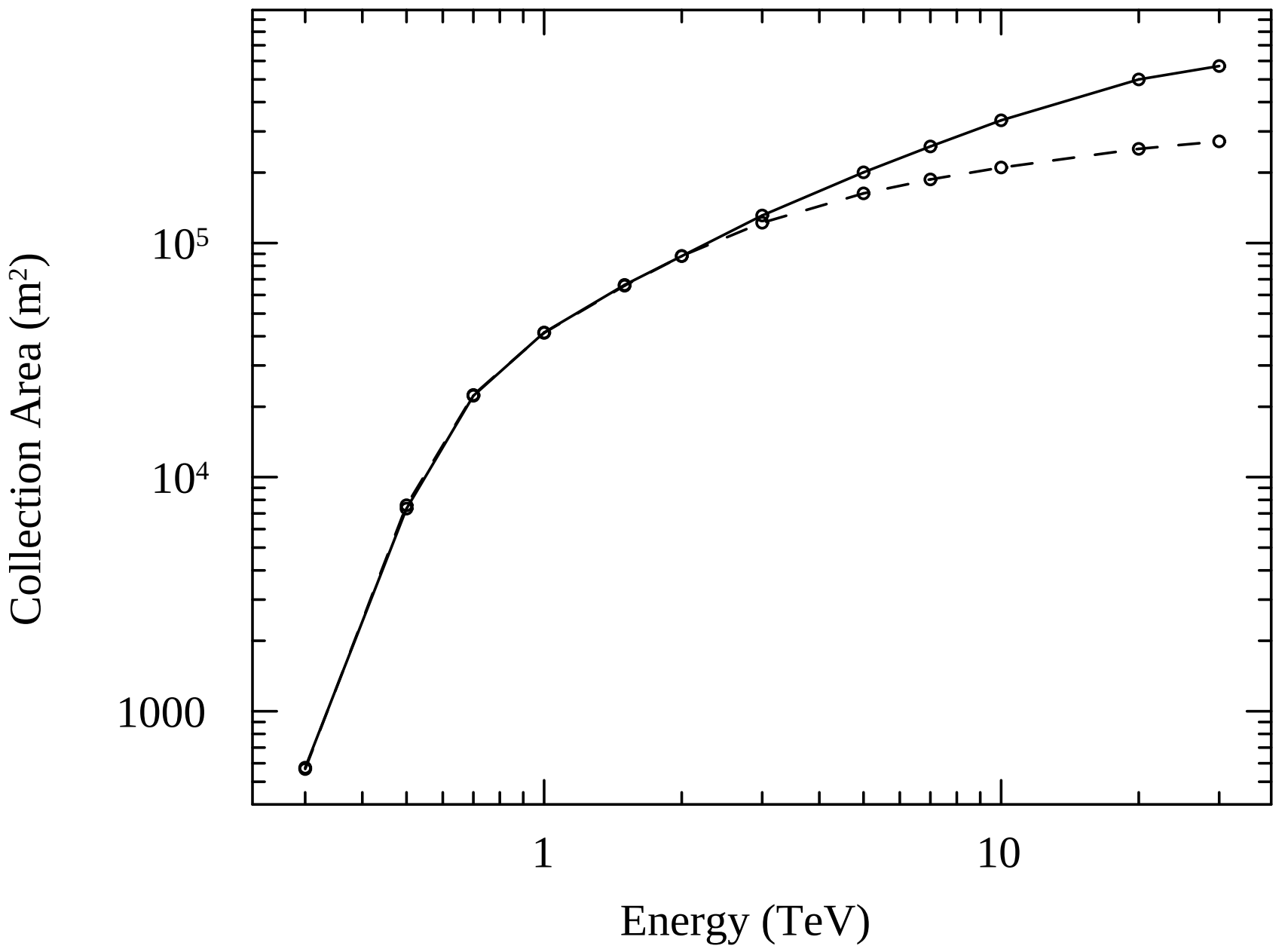

Figure 1: Collecting area of the GAW telescope array vs energy for on-axis Gamma Ray events. Solid line refers to a focal plane detector with a large FoV of 24° × 24° while the dashed line refers to a small FoV of $\sim$ 5° × 5°.

Another important difference of GAW with respect to traditional Čerenkov telescopes is the detector working mode. The focal plane detector of each telescope consists of a grid of 40 × 40 Multi-Anode Photomultipliers Tubes (MAPMT), with 64 anodes each, arranged in 8 × 8 matrix, operated in sin-gle photoelectron counting mode [2] instead of the charge integration method widely used in the IACT experiments. The active channels, 102400 for each telescope, will record the Čerenkov binary image with high granularity, which is fundamental to minimize the probability of photoelectron pile-up within intervals shorter than the sampling time of 10 ns. In such working mode, the electronic noise and the PMT gain differences are negligible, allowing to lower the photoelectron trigger threshold and, as a consequent result, to achieve a low telescope energy threshold in spite of the relatively small dimension of the Čerenkov light-collector. The stereoscopic observational approach will guarantee the sensitivity necessary to accomplish its feasibility and scientific goals.

GAW will be erected in the Calar Alto Observatory (Sierra de Los Filabres - Andalucía, Spain), at the altitude of 2150 m a.s.l..

Two phases are foreseen for the project:

In the first phase, only part of the GAW focal detector will be implemented to cover a FoV of $5° \times 5°$. The detector will be mounted on a rack frame and it can be moved to cover the entire focal plane. The instrumental sensitivity in this configuration will be tested observing the Crab Nebula on- and off-axis up to 12°.

In the second phase the focal plane detector will be enlarged to cover a FoV of $24° \times 24°$. We plan to survey a sky belt with an extension of 60° in the North-South direction.

## 3 GAW expected performances

The GAW expected performances were evaluated with a complete end-to-end simulation. Physical processes involved in the interaction of a $\gamma$-ray or a proton in the atmosphere, shower production and development, generation of Čerenkov light and effects of the atmospheric absorption were simulated using the CORSIKA code [3]. The collection of the Čerenkov light by the stereoscopic array, the optics transmission, the angular spread of the Fresnel lens, the focal plane detector geometry, the quantum efficiency of the MAPMTs and the trigger electronics were simulated by a proper code. The image analysis and event reconstruction were performed with an "ad hoc" procedure on the Čerenkov images [4].

Fig. 1 shows the collection area vs. energy for mono-energetic $\gamma$-ray events coming from a cosmic on-axis source at the zenith and with the constraints of trigger coincidence in all three telescopes. The solid line represents the collection area with a FoV of $24° \times 24°$, while the dashed line shows, for comparison, the collection area with the reduced FoV of $\sim 5° \times 5°$. The advantage of a large FoV is even more striking at higher energies. The detection trigger rate for a Crab-like spectrum peaks at 0.7 TeV. The main performance of GAW is summarized by its integrated flux sensitivity as function of the energy. Fig. 2 shows the sensitivity for a Crab-like point source in 50 hours observation with 5 sigma detection limit. For comparison, the flux of the Crab Nebula and the sensitivity of other TeV experiments are also shown.

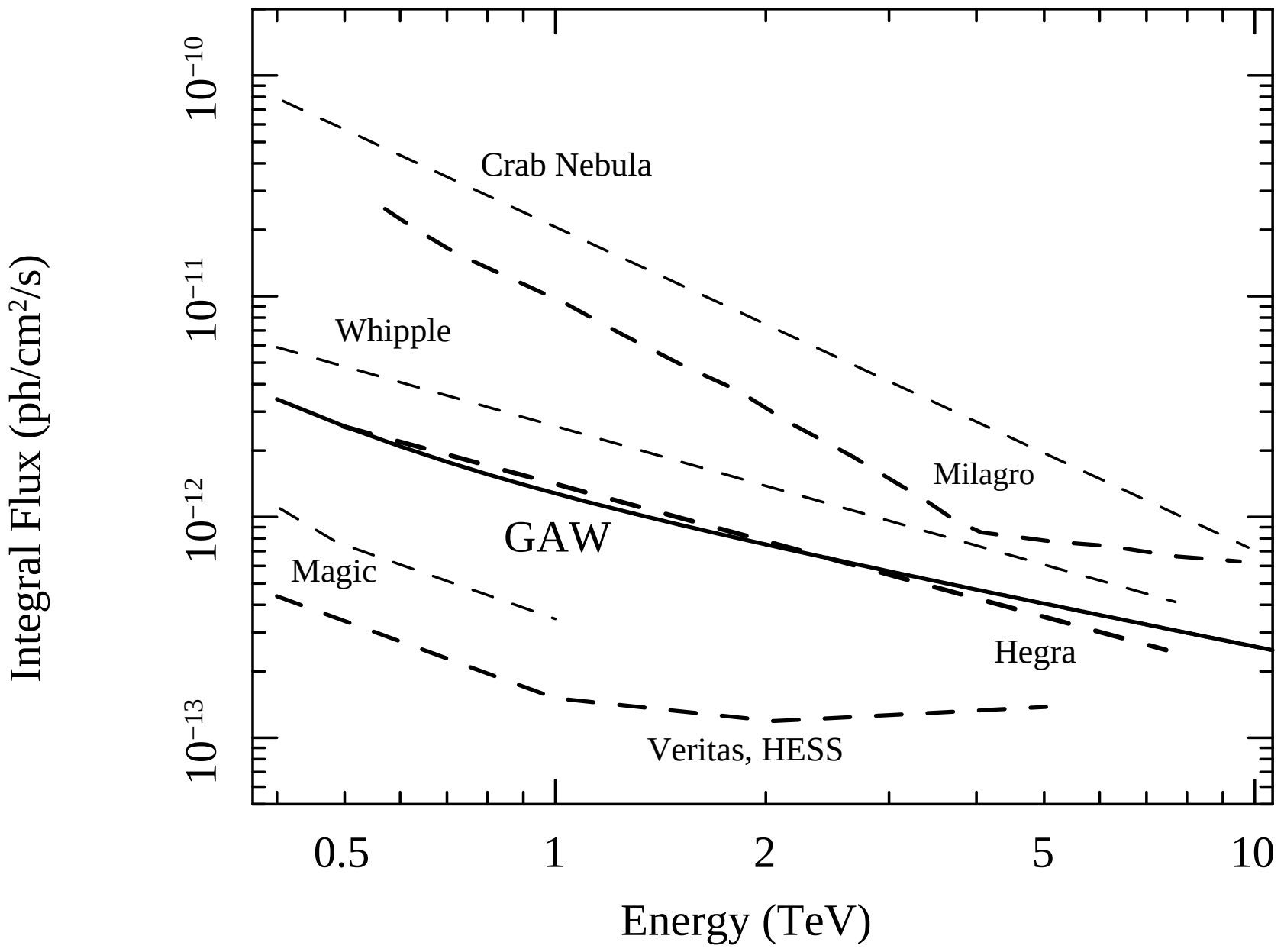

Figure 2: The sensitivity limit ($5\sigma$) detection of GAW (solid line) for Crab-like point sources in 50 hours observations. The flux of the Crab Nebula and the sensitivity of other TeV experiments (dashed lines) are also shown for comparison. The Milagro sensitivity is for 1 year of observation.

## 4 Summary

IACTs with large FoV will offer two important advantages: they will survey the sky for serendipitous TeV detections and, at the same time, will increase the IACT collection area, triggering events whose core is far away from the telescope axis and therefore improving the statistics of the high energy tail of the source spectra.

Presently, GAW is a R&D experiment made up of an array of three identical relatively small Čerenkov telescopes that will test the feasibility of a new generation IACT that joins large FoV and high flux sensitivity. Large FoV will be achieved by using refractive optics made of single side flat Fresnel lens of moderate size (2.13 m diameter). The focal camera will use the single photon counting mode instead of the charge integration mode widely used in the present IACT experiments. This detector working mode will allow us to operate with a very low photoelectron threshold and a consequent lowering of the energy threshold. The stereoscopic observational approach will improve the angular resolution, the cability of identifying $\gamma$-ray induced showers and a good determination of the primary photon energy. In this way it will reach the necessary sensitivity to accomplish the foreseen scientific goals.

GAW is a collaboration effort of Research Institutes in Italy, Portugal and Spain. It will be erected in the Calar Alto Observatory (Sierra de Los Filabres - Andalucía, Spain). The first telescope will be settled within Autumn 2007.

## 5 Acknowledgements

The GAW collaboration acknowledge the support by the CSIC, the grant by the Junta de Andalucía under project numbers FQM160 and P06-FQM-01392 (Spain) and the support of FCT - Fundacão para a Ciência e Tecnologia (Portugal).

# TENTEN: A NEW IACT ARRAY FOR MULTI-TEV GAMMA-RAY ASTRONOMY

Gavin Rowell [a], Victor Stamatescu [a], Roger Clay [a], Bruce Dawson [a], Ray Protheroe [a], Andrew Smith [a], Greg Thornton [a], Neville Wild [a]

[a] *School of Chemistry & Physics, University of Adelaide, Adelaide 5005, Australia*

**Abstract**

The H.E.S.S. results of recent years suggest a population of $\gamma$-ray sources at energies E>10 TeV, opening up future studies and new discoveries in the multi-TeV energy range. This energy range addresses the origin of PeV cosmic-rays (CRs) and the astrophysics of multi-TeV $\gamma$-ray production in a growing number of astrophysical environments. Here, we outline the *TenTen* project — a proposed stereoscopic array of modest-sized (10 to 30m$^2$) Cherenkov imaging telescopes optimised for the E$\sim$10 to 100 TeV range. The telescopes will operate with a wide field of view ($8^\circ$ to $10^\circ$ diameter) and the array is expected to achieve an effective area of $\sim$10 km$^2$ at energies above 10 TeV. A summary of the motivation for *TenTen* and key performance parameters are given.

## 1 Introduction

Ground-based $\gamma$-ray astronomy operating in the $\sim$0.1 to $\sim$10 TeV range has become a mainstream astronomical discipline due to the exciting results from H.E.S.S. [21] in the Southern Hemisphere over recent years. In the Northern Hemisphere the MAGIC [27] and MILAGRO [28] telescopes are revealing new sources and similar results can be expected from VERITAS [38]. The TeV source catalogue extends to over 30 individual sources, and we are now able to perform detailed studies of extreme environments capable of CR particle acceleration. Several lessons can be gained from the experience of H.E.S.S. and others, which motivate development of a new dedicated instrument for studies at multi-TeV ($E >$few TeV) energies (see also [34, 6]):

**Increasing variety of TeV Sources:** The number of environments established as sources of gamma radiation (to energies exceeding $\sim$ 10 TeV in many cases) is growing. In the case of Galactic sources we have shell-type supernova remnants (SNRs), pulsar-wind-nebulae (PWN), compact binaries and/or X-ray binaries, young stellar systems/clusters and molecular clouds acting as targets for CRs in their vicinity. A growing number of Galactic sources (all extended) remain unidentified, essentially without any clear counterpart at other energies. This latter issue has opened up the new sub-field of *dark* TeV $\gamma$-ray sources.
**Hard Photon Spectra:** The majority of new Galactic sources exhibit hard power law photon spectra $dN/dE \sim E^{-\Gamma}$ where $\Gamma < 2.5$ without indication of cutoffs, suggesting that their emission extends beyond 10 TeV.
**Extended Sources Require Large Fields of View:** The majority of Galactic sources are found to be extended up to several degrees. Morphology studies, coupled with multiwavelength information, are allowing us to probe TeV $\gamma$-ray production and transport processes. This activity requires large fields of view, not only to encompass sources of interest but to also allow selection of regions for CR background estimation and energy spectra determination. The 5° diameter FoV of the H.E.S.S. cameras has permitted highly successful surveys of the inner Southern Galactic Plane within just a few years [4, 7], and also the establishment of degree-scale morphology in several strong sources.
**Present Instruments have Limited Multi-TeV Sensitivity:** Current intruments operate with a $\sim$0.1 TeV threshold energy and effective collection area ($E >$ 10 TeV) of less than 1 km$^2$. The fluxes of the new TeV sources are in the few to $\sim$15% Crab flux range, reflecting the intrumental sensitivities. Observational opportunities for detailed multi-TeV studies are limited due to the ever growing source catalogues at $E >$0.1 TeV spread across different astrophysical programmes. Fig. 1 illustrates an example of this where a new weak TeV source is revealed to the north of the pulsar wind nebula HESS J1825−137 after only deep ($>$ 50 hr) observations. Further, detailed studies of this weak source, motivated perhaps by its relatively rare (at present) coincidence with a MeV/GeV EGRET source, would not be practical with H.E.S.S. Such studies would require >100 hr of observations and face stiff competition from other source programmes.
***E* > 10 TeV Sources Already Exist:** Several Galactic sources (two shell-type SNRs and several PWN [5, 8, 11, 9, 10]) with strong fluxes ($>$ 15% Crab) and/or deep observation times ($\geq$ 50 hr), have established photon spectra reaching $\sim$50 TeV or greater, demonstrating that particle acceleration to energies exceeding 100 TeV is occurring in these types of objects. MILAGRO has also recently revealed degree-scale emission (with total flux exceeding 1 Crab) at energies above 10 TeV in the Cygnus and other regions of the Northern Galactic Plane [1, 2], further highlighting the potential of future large-FoV telescopes in the multi-TeV range.

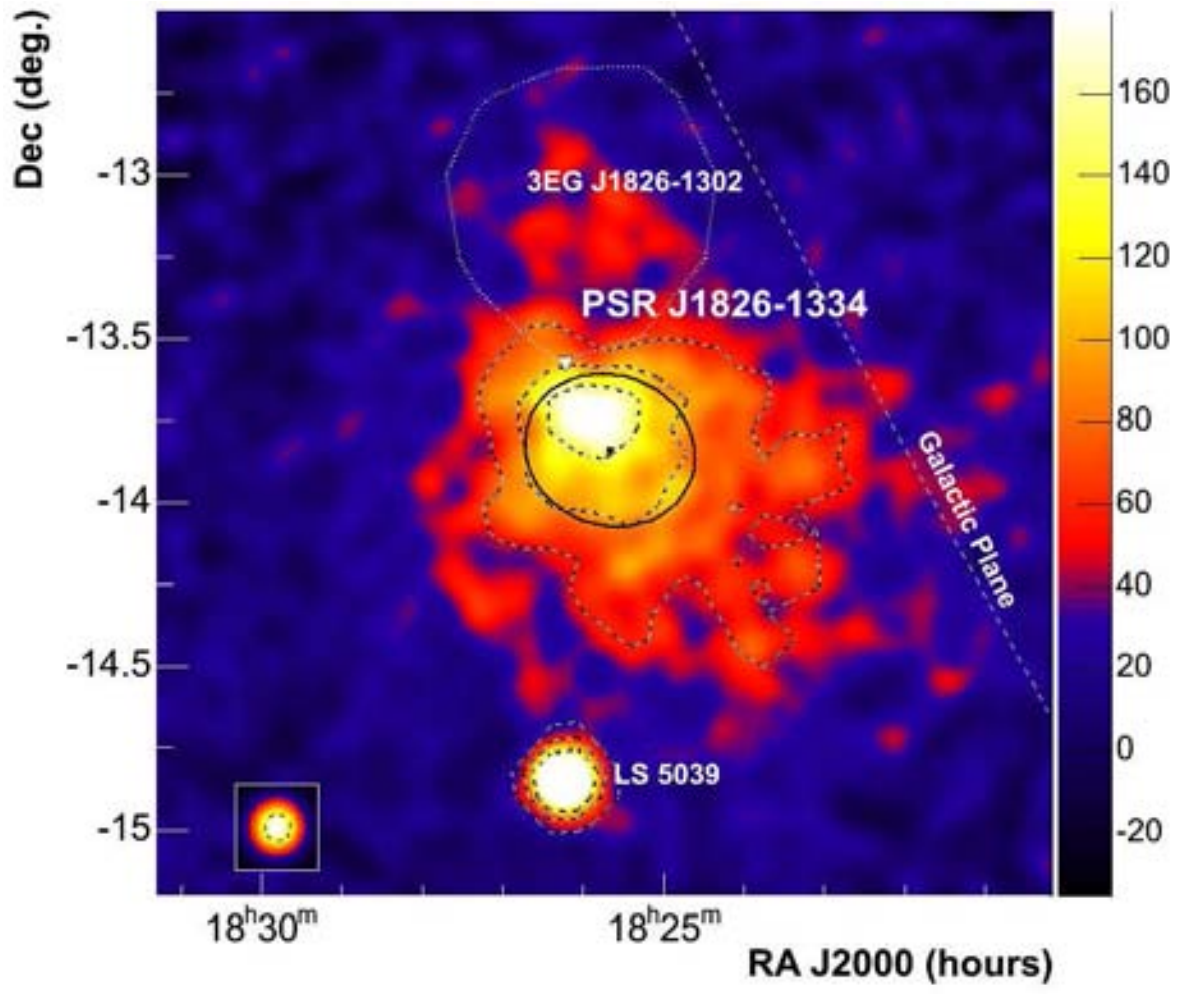

Figure 1: H.E.S.S. image of the pulsar wind nebula HESS J1825−137 [9] after 52 hr observations. A weaker source is found to the north of HESS J1825−137, and may be a counterpart to the EGRET source 3EG J1826−1302. This weaker source is unlikely to be studied further by H.E.S.S.

There are also clear theoretical grounds for pushing deep into the multi-TeV domain:

**Particle Acceleration to the *knee* and beyond:** The desire to understand particle acceleration to the CR *knee* ($E \sim 1$ PeV) energy and beyond remains a key motivation for multi-TeV studies. Although it is generally accepted that CRs can be accelerated in shell-type SNRs [19] to energies $E_{\rm max}$ $\sim$few$\times 10^{14}$ eV [23] (via the diffusive shock acceleration process), there is considerable uncertainty as to how particles can reach the *knee* energy and beyond (eg. [26, 16]) in so-called *Pevatrons*. Several ideas have been put forward, for example: strong amplification of pre-shock magnetic fields [12]; local Gamma-Ray-Bursts (GRBs) [37, 15]; and superbubbles which combine the effects of many SNRs and maybe Wolf-Rayet/OB stellar winds [17, 14, 30, 13]. Extragalactic sources with large-scale kpc shocks such as galaxy clusters (eg. [39]) and AGN jets and giant lobes (eg. [22]) could also contribute. Only with observations at around $\geq$100 TeV can we begin to solve the mystery of PeV CR acceleration.

***E* > 10 TeV — Easier Separation of Hadronic & Electronic Components:** A major complication in interpreting present results in the 0.1 to $\sim$10 TeV range concerns the separation of $\gamma$-ray components from accelerated hadrons (from secondary $\pi^{\circ}$-decay) and those from accelerated electrons (usually from inverse-Compton scattering). Multiwavelength information at radio and X-ray energies for example, can provide constraints on these components but often one requires model-dependent assumptions to decide the nature of the parent particles. At energies $E$ >10 TeV the electronic component can be suppressed due to strong radiative synchrotron energy losses suffered by electrons in magnetised post-shock environments, such as that in shell-type SNRs. In addition, the Klein-Nishina effect on the inverse-Compton cross-section can significantly reduce the efficiency of this process. Except in those cases where a strong source of electrons exists, such as in PWN, interpretation of $E > 10$ TeV spectra may therefore be much more confidently interpreted as arising from accelerated hadrons. Such hadronic/electronic separation also allows us to study in detail the astrophysics of accelerated electrons in multi-TeV sources.
**Probing Local Intergalactic/Interstellar Photon Fields:** $E > 10$ TeV photons indirectly allow us to probe ambient soft photon fields. In the ∼10 to ∼100 TeV energy range, absorption on the cosmic infra-red background (CIB) in the 10 to 100 $\mu$m range dominates with mean free paths extending beyond 1 Mpc. Constraints on the (nearby) intergalactic CIB via $\gamma$-ray spectral studies of nearby extragalactic sources, such as M 87 (an established TeV source) can yield important information concerning star and galaxy formation in our local intergalactic neighbourhood [3]. Constraints on the interstellar CIB may also be possible via $E > 10$ TeV spectral studies of multi-TeV Galactic source populations [29].

## 2 TenTen: Initial Simulation Study & Performance

Given that source fluxes rapidly decrease with energy, any dedicated instrument operating in the multi-TeV energy domain must have a very large effective collection area $A_{\rm eff}$ ∼10 km$^2$ or greater. While there are several ways to achieve 10 km$^2$ using ground-based techniques, earlier simulations [31] have shown that a proven, technically straightforward, and highly sensitive method would employ stereoscopy in an array of 30 to 50 modest-sized imaging atmospheric Cherenkov telescopes (IACTs) in a cell-based approach. Each telescope would have mirror area 10 to 30 m$^2$, field of view (FoV) 5° to 10°, and inter-telescope spacing within a single cell ≥200 metres (compared to ∼100 m employed by arrays such as H.E.S.S.). The large FoV, limited practically by optical aberrations, allows events to trigger out to core distances ≥200 m, thereby increasing the effective collection area of a cell. We propose here such an array, known as *TenTen*, which stands for 10 km$^2$ above 10 TeV. Similar and other ideas for 100 TeV studies have also been suggested [25, 40, 24].

Our initial simulation study [33] looked at the performance of a single cell of 5 telescopes, each with mirror area 23.8 m$^2$ (84x60 cm diameter spherical mirror facets), f/1.5 optics, and 1024 pixel camera spanning 8.2° diameter (with pixel diameter 0.25°). The optics were based on an elliptical dish profile outlined in [35], and provide an 80% containment diameter $\leq 0.25°$ out to $\sim 4°$ off-axis. The layout of the cell has the outer four telescopes arranged in a square of side length $L$ with a single telescope at the centre (see Fig. 2 Left panel), not unlike the HEGRA IACT-System layout [32]. Gamma-ray and proton extensive air shower simulations (30° zenith — with CORSIKA v6.204 [20] and SIBYLL [18]) coupled with telescope responses (based on [36]) were used to investigate basic performance parameters of the cell. An observation altitude 200 m a.s.l. was chosen since we are investigating sites in Australia. For $E >$ 10 TeV, low-altitude sites are beneficial in terms of collection area compared to mid/high altitudes due to the larger distances between telescopes and shower maxima. Using a fixed ADC gate of 20 ns (similar to H.E.S.S. electronics) and a conservative trigger setup (pixel threshold of 12 photoelectrons in $\geq$2 pixels, image $size \geq 60$ photoelectrons, image $dis < 3.5°$ and stereo trigger of $\geq$2 telescopes) we found that for a cell with side length $L = 300$ m, an on-axis $\gamma$-ray effective collection area $A_{\rm eff}$ exceeding 1 km$^2$ for $E > 30$ TeV can be achieved in a single cell (see Fig. 3). The large FoV is a key factor in allowing events to be triggered out to core distances approaching 800 m from the cell centre (Fig. 2 Right panel). In addition, similar cosmic-ray background rejection power (based on scaled width) and $\sim$6 arc-minute angular resolution is achieved in the $E > 10$ TeV range as H.E.S.S. and the HEGRA IACT-System achieve in their respective energy ranges. An energy threshold in the 1 to few TeV range is also indicated, depending on the cell side length $L$. This encouraging result suggests that expanding the array to (for example) ~10 cells sufficiently spaced so that there are no common events between cells could yield collection areas ~10 km$^2$, exceeding that of H.E.S.S. by factors approaching 50 at 100 TeV. The approximate flux sensitivity (based on the improvement in collection area for 10 cells over H.E.S.S.) and energy coverage of a 10 cell *TenTen* array is depicted in Fig. 4. Flux sensitivities for large extended sources such as the 2° diameter shell type SNR RX J0852.0−4622 would be $\sim 10^{-12}$ erg cm$^{-2}$ s$^{-1}$ at 10 TeV. For point-like sources, the accessible fluxes would be a factor of 10 to 20 lower again (less than $\sim 10^{-13}$ erg cm$^{-2}$ s$^{-1}$ above 10 TeV). A 1 TeV to a few TeV energy threshold would also allow detailed studies of existing $\gamma$-ray sources that are at the flux threshold of detection for H.E.S.S. Even though our simulations were limited to $E \leq 100$ TeV, we also expect a high collection area above this energy.

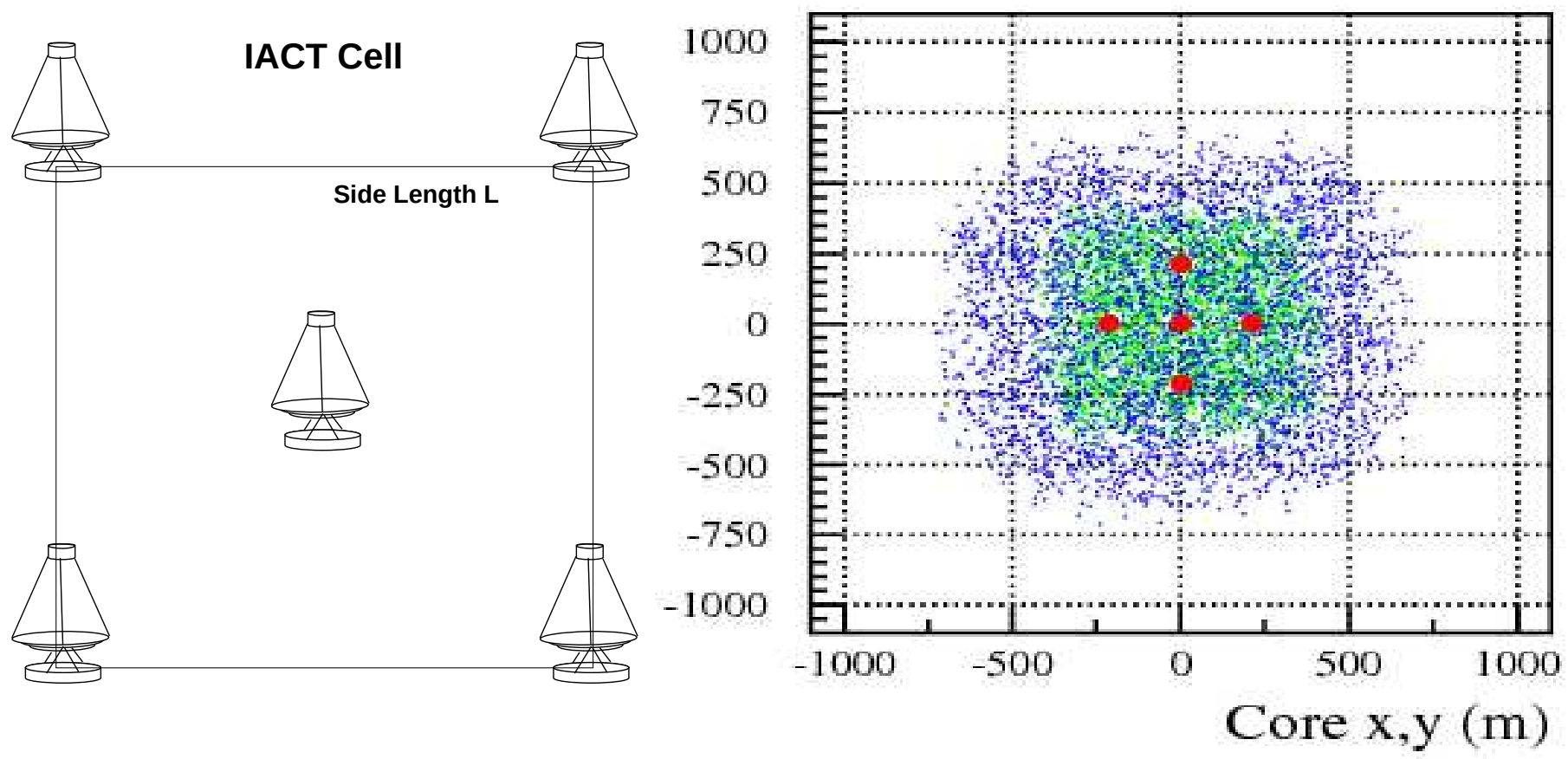

Figure 2: **Left:** Schematic of an IACT cell of 5 telescopes as simulated for TenTen. A cell side length $L$ in the range 200 to 500 metres has so far been simulated in our initial study. **Right:** True core locations of triggered events for a cell of side length $L$=300 m. Trigger conditions are given in text. Telescope positions are represented by the large red dots (the array is rotated 90° with respect to the left panel) with 1-10 TeV events shown as small green dots, and 10-100 TeV events shown as small blue dots.

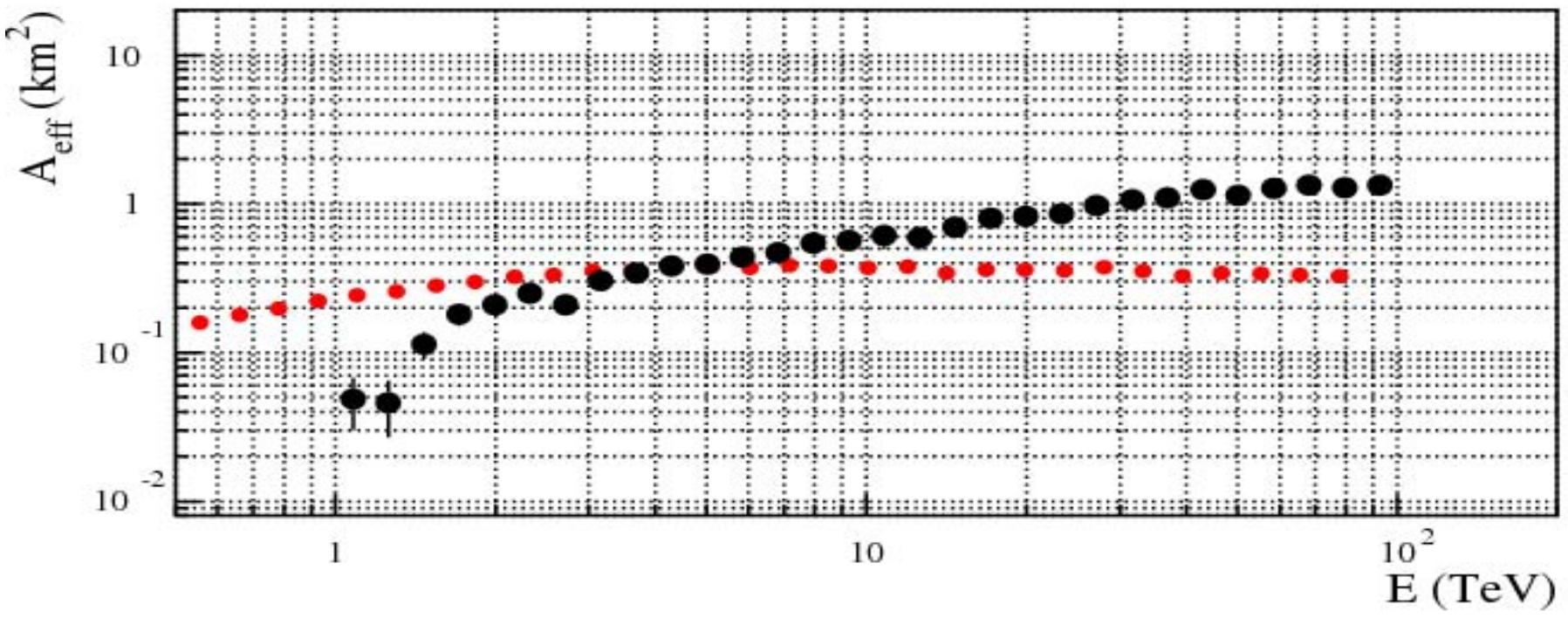

Figure 3: Effective area (km$^2$) vs energy $E$ (TeV) for a 5-telescope cell of side length $L$=300 m, at 200 m above sea level (Large black filled cirles). Telescopes of total mirror area 23.8 m$^2$, coupled to a focal plane array of 1024 pixels spanning $\sim 8°$ field of view were used, along with trigger conditions described in text. This can be compared to the effective area achieved by H.E.S.S. (small red solid circles).

## 3 Summary & Conclusions

We have described the motivation and some important performance parameters for a new array of IACTs achieving 10 km$^2$ at $E > 10$ TeV. This proposed array, known as *TenTen*, will be dedicated to multi-TeV astronomy, and based on results from H.E.S.S., is expected to yield new insights into PeV particle acceleration and multi-TeV astrophysics. Studies are currently underway to further optimise individual telescopes (optics, electronics, camera design), array

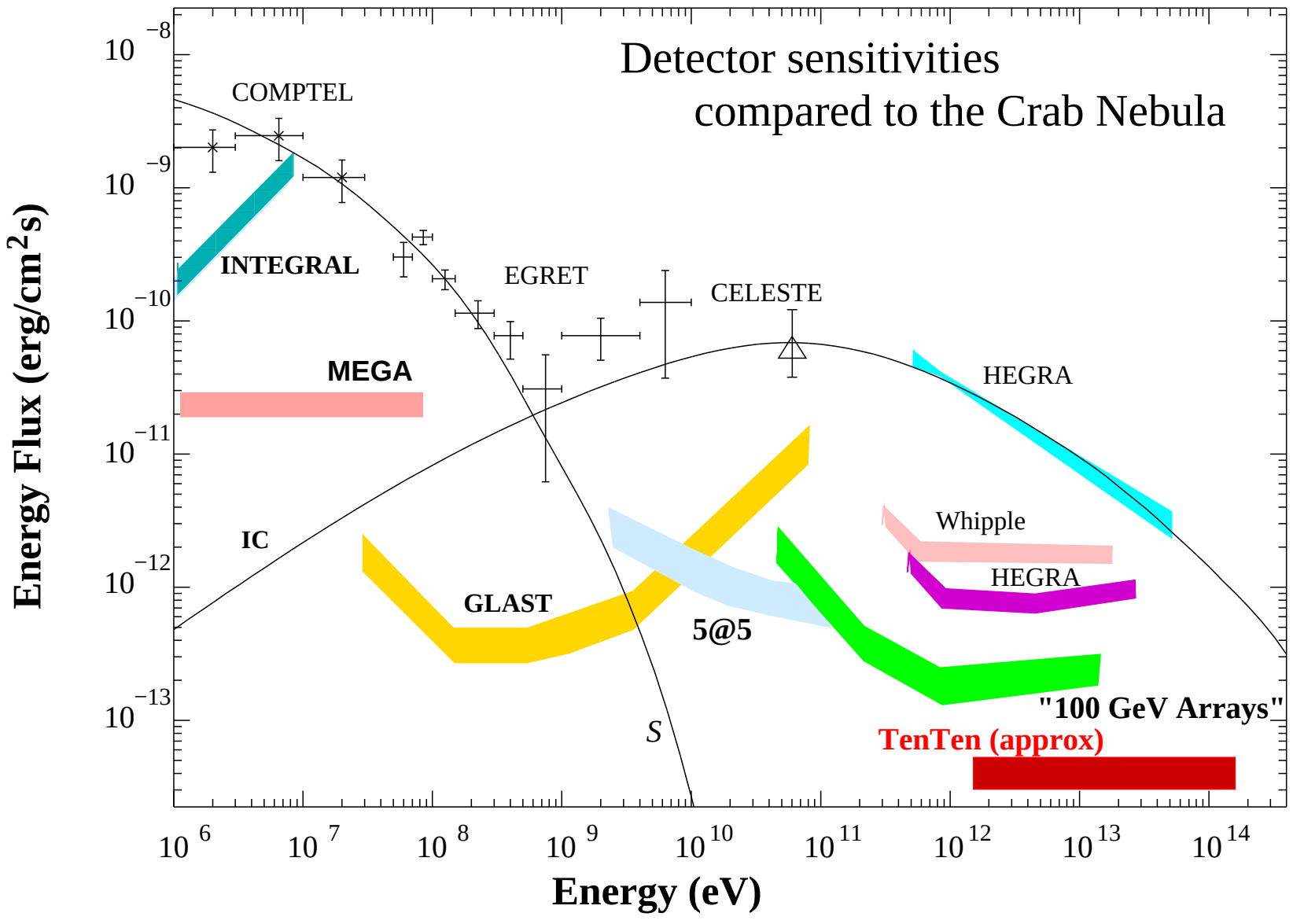

Figure 4: Approximate point source energy flux sensitivity for *TenTen* (assuming 10 cells of 5 telescopes discussed in text) in comparison with other $\gamma$-ray instruments (space-borne and ground-based) and the flux from the Crab Nebula. An observation time of 50 hr and signal significance of $5\sigma$ is required. The *TenTen* sensitivity is estimated from its collection area improvement over H.E.S.S. (represented by "100 GeV arrays") from our initial simulation study as discussed in text.

layout, and potential sites in Australia.

# DARK MATTER SPIKES AND MINI-SPIKES

Gianfranco Bertone

*Institut d'Astrophysique de Paris, UMR 7095-CNRS, Université Pierre et Marie Curie, 98bis Bd Arago, 75014 Paris, France*

**Abstract**

Spikes are strong Dark Matter overdensities around Massive Black Holes. Although they are unlikely to survive over cosmological timescales at the center of galactic haloes, due to dynamical processes such as scattering off stellar cusps, they can form and survive around intermediate mass black holes. We review the formation scenarios of spikes and mini-spikes and discuss the implications for indirect Dark Matter searches with existing and upcoming experiments such as GLAST and PAMELA.

## 1 Introduction

Although many astrophysical and cosmological observations provide convincing evidence for the existence of a "dark" component in the matter density of the Universe, the nature of this *dark matter* (DM) remains unkown. It is commonly assumed that DM is made of new, as yet undiscovered, particles, associated with theories beyond the Standard Model of Particle Physics. Among the most widely studied candidates are the supersymmetric neutralino and candidates arising in theories with extra-dimensions, which appear difficult to constrain with direct searches (i.e. by looking for nuclear recoils due to DM particles scattering off nuclei) and whose prospects of discovery at future accelerators strongly depend on the details of the particle physics setup (for recent reviews see e.g. Refs. [1, 2]). Indirect searches via the detection of annihilation radiation may provide an interesting alternative, but they are usually affected by large astrophysical and cosmological uncertainties. Furthermore, in many cases, the detection of an annihilation signal may be difficult to distinguish from less exotic astrophysical sources. Recenty, new scenarios have been discussed in literature, where the formation of Massive Black Holes leads to strong DM overdensities called "spikes", that might be observed as point sources of gamma-rays [16] and neutrinos [22]. We show here that spikes and mini-spikes may lead to a dramatic enhancement of gamma-ray and antimatter fluxes, bringing them within the reach of current and upcoming experiments.

## 2 Spikes

The effect of the formation of a central object on the surrounding distribution of matter has been investigated in Refs. [4] and for the first time in the framework of DM annihilations in Ref. [5]. It was shown that the *adiabatic* growth of a massive object at the center of a power-law distribution of DM, with index $\gamma$, induces a redistribution of matter into a new power-law (dubbed "spike") with index $\gamma_{sp} = (9 - 2\gamma)/(4 - \gamma)$. This formula is valid over a region of size $R_{sp} \approx 0.2\, r_{BH}$, where $r_{BH}$ is the radius of gravitational influence of the black hole, defined implicitly as $M(< r_{BH}) = M_{BH}$, where $M(< r)$ denotes the mass of the DM distribution within a sphere of radius $r$, and where $M_{BH}$ is the mass of the Black Hole [6]. The process of adiabatic growth is in particular valid for the SMBH at the galactic center. A critical assessment of the formation *and survival* of the central spike, over cosmological timescales, is presented in Refs. [8, 7] – see also references therein. We limit ourselves here to note that adiabatic spikes are rather fragile structures, that require fine-tuned conditions to form at the center of galactic halos [9], and that can be easily destroyed by dynamical processes such as major mergers [10] and gravitational scattering off stars [11, 8].

It was recently shown that a $\rho \propto r^{3/2}$ DM overdensity can be predicted in any halo at the center of any galaxy old enough to have grown a power-law density cusp *in the stars* via the Bahcall-Wolf mechanism [12]. Collisional generation of these DM "crests" – Collisionally REgenerated STructures – was demonstrated even in the extreme case where the DM density was lowered by slingshot ejection from a binary supermassive black hole. However, the enhancement of the annihilation signal from a DM crest is typically much smaller than for adiabatic spikes [12].

## 3 Mini-Spikes

If intermediate-mass black holes (IMBHs), with a mass ranging between $10^2$ and $10^6 M_\odot$ (e.g. [3]), exist in the Galaxy, their adiabatic growth would have modified the DM distribution around them, leading to the formation of "mini-spikes". The DM annihilation rate being proportional to the square of the number density of DM particles, these mini-spikes would be bright gamma-ray sources, distributed in a roughly spherically-symmetric way about the galactic center, and well within the observational reach of the next-generation gamma-ray experiments. Their brightness and isotropy make them ideal targets of large field-of-view gamma-ray experiments such as GLAST [15]. In case of a positive detection, Air Cherenkov Telescopes such as CANGAROO [17], HESS [18], MAGIC [19] and VERITAS [20] could extend the observations to higher energies and improve the angular resolution. Mini-spikes could also be detectable with neutrino experiments such as Antares and IceCube. Furthermore they may also lead to strong enhancements of anti-matter fluxes, within the reach of experiments such as PAMELA. The observation of numerous (up to $\sim 100$) point-like gamma-ray sources with identical cut-offs in their energy spectra, at an energy equal to the mass of the DM particle, would provide smoking-gun evidence for DM particles.

Mini-spikes result from the reaction of DM mini-halos to the formation or growth of IMBHs. To make quantitative predictions, we focus on a specific IMBHs formation scenario [13], representative of a class of models where these objects form directly out of cold gas in early-forming DM halos, and are characterized by a large mass scale, of order $10^5 M_\odot$ (see also Ref. [14] and references therein). In Fig. 1 we show the distribution of these IMBHs as obtained in

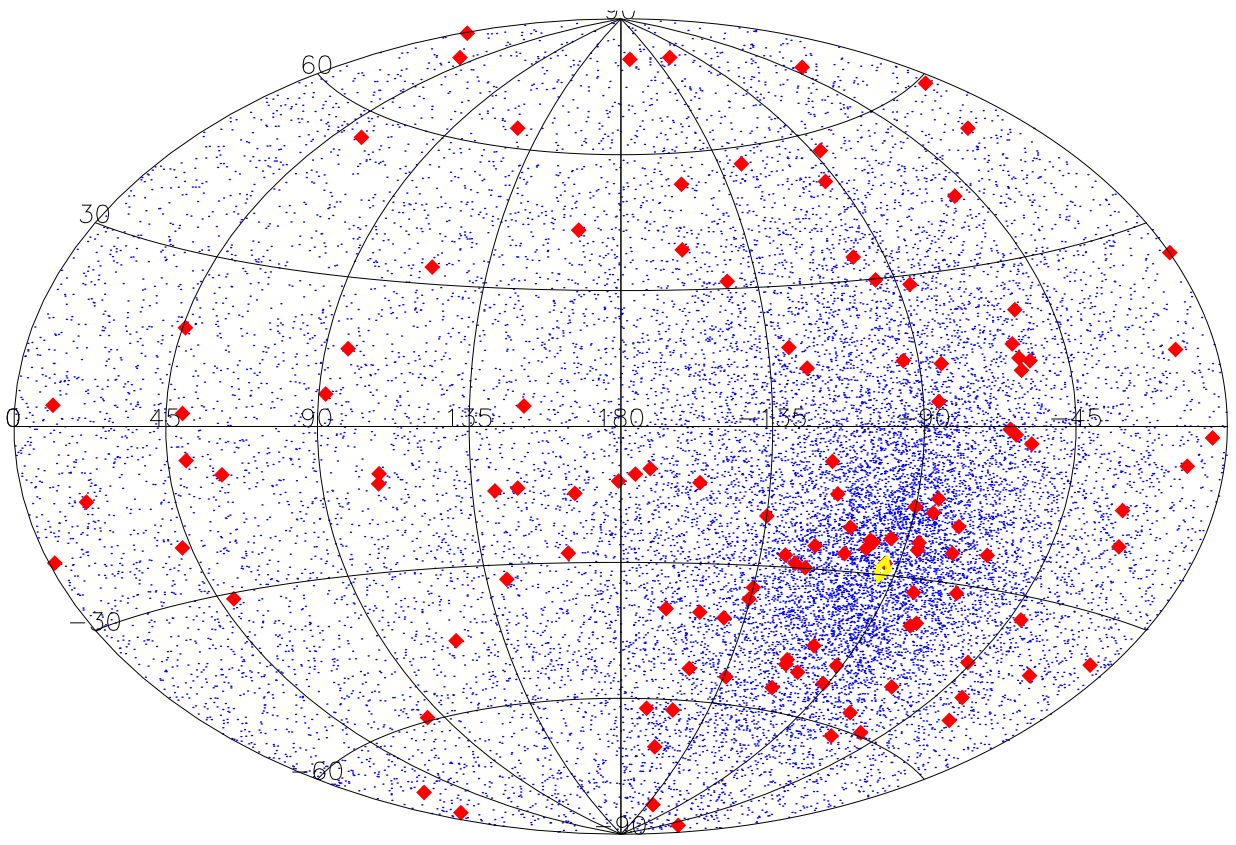

Figure 1: Sky map in equatorial coordinates showing the position of Intermediate Mass Black Holes in one random realization of a Milky-Way like halo (red diamonds), and in all 200 realizations (blue dots). The concentration at negative declinations corresponds to the position of the Galactic center (black open diamond). From Ref. [22]

Ref. [22], which is based on the Monte Carlo halo evolution procedure given in Ref. [23]. The method for populating black holes at high-redshift are described in detail in Refs. [24, 16].

Mini-spikes would be copious sources of Dark Matter annihilation products, such as gamma-rays, neutrinos, anti-protons and positrons , which can be produced either directly, or through fragmentation and decay of secondary particles such as quarks, leptons and gauge bosons. If $N_i(E)$ is the spectrum of secondary particles of species $i$ per annihilation, the flux from an individual mini-spike can be expressed as [16]

$$\Phi^0_{\nu_\ell}(E) = \phi_0 m_{\chi,100}^{-2} (\sigma v)_{26} D_{\rm kpc}^{-2} L_{\rm sp} N_i(E) \qquad (1)$$

with $\phi_0 = 9 \times 10^{-10}\text{cm}^{-2}\text{s}^{-1}$. The first two factors depend on the particle physics parameters, viz. the mass of the DM particle in units of 100 GeV $m_{\chi,100}$, and its annihilation cross section in units of $10^{-26}\text{cm}^3/\text{s}$, $(\sigma v)_{26}$, while the third factor accounts for the flux dilution with the square of the IMBH distance to the Earth in kpc, $D_{\rm kpc}$. Finally, the normalization of the flux is fixed by an adimensional *luminosity factor* $L_{\rm sp}$, that depends on the specific properties of individual spikes. In the case where the DM profile *before* the formation of the IMBH follows the commonly adopted Navarro, Frenk and White profile [25], the final DM density $\rho(r)$ around the IMBH will be described by a power law $r^{-7/3}$ in a region of size $R_s$ around the IMBHs. Annihilations themselves will set an upper limit to the DM density $\rho_{\max} \approx m_\chi/[(\sigma v)t]$, where $t$ is the time elapsed since the formation of the mini-spike, and we denote with $R_{\rm c}$ the "cut" radius where $\rho(R_{\rm c}) = \rho_{\max}$. With these definitions, the intrinsic luminosity factor in Eq. 1 reads

$$L_{\rm sp} \equiv \rho_{100}^2(R_{\rm s}) R_{\rm s,pc}^{14/3} R_{\rm c,mpc}^{-5/3} \qquad (2)$$

where $R_{\rm s,pc}$ and $R_{\rm c,mpc}$ denote respectively $R_{\rm s}$ in parsecs and $R_{\rm c}$ in units of $10^{-3}$pc, $\rho_{100}(r)$ is the density in units of 100GeV cm$^{-3}$. Typical values of $L_{\rm sp}$ lie in the range $0.1 - 10$ [16]. To estimate the flux, we need now to specify the gamma-ray spectrum per annihilation $\mathrm{d}N/\mathrm{d}E$, which depends on the nature of the DM particle. In most scenarios, direct annihilation in two photons is severely suppressed, but a continuum spectrum is expected from the decay of secondary neutral pions.

In Fig. 2, we show the (average) integrated luminosity function of IMBHs in one of the scenarios discussed in Ref. [16]. We define the integrated luminosity function as the number of black holes producing a gamma-ray flux larger than $\Phi$, as a function of $\Phi$. The upper (lower) line corresponds to $m_\chi = 100$ GeV, $\sigma v = 3{\times}10^{-26}$ cm$^3$s$^{-1}$ ( $m_\chi = 1$ TeV, $\sigma v = 10^{-29}$ cm$^3$s$^{-1}$). In a practical sense, the plot shows the number of IMBHs that can be detected with experiments with point source sensitivity $\Phi$ above 1 GeV. We show for comparison the point source sensitivity above 1 GeV for EGRET and GLAST, corresponding roughly to the flux for a $5\sigma$ detection of a high-latitude point-source in an observation time of 1 year [21]. Note that the fluxes scale in this case as $\sim (\sigma v)^{2/7} m_\chi^{-9/7}$, due to the implicit dependence of $r_{cut}$ on $\sigma v$ and m.

The number of detectable sources is very high, even in the pessimistic case, and either strong constraints on a combination of the astrophysics and particle physics of this scenario, or an actual detection, should be possible within the first year of operation of GLAST, which is expected to be launched in 2007.

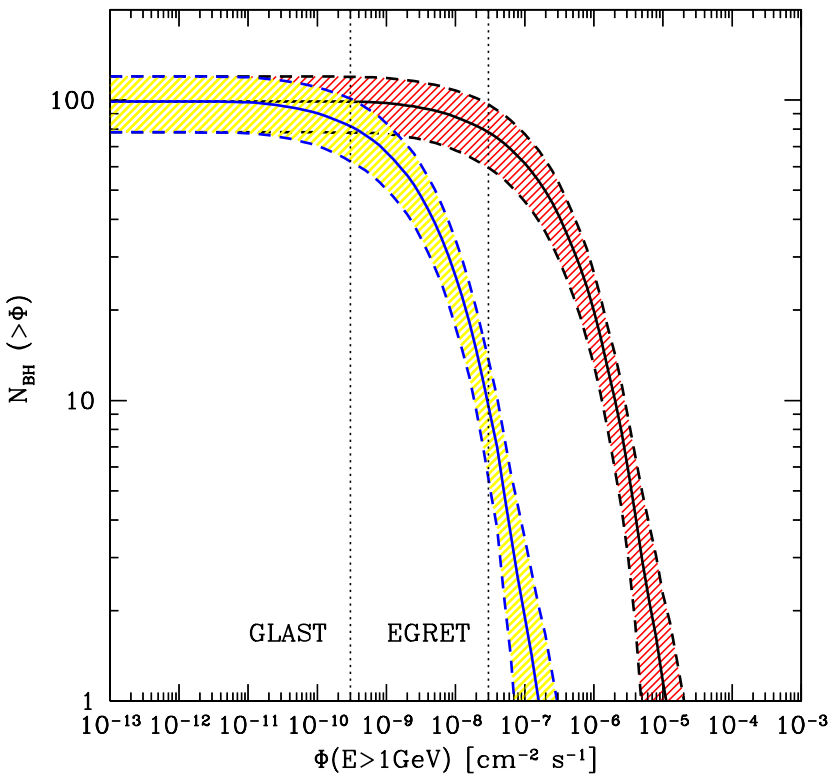

Figure 2: IMBHs integrated luminosity function, i.e. number of black holes producing a gamma-ray flux larger than a given flux, as a function of the flux. The upper (lower) line corresponds to $m_\chi = 100$ GeV, $\sigma v = 3 \times 10^{-26}$ cm$^3$ s$^{-1}$ ($m_\chi = 1$ TeV, $\sigma v = 10^{-29}$ cm$^3$ s$^{-1}$). For each curve we also show the 1-$\sigma$ scatter among different realizations of Milky Way-sized host DM halos. We show for comparison the $5\sigma$ point source sensitivity above 1 GeV of EGRET and GLAST (1 year). From Ref. [16].

The prospects for detecting high energy neutrinos from mini-spikes are also interesting. The results are shown in Fig. 3, where we plot the number of black holes producing a rate of events $R$, or larger, in ANTARES and IceCube, assuming $m_\chi = 1$ TeV, and $\sigma v = 10^{26}$ cm$^3$ s$^{-1}$. For comparison, we show the rate of atmospheric neutrino events in a search cone of size 1° around the source, for Antares and for a kilometer-scale telescope. For the atmospheric background in Antares we have adopted the value derived in a dedicated study [26], relative to a point source coincident with the Galactic center, while for kilometer scale telescopes, we have used the so-called Bartol flux [27], and a source at the same declination.

The presence of mini-spikes may have important consequences for indirect DM searches based on exotic contributions to anti-matter fluxes, Monte-Carlo simulations involving $10^6$ realizations of the IMBHs population have been per-formed in Ref. [30]. The distribution of the resulting boost factors (for positron flux) at the Earth is sketched in Fig. 4 as a function of energy. The shaded area show the region where the boost factor is expected to lie, with a 1-$\sigma$ and 2-$\sigma$ confidence level. The yellow (grey) areas correspond to the 1 $\sigma$ region, the lighter one being obtained by fixing the annihilation volumes $\xi_i$ to their mean expected value, while the darker one corresponds to the general case for which mini-spikes have different values of $\xi_i$. In both cases, the dot-dashed curves stand for the 1 $\sigma$ contours obtained analytically. These curves are in good agreement with the ones obtained from the Monte Carlo, the small increase of the variance for the Monte Carlo with respect to the analytical expectation is due to the number of black holes that vary from one Monte Carlo realization to another, while this effect is not implemented in the analytical determination of the boost factor. The fact that these curves are fairly close to each other confirms that the dispersion of the number $N_{\rm BH}$ of IMBHs in the Milky Way poorly influences the final dispersion of the boost. This figure also shows that the boost factor can be very large, the expected value being of order 8000 (for a DM particle mass of 1 TeV).

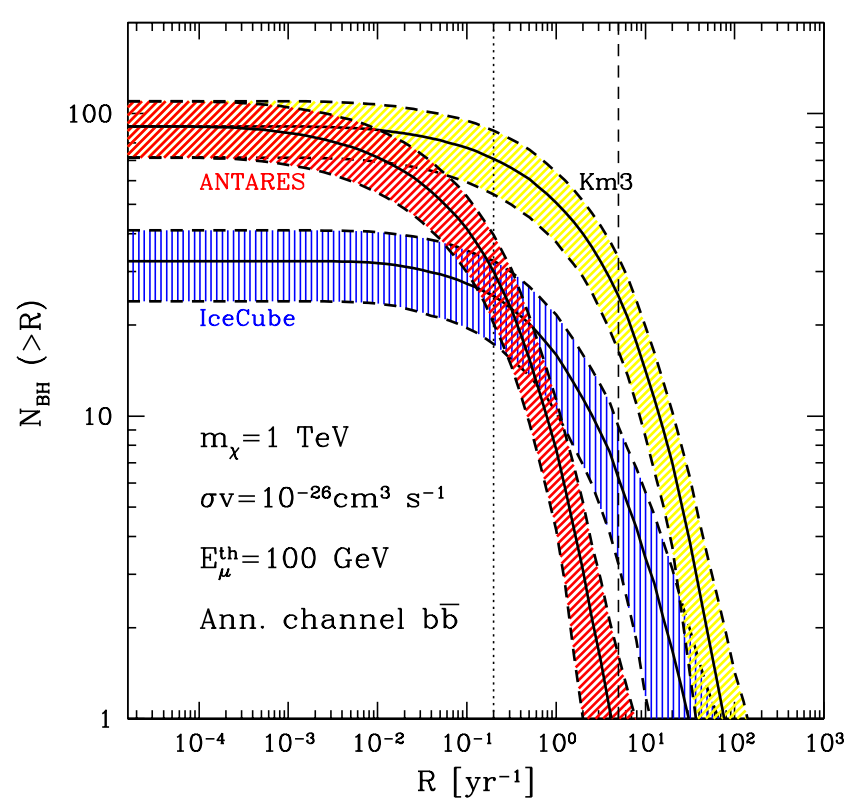

Figure 3: Number of IMBHs producing a rate $R$, or larger, in Antares, IceCube and Km3. Solid lines corresponds to the average over all realizations, while dashed line denote the 1-$\sigma$ scatter from one simulation to another. For reference we show the rate induced by atmospheric neutrinos in Antares (vertical dotted) and Km3 (vertical dashed) in a 1° cone around a source located at the Galactic center.

Finally, we stress that the identification of the exotic nature of mini-spikes as gamma-ray sources would be made easier if these object were detected in the Adromeda Galaxy [31].

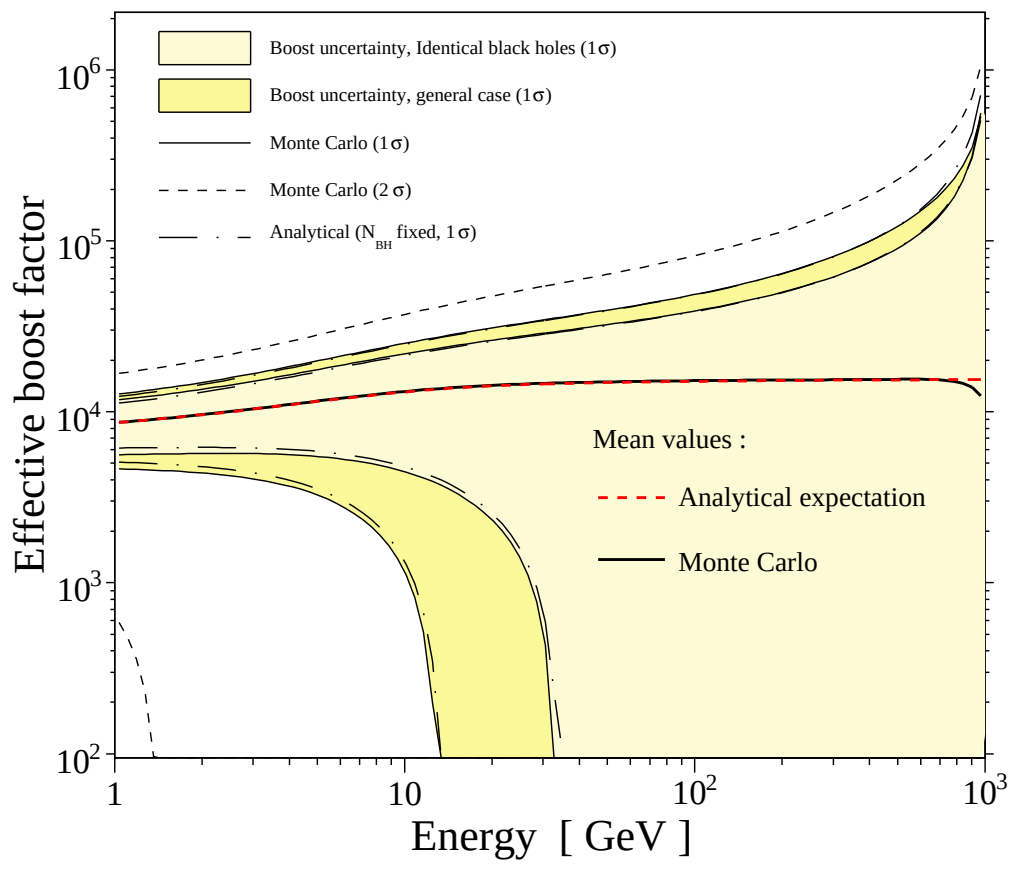

Figure 4: Results from the Monte-Carlo simulations of the IMBHs population inside the Milky Way are compared to the analytical computations of the effective boost factor and its dispersion, for $m_\chi = 1$ TeV. From Ref. [30].

.

**List of Participants**

| | | |
|---|---|---|
| Aharonian Felix | DIAS (Dublin) and MPIK (Heidelberg) | felix.aharonian@mpi-hd.mpg.de |
| Antonelli L. Angelo | INAF-OAR and ASDC | antonelli@mporzio.astro.it |
| Bambi Cosimo | University of Ferrara | bambi@fe.infn.it |
| Barbiellini Guido | University and INFN Trieste | barbiellini@ts.infn.it |
| Bassan Massimo | University and INFN Roma Tor Vergata | massimo.bassan@roma2.infn.it |
| Bastieri Denis | University and INFN Padova | denis.bastieri@pd.infn.it |
| Bellazzini Ronaldo | INFN Pisa | ronaldo.bellazzini@pi.infn.it |
| Bernabei Rita | University and INFN Roma Tor Vergata | bernabei@roma2.infn.it |
| Bertone Gianfranco | IAP Paris | bertone@iap.fr |
| Bonamente Emanuele | University of Perugia | emanuele.bonamente@pg.infn.it |
| Bringmann Torsten | SISSA | bringman@sissa.it |
| Buttaro Vincenzo | University and INFN Roma Tor Vergata | buttaro@roma2.infn.it |
| Camarri Paolo | University and INFN Roma Tor Vergata | paolo.camarri@roma2.infn.it |
| Caraveo Patrizia | IASF/INAF | pat@iasf-milano.inaf.it |
| Carbone Rita | INFN Napoli | carbone@na.infn.it |
| Casanova Sabrina | Max Planck fuer Kernphysik | Sabrina.Casanova@mpi-hd.mpg.de |
| Cavaliere Alfonso | University of Roma Tor Vergata | cavaliere@roma2.infn.it |
| Cavazzuti Elisabetta | ASI | elisabetta.cavazzuti@asi.it |
| Cecchi Claudia | University and INFN Perugia | claudia.cecchi@pg.infn.it |
| Colafrancesco Sergio | ASI-ASDC and INAF | cola@mporzio.astro.it |
| Conrad Jan | Stockholms Universitet | conrad@particle.kth.se |

| | |
|---|---|
| Costa Enrico | IASF-Roma, INAF<br>enrico.costa@iasf-roma.inaf.it |
| Cuoco Alessandro | University of Napoli<br>cuoco@na.infn.it |
| Cusumano Giancarlo | IASF-Pa - INAF<br>cusumano@ifc.inaf.it |
| Cutini Sara | ASI-ASDC and and University of Perugia<br>sara.cutini@asdc.asi.it |
| D'Armiento Daniele | University of Roma La Sapienza<br>danieled@quipo.it |
| De Angelis Alessandro | University of Udine and INFN Trieste<br>deangelis@fisica.uniud.it |
| De Lotto Barbara | University of Udine and INFN Trieste<br>delotto@fisica.uniud.it |
| Del Monte Ettore | INAF IASF Roma<br>ettore.delmonte@iasf-roma.inaf.it |
| Dell'Agnello Simone | INFN-LNF<br>simone.dellagnello@lnf.infn.it |
| Di Ciaccio Anna | University and INFN Roma Tor Vergata<br>anna.diciaccio@roma2.infn.it |
| Di Falco Stefano | INFN Pisa<br>stefano.difalco@pi.infn.it |
| do Couto e Silva | Eduardo KIPAC/SLAC<br>eduardo@slac.stanford.edu |
| Donato Fiorenza | University and INFN Torino<br>donato@to.infn.it |
| Doro Michele | University and INFN Padova<br>michele.doro@pd.infn.it |
| Dubois Richard | Stanford Linear Accelerator Center<br>richard@slac.stanford.edu |
| Enrico Costa | IASF-Roma, INAF<br>enrico.costa@iasf-roma.inaf.it |
| Fargion Daniele | University and INFN Roma La Sapienza<br>daniele.fargion@roma1.infn.it |
| Feroci Marco | IASF-Roma, INAF<br>marco.feroci@iasf-roma.inaf.it |
| Fuschino Fabio | INAF/IASF-Bologna<br>fuschino@iasfbo.inaf.it |
| Fusco Piergiorgio | University and INFN Bari<br>piergiorgio.fusco@ba.infn.it |
| Fusco-Femiano Roberto | IASF-Roma/INAF<br>roberto.fuscofemiano@iasf-roma.inaf.it |
| Gaggero Daniele | University of Pisa<br>daniele.gaggero@pi.infn.it |

| | | |
|---|---|---|
| Galli Alessandra | INAF/IASF-Roma | alessandra.galli@iasf-roma.inaf.it |
| Gasparrini Dario | ASDC/INAF/University of Perugia | gasparrini@asdc.asi.it |
| Gaug Markus | INFN Padova | gaug@pd.infn.it |
| Germani Stefano | INFN Perugia | germani@pg.infn.it |
| Giebels Berrie | LLR Ecole Polytechnique | berrie@poly.in2p3.fr |
| Giglietto Nicola | University and INFN Bari | giglietto@ba.infn.it |
| Giommi Paolo | ASI | paolo.giommi@asi.it |
| Glicenstein Jean-Francois | DAPNIA CEA Saclay | glicens@cea.fr |
| Grasso Dario | INFN Pisa | dario.grasso@pi.infn.it |
| Lamanna Giovanni | LAPP - CNRS/IN2P3 | lamanna@lapp.in2p3.fr |
| Lanciano Orietta | IASFC | orietta.lanciano@gmail.com |
| Lavalle Julien | CPPM CNRS-Université de la Mediterranée | lavalle@in2p3.fr |
| Lionetto Andrea | INFN Roma Tor Vergata | lionetto@roma2.infn.it |
| Longo Francesco | University and INFN Trieste | francesco.longo@ts.infn.it |
| Lubrano Pasquale | INFN Perugia | pasquale.lubrano@pg.infn.it |
| Maccione Luca | SISSA | maccione@sissa.it |
| Malvezzi Valeria | University and INFN Roma Tor Vergata | valeria.malvezzi@roma2.infn.it |
| Marcelli Laura | University and INFN Roma Tor Vergata | laura.marcelli@roma2.infn.it |
| Marisaldi Martino | INAF-IASF | marisaldi@iasfbo.inaf.it |
| Massaro Francesco | University Roma Tor Vergata | massaro@roma2.infn.it |
| Minori Mauro | INFN Roma Tor Vergata | minori@roma2.infn.it |
| Moiseev Alexander | CRESST/NASA/GSFC | moiseev@milkyway.gsfc.nasa.gov |

| | |
|---|---|
| Monte Claudia | INFN Bari<br>claudia.monte@ba.infn.it |
| Morselli Aldo | INFN and University Roma Tor Vergata<br>aldo.morselli@roma2.infn.it |
| Moskalenko Igor | Stanford University<br>imos@stanford.edu |
| Muñoz Carlos | Universidad Autonoma de Madrid and IFT<br>carlos.munnoz@uam.es |
| Nichelli Elisa | Dublin Institute for Advanced Studies<br>enichelli@cp.dias.ie |
| Oliva Pietro | University of Roma La Sapienza<br>pietro.oliva@roma1.infn.it |
| Omodei Nicola | INFN Pisa<br>nicola.omodei@pi.infn.it |
| Orsi Silvio | INFN Roma Tor Vergata<br>silvio.orsi@roma2.infn.it |
| Picozza Piergiorgio | University and INFN Roma Tor Vergata<br>piergiorgio.picozza@roma2.infn.it |
| Pieri Lidia | INAF-OAPD and INFN<br>lidia.pieri@oapd.inaf.it |
| Pittori Carlotta | INAF-ASDC<br>carlotta.pittori@asdc.asi.it |
| Pizzella Alessandro | University of Padova<br>alessandro.pizzella@unipd.it |
| Pizzella Guido | University and INFN Roma Tor Vergata<br>guido.pizzella@roma2.infn.it |
| Protheroe Raymond | University of Adelaide<br>rprother@physics.adelaide.edu.au |
| Regis Marco | S.I.S.S.A.<br>regis@sissa.it |
| Ricci Marco | INFN - LNF<br>marco.ricci@lnf.infn.it |
| Roncadelli Marco | INFN Pavia<br>marco.roncadelli@pv.infn.it |
| Ronga Francesco | INFN - LNF<br>francesco.ronga@lnf.infn.it |
| Rowell Gavin | University of Adelaide<br>growell@physics.adelaide.edu.au |
| Schelke Mia | INFN Torino<br>schelke@to.infn.it |
| Schlenstedt Stefan | DESY<br>stefan.schlenstedt@ifh.de |
| Sellerholm Alexander | Physics, HEAC<br>sellerholm@physto.se |
| Sparvoli Roberta | University and INFN Roma Tor Vergata<br>roberta.sparvoli@roma2.infn.it |
| Tavani Marco | INAF and University Roma Tor Vergata<br>tavani@iasf-roma.inaf.it |
| Tibolla Omar | University and INFN Padova<br>Omar.Tibolla@pd.infn.it |
| Tosti Gino | INFN Perugia<br>tosti@pg.infn.it |
| Ullio Piero | SISSA<br>ullio@sissa.it |
| Vernetto Silvia | IFSI /INAF Torino<br>vernetto@to.infn.it |
| Verrecchia Francesco | ASI Science Data Center<br>verrecchia@asdc.asi.it |
| Vitale Vincenzo | University of Udine<br>vitale@fisica.uniud.it |
| Wischnewski Ralf | DESY<br>wischnew@ifh.de |

# Index